\documentclass[11pt]{book}

\usepackage[utf8]{inputenc}		
\usepackage[T1]{fontenc}

\usepackage[frenchb]{babel}
\usepackage{lmodern}
\usepackage{ae,aecompl}										
\usepackage[upright]{fourier}

\usepackage{sectsty}

\usepackage[top=2.5cm, bottom=2cm, left=3cm, right=2.5cm,
			headheight=15pt]{geometry} 
\usepackage{fancyhdr}			
\usepackage{enumerate}
\usepackage{enumitem}
\usepackage[section]{placeins}	
\usepackage{epigraph}
\usepackage[font={small}]{caption}
\usepackage[francais]{minitoc}		
\usepackage{pdflscape}				

\usepackage{amsmath}			
\usepackage{amssymb}			
\usepackage{amsfonts}			
\usepackage{nicefrac}
\usepackage{upgreek}			
\usepackage[boxed,ruled,linesnumbered,french]{algorithm2e}

\usepackage{multirow}
\usepackage{booktabs}
\usepackage{colortbl}
\usepackage{tabularx}
\usepackage{multirow}
\usepackage{threeparttable}
\usepackage{etoolbox}
	\appto\TPTnoteSettings{\footnotesize}
\addto\captionsfrench{}	

\usepackage{graphicx}			
\usepackage{subcaption}
\usepackage{pdfpages}
\usepackage{rotating}
\usepackage{pgfplots}
	\usepgfplotslibrary{groupplots}
\usepackage{tikz}
	\usetikzlibrary{backgrounds,automata}
	\pgfplotsset{width=7cm,compat=1.3}
	\tikzset{every picture/.style={execute at begin picture={
   		\shorthandoff{:;!?};}
	}}
	\pgfplotsset{every linear axis/.append style={
		/pgf/number format/.cd,
		use comma,
		1000 sep={\,},
	}}
\usepackage{eso-pic}
\usepackage{import}
\usepackage{cclicenses}
\usepackage{pgfplotstable}
\usepackage{verbatim}

\makeatletter
\patchcmd{\BR@backref}{\newblock}{\newblock(page~}{}{}	
\patchcmd{\BR@backref}{\par}{)\par}{}{}
\makeatother
\usepackage{breakcites}	
\usepackage{bibentry}
\nobibliography*
\usepackage[pdftex,pdfborder={0 0 0},
			colorlinks=false,
			linkcolor=blue,
			citecolor=blue,
			pagebackref=true,
			]{hyperref} 

\usepackage{silence}
\chapterfont{\centering}
\chapternumberfont{\centering}

\newenvironment{figureth}{
		\begin{figure}[htbp]
			\centering
	}{
		\end{figure}
		}

\newenvironment{tableth}{
		\begin{table}[htbp]
			\centering
	}{
		\end{table}
		}

\newenvironment{subfigureth}[1]{
	\begin{subfigure}[t]{#1}
	\centering
}{
	\end{subfigure}
}

\DeclareMathOperator*{\argmin}{arg\,min}
\DeclareMathOperator*{\argmax}{arg\,max}
\newcommand\floor[1]{\lfloor#1\rfloor}
\newcommand\ceil[1]{\lceil#1\rceil}

\newtheorem{defi}{Définition}
\newtheorem{prop}{Proposition}
\newtheorem{theo}{Théorème}
\newtheorem{lem}{Lemme}
\newtheorem{preuve}{Preuve}

\newtheorem{note}{Remarque}

\title{}     
\date{} 
\author{}
\begin{document}

\thispagestyle{empty}

\begin{figure}[!tbp]
  \centering
  \begin{minipage}[b]{0.4\textwidth}
    \includegraphics[width=\textwidth]{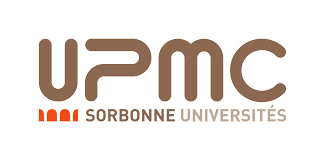}
  \end{minipage}
  \hfill
  \begin{minipage}[b]{0.4\textwidth}
    \includegraphics[width=\textwidth]{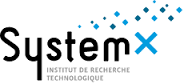}
  \end{minipage}
\end{figure}

{\large

\vspace*{1cm}

\begin{center}

{\bf TH\`ESE DE DOCTORAT DE \\ l'UNIVERSIT\'E PIERRE ET MARIE CURIE}

\vspace*{0.5cm}

Sp\'ecialit\'e \\ [2ex]
{\bf Informatique}\ \\ 

\vspace*{0.5cm}

Ecole doctorale Informatique, Télécommunications et Electronique (Paris)

\vspace*{1cm}

Pr\'esent\'ee par \ \\

\vspace*{0.5cm}

{\Large {\bf Thibault GISSELBRECHT}}

\vspace*{1cm}
Pour obtenir le grade de \ \\[1ex]
{\bf DOCTEUR de l'UNIVERSIT\'E PIERRE ET MARIE CURIE} \ \\

\vspace*{0.5cm}

\end{center}

\flushleft{Sujet de la th\`ese :}\ \\
\ \\
{\Large {\bf Algorithmes de bandits pour la collecte d’informations en temps réel dans les réseaux sociaux \\ }}

\vspace*{0.5cm} 
\flushleft{Soutenue le vendredi 24 mars 2017}\\[2ex]
\flushleft{Devant le jury composé de :  }\\[1ex]
\flushleft{\begin{tabular}{r@{\ }lll}
  & M. Philippe  {\sc Preux} & Professeur & Rapporteur  \\
  & M. Liva {\sc Ralaivola} & Professeur & Rapporteur  \\
  & Mme Michèle {\sc Sebag} & Directrice de recherche & Examinatrice  \\
  & M. Olivier  {\sc Sigaud} & Professeur & Examinateur  \\
  & M. Sylvain {\sc Lamprier} & Maître de conférence & Co-Encadrant \\
  & M. Patrick {\sc Gallinari} & Professeur & Directeur de thèse\\
\end{tabular}}

}


			\dominitoc
			\pagenumbering{roman}
			\dominitoc		
			\setcounter{tocdepth}{2}	
            \vspace{2cm}
\textbf{ \large{Résumé: }}
Dans cette thèse, nous nous intéressons au problème de la collecte de données en temps réel dans les médias sociaux. En raison des différentes limitations imposées par ces médias, mais aussi de la quantité très importante de données, il n'est pas envisageable de collecter la totalité des données produites par des sites tels que \textit{Twitter}. Par conséquent, pour être en mesure de récolter des informations pertinentes, relativement à un besoin prédéfini, il est nécessaire de se focaliser sur un sous-ensemble des données existantes. Dans ce travail, nous considérons chaque utilisateur d'un réseau social comme une source de données pouvant être écoutée à chaque itération d'un processus de collecte, en vue de capturer les données qu'elle produit. Ce processus, dont le but est de maximiser la \textit{qualité} des informations récoltées, est contraint à chaque pas de temps par le nombre d'utilisateurs pouvant être écoutés simultanément. Le problème de sélection du sous-ensemble de comptes à écouter au fil du temps constitue un problème de décision séquentielle sous contraintes, que nous formalisons comme un problème de bandit avec sélections multiples. Dans cette optique, nous proposons plusieurs modèles visant à identifier en temps réel les utilisateurs les plus pertinents. Dans un premier temps, le cas du bandit dit stochastique, dans lequel chaque utilisateur est associé à une distribution de probabilité stationnaire, est étudié. Par la suite, nous étudions deux modèles de bandit contextuel, l'un stationnaire et l'autre non stationnaire, dans lesquels l'utilité de chaque utilisateur peut être estimée de façon plus efficace en supposant une certaine structure, permettant ainsi de mutualiser l'apprentissage. En particulier, la première approche introduit la notion de profil, qui correspond au comportement moyen de chaque utilisateur. La seconde approche prend en compte l'activité d'un utilisateur à un instant donné pour prédire son comportement futur. Pour finir, nous nous intéressons à des modèle permettant de prendre en compte des dépendances temporelles complexes entre les utilisateurs, grâce à des transitions entre états cachés du système d'une itération à la suivante. Chacune des approches proposées est validée sur des données artificielles et réelles.

\vspace{0.5cm}
\textbf{ \large{Abstract: }}
In this thesis, we study the problem of real time data capture on social media. Due to the different limitations imposed by those media, but also to the very large amount of information, it is not possible to collect all the data produced by social networks such as \textit{Twitter}. Therefore, to be able to gather enough relevant information related to a predefined need, it is necessary to focus on a subset of the information sources. In this work, we focus on user-centered data capture and consider each account of a social network as a source that can be listened to at each iteration of a data capture process, in order to collect the corresponding produced contents. This process, whose aim is to maximize the quality of the information gathered, is constrained at each time step by the number of users that can be monitored simultaneously. The problem of selecting a subset of accounts to listen to over time is a sequential decision problem under constraints, which we formalize as a bandit problem with multiple selections. Therefore, we propose several bandit models to identify the most relevant users in real time. First, we study of the case of the so-called stochastic bandit, in which each user corresponds to a stationary distribution. Then, we introduce two contextual bandit models, one stationary and the other non stationary, in which the utility of each user can be estimated more efficiently by assuming some underlying structure in the reward space. In particular, the first approach introduces the notion of profile, which corresponds to the average behavior of each user. On the other hand, the second approach takes into account the activity of a user at a given instant in order to predict his future behavior. Finally, we are interested in models that are able to take into account complex temporal dependencies between users, with the use of a latent space within which the information transits from one iteration to the other. Moreover, each of the proposed approaches is validated on both artificial and real datasets.

            \chapter*{Remerciements}

Je remercie chaleureusement toutes les personnes du LIP6 qui m’ont aidé pendant l’élaboration de ma thèse et notamment mon directeur Monsieur Patrick Gallinari, pour son intérêt et ses conseils tout au long de ses trois années. Je tiens en particulier à dire merci mon encadrant, Monsieur Sylvain Lamprier pour son soutien et sa grande disponibilité. Ce travail est le fruit d'une collaboration étroite entre nous. Il a toujours été à l’écoute, et s’est toujours beaucoup impliqué dans mes travaux. Les nombreuses discussions que nous avons eues ainsi que ses conseils sont pour beaucoup dans le résultat final de ce travail. Enfin, ses relectures et ses corrections ont été très appréciables. Je souhaite également remercier l'equipe MLIA et en particulier Ludovic Denoyer pour son aide dans la rédaction de divers articles.

Ce travail n’aurait pas été possible sans le soutien de l'Institut de Recherche Technologique SystemX et du projet au projet IMM, qui m’ont permis, grâce aux aides financières, de me consacrer sereinement à l’élaboration de ma thèse. Je pense à tous les intervenants avec qui j'ai eu l'honneur de travailler et souhaite en particulier remercier Olivier Mesnard, chef du projet IMM, de m'avoir apporté son soutien. Je remercie également Nicolas, Jérémy, Jean-Alexis, Pierre et Romain pour leur aide scientifique et technique, mais aussi pour les nombreuses échanges que nous avons eus.

Je voudrais aussi remercier Philippe Preux et Liva Ralaivola d’avoir accepté de relire cette thèse et d’en être rapporteurs, ainsi que Michèle  Sebag et Olivier Sigaud d’avoir accepté d’être membres du jury.

Au terme de ce parcours, je remercie enfin celles et ceux qui me sont chers et que j’ai quelque peu délaissé ces derniers mois pour achever cette thèse. Leurs attentions et encouragements m’ont accompagné tout au long de ces années. Merci à mes parents, pour leur soutien moral et matériel et leur confiance, et à ma tante pour sa relecture du manuscrit.

			\tableofcontents
			\renewcommand*\listfigurename{Liste des figures}
			\listoffigures
			\listoftables
			\listofalgorithms
            
	\pagenumbering{arabic}

	\chapter{Introduction}

\section{Les réseaux sociaux aujourd'hui}
Depuis leur apparition il y a une quinzaine d'années, les médias sociaux en ligne sont rapidement devenus un vecteur d'information incontournable, mettant en jeu des dynamiques complexes de communication entre utilisateurs. Le premier réseau social, nommé Classmates, dont le but était de remettre en contact d’anciens camarades de classe, fut créé en 1995 par Randy Conrads. Il fut largement détrôné par l'arrivée des géants comme Facebook en 2004 ou Twitter en 2006 qui comptent aujourd'hui respectivement plus d'un milliard et 300 millions d'utilisateurs. Ils ont ensuite été suivis par une multitude d’autres réseaux plus spécifiques : Instagram, Vine, Periscope, Pinterest, Snapchat... A l'heure actuelle, pour beaucoup d’internautes, utiliser ces sites est considéré comme une activité sociale à part entière. Ces nouveaux moyens de communication ont entraîné la création d'une nouvelle génération de consommateurs de l'information, toujours plus désireux de savoir vite, et plus. 
Dans le monde ultra connecté de 2016, la communication via les réseaux sociaux est donc devenue un enjeu majeur, si bien que la compréhension des dynamiques sous-jacentes constitue une question clé pour de nombreux acteurs, industriels ou académiques. 
D'un point de vue industriel, les entreprises ont rapidement compris le potentiel des médias sociaux. Il s'agit en effet d'un outil permettant de toucher un public très large et de façon quasi instantanée, si bien que de nombreuses campagnes de publicité ont désormais lieu sur les réseaux sociaux. Les réseaux sociaux leur permettent de gérer leur image, de se développer financièrement et d’enrichir leur expérience sur un marché du web offrant de nouvelles opportunités, auparavant inexistantes. L'apparition de nouveaux métiers tels que \textit{community manager}, \textit{content manager} et autres \textit{social media planner} témoigne de l'intérêt porté à ces moyens de communication. D'autre part, dans le monde de la recherche, de nombreux scientifiques se sont intéressés à rendre compte des mécaniques à l'œuvre sur les réseaux sociaux. En effet, la mise à disposition d'une quantité d'informations considérable permet d'étudier des phénomènes auparavant impalpables si bien que le domaine de l'analyse des réseaux sociaux a fait des progrès considérables. Cette branche des sciences mêle sociologie, théorie des graphes et statistiques au sens large. A titre illustratif, de nombreux aspects tels que la modélisation du phénomène de bouche à oreille, la détection de communautés, la détection de source et beaucoup d'autres sont désormais des sujets à part entière. Toutes ces études sont rendues possibles par la présence de données sociales sur le web, dont la collecte représente une étape clé. Or, l'accès à la donnée des réseaux sociaux est contraint par différents facteurs qu'il convient de prendre en compte. D'une part, cette dernière soulève de nombreuses questions liées au respect de la vie privée et les débats à ce sujet ont été nombreux ces dernières années. La plupart des réseaux sociaux autorisent désormais chaque utilisateur à personnaliser les conditions d'accès à son profil par des tiers. D'autre part, même sur les réseaux sociaux les plus ouverts comme Twitter, l'accès à la donnée est restreint par les médias eux-mêmes. En effet, conscients de la valeur financière de celle-ci, il est souvent impossible d'avoir accès à la totalité des contenus. La collecte des données produites par les réseaux sociaux constitue donc un enjeu majeur en amont de toute étude. Nous proposons dans cette thèse de s'intéresser à cette problématique.

\section{La collecte d'information}

Afin de permettre le suivi de l'activité de leurs utilisateurs sur leur système, la plupart des médias sociaux actuels proposent un service de capture de données via des API (Application Programming Interface). D'une façon générale, il existe deux types d'API - sur lesquels nous reviendrons en détail par la suite - permettant d'accéder aux données: des API donnant accès à des données historiques stockées en base et des API fournissant les données en temps réel, au fur et à mesure qu'elles sont produites. Alors que la navigation dans les bases de données historiques peut s'avérer difficile et coûteuse, l'accès temps réel permet en outre de s'adapter aux dynamiques des réseaux étudiés. Dans ce manuscrit, nous étudierons ce second moyen d'accès aux données permettant l'acquisition en temps réel des flux produits sur le média social considéré. Néanmoins, l'utilisation d'un tel service peut se heurter à diverses contraintes, aussi bien techniques que politiques. Tout d'abord, les ressources de calcul disponibles pour le traitement de ces données sont souvent limitées. D'autre part, des restrictions sont imposées par les médias sociaux sur l'utilisation des API, dites de \textit{streaming}, qu'ils mettent à disposition pour permettre un traitement en temps réel de leur contenu. 
Bien souvent, comme c'est le cas sur Twitter, seules les données relatives à un nombre limité d'indicateurs (auteurs ou mots-clés contenus par exemple) peuvent être considérées simultanément, restreignant alors considérablement la connaissance du réseau à un sous-ensemble limité de son activité globale. La collecte en temps réel de la totalité des données produites sur un média social est donc bien souvent impossible et il s'agit alors d'échantillonner les données collectées, en définissant des méthodes automatiques de collecte. Une stratégie consiste à définir des filtres permettant d'orienter la collecte vers des données correspondant à un besoin particulier. Il s'agit de sélectionner les sources de données à écouter les plus susceptibles de produire des données pertinentes pour le besoin défini.

 
Dans ce contexte, définir un besoin de données / informations peut s'avérer une tâche complexe : comment définir un ensemble d'indicateurs permettant une collecte efficace, alors que l'on ne connaît pas la distribution des données pertinentes sur le réseau ? D'autant plus dans un contexte dynamique ? Si une collecte concernant une thématique particulière peut se faire en définissant une liste de mots-clés spécifiques que doivent contenir les messages à récupérer, les données obtenues via cette méthode sont souvent très bruitées ou hors sujet du fait du trop grand nombre de réponses ou d'interférences entre divers événements. Les entreprises qui vendent des solutions d'accès aux données des réseaux sociaux connaissent bien cette problématique et beaucoup d'entre elles ont recours à l’intervention d'un opérateur humain pour définir et modifier les indicateurs permettant de filtrer les données à collecter, ce qui est onéreux et n'est pas envisageable à grande échelle. 

Ceci nous amène à considérer le problème général de la collecte de données dans un réseau social pour un besoin spécifique lorsque le nombre de sources simultanément observables est restreint. Plus formellement, considérons un système d'écoute qui, compte tenu d'un ensemble d'utilisateurs sources à écouter, fournit le contenu produit par ces derniers pendant une période de temps donnée. Etant donné une fonction de qualité spécifique à la tâche en question, permettant d'évaluer la pertinence du contenu délivré par une source pour un besoin particulier, nous proposons une solution à ce problème d'échantillonnage de sources, basée sur une méthode d'apprentissage automatique. Nous utilisons pour cela le formalisme du problème du bandit manchot qui, comme nous le verrons, s'adapte bien à cette tâche de collecte orientée. En effet, le problème du bandit traite du compromis entre exploration et exploitation dans un processus de décision séquentiel où, à chaque pas de temps, un agent doit choisir une action parmi un ensemble d'actions possibles puis reçoit une récompense traduisant la qualité de l'action choisie. 
Dans notre cas, en optimisant à chaque pas de temps la fonction de qualité, les algorithmes de bandits nous permettront d'explorer, d'évaluer et de redéfinir l'ensemble des sources à considérer à chaque pas de temps. Cela permettra en particulier d'apprendre progressivement à se concentrer sur les sources d'information les plus pertinentes du réseau, sous les contraintes spécifiées (relatives à la capacité d'écoutes simultanées). Cette méthode a l'avantage de fonctionner pour n'importe quel besoin, sous réserve que ce dernier puisse être exprimé sous la forme d'une fonction de qualité associant une note à un contenu. Cette dernière peut prendre diverses formes et peut être utilisée par exemple pour collecter des messages d'actualité, identifier des influenceurs thématiques ou capturer des données qui tendent à satisfaire un panel d'utilisateurs finaux donnés.

La partie \ref{partEtatArt} de ce manuscrit est dédiée à l'état de l'art. En particulier, dans le chapitre \ref{EtatArtRI}, nous proposons un état des lieux des principales tâches et méthodes liées à la collecte d'informations sur le web. Nous commencerons par nous intéresser aux modèles traditionnels de recherche d'information, fondés sur une approche statique de la donnée disponible. Nous verrons par la suite comment l'évolution du web, avec en particulier la présence de flux de données évoluant en continu dans les médias sociaux, a entraîné l'apparition de nouvelles tâches - et donc de nouvelles méthodes - devant prendre en compte les contraintes propres à ce nouveau cadre. En particulier, les enjeux liés à l'exploration de l'environnement nous permettront de faire un premier pas dans le monde des bandits, auxquels nous dédions tout le chapitre \ref{EtatArt}. Nous verrons que ce problème de décision séquentielle peut prendre un grand nombre de formes et détaillerons les plus célèbres. La partie \ref{partContributions} de ce manuscrit sera consacrée aux différents modèles que nous avons proposés, dont nous décrivons les grandes lignes ci-après.

\section{Contributions}

\subsection{\nameref{FormalisationTache} (chapitre \ref{FormalisationTache})}
Dans cette première contribution, nous nous intéressons à la tâche de collecte d'informations en temps réel dans les médias sociaux, que nous formalisons comme un problème de bandit. Supposons que l'on dispose d'un processus permettant de collecter l'information produite par un utilisateur pendant une fenêtre de temps, et d'une fonction de qualité donnant un score aux contenus récupérés (dans la pratique, une API de \textit{streaming} nous autorise à faire cela). Notre but est de maximiser la qualité de l'information récoltée au cours du temps. Etant donné que l'on est restreint sur le nombre d'utilisateurs que l'on peut écouter simultanément, il est nécessaire de définir une stratégie de positionnement des capteurs à chaque instant. Il s'agit donc d'un problème de bandit avec sélection multiple. Il s'agira en particulier de décrire un modèle mathématique de la tâche et de faire le lien avec les problèmes de bandits décrits auparavant. Nous présenterons au passage les différents jeux de données utilisés tout au long du manuscrit, ainsi que certaines fonctions de qualité utilisées lors des différentes expérimentations.

\subsection{\nameref{ModeleStationnaire} (chapitre \ref{ModeleStationnaire})}

La première approche proposée consiste en un modèle de bandit stationnaire pour la collecte d'informations. Dans ce modèle, chaque utilisateur est associé à une distribution de récompense stationnaire, dont l'espérance reflète la qualité des informations qu'il produit, relativement à une fonction de qualité mesurant la pertinence de ses contenus. Dans cette optique, nous proposons une extension d'un algorithme existant adapté à notre tâche ainsi qu'une borne de convergence théorique. Les résultats obtenus à partir de ce modèle montrent que cette approche est pertinente pour la tâche de collecte de données sur Twitter.

\textbf{Publications associées :}
\begin{itemize}
\item \bibentry{icwsm15}
\item \bibentry{dn15}
\end{itemize}

\subsection{\nameref{ModeleContextuelFixe} (chapitre \ref{ModeleContextuelFixe})}

Dans une seconde contribution, nous proposons un modèle de bandit contextuel basé sur des profils utilisateurs. L'hypothèse est que l'utilité espérée des différents utilisateurs dépend de profils constants définis sur chacun d'entre eux, et que leur utilisation peut permettre une meilleure exploration des utilisateurs. Nous considérons néanmoins le cas, correspondant aux contraintes de notre tâche, où ces profils ne sont pas directement visibles, seuls des échantillons bruités de ces profils (les messages postés par les utilisateurs) sont obtenus au cours de la collecte. Nous proposons un nouvel algorithme, avec garanties théoriques de convergence, permettant d'améliorer l'efficacité de la collecte dans diverses configurations expérimentales réalistes.

\textbf{Publication associée :}

\begin{itemize}
\item \bibentry{ecml16}
\end{itemize}

\subsection{\nameref{ModeleContextuelVariable} (chapitre \ref{ModeleContextuelVariable})}

Dans les contributions précédentes, des hypothèses de stationnarité nous ont permis de définir des algorithmes efficaces avec des garanties de convergences théoriques. Néanmoins, cette stationnarité n'est pas toujours vérifiée. Nous proposons alors un modèle contextuel avec contextes variables, permettant d'appréhender les variations d'utilité espérée des utilisateurs. En supposant que l'utilité espérée d'un utilisateur à un pas de temps dépend du contenu qu'il a diffusé au pas de temps précédent, l'idée est d'exploiter ces corrélations pour améliorer les choix d'utilisateurs successifs. Cependant, dans notre cas, du fait des restrictions imposées par les API des médias considérés, seule une partie des contextes est visible à chaque pas de temps. On propose un algorithme adapté à ce genre de contraintes, ainsi que des expérimentations associées.

\textbf{Publications associées :}
\begin{itemize}
\item \bibentry{icwsm16}
\item \bibentry{dn16}
\end{itemize}

\subsection{\nameref{ModeleRelationnel} (chapitre \ref{ModeleRelationnel})}

Enfin, ce dernier chapitre propose un nouveau type de bandit, dans lequel il existe des dépendances entre les distributions d'utilité des périodes de temps successives. Pour notre tâche de collecte, il s'agit alors d'émettre des hypothèses sur la manière dont transite l'information pertinente afin d'estimer les meilleurs utilisateurs à écouter à chaque période de temps. Deux modèles basés sur l'algorithme de Thompson sampling et des approximations variationnelles ont été proposées.

	\part{Etat de l'art}
\label{partEtatArt}

\chapter[De la RI traditionnelle à l'exploitation en ligne des réseaux sociaux]{De la recherche d'information traditionnelle à l'exploitation en ligne des réseaux sociaux}
  \minitoc
  \newpage

\label{EtatArtRI}

Depuis l'apparition des réseaux sociaux, le mode de consommation de l'information sur internet a beaucoup évolué. Traditionnellement, le web d'une façon générale est considéré comme un répertoire géant, plus ou moins statique, dans lequel un ensemble de pages ou documents sont référencés dans un index. Pour accéder à un contenu particulier, un utilisateur d'internet doit traduire son besoin sous la forme d'une requête, à l'image d'une recherche sur le célèbre moteur Google. Aujourd'hui, grâce aux réseaux sociaux l'information nous arrive en temps réel, de façon continue, et provient de multiples sources. Ces trois facteurs ont révolutionné notre façon d'appréhender le web, dans lequel une multitude de "mini-sources" peuvent maintenant créer des contenus et les partager avec les autres utilisateurs en une fraction de seconde. La vision du web comme un répertoire statique semble alors dépassée. Nous proposons dans la suite de ce chapitre quelques éléments associés à cette conception, ce qui permettra de souligner en quoi ceux-ci ne sont pas adaptés au mode de fonctionnement du web actuel et des réseaux sociaux. Nous verrons par la suite que ces nouveaux médias requièrent la définition de nouveaux enjeux et par conséquent de nouvelles tâches de recherche, dont nous présenterons certains aspects. Nous positionnerons le problème de la collecte en temps réel traité dans cette thèse au fur et à mesure du discours.


\section{Traitement de l'information hors-ligne}

\subsection{Modèle classique}
             
  Depuis sa création au début des années 1990, Internet a connu une croissance exponentielle. De quelques centaines de sites, sa taille estimée dépasse aujourd’hui plusieurs dizaines de milliards de pages. Progressivement, les données qu'il contient sont devenues extrêmement intéressantes pour les entreprises, mais aussi les particuliers. Naturellement, la question de la récupération et de l'exploitation de cette source quasi inépuisable de données est apparue. C’est dans le cadre de cette problématique qu'a émergé le concept de \textit{crawling}, outil majeur de la collecte de données sur le web. Originellement, le \textit{crawling} consiste à parcourir et indexer le web afin d’en établir la cartographie, le but final étant de permettre à un moteur de recherche de trouver les documents les plus pertinents pour une requête donnée. Les trois principales étapes d'un processus de recherche d'information sont les suivantes :
    
    \begin{enumerate}
    \item Le \textit{crawl} - ou \textbf{exploration} - est effectué par un robot d’indexation  qui parcourt récursivement tous les hyperliens qu’il trouve. Cette exploration est lancée depuis un nombre restreint de pages web (\textit{seeds});
  \item L’\textbf{indexation} des ressources récupérées consiste à extraire les termes considérés comme significatifs du corpus de documents exploré. Diverses structures - tel que le dictionnaire inverse - permettent alors de stocker les représentations des documents;
    \item Finalement, la fonction d'\textbf{appariement} permet d'identifier dans le corpus documentaire (en utilisant l’index) les documents qui correspondent le mieux au besoin exprimé dans la requête afin de retourner les résultats par ordre de pertinence.
    \end{enumerate}

En partant de ce principe générique, un grand nombre de méthodes existe pour effectuer chacune des sous-tâches en questions.

\subsubsection{Crawling} 

Un \textit{crawler} est un programme qui explore automatiquement le web afin  de collecter les ressources (pages web, images, vidéos, etc.) dans le but de les indexer. 
 L'hypothèse sous-jacente est que les contenus du web évoluent de façon relativement lente, autorisant les crawlers à se rafraîchir de façon plus ou moins régulière selon les sites pour maintenir l'index à jour. D'un côté, les sites de nouvelles, dont les contenus évoluent rapidement, sont visités très régulièrement, alors que d'un autre, une page personnelle par exemple peut avoir une inertie de quelques jours. Par ailleurs, étant donné la taille du réseau, il est impossible, même pour les plus grands moteurs de recherche de couvrir la totalité des pages publiques. Une étude datant de 2005 \cite{studyWebCrawl} a montré que ceux-ci ne sont en mesure d'indexer qu'entre 40\% et 70\% du web. Dans cette optique, il est souhaitable que la portion de pages visitées  contienne les pages les plus pertinentes et pas seulement un échantillon aléatoire. Ceci amène à considérer la nécessité d'une métrique permettant de hiérarchiser les pages web par ordre d'importance, celle-ci étant souvent définie comme une fonction du contenu et de la popularité en termes de liens ou de visites. Dans ce contexte, les algorithmes de \textit{crawling} sont très nombreux et se différencient entre autres par l'heuristique permettant de donner un score aux différentes adresses visitées, ou à visiter, dans le but de prioriser les visites les plus utiles (en fonction de leur importance et/ou de leur fréquence de modification). Parmi les nombreux algorithmes de \textit{crawling} nous citerons \texttt{Breadth first} \cite{breathfirst}, \texttt{Depth first} \cite{depthfirst}, \texttt{OPIC} \cite{OPIC}, \texttt{HITS} \cite{HITS} ou encore \texttt{Page Rank} \cite{PageRank}. A titre illustratif, l'algorithme \texttt{PageRank} de Google attribue à chaque page une valeur  proportionnelle au nombre de fois que passerait par cette page un utilisateur parcourant le web en cliquant aléatoirement sur un des liens apparaissant sur chaque page. Ainsi, une page a un score d'autant plus important qu'est grande la somme des scores des pages qui pointent vers elle. Nous orientons le lecteur vers le travail de \cite{surveyCrawl} pour une étude des différentes possibilités. 
 Dans un contexte où le nombre pages est de plus en plus grand, et pour répondre à un besoin de collecte d'information plus ciblée, le concept de  \textit{focused crawling} est apparu. Au lieu de réaliser une exploration du web uniquement basé sur les liens entre les différentes pages, ce dernier permet de cibler particulièrement des pages présentant certaines caractéristiques. On peut par exemple s'intéresser uniquement aux pages parlant de tel ou tel sujet, rendant ainsi la tâche d'exploration moins coûteuse, le nombre de pages à visiter étant restreint. Le \textit{crawling} ciblé fut introduit dans \cite{Chakrabarti1999}, où les auteurs définissent une méthode permettant de décider si oui ou non une page doit être visitée en fonction des informations dont on dispose - par exemple selon ses métadonnées - et de ses liens vers d'autres pages considérées pertinentes relativement à un sujet prédéfini. De nombreuses variantes ont été proposées par la suite, par exemple dans \cite{Micarelli2007} où une stratégie de \textit{crawling } adaptatif est élaborée.

\subsubsection{Indexation} 
D'autre part, le processus d'indexation consiste à établir une représentation des différents documents, sur laquelle la fonction d'appariement pourra se baser pour trouver des scores de pertinence de manière efficace. Le modèle le plus simple, nommé sac de mots, fut évoqué pour la première fois dans \cite{bagOfWord}. Il s'agit de représenter les corpus sous forme d'une matrice creuse dans laquelle chaque document est encodé selon les termes qui le composent (matrice indexée par les termes dans le cas du dictionnaire inversé). Des modèles plus élaborés permettent de prendre en compte l'importance relative des termes dans un corpus afin d'augmenter ou de diminuer leur pouvoir discriminant, comme c'est le cas du schéma de pondération TF-IDF. Ce type de modèle utilise en particulier les notions de fréquence de terme (TF) et de fréquence inverse de document (IDF). Pour un terme et un document donné, la première grandeur correspond simplement au nombre d'occurrences du terme dans le document considéré (on parle de "fréquence" par abus de langage) tandis que la seconde grandeur est une mesure d'importance du terme dans l'ensemble du corpus. Chaque document peut donc être représenté par un vecteur composé des scores TF-IDF de chacun de ses termes, où TF-IDF peut correspondre à différentes combinaisons des deux grandeurs TF et IDF. Notons qu'il est également possible d'ajouter un prétraitement des données visant en particulier à enlever les mots les plus courants (\textit{stop words}), procéder à une racinisation (\textit{stemming}) ou encore une lemmatisation. Des modèles plus complexes tels LSA (\textit{Latent Semantic Analysis}), PLSA (\textit{Probabilisitc Latent Semantic Analysis}) \cite{PLSA}  ou encore LDA  (\textit{Latent Dirichlet Allocation}) \cite{LDA} ont ensuite permis de réduire la dimension de l'espace des représentations en vue d'obtenir des formes plus concises des différents documents. Leur objectif est d'obtenir une meilleure généralisation en passant d'une représentation dans l'espace des mots à une représentation dans un espace de concepts latents. La méthode LSA consiste en la factorisation de la matrice termes-documents du modèle sac de mots via une décomposition en valeurs singulières. Les méthodes PLSA et LDA quant à elles ne sont pas purement algébriques contrairement à LSA. Elles contiennent une dimension probabiliste permettant de modéliser chaque document comme un tirage échantillonné selon un modèle plus ou moins complexe. Plus récemment dans \cite{word2vec}, le très populaire modèle \textit{word2vec} définit une représentation distribuée des termes dans un espace de dimension restreinte, de manière à encoder de façon compacte différents types de dépendances, grâce à des techniques basées sur  des architectures de réseaux de neurones.

\subsubsection{Recherche}  
Etant donné une requête - qui peut elle-même être considérée comme un document de petite taille - et un ensemble de documents indexés, le but est d'ordonnancer ces derniers en fonction de leur pertinence relativement à un besoin d'information exprimé dans la requête. D'une façon générale, il s'agit d'attribuer un score numérique à chaque document, dont le calcul peut être effectué de différentes manières. Les modèles les plus simples reposent directement sur le modèle sac de mots. Par exemple, il peut s'agir d'un score binaire traduisant la présence ou non des termes de la requête dans un document (modèle booléen). D'autre part, le populaire modèle vectoriel mesure la similarité entre requête et document en considérant l'angle séparant leurs vecteurs dans l'espace des représentations (TF-IDF, \textit{word2vec} ou autre). Des scores plus complexes peuvent aussi être pris en compte, par exemple  BM25 \cite{BM25} et ses extensions BM25F \cite{BM25F} et BM25$+$ \cite{BM25Plus} qui sont encore considérés comme des méthodes à l'état de l'art dans le domaine. Enfin, d'autres approches consistent en l'élaboration d'un modèle de lanque par document. La pertinence de chaque document est alors mesurée en fonction de la probabilité que la requête ait été générée selon le modèle de langue correspondant.  Les premiers modèles de langues furent introduits dans les années 80 et considèrent des distributions de probabilité sur des séquences de mots (modèle de type $n$-grammes par exemple) \cite{surveyLangModel}. En outre, si des méthodes d'indexation plus complexes sont utilisées en amont (voir plus haut), on peut effectuer des comparaisons de documents dans un espace de dimension réduite. D'une façon générale, afin d'améliorer la pertinence des résultats renvoyés, il est possible d'utiliser des informations additionnelles comme le fait notamment Google en utilisant \texttt{PageRank} pour donner un score d'importance \textit{a priori} à chaque page. Finalement, l'ordonnancement  des résultats d'une requête peut être considéré comme un problème d'apprentissage  (\textit{learning to rank}) \cite{Letor}. Les différentes approches sont divisées en trois catégories, selon la fonction de coût à optimiser: \textit{pointwise ranking}, \textit{pairwise ranking} et \textit{listwise ranking}. Les détails de chacune des approches sont décrits dans \cite{learningToRank}.

\subsection{Le cas particulier des réseaux sociaux}

De par le nombre d'utilisateurs y étant actifs et la rapidité avec laquelle les contenus y sont produits, il apparaît que les méthodes de \textit{crawling} et d'indexation, jusqu'alors considérées comme des processus hors-ligne effectués en arrière-plan, et donc avec une inertie plus ou moins importante, ne sont pas toujours adaptés au cas des réseaux sociaux. En effet, si l'on souhaitait utiliser ces dernières sans modification, cela nécessiterait de référencer en continu toutes les publications de tous les utilisateurs dans chaque réseau social, soit plusieurs dizaines de milliers par seconde. De plus, l'aspect éphémère de l'information, sur laquelle nous reviendrons, pose des difficultés supplémentaires.

Les modèles permettant l'accès à l'information sur les réseaux sociaux doivent donc être adaptés par rapport au cas traditionnel. Sur ces derniers, les utilisateurs sont acteurs et ont la possibilité d'interagir directement avec les autres sans aucun besoin de passer par un quelconque moteur de recherche intermédiaire. L'accès à l'information en devient beaucoup plus rapide et plus libre, mais aussi moins contrôlé si bien que l'étude des phénomènes sous-jacents est beaucoup plus complexe. Afin d'avoir une compréhension plus fine des mécanismes à l'œuvre, un certain nombre de tâches ont été étudiées dans la littérature ces dernières années. En voici quelques-unes à titre illustratif :
\begin{itemize}
\item Diffusion de l'information : ce champ d'étude cherche à modéliser la façon dont une information se propage d'utilisateur en utilisateur dans un réseau social. De nombreux modèles ont été proposés dans la littérature pouvant utiliser des modèles de graphe connu ou pas \textit{a priori}  \cite{ IC,NetRate}, mais aussi des modèles de représentation dans des espaces latents \cite{heatDiffSimon};
\item Détection de source : il s'agit de retrouver le ou les utilisateurs étant source d'une rumeur sur un réseau social \cite{sourceDetection,sourceDetectionSimon,sourceDetectionSIR};
\item Détection de communauté : il s'agit de partitionner les utilisateurs en sous-ensembles dans lesquels les différents membres ont des caractéristiques communes, par exemple un centre d'intérêt \cite{communityDetection};
\item Maximisation de l’influence : le but est ici de trouver un ensemble de comptes tels que la propagation d'une information provenant de ces derniers soit maximale sur le réseau \cite{Kempe2003}.
\end{itemize}

Chacune de ces tâches est propre aux réseaux sociaux et peut, bien que l'aspect dynamique de ces derniers soit inhérent, être traitée avec des modèles hors-ligne et statiques (en utilisant les hypothèses adéquates). Par exemple, on peut considérer que l'information se diffuse via un graphe figé dans le temps et que les liens entre utilisateurs sont fixes.

Ainsi, beaucoup de travaux ont été effectués sur l'idée qu'un réseau social pouvait être considéré comme un graphe du web, où les nœuds correspondraient aux utilisateurs, et les liens aux relations sociales existantes entre eux. Avec un tel modèle, il est possible d'appliquer les approches de \textit{crawling} présentées plus haut. Sur ce principe, des systèmes personnalisés et adaptés à chaque utilisateur ont vu le jour pour permettre d'orienter les choix des utilisateurs vers des sources pertinentes dans le réseau. En particulier, dans \cite{recoFollow} et \cite{WTF}, les auteurs construisent des systèmes de recommandation, appelées respectivement \textit{Twittomender} et \textit{Who to Follow}. Dans le premier, des méthodes de filtrage collaboratif sont utilisées pour trouver les comptes Twitter qui sont susceptibles d'intéresser un utilisateur cible. Dans le second, une approche appelée \textit{circle of trust} (correspondant au résultat d'une marche aléatoire égocentrique similaire au  \texttt{personalized PageRank} \cite{personalizedPageRank}) est adoptée. Cependant, ce genre d'approche nécessite d'avoir en entrée le graphe des relations entre entités considérées, ce qui n'est pas évident. En effet, d'une façon générale, les relations entres les utilisateurs ne sont pas connues. Or, si dans le cas classique, les liens entre pages sont découverts à mesure que le robot d'indexation parcourt les pages web, dans le cas des réseaux sociaux, seul un nombre limité d'arcs du graphe social peuvent être obtenus via les APIs mises à disposition par les réseaux sociaux. 
Dans cette optique, les auteurs de \cite{Boanjak2012} ont proposé un système distribué permettant de faire du \textit{crawling} sur Twitter en interrogeant l'API depuis de nombreux clients distincts pour éviter les restrictions. Dans cette veine, les travaux proposés dans \cite{crawlingFB} et \cite{walkingFB} s'intéressent à l'exploration du réseau social Facebook. Le point commun à tous ces systèmes est qu'ils doivent prendre en compte le fait que l'acquisition de données est à la fois limitée et coûteuse. 
 Cependant, les politiques des médias sociaux étant de plus en plus restrictives (en termes de requêtes) à l'heure actuelle, une telle approche ne semble pas viable. En effet si l'on prend l'exemple de Twitter, un maximum de 180 requêtes peut être effectué par tranche de 15 minutes pour récolter les informations dans leur base de données, ce qui peut rapidement s'avérer contraignant lorsque l'on souhaite récupérer un graphe d'utilisateurs de façon récursive par exemple. 

Toutes ces approches classiques ont donc des applications limitées dans un cadre réel pour deux raisons principales : premièrement, l'apprentissage d'un modèle hors-ligne n'est pas compatible avec la dynamique des réseaux sociaux dans lesquelles les conditions peuvent évoluer très rapidement. Deuxièmement, comme nous l'avons déjà évoqué, quand bien même ce ne serait pas le cas, des problématiques d'accès à l'information se posent. En effet, dans la plupart des modèles appris hors-ligne, le processus d'apprentissage nécessite l'accès à beaucoup d'information pour s'effectuer. Il peut s'agir par exemple d'un graphe d'utilisateurs, qui dans la pratique n'est pas accessible à cause des restrictions imposées par les médias sociaux. Dans tous les cas, ces modèles nécessitent en amont un ensemble de données pour être appris, dont la collecte est souvent considérée comme une étape préparatoire. 

La plupart des médias sociaux proposent à leurs utilisateurs deux moyens d'accéder aux informations qu'ils contiennent. La première consiste à interroger une base dans laquelle toutes les données passées sont enregistrées\footnote{Documentation Twitter disponible à l'URL: \url{https://dev.twitter.com/rest/public}}. Cette approche peut être pertinente dans certaines situations, par exemple pour étudier des phénomènes \textit{a posteriori}, ou valider certaines hypothèses, mais possède l'inconvénient d'être extrêmement contraint. En effet, comme il est précisé plus haut, le nombre de requêtes est très limité, les médias sociaux ayant rapidement compris la valeur des données qu'ils détiennent. La seconde approche consiste à récupérer en temps réel les données produites par les utilisateurs, il s'agit de la collecte en ligne, qui nous intéresse particulièrement puisqu'elle permet de s'adapter à la dynamique des réseaux sociaux. 
La partie suivante est dédiée aux modèles d'exploitation en temps réel de l'information, ce qui nous permettra au passage de positionner les travaux effectués dans cette thèse de façon plus précise.

\section{Exploitation en temps réel de l'information}

De nombreuses applications nécessitent aujourd’hui le traitement d'informations en continu. Or dans la majorité des cas les méthodes hors-ligne ne sont pas transposables directement pour une utilisation en ligne. Cela implique l'élaboration de nouvelles méthodes pour exploiter les flux de données en temps réel. Ces modèles, en incorporant de nouvelles connaissances en continu, ont l'avantage d'être en mesure de s'adapter à des environnements changeants plus ou moins rapidement. Nous proposons dans la suite de faire un état des lieux des principales problématiques liées à ce champ d'études spécifique. Nous parlerons ensuite des diverses applications dans les réseaux sociaux.


\subsection{La fouille de flux de données: un besoin de méthodes adaptées}

Face à l'augmentation constante des données produites sur le web, d'importants efforts ont été déployés pour proposer des méthodes efficaces permettant de les exploiter. En particulier, des progrès considérables ont été réalisés dans le domaine du stockage de la donnée, de la parallélisation et du calcul distribué. Cependant, le rythme de production de certaines sources de données telles que les réseaux sociaux ne cessant d'augmenter, de nouvelles méthodes permettant de ne pas avoir à stocker l'intégralité des données se sont développées. Ces dernières se consacrent à l'étude de flux de données en ligne, qui contrairement aux approches classiques ne nécessitent pas - ou peu - de stockage de l'information. En effet, on considère dans ces approches qu'une donnée n'est lisible qu'une fois - ou un nombre de fois très faible - dans un système limité en capacité de stockage. Les flux sont continus, potentiellement illimités et arrivent rapidement. De plus, leurs distributions ne sont pas stationnaires. En effet dans beaucoup de situations, la distribution qui sous-tend les données ou les règles sous-jacentes peuvent changer avec le temps. Ainsi, les algorithmes de fouille de données (classification, clustering...) doivent prendre en compte ce facteur pouvant potentiellement modifier leurs sorties. Ce phénomène est appelé dérive conceptuelle (\textit{concept drift}). Par exemple, le trafic réseau, les contenus postés sur Twitter, les conversations téléphoniques, etc. sont des flux de données où l'on peut observer des évolutions au cours du temps et dans lesquels il s'agit de réussir à capturer les dynamiques.


Différents types de solutions permettant de traiter les flux de données ont été développés. Nous présentons ci-dessous les méthodes classiques permettant de traiter les flux de données, soit en n'utilisant qu'un sous-ensemble des données, soit en cherchant à en extraire des descripteurs plus concis:




\begin{itemize}
\item \textbf{Echantillonnage} : Il s'agit de définir un processus probabiliste permettant de sélectionner les exemples à traiter ou non dans l'apprentissage, tout en conservant suffisamment d'information utile. Par exemple, les arbres de Hoeffding, qui constitue une extension des arbres de décision classiques, utilisent l'idée selon laquelle un petit échantillon de données est suffisant pour effectuer l'apprentissage d'un modèle de classification. Ceux-ci sont en mesure de déterminer, avec une forte probabilité, le nombre minimal d'exemples nécessaires pour effectuer un apprentissage de façon efficace. Les algorithmes \texttt{VFDT} \cite{hoeffdingTree} et \texttt{CVFDT} \cite{hoeffdingTree2} reposent sur ce principe. Cette technique a l'avantage d'être simple et rapide, mais comporte un certain nombre de désavantages. En effet, comme l'évoque \cite{streamReview}, elle ne prend pas en compte les fluctuations possibles dans la quantité d'information présente dans le flot de données.

\item\textbf{Load shedding} : Proche de l'échantillonnage, cette méthode consiste à directement écarter une partie des données à traiter, souvent selon un processus aléatoire \cite{dataBasedShedding}. Elle a été appliquée avec succès pour effectuer des requêtes dans un flux de données dans \cite{dataBasedSheddingApp}. En outre, cette approche possède les mêmes inconvénients que l'échantillonnage présenté ci-dessus.

\item \textbf{Sketching} : Cette méthode fut proposée à l'origine dans \cite{dataBasedSketching} et consiste à projeter les données dans un espace de dimension réduite pour permettre de résumer le flot de données à l'aide d'un petit nombre de variables aléatoires. On distingue deux types de projection: le premier type, dit \textit{data-oblivious}, consiste à utiliser des projections aléatoires, ou apprises sur en ensemble de données hors-ligne. Dans ce cas, les projections sont fixées à l'avance et ne peuvent évoluer en fonction des données du flux. Les méthodes LSH \cite{LSH}, MDSH \cite{MDSH} et KLSH \cite{KLSH} font partie de cette famille. 
Le second type de projection, nommé \textit{data-dependent}, utilise directement les données du flux et sont apprises en ligne. Ces méthodes, dont \textit{PCA} \cite{dataBasedSketchingPCA}, \textit{Laplacian Eigenmaps} \cite{LaplacianEigen} et \textit{Spherical Hashing} \cite{SphericalHashing} font partie, sont souvent plus performantes que les méthodes \textit{data-oblivious}, mais sont plus coûteuses à mettre en œuvre. 

\item \textbf{Agrégation} : Il s'agit de calculer un certain nombre de valeurs statistiques telles que la moyenne ou la variance, de les agréger et de les utiliser comme entrée de l'algorithme d'apprentissage. Ces méthodes ont l'avantage d'être simples à mettre en oeuvre, mais atteignent leurs limites lorsque les flux de données sont hautement non-stationnaires. Elles ont été appliquées avec succès dans 
\cite{dataBasedAggreg} et 
\cite{dataBasedAggreg2} respectivement pour des tâches de \textit{clustering} et de classification dans les flux de données. 

\item \textbf{Ondelettes}: Cette technique consiste à reconstruire un signal en le projetant sur une base orthogonale de fonctions particulières. Les vecteurs de base les plus utilisés sont les ondelettes de Haar, qui sont des fonctions constantes par morceaux, ce qui en fait les ondelettes les plus simples à comprendre et à implémenter. Grâce à ces dernières, la reconstruction du signal est la meilleure approximation selon la norme $l^{2}$. 
On citera le travail de \cite{classifApp1}, dans lequel les auteurs spécifient l'algorithme de classification  \texttt{AWESOME}, utilisant la technique des ondelettes pour rendre plus compacte la représentation des informations. 

\item \textbf{Histogrammes}:
 Il s'agit de structures qui agrègent les données en les découpant selon une règle prédéfinie, ayant pour but d'approximer les distributions des données dans le flux considéré. La technique de référence dans ce domaine consiste à utiliser les histogrammes dits \textit{V-Optimal} \cite{dataBasedHisto3}, permettant d'approximer la distribution d'un ensemble de valeurs par une fonction constante par morceaux, de façon à minimiser l'erreur quadratique. 





\item \textbf{Fenêtre glissante} : La méthode des fenêtres glissantes consiste à restreindre l'analyse aux éléments les plus récents du flux de données. Cette technique a notamment été étudiée théoriquement dans \cite{taskBasedSliding}.  Elle possède l'avantage d'être aisée à implémenter, relativement facile à analyser de par sa nature déterministe,  mais peut s'avérer trop simpliste dans certaines applications nécessitant de prendre en compte le passé de façon plus structurée.

\end{itemize}


Toutes les approches présentées ci-dessus constituent une étape préalable pour résoudre des tâches d'apprentissage classique (régression, classification, etc.) dans lesquelles les exemples arrivent au fil de l'eau, de façon potentiellement rapide, et sans pouvoir être réutilisés. Grâce aux avancées technologiques de ces dernières années, les capteurs qui nous entourent sont de plus en plus nombreux et variés. Le terme de \textit{sensor network}, qui désigne un ensemble de capteurs permettant de mesurer toutes sortes de conditions dans des environnements divers (conditions météo, trafic routier, flux de messages, etc.), prend de plus en plus d'importance aujourd'hui, en particulier avec l'avènement  des objets connectés. Dans cette optique, chaque capteur peut être considéré comme source d'un flux de données particulier. Si l'exploitation d'un seul flux de données représente déjà un défi, l'exploitation de plusieurs flux de données est une tâche encore plus difficile, nécessitant d'adapter les techniques présentées plus haut (voir par exemple les travaux de \cite{multipleStreamClust} qui étendent une  méthode pour du \textit{clustering} sur les données d'un flux unique, à une méthode pour du \textit{clustering} sur les données d'un grand nombre de flux).

Un autre champ d'étude s'intéresse directement à l'étude des flux de données multiples en tant qu'objets. Dans cette optique, chaque flux de données  est considéré comme une séquence de mesures évoluant au cours du temps. Ainsi, diverses sous-tâches de prédiction (modèles autorégressifs par exemple), d'analyse de tendance, de classification de séquences ou de recherche de similarité entre séries temporelles existent. Par exemple les travaux de \cite{timeSeriesApp3} étudient le \textit{clustering} de séries temporelles provenant de téléphones mobiles (accélération, niveau sonore, luminosité, etc.) afin d'inférer un contexte utilisateur pour proposer divers services personnalisés. Par ailleurs dans \cite{timeSeriesApp4}, les auteurs ont développé une approche générique de détection d'événements dans les séries temporelles, qui constitue aujourd'hui un domaine à part entière (voir plus bas dans la partie sur les réseaux sociaux). En outre, dans le domaine de la santé, le sujet de la classification de séries provenant d'électrocardiogrammes constitue un sujet très étudié à l'heure actuelle \cite{ECGClassif}.

Finalement, dans un contexte plus proche des travaux effectués dans ce manuscrit, l'étude de la sélection de capteurs constitue un domaine dans lequel on considère des contraintes sur le nombre de flux auxquels il est possible d'accéder. Il s'agit donc de sélectionner le meilleur sous-ensemble de $k$ capteurs, dans le but d'estimer au mieux un certain nombre de grandeurs \cite{sensorSelection}. Par exemple, le déploiement de caméras de vidéosurveillance dans une ville constitue un problème de sélection de capteurs, dans lequel on cherche à avoir une vision de la ville la plus complète, tout en minimisant le nombre de caméras utilisées.  
Ce problème est d'autant plus complexe dans un cadre dynamique, c'est-à-dire lorsque les capteurs sélectionnés sont modifiés au cours du temps. Par exemple dans \cite{sensorSelectionDyna}, les auteurs étudient la sélection d'un sous-ensemble de flux d'images en temps réel, la totalité de ceux-ci n'étant pas exploitable, en vue de maximiser une fonction objectif. Nous verrons par la suite que le problème de la sélection de capteurs a été appliqué à la collecte d'information dans les réseaux sociaux, qui nous intéresse particulièrement ici.

Dans la section suivante, on s'intéresse au cas particulier des flux de données produits par les médias sociaux.

\subsection{Applications en temps réel dans les médias sociaux}
\label{twitterAPI}

Aujourd'hui, les plus importants flux de données sur le web proviennent des médias sociaux. A titre illustratif, sur Twitter on peut atteindre un taux de 7000 messages par secondes. Pour permettre de traiter cette quantité de données arrivant en flux continus, les algorithmes en ligne ont rapidement suscité un intérêt très particulier. Afin d'offrir la possibilité à des applications tierces d'utiliser ces données, les réseaux sociaux mettent la plupart du temps des API dites de \textit{streaming} à disposition. Ces dernières permettent de récupérer les flux de données produits en temps réel sur un média social. Cependant, les conditions d'accès aux données produites sur les réseaux sociaux sont souvent très restrictives. Par exemple pour Twitter, seulement 1\% du trafic total est accessible en temps réel.

Dans la section précédente, nous nous sommes concentrés sur le fait que la totalité des données ne pouvait pas être stockée et/ou traitée. Ainsi, des méthodes intervenant directement au niveau des données ou au niveau des algorithmes ont été développées pour répondre à ce challenge. Entre autres, des stratégies de sous-échantillonnage ont été établies en supposant que toute la donnée était potentiellement accessible. Il apparaît désormais de façon claire qu'une seconde contrainte vient se poser dans le cadre particulier des médias sociaux. En effet, notre accès aux flux de données est fortement restreint par les propriétaires de ces médias. Au regard des différents éléments présentés jusqu'ici, les spécificités de la tâche de collecte d'information en temps réel dans les médias sociaux qui fait l'objet de cette thèse commencent à apparaître. Rappelons que le but est le suivant: étant donné un ensemble d'utilisateurs postant des contenus en temps réel et un système permettant l'écoute d'un nombre restreint d'entre eux, notre objectif est de définir des stratégies permettant de choisir lesquels écouter à chaque itération afin de maximiser la pertinence des contenus récoltés. A cause de la fonction que l'on souhaite maximiser, nous verrons que des stratégies d'échantillonnage aléatoires ne sont pas adaptées et nous devrons utiliser des approches sélectionnant les sources de façon active, en effectuant un apprentissage. Un certain nombre de recherches se sont déjà intéressées aux flux de données en temps réel dans les médias sociaux ainsi qu'à des cas d'applications. Avant d'étudier le cas particulier de la collecte, nous présentons un certain nombre de cas d'applications.

\subsubsection{Cas d'applications courants}

\begin{itemize}
\item \textbf{Topic Detection and Tracking}: Une très large littérature s'est intéressée à l'identification et le suivi d'événements dans les flux de données. Le domaine du Topic Detection and Tracking (TDT), introduit dans \cite{Allan98}, propose des modèles pour traiter cette problématique. Cela inclut en particulier le découpage du flux, la détection d’événements et l'assemblage d'éléments d'une même histoire provenant de sources différentes dans le flux de données. Bien que cela puisse paraître très proche de notre tâche, car traitant des flux de données textuelles, le TDT considère que tous les flux sont disponibles. Dans notre cas, non seulement les flux ne sont pas tous disponibles, mais le but est de chercher à sélectionner les meilleurs, relativement à une fonction de qualité définie \textit{a priori}. Les travaux sur la détection de \textit{trending topics} se situent dans cette lignée. Il s'agit de détecter en temps réel, via l'analyse des flux de données, des sujets tendance sur un réseau social à un moment précis. Par exemple dans \cite{Cataldi2010}, les auteurs proposent une approche traitant en continu des messages provenant de Twitter afin de découvrir des mots-clés correspondant à des sujets en vogue sur le réseau. La question de la collecte des données n'est cependant pas traitée dans ce travail, qui considère uniquement le flux aléatoire de Twitter renvoyant en temps réel 1\% des contenus publics postés sur le réseau. Plus récemment dans \cite{Colbaugh2011}, la blogosphère est modélisée comme un réseau où chaque nœud correspond à un blog. L'objectif est d'identifier les blogs qui font suivre les contenus émergents du moment. En ce sens, l'approche proposée oriente la collecte vers des sources de données utiles, mais cela se fait de manière statique, sur des données d'apprentissage collectées sur tous les nœuds du réseau, ce qui n'est possible que lorsque le réseau est suffisamment petit. Finalement, le travail présenté dans \cite{Li2013} propose un processus de réorientation de la collecte en utilisant l'API \textit{streaming} de Twitter permettant de suivre l'emploi de certains mots en temps réel. Dans ce travail, les auteurs redéfinissent l'ensemble complet des mots à suivre à chaque étape, en sélectionnant les termes les plus employés du moment à partir d'observations faites depuis le flux aléatoire proposé par Twitter. Le fait de sonder le réseau à partir de mots-clés peut poser un certain nombre de problèmes dans le cadre de la collecte de données. En effet, si une collecte relative à un sujet peut bien être effectuée à partir d'une liste de mots-clés, les données récoltées avec cette approche peuvent s'avérer très bruitées ou bien hors sujet, en raison du faible pouvoir expressif d'une telle formulation du besoin d'une part et du grand nombre de messages retournés d'autre part.


\item \textbf{Analyse de sentiment}: l’objectif est d'analyser de grandes quantités de données afin d'en déduire les différents sentiments qui y sont exprimés. Les sentiments extraits peuvent ensuite faire l'objet d'études sur le ressenti général d'une communauté. Ce domaine a connu un succès grandissant dû à l'abondance de données provenant des réseaux sociaux, notamment celles fournies par Twitter. Ce type d'outils est aujourd'hui très utilisé dans de nombreux domaines. Ainsi, en politique il est possible d'extraire le sentiment des citoyens sur certaines thématiques/personnalités, en marketing de se faire une idée de l'opinion des clients sur un produit, etc. Le travail effectué dans \cite{opinionMiningSurvey} fait un état des lieux des nombreuses méthodes d'apprentissages permettant de traiter cette tâche. En particulier lorsque les données arrivent en continu sur un réseau social, il s'agit d'un problème de classification en ligne, pour lequel les auteurs de  \cite{realTimeOpinionMining} proposent une architecture. D'autre part dans  \cite{opinionMiningRealTimeUSElections}, un système d'analyse de sentiment en temps réel sur Twitter au sujet des candidats à l'élection présidentielle de 2012 est décrit. De nombreux autres travaux, comme ceux de  \cite{opinionMiningRealTimeTwitter} ou encore \cite{opinionMiningRealTimeTwitter2Hadoop}, ont été effectués à ce sujet.

\item \textbf{Analyse de la propagation d'informations en temps réel}: le phénomène de propagation de l'information dans les réseaux sociaux a beaucoup été étudié dans une configuration hors-ligne. Cependant, la croissance rapide des données sociales a fait émerger le besoin d'inventer des modèles plus réalistes, appris en ligne directement. Dans cette optique, les travaux décrits dans \cite{diffusionRealTime,diffusionRealTime2} proposent des approches en temps réel pour traiter cette tâche. Leur approche se base sur le graphe social (\textit{follower/followees}) pris à un instant donné, et ne  permet donc pas de prendre en compte la dynamique des relations entre les utilisateurs. 
\end{itemize}

          \subsubsection{La collecte orientée en ligne}
        
        Comme le font remarquer les travaux récents de \cite{dataCapture}, la collecte de donnée dans les réseaux sociaux est souvent vue comme une étape préliminaire, mais n'a pas beaucoup été étudiée. \textit{A fortiori}, le sujet plus complexe de la collecte en temps réel dans ces médias, qui comporte des contraintes spécifiques, représente un champ d'étude peu exploré. Il est important de noter que dans ce cas, si une information n'est pas récupérée au moment où elle est produite alors elle est perdue, ce qui n'est pas le cas dans les approches traditionnelles. Ainsi il est primordial pour le processus de collecte de réussir à bien s'orienter dans les différents flux de données afin de minimiser l'information perdue, ou de façon équivalente, de maximiser les contenus pertinents récoltés. Les travaux décrits dans \cite{deployMonitPoint} s'intéressent à cette problématique. Dans leur cas, il s'agit de déployer un ensemble de capteurs sur le réseau dans un contexte où il est impossible de collecter la donnée en positionnant un capteur sur chaque compte (restrictions des APIs). Dans ce travail, les auteurs considèrent que les relations sociales peuvent être exploitées de la façon suivante: si un utilisateur A suit (\textit{follow}) un utilisateur B alors, lorsque l'utilisateur B poste un contenu, cette information parvient à A. Il n'est ainsi pas nécessaire de placer un capteur sur B, puisque toutes les interactions de B avec le réseau social seront notifiées à A. En se basant sur ce principe, les auteurs utilisent une approche de couverture par ensembles (\textit{Set Cover problem}), dont le but est de trouver comment positionner les $k$ capteurs disponibles de manière à couvrir au mieux les informations produites sur un réseau social, sachant que selon leur positionnement, les capteurs permettent de collecter des informations sur des zones plus ou moins vastes du réseau. Cette approche possède deux inconvénients majeurs: premièrement, elle nécessite la connaissance du graphe social, dont l'acquisition constitue un problème à part entière. Par ailleurs, cette approche est stationnaire et considère le problème de la couverture du réseau à un instant donné, ne prenant ainsi pas en compte la dynamique des comportements des utilisateurs.

       Dans cette thèse, nous proposons une autre approche pour la collecte d'information en ligne, plus réaliste, dans laquelle le graphe social n'est pas disponible. De plus, la méthode que nous utilisons considère des sources potentiellement instationnaires, dont l'utilité peut évoluer au cours du temps. Ainsi, dans notre scénario, les positions de l'ensemble des capteurs peuvent être reconsidérées à chaque itération d'un processus de collecte dynamique. N'ayant pas d'information \textit{a priori} sur le comportement des utilisateurs, le processus de collecte doit être en mesure d'apprendre au fur et à mesure à s'orienter dans le flux de données. Le problème du choix du positionnement des capteurs se pose alors. En effet, imaginons que notre système, à un instant donné, repère une source de bonne qualité après y avoir positionné un capteur pendant un certain temps. Etant donné que cette source mobilise un capteur et que le nombre de capteurs est restreint, la question est de savoir si le système doit exploiter cette ressource ou alors tenter d'en explorer d'autres, potentiellement meilleures. Il s'agit donc d’un processus de décision séquentielle devant faire face au célèbre dilemme exploitation/exploration, auquel les problèmes de bandits sont spécifiquement dédiés. Dans le chapitre \ref{EtatArt} qui suit, nous proposons un état de l’art relatif aux problèmes de bandits. Nous formalisons ensuite au chapitre \ref{FormalisationTache} la tâche de collecte en ligne comme un problème de bandit, ce qui fournira les bases nécessaires aux différentes études menées dans cette thèse.
	
    \chapter{Problèmes de bandits et algorithmes}
	\minitoc
	\newpage

\label{EtatArt}


\label{EtatArtBandit}
Dans ce chapitre, nous introduisons le problème du bandit multibras (MAB en abrégé), que nous considérerons dans l'ensemble de cette thèse pour notre tâche de collecte de données en temps réel dans les réseaux sociaux. Nous commençons par une présentation générale du problème et des notations qui seront utilisées dans la suite. Nous nous intéressons ensuite à diverses instances particulières. 
Dans un premier temps, on présente le  bandit stochastique, qui constitue le cas le plus étudié dans la littérature. Le bandit contextuel, qui considère la présence d'un contexte décisionnel, est ensuite présenté. On considère également l'extension de ces deux modèles au cas des sélections multiples. Nous présenterons ensuite brièvement les problèmes de bandit dans les graphes et ceux de bandit non stationnaire. Pour chacune de ces configurations, nous définissons la notion centrale de regret et  présentons les principaux algorithmes de l'état de l'art. Notons que nous nous restreignons aux approches les plus populaires qui ne représentent pas l'intégralité de la multitude de variantes existantes.

	\section{Problème générique et notations}
    
		\subsection{Position du problème}
        
Le problème du bandit traite du compromis entre exploration et exploitation dans un processus de décision séquentielle dans lequel, à chaque pas de temps, un agent doit choisir une action parmi un ensemble de $K$ actions disponibles. A l'issue de ce choix, l'agent reçoit une récompense qui quantifie la qualité de l'action choisie. Le but de l'agent est de maximiser ses gains cumulés au cours du temps. Le terme "bandit" provient du monde des jeux de casino, dans lequel il désigne une machine à sous. Il modélise un joueur en face d'un certain nombre de machines, ayant la possibilité d'en jouer une chaque itération. Son but est évidemment de maximiser la somme de ses gains. Pour cela, le joueur doit trouver compromis entre l'\textbf{exploitation} des bonnes machines à sous et l'\textbf{exploration} de nouvelles, ou peu connues. Ce compromis représente le point central de tous les algorithmes de bandit. Plus précisément, il s'agit de trouver un équilibre entre:

\begin{itemize}
\item \textbf{Exploitation: } choisir les actions offrant des récompenses empiriques élevées;
\item \textbf{Exploration: } choisir les actions moins connues afin de ne pas passer à côté d'actions potentiellement de bonne qualité.
\end{itemize}

En partant de ce principe générique, un grand nombre d'instances du problème de bandit ont été proposées et se distinguent de par les hypothèses faites sur les distributions des différentes récompenses. Le premier cas, le plus étudié dans la littérature, est nommé bandit stochastique et considère que les récompenses observées pour une action donnée proviennent d'une loi stationnaire et sont tirées de façon indépendante. L'étude de ce problème consitue cœur de la section \ref{SectionBanditStochastique}. D'autre part, dans le cas du bandit contextuel présenté en section \ref{SectionBanditContextuel}, les distributions de récompenses des différentes actions sont liées aux valeurs d'un vecteur de contexte décisionnel observé. Il est également possible de structurer les différentes actions à l'aide d'un graphe (voir section \ref{graphBanditEtatArt}) ou encore de se placer dans le cadre très large des récompenses non stationnaires (voir section  \ref{nonstatBanditEtatArt}). Finalement, notons qu'un champ d'étude à part entière, nommé bandit adverse, concerne le cas où les récompenses sont choisies par un agent extérieur (adversaire). Plus de détails à ce sujet peuvent être trouvés dans \cite{nonStoBandit}, \cite{banditMOSS} et \cite{Bubeck2012}.


		\subsection{Notations}
Dans cette partie, nous introduisons un ensemble de notations que nous utiliserons dans toute la suite du document. 

			\begin{itemize}
				\item $\mathcal{K}$ désigne l'ensemble des $K$ actions disponibles à chaque instant;
				\item $i_{t}$ désigne l'action choisie à l'itération courante $t$; 
				\item $r_{i,t}$ désigne la récompense produite par le bras (ou action) $i$ au temps $t$;
				\item $r_{t}$ est une notation simplifiée pour la récompense obtenue à l'instant $t$, c.-à-d. nous avons: $r_{t}=r_{i_{t},t}$;
			\end{itemize}

De façon générique, un problème de bandit est un processus séquentiel dans lequel à chaque instant $t=1,2...,T$, un agent doit:

\begin{enumerate}
\item Choisir une action $i_{t} \in \mathcal{K}$ en fonction des choix passés;
\item Recevoir la récompense $r_{t}$;
\item Améliorer la stratégie de sélection grâce aux nouvelles observations.
\end{enumerate}

		\subsection{Applications courantes}
  
  Les algorithmes de bandits trouvent de nombreuses applications dans des problèmes pratiques. En voici une liste non exhaustive:
  
  \begin{itemize}
  \item Placement de publicité en ligne: dans ce cas, chaque publicité correspond à une action. Le but est de sélectionner la publicité à présenter à chaque utilisateur d'un site internet en vue de maximiser le nombre de clics. Les bandits ont été appliqués à cette tâche avec succès dans \cite{banditAd} et \cite{UnbiasedEvaluationBanditCtxt}.
  \item Routage: considérons un réseau représenté par un graphe dans lequel on doit envoyer une séquence de paquets d'un nœud à un autre. Pour chaque paquet, on choisit un chemin à travers le graphe et le paquet en question met un certain temps à arriver à destination. En fonction du trafic, les temps de parcours peuvent changer et la seule information disponible est le temps de parcours sur le chemin choisi à un instant donné. L'objectif est de minimiser ce délai total pour la séquence de tous les paquets de données. Ce problème peut être vu comme un bandit, l'ensemble des bras étant l'ensemble des chemins entre les deux sommets que chaque paquet peut emprunter (cf. \cite{banditRouting}).
  \item Monte Carlo tree search: une application importante des algorithmes de bandits est le programme MoGo de \cite{banditUCT2} permettant de jouer au Go contre des humains professionnels. Avec une méthode de Monte-Carlo pour évaluer la valeur d'une position, il est possible de voir le problème du jeu de Go comme des bandits dans un arbre. Les auteurs abordent ce problème avec la célèbre stratégie \texttt{UCT} décrite dans \cite{banditUCT}.
  \item Allocation de chaîne dans les réseaux de téléphonie: lors d'une communication entre deux téléphones mobiles, l'opérateur peut changer le canal de communication plusieurs fois. Ici, l'ensemble des bras correspond à l'ensemble des canaux possibles et un pas de temps représente un intervalle de temps où le canal est fixe. Ainsi le but est de choisir les meilleurs canaux tout au long de la communication (cf. \cite{banditAllocation}).
      \item Optimisation de fonction sous contraintes: les travaux de \cite{banditOptimPreux} proposent d'appliquer le formalisme du bandit à l'optimisation de fonctions mathématiques. Dans ce cas, l'ensemble des actions possibles correspond au domaine de définition de la fonction à optimiser (une action=un point de l'espace) et la récompense associée correspond à  la valeur de la fonction en ce point, le but étant de touver le point où la fonction à la valeur la plus élevée.
  \item Optimisation des paramètres d'un algorithme: considérons un algorithme avec un paramètre à régler, et dont la performance peut facilement être évaluée à un bruit aléatoire près. En considérant ce problème comme un problème de bandit stochastique, où l'ensemble des bras correspond à l'ensemble des valeurs de paramètres possibles, on peut construire des stratégies pour ajuster automatiquement le paramètre au cours du temps. 

  \end{itemize}

Dans le contexte particulier des médias sociaux, les algorithmes de bandit ont également des applications, dont on présente certaines ci-dessous:

\begin{itemize}
\item Le travail effectué dans \cite{Lage2013} s'intéresse à une tâche de maximisation de l'audience dans un réseau social. Dans ce contexte, un algorithme de bandit est utilisé pour sélectionner quels messages un utilisateur doit publier s'il souhaite maximiser le nombre de réactions (\textit{re-posts} par exemple).
\item Dans \cite{banditSNApp}, les auteurs proposent une approche générique permettant d'orienter un robot d'indexation dans réseau social, visant à assurer un bon équilibre entre exploration et exploitation grâce à un algorithme de bandit.
\item Dans \cite{gangOfBandit}, il est question d'utiliser la structure de graphe sous-jacente d'un réseau social pour améliorer les performances des algorithmes de recommandation. Cette approche, nommée \textit{Gang of Bandit}, fait l'hypothèse que des utilisateurs connectés entre eux auront des préférences similaires. Elle est évaluée empiriquement sur divers réseaux tels que \textit{Last.fm} ou  \textit{Delicious}.
\item Dans \cite{clusteringOfBandit} on s'intéresse également au problème de la recommandation dans un réseau social, mais contrairement à l'approche précédente, la structure de graphe n'est pas connue au préalable. Cette approche, plus réaliste, est également testée sur des données réelles.
\item Dans \cite{banditSNAppInfMax}, les auteurs abordent le problème de la maximisation de l'influence sous la forme d'un problème de bandit. Cette tâche consiste à sélectionner un ensemble d'utilisateurs source auxquels partager une information ou un produit dans le but de maximiser sa diffusion dans un réseau social. La validité de cette approche est évaluée empiriquement sur des données du  un site de partage de photographies et de vidéos  \textit{Flickr}.
\end{itemize}

Nous verrons par la suite et tout au long de ce manuscrit que les approches de bandits peuvent aussi être appliquées à la collecte d'information ciblée dans un réseau social.

	\section{Bandit stochastique}

     \label{SectionBanditStochastique}
    
    	\subsection{Problème et notations}
        
	Comme indiqué précédemment, le problème du bandit stochastique consiste à choisir une action parmi $K$, en considérant des distributions de récompenses  stationnaires et des tirages indépendants d'un pas de temps à l'autre. Nous utilisons les notations suivantes dans la suite du chapitre:
    
    \begin{itemize}
        	\item $\nu_{i}$ représente la distribution de la récompense du bras $i$, c.-à-d. $\forall t\geq 1, r_{i,t} \underset{iid}{\sim} \nu_{i}$;
			\item $\mu_{i}$ désigne l'espérance de la récompense de l'action $i$, c.-à-d. $\mu_{i}=\mathbb{E}\left[\nu_{i}\right]$;
            \item $\mu^{\star}=\max\limits_{i\in \left\{1,2,...,K\right\}}\quad \mu_{i}$ la moyenne la plus élevée;
            \item $i^{\star}$ l'action correspondant à la plus grande moyenne, c.-à-d. l'action $i$ telle que $\mu^{\star}=\mu_{i}$. Notons que cet indice n'est pas forcément unique dans le cas général;          
            \item $\Delta_{i}=\mu^{\star}-\mu_{i}$ la différence des moyennes entre l'action optimale et l'action $i$;
			\item $N_{i,t}$ représente le nombre de fois que l'action $i$ a été choisie jusqu'au temps $t$, c.-à-d. $N_{i,t}=\sum\limits_{s=1}^{t}\mathbb{1}_{\left\{i_{t}=i\right\}}$;
			\item $\hat{\mu}_{i,t}$ représente la moyenne empirique de la récompense du bras $i$ jusqu'au temps $t$, c.-à-d. $\hat{\mu}_{i,t}=\dfrac{1}{N_{i,t}}\sum\limits_{s=1}^{t}\mathbb{1}_{\left\{i_{t}=i\right\}}r_{i,s}$;
			\item $V_{i,t}$ représente la variance empirique de la récompense du bras $i$ jusqu'au temps $t$, c.-à-d. $V_{i,t}=\dfrac{1}{N_{i,t}}\sum\limits_{s=1}^{t}(\mathbb{1}_{\left\{i_{t}=i\right\}}r_{i,s}-\hat{\mu}_{i,t})^{2}$;
			\end{itemize}
 
Dans ce qui suit, sauf indication contraire explicite, on considère des récompenses positives et bornées, autrement dit: $\forall t \in \left\{1,2,...,T\right\}, \forall i \in \left\{1,2,...,K\right\}: r_{i,t}\in \left[0,b\right]$, avec $b$ une constante strictement positive.
  
		\subsection{Regret}

    Le but d'un algorithme de bandit est de maximiser la somme des récompenses collectées au cours du processus décisionnel d'horizon $T$. Afin d'analyser les performances des algorithmes, il est d'usage de se comparer à une politique optimale (inconnue) qui choisirait l'action possédant la plus haute somme de récompenses.
    

Plus précisément, on définit la notion de regret de la façon suivante:

\begin{defi}
    	\label{regretCumulto} 
        Le regret est défini par:
    	\begin{equation}
    		R_{T}=\max\limits_{i\in \left\{1,2,...,K\right\}}\sum\limits_{t=1}^{T}r_{i,t} - \sum\limits_{t=1}^{T}r_{i_{t},t}
    	\end{equation}
    	
\end{defi}

Cependant, les récompenses, mais aussi la politique permettant de choisir $i_{t}$, étant des variables aléatoires, on s'intéresse plus souvent à la moyenne du regret, dont on définit deux notions dans ce qui suit: le regret moyen et le pseudo-regret  (cette nomenclature est tirée de l'article de \cite{Bubeck2012}).

 \begin{defi}
    \label{regretCumultoEsp}   
     Le regret moyen est défini par:
     	 \begin{equation}	\mathbb{E}[R_{T}]=\mathbb{E}\left[\max\limits_{i\in \left\{1,2,...,K\right\}}\sum\limits_{t=1}^{T}r_{i,t} - \sum\limits_{t=1}^{T}r_{i_{t},t}\right]
    	\end{equation}
\end{defi}

\begin{defi}
    \label{regretCumultoPseudo}    
    Le pseudo-regret est défini par:
     \begin{equation}	
     \hat{R}_{T}=\max\limits_{i\in\left\{1,2,...,K\right\}}\mathbb{E}\left[\sum\limits_{t=1}^{T}r_{i,t} - \sum\limits_{t=1}^{T}r_{i_{t},t}\right]
    	\end{equation}

\end{defi}

Dans les deux cas, l'espérance est calculée selon la politique et les distributions de récompenses. On remarque que le pseudo-regret est une notion moins forte que le regret moyen. En effet, le pseudo-regret utilise la meilleure action seulement en moyenne tandis que le regret moyen prend en compte la séquence des récompenses. Formellement, on a $\hat{R}_{T}\leq \mathbb{E}[R_{T}]$. En pratique on cherche à minimiser le pseudo-regret, qui peut se réécrire de la façon suivante, en faisant intervenir le nombre de fois que les bras sous optimaux sont choisis par l'algorithme à étudier:

\begin{align*}
\hat{R}_{T} &=\max\limits_{i\in\left\{1,2,...,K\right\}}\mathbb{E}\left[\sum\limits_{t=1}^{T}r_{i,t} - \sum\limits_{t=1}^{T}r_{i_{t},t}\right]
\\
&= \max\limits_{i\in\left\{1,2,...,K\right\}}\mathbb{E}\left[\sum\limits_{t=1}^{T}r_{i,t}\right] - \mathbb{E}\left[\sum\limits_{t=1}^{T}r_{i_{t},t}\right] \\
&= \max\limits_{i\in\left\{1,2,...,K\right\}}\sum\limits_{t=1}^{T}\mathbb{E}\left[r_{i,t}\right] - \sum\limits_{t=1}^{T}\mathbb{E}\left[r_{i_{t},t}\right] \\
&=\sum\limits_{t=1}^{T} \mu^{\star}-\sum\limits_{t=1}^{T}\mathbb{E}\left[\mu_{i_{t}}\right] \\
&=T\mu^{\star}-\sum\limits_{i=1}^{K}\mu_{i}\mathbb{E}\left[N_{i,T}\right]\\
&=\sum\limits_{i=1}^{K}\Delta_{i}\mathbb{E}\left[N_{i,T}\right]\\
&=\sum\limits_{i: \Delta_{i}>0}\Delta_{i}\mathbb{E}\left[N_{i,T}\right]
\end{align*}

On peut montrer qu'une politique purement aléatoire possède un regret linéaire en $T$. En effet si chaque action est sélectionnée uniformément, alors $\mathbb{E}\left[N_{i,T}\right]$ est de l'ordre de $T/K$, d'où, d'après l'équation précédente, un regret linéaire en temps. On s'intéresse donc naturellement à des politiques a minima meilleures qu'une politique purement aléatoire. On appelle ainsi politique admissible, une politique dont le pseudo-regret est sous-linéaire, c'est-à-dire telle que $\hat{R}_{T}=o(T^{\beta})$ avec $0<\beta < 1$. En utilisant la formule ci-dessus, on peut montrer que la condition $\mathbb{E}[N_{i,t}]=o(T^{\beta})$ pour tout $i\neq i^{\star}$ est suffisante pour obtenir un regret sous-linéaire. En montrant qu'il existe des politiques dont le pseudo-regret est majoré par un terme de la forme $C\log (T)$, où $C$ est une constante, les auteurs de \cite{LaiRobbins} ont prouvé l'existence de politiques admissibles. On définit ci-dessous la divergence de Kullback-Leibler, qui permettra ensuite de définir une borne inférieure du pseudo-regret pour  n'importe quelle politique de bandit. 

 \begin{defi}[Kullback-Leibler]
 
    \label{KLDiv}   
	La divergence de Kullback-Leibler entre deux distributions discrètes $P$ et $Q$ avec pour support $\mathcal{S}$ est définie comme:
    	\begin{equation}
			KL(P,Q)=\sum\limits_{j\in \mathcal{S} }P_{j}\log\left(\dfrac{P_{j}}{Q_{j}}\right)
		\end{equation}
        
   La divergence de Kullback-Leibler entre deux distributions continues $P$ et $Q$ ayant pour densité respectivement $p$ et $q$ est définie comme:  
        \begin{equation}
			KL(P,Q)=\int_{-\infty}^{+\infty} p(x) \log\left(\dfrac{p(x)}{q(x)}\right) \, \mathrm{d}x
		\end{equation}
\end{defi}


\begin{theo}[Borne inférieure]
	\label{infRegSto}
	D'après \cite{LaiRobbins}, pour toute politique telle que pour tout $i\neq i^{\star}$, $\mathbb{E}[N_{i,t}]=o(T^{\beta})$ avec $0<\beta < 1$, on a:
		\begin{equation}
        \label{bonreInfnbPlayes}
			\liminf\limits_{T\rightarrow\infty} \dfrac{\mathbb{E}[N_{i,T}]}{\log(T)} \geq  \dfrac{1}{KL_{inf}(i,i^{\star})}
		\end{equation}
        Avec $KL_{inf}(i,i^{\star})=\inf\left\{KL(\nu_{i},\nu):~\mathbb{E}[\nu] > \mu_{i^{\star}}\right\}$.
        
         Par conséquent:
         \begin{equation}
         \label{bonreInfReg}
			\liminf\limits_{T\rightarrow\infty} \dfrac{\hat{R}_{T}}{\log(T)} \geq \sum\limits_{i: \Delta_{i}>0} \dfrac{\Delta_{i}}{KL_{inf}(i,i^{\star})}
		\end{equation}
\end{theo}

Le théorème précédent établit une limite inférieure du pseudo-regret de la meilleure politique atteignable. Une politique est dite optimale si la borne supérieure de son pseudo-regret tend vers à la borne inférieure lorsque $T$ augmente. Pour étudier les performances d'un algorithme de bandit, il est donc de coutume d'établir une borne supérieure de son pseudo-regret. Nous présentons dans ce qui suit différents algorithmes ainsi que la borne supérieure du regret associé.

		\subsection{Algorithmes}

    Nous présentons dans ce paragraphe les trois différents types de stratégies suivantes:
    
    \begin{itemize}
		\item \textbf{Algorithmes de type $\epsilon$-greedy :} il s'agit de la stratégie la plus simple dans laquelle on fait varier la probabilité d'exploration.
        \item \textbf{Algorithmes de type UCB :} cette méthode utilise la borne supérieure de l'intervalle de confiance sur les moyennes des récompenses. Cette borne est maintenue pour chaque bras à chaque itération. Ainsi, à chaque pas de temps, l'agent choisit l'action ayant la borne supérieure la plus élevée. Ce type d'algorithme est dit optimiste car il considère pour chaque bras la meilleure utilité possible étant donné son intervalle de confiance.
        \item \textbf{Algorithmes de Thompson sampling :} ce type d'algorithme choisit l'action à jouer en fonction de sa probabilité \textit{a posteriori} d'être la meilleure en mettant à jour un modèle bayésien à chaque itération et pour chaque bras.
	\end{itemize}
    
			\subsubsection{Algorithmes de type $\epsilon$-greedy}

                   Imaginons une politique qui, après avoir initialisé chaque action en la jouant une fois, sélectionne l'action ayant la plus forte moyenne empirique à chaque itération. 
 En raison de l'aspect aléatoire des récompenses, il est possible que ce genre de politique s'enferme dans un séquence de décisions sous-optimales, sans pouvoir reconsidérer les meilleures actions. A titre d'exemple, considérons deux actions $A$ et $B$ suivant des lois de Bernouilli de moyennes $\mu_{A}$ et $\mu_{B}$, avec $\mu_{A}>\mu_{B}$. Imaginons qu'à l'initialisation, la récompense associée à $A$ soit de 0 et celle associée à $B$ soit de 1. Une politique se basant uniquement sur la moyenne empirique sélectionnerait indéfiniment l'action $B$, malgré le fait que $\mu_{A}>\mu_{B}$. Le pseudo-regret  à l'instant $T$ vaudrait donc à $(\mu_{A}-\mu_{B})(T-1)$, qui est une fonction linéaire. Il apparaît donc que les politiques uniquement basées sur l'exploitation ne peuvent pas, d'une façon générale, conduire à une borne supérieure du regret sous-linéaire. 
Pour éviter ce genre de focalisation sur des actions sous-optimales, les politiques de type $\epsilon$-greedy consistent à exploiter (c-à-d sélectionner l'action ayant la meilleure moyenne empirique) avec une probabilité $\epsilon \in \left[ 0,1\right]$ et à explorer (c-à-d sélectionner une action au hasard) avec une probabilité $1-\epsilon$. Cette méthode, fut introduite par \cite{Sutton}. Cependant, comme le montre \cite{banditUCB}, l'utilisation d'un terme d'exploration $\epsilon$ constant conduit également à un regret linéaire. Pour remédier à cela, ces mêmes auteurs proposent de faire décroître le terme d'exploration, noté $\epsilon_{t}$, avec le temps. Cette politique, nommée $\epsilon_{t}$-\texttt{greedy} est décrite dans l'algorithme \ref{epsilonreedy}, pour une suite de valeurs du paramètre $\epsilon_{t}$ définies à l'avance. 
                   

					\begin{algorithm}[h!]
						\KwIn{$\epsilon_1$,$\epsilon_2$,...,$\epsilon_T$}
 						\For{$t=1..K$}{
							Sélectionner $i_{t} =t $ \\
							Recevoir récompense $r_{i_{t}}$ et mettre à jour $\hat{\mu}_{i_{t},t}$\\
 						}
 						\For{$t=K+1..T$}{

 						 	$
    							i_{t}=\begin{cases}
      									\argmax\limits_{i \in \left\{1,...,K \right\}} \hat{\mu}_{i,t-1} & \text{avec probabilité~} 1-\epsilon_t \\
     									\text{Uniform}(\left\{1,...,K \right\}) & \text{avec probabilité~} \epsilon_t
    								  \end{cases}
  							$\\
							Recevoir récompense $r_{i_{t}}$ et mettre à jour $\hat{\mu}_{i_{t},t}$\\
 						}
						\caption{$\epsilon_{t}$-\texttt{greedy} \cite{banditUCB}}
						\label{epsilonreedy}
					\end{algorithm}

Le théorème suivant établit une borne supérieure du regret de l'algorithme $\epsilon_{t}$-\texttt{greedy}, lorsque les valeurs des $\epsilon_{t}$ possèdent une forme particulière.

					\begin{theo}
					\label{RegEpsilongreedy}
					\cite{banditUCB} Soit $c>0$ et $0<d\leq\min\limits_{i\in \left\{1,...,K \right\}, i\neq i^{\star}}\Delta_{i}$. Le pseudo-regret  de l'algorithme $\epsilon_{t}$-\texttt{greedy}, avec $\epsilon_{t}=\min\left(1,\dfrac{cK}{d^{2}t}\right)$, est majoré pour tout $t>\dfrac{cK}{d}$ comme suit:
						\begin{equation}
							\hat{R}_{T} \leq  \sum\limits_{i: \Delta_{i}>0} \Delta_{i}  \left(    \dfrac{cd^{2}}{T} + 2\left(\dfrac{c}{d^{2}}\log \left(\dfrac{(T-1)d^{2}e^{1/2}}{cK} \right)\right)\left(\dfrac{cK}{(T-1)d^{2}e^{1/2}}\right)^{c/(5d^{2})} +\dfrac{4e}{d^{2}} \left(\dfrac{cK}{(T-1)d^{2}e^{1/2}}\right)^{c/2} \right)
						\end{equation}
					\end{theo}

     Une étude fine des différents termes de la formule ci-dessus montre que le regret est bien sous-linéaire. Cependant, la valeur du paramètre de l'algorithme $d$ dépend des paramètres inconnus du problème. En effet, $d$ doit être fixé de façon à avoir $0<d\leq\min\limits_{i\in \left\{1,...,K \right\}, i\neq i^{\star}}\Delta_{i}$, or, la valeur de $\min\limits_{i\in \left\{1,...,K \right\}, i\neq i^{\star}}\Delta_{i}$ n'est généralement pas connue à l'avance. Cette politique est donc en pratique impossible à mettre en œuvre sous cette forme exacte, et $d$ doit être fixé à la main, ce qui ne permet alors pas d'atteindre une borne du regret sous-linéaire. Nous proposons dans la section suivante d'étudier les algorithmes de type UCB qui, eux, ne nécessitent pas de connaissance \textit{a priori} sur les paramètres du problème, en dehors de la valeur de $b$, la borne supérieure des récompenses.
                    
			\subsubsection{Algorithmes optimistes}
            
            Ce type d'algorithme traduit le concept d'optimisme face à l'incertain. En dépit du manque de connaissance exacte de l'action optimale, une estimation optimiste de la qualité de chacune est établie, puis l'action ayant la plus élevée est sélectionnée. Si cette estimation est erronée alors la supposition optimiste  décroît et  permet de sélectionner une action différente aux itérations suivantes. En revanche s'il s'avère que le choix effectué est bon, l'algorithme est en mesure d'exploiter cette action et d'atteindre un regret faible. Le compromis exploitation / exploration est abordé de cette façon. Concrètement, cela se traduit par la construction d'un intervalle de confiance  pour chaque action, dans lequel la récompense moyenne associée est comprise avec une forte probabilité. Il est important de noter que chaque action possède son propre intervalle de confiance, mais que ces intervalles doivent tous correspondre à une même probabilité de confiance. Le principe optimiste revient alors à sélectionner à chaque itération  l'action ayant la borne supérieure de l'intervalle de confiance (UCB pour Upper Confidence Bound) la plus élevée. Ce principe est illustré dans la figure \ref{fig:UCBConfInt}, où l'on dispose de quatre actions, avec pour chacune une estimation de la moyenne (en trait plein) ainsi qu'une borne supérieure de l'intervalle de confiance à 95$\%$ (en pointillé). Il apparaît que, même si l'action située en bas possède la moyenne empirique la plus faible, son incertitude lui confère la meilleure borne supérieure. Un algorithme de type UCB, basé sur l'intervalle de confiance en question, sélectionnerait alors cette action. Dans cet exemple, de simples fonctions gaussiennes sont utilisées, cependant de façon générale diverses inégalités de concentration sont utilisées pour dériver les algorithmes de type optimiste. Nous proposons des exemples d'algorithmes dans la suite.

	\begin{figureth}		\includegraphics[width=0.9\linewidth]{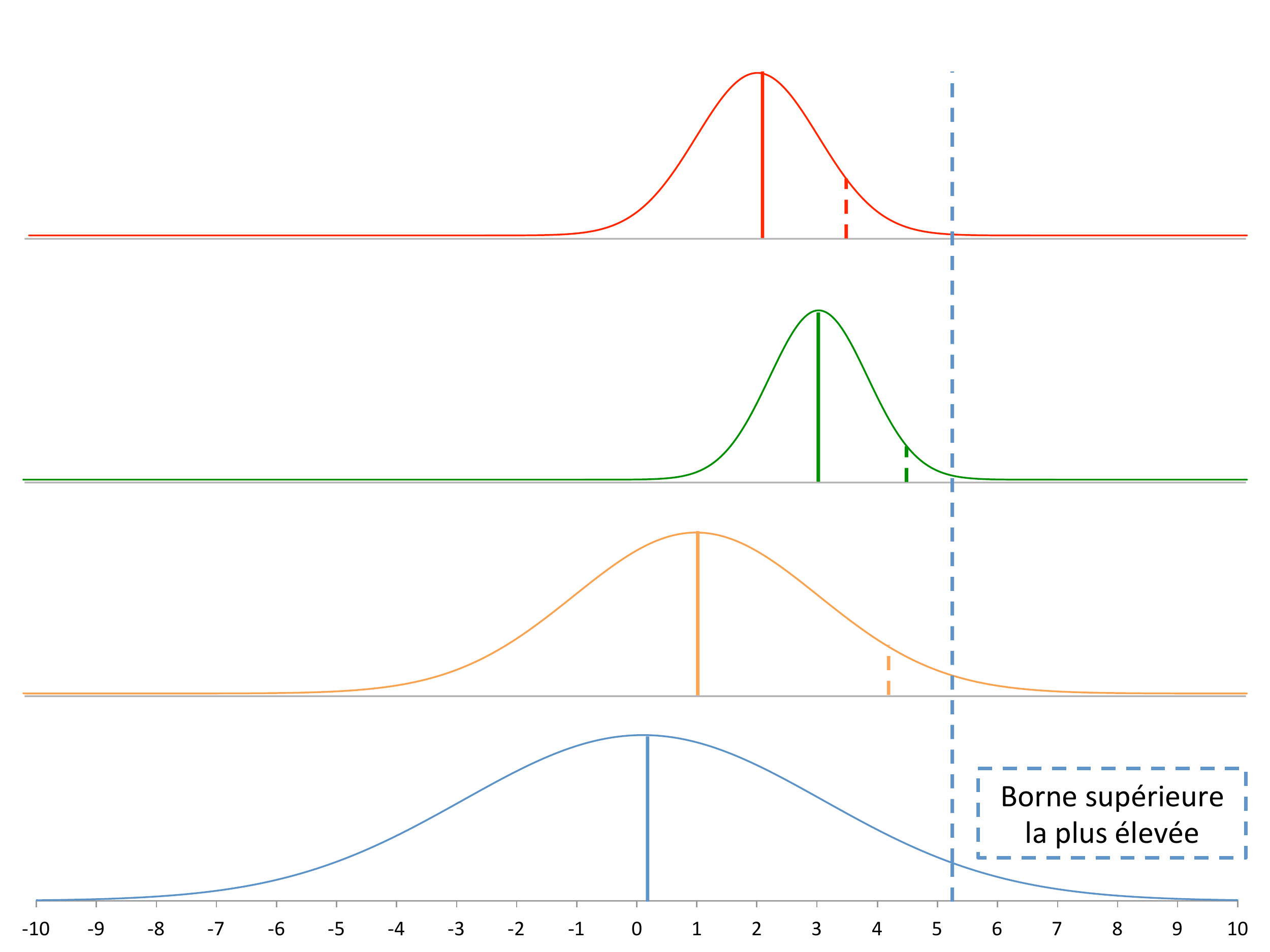}
			\caption[Principe des algorithmes optimistes.]{Principes des algorithmes optimistes.}		
			\label{fig:UCBConfInt}
	\end{figureth}

				\paragraph{Algorithme \texttt{UCB}}

La politique \texttt{UCB}, initialement proposée par \cite{banditUCB} et décrite dans l'algorithme \ref{UCB}, utilise l'inégalité de Hoeffding \cite{Hoeffding}, qui se traduit de la façon suivante dans notre cas.

			\begin{theo}
            \label{Hoeffding}
            \cite{Hoeffding}
				Soit $\epsilon>0$, on a, $\forall i \in \left\{1,...,K \right\}, \forall t>K$:
                \begin{align}
                	\mathbb{P}(\hat{u}_{i,t}-\mu_{i} \geq \epsilon) & \leq e^{-2\epsilon^{2}N_{i,t}/b^{2}} \\
                    \mathbb{P}(\hat{u}_{i,t}-\mu_{i} \leq -\epsilon) & \leq e^{-2\epsilon^{2}N_{i,t}/b^{2}} \\
                	\mathbb{P}(|\hat{u}_{i,t}-\mu_{i}| \geq \epsilon) & \leq 2e^{-2\epsilon^{2}N_{i,t}/b^{2}}
                \end{align}
			\end{theo}

Avec $\epsilon=\epsilon_{i,t-1}=\sqrt{\dfrac{2b^{2}\log(t)}{N_{i,t-1}}}$, on a $\mathbb{P}(\mu_{i} \leq \hat{u}_{i,t-1} + \epsilon_{i,t-1} ) \geq 1-\dfrac{1}{t^{4}}$. 
Ainsi, à l'instant $t$, pour chaque $i$, le terme  $\hat{u}_{i,t-1} + \epsilon_{i,t-1}$ est une borne supérieure de l'intervalle de confiance de la récompense associée, qui constitue le score associé à chaque action de l'algorithme \texttt{UCB}. Ce score est constitué d'un terme d'exploitation $\hat{u}_{i,t-1}$ favorisant les actions ayant de bonnes moyennes empiriques d'une part, et d'un terme d'exploration $\epsilon_{i,t-1}$ d'autre part. Le terme $\epsilon_{i,t-1}$ favorise bien l'exploration des actions les moins connues puisqu'il est d'autant plus élevé que le terme $N_{i,t-1}$ est faible. Le choix de la forme de ce terme d'exploration permet d'assurer un regret logarithmique, comme le montre le théorème suivant.

					\begin{algorithm}[h!]
 						\For{$t=1..K$}{
							Sélectionner $i_{t} =t $ \\
							Recevoir récompense $r_{i_{t}}$ et mettre à jour $\hat{\mu}_{i_{t},t}$\\
 						}
 						\For{$t=K+1..T$}{

 						 	$
    							i_{t} = \argmax\limits_{i \in \left\{1,...,K \right\}} \hat{\mu}_{i,t-1}  +\sqrt{2\dfrac{b^{2}\log(t)}{N_{i,t-1}}}
  							$\\
							Recevoir récompense $r_{i_{t}}$\\
							Mettre à jour $\hat{\mu}_{i_{t},t}$ \\
							$N_{i_{t},t}=N_{i_{t},t-1}+1$\\
   
 						}
						\caption{\texttt{UCB} \cite{banditUCB}}
						\label{UCB}
					\end{algorithm}

\begin{note}
Dans les algorithmes présentés, nous explicitons uniquement les calculs des paramètres $\hat{\mu}_{i_{t},t}$ et $N_{i_{t},t}$ pour l'action $i_{t}$ sélectionnée au temps $t$. Pour les bras non sélectionnés, la valeur de ces paramètres est la même qu'au temps précédent, autrement dit à un instant $t$,  $\forall i \in \left\{1,...,K\right\}$ tels que $i\neq i_{t}$, on a $\hat{\mu}_{i,t}=\hat{\mu}_{i,t-1}$ et $N_{i,t}=N_{i,t-1}$. Nous n'explicitons volontairement pas cette mise à jour triviale afin de ne pas alourdir les algorithmes. 
\end{note}

					\begin{theo}
					\label{RegUCB}
					\cite{banditUCB} Le pseudo-regret moyen de l'algorithme \texttt{UCB} est majoré par: 
						\begin{equation}
							\hat{R}_{T} \leq 8 \left(\sum\limits_{i: \Delta_{i}>0} \dfrac{b^{2}}{\Delta_{i}}\right) \log(T) +\left(1+\dfrac{\pi^{2}}{3}\right)\sum\limits_{i: \Delta_{i}>0} \Delta_{i}
						\end{equation}
					\end{theo}

\textbf{Schéma de preuve: }Nous proposons ici de prouver le résultat du théorème précédent. Non seulement ceci donnera une justification plus claire de la pertinence d'utiliser l'inégalité de Hoeffding mais aussi une idée de la méthode à utiliser. Comme nous l'avons vu précédemment, trouver une borne supérieure du regret dans le cadre du bandit stochastique revient à majorer l'espérance $\mathbb{E}[N_{i,T}]$ du nombre de fois où chaque bras sous optimal a été choisi jusqu'à l'instant $T$. Dans la suite, on prend, sans perte de généralité $b=1$.  D'après l'inégalité de Hoeffding, pour tout bras $i$ à l'instant $t$, on a:

$P(\hat{\mu}_{i,t-1} + \sqrt{\dfrac{2\log(t)}{N_{i,t-1}}}\leq \mu_i) \leq t^{-4} $ \text{ et } 
				$P(\widehat{\mu}_{i,t-1} -  \sqrt{\dfrac{2\log(t)}{N_{i,t-1}}} \geq \mu_i) \leq t^{-4} $
                
Ce qui définit l'intervalle de confiance de niveau $1-t^{-4}$ suivant:

$ \hat{\mu}_{i,t-1}- \sqrt{\dfrac{2\log(t)}{N_{i,t-1}}}  \leq^{(a)} \mu_{i}  \leq^{(b)} \hat{\mu}_{i,t-1} + \sqrt{\dfrac{2\log(t)}{N_{i,t-1}}} $. 

Lorsque l'algorithme \texttt{UCB} sélectionne une action sous-optimale $i$ à un instant $t$, on a forcément:

$\hat{\mu}_{i,t-1} +\sqrt{\dfrac{2\log(t)}{N_{i,t-1}}} \geq \hat{\mu}_{i^{\star},t-1} + \sqrt{\dfrac{2\log(t)}{N_{i^{\star},t-1}}} $

Ainsi, les deux possibilités suivantes existent:

		\begin{itemize}
			\item Si on est dans l'intervalle de confiance, on a alors:
		$\mu_{i} + 2\sqrt{\dfrac{2\log(t)}{N_{i,t-1}}} \geq \mu^{\star}$, ce qui implique : $N_{i,t-1} \leq \dfrac{8 \log{t}}{\Delta_i^2}$ ;
		
			\item Sinon, c'est que l'une des inégalités (a) ou (b) n'est pas vérifiée pour $i$ ou $i^{\star}$.
		\end{itemize}
     
     Par ailleurs, en notant $s=N_{i,t-1}$ et $s^{\star}=N_{i^{\star},t-1}$, pour tout $u \geq 0$ on a: 

	$N_{i,T}\leq u + \sum\limits_{t=u+1}^{T} \mathbb{1}_{\left\{i_{t}=i, N_{i,t}>u\right\}}$

     $N_{i,T}\leq u + \sum\limits_{t=u+1}^{T} \mathbb{1}_{\left\{\exists s: u < s \leq t, \exists s^{\star}: 1 \leq s^{\star} \leq t,  \hat{\mu}_{i,t-1} + \sqrt{2\log(t)/s}  \geq \hat{\mu}_{i^{\star},t-1} + \sqrt{2\log(t)/s^{\star} }\right\}}$

        En choisissant $u=\dfrac{8 \log{T}}{\Delta_i^2}$, on sait alors qu'un bras sous-optimal est choisi seulement si (a) ou (b) n'est pas vérifiée. Or: 
		\begin{itemize}
			\item (a) n'est pas vérifiée avec une probabilité de $t^{-4}$ 
			\item (b) n'est pas vérifiée avec une probabilité de $t^{-4}$
		\end{itemize}
Donc : 
		$\mathbb{E}[N_{i,T}] \leq \dfrac{8 \log{T}}{\Delta_i^2} + \sum\limits_{t=u+1}^{T} \left[\sum\limits_{s=u+1}^t t^{-4} + \sum\limits_{s=1}^t t^{-4}\right]  \leq  \dfrac{8 \log{T}}{\Delta_i^2} + 1 + \dfrac{\pi^2}{3} $, 
        ce qui conclut la preuve.
        
				\paragraph{Algorithme \texttt{UCBV}}

			La politique \texttt{UCBV} (que nous décrivons  dans l'algorithme  \ref{UCBV}) a été proposée par \cite{banditUCBV} et utilise la variance empirique des actions pour le calcul de l'intervalle de confiance (inégalité de Bernstein). Initialement, l'idée d'utiliser la variance fut introduite par \cite{banditUCB} avec l'algorithme \texttt{UCB-Tuned}. Bien qu'offrant des performances empiriques élevées, aucune garantie théorique de convergence n’a pu être démontrée pour \texttt{UCB-Tuned}, contrairement à \texttt{UCBV} (voir le théorème ci-après). L'idée de cet algorithme est de donner un bonus d'exploration aux actions ayant une variance empirique élevée. Cet algorithme fait intervenir deux paramètres $a$ et $c$ permettant de régler l'exploration.

					\begin{algorithm}[h!]
						\KwIn{$c>0, a>0$}
 						\For{$t=1..K$}{
							Sélectionner $i_{t} =t $ \\
							Recevoir récompense $r_{i_{t}}$ et mettre à jour $\hat{\mu}_{i_{t},t}$\\
 						}
 						\For{$t=K+1..T$}{

 						 	$
    							i_{t} = \argmax\limits_{i \in \left\{1,...,K \right\}} \hat{\mu}_{i,t-1}  +\sqrt{\dfrac{2aV_{i,t-1}\log(t)}{N_{i,t-1}}}+3cb\dfrac{\log(t)}{N_{i,t-1}}
  							$\\
							Recevoir récompense $r_{i_{t}}$\\
							Mettre à jour $\hat{\mu}_{i_{t},t}$ et $V_{i,t}$\\
							$N_{i_{t},t}=N_{i_{t},t-1}+1$ \\
 						}
						\caption{\texttt{UCBV} \cite{banditUCBV}}
						\label{UCBV}
					\end{algorithm}

					\begin{theo}
						\label{RegUCBV}
						\cite{banditUCBV}
							En prenant $c=1$ et $a=1.2$, le pseudo-regret de l'algorithme \texttt{UCBV} est majoré par: 
							\begin{equation}
								\hat{R}_{T}\leq 10 \left(\sum\limits_{i: \Delta_{i}>0} \dfrac{\sigma_{i}^{2}}{\Delta_{i}}+2b\right) \log(T)
							\end{equation}
                            Où $\sigma_{i}^{2}$ est la variance de l'action $i$.
					\end{theo}

			Comparée à la borne pour l'algorithme \texttt{UCB} proposé dans le théorème \ref{RegUCB}, cette dernière ne varie pas avec $b^{2}$ dans la somme, mais en $\sigma_{i}^{2}$. Nous observons par ailleurs qu'il existe toujours une dépendance linéaire en $b$, qui d'après l'étude de \cite{banditUCBV} est inévitable. Ainsi, selon les distributions de récompenses considérées, cette borne peut être plus large ou plus serrée que la borne de l'algorithme \texttt{UCB}. 
            

            \paragraph{Algorithme \texttt{UCB}-$\delta$}
	
			La stratégie \texttt{UCB}-$\delta$, décrite dans l'algorithme \ref{UCBDelta}, fut proposée par \cite{OFUL} et dérivée à partir d'une instance spécifique d'un problème de bandit contextuel (c.f. section \ref{SectionBanditContextuel}). Les auteurs utilisent une inégalité de concentration basée sur la théorie des processus auto normalisés \cite{PenaLaiShao2009}, permettant de déterminer une borne supérieure de l'intervalle de confiance de chaque action.

					\begin{algorithm}[h!]
						\KwIn{$0<\delta<1$}
 						\For{$t=1..K$}{
							Sélectionner $i_{t} =t $ \\
							Recevoir récompense $r_{i_{t}}$ et mettre à jour $\hat{\mu}_{i_{t},t}$\\
 						}
 						\For{$t=K+1..T$}{

 						 	$
    							i_{t} = \argmax\limits_{i \in \left\{1,...,K \right\}} \hat{\mu}_{i,t-1}  +\sqrt{\dfrac{1+N_{i,t-1}}{N_{i,t-1}^{2}}\left(1+2\log\left(\dfrac{K(1+N_{i,t-1})^{1/2}}{\delta}\right)\right)}
  							$\\
							Recevoir récompense $r_{i_{t}}$\\
							Mettre à jour $\hat{\mu}_{i_{t},t}$\\
							$N_{i_{t},t}=N_{i_{t},t-1}+1$ \\
 						}
						\caption{\texttt{UCB}-$\delta$ \cite{banditUCBV}}
						\label{UCBDelta}
					\end{algorithm}

					\begin{theo}
						\label{RegUCBdelta}
						\cite{OFUL}
							Soit $0<\delta<1$, alors avec une probabilité au moins égale à $1-\delta$, pour l'algorithme \texttt{UCB}-$\delta$, on a:
							\begin{equation}
								\sum\limits_{i: \Delta_{i}>0}\Delta_{i}N_{i,T} \leq \sum\limits_{i: \Delta_{i}>0} \left(3\Delta_{i}+  \dfrac{16}{\Delta_{i}}\log\left(\dfrac{2K}{\delta\Delta_{i}}\right)\right)
							\end{equation}
					\end{theo}

	On remarque qu'il ne s'agit pas exactement d'une borne du pseudo-regret, car c'est directement $N_{i,T}$ qui apparaît et non son espérance $\mathbb{E}[N_{i,T}]$ dans la grandeur qui est majorée. La borne supérieure proposée dépend de $\delta$ et est valide avec une certaine probabilité (dépendant aussi de $\delta$). Notons qu'en prenant $\delta=1/T$, on a bien une dépendance logarithmique en $T$, vraie avec une probabilité tendant vers 1 quand l'horizon est suffisamment grand. Le lecteur intéressé par une borne du pseudo-regret peut se référer à l'article de \cite{OFUL} qui précise qu'une borne du même type que  celle de \cite{banditUCB} peut être dérivée en utilisant une approche similaire. Cependant, contrairement à  l'algorithme \texttt{UCB} classique qui est conçu à l'origine pour des récompenses bornées, l'algorithme \texttt{UCB}-$\delta$ fait l'hypothèse de récompenses sous-gaussiennes. Nous reviendrons par la suite sur cette notion dans la section sur les problèmes de bandits contextuels. Nous testerons les performances de cet algorithme au cours de ce manuscrit.

				\paragraph{Algorithme \texttt{MOSS}}

					La politique \texttt{MOSS}, proposé par \cite{banditMOSS}, est décrite dans l'algorithme \ref{MOSS}. Dans ce travail, les auteurs étudient le problème du bandit adverse pour lequel ils proposent une nouvelle caractérisation du regret. Ils considèrent par la suite le problème du bandit stochastique (qui nous intéresse ici) et proposent un algorithme de type UCB. Dans cet algorithme, le score de chaque action ayant été sélectionnée plus de $T/K$ fois correspond exactement à sa moyenne empirique. Pour les autres, il s'agit d'une borne supérieure de la récompense moyenne associée, provenant de l'inégalité de Hoeffding. Notons par ailleurs que cet algorithme nécessite la connaissance \textit{a priori} de l'horizon $T$, ce qui selon les applications peut ne pas être le cas.

					\begin{algorithm}[h!]
 						\For{$t=1..K$}{
							Sélectionner $i_{t} =t $ \\
							Recevoir récompense $r_{i_{t}}$ et mettre à jour $\hat{\mu}_{i_{t},t}$\\
 						}
 						\For{$t=K+1..T$}{

 						 	$
    							i_{t} = \argmax\limits_{i \in \left\{1,...,K \right\}} \hat{\mu}_{i,t-1}  +\sqrt{\dfrac{\max\left(\log\left(\dfrac{T}{KN_{i,t-1}}\right),0\right)}{N_{i,t-1}}}
  							$\\
							Recevoir récompense $r_{i_{t}}$\\
							Mettre à jour $\hat{\mu}_{i_{t},t}$\\
							$N_{i_{t},t}=N_{i_{t},t-1}+1$ \\						}
						\caption{\texttt{MOSS} \cite{banditMOSS}}
						\label{MOSS}
					\end{algorithm}

						\begin{theo}
							\label{RegMOSS}
							\cite{banditMOSS}
                           Le pseudo-regret de l'algorithme \texttt{MOSS} est majoré par: 
							\begin{equation}
								\hat{R}_{T} \leq 23K \sum\limits_{i: \Delta_{i}>0} \dfrac{\max\left(\log\left(\dfrac{n\Delta_{i}^{2}}{K}\right),1\right)}{\Delta_{i}}
							\end{equation}
							\end{theo}

La preuve de cette borne fait appel à de nombreux outils mathématiques, que nous ne détaillerons pas ici. D'une façon générale, il est nécessaire d'introduire des variables intermédiaires ayant pour but de découpler les  dépendances entre bras, puis de majorer les différents termes en utilisant l'inégalité de Hoeffding. A nouveau, on ne peut pas comparer directement cette borne avec celles des autres algorithmes dans un cas général. Nous verrons lors de diverses expérimentations que selon les situations, on peut obtenir des performances plus ou moins élevées. 

				\paragraph{Algorithme \texttt{kl-UCB} et \texttt{Empirical KL-UCB}}
	Les algorithmes précédemment présentés ont une borne supérieure du pseudo-regret logarithmique en $T$. En effet, que ce soit pour \texttt{UCB} \cite{banditUCB},   \texttt{UCBV} \cite{banditUCBV},\texttt{UCB}-$\delta$ \cite{OFUL} ou \texttt{MOSS} \cite{banditMOSS}, le nombre $\mathbb{E}[N_{i,T}]$ de fois où une action sous-optimale $i\neq i^{\star}$ est sélectionnée peut être majoré par un terme de la forme:
    
    \begin{equation}
   	\mathbb{E}[N_{i,T}] \leq \dfrac{K_{1}}{\Delta_{i}^{2}}\log(T)+K_{2}
    \end{equation}
    
    Pour étudier l'optimalité d'un algorithme, au sens où la borne supérieure et la borne inférieure sont identiques, ce terme est à comparer avec la borne inférieure de l'équation \ref{bonreInfnbPlayes}, à savoir:
    
        \begin{equation*}
			\liminf\limits_{T\rightarrow\infty} \dfrac{\mathbb{E}[N_{i,T}]}{\log(T)} \geq  \dfrac{1}{KL_{inf}(i^{\star},i)}
		\end{equation*}
    
    Or, l'inégalité de Pinsker, utilisée en particulier par \cite{banditKLUCBBis}, spécifie que le terme $KL_{inf}(i^{\star},i)$ peut être bien plus grand $\Delta_{i}^{2}/2$. Par conséquent, les bornes précédemment exposées ne sont pas optimales car le terme multiplicatif devant $\log(T)$ ne correspond pas à celui de la borne inférieure. 
    
    Dans le but de s'approcher de la borne inférieure, une famille d'algorithmes utilisant directement la divergence de Kullback-Leibler  dans le calcul de l'intervalle de confiance des récompenses est présentée en détail dans \cite{banditKLUCBBounded}. L'algorithme générique proposé est nommé \texttt{KL-UCB}. La borne supérieure de l'intervalle de confiance utilisée dans ce dernier fait appel à un problème d'optimisation mathématique, qui n'est pas soluble dans un cadre général. De plus, la borne supérieure du regret associé n'est optimale que sous certaines conditions. Ainsi, les auteurs proposent l'étude de différents scénarios répondant à diverses hypothèses dans le but de concrétiser l'algorithme, mais aussi de prouver son optimalité:
    
    \begin{enumerate}
    \item \textbf{Récompenses dans la famille exponentielle}: dans ce cas, chaque distribution de récompense est supposée appartenir à la famille exponentielle canonique. L'algorithme associé (algorithme \ref{klUCB}), nommé \texttt{kl-UCB} est optimal au sens définit précédemment. La fonction non décroissante $f$ nécessaire à l'implémentation de l'algorithme est en pratique prise égale au logarithme. Le calcul du score de chaque action revient à trouver le zéro d'une fonction croissante et convexe, ce qui peut se faire par de nombreux processus itératifs (dichotomie, méthode de Newton, etc.). D'autre part, pour initialiser la recherche de façon pertinente, il est possible de partir d'un intervalle de confiance de type Hoeffding. Dans  \cite{banditKLUCBGeneral}, les auteurs effectuent les calculs de façon explicite pour les distributions binomiales, de Poisson, binomiales négatives, gaussiennes et Gamma. 
    \item \textbf{Récompenses bornées à support fini}: dans ce cas, les distributions de récompenses sont supposées à la fois bornées et à support fini. Les auteurs montrent alors que le score associé à chaque bras admet une forme pouvant être explicitement calculée. L'algorithme correspondant est nommé \texttt{Empirical KL-UCB} et est décrit dans l'algorithme \ref{EmpKLUCB}. 
    Son optimalité est également démontrée pour un choix judicieux de la fonction $f$\footnote{$f(t)=\log(t)+\log(\log (t))$ pour $t\geq 2$}. 
     \item \textbf{Récompenses bornées}: ici, les récompenses sont uniquement supposées bornées, comme dans le cadre étudié par \cite{banditUCB} (algorithme \texttt{UCB}). Bien qu'aucun algorithme ne soit spécifiquement créé pour ce cas, les auteurs montrent que l'application de l'algorithme \texttt{kl-UCB}, en supposant des distributions de Bernouilli offre de meilleures performances théoriques que l'algorithme \texttt{UCB}. Les auteurs évoquent par ailleurs l'utilisation directe de l'algorithme \texttt{Empirical KL-UCB} dans ce cas. Bien que l'optimalité ne puisse pas être démontrée, une analyse préliminaire permettant de justifier les bonnes performances expérimentales est conduite. 
    \end{enumerate}

    \begin{note}
    Nous ne détaillons volontairement pas les différentes notations ensemblistes présentes dans le calcul des scores de chaque action en raison de leur complexité, ces dernières sont présentées de façon détaillée dans \cite{banditKLUCBGeneral}.
    \end{note}

					\begin{algorithm}[h!]
                    \KwIn{Une fonction non décroissante $f:\mathbb{N}\rightarrow\mathbb{R}$}
 						\For{$t=1..K$}{
							Sélectionner $i_{t} =t $ \\
							Recevoir récompense $r_{i_{t}}$ \\
 						}
 						\For{$t=K+1..T$}{
 						 	$
    							i_{t} = \argmax\limits_{i \in \left\{1,...,K \right\}} \sup\left(\mu \in \bar{I}: d(\hat{\mu}_{i,t},\mu)\leq \dfrac{f(t)}{N_{i,t-1}}\right)
  							$\\
							Recevoir récompense $r_{i_{t}}$\\			}
						\caption{\texttt{kl-UCB} \cite{banditKLUCBBounded}}
						\label{klUCB}
					\end{algorithm}

                   \begin{algorithm}[h!]
                   \KwIn{Une fonction non décroissante $f:\mathbb{N}\rightarrow\mathbb{R}$}
 						\For{$t=1..K$}{
							Sélectionner $i_{t} =t $ \\
							Recevoir récompense $r_{i_{t}}$ et mettre à jour $\hat{\mu}_{i_{t},t}$\\
 						}
 						\For{$t=K+1..T$}{

 						 	$
    							i_{t} = \argmax\limits_{i \in \left\{1,...,K \right\}} \sup\left(\mathbb{E}[\nu]: \nu\in \mathcal{M}_{1}\left(Supp(\hat{\nu}_{i,t})\cup \left\{1\right\}\right) et \leq   KL(\hat{\nu}_{i,t},\nu) \leq \dfrac{f(t)}{N_{i,t}} \right)
  							$\\
							Recevoir récompense $r_{i_{t}}$\\
                            }
						\caption{\texttt{Empirical KL-UCB} \cite{banditKLUCBGeneral}}
						\label{EmpKLUCB}
					\end{algorithm}

			\subsubsection{Algorithmes de \texttt{Thompson sampling}}
            
            L'algorithme du  \texttt{Thompson sampling} fut introduit par \cite{Tho33}. Il s'agit d'une stratégie stochastique permettant d'aborder le compromis exploitation / exploration par tirages successifs à partir de distributions des moyennes des récompenses estimées pour chaque bras. D'une façon générale, il s'agit de maintenir une distribution \textit{a posteriori} sur les moyennes de chaque action, à mesure que les observations de récompenses arrivent. Formellement, les éléments suivants sont manipulés:
            
            \begin{enumerate}
            	\item Une distribution \textit{a priori} pour chaque bras $i$ sur le paramètre $\mu_{i}$ notée  $p(\mu_{i})$ ;
                \item L'ensemble des observations passées, noté $\mathcal{H}_{t-1}=\left\{(i_{s},r_{s})\right\}_{ s=1..t-1}$, avant le temps  $t$;
                \item Une vraisemblance $p(r_{i,t}|\mu_{i})$ pour la récompense de chaque bras $i$ à l'instant $t$;
                \item Grâce au théorème de Bayes, on peut calculer la distribution  \textit{a posteriori} $p(\mu_{i}|\mathcal{H}_{t-1})\propto p(\mu_{i})p(\mathcal{H}_{t-1}|\mu_{i})$ pour chaque $i$ à l'instant $t$;
			\end{enumerate}            

La stratégie du \texttt{Thompson sampling} consiste à sélectionner l'action selon sa probabilité \textit{a posteriori} d'être optimale. En pratique, on effectue un tirage $\tilde{\mu}_{i}$ suivant la loi  $p(\mu_{i}|\mathcal{H}_{t-1})$ et on choisit l'action $i_{t}=\argmax\limits_{i \in \left\{1,...,K \right\}} \tilde{\mu}_{i}$.  
Si la distribution \textit{a posteriori} est de la même forme que la distribution \textit{a priori}, la distribution \textit{a priori} est appelé \textit{prior conjugué} de la \textit{vraisemblance}. Cela permet en particulier de garantir que les distributions \textit{a posteriori} possèdent une forme analytique et sont échantillonnables facilement.  Ce type de \textit{prior conjugué} existe pour de nombreuses distributions, en particulier pour les distributions de la famille exponentielle.

Le Thompson sampling, bien que découvert en 1933 fut longtemps ignoré de par le manque de garanties théoriques. Cependant, à la suite de la publication de l'article de \cite{banditThompsonCtxtEmpirical} mettant en avant les performances empiriques ainsi que la simplicité d'implémentation de ce dernier, un regain d'intérêt pour cet algorithme fut observé. Des garanties théoriques sur le regret ont été démontrées dans de nombreux articles, on citera en particulier \cite{banditBernThompsonAnalysis} qui ont démontré l'optimalité dans le cas où les récompenses suivent des lois de Bernoulli et \cite{banditSimpleThompsonAnalysis} qui ont montré une borne supérieure du regret (non optimale) pour le cas plus général des récompenses bornées. Par la suite, les travaux de \cite{banditThompsonPriorFree} et \cite{banditSimpleThompsonAnalysis2} on démontré la convergence pour des cas plus généraux.

\begin{note}[Regret fréquentiste et regret bayésien] Etant donné que les paramètres des distributions sont ici considérés comme des variables aléatoires, il est possible de définir une nouvelle notion du regret, prenant l'espérance sur ces paramètres. Ce regret est nommé regret bayésien, en opposition au regret dit fréquentiste que nous avons étudié dans les sections précédentes, et qui considère les paramètres des distributions comme des constantes. Dans la suite, lorsque nous parlerons de regret, nous ferons référence au regret fréquentiste. L'étude du regret bayésien constitue un champ de recherche spécifique, hors du scope de ce manuscrit. Le lecteur intéressé pourra se référer aux travaux présentés dans  \cite{banditTSbayesianReg2} ou \cite{banditTSbayesianReg}.
\end{note}

Dans la suite, on présente des algorithmes considérant d'une part des récompenses binaires et d'autre part des récompenses gaussiennes.

                \paragraph{Récompenses binaires}
                
     L'algorithme proposé dans cette section est le plus étudié dans la littérature. Il traite le cas ou les récompenses des actions sont binaires (0 ou 1) et suivent des lois de Bernoulli, c.-à-d. $\forall i, \forall t: p(r_{i,t}=1|\mu_{i})=\mu_{i}$ avec $0<\mu_{i}<1$. Afin d'obtenir une distribution \textit{a posteriori} soluble analytiquement pour les moyennes des récompenses des différentes actions, on utilise le \textit{prior} conjugué de la distribution de Bernoulli, à savoir la loi Beta. Soient $\alpha$ et $\beta$ les paramètres de la loi \textit{a priori} de $\mu_{i}$: $\mu_{i} \sim$ Beta$(\alpha,\beta)$. Si l'on note $S_{i,t}$ le nombre de 1 et $F_{i,t}$ le nombre de 0 obtenu en tirant le bras $i$, alors la distribution \textit{a posteriori} de $\mu_{i}$ est  une loi Beta$(\alpha+S_{i,t},\beta+F_{i,t})$. L'algorithme  \ref{TSBetaBern} décrit la politique de Thompson sampling associée (notons qu'on prend $\alpha=\beta=1$, ce qui correspond à la loi uniforme sur $\left[0,1\right]$).
     
     				\begin{algorithm}[h!]
                    	\For{$i=1..K$}{
                        	$S_{i,0}=0$ \\
                            $F_{i,0}=0$ \\
                        }
 						\For{$t=1..T$}{
                        	        \For{$i=1..K$}{
                        				Échantillonner $\tilde{\mu}_{i} \sim$ Beta$(S_{i,t-1}+1,F_{i,t-1}+1)$ \\
                        				}
 						 	$
    							i_{t} = \argmax\limits_{i \in \left\{1,...,K \right\}} \tilde{\mu}_{i}
  							$\\
							Recevoir récompense $r_{i_{t}}$\\
                            \eIf{$r_{i_{t}}=1$}{
                            		 $S_{i_{t},t}=S_{i_{t},t-1}+1$ \\
                            	}
                            				{
                                     $F_{i_{t},t}=F_{i_{t},t-1}+1$ \\
                            	}

 						}
						\caption{\texttt{Thompson sampling} récompenses binaires \cite{banditBernThompsonAnalysis}}
							\label{TSBetaBern}
							\end{algorithm}

\textbf{Extension au cas des récompenses bornées:}  Plutôt que de considérer des récompenses binaires, l'algorithme \ref{TSBetaGen} est une extension de l'algorithme \ref{TSBetaBern} dans le cas plus général des distributions bornées dans $\left[0,1\right]$. Notons que ce cas entre bien dans notre cadre (notre hypothèse étant que les récompenses sont dans $\left[0,b\right]$) à une normalisation par $b$ près. Cet algorithme fait intervenir une variable aléatoire extérieure $X$ que l'on échantillonne de façon à ce que sa moyenne soit la même que la moyenne du bras choisi. En effet, soit $p(r_{i,t})$ la densité de probabilité de la variable $r_{i,t}$ et soit X une variable aléatoire telle que $X \sim \text{Ber}(r_{i,t})$, on a $p(X=1)=\int_{0}^{1} p(X=1|r_{i,t}) p(r_{i,t})  \, \mathrm{d}r_{i,t} =\int_{0}^{1} r_{i,t} p(r_{i,t})  \, \mathrm{d}r_{i,t}=\mathbb{E}[r_{i,t}]=\mu_{i}$.

                     	\begin{algorithm}[h!]
                    	\For{$i=1..K$}{
                        	$S_{i,0}=0$ \\
                            $F_{i,0}=0$ \\
                        }
 						\For{$t=1..T$}{
                        	        \For{$i=1..K$}{
                        				Échantillonner $\tilde{\mu}_{i} \sim$ Beta$(S_{i,t-1}+1,F_{i,t-1}+1)$ \\
                        				}
 						 	$
    							i_{t} = \argmax\limits_{i \in \left\{1,...,K \right\}} \tilde{\mu}_{i}
  							$\\
							Recevoir récompense $r_{i_{t}}$\\
                            Échantillonner $X \sim \text{Ber}(r_{i_{t}})$ \\
                            \eIf{$X=1$}{
                            		 $S_{i_{t},t}=S_{i_{t},t-1}+1$ \\
                            	}
                            				{
                                     $F_{i_{t},t}=F_{i_{t},t-1}+1$ \\
                            	}

 						}
						\caption{\texttt{Thompson sampling} récompenses bornées \cite{banditSimpleThompsonAnalysis}}
						\label{TSBetaGen}
					\end{algorithm}

                         \begin{theo}
							\label{RegTSBernBeta}
							\cite{banditBernThompsonAnalysis}
                           Le pseudo-regret de l'algorithme  \texttt{Thompson sampling} est optimal dans le cas ou les récompenses suivent des lois de Bernouilli.
						\end{theo}

				\paragraph{Récompenses gaussiennes}
                
                Nous présentons dans ce paragraphe un cas où les récompenses  sont gaussiennes \cite{banditSimpleThompsonAnalysis2}. On utilise pour cela des ditributions \textit{a priori} gaussiennes. Formellement, $\forall i, \forall t: p(r_{i,t}|\mu_{i})=\mathcal{N}\left(r_{i,t};\mu_{i},\sigma^{2}\right)$ et $p(\mu_{i})=\mathcal{N}\left(\mu_{i};a,b\right)$ où $\mathcal{N}\left(x;a,b\right)$ est la densité de probabilité d'une gaussienne de moyenne $a$ et de variance $b$. 
 Ainsi, après $t$ pas de temps, la distribution \textit{a posteriori} pour l'action $i$ s'écrit :
 
 $p(\mu_{i})=\mathcal{N}\left(\mu_{i};\left(\dfrac{a}{b}+\dfrac{S_{i,t}}{\sigma^{2}}\right)\left(\dfrac{N_{i,t}}{\sigma^{2}}+\dfrac{1}{b}\right)^{-1},\left(\dfrac{N_{i,t}}{\sigma^{2}}+\dfrac{1}{b}\right)^{-1}\right)$, où $S_{i,t}$ est la somme des récompenses du bras $i$ jusqu'au temps $t$. Il est ainsi possible pour chaque action $i$ d'effectuer un échantillonnage de sa moyenne \textit{a posteriori}. On présente cette politique dans l'algorithme \ref{TSGauss}. 
                
                     \begin{algorithm}[h!]
                     \KwIn{a,b}
                    	\For{$i=1..K$}{
                        	$S_{i,0}=0$ \\
                            $N_{i,0}=0$ \\
                        }
 						\For{$t=1..T$}{
                        	        \For{$i=1..K$}{
                        				Échantillonner $\tilde{\mu}_{i} \sim \mathcal{N}\left(\mu_{i};\left(\dfrac{a}{b}+\dfrac{S_{i,t-1}}{\sigma^{2}}\right)\left(\dfrac{N_{i,t-1}}{\sigma^{2}}+\dfrac{1}{b}\right)^{-1},\left(\dfrac{N_{i,t-1}}{\sigma^{2}}+\dfrac{1}{b}\right)^{-1}\right)$ \\
                        				}
 						 	$
    							i_{t} = \argmax\limits_{i \in \left\{1,...,K \right\}} \tilde{\mu}_{i}
  							$\\
							Recevoir récompense $r_{i_{t}}$\\
                            $S_{i_{t},t}=S_{i_{t},t-1}+r_{i_{t}}$ \\
                            $N_{i_{t},t}=N_{i_{t},t-1}+1$ \\

 						}
						\caption{\texttt{Thompson sampling} récompenses gaussiennes \cite{banditSimpleThompsonAnalysis2}}
					\label{TSGauss}
				\end{algorithm}

            \subsubsection{Autres algorithmes notables}
            
            Proposé par \cite{bayesUCBunifyingModel}, l'algorithme \texttt{Bayes-UCB} se situe à la croisée des chemins entre l'approche purement bayésienne du Thompson sampling et l'approche UCB. Il s'agit du premier algorithme bayésien pour lequel l'optimalité a été démontrée. Ce dernier utilise la notion de quantile pour calculer les scores de chaque action. D'autre part, l'algorithme \texttt{DMED} (Deterministic Minimum Empirical Divergence) fut proposé par \cite{banditDMED,banditDMED2}. Ce dernier utilise également la divergence de Kullbak-Leibler et il est prouvé que le regret associé est asymptotiquement optimal. Pour plus de détails sur ces deux dernières méthodes le lecteur intéressé pourra se référer respectivement à \cite{bayesUCBunifyingModel} et \cite{banditDMED,banditDMED2}.

	\section{Bandit contextuel}
	
    \label{SectionBanditContextuel}

   On peut montrer que le regret des algorithmes précédents augmente en $\sqrt{K}$, ce qui, lorsque le nombre d'actions devient grand, peut s'avérer problématique. Lorsque des informations auxiliaires associées aux actions sont disponibles, il est possible d'exploiter une notion de structure entre les actions, afin que l'apprentissage de la qualité des unes fournisse des informations sur les autres. Cela permet de réduire l'exploration des mauvaises actions et d'augmenter l'exploitation des bonnes. Le problème du bandit contextuel fait l'hypothèse qu'il existe un contexte décisionnel (nous préciserons sa forme par la suite) dont l'utilisation peut permettre à l'agent  d'améliorer sa stratégie de sélection. D'autre part, si ce contexte varie au cours du temps, alors sa prise en compte peut en particulier mener à une meilleure appréhension de la non-stationnarité des différentes récompenses. La suite de cette section est dédiée dans un premier temps à la formalisation du problème, puis à l'étude d'un certain nombre d'algorithmes de l'état l'art permettant d'exploiter la notion de contexte.


		\subsection{Problème et notations}

    D'une façon générale, dans le problème du bandit contextuel, on fait l'hypothèse qu'à chaque itération, un contexte est présenté à l'agent décisionnel avant qu'il prenne sa décision. L'agent dispose  d'informations supplémentaires lui permettant d'améliorer sa stratégie en apprenant une fonction - supposée linéaire dans le bandit contextuel classique étudié ici - liant contextes et récompenses. La structure induite dans l'espace des récompenses permet de mutualiser l'apprentissage et de s'orienter plus rapidement vers les actions les plus prometteuses à un instant donné. Dans un scénario de publicité en ligne, ce contexte peut par exemple représenter les caractéristiques  d'un utilisateur (son genre, son âge, son historique...) à qui on souhaite présenter la publicité la plus adaptée. Chaque publicité correspondant à une action, il est possible de modifier l'espérance de clic sur chacune selon les caractéristiques de l'utilisateur considéré, permettant ainsi d'améliorer la pertinence des publicités proposées.

    Le bandit contextuel peut se décliner sous différentes instances : 

\begin{itemize}

  \item Avec contexte global (état général du monde au moment de la décision) ou individuel (chaque action a des caractéristiques qui lui sont propres) 

  \item Avec contexte fixe (stationnarité des contextes)   ou variable (le contexte décisionnel ou les caractéristiques de chaque action ont changé).

  \item Avec prise en compte globale (paramètres de mise en relation entre contexte et utilité espérée partagés) ou individuelle (chaque bras - ou plutôt son utilité espérée - réagit différemment à un changement de contexte décisionnel et/ou propriétés individuelles)

\end{itemize}

La variabilité des contextes (qu'ils soient communs ou individuels) peut permettre d'appréhender des variations d'utilité espérée pour chaque action. En outre, l'observation de contextes individuels (même fixes) peut permettre de mieux explorer les actions en les considérant disposés dans un espace de caractéristiques (exploitation de la structure). A noter cependant que le cas du contexte global fixe n'a pas d'intérêt (même contexte pour tout le monde, donc ne permet pas de discrimination entre les actions, et stationnaire, donc ne permet pas d'appréhender des variations d'utilité espérée). Par ailleurs, le cas du contexte global variable avec prise en compte commune ne présente pas d'intérêt non plus, puisque les utilités espérées sont dans ce cas les mêmes pour tous les bras. Le tableau \ref{tab:configBanditCtxt} référence les différentes possibilités, en donnant pour chacune la manière dont est déterminée l'espérance de la récompense de chaque action dans le cas linéaire (bien que bien d'autres hypothèses pourraient être envisagées). 

Différentes hybridations de ces instances peuvent également être considérées. Par exemple, \cite{banditCtxtLinUCB} proposent d'utiliser un algorithme de bandit contextuel pour un problème de recommandation d'articles à des visiteurs d'un site de nouvelles, en considérant des vecteurs de contexte incluant à la fois des informations sur l'utilisateur considéré à l'instant $t$ (contexte commun variable) et des informations sur les articles à présenter (propriétés fixes propres à chaque article). Les auteurs proposent également une hybridation au niveau de la prise en compte des contextes, en considérant à la fois des paramètres individuels et des paramètres partagés par l'ensemble des actions.

    \begin{tableth}
        \setlength\extrarowheight{5pt}
				\begin{tabular}{|c|c||c|c|}
                	\hline
               \multicolumn{2}{|c||}{\cellcolor{black}} & \multicolumn{2}{|c|}{\textbf{\Large{Paramètres}}} \\
                	\hline 
                    \multicolumn{2}{|c||}{\Large{\textbf{Contextes}}} & \textbf{Globaux: $\beta$} & \textbf{Individuels: $(\theta_{1},...,\theta_{K})$}  \\ 
                	\hline\hline
               \textbf{Constants} &   \textbf{Individuels: $x_{i}$} &$x_{i}^{\top}\beta$ & $x_{i}^{\top}\theta_{i}$\\
					\hline 
                \multirow{2}{*}{\textbf{Variables}} & \textbf{Communs: $x_{t}$ } & \cellcolor{gray} & $x_{t}^{\top}\theta_{i}$  \\ \cline{2-4}
									      & \textbf{Globaux: $x_{i,t}$} & $x_{i,t}^{\top}\beta$ & $x_{i,t}^{\top}\theta_{i}$ \\ 
					\hline
					\end{tabular}
		\caption[Configurations bandit contextuel]{Différentes instances possibles du bandit contextuel linéaire à $K$ bras.}
		\label{tab:configBanditCtxt}
	\end{tableth}

Dans ce qui suit,  nous étudions le cas où les contextes de chaque action sont individuels et variables au cours du temps, et dont les paramètres de prise en compte sont partagés. Formellement, nous faisons l'hypothèse de linéarité suivante :
        	\begin{equation}
         		\exists \beta \in \mathbb{R}^{d} \text{ tel que } r_{i,t} = x_{i,t}^{\top}\beta + \eta_{i,t}
        	\end{equation}
            
Où $\eta_{i,t}$ est un bruit sous-gaussien de moyenne nulle et de constante caractéristique $R$, c'est à dire: $\forall \lambda \in \mathbb{R}: \mathbb{E}[e^{\lambda\eta_{i,t}}|\mathcal{H}_{t-1}] \leq e^{\lambda^{2}R^{2}/2}$ avec $\mathcal{H}_{t-1}=\left\{(i_{s},x_{i_{s},s},r_{i_{s},s})\right\}_{s=1..t-1}$. Soulignons qu'un bruit gaussien $\mathcal{N}(0,\sigma^{2})$ est sous-gaussien de constante $\sigma$ et qu'une variable uniforme dans l'intervalle $[-a,a]$ est aussi sous-gaussienne de constante $a$ (voir \cite{subGaussian} pour plus de résultats sur les variables aléatoires sous-gaussiennes).

            Ainsi, étant donné un ensemble $\mathcal{K}$ de $K$ actions, le problème du bandit contextuel procède de la façon suivante à chaque itération $t \in \left\{ 1, 2, 3,\ldots, T\right\}$:

\begin{enumerate}
	\item Pour tout $i \in \left\{1,...,K\right\}$, observer $x_{i,t}\in \mathbb{R}^{d}$, le vecteur de contexte de chaque action $i$;
	\item En utilisant les observations historiques, sélectionner l'action $i_{t}$ et recevoir la récompense $r_{i_{t},t}$;
	\item Améliorer la stratégie de décision en considérant la nouvelle observation $(i_{t},x_{i_{t},t},r_{i_{t},t})$. 
\end{enumerate}

           \begin{note}
      Dans cette thèse, on s'intéresse principalement à une structure linéaire entre les récompenses des actions, mais d'autres possibilités existent. En particulier, l'utilisation d'un graphe sous jacent peut également permettre de structurer les actions entre elle et améliorer l'exploration (voir la section \ref{graphBanditEtatArt}). D’autre part, le lecteur intéressé pourra se référer au travail de \cite{banditUnimodal} lorsqu'on fait l'hypothèse d'une structure unimodale et à celui de \cite{banditXArmed} lorsqu'il s'agit d'une structure lipchitzienne. Notons qu'il est possible de travailler dans des espaces d'actions continues, par conséquent avec un nombre d'actions infini (nous faisons un point sur ce sujet dans la section \ref{sectionOFUL}). 
   \end{note}
    
    Dans la partie suivante, on définit la notion de regret pour le problème du bandit contextuel, puis nous présenterons un certain nombre d'algorithmes de l'état de l'art permettant de tirer profit des contextes décisionnels observés.

        \subsection{Regret}
        
        Tout comme dans le cadre du bandit stochastique, on définit la notion de regret, qui quantifie la qualité d'un algorithme par rapport à une stratégie optimale. Ici, la stratégie optimale est définie par un oracle ayant connaissance du paramètre $\beta$ et qui choisirait à chaque instant l'action $ i_{t}^{\star}=\argmax\limits_{i\in \left\{ 1,..., K\right\}} x_{i,t}^{\top}\beta$.
        
        \begin{defi}
        	\label{defRegCtxt}
            Dans le cadre du bandit contextuel le regret est défini par:
            \begin{equation}
            R_{T}=\sum\limits_{t=1}^{T} r_{i_{t}^{\star},t}-r_{i_{t},t}
            \end{equation}
        \end{defi}
      
          \begin{defi}
        	\label{defPseudoRegCtxt}
            On définit également le pseudo-regret, calculé en prenant l'espérance du regret:
            \begin{equation}
            \hat{R}_{T}=\sum\limits_{t=1}^{T} x_{i_{t}^{\star},t}^{\top}\beta - x_{i_{t},t}^{\top}\beta
            \end{equation}
        \end{defi}
        
        Dans la pratique, on cherchera à borner le pseudo-regret d'un algorithme avec une forte probabilité pour en  analyser les performances théoriques. Pour deux fonctions $f$ et $g$ à variable entière, on utilisera la notation $f=\mathcal{O}(g)$ s'il existe une constante $C$ et un entier $M$ tels que $f(n)\leq Cg(n), \forall n\geq M$.
        
        \begin{theo}[Borne inférieure]
        	D'après \cite{banditCtxtLinUCBpreuve}, pour tout algorithme de bandit contextuel linéaire il existe une constante $\gamma>0$, telle que pour $T \geq d^{2}$:
			\begin{equation}
            \label{borneInfRegCtxt}
            	\hat{R}_{T} \geq \gamma \sqrt{dT}
            \end{equation}
        \end{theo}
        
        Ce théorème décrit une borne inférieure du pseudo-regret de tout algorithme de bandit contextuel linéaire, définissant les performances de la meilleure politique atteignable.

		\subsection{Algorithmes}

			Pour les trois algorithmes que nous allons décrire, nous utiliserons les notations matricielles suivantes:
				\begin{itemize}
					\item $D_{t}$ la matrice de taille $t \times d$ des contextes correspondant aux actions sélectionnées jusqu'au temps $t$:  $D_{t}=\left(x_{i_{s},s}^{\top}\right)_{s=1..t}$;
                    \item $c_{t}$ le vecteur de taille $t$ des récompenses reçues jusqu'au temps $t$: $c_{t}=\left(r_{i_{s},s}\right)_{s=1..t}$;
                    \item $V_{t}=D_{t}^{\top}D_{t}+\lambda I$, avec $I$ la matrice identité de taille $d$;
                    \item $b_{t}=D_{t}^{\top}c_{t}$ (vecteur de taille $d$).
				\end{itemize}
            
				\subsubsection{Algorithme \texttt{LinUCB}}
          
          L'algorithme \texttt{LinUCB} fut introduit par \cite{banditCtxtLinUCB}. Nous présentons ici une formulation légèrement différente, à savoir celle de \cite{banditCtxtLinUCBpreuve}, qui correspond à notre instance avec contextes individuels et paramètres communs. Cet algorithme fait usage des paires (contexte,récompense) observées dans le passé pour apprendre un estimateur  $\hat{\beta}_{t}$ du paramètre $\beta$, avec un certaine confiance, permettant de dériver un intervalle de confiance pour chaque action. De cette façon, il est possible de définir une politique optimiste sélectionnant l'action ayant la borne supérieure de l'intervalle de confiance la plus élevée. 
          
          
          On note $\hat{\beta}_{t}$ la solution du problème de régression linéaire $l^{2}$-régularisé  avec coefficient de régularisation $\lambda >0$ jusqu'au temps $t$:
          
                \begin{align*}
 \hat{\beta}_{t}&=\argmin\limits_{\beta}\sum\limits_{s=1}^{t} (r_{i_{s},s}-x_{i_{s},s}^{\top}\beta)^{2} +\lambda||\beta||^{2}\\
 				&=\argmin\limits_{\beta} ||c_{t}-D_{t}\beta||^{2}+\lambda||\beta||^{2} \\
                &= \left(D_{t}^{\top}D_{t}+\lambda I\right)^{-1}D_{t}^{\top}c_{t} \\
                &= V_{t}^{-1}b_{t}
                \end{align*}

  Le théorème suivant définit l'intervalle de confiance associé à chaque action $i$ au temps $t$ (avant d'effectuer la sélection de $i_{t}$) en fonction des paramètres de l'estimateur de $\beta$.
  
  \begin{prop}
  		\cite{banditCtxtLinUCB}
     Soit $0<\delta<1$, alors avec une probabilité au moins égale à $1-\delta$ on a pour toute action $i$ et à chaque instant $t \geq 1$:    
     \begin{equation}
     	|x_{i,t}^{\top}\hat{\beta}_{t-1}-x_{i,t}^{\top}\beta|\leq \alpha \sqrt{x_{i,t}^{\top}V_{t-1}^{-1}x_{i,t}}
     \end{equation}
     Avec $\alpha=1+\sqrt{\dfrac{\log(2/\delta)}{2}}$, en utilisant les conventions $A_{0}=\lambda I$ et $b_{0}=0$ le vecteur nul de taille $d$.
  \end{prop}

 Ainsi, on peut définir une borne supérieure de l'intervalle de confiance avec une probabilité $\delta$, notée $s_{i,t}$, de l'action $i$ à l'instant $t$ telle que: $s_{i,t}=x_{i,t}^{\top}\hat{\beta}_{t-1}+\alpha \sqrt{x_{i,t}^{\top}V_{t-1}^{-1}x_{i,t}}$. Le paramètre  $\alpha$, qui dépend de $\delta$ permet de régler l'exploration. Si l'on souhaite un algorithme dont la composante exploratoire est élevée, il faut choisir une valeur de $\delta$ faible, correspondant à un intervalle de confiance large, et donc un paramètre $\alpha$ élevé. La politique finale est décrite dans l'algorithme \ref{LinUCB}.
  
                   \begin{algorithm}[h!]
                   \KwIn{$\alpha >0$, $\lambda>0$}
                   $V=\lambda I$\\
                   $b=0$\\
 						\For{$t=1..T$}{
                        			$\hat{\beta}=V^{-1}b$\\
                        	        \For{$i=1..K$}{
                        				Observer contexte $x_{i,t}$ \\
                                        Calculer $s_{i,t}=x_{i,t}^{\top}\hat{\beta}+\alpha \sqrt{x_{i,t}^{\top}V^{-1}x_{i,t}}$ \\
                        				}
                                        Sélectionner 
 						 	$
    							i_{t} = \argmax\limits_{i \in \left\{1,...,K \right\}} s_{i,t}
  							$\\
							Recevoir récompense $r_{i_{t}}$\\
							$V	=V+x_{i_{t},t}x_{i_{t},t}^{\top}$\\
            				$b=b+r_{i_{t},t}x_{i_{t},t}$
 						}
						\caption{\texttt{LinUCB}  \cite{banditCtxtLinUCBpreuve}}
					\label{LinUCB}
					\end{algorithm}
    
    \begin{note}
    	La mise à jour des paramètres de l'algorithme d'un temps à un autre ne nécessite pas de conserver en mémoire l'intégralité des contextes et récompenses passés. En effet, les matrices $V_{t}$ et le vecteur $b_{t}$ sont mis à jour grâce aux formules  de récurrence suivantes:
        \begin{itemize}
         \item $V_{t}=V_{t-1}+x_{i_{t},t}x_{i_{t},t}^{\top}$ ;
         \item $b_{t}=b_{t-1}+r_{i_{t},t}x_{i_{t},t}$.
    \end{itemize}
        
    \end{note}

                Bien que l'algorithme \texttt{LinUCB} classique soit extrêmement performant et rapide d'un point de vue expérimental \cite{banditCtxtLinUCB}, son analyse théorique pose des difficultés techniques. En effet, comme le font remarquer les auteurs dans \cite{banditCtxtLinUCBpreuve}, pour pouvoir utiliser l'inégalité de Azuma-Hoeffding dans l'analyse de son regret, il faut que les estimateurs de récompenses à chaque itération proviennent d'une combinaison linéaire de récompenses passées indépendantes. Or, l'algorithme \texttt{LinUCB} calcule un estimateur de $\beta$ qui dépend des choix précédents.
                 Pour outrepasser cette difficulté, les auteurs proposent une modification de \texttt{LinUCB}, appelée \texttt{SupLinUCB}, garantissant une estimation des paramètres indépendante des décisions passées, en maintenant différents ensembles d'apprentissage en parallèle afin d'isoler la constitution de ces ensembles du processus de sélection. Bien que bien moins performant que \texttt{LinUCB} du fait de la dispersion des observations sur les différents ensembles d'apprentisage, \texttt{SupLinUCB} permet de garantir la borne supérieure du regret sous-linéaire présentée ci-dessous.
                

                    \begin{theo}
                    	\cite{banditCtxtLinUCBpreuve} Soit $0<\delta<1$. Si $\forall i, \forall t: ||x_{i,t}|| \leq 1$, $0\leq||r_{i,t}||\leq 1$ et $||\beta || \leq 1$  alors avec une probabilité au moins égale à $1-\delta$, le pseudo-regret de l'algorithme \texttt{SupLinUCB} est tel que:
                        \begin{equation}
                        	\hat{R}_{T} = \mathcal{O}\left(\sqrt{dT\log^{3}\left(\dfrac{KT\log(T)}{\delta}\right)}\right)
                        \end{equation}
                    \end{theo}
                    
                Cette borne est dite optimale à un facteur logarithmique près.

				\subsubsection{Algorithme \text{OFUL}}
                \label{sectionOFUL}

                L'algorithme \texttt{OFUL} (Optimism in the Face of Uncertainty Linear) présenté dans \cite{OFUL} est un algorithme optimiste traitant le problème du bandit contextuel linéaire, dans lequel l'espace des actions peut être continu. Cette instance spécifique du bandit contextuel fut auparavant étudiée par \cite{Dani2008} puis par \cite{Tsitsiklis2008}. Nous verrons par la suite que l'on peut aisément adapter cet algorithme au cas discret. Commençons par présenter le problème sous sa forme générique.

                A chaque instant $t$, l'agent décisionnel doit choisir une action $x_{t}$ dans un espace $\mathcal{D}_{t} \subset \mathbb{R}^{d}$. La récompense associée à cette action est toujours égale à $r_{t}=x_{t}^{\top}\beta+\eta_{t}$, avec  $\beta \in \mathbb{R}^{d}$ et $\eta_{t}$ un bruit sous-gaussien de constante $R>0$. 
                Soit $\hat{\beta}_{t}$ la solution du problème de régression linéaire $l^{2}$-régularisé  avec coefficient de régularisation $\lambda >0$ jusqu'au temps $t$. Comme précédemment on a: $\hat{\beta}_{t}=\left(D_{t}^{\top}D_{t}+\lambda I\right)^{-1}D_{t}^{\top}c_{t}$.  \cite{OFUL} montre qu'à un instant $t$, le véritable paramètre appartient à un ellipsoïde centré sur l'estimateur $\hat{\beta}_{t}$, noté $\mathcal{C}_{t}$ avec une certaine probabilité, selon la théorie des processus auto normalisés de \cite{PenaLaiShao2009}. Dans la suite, on note $||x||_{A}=\sqrt{x^{\top}Ax}$ la norme induite par une matrice définie positive $A$.

                \begin{theo}
                \label{ellipOFUL}
                	Soit $\delta > 0$ et $\lambda >0$. Si $||\beta|| \leq S$, alors avec une probabilité au moins égale à $1-\delta$:
                	\begin{equation}
                    \label{ellipOFULEq}
						\beta \in \mathcal{C}_{t} = \left\{\tilde{\beta} \in \mathbb{R}^{d}:  ||\hat{\beta}_{t}-\tilde{\beta} ||_{V_{t}} \leq R\sqrt{2\log\left(\dfrac{det(V_{t})^{1/2}det(\lambda I)^{-1/2}}{\delta}\right )}+\lambda^{1/2}S \right\}
                	\end{equation}	 
                \end{theo}
                 
				Cet ellipsoïde de confiance est plus serré que celui trouvé précédemment par \cite{Dani2008} ou \cite{Tsitsiklis2008}, pour cette raison nous ne présentons pas les algorithmes associés à ces deux publications. Le lecteur intéressé pourra se référer à \cite{OFUL} pour une comparaison exhaustive.

				A un instant $t$, l'agent décisionnel est en mesure de calculer $\mathcal{C}_{t-1}$ en fonction des choix passés. L'idée de l'algorithme est de calculer un estimateur optimiste $\tilde{\beta}_{t-1}=\argmax\limits_{\tilde{\beta}\in \mathcal{C}_{t-1}}(	\max\limits_{x\in \mathcal{D}_{t} } (x^{\top}\tilde{\beta}))$ puis de choisir $x_{t}=\argmax\limits_{x\in \mathcal{D}_{t}} x^{\top}\tilde{\beta}_{t-1}$. Dans l'article original, les auteurs proposent la notation compacte suivante : $(x_{t},\tilde{\beta}_{t-1})=\argmax\limits_{(x,\tilde{\beta})\in \mathcal{D}_{t}\times \mathcal{C}_{t-1}} x^{\top}\tilde{\beta}$. La politique associée est détaillée dans l'algorithme \ref{OFUL}.

                      \begin{algorithm}[h!]
                      	Initialiser $\mathcal{C}_{0}$ (équation \ref{ellipOFULEq})\\
 						\For{$t=1..T$}{
                        	     $(x_{t},\tilde{\beta}_{t-1})=\argmax\limits_{(x,\tilde{\beta})\in \mathcal{D}_{t}\times \mathcal{C}_{t-1}} x^{\top}\tilde{\beta}$
								\\
								Sélectionner $x_{t}$\\
								Recevoir récompense $r_{t}$\\ 
								Mettre à jour $\mathcal{C}_{t}$ (équation \ref{ellipOFULEq})
												}
						\caption{\texttt{OFUL}  \cite{OFUL}}
					\label{OFUL}
					\end{algorithm}

  Bien que dans le cas continu, où l'espace $\mathcal{D}_{t}$ peut être complexe, la résolution du problème d'optimisation $(x_{t},\tilde{\beta}_{t-1})=\argmax\limits_{(x,\tilde{\beta})\in \mathcal{D}_{t}\times \mathcal{C}_{t-1}} x^{\top}\tilde{\beta}$ puisse s'avérer très compliquée, dans le cas qui nous intéresse dans cette thèse, c.-à-d. le cas discret ($\mathcal{D}_{t}=\left\{x_{1,t},...,x_{K,t}\right\}$), on a:

  		$$i_{t} = \argmax\limits_{i \in \left\{1,...,K\right\}} \left( x_{i,t}^{\top}\hat{\beta}_{t-1}+\left(R\sqrt{2\log\left(\dfrac{det(V_{t-1})^{1/2}det(\lambda I)^{-1/2}}{\delta}\right )}+\lambda^{1/2}S\right)||x_{i,t}||_{V_{t-1}^{-1}}\right)$$





  Si l'on note $\alpha_{t-1}=R\sqrt{2\log\left(\dfrac{det(V_{t-1})^{1/2}det(\lambda I)^{-1/2}}{\delta}\right )}+\lambda^{1/2}S$, l'algorithme \texttt{OFUL} consiste à sélectionner l'action ayant le score $s_{i,t}=x_{i,t}^{\top}\hat{\beta}_{t-1}+\alpha_{t-1} \sqrt{x_{i,t}^{\top}V_{t-1}^{-1}x_{i,t}}$ le plus élevé. Ce score ressemble fortement au score utilisé dans \texttt{LinUCB}, sauf qu'ici, le paramètre d'exploration $\alpha_{t-1}$ varie au cours du temps. 

                      \begin{theo}
                    	\cite{OFUL} Soit $0<\delta<1$. Si l’on fait l'hypothèse que $||\beta|| \leq S$ et que $|x^{\top}\beta|\leq 1$,  alors avec une probabilité au moins égale à $1-\delta$, le pseudo-regret de l'algorithme \texttt{OFUL} est tel que:
                        \begin{equation}
                        	\hat{R}_{T} = \mathcal{O}\left(  d\log(T)\sqrt{T}+\sqrt{dT\log\left(\dfrac{T}{\delta}\right)}\right)
                        \end{equation}
                    \end{theo}
                
Cette borne a l'avantage d'être relativement serrée, avec un algorithme  relativement simple à implémenter. En effet il possède la même complexité que \texttt{LinUCB}, pour une borne du même ordre que \texttt{SupLinUCB}, beaucoup plus lourd à mettre en œuvre.

				\begin{note}
					Il existe une version allégée de l'algorithme \texttt{OFUL}, nommée \texttt{Rarely Switching OFUL} ne nécessitant qu'un nombre restreint de mises à jour du paramètre $\tilde{\beta}_{t}$. Ce dernier n'est alors recalculé que lorsque le réel $det(V_{t})$ augmente d'un facteur $(1+C)$ où $C$ est un paramètre de l'algorithme. Cet algorithme possède une borne supérieure du regret du même ordre que l'algorithme \texttt{OFUL} pour un coût bien moindre.
				\end{note}


\subsubsection{\texttt{Thompson sampling} contextuel}
            \label{TSLinBanditEtatArt}
            		
            		Sur le même principe que l'algorithme de \texttt{Thompson sampling} dans le cas du bandit stochastique, il s'agit d'un algorithme bayésien qui fut analysé par \cite{banditThompsonCtxtPreuve} pour le bandit contextuel linéaire. Ce type d'approche fut dans un premier temps remarqué grâce au travail de  \cite{banditThompsonCtxtEmpirical}, dans lequel les auteurs en ont montré les performances empiriques dans un cas concret d'application concernant la publicité en ligne. Comme dans le cas stochastique, le \texttt{Thompson sampling} nécessite de spécifier deux distributions:

            		\begin{itemize}
            			\item Une distribution \textit{a priori} sur le paramètre inconnu $\beta$: $p(\beta)$;
            			\item Une vraisemblance des récompenses $r_{i,t}$: $p(r_{i,t}|\beta)$.
            		\end{itemize}

            		Cela permet, grâce au théorème de Bayes, de maintenir une distribution \textit{a posteriori} sur le paramètre $\beta$ à mesure que les exemples d'apprentissages arrivent, c.-à-d à un instant $t$: $p(\beta|\mathcal{H}_{t-1})\propto \prod\limits_{s=1}^{t-1}p(r_{i_{s},s}|\beta)p(\beta)$. 

            		
Soient les distributions suivantes: 

            		\begin{itemize}
            			\item $p(\beta)=\mathcal{N}\left(\beta;0,\sigma^{2} I\right)$; 
            			\item $p(r_{i,t}|\beta)=\mathcal{N}\left(r_{i,t};x_{i,t}^{\top}\beta,\sigma^{2}\right)$.
            		\end{itemize}
            		  
On peut alors déterminer $p(\beta|\mathcal{H}_{t-1})$ en fonction des observations réalisées jusqu'au temps $t-1$:

            		  \begin{align*}
            		  	p(\beta|\mathcal{H}_{t-1}) & \propto \prod\limits_{s=1}^{t-1} \exp\left(-\dfrac{1}{2\sigma^{2}}(r_{i_{s},s}-x_{i_{s},s}^{\top}\beta)^{2}\right) \exp\left(-\dfrac{\beta^{\top}\beta}{2\sigma^{2}}\right)\\
            		  	&=\exp\left(-\dfrac{1}{2\sigma^{2}}\left( \sum\limits_{s=1}^{t-1}(r_{i_{s},s}-x_{i_{s},s}^{\top}\beta)^{2} +  \beta^{\top}\beta   \right)    \right)\\
            		  	&= \exp\left(-\dfrac{1}{2\sigma^{2}}\left( (c_{t-1}-D_{t-1}\beta)^{\top}(c_{t-1}-D_{t-1}\beta) +  \beta^{\top}\beta   \right)    \right)
            		  \end{align*}

					En remaniant les termes présents dans cette formule et en utilisant le fait que le terme $c_{t-1}^{\top}c_{t-1}$ ne dépend pas de $\beta$, on arrive à: 
            		  $p(\beta|\mathcal{H}_{t-1})\propto \exp\left(-\dfrac{1}{2\sigma^{2}}\left( (\beta - V_{t-1}^{-1}b_{t-1})^{\top}V_{t-1}(\beta - V_{t-1}^{-1}b_{t-1})  \right)    \right)$, avec $V_{t-1}=D_{t-1}^{\top}D_{t-1}+ I$ et $b_{t-1}=D_{t-1}^{\top}c_{t-1}$. Ceci correspond à une densité de probabilité représentant une loi gaussienne $\mathcal{N}(\beta; V_{t-1}^{-1}b_{t-1},\sigma^{2}V_{t-1}^{-1})$. Remarquons que la matrice $V_{t-1}$ est toujours définie positive et par conséquent la densité est toujours bien définie. La politique finale, décrite dans l'algorithme \ref{TSContextLin}, échantillonne la paramètre $\tilde{\beta}$ selon cette loi gaussienne  et sélectionne l'action $i_{t}=\argmax\limits_{i \in \left\{1,...,K \right\}} x_{i,t}^{\top}\tilde{\beta}$

                       \begin{algorithm}[h!]
                   		$V=I$\\
                   		$b=0$\\
 						\For{$t=1..T$}{
                        			$\hat{\beta}=V^{-1}b$\\
                        			Échantillonner $\tilde{\beta}\sim \mathcal{N}(\hat{\beta}, \sigma^{2}V^{-1})$\\ 
                        	        \For{$i=1..K$}{
                        				Observer contexte $x_{i,t}$ \\
                        				}

                                        Sélectionner 
 						 	$
    							i_{t} = \argmax\limits_{i \in \left\{1,...,K \right\}} x_{i,t}^{\top}\tilde{\beta}
  							$\\
							Recevoir récompense $r_{i_{t}}$\\
							$V=V+x_{i_{t},t}x_{i_{t},t}^{\top}$\\
            				$b=b+r_{i_{t},t}x_{i_{t},t}$
 						}
						\caption{\texttt{Thompson sampling} contextuel linéaire  \cite{banditThompsonCtxtPreuve}}
					\label{TSContextLin}
					\end{algorithm}

					\begin{note}
				Finalement pour chaque action $i$ à l'instant $t$, une fois son contexte observé, la variable aléatoire $\mathbb{E}[r_{i,t}]=x_{i,t}^{\top}\beta$ suit une loi gaussienne unidimensionnelle $\mathcal{N}(x_{i,t}^{\top}V_{t-1}^{-1}b_{t-1},\sigma^{2}x_{i,t}^{\top}V_{t-1}^{-1}x_{i,t})$, d'où l'on peut directement retrouver l'intervalle de confiance utilisé dans l'algorithme \texttt{LinUCB}.
					\end{note}

                      \begin{theo}
                    	\cite{banditThompsonCtxtPreuve} Soit $0<\delta<1$, $0<\epsilon<1$ et $\sigma^{2}=R\sqrt{\dfrac{24}{\epsilon}d\log\left(\dfrac{1}{\delta}\right)}$. Si $\forall i, \forall t: ||x_{i,t}|| \leq 1$ et $||\beta || \leq 1$ alors avec une probabilité au moins égale à $1-\delta$, le pseudo-regret de l'algorithme \texttt{Thompson sampling} contextuel linéaire est tel que:
                        \begin{equation}
                        	\hat{R}_{T} = \mathcal{O}\left(\dfrac{d^{2}}{\epsilon}\sqrt{T^{1+\epsilon}}\log(dT)\log\left(\dfrac{1}{\delta}\right)\right)
                        \end{equation}
                    \end{theo}

Ce théorème fut le premier résultat théorique établi sur les \texttt{Thompson sampling} contextuel. La borne proposée, bien que non optimale au sens où elle n'est pas exactement égale à la valeur de la borne inférieure, s'en rapproche à un terme logarithmique près.

	\section{Bandit avec sélections multiples}
    \label{banditMultSel}
    
		Le problème du bandit avec sélections multiples est une extension du problème du bandit traditionnel dans le cas où l'agent sélectionne plus d'une action à chaque itération. On notera $\mathcal{K}_{t} \subset \mathcal{K}$ l'ensemble des actions sélectionnées et $k$ le nombre d'éléments qu'il contient ($k=|\mathcal{K}_{t}|$), c'est-à-dire le nombre d'actions à sélectionner. Notons que $k$ est un paramètre du problème fixé à l'avance dans notre cas. Le problème se décline sous les mêmes formes que dans le cas traditionnel, à savoir le cas stochastique et le cas contextuel, que l'on propose de détailler ci-dessous. Finalement, nous nous restreignons ici au cas où la récompense associée à $k$ actions est la somme des récompenses individuelles. Cette instance spécifique est incluse dans un champ de recherche désignée sous l'appellation anglophone \textit{Combinatorial optimization with semi-bandit feedback} \cite{combOptSemiBandit}. Notons toutefois que des structures de récompenses plus complexes peuvent également être considérées. Le cas générique où la récompense de  $k$ actions est une fonction lipschitzienne des $k$ récompenses individuelles est présenté dans \cite{banditCombinatoire} pour le bandit stochastique et dans \cite{banditCtxtCombLinUCB}  pour le bandit contextuel.

		\subsection{Cas stochastique}
        
        Le bandit stochastique avec sélections multiples fut dans un premier temps introduit par \cite{banditThompsonCombinatorialLowerBound}. Dans ce travail, les auteurs proposent une borne inférieure du regret ainsi qu'un algorithme permettant d'atteindre cette borne. Cependant, leur méthode très complexe d'un point de vue calculatoire n'est pas aisée à implémenter de façon efficace. Plus récemment, \cite{banditCombinatoire} et \cite{banditCombinatoireESCB} ont proposé respectivement \texttt{CUCB} et \texttt{ESCB}, des algorithmes de type UCB permettant une implémentation efficace et rapide. D'autre part, \cite{banditThompsonComplexProblem} ont étendu la méthode de Thompson sampling à une classe plus large de problèmes, incluant celui qui nous intéresse ici. Dans les deux cas, les auteurs montrent une borne supérieure du regret pour l'algorithme associé, mais cette dernière n'est pas optimale. Finalement, dans les travaux récents de \cite{banditThompsonCombinatorialProof}, un algorithme de Thompson Sampling spécifique au cas du bandit stochastique avec sélections multiples est proposé et son optimalité est également démontrée lorsque les récompenses suivent des lois de Bernoulli. 
        
        On suppose que les moyennes des actions sont distinctes et ordonnées de la façon suivante, c.-à-d. $\mu_{1} > \mu_{2} >...> \mu_{K}$ et on note $\mathcal{K}^{\star}$ l'ensemble des $k$ actions optimales $\left\{1,,...,k\right\}$. Évidemment les algorithmes n'ont pas connaissance de cette relation d'ordre. 
        
        \begin{defi}
   			 \label{regretCumultoPseudoMP}    
   			 Le pseudo-regret dans le cas du bandit stochastique avec sélections multiples est défini par:
     		\begin{equation}	
     			\hat{R}_{T}= \mathbb{E}\left[\sum\limits_{t=1}^{T}  \left( \sum\limits_{i\in \mathcal{K}^{\star}} \mu_{i} -  \sum\limits_{i\in \mathcal{K}_{t}} \mu_{i}\right) \right]
    		\end{equation}
		\end{defi}
        
        Contrairement au cas classique, le regret ici ne dépend pas uniquement du nombre de fois ou les actions sous-optimales ont été sélectionnées. En effet, ce dernier dépend fortement de la structure combinatoire des récompenses. Pour illustrer cela nous présentons ci-dessous un exemple évoqué dans \cite{banditThompsonCombinatorialProof}: on prend $K=4$, $k=2$, $T=2$ et $(\mu_{1},\mu_{2},\mu_{3},\mu_{4}) =(0.10,0.09,0.08,0.07)$ et on propose de comparer deux scénarios. Dans le premier on prend $\mathcal{K}_{1}=\left\{1,2\right\}$, $\mathcal{K}_{2}=\left\{3,4\right\}$ et dans le second $\mathcal{K}_{1}=\left\{1,3\right\}$, $\mathcal{K}_{2}=\left\{1,4\right\}$. Ainsi le regret du premier vaut $\hat{R}_{2}=0.04$ et dans le second $\hat{R}_{2}=0.03$. Or dans les deux cas on a $N_{3,2}=1$ et $N_{4,2}=1$ et donc bien que les actions sous-optimales aient été jouées le même nombre de fois le regret est différent. Par conséquent pour atteindre un regret optimal, un algorithme ne peut se limiter à sélectionner les actions sous-optimales le moins de fois possibles. En se basant sur ce constat \cite{banditThompsonCombinatorialLowerBound} ont montré que le pseudo-regret est minoré par une fonction dépendant du nombre de fois où les actions sous optimales ont été sélectionnées: $\hat{R}_{T}\geq \sum\limits_{i\in \mathcal{K}\backslash \mathcal{K}^{\star}} (\mu_{k}-\mu_{i})\mathbb{E}[N_{i,T}]$, où $\mu_{k}$ est l'espérance de la moins bonne action optimale. Le théorème suivant établit une borne inférieure de $\mathbb{E}[N_{i,T}]$ pour chaque action $i$, d'où l'on peut déduire une borne inférieure du regret grâce à l'étude précédente.

\begin{theo}[Borne inférieure]
	\label{infRegStoMP}
	D'après \cite{banditThompsonCombinatorialLowerBound}, pour toute politique telle que $\mathbb{E}[N_{i,T}]=o(T^{\beta})$ avec $0<\beta < 1$, on a:
		\begin{equation}
        \label{bonreInfnbPlayesMP}
			\liminf\limits_{T\rightarrow\infty} \dfrac{\mathbb{E}[N_{i,T}]}{\log(T)} \geq  \dfrac{1}{KL_{inf}(i_{k},i)}
		\end{equation}
        Où l'on rappelle que $i_{k}$ désigne l'indice de la moins bonne action optimale.
        
         Ceci se traduit sur la borne inférieure du regret par:
         \begin{equation}
         \label{bonreInfRegMP}
			\liminf\limits_{T\rightarrow\infty} \dfrac{\hat{R}_{T}}{\log(T)} \geq \sum\limits_{i: \Delta_{i,k}>0} \dfrac{\Delta_{i,k}}{KL_{inf}(i_{k},i)}
		\end{equation}
        où $\Delta_{i,k}=\mu_{k}-\mu_{i}$
\end{theo}

	Pour un descriptif précis des algorithmes cités plus haut, nous orientons le lecteur vers les références correspondantes. Cependant, notons que dans le cas qui nous intéresse ici, à savoir celui où la récompense d'un ensemble de bras est égale à la somme des récompenses individuelles de chaque bras, ces algorithmes prennent une forme simplifiée. En effet, \texttt{CUCB} consiste à sélectionner les $k$ meilleurs bras, ordonnés selon un score \texttt{UCB} classique, tandis que \texttt{Thompson sampling} revient à échantillonner chaque bras indépendamment puis à sélectionner les $k$ meilleurs. 

		\subsection{Cas contextuel}

Le problème du bandit contextuel avec sélections multiples est une extension naturelle du cas contextuel traditionnel dans lequel à chaque itération $t$ et pour chaque action $i$, un vecteur de contexte $x_{i,t}$ est fourni à l'agent avant qu'il effectue sa sélection. Ce problème a été étudié et appliqué à un scénario de recommandation dans \cite{banditCtxtCombLinUCB}. Dans ce travail, les auteurs proposent l'algorithme \texttt{C$^{2}$UCB} (Contextual Combinatorial UCB) et une borne (non optimale) associée. Il s'agit d'une extension de l'algorithme \texttt{OFUL} au cas des sélections multiples. Dans notre cas il s'agit de sélectionner à chaque instant les $k$ actions ayant les meilleurs scores. On peut également étendre les autres algorithmes de bandit contextuel (\texttt{LinUCB}, \texttt{Thompson Sampling} etc.) à cette situation. Le pseudo-regret est défini de la façon suivante.

         \begin{defi}
        	\label{defPseudoRegCtxtMP}
            Le pseudo-regret dans le cas du bandit contextuel avec sélections multiples est défini par:
            \begin{equation}
            \hat{R}_{T}=\sum\limits_{t=1}^{T} \left(\sum\limits_{i\in \mathcal{K}_{t}^{\star}}  x_{i,t}^{\top}\beta - \sum\limits_{i\in \mathcal{K}_{t}} x_{i,t}^{\top}\beta\right)
            \end{equation}
            Où  $\mathcal{K}_{t}^{\star}$ est  l'ensemble des actions optimales c.-à-d. les $k$ actions ayant la plus grande valeur de $x_{i,t}^{\top}\beta$.
        \end{defi}

			\section{Bandit dans les graphes}
            		\label{graphBanditEtatArt}
            Plusieurs travaux se sont intéressés à des problèmes de bandit dans lesquels une structure de graphe sous-jacente existe. Nous proposons dans cette section un aperçu de certains travaux, ce qui nous permettra par la suite de situer notre travail, en particulier dans le chapitre \ref{ModeleRelationnel}. Dans ce type de problème, l'agent décisionnel a la possibilité de tirer parti de cette structure, ce qui lui permet d'apprendre plus rapidement comparé au cas du bandit stochastique, où l'apprentissage est effectué sur chaque nœud séparément. En raison des nombreuses applications possibles, en particulier dans le marketing et la publicité en ligne, cette formulation du problème de bandits a connu un fort intérêt ces dernières années. Typiquement, le graphe peut être une représentation d'un réseau social dans lequel un nœud correspond à un utilisateur et une arrête à un lien social. Nous proposons ci-dessous de référencer les principales instances de ce problème:

            \begin{itemize}
            \item Les problèmes de bandits dans les graphes furent introduits dans  \cite{banditNetworfFirst} pour le cas du bandit adverse puis dans \cite{banditNetwork1UCBN}  pour le bandit stochastique. L'hypothèse sous-jacente est que lorsque l'on sélectionne une action, on reçoit la récompense associée, ainsi que celles de toutes les actions voisines (dans le graphe), permettant ainsi d'accélérer considérablement l'apprentissage. Ainsi, les algorithmes  optimistes \texttt{UCB-N} et \texttt{UCB-MaxN}, proches de la politique \texttt{UCB} classique sont étudiés dans \cite{banditNetwork1UCBN}. Dans le premier, on utilise  simplement le fait que l'on accumule plus de connaissances à chaque itération, nous permettant ainsi d'avoir un score UCB plus fin et de converger plus rapidement vers la solution optimale. Dans le second, on choisit dans un premier temps un nœud via un score UCB classique, et on sélectionne ensuite l'action ayant la meilleure moyenne empirique dans le voisinage de ce nœud. Ces méthodes permettent d'améliorer les garanties de convergence et en particulier d'ôter la dépendance explicite du regret en $K$, ce qui est bénéfique lorsque le nombre d'actions est important. Ces résultats ont ensuite été améliorés par \cite{banditNetwork2UCBLP} grâce à l'algorithme \texttt{UCB-LP}.
            \item  D'autre part, le problème nommé \textit{spectral bandits} fut étudié dans \cite{spectralBandit} et \cite{spectralBandit2}. Dans cette configuration, on suppose une hypothèse de régularité sur le graphe, les nœuds connectés du réseau étant supposés  se comporter de manière similaire. Les nœuds connectés ont alors tendance à posséder des distributions d'utilité proches, ce qui peut être exploité pour définir des stratégies efficaces : l'observation d'une récompense sur un noeud du réseau permet de faire des suppositions sur le niveau de récompense (que l'on n'a pas observé) pour tous les noeuds de son voisinage et ainsi accroître la vitesse d'apprentissage.
            \item Dans un autre cadre, les bandits contextuels ont également été étudiés lorsqu'une structure de graphe est présente. En particulier dans \cite{gangOfBandit}  et \cite{clusteringOfBandit}, chaque nœud correspond à un bandit contextuel et les arêtes du graphe sont utilisées pour définir les similarités entre les poids des différents bandits. Dans  \cite{gangOfBandit} (algorithme \texttt{Gob.Lin}) les poids des arêtes sont connus à l'avance et sont directement utilisés dans la stratégie de sélection. Dans \cite{clusteringOfBandit} (algorithme \texttt{CLUB}), les actions sont supposées appartenir à différents \textit{clusters}, inconnus à l'avance, caractérisant des similarités entre elles. Plus récemment, des approches  utilisant les bandits dans des environnements collaboratifs ont été développées pour des problèmes de recommandation. On citera en particulier les travaux  de \cite{banditCollaborative2} et \cite{banditCollaborative}, dans lesquels les algorithmes \texttt{CoLinUCB} et \texttt{COFIBA} sont développés.
            \item Dans un autre registre, les travaux récents de \cite{banditInfluence} traitent du problème de la maximisation de l'influence locale (c'est à dire chaque noeud ne peut influencer que ses voisins directes), dans lequel aucune connaissance du graphe n'est supposée \textit{a priori}.  Une des applications de ce modèle concerne  le marketing dans les réseaux sociaux, où les marques souhaitent tirer profit des utilisateurs les plus influents pour diffuser leurs produits. Le modèle suppose l'existence d'un graphe sous-jacent dont les poids des arcs modélise les probabilités d'influence des nœuds entre eux. A chaque itération, l'agent sélectionne un nœud, observe l'ensemble des nœuds influencés, et la récompense associée  correspond au nombre de nœuds ayant été influencés. Les auteurs proposent une politique nommée \texttt{BARE} permettant à de maximiser les récompenses récoltées. Cet algorithme, dans lequel l'horizon $T$ doit être connu, extrait un sous-ensemble de nœuds après une phase d'exploration pure, puis un algorithme de bandit classique est appliqué à ce sous-ensemble pour trouver le plus influent. Notons que cette approche fait une hypothèse de stationnarité des récompenses.
            \end{itemize}

Nous nous sommes volontairement limités au cadre du bandit stochastique et contextuel. Cependant, de nombreux travaux sur les problèmes de bandit adverse dans les graphes ont été formalisés. Nous orientons le lecteur intéressé vers les lectures suivantes : \cite{banditNetworfFirst}, \cite{advBanditGraph}, \cite{advBanditGraph2}, \cite{banditGraph}, \cite{advBanditGraph3}.

			\section{Bandit non stationnaire}
            \label{nonstatBanditEtatArt}
            
Dans un certain nombre d'applications, des changements dans les distributions de récompense au cours du temps sont inhérents et doivent être pris en compte dans le modèle. En réalité, l'hypothèse de stationnarité, bien que permettant de caractériser très finement le regret, n'est souvent pas réaliste, comme le soulignent les auteurs dans \cite{nonStationaryBanditBound}. Ainsi, un champ de la littérature au sujet des bandits s'intéresse au cas où les récompenses sont non-stationnaires. Dans ce cas pour chaque action $i$, on note $\mu_{i}(t)$ la valeur moyenne de la récompense associée à un instant $t$. Si dans un problème de bandit stochastique traditionnel les performances d'un algorithme sont comparées à celles d'un oracle statique choisissant toujours la même action, ceci n'est plus pertinent dans le scénario non stationnaire. En effet, il convient de définir une autre notion de regret, mesurant les performances d'un algorithme par rapport à celle d'un oracle dynamique.

\begin{defi}
\label{defregNonStat}
En notant $i_{t}^{\star}=\argmax\limits_{i\in \left\{1,...,K\right\}}$ l'action ayant la plus forte espérance à un instant $t$, le regret cumulé au temps $T$ est défini par:
\begin{equation}
\label{defregNonStatFormule}
\hat{R}_{T}=\mathbb{E}\left[\sum\limits_{t=1}^{T} \mu_{i_{t}^{\star}}(t) -  \mu_{i_{t}}(t)\right]
\end{equation}
Où $\mathbb{E}[.]$ est l'espérance prise selon la politique de sélection considérée.
\end{defi}

L'expression "non stationnaire" étant relativement vague, il convient lors de chaque étude d'effectuer des hypothèses supplémentaires afin de cadrer le problème. Nous proposons ci-dessous différents scénarios ayant été étudiés concernant la façon dont cette instationnarité se manifeste:

\begin{itemize}
\item La première instance du bandit non stationnaire a été introduite dans \cite{restlessBandit}, où le terme \textit{restless bandit} fut introduit. Dans ce modèle l'état de chaque action est modifié à chaque itération selon un processus aléatoire\footnote{Lorsque seul l'état de l'action sélectionnée change à chaque itération, le problème et appelé \textit{rested bandit}.}. En partant de cette hypothèse, de nombreux processus ont ensuite été étudiés. Par exemple dans \cite{nonStationaryBanditBrown} les récompenses espérées de chaque action évoluent selon des mouvements browniens indépendants. D'autre part dans, \cite{restlessBandit2} et  \cite{restlessBandit3}, un processus de Markov (MDP) discret régit l'évolution des différents états. Ainsi, pour chaque action, on suppose l'existence de probabilités de transition entre états (inconnues de l'agent), chaque état correspondant à une espérance de récompense différente (inconnue de l'agent). Notons que dans cette dernière approche, les états des différentes actions évoluent de façon indépendante. Par ailleurs, les travaux de \cite{restlessBanditLiva} étudient un autre instance du \textit{restless bandit}, dans le cas où les actions sont associées à des processus de mélange ($\phi$\textit{-mixing process}), et où, par conséquent, des dépendances entre les récompenses d'une même action à différents pas de temps existent.

\item Dans un autre contexte, le cas où l'espérance des récompenses varie de façon abrupte au cours du temps a été étudié dans \cite{nonStationaryBanditAbrupt}. Dans ce cas, on fait cependant l'hypothèse que le nombre de changements est limité. Si l'on note $\gamma_{T}$ le nombre de changements sur un horizon de taille $T$, alors les auteurs démontrent que la borne inférieure du regret associé est en $\mathcal{O}(\sqrt{\gamma_{T}T})$. Les auteurs étudient deux algorithmes permettant de traiter ce problème:
		\begin{itemize}
			\item Le premier, nommé \texttt{Dicounted UCB}, utilise un facteur de \textit{discount} pour estimer l'espérance des récompenses à un instant donné. Il est ainsi possible de régler l'importance que l'on donne aux exemples récents relativement aux exemples les plus anciens. Si l'on choisit $\gamma=1$ comme facteur de \textit{discount} alors cet algorithme est équivalent à l'algorithme \texttt{UCB} classique. Si le nombre de changements $\gamma_{T}$ est connu à l'avance, alors les auteurs démontrent que le regret est en $\mathcal{O}(\sqrt{\gamma_{T}T}\log(T))$.
            \item Le second, nommé \texttt{Sliding Window UCB}, utilise une fenêtre glissante permettant de ne prendre en compte les récompenses passées que sur un intervalle de taille fixe. Ainsi à un instant donné, au lieu d'utiliser un facteur de \textit{discount} comme dans \texttt{Dicounted UCB}, on  calcule les scores de sélection de la même façon que dans un \texttt{UCB}, mais en utilisant uniquement les $\tau$ derniers pas de temps. Les auteurs montrent également que si l'on connaît $\gamma_{T}$ à l'avance, le regret est également en $\mathcal{O}(\sqrt{\gamma_{T}T}\log(T))$. 

		\end{itemize}

\item Plus récemment dans \cite{nonStationaryBanditBound}, les auteurs introduisent une notion de budget total de variation contraignant également les changements des valeurs espérées des récompenses. Cependant, ces contraintes sont moins fortes que dans les deux cas précédents, qui sont englobés dans le formalisme proposé. Dans ce modèle, les récompenses peuvent changer un nombre de fois arbitraire, mais c'est la variation totale des valeurs espérées, définie par $\sum\limits_{t=1}^{T-1}\sup\limits_{i\in \mathcal{K}} |\mu_{i}(t) - \mu_{i}(t+1) |$, qui est bornée. L'algorithme proposé dans ce travail est nommé \texttt{Rexp3}.
\end{itemize}

\begin{note}
Le bandit contextuel permet de capter des variations d'utilités des récompenses en fonction d'un contexte décisionnel pouvant évolué dans le temps. Dans ce sens, il modélise des récompenses non stationnaires. Cependant, on considère généralement que les paramètres du modèle, c'est-à-dire les poids définissant la relation entre contextes et récompenses, sont constants au cours du temps. Des poids évoluant avec le temps pourraient également être considérés. Dans ce cadre, une démarche similaire à celle de  l'algorithme \texttt{Sliding Window UCB} pourrait être appliquée, en restreignant l'apprentissage des poids du modèle à une fenêtre de temps. Cependant, les garanties théoriques de convergence ne seraient plus assurées a priori.
\end{note}


    \part{Contributions}
\label{partContributions}

\chapter{La collecte vue comme un problème de bandit}
  \minitoc
  \newpage

\label{FormalisationTache}

Ce premier chapitre  constitue le scocle commun à toutes les approches que nous proposons tout au long de ce manuscrit. Nous définissons dans un premier temps le processus de collecte d'information dynamique, puis nous le formalisons comme un instance particulière d'un problème de bandit. Nous présentons ensuite les différents jeux de données et modèles de récompense qui permettront d'évaluer les performances du processus de collecte.

  \section{Processus de collecte dynamique}
    
    Dans cette thèse, nous nous intéressons donc au problème de collecte de données en temps réel sur les réseaux sociaux, dans un cadre générique, tel que les approches définies puissent être valides pour divers types de collectes plus ou moins complexes. Comme mentionné précédemment, la plupart des grands médias sociaux mettent à disposition des APIs dites de {\it streaming}, fournissant une portion plus ou moins restreinte de l'activité de leurs utilisateurs en temps réel. Dans cette thèse, les expérimentations sont menées dans le cadre du réseau Twitter, qui propose les APIs de {\it streaming} suivantes\footnote{Documentation Twitter disponible à l'URL: \url{https://dev.twitter.com/streaming/public}}:

\begin{itemize}
\item API \textit{Sample Streaming}: renvoie aléatoirement 1\% des contenus en temps réel.
\item API \textit{Follow Streaming}: permet de suivre en temps réel les contenus produits par 5000 utilisateurs, dans la limite de 1\% du volume total des contenus produits;
\item API \textit{Track Streaming}: permet de suivre en temps réel les publications contenant au moins un mot parmi une liste de 400 mots-clés, dans la limite de 1\% du volume total des contenus produits;
\item API \textit{Geo Streaming}: permet de suivre en temps réel les publications géolocalisées dans 25 zones, dans la limite de 1\% du volume total des contenus produits;

\end{itemize}   
Ainsi, l'API {\it Sample Streaming}  est très limitée et son utilisation seule pour une recherche de données particulière peut nous amener à être noyé sous les messages collectés, tout en passant complètement à côté des messages potentiellement pertinents (puisque l'on ne collecte que 1\% de l'activité globale, rien ne garantit que ne serait-ce qu'un seul message  collecté soit pertinent,  malgré la publication de multiples messages répondant à nos besoins sur le réseau). Aucun suivi d'activité ciblé n'est possible avec cette API. Pour des recherches simples que l'on peut exprimer sous la forme d'une liste de mots-clés, l'API {\it Track Streaming} pourrait paraître le meilleur choix de prime abord, mais son utilisation se heurte à diverses limitations. Premièrement, en raison de son mode de filtrage disjonctif (modèle booléen : tous les messages contenant au moins un terme de la liste de filtrage peuvent être retournés), la présence d'un seul terme trop générique dans la liste amène à n'obtenir quasiment que des messages contenant ce terme uniquement, et donc risque  de nous faire passer à côté de tous les messages réellement intéressants. Les messages contenant tous les termes de la liste ne sont en effet pas prioritaires par rapport aux autres, comme cela serait le cas dans des modèles d'ordonnancement définis dans le cadre de la recherche d'information classique. Rien ne garantit alors d'avoir des messages pertinents dans l'ensemble retourné, car il est très probable que la taille de l'ensemble des messages candidats dépasse largement les 1\% de l'activité globale si la liste de filtrage comporte un mot fréquent. Une méthode automatique qui sélectionnerait les mots à suivre selon cette API risquerait de se heurter très rapidement à ce problème, en choisissant probablement des mots génériques trop fréquemment. D'autre part, l'ensemble des recherches considérées dans cette thèse ne concerne pas uniquement des besoins pouvant s'exprimer sous la forme de listes de termes, et pour des recherches plus complexes cette API de filtrage par mots-clés peut paraître peu pertinente. Est-il possible de définir une liste de mots-clés présents dans la majorité de messages pertinents? Alors qu'une recherche dynamique selon l'API {\it Follow Streaming} peut permettre d'identifier les utilisateurs du réseau les plus susceptibles de produire du contenu pertinent, une recherche par mots-clés se contenterait d'identifier les termes les plus souvent présents dans les messages pertinents, ce qui paraît bien plus limité. En outre, dans le cadre de problèmes de collecte où l'on souhaite obtenir des données relationnelles relativement denses, centrées sur une communauté d'utilisateurs retreinte afin d'en extraire diverses statistiques, ce genre d'API par filtrage sur mots-clés n'est pas du tout adapté. Il s'agit alors de s'intéresser aux producteurs de contenu, plutôt qu'à des marqueurs particuliers de pertinence. L'API {\it Follow Streaming} paraît alors bien plus adaptée, c'est l'approche étudiée dans cette thèse pour la collecte de données dynamique (bien que l'ensemble des approches définies dans cette thèse pourrait être également appliquées dans le cadre de recherches par mots-clés). Notons enfin que chercher le contenu pertinent par sélection d'utilisateurs à suivre permet par ailleurs d'être plus générique par rapport au réseau social considéré, puisque, alors que des APIs de {\it streaming} filtrées selon des mots-clé ne sont pas toujours disponibles, les médias sociaux se voient dans l'obligation, en vue d'être promus par des acteurs externes et utilisés dans divers contextes à travers le Web, de proposer des mécanismes permettant à des applications tierces de récupérer l'activité en temps réel d'utilisateurs ciblés. Par exemple, Facebook ne propose pas d'API de {\it streaming} filtrant les informations selon les mots employés, mais met à disposition un système d'abonnement à des pages d'utilisateurs, fournissant des rapports temps-réel de leur activité similaires à ce que produit l'API {\it Follow Streaming} de Twitter.

Le problème de la collecte dynamique de données sur les réseaux sociaux peut alors  se voir comme un problème de sélection de sources à écouter: n'ayant pas la capacité de considérer la totalité de l'activité du réseau à chaque instant de la collecte (pour des raisons techniques liées aux ressources à disposition ou bien en raison des limitations), l'objectif est de définir un sous-ensemble d'utilisateurs susceptibles de produire du contenu pertinent pour le besoin spécifié. On se focalise ici sur la sélection de sources pertinentes dans un réseau social lorsque le nombre de comptes d'utilisateurs pouvant être écoutés simultanément est limité. 
Nous proposons de considérer ce problème dans le contexte des réseaux sociaux de grande taille, où le choix des sources ne peut être effectué manuellement en raison du trop grand volume de données produites et du trop grand nombre de sources à envisager. Bien que notre approche puisse être appliquée à bien d'autres réseaux sociaux (tous ceux qui offrent un accès temps réel à un sous-ensemble de leurs données), nous nous intéressons à la collecte de données sur Twitter,  où le nombre d'utilisateurs total est supérieur à $313$ millions, pour une limite de 5000 utilisateurs pouvant être écoutés simultanément. Dans ce contexte, choisir quel sous-ensemble d'utilisateurs suivre à chaque instant est une question complexe, qui requiert la mise en œuvre de techniques d'apprentissage permettant une exploration intelligente de l'ensemble des utilisateurs du réseau pour déterminer de manière efficace les meilleurs candidats à écouter pour la tâche considérée.

Étant donné un processus qui, à chaque instant de la collecte, permet de récupérer les contenus produits par un sous-ensemble d'utilisateurs d'un réseau social, et une fonction de qualité permettant de mesurer la pertinence d'un contenu publié par un utilisateur écouté dans un intervalle de temps, il s'agit de définir une stratégie de décision permettant au processus de se concentrer sur les utilisateurs les plus intéressants à mesure que le temps s'écoule. Cette stratégie, apprise de façon incrémentale, est définie de façon à maximiser le gain cumulé - évalué via la fonction de qualité -  par les utilisateurs écoutés au cours du temps. Dans ce qui suit, cette stratégie de décision sera appelée de façon équivalente \textit{politique de sélection}. Sur Twitter par exemple, l'information capturée pourrait correspondre à l'ensemble des \textit{tweets} publiés par les utilisateurs écoutés durant la période de temps considérée, et la fonction de qualité pourrait correspondre à un score évaluant le contenu par rapport à une thématique donnée ou bien sa popularité. Nous reviendrons sur les différentes fonctions de qualité que nous utilisons par la suite.

Dans notre contexte, une difficulté majeure provient du fait que l'on ne connaît rien \textit{a priori}  des utilisateurs du réseau ni des relations qui peuvent les relier: récupérer des profils utilisateurs ou la liste des utilisateurs amis d'un utilisateur donné n'est bien souvent pas envisageable, car cela requiert le questionnement d'une API coûteuse ou restreinte (fréquence des requêtes possibles souvent très limitée) et se heurte parfois à des problèmes de confidentialité des données. Cela implique entre autres que nous ne pouvons pas nous baser sur le graphe du média social pour explorer les utilisateurs (ce qui interdit l'emploi de techniques utilisées pour des problèmes connexes de \textit{Crawling}) et que nous ne connaissons pas \textit{a priori} l'ensemble complet des utilisateurs du réseau. Ainsi, en plus de sélectionner les utilisateurs les plus pertinents, à chaque itération, le processus doit alimenter l'ensemble des utilisateurs potentiellement écoutables. Par exemple, un nouvel utilisateur peut être référencé dans un message si l'utilisateur à l'origine du message le mentionne dans son contenu. Typiquement, sur Twitter, les utilisateurs peuvent se répondre entre eux ou republier des messages d'autres utilisateurs, ce qui nous offre la possibilité de découvrir de nouvelles sources. Le processus général de collecte, présenté dans la figure \ref{fig:process}, opère de la façon suivante à chaque itération:

\begin{enumerate}
\item Sélection d'un sous-ensemble d'utilisateurs relativement à la politique de sélection considérée ;
\item Ecoute de ces utilisateurs pendant une fenêtre de temps donné ;
\item Alimentation de l'ensemble des utilisateurs potentiellement écoutable  (par exemple en fonction des nouveaux utilisateurs référencés
dans les messages enregistrés) ;
\item Evaluation des données collectées en fonction de leur pertinence pour 
la tâche à résoudre;
\item Mise à jour de la politique de sélection en fonction des scores  obtenus.
\end{enumerate}

  \begin{figureth}
      \includegraphics[width=0.9\linewidth]{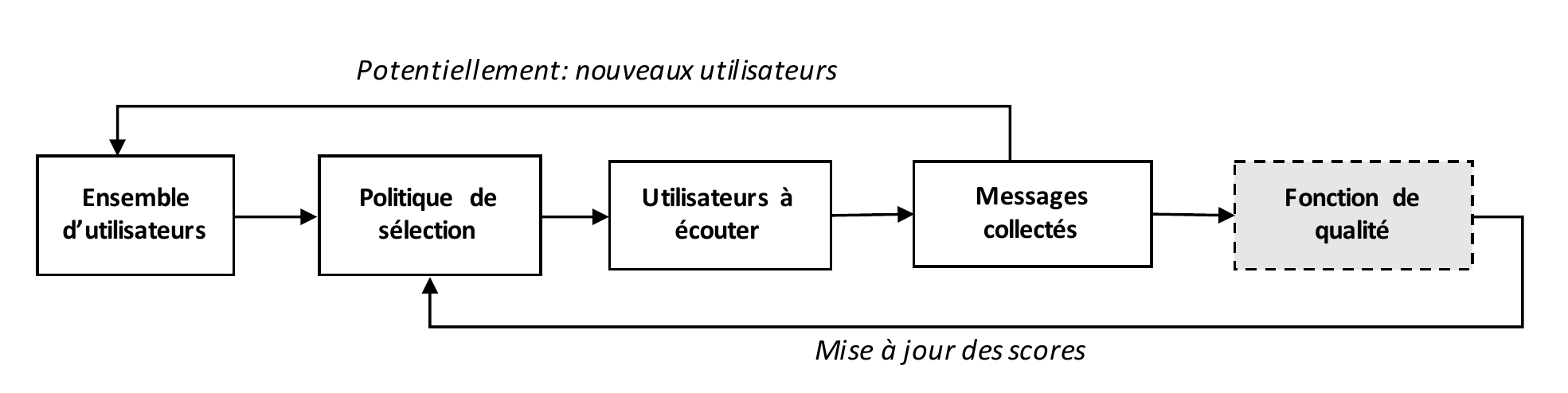}
      \caption[Processus général de la capture de données]{Processus général de la capture de données.}   
      \label{fig:process}
  \end{figureth}

    \section{Un problème de bandit}


Formellement, en notant $\mathcal{K}$ l'ensemble des $K$ utilisateurs du réseau, la tâche de collecte présentée ci-dessus revient à sélectionner à chaque itération $t \in\left\{1,...,T\right\}$ un sous-ensemble noté ${\cal K}_t$ de $k$ profils à suivre parmi l'ensemble des utilisateurs (${\cal K}_t \subset {\cal K}$), selon leur propension à poster des tweets pertinents par rapport au besoin exprimé. En notant $\omega_{i,t}$ le contenu produit par l'utilisateur $i$ pendant la fenêtre de temps $t$ et $r_{i,t}$ la note associée, le but est de sélectionner le sous-ensemble d’utilisateurs maximisant la somme des scores de pertinence tout au long du processus de collecte. Évidemment, les scores de pertinence $r_{i,t}$ sont seulement observés pour les utilisateurs suivis à l'itération $t$ (c.-à-d. pour les $i \in {{\cal K}_t}$). On peut donc voir la tâche de collecte comme un problème d'optimisation en ligne:

\begin{equation}
\max\limits_{({\cal K}_t)_{t=1..T}} \quad \sum\limits_{t=1}^{T} \sum\limits_{i \in {\cal K}_t}  r_{i,t}
\label{sum}
\end{equation}   

Les scores de pertinence considérés dépendent d'un besoin d'information, défini par l’utilisateur du système, pouvant prendre des formes variées. Par exemple, l'utilisateur pourrait souhaiter suivre des profils actifs sur des thématiques précises, ou bien des profils influents (au sens où leurs messages sont souvent repris par d'autres utilisateurs). Ces scores peuvent soit être assignés manuellement, relativement à une évaluation humaine des contenus, ou bien automatiquement, par exemple à l'aide d'un classifieur, comme nous le verrons dans la section suivante.

Ainsi, en considérant l'écoute de l'utilisateur $i$ dans la fenêtre de temps $t$ comme une action, il apparaît que notre tâche entre dans le formalisme du bandit avec sélection multiple présenté dans le chapitre \ref{EtatArt}, dans laquelle on cherche à maximiser la somme des récompenses récoltées. D'un point de vue pratique, la maximisation de la fonction définie dans la formule \ref{sum} est contrainte par différents facteurs  liés aux API de \textit{streaming} que nous utilisons pour la collecte des données, en particulier celle de l'API \textit{Follow} \textit{streaming}, qui permet d'obtenir en temps réel les contenus produits par 5000 utilisateurs définis. Nous utilisons cette API pour capturer les données produites par les utilisateurs sélectionnés par notre système. Plusieurs possibilités s'offrent à nous pour alimenter l'ensemble des utilisateurs potentiellement écoutable au fur et à mesure. La première option consiste à utiliser les comptes mentionnés par un utilisateur étant écouté. En effet lorsqu'un utilisateur interagit avec un autre (\textit{Retweet} ou \textit{Reply}  par exemple), il est possible de récolter les noms des personnes en interactions et ainsi d'alimenter notre liste. Par exemple si l'utilisateur est écouté, et que ce dernier répond à un autre utilisateur non connu, alors ce dernier sera ajouté à l'ensemble des profils potentiellement écoutables à la prochaine itération. Notons que ceci nécessite la définition d'un ensemble d'utilisateurs initial afin que le processus puisse démarrer. La seconde possibilité consiste à utiliser l'API \textit{Random streaming} permettant de récolter 1$\%$ des contenus produits sur le réseau social, et par conséquent les profils associés.

Suivant les hypothèses faites quant aux distributions de récompenses, on se situera tantôt dans un modèle de bandit stochastique tantôt dans un modèle de bandit contextuel. Dans la suite du manuscrit, on étudie différentes modélisations de ce problème. Dans un premier temps, au chapitre \ref{ModeleStationnaire}, on s'intéresse à un modèle de récompense stationnaire. Nous utiliserons des modèles contextuels, où les contextes seront considérés comme constants dans le chapitre \ref{ModeleContextuelFixe} et variables dans le chapitre \ref{ModeleContextuelVariable}. Nous verrons que dans ces deux derniers cas, nous ne pourrons pas utiliser des algorithmes existants à cause des contraintes présentes sur l'observation des contextes dues aux restrictions des API. Finalement dans le chapitre \ref{ModeleRelationnel} nous modéliserons les relations pouvant exister entre les utilisateurs d'un pas de temps à l'autre.

\begin{note}
Contrairement à la majorité des tâches classiques en recherche information, nous ne nous préoccupons pas ici de la précision des informations collectées. L'objectif est de maximiser le volume de messages pertinents collectés, un filtrage pouvant être éventuellement appliqué \textit{a posteriori}. Cet aspect diffère des tâches habituelles en recherche d'information où l'on s'intéresse à un compromis précision-rappel.
\end{note}

    \section{Modèles de récompenses utilisés}
        \label{chap0ModelesRec}
                Sur Twitter, il existe différentes façons pour un utilisateur d'apparaître sur le réseau. D'une part de façon active, lorsqu'un utilisateur poste un \textit{Tweet}, c'est-à-dire publie un nouveau message directement depuis son compte. Un utilisateur peut aussi effectuer des reprises de contenu produit par d'autres comptes, appelé  \textit{Retweet}. Il est également possible de répondre à des contenus publiés par d'autres profils,  appelés  \textit{Reply}. D'autre part, un utilisateur peut apparaître de façon passive, lorsque d'autres utilisateurs effectuent des \textit{Retweet} ou des \textit{Reply} sur des messages qu'il a publiés. Lorsqu'un utilisateur est écouté pendant une période donnée, toutes ses apparitions sur le réseau - actives ou passives - sont captées. Ceci est résumé dans le tableau \ref{tab:interactions}, dans lequel on introduit les notations pour les événements associés à un utilisateur $i$ pendant l'itération $t$. Dans les cinq cas, $k\in \left\{0,1,2,3,4\right\}$, $\omega_{i,t}^{k}$ désigne le contenu associé à l'événement $k$ pendant la période $t$ pour l'utilisateur $i$.

      \begin{tableth}
        \setlength\extrarowheight{5pt}
        \begin{tabular}{|c|c|c||c|c|}
                  \hline
               \multicolumn{3}{|c||}{\textbf{Actif}} & \multicolumn{2}{c|}{\textbf{Passif}} \\
                  \hline 
                   Tweet & Retweet & Reply &Retweet & Reply \\
                  \hline
                  $\omega_{i,t}^{0}$ & $\omega_{i,t}^{1}$ & $\omega_{i,t}^{2}$ &$\omega_{i,t}^{3}$ & $\omega_{i,t}^{4}$ \\    
          \hline
          \end{tabular}
    \caption[Différents types d'interactions d'un utilisateur avec Twitter.]{Différents types d'interactions d'un utilisateur avec Twitter.}
    \label{tab:interactions}
  \end{tableth}
    
        Bien que d'autres types de fonction récompense auraient pu être envisagés, nous décidons dans nos expérimentations de nous intéresser à des modèles de collecte visant à récompenser les messages ayant un fort impact sur une thématique donnée . Trois thématiques sont considérées (correspondant à trois différents modèles de récompense) : \textit{politique}, \textit{religion} et \textit{science}. L'appartenance à chacune de ces trois thématiques se fait par l'application d'un classifieur entraîné sur le corpus \textit{20 Newsgroups}\footnote{\url{http://qwone.com/jason/20Newsgroups/}}, constitué de 18846 documents et d'un vocabulaire de 61188 termes. Le classifieur utilisé est un SVM à noyau linéaire\footnote{Avec un coefficient de pénalité fixé à 1.}, dont les bonnes performances pour la classification de textes ont, entre autres, été évaluées dans \cite{SVMtextClassif}. Les caractéristiques utilisées pour chaque document correspondent au poids TF-IDF de chaque terme le composant. On rappelle qu'étant donnée un document $d_j$, la fréquence d'un  terme $t_i$ (notée $TF_{t_i,d_j}$) est égale au nombre d'occurrences de $t_i$ dans le document $d_j$, et permet de mesurer l'importance d'un terme au sein d'un document. De plus, étant donné un corpus $\mathcal{D}={d_{1},...,d_{D}}$ de $D$ documents, la fréquence inverse de document  d'un terme $t_{i}$ (notée $IDF_{t_i}$) est une mesure de l'importance du terme dans l'ensemble du corpus: $IDF_{t_i}=\log\left(D/|\left\{d_{j} \text{ tel que } t_{i}\in d_{j}\right\}|\right)$, avec $|\left\{d_{j} \text{ tel que } t_{i}\in d_{j}\right\}|$ le nombre de documents où le terme $t_{i}$ est présent. Dans le schéma TF-IDF, cette mesure vise à donner un poids plus important aux termes les moins fréquents, considérés comme plus discriminants. Finalement, le score TF-IDF d'un terme $t_{i}$ dans un document $d_{j}$ est égal au produit des deux score: $TF_{t_{i},d_{j}}IDF_{t_{i}}$.

        Le classifieur utilisé a été entraîné pour chaque classe "contre les autres" (\textit{one-vs.rest}) en considérant l'ensemble des labels présents dans le corpus (au nombre de 20). Ainsi, certains \textit{tweets} peuvent n'appartenir à aucune des classes \textit{politique}, \textit{religion} ou \textit{science}. En revanche, un \textit{tweet} ne peut appartenir qu'à une seule classe, celle associée à la valeur de la fonction de décision la plus élevée.

        Pour un contenu $\omega_{i,t}^{k}$, avec $k \in \left\{0,1,2,3,4\right\}$ suivant le type de message considéré, (voir tableau \ref{tab:interactions}), on note $g(\omega_{i,t}^{k})$ la valeur associée au résultat de ce classifieur pour une thématique donnée:
        
        \begin{equation*}
      g(\omega_{i,t}^{k}) = \begin{cases}
        1 \text{ si le contenu appartient à la thématique}\\
        0 \text{ sinon}
          \end{cases}
    \end{equation*}
        
        Conscient du fait que les \textit{tweets} bruts peuvent être bruités par la présence d'URL, ou de mention d'autres utilisateurs, nous prenons le soin de nettoyer chaque message avant de lui appliquer le classifieur. 
        
        De plus, pour un utilisateur, on peut collecter plusieurs contenus d'un même type durant une fenêtre d'écoute. Ainsi pour un utilisateur $i$ à l'instant $t$, on note $p_{i,t}^{k}$ le nombre de contenus collectés de type $k$. Ainsi d'une façon générique, la récompense d'un utilisateur $i$ au temps $t$, pour une thématique donnée, est calculée comme une fonction de la somme pondérée des valeurs de la fonction $g$ sur les contenus collectés:

                
        
        \begin{equation}
        \label{defiRwd}
        r_{i,t}=\tanh\left(\sum\limits_{k=0}^{4} \alpha_{k}\sum\limits_{p=0}^{p_{i,t}^{k}} g(\omega_{i,t}^{k,p})\right)
        \end{equation}
        
                Où $\alpha_{k}$ correspond au poids que l'on souhaite donner au type de contenu $k$ et $\omega_{i,t}^{k,p}$ représente le $p^{\text{ième}}$ contenu de type $k$ de l'utilisateur $i$ à l'itération $t$, et $\tanh$ désigne la fonction tangente hyperbolique. La fonction tangente hyperbolique permet de ramener les récompenses entre 0 et 1 (car son argument est ici toujours positif), et les différents coefficients permettent de jouer sur la pente plus ou moins abrupte de cette fonction. Par exemple, une valeur de $\alpha_{0}$ élevée aura tendance à privilégier les utilisateurs étant à la source de nombreux messages originaux tandis qu'un $\alpha_{3}$ élevé favorisera les utilisateurs étant beaucoup repris par les autres.

        Dans la pratique, nous avons décidé de nous concentrer sur la recherche d'utilisateur étant soit à l'origine de contenu, soit faisant l'objet de nombreuses réactions vis-à-vis des autres utilisateurs, c'est à dire étant "Retweeté" ou suscitant des "Replies": $\alpha_{0}=\alpha_{3}=\alpha_{4}=0.1$ et $\alpha_{1}=\alpha_{2}=0$. Dans la suite du manuscrit, nous désignons ce modèle de récompense par \textit{Topic+Influence}. Encore une fois, ce type de schéma de récompense a uniquement été imaginé à des fins expérimentales, pour démontrer les performances des approches proposées. Bien d'autres schémas peuvent être employés dans le même cadre applicatif, en utilisant les modèles présentés dans les chapitres suivants.

        \section{Représentations des messages} 
        \label{repMessage}
        
        Certains modèles proposés dans cette thèse nécessitent l'utilisation d'une représentation vectorielle des différents messages (voir chapitres  \ref{ModeleContextuelFixe} et \ref{ModeleContextuelVariable}). Pour cela, une première possibilité consiste à utiliser une représentation de type sac de mots. Dans ce cas, étant donné un dictionnaire de taille $m$, un message correspond à un vecteur dans $\mathbb{R}^{m}$, dont chaque composante est égale au poid TF-IDF des termes qui le composent. Dans l'application que nous proposons, nous appliquons préalablement un algorithme de  racinisation (\textit{Porter stemmer}) aux différents messages. Notons qu'en  linguistique, la racine d’un mot correspond à la partie du mot restante une fois que l’on a supprimé ses préfixes et ses suffixes. Ceci permet d'unifier les variantes d'un terme sous une même forme syntaxique. Nous construisons le dictionnaire final en sélectionnant les 2000 ($m=2000$) termes ayant les scores TF-IDF les plus élevés.

        Comme nous le verrons dans le manuscrit, certains algorithmes nécessiteront de manipuler (et en particulier d'inverser) des matrices de taille égale à la dimension des vecteurs représentant les documents, soit $2000 \times 2000$. En pratique cela peut s'avérer très complexe, et ralentir fortement l'exécution des différents algorithmes. En vue de réduire la dimension de l'espace des représentations des messages et d'obtenir des descripteurs plus concis, nous proposons d'utiliser une transformation LDA (Latent Dirichlet Allocation) \cite{LDA}. Il s'agit d'un modèle génératif, souvent utilisé en recherche d'information, qui modélise chaque message comme un mélange probabiliste de $d$ sujets (ou concepts), où chaque sujet est caractérisé par une distribution sur les termes du dictionnaire. Cependant, le format des messages sur Twitter étant très court, il s'avère que l'apprentissage direct du modèle LDA sur des messages individuels n'est pas pertinent. Nous privilégions ainsi l'approche spécifiquement créée par \cite{LDATwitter2}, dans laquelle les messages d'un même utilisateur sont agrégés en un seul document lors de l'apprentissage du modèle. Le nombre $d$ de sujets est un paramètre du modèle que nous fixons à 30 dans les  expérimentations à venir. Ce modèle est entraîné sur un ensemble de \textit{tweets} collectés via l'API \textit{Random Streaming} pendant une durée de trois jours. La méthode employée, qui consiste à entraîner un modèle de projection fixe sur un corpus hors-ligne rentre dans le cadre du \textit{data-oblivious sketching} présenté au chapitre \ref{EtatArtRI}.
                
        \section{Jeux de données}
        \label{datasets}
        
        Tout au long de ce manuscrit, en plus des expérimentations en ligne pour tester les approches en conditions réelles, nous effectuerons des expérimentations sur trois ensembles de données enregistrées, ce qui permet de simuler les processus de collecte plusieurs fois et de comparer divers algorithmes. Concrètement, chaque jeu de données est constitué de l'activité de 5000 utilisateurs, enregistrée pendant un période définie. Ceci permet ensuite de jouer divers scénarios de collecte dans un mini réseau social de $K=5000$ personnes.
        
        Les jeux de données collectés sont les suivants:

\begin{itemize} 
\item Le premier jeu de données, appelé \textit{USElections}, correspond à un ensemble de messages récoltés grâce à l'API Twitter en suivant 5000 comptes d'utilisateurs sur une période de 10 jours précédant les élections américaines de 2012. Nous avons choisi ces comptes en prenant les 5000 premiers à avoir employé les mots-clés "Obama", "Romney" ou le \textit{hashtag} "\#USElections". Il contient un total de 3 587 961 messages. 
\item Le second jeu de données, appelé \textit{OlympicGames}, fut collecté en août 2016 pendant une période de trois semaines sur la thématique des Jeux olympiques d'été de Rio. Les 5000 comptes écoutés furent sélectionnés en prenant les 5000 comptes à avoir le plus utilisé les  \textit{hashtags} "\#Rio2016", "\#Olympics", "\#Olympics2016" ou "\#Olympicgames" dans une période antérieure de trois jours avant le début des JO. Il contient un total de 15 010 322 messages. 
\item Le troisième jeu de données, appelé \textit{Brexit}, fut collecté en octobre 2016 sur la thématique du Brexit pendant une période d'une semaine. Les 5000 comptes écoutés furent sélectionnés en prenant les 5000 premiers à avoir employé le \textit{hashtag} "\#Brexit". Il contient un total de 2 118 235 messages. 
\end{itemize}

D'autres caractéristiques de ces jeux de données sont présentées dans le tableau \ref{tab:dataset}. Les différentes valeurs exposées dans ce tableau sont définies de la façon suivante:

\begin{itemize} 
\item \textbf{Tweet} ($k=0$): nombre de \textit{Tweets} dont l'origine vient de l'un des 5000 comptes écoutés lors de la collecte. Il s'agit de messages originaux postés directement par des utilisateurs écoutés;
\item \textbf{Retweet} ($k=1$): nombre de \textit{Retweets} dont l'origine vient de l'un des 5000 comptes écoutés lors de la collecte. Il s'agit de reprises par les utilisateurs écoutés de messages  d'autres utilisateurs (écoutés ou non);
\item \textbf{Reply} ($k=2$): nombre de \textit{Replies} dont l'origine vient de l'un des 5000 comptes écoutés lors de la collecte. Il s'agit de réponses par les utilisateurs écoutés à des messages d'autres utilisateurs (écoutés ou non);
\item \textbf{Retweet} ($k=3$): nombre de \textit{Retweets} dont l'origine ne vient pas de l'un des 5000 comptes écoutés lors de la collecte. Il s'agit de reprises  par des utilisateurs non écoutés de messages postés par des utilisateurs écoutés;
\item \textbf{Reply} ($k=4$): nombre de \textit{Replies} dont l'origine ne vient pas de l'un des 5000 comptes écoutés lors de la collecte. Il s'agit de réponses d'utilisateurs non écoutés à des utilisateurs écoutés;
\end{itemize}

  \begin{tableth}
    \begin{tabular}{|c|c|c|c|c|c|}
          \hline
      \textbf{Jeu de données}  & \textbf{Tweet} ($k=0$) & \textbf{Retweet} ($k=1$) & \textbf{Reply} ($k=2$) & \textbf{Retweet}: ($k=3$) & \textbf{Reply} ($k=4$) \\
      \hline 
      \textit{USElections}   & 1 057 622 & 770 062 & 295 128 & 1 100 255 & 364 894\\
            \hline
            \textit{OlympicGames}  & 3 241 413 & 3 756 089 & 341 734 & 6 807 064 & 864 022 \\
            \hline 
            \textit{Brexit}  & 406 776 & 1 068 578 & 178 234 & 324 150 & 140 497 \\
            \hline 
    \end{tabular}
    \caption[Caractéristiques des jeux de données.]{Caractéristiques des jeux de données.}
    \label{tab:dataset}
  \end{tableth}

Ainsi, chaque jeu de données possède des caractéristiques qui lui sont propres, avec une part plus ou moins importante des différents contenus. On note que d'une façon générale, les réponses constituent un événement plus rare que les reprises. 

Enfin, dans toutes les expérimentations hors-ligne sur ces jeux de données  que nous effectuerons au cours de la thèse, le processus d'alimentation de l'ensemble d'utilisateurs  doit être défini. Etant donné que nous utilisons un ensemble d'utilisateurs prédéfini  il n'est pas possible d'utiliser la méthode évoquée plus haut qui consiste à partir d'un petit ensemble de source, puis à alimenter l'ensemble des utilisateurs grâce aux divers interactions (réponses ou reprise de messages) entre ces sources et des utilisateurs non connus. En effet, les chances pour que les 5000 utilisateurs écoutés pendant la collecte des divers jeux de données aient interagit sont faibles. Nous choisissons donc, pour les expérimentations hors-ligne, d'ajouter un utilisateur dès lors qu'il apparaît actif pour la première fois. Bien sûr, un système permettant cela n'est pas envisageable dans une expérimentation en ligne en conditions réelles, ce qui nous conduira à choisir un autre processus d'alimentation de la base de comptes lors des expérimentations à venir.

\section{Conclusion}

Le problème de la collecte de données en temps réel est désormais formalisé comme un problème de bandit. Les corpus d'étude et les modèles de récompenses sont eux aussi définis. Les éléments donnés dans ce chapitre serviront de base à l'ensemble des approches considérées dans la suite du manuscrit.


\chapter{Modèle stationnaire stochastique}
  \minitoc
  \newpage
    
\label{ModeleStationnaire}

Dans ce chapitre, nous proposons de modéliser la récompense de chaque utilisateur par une distribution stationnaire afin de se placer dans le cadre du bandit stochastique. L'objectif de cette première approche est de montrer l'intérêt des algorithmes de bandits pour traiter notre tâche de collecte d'information en temps réel dans un média social. Dans cette optique, nous verrons que les algorithmes existants peuvent être adaptés à notre cas, et nous proposerons également un nouvel algorithme, dont nous testerons les performances dans une partie expérimentale.
    
\section{Modèle et algorithmes}
\label{chap1Modele}

      Rappelons que l'on considère un ensemble de $K$ utilisateurs noté $\mathcal{K}$. A chaque itération $t$ du processus de collecte, l'agent décisionnel, autrement dit la politique de sélection, doit choisir un sous-ensemble noté $\mathcal{K}_{t} \subset \mathcal{K}$ de $k$ utilisateurs à écouter. La récompense $r_{i,t}$ associée à un utilisateur $i$ suivi pendant la fenêtre d'écoute $t$ est immédiatement évaluée selon l'un des modèles proposés dans la partie \ref{chap0ModelesRec}. Notre but est de collecter un maximum d'information pertinente - relativement à notre modèle de récompense - tout au long du processus composé de $T$ itérations, ce qui correspond à la maximisation de la somme des récompenses récoltées au cours du temps. Nous supposons que chaque utilisateur $i$ est associé à une distribution de récompense stationnaire $\nu_{i}$ de moyenne $\mu_{i}$. Ainsi, à chaque temps $t$, la récompense émise par un profil $i$ correspond à un échantillon de la loi $\nu_{i}$, c'est à dire $r_{i,t} \sim \nu_{i}$. Afin de rester dans le cadre des bandits stationnaires, on suppose également que tous les échantillons sont indépendants entre eux.




Les hypothèses de stationnarité utilisées dans ce chapitre nous positionnent dans le cadre du bandit avec sélections multiples décrit dans l'état de l'art (voir section \ref{banditMultSel}), pour lequel l'algorithme \texttt{CUCB} a été proposé dans \cite{banditCombinatoire}. Dans le cas qui nous intéresse, c'est-à-dire lorsque la récompense d'un ensemble de bras est égale à la somme des récompenses individuelles des bras qui le composent, l'algorithme \texttt{CUCB} correspond à une extension de l'algorithme  \texttt{UCB} \cite{banditUCB}. Pour effectuer la sélection de $k$ actions à chaque itération, cet algorithme associe à chaque action $i$ et à chaque instant $t$ un score noté $s_{i,t}$ correspondant à une borne supérieure de l'intervalle de confiance de la récompense associée. Cette politique est dite optimiste, car elle suppose que pour chaque utilisateur, la récompense associée est la meilleure de ce qu'elle pourrait être selon l'intervalle de confiance considéré.

Selon l'algorithme \texttt{CUCB}, et pour le cas qui nous intéresse où toutes les actions ne sont pas connues \textit{a priori}, le score de chaque action connue $i$ à l'instant $t$ s'écrit:

\begin{equation}
\label{scoreUCBGenerique}
            s_{i,t}=\begin{cases}
                        \hat{\mu}_{i,t-1} +B_{i,t} \text{ si } N_{i,t-1}>0\\
                       +\infty \text{ si } N_{i,t-1}=0
                      \end{cases}
\end{equation}

 Où $N_{i,t-1}=\sum\limits_{s=1}^{t-1}\mathbb{1}_{\left\{i\in \mathcal{K}_{s}\right\}}$ est égal au nombre de de fois où l'action $i$ a été sélectionnée jusqu'au temps $t-1$, $\hat{\mu}_{i,t-t}=\dfrac{1}{N_{i,t-1}}\sum\limits_{s=1}^{t-1}\mathbb{1}_{\left\{i\in \mathcal{K}_{s}\right\}}r_{i,s}$ correspond à la moyenne empirique de l'action $i$ et $B_{i,t}=\sqrt{\dfrac{2\log(t)}{N_{i,t-1}}}$ est un terme exploratoire. Le score $s_{i,t}$ représente bien un compromis entre exploitation et exploration puisqu’il s’agit de la somme d’un premier terme estimant la qualité d’une action $i$ et d’un second terme décroissant avec le nombre de fois où l’action i est choisie. De plus, étant donné que l'on ne connaît pas tous les utilisateurs à l'instant initial, le score des utilisateurs non écoutés au moins une fois (c.-à-d. $N_{i,t-1}=0$) est initialisé à $+\infty$ afin de forcer le système à les sélectionner.  Avec ceci, nous pouvons donc   directement appliquer l'algorithme \texttt{CUCB} à notre cas. 
Le processus de collecte générique associé est détaillé dans l'algorithme  \ref{AlgoCollecteStat}, dans lequel on associe un score $s_{i,t}$ à chaque utilisateur pour effectuer la sélection.

                      \begin{algorithm}[h!]
                        \KwIn{$\mathcal{K}_{init}$}
            \For{$t=1..T$}{
                           \For{$i \in \mathcal{K} $}{
                              Calculer $s_{i,t}$ selon l'équation \ref{scoreUCBGenerique}
                             }
                              Ordonner les utilisateurs par ordre décroissant selon $s_{i,t}$\;
                              Sélectionner les $k$ premiers pour fixer $\mathcal{K}_{t}$\;
                              Ecouter en parallèle tous les utilisateurs $i\in \mathcal{K}_{t}$ et observer $\omega_{i,t}$ \;
                        \For{$i \in \mathcal{K}_{t} $}{
                                  
                                    Recevoir la récompense associée $r_{i,t}$ \;
                                    Alimenter $\mathcal{K}$ avec les nouveaux utilisateurs $j$, $j\notin\mathcal{K} $
                        }
                        }
            \caption{Algorithme de collecte - hypothèse stationnaire}
          \label{AlgoCollecteStat}
          \end{algorithm}

 \begin{note}
Etant donné que tous les utilisateurs ne sont pas connus à l'initialisation, ce problème problème entre dans le cadre du  \textit{sleeping} bandit \cite{sleepingBandit}, dans lequel l'ensemble des actions disponibles à chaque itération change d'une itération à l'autre. Il est alors possible d'appliquer les algorithmes de bandit stationnaire classiques à la différence près qu'au lieu de sélectionner une fois chaque action en début de processus pour initialiser les moyennes empiriques, chaque action est jouée une fois lorsqu'elle apparaît pour la première fois dans l'ensemble des actions disponibles.
 \end{note}

Pour la tâche de collecte d'information définie, la récompense de chaque utilisateur est basée sur la pertinence du contenu produit pendant une période finie. Cependant, de fortes variations peuvent être observées sur la fréquence de publication des utilisateurs écoutés. Typiquement, la plupart du temps, les utilisateurs ne produisent aucun contenu. Avec la politique \texttt{CUCB} présentée précédemment (c.-à-d., avec $B_{i,t}=\sqrt{\frac{2\ln(t)}{N_{i,t-1}}}$), le score $s_{i,t}$ peut tendre à pénaliser les utilisateurs produisant peu de contenus les premières fois qu'ils sont écoutés. De plus, aucune différence ne peut être faite entre un utilisateur produisant beaucoup de contenus de qualité moyenne et un utilisateur produisant peu de contenus, mais d'une grande qualité. En vue de prendre en compte cette forte variabilité dans le comportement des utilisateurs, nous proposons un nouvel algorithme de bandit à sélections multiples, que nous appelons \texttt{CUCBV}, et qui considère la variance des récompenses récoltées. L'algorithme \texttt{CUCBV} est une extension de l'algorithme \texttt{UCBV} proposé dans \cite{banditUCBV}. 
Cet algorithme associe un score $s_{i,t}$ à chaque action $i$ à l'instant $t$ de la forme proposée dans l'équation \ref{scoreUCBGenerique}, mais utilise la variance dans le terme d'exploration, ce qui semble mieux adapté à notre tâche. 
Le terme d'exploration $B_{i,t}$ de l'algorithme \texttt{CUCBV} est défini par:

 \begin{equation}
 \label{scoreCUCBV}
 B_{i,t}= \sqrt{\dfrac{2aV_{i,t-1}\log(t)}{N_{i,t-1}}}+3c\dfrac{\log(t)}{N_{i,t-1}}
 \end{equation}

Où $c$ et $a$ sont des paramètres de l'algorithme permettant de contrôler l'exploration, et $V_{i,t-1}=\dfrac{1}{N_{i,t-1}}\sum\limits_{s=1}^{t-1}(\mathbb{1}_{\left\{i_{t}=i\right\}}r_{i,s}-\hat{\mu}_{i,t-1})^{2}$ est la variance empirique de l'utilisateur $i$. 

Avec un tel facteur d’exploration, la politique tend à plus explorer les utilisateurs ayant une grande variance, puisque plus d’informations sont nécessaires pour avoir une bonne estimation de leur qualité. Dans la section suivante, on discute des garanties de convergence théoriques de l'algorithme \texttt{CUCBV} que nous proposons.

  \section{Etude du regret}

    Dans ce qui suit nous supposons, sans perte de généralité que les actions sont ordonnées de la façon suivante: $\forall i, \mu_{i} > \mu_{i+1}$. De plus, on utilise les notations supplémentaires suivantes:

\begin{itemize}
\item $\sigma_{i}$ l'écart type de la récompense associée à l'action $i$;
\item $\mathcal{K}^{\star}$ est l'ensemble des $k$ actions ayant la plus forte espérance: $\mathcal{K}^{\star}=\left\{1,..,k\right\}$; 
\item $\overline{\mu}^{\star}$ est la moyenne des espérances des actions dans $\mathcal{K}^{\star}$, $\overline{\mu}^{\star}=\dfrac{1}{k}\sum\limits_{i=1}^{k}\mu_{i}$
\item $\Delta_i$ est défini comme la différence entre $\overline{\mu}^{\star}$, la moyenne des espérances des actions dans $\mathcal{K}^{\star}$, et $\mu_{i}$: $\Delta_i = \overline{\mu}^{\star}-\mu_{i}$; 
\item $\underline{\mu}^{\star}$ est la plus faible espérance dans $\mathcal{K}^{\star}$: $\underline{\mu}^{\star}=\min\limits_{i\in\mathcal{K}^{\star} }\mu_{i}=\mu_{k}$ 
\item $\delta_{i}$ correspond à la différence entre $\underline{\mu}^{\star}$, la plus faible espérance dans $\mathcal{K}^{\star}$, et $\mu_i$: $\delta_{i}=\underline{\mu}^{\star} - \mu_i$;
\end{itemize}

          La performance d'un algorithme de bandit est habituellement mesurée par la notion de regret, qui correspond à la perte de récompense qu’un agent est susceptible de subir en choisissant une action $i$ au lieu d’une action optimale (au sens de la moyenne). De plus, il est usuel d'étudier théoriquement des garanties sur le pseudo-regret, qui correspond, conformément à la définition  \ref{regretCumultoPseudoMP} (voir section \ref{banditMultSel}), à:
            
        \begin{equation}
          \hat{R}_{T}= \mathbb{E}\left[\sum\limits_{t=1}^{T}  \left( \sum\limits_{i\in \mathcal{K}^{\star}} \mu_{i} -  \sum\limits_{i\in \mathcal{K}_{t}} \mu_{i}\right) \right]
        \end{equation}


\begin{prop}
\label{defAlternativeRegComb}
Pour un algorithme de bandit stationnaire avec sélection multiple, le pseudo-regret peut se réécrire de la façon suivante:
        \begin{equation}
          \hat{R}_{T}=\sum\limits_{i=1}^{K} \mathbb{E}[N_{i,T}]\Delta_{i}
        \end{equation}
\end{prop}

\begin{preuve}
Disponible en annexe \ref{PreuveRegCUCBVFirstPart}.
\end{preuve}

Lorsque $k=1$, cette écriture revient à la définition du regret du bandit stochastique, car $\Delta_i=\mu^{\star}-\mu_{i}$. Ainsi seuls les bras sous-optimaux ont une contribution non nulle (et forcément positive) dans le regret. Dans le cas où $k>1$, la présence de plusieurs bras optimaux complexifie la façon dont le regret se comporte, car toutes les actions, qu'elles soient optimales ou non, jouent un rôle dans le regret. Nous remarquons que, bien que par définition le regret total ne puisse pas être négatif, certaines contributions peuvent ponctuellement être négatives. En effet, pour certains bras optimaux $i\in \mathcal{K}^{\star}$, on aura $\Delta_i <0$ selon qu'on est au-dessus ou en dessous de la moyenne. Afin de nous affranchir de cette difficulté combinatoire, nous proposons ci-dessous une première majoration du regret, ne faisant intervenir que les contributions des actions sous-optimales.

\begin{prop}
\label{defAlternativeRegCombBis}
Pour un algorithme de bandit stationnaire avec sélection multiple, le pseudo-regret est majoré par:
        \begin{equation}
          \hat{R}_{T}\leq \sum\limits_{i \notin \mathcal{K}^{\star}}^{K} \mathbb{E}[N_{i,T}]\Delta_{i}
        \end{equation}
\end{prop}

\begin{preuve}
En remarquant que le regret peut s'écrire \\
$\hat{R}_{T}=\sum\limits_{i=1}^{k} \mathbb{E}[N_{i,T}] \Delta_i + \sum\limits_{i=k+1}^{K} \mathbb{E}[N_{i,T}] \Delta_i$, que pour toute action $\mathbb{E}[N_{i,T}]\leq T$ et que $\sum_{i=1}^{k}\Delta_i = 0$ on obtient le résultat proposé.
\end{preuve}

\begin{theo}
\label{theoRegCUCBV}
En considérant l'ensemble complet des actions $\mathcal{K}$ connu \textit{a priori}, en choisissant $a=c=1$ (paramètres de réglages de l'algorithme), le pseudo-regret de l'algorithme \texttt{CUCBV} est majoré par:
        \begin{equation}
          \hat{R}_{T}\leq \ln(T)\sum\limits_{i \notin \mathcal{K}^{\star}} \left( C+8\left(\dfrac{\sigma_{i}^{2}}{\delta_{i}^{2}} + \dfrac{2}{\delta_{i}}\right)\right)\Delta_i + D
        \end{equation}
Où $C$ et $D$ sont des constantes, $\Delta_i$ est la différence entre $\overline{\mu}^{\star}$, la moyenne des moyennes des récompenses dans $\mathcal{K}^{\star}$, et $\mu_{i}$, la moyenne de $i$: $\Delta_i = \overline{\mu}^{\star}-\mu_{i}$, $\delta_{i}$ est la différence entre $\underline{\mu}^{\star}$, la moyenne associée à la moins bonne action de $\mathcal{K}^{\star}$, et $\mu_i$: $\delta_i = \underline{\mu}^{\star}-\mu_{i}$, $\sigma_{i}^{2}$ est la variance de l'action $i$.
\end{theo}

\begin{preuve}
Disponible en annexe \ref{PreuveRegCUCBVSecondPart}.
\end{preuve}

A une constante additive près, ce résultat nous permet de garantir une convergence logarithmique. 
Bien que ce résultat ne soit valide que pour le cas où l’ensemble complet des utilisateurs est connu \textit{a priori}, il nous permet d’affirmer que lorsqu’un utilisateur optimal (avec une moyenne parmi les $k$ meilleures) entre dans l’ensemble $K$, le processus converge de façon logarithmique vers une politique le reconnaissant comme tel. En revanche, cette borne n'est pas optimale au sens ou le facteur multiplicatif devant le logarithme n'est pas égal à celui de la borne inférieure du regret de tout algorithme de bandit à sélections multiples (la formule de cette borne inférieure est détaillée dans le théorème \ref{infRegStoMP} de l'état de l'art). Ceci est dû au fait, d'une part que l'algorithme \texttt{UCBV} avec une seule action sélectionnée à chaque pas de temps n'est déjà pas optimal, et d'autre part à la majoration effectuée dans la proposition \ref{defAlternativeRegCombBis}. 

Rappelons finalement que nous considérons ici le cas des distributions de récompenses stationnaires. Bien que cette hypothèse ne soit pas toujours vérifiée dans notre contexte de réseaux sociaux, cette preuve de convergence nous indique que si des sources sont bonnes pendant une période de temps suffisamment longue, notre algorithme est en mesure de les détecter. Dans la section suivante, nous proposons diverses expérimentations de cet algorithme pour notre tâche de collecte de données.

\section{Expérimentations}

Nous expérimentons les algorithmes de bandits étudiés précédemment dans le cadre de la collecte orientée, d'une part sur des données hors ligne et d'autre part dans une expérimentation en ligne.

    \subsection{Hors ligne}

      \subsubsection{Protocole}
      \label{protocole}

Dans le but de tester différents algorithmes, nous réalisons dans un premier temps des expérimentations sur les bases de données collectées  \textit{USElections}, \textit{OlympicGames} et \textit{Brexit}, dont les caractéristiques ont été détaillées dans la section \ref{datasets}. On rappelle que ces bases contiennent chacune l'activité de $K=5000$ utilisateurs sur une certaine période, ce qui nous permet de simuler un réseau social. Afin de se placer dans le cadre de notre tâche, chaque période est divisée artificiellement en $T$ itérations.  Dans les trois cas nous avons fixé la taille d'une itération à 100 secondes. Ainsi la base \textit{USElections} comporte $T=8070$ itérations, la base \textit{OlympicGames} en comporte $T=16970$ et la base \textit{Brexit} en comporte $T=4334$. La valeur de $k$, le nombre d'utilisateurs que l'on est autorisé à écouter à chaque itération est fixé à $k=100$ \footnote{Nous testerons d'autres valeurs par la suite.}. Finalement, on utilise le modèle de récompense \textit{Topic+Influence} présenté dans le  chapitre \ref{FormalisationTache} (section \ref{chap0ModelesRec}).

Le but de ces expérimentations est double. Premièrement, il s'agit de vérifier que les algorithmes de bandit permettent effectivement de collecter la donnée de façon intelligente en s'orientant vers des utilisateurs pertinents. Dans cette optique, nous nous comparerons à une politique aléatoire, nommée \texttt{Random}, sélectionnant les utilisateurs uniformément à chaque instant. De plus, il s'agit de montrer que l'algorithme \texttt{CUCBV} que nous avons proposé dans la partie précédente est plus adapté que ses concurrents dans le cas spécifique de la collecte de données. Dans cette optique, on se comparera également à l'algorithme \texttt{CUCB}, mais aussi aux algorithmes \texttt{UCB}-$\delta$ et \texttt{MOSS} décrits dans le chapitre \ref{EtatArtBandit} et proposés respectivement dans \cite{OFUL} et \cite{banditMOSS}. On adapte ces derniers à notre cas (sélection multiple) en considérant les scores des politiques  respectives (voir algorithme \ref{MOSS} pour  \texttt{MOSS} et algorithme \ref{UCBDelta} pour \texttt{UCB}-$\delta$) et en sélectionnant à chaque itération les $k$ utilisateurs ayant les scores les plus élevés. Notons que \texttt{UCB}-$\delta$ possède un paramètre $\delta$ permettant de régler l'exploration. Nous fixons ce paramètre à 0.05, qui est une valeur souvent utilisée dans la littérature. De plus, la politique \texttt{MOSS} nécessite la connaissance de l'horizon $T$ et du nombre total d'utilisateurs $K$ en paramètres. Dans la pratique, nous pouvons seulement fixer une borne supérieure pour $K$, le nombre total d'utilisateurs étant inconnu à l'avance. En outre, la présence de $T$ peut également poser des difficultés dans le cas où l'horizon n'est pas connu au préalable. Cependant, dans les expérimentations hors ligne que nous effectuons ici, il est possible de fixer des valeurs optimales pour \texttt{MOSS}, ce qui n'est pas possible en pratique et favorise cet algorithme pour ces expérimentations. 

    \subsubsection{Résultats}

         Les figures \ref{fig:XPStatUSRT}, \ref{fig:XPStatOGRT} et \ref{fig:XPStatBrexitRT} contiennent l'évolution du gain cumulé en fonction du temps respectivement pour les bases \textit{USElections}, \textit{OlympicGames} et \textit{Brexit} avec le modèle de récompense \textit{Topic+Influence} pour les trois thématiques. On remarque que les trois stratégies proposées offrent de meilleurs résultats que la stratégie \texttt{Random}, ce qui confirme que l'application d'algorithmes de bandits pour la capture de données sur un réseau social est pertinente. On remarque dans un premier temps que d'une façon générale, les politiques \texttt{CUCB} et \texttt{UCB}-$\delta$ offrent des performances similaires. Selon les cas, l'algorithme \texttt{MOSS} est plus performant que ces deux derniers. Comme nous pouvons le voir, cela dépend de la récompense et du jeu de données utilisé. Remarquons qu'étant donné les valeurs de $K$ et $T$ associées à nos trois bases de données, l'algorithme \texttt{MOSS} possède un terme d'exploration très faible, ce qui le conduit à favoriser l'exploitation au détriment de l'exploration. En effet pour chaque action $i$ à l'instant $t$ ce dernier vaut $\sqrt{\max(\log(T/(KN_{i,t-1})),0)/N_{i,t-1}}$ qui devient nul dès que $N_{i,t-1}\geq T/K$. 
         
        Dans toutes les configurations, l'algorithme \texttt{CUCBV} offre de meilleurs résultats que \texttt{CUCB}, \texttt{UCB}-$\delta$ et \texttt{MOSS}. Le fait de considérer la variance empirique des utilisateurs dans le terme d'exploration conduit à une meilleure estimation des comptes ayant une forte variabilité, en autorisant l'algorithme à les sélectionner plus souvent. Dans ce scénario de capture de données, ceci semble particulièrement pertinent puisque la récompense obtenue dépend grandement de la fréquence de publication des utilisateurs. Par exemple, il est possible d'observer des récompenses nulles pendant plusieurs itérations pour certains utilisateurs uniquement par malchance (ils n'ont pas parlé à cet instant précis), sans pour autant qu'ils soient intrinsèquement non pertinents. Le fait d'augmenter le terme d'exploration pour ces profils à haute variabilité autorise le processus à les reconsidérer plus souvent s'il leur est déjà arrivé d'apporter de bonnes récompenses par le passé. De plus, certains profils, peut-être moins actifs que d'autres, mais possédant une forte utilité relativement au modèle de récompense en question - par exemple parce qu'ils sont actifs uniquement sur la thématique recherchée - auront plus de chances d'être réécoutés.

  \begin{figureth}
    \begin{subfigureth}{0.5\textwidth}
      \includegraphics[width=\linewidth]{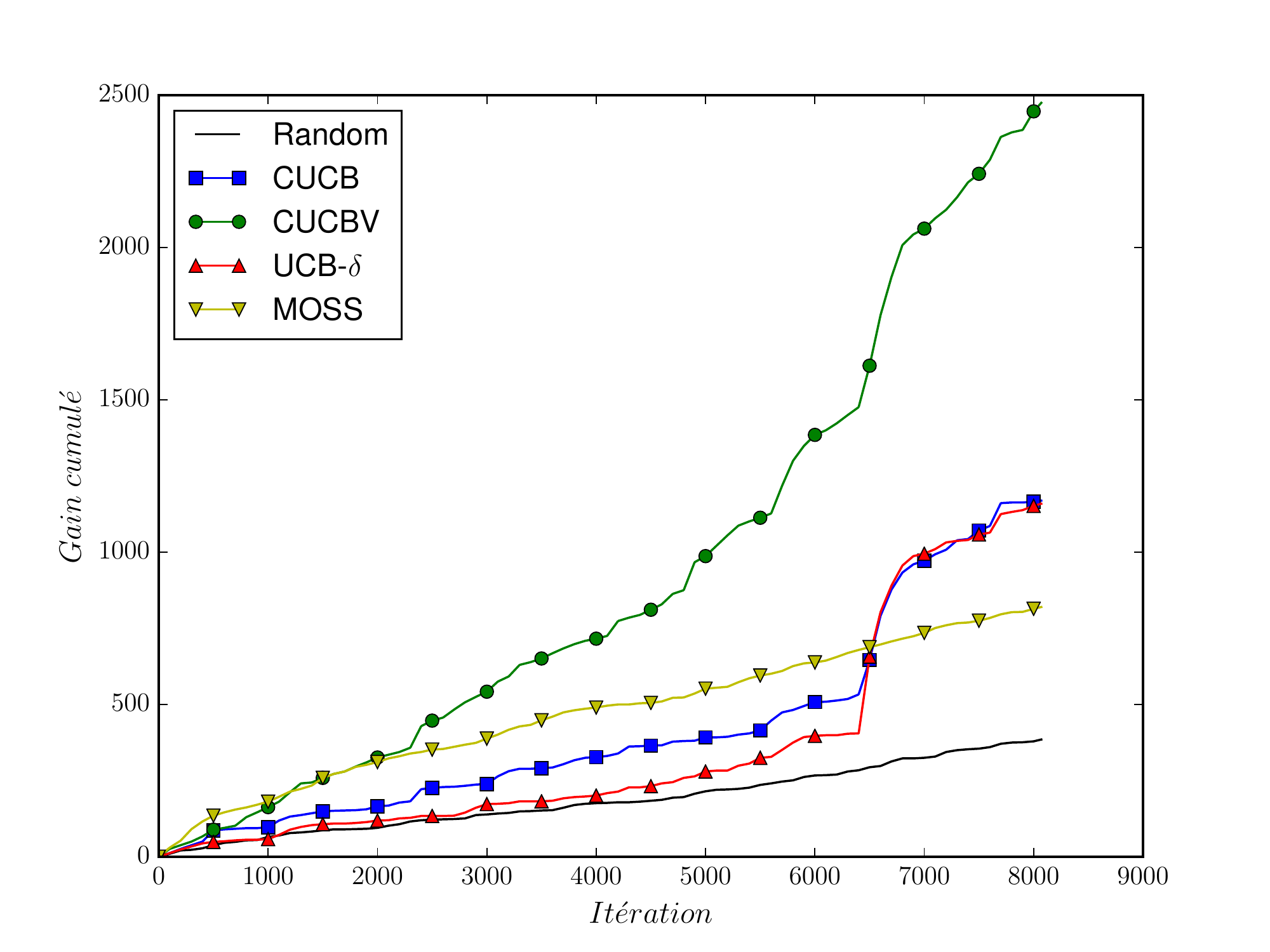}
      \caption{Politique}
    \end{subfigureth}
    \begin{subfigureth}{0.5\textwidth}
      \includegraphics[width=\linewidth]{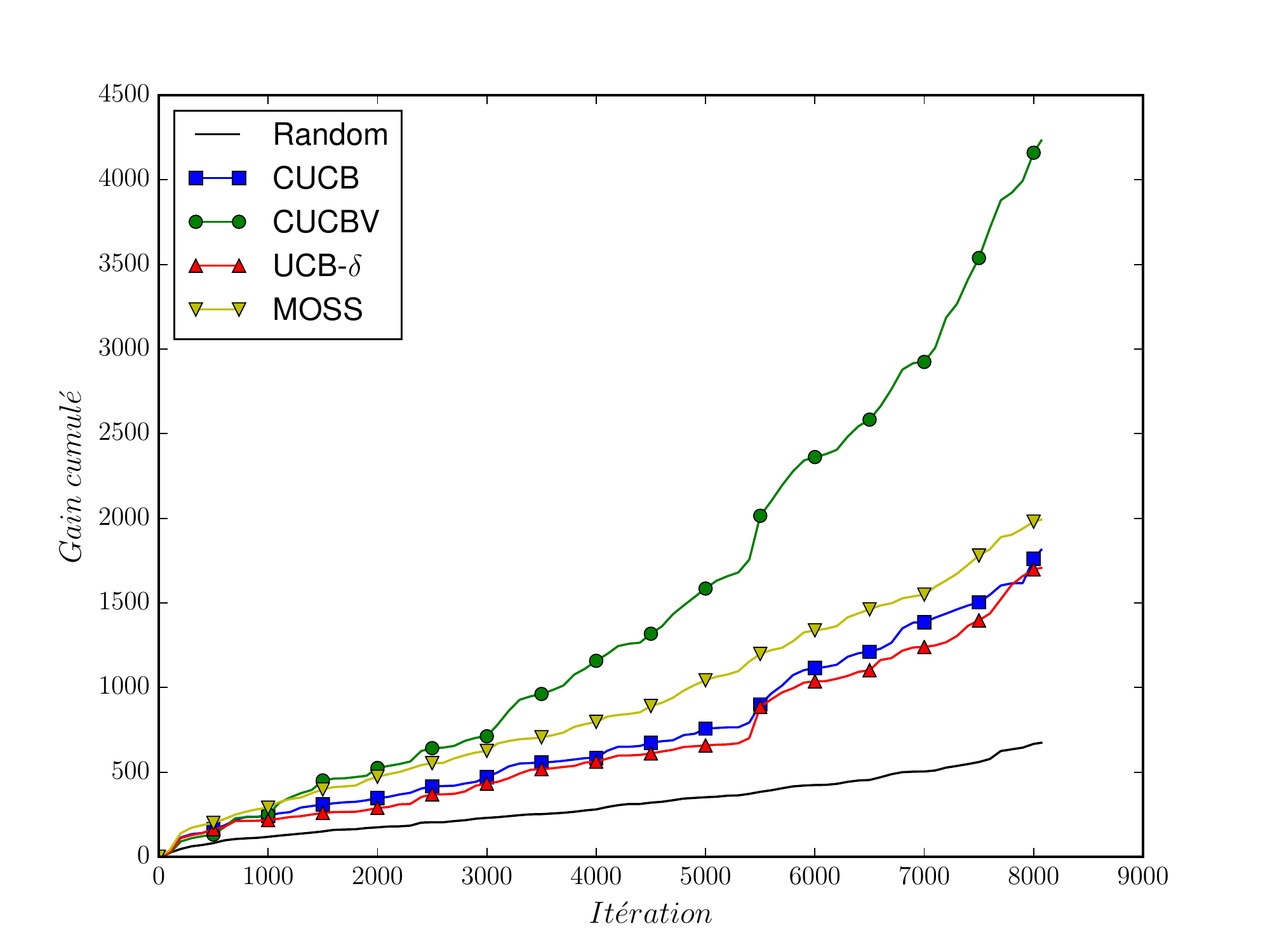}
      \caption{Religion}
    \end{subfigureth}
        \begin{subfigureth}{0.5\textwidth}
      \includegraphics[width=\linewidth]{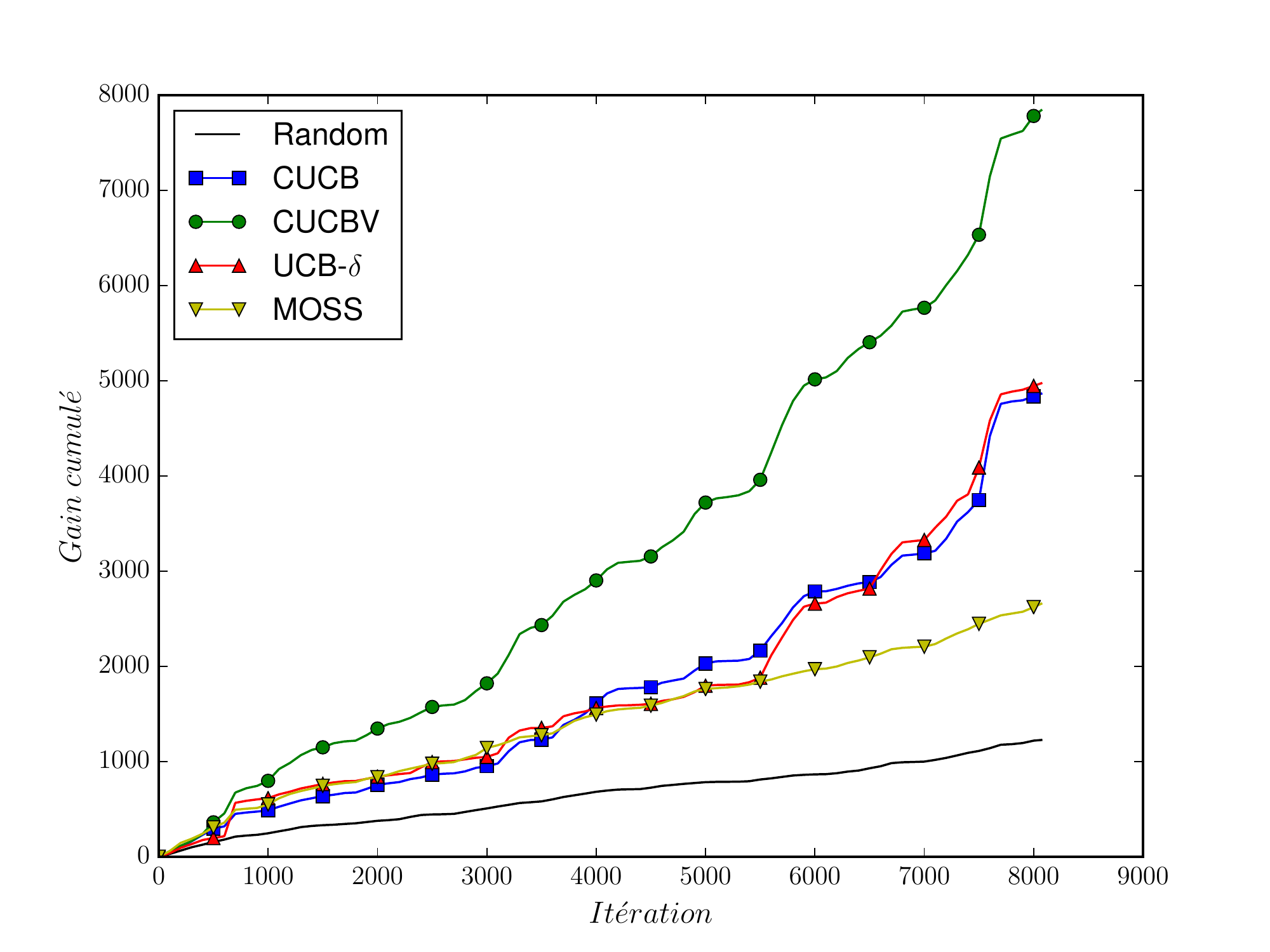}
      \caption{Science}
    \end{subfigureth}
    \caption[Gain vs temps base \textit{USElections} modèle statique]{Evolution de la récompense cumulée en fonction du temps sur la base \textit{USElections} pour différentes politiques et le modèle \textit{Topic+Influence} avec différentes thématiques.} 
      \label{fig:XPStatUSRT}
  \end{figureth}

  \begin{figureth}
    \begin{subfigureth}{0.5\textwidth}
  \includegraphics[width=\linewidth]{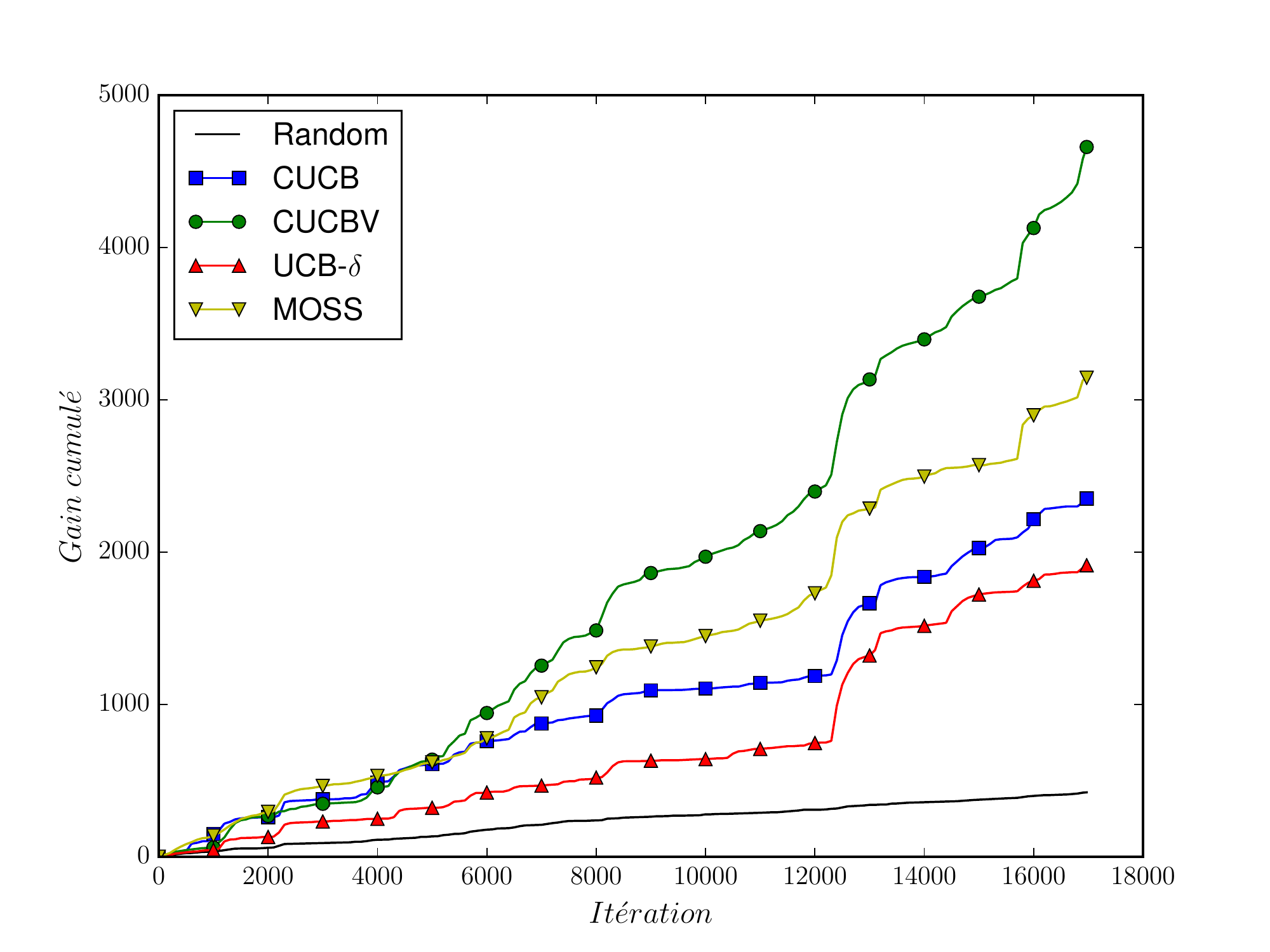}
      \caption{Politique}
    \end{subfigureth}
    \begin{subfigureth}{0.5\textwidth}
      \includegraphics[width=\linewidth]{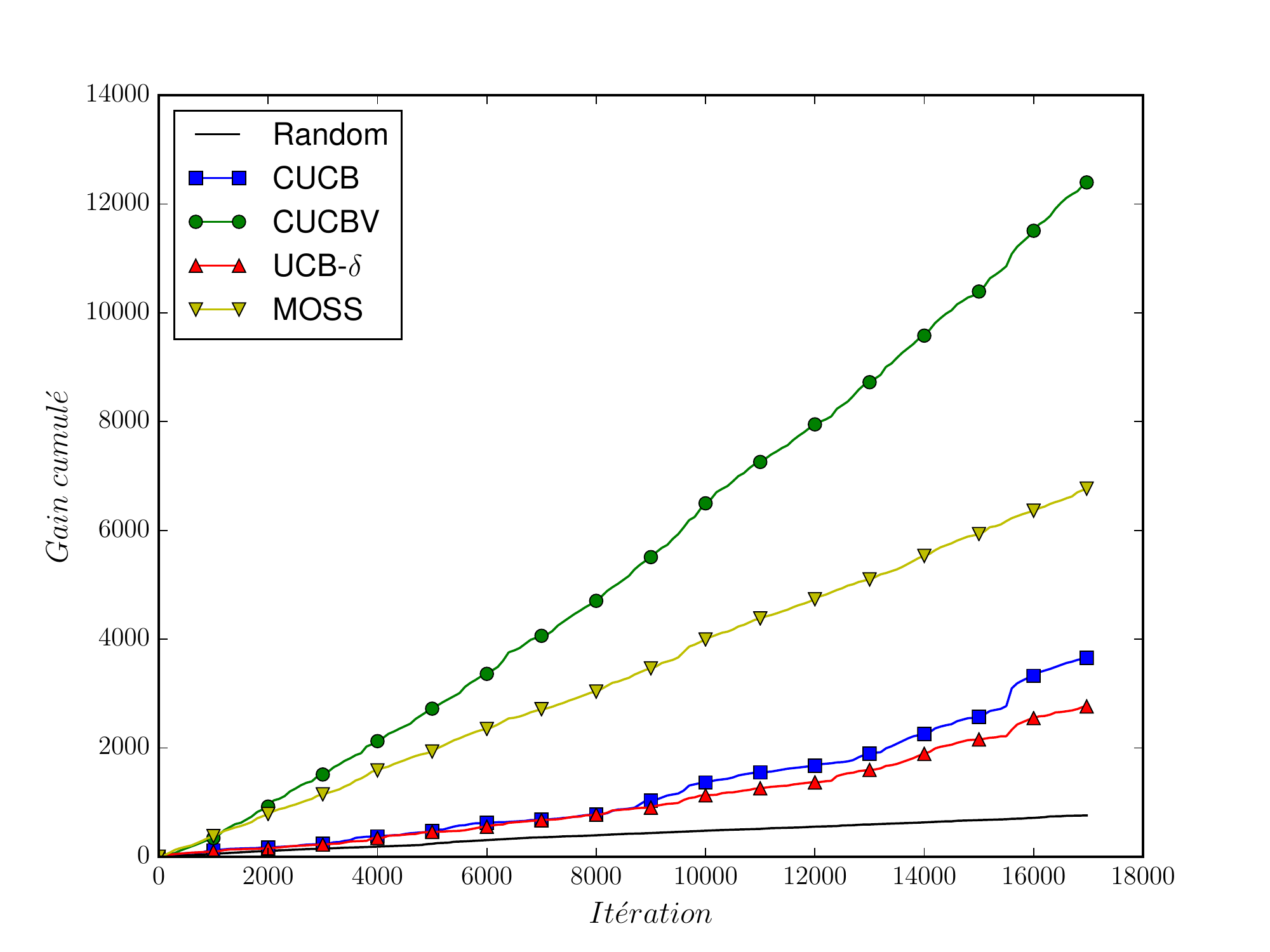}
      \caption{Religion}
    \end{subfigureth}
        \begin{subfigureth}{0.5\textwidth}
      \includegraphics[width=\linewidth]{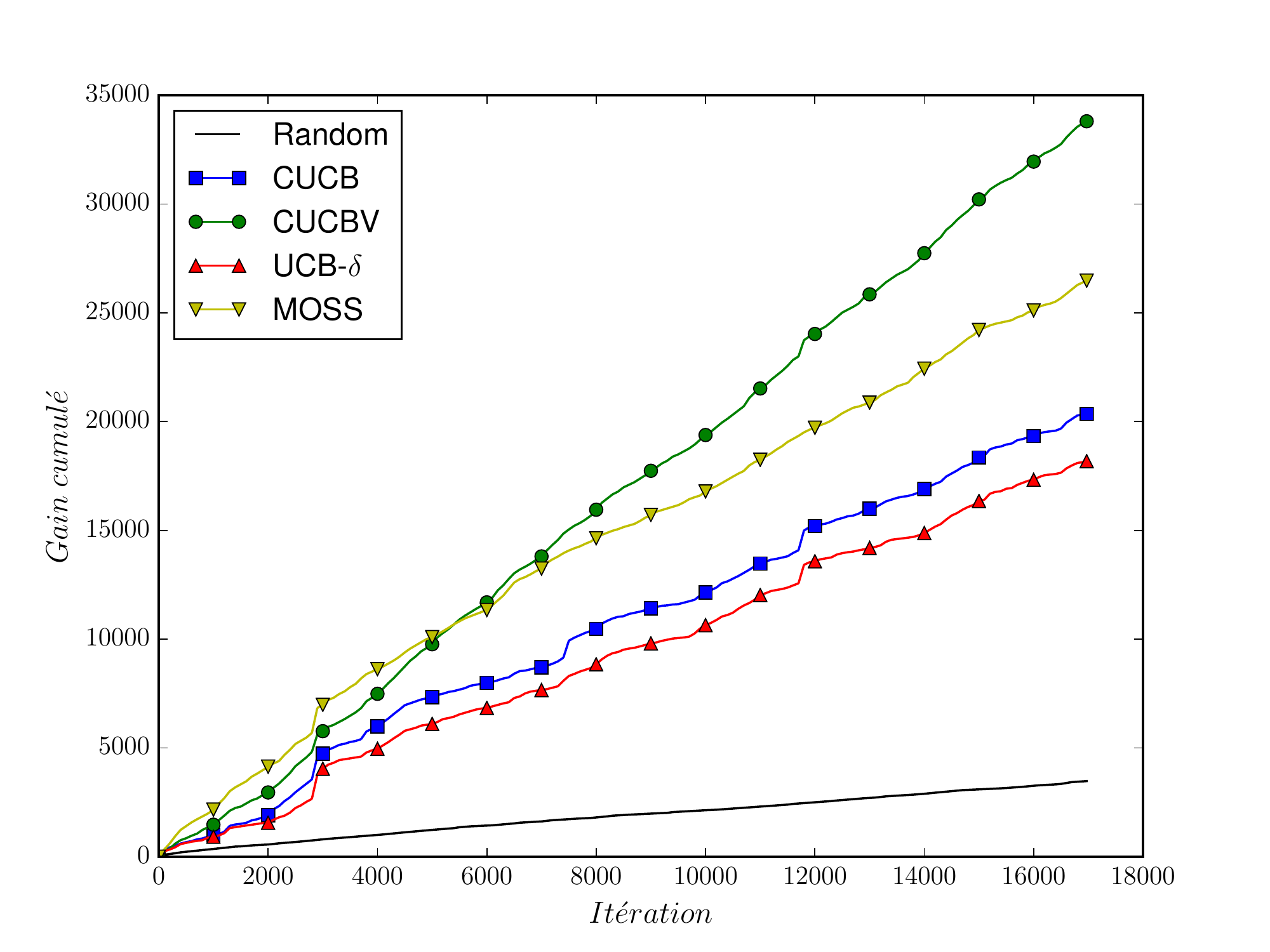}
      \caption{Science}
    \end{subfigureth}
    \caption[Gain vs temps base \textit{OlympicGames} modèle statique]{Evolution de la récompense cumulée en fonction du temps sur la base \textit{OlympicGames} pour différentes politiques et le modèle \textit{Topic+Influence} avec différentes thématiques.} 
      \label{fig:XPStatOGRT}
  \end{figureth}

  \begin{figureth}
    \begin{subfigureth}{0.5\textwidth}
  \includegraphics[width=\linewidth]{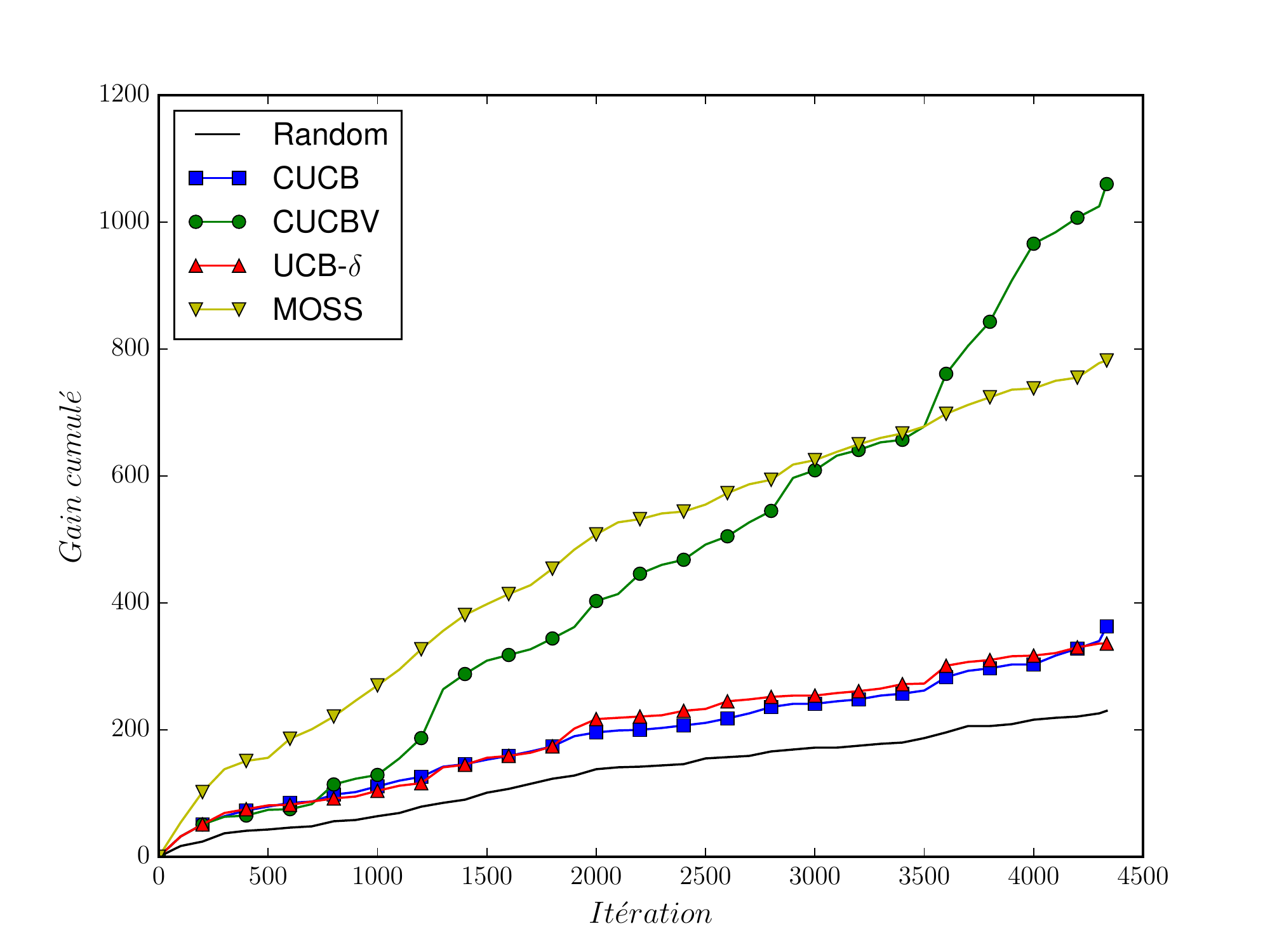}
      \caption{Politique}
    \end{subfigureth}
    \begin{subfigureth}{0.5\textwidth}
      \includegraphics[width=\linewidth]{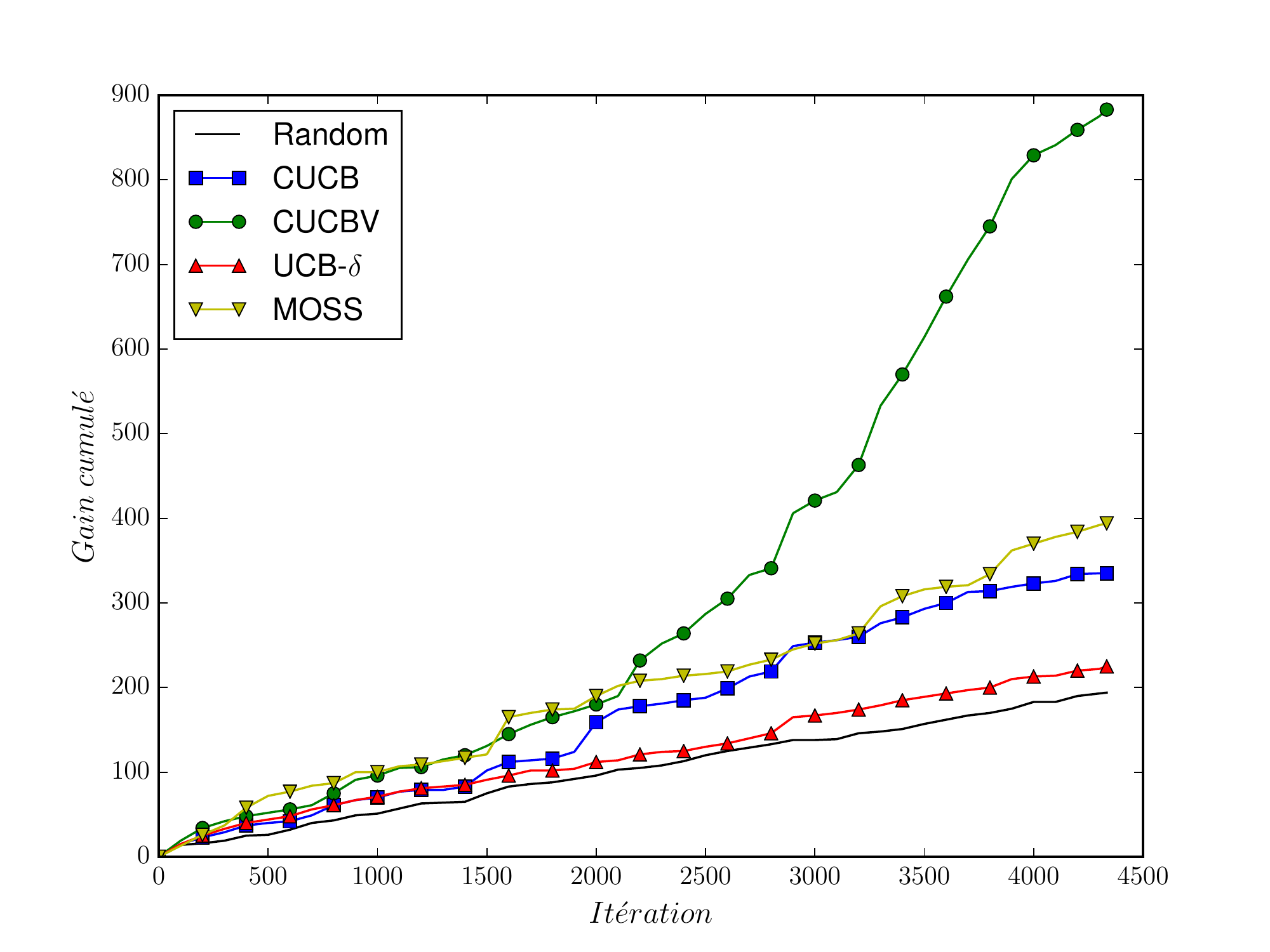}
      \caption{Religion}
    \end{subfigureth}
        \begin{subfigureth}{0.5\textwidth}
      \includegraphics[width=\linewidth]{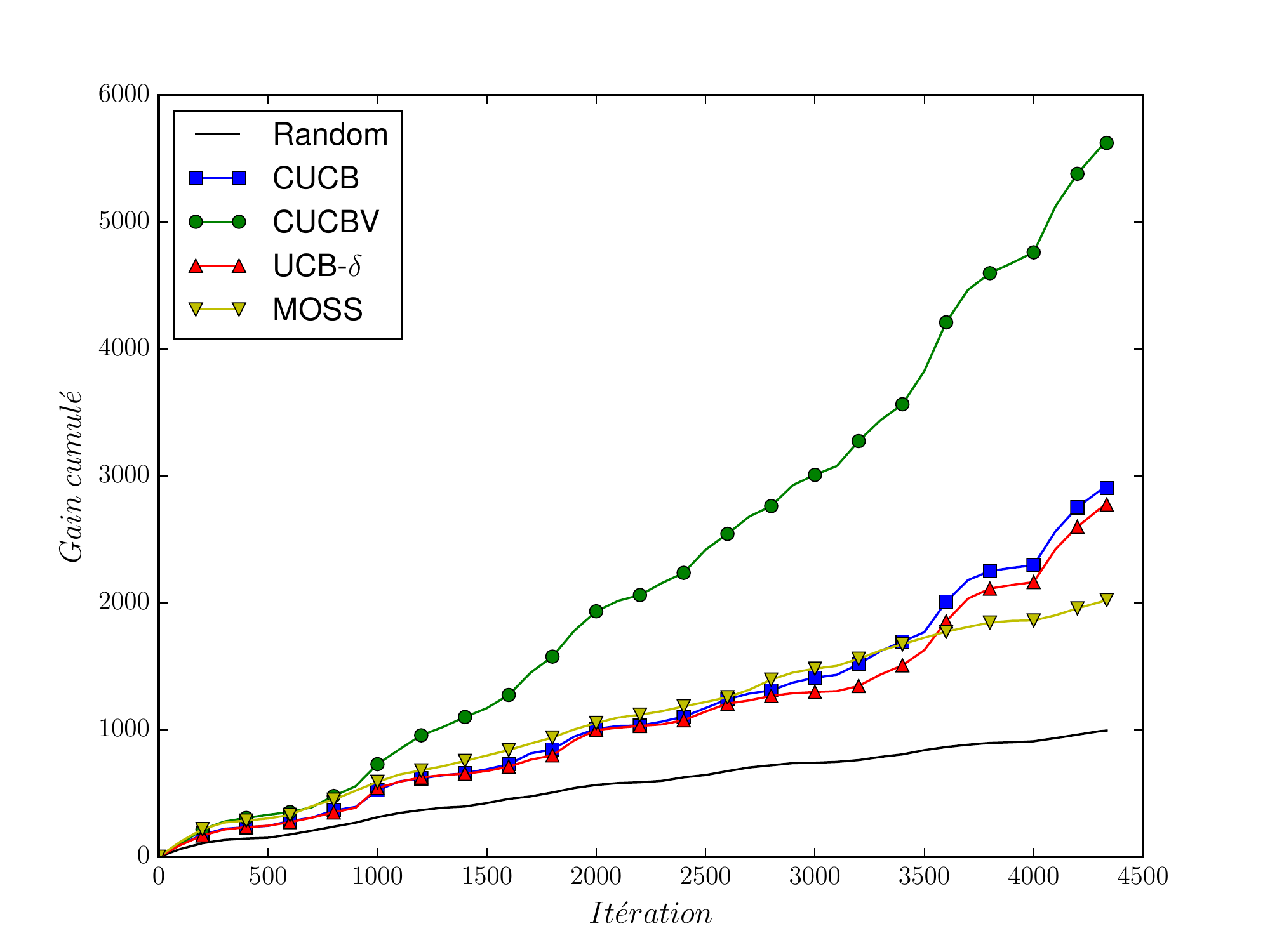}
      \caption{Science}
    \end{subfigureth}
    \caption[Gain vs temps base \textit{Brexit} modèle statique]{Evolution de la récompense cumulée en fonction du temps sur la base \textit{Brexit} pour différentes politiques et le modèle \textit{Topic+Influence} avec différentes thématiques.} 
      \label{fig:XPStatBrexitRT}
  \end{figureth}

      Finalement, la figure \ref{fig:XPStat3D} représente le gain cumulé pour les politiques \texttt{Random}, \texttt{CUCB} et \texttt{CUCBV} en fonction du temps et du nombre d'utilisateurs $k$ sélectionnés à chaque itération, sur le jeu de données \textit{USElections} et pour le modèle \textit{Topic+Influence} avec la thématique \textit{Science}. Nous remarquons que les trois surfaces ne se coupent jamais et que celle représentant \texttt{CUCBV} est toujours au-dessus de celle de \texttt{CUCB}, elle-même au-dessus de \texttt{Random}. Cela montre que peu importe le nombre d'utilisateurs écoutés simultanément, la politique \texttt{CUCBV} semble être la meilleure option.

    \begin{figureth}
      \includegraphics[width=0.7\linewidth]{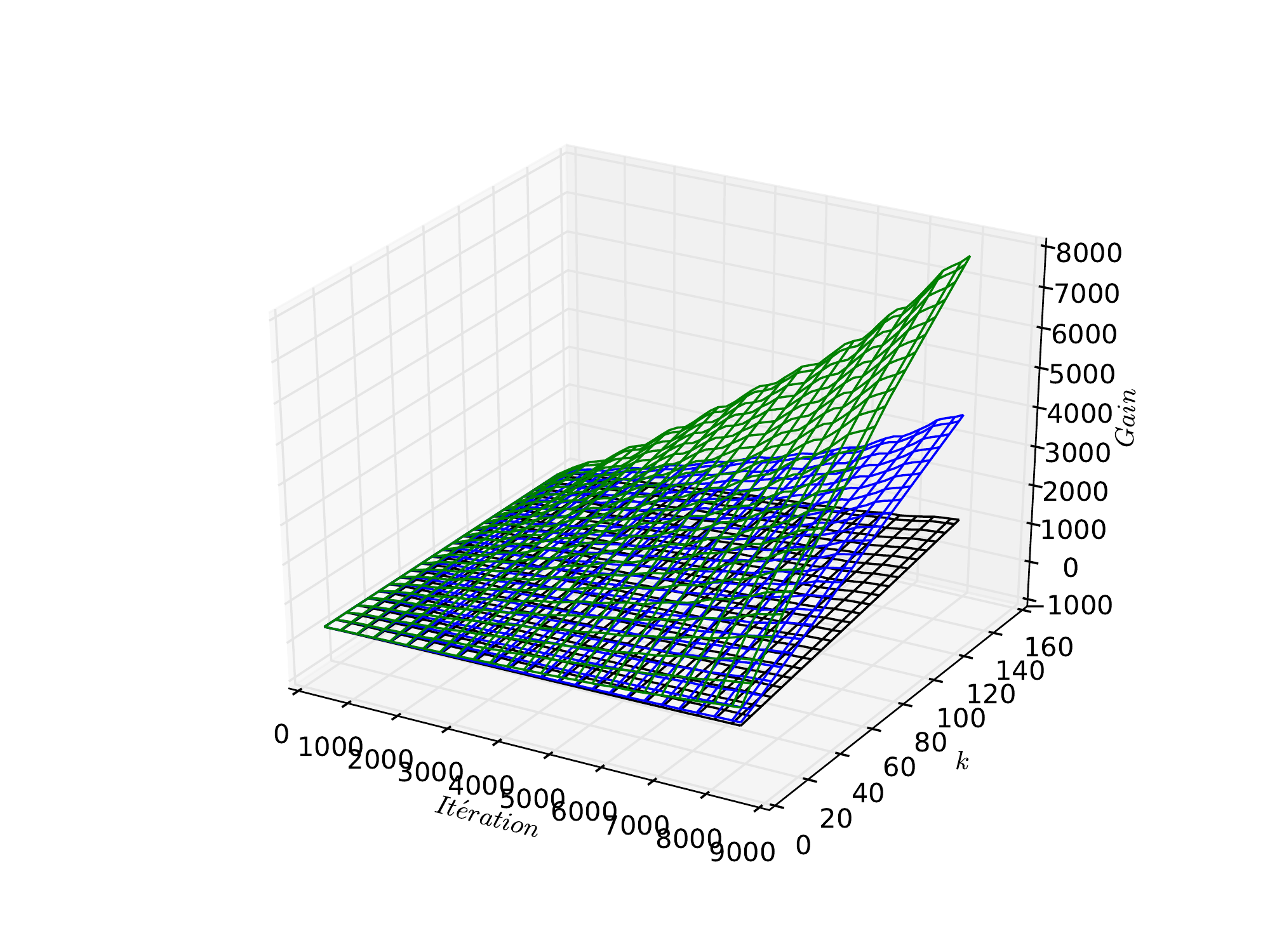}
        \caption[Gain vs temps vs $k$ base \textit{USElections} modèle statique]{Evolution de la récompense cumulée en fonction du temps sur la base \textit{USElections} pour différentes politiques et le modèle \textit{Topic+Influence} avec la thématique \textit{Science} pour différentes valeurs de $k$. La courbe du dessus représente la politique \texttt{CUCBV}, celle du milieu \texttt{CUCB} et celle du dessous \texttt{Random}.}
      \label{fig:XPStat3D}
  \end{figureth}

    \subsection{En ligne}
        
                \subsubsection{Protocole}

Nous proposons maintenant de tester nos modèles sur une expérience grandeur nature, en utilisant l'API de \textit{Follow Streaming} de Twitter, afin d'en confirmer l'efficacité lorsqu'il s'agit d'un réseau social entier avec la contrainte de l'API limitant le nombre d'utilisateurs pouvant être écoutés en même temps. On rappelle que la limite maximale d'utilisateurs pouvant être écoutés en même temps est de 5000. Étant donné que le nombre de connexions simultanées à l'API \textit{Streaming} de Twitter est limité, nous nous concentrons sur  l'algorithme \texttt{CUCBV}, qui est apparu comme le plus performant dans les expériences hors ligne. Nous le comparons à une politique \texttt{Random}. 

Il est nécessaire de définir un processus d'alimentation de la base d'utilisateurs prenant en compte les contraintes réelles des API. Nous choisissons ici d'utiliser les utilisateurs référencés dans les messages des comptes écoutés à chaque instant. Concrètement, il est nécessaire de sélectionner un ensemble de comptes initiaux, puis l'ensemble des comptes $\mathcal{K}$ est alimenté à chaque itération $t$ par les utilisateurs ayant \textit{retweeté} ou répondu aux utilisateurs de $\mathcal{K}_t$. Notons que d'autres possibilités existent et seront présentées dans le prochain chapitre.

Nous considérons le modèle \textit{Topic+Influence} avec la thématique \textit{Politique}. Pour cette expérience, le nombre d'utilisateurs écoutés simultanément est fixé à $k=5000$ (maximum autorisé par l'API) et la durée de chaque itération est de 5 minutes. Notre collecte s'est déroulée sur une période de 75 heures soit un total de $T=900$ itérations, en utilisant les comptes initiaux suivants:\textit{BBC}, \textit{CNN} et \textit{FoxNews}.

Etant donné que le nombre de nouveaux utilisateurs rencontrés à chaque itération peut être très élevé, nous avons limité le nombre maximal de nouveaux comptes à 1000 utilisateurs à chaque itération. Cela permet d'éviter le cas où ce nombre atteindrait $k$, ce qui aurait pour conséquence de ne permettre aucune exploitation.

      \subsubsection{Résultats}
            
            A nouveau, la figure  \ref{fig:XPStatLive} représente la récompense cumulée en fonction du temps pour les deux algorithmes testés. Elle met en valeur les bonnes performances de la méthode proposée, car la courbe \texttt{CUCBV} augmente significativement plus rapidement que la courbe \texttt{Random}. A la fin du processus, \texttt{CUCBV} obtient un gain cumulé 3 fois plus élevé que \textit{Random}. 
Finalement, notons que nous avons terminé avec un nombre total d'utilisateurs  beaucoup plus large pour \texttt{CUCBV} que pour \texttt{Random}, ce qui est cohérent avec le fait que l'on s'oriente mieux vers des utilisateurs populaires, plus souvent \textit{retweetés}, étant donné le processus d'alimentation de la base d'utilisateurs et le modèle de récompense choisi, qui favorise les utilisateurs dont les messages sont repris.

          \begin{figureth}
      \includegraphics[width=0.7\linewidth]{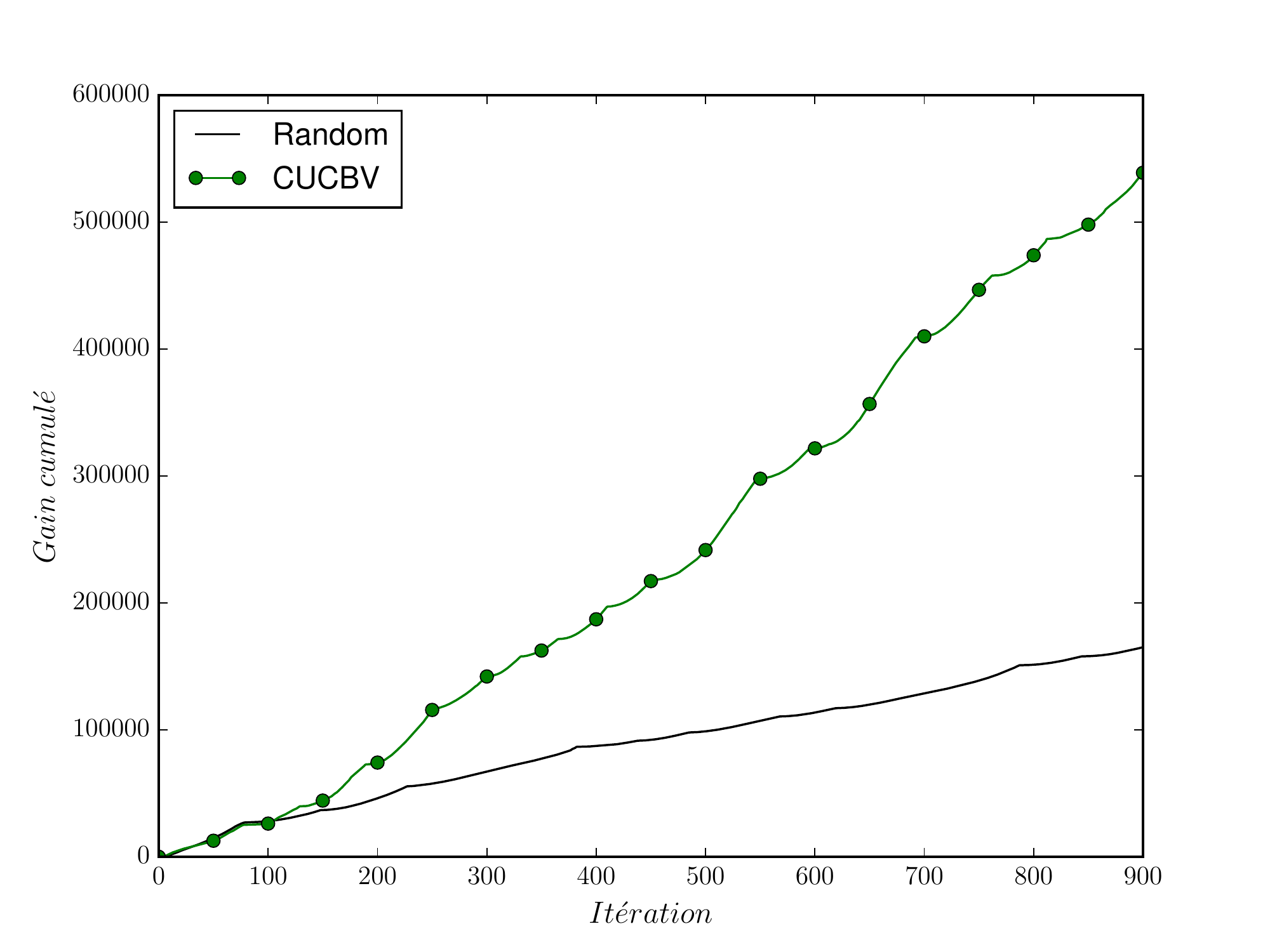}
        \caption[Gain vs temps XP en ligne modèle statique]{Evolution de la récompense cumulée en fonction du temps pour l'expérience en ligne pour différentes politiques.}  
      \label{fig:XPStatLive}
  \end{figureth}

\section{Conclusion}

Dans ce chapitre, nous avons proposé une première approche de bandit à sélection multiple pour résoudre notre tâche de collecte d'information ciblée dans un réseau social. En modélisant chaque distribution de récompense par une distribution stationnaire de moyenne et de variance inconnues, nous avons été en mesure d'appliquer des algorithmes de bandits classiques, adaptés à ce cas. En outre, face à la forte variabilité des récompenses obtenues dans notre contexte, nous avons défini l’algorithme \texttt{CUCBV}, prenant en compte la variance dans sa stratégie d’exploration des récompenses. Après avoir prouvé certaines garanties de convergence, nous l’avons expérimenté dans le contexte des médias sociaux. Les expériences sur données réelles, aussi bien hors ligne qu’en ligne, ont montré la capacité de l’algorithme à s’orienter automatiquement vers des sources pertinentes relativement à une fonction de récompense donnée.

Ce premier travail, basé sur des hypothèses de stationnarité, et ne faisant usage d'aucune information relative aux différents  utilisateurs autres que les récompenses qu'ils ont produites, ouvre  diverses perspectives quant à l'application des algorithmes de bandit à la collecte d'information en temps réel dans les médias sociaux. En particulier, l'utilisation de modèles plus complexes, modélisant plus finement le comportement des utilisateurs du réseau, semble être une voie prometteuse, que nous proposons d'explorer dans les chapitres suivants.

%



	\chapter{Modèle stationnaire avec profils constants}
	\minitoc
	\newpage

\label{ModeleContextuelFixe}

Dans ce chapitre, nous faisons l'hypothèse qu'un profil associé à chaque utilisateur sous-tend son utilité espérée. Une hypothèse de linéarité mettant en relation les profils et un paramètre inconnu, commun à tous les utilisateurs, nous permet de définir une stratégie  d'exploration de l'espace des actions plus efficace que dans le chapitre précédent. Nous verrons que ce modèle peut se traduire par un problème de bandit contextuel avec des contextes (que nous appelons ici profils) constants. Cependant, pour notre tâche de collecte d'informations sous-contrainte, deux principaux facteurs nous empêchent d'utiliser des algorithmes existants: premièrement, les vecteurs de profil ne sont pas visibles directement. En effet seulement des échantillons de vecteurs, centrés sur les profils, sont accessibles à l'agent décisionnel. Deuxièmement, ces échantillons ne sont accessibles que pour un sous-ensemble d'utilisateurs à chaque itération. Dans cette optique, nous proposons à la fois un nouveau problème de bandit contextuel et un algorithme associé, dont nous proposons une borne supérieure du regret. Enfin, nous terminons par des expérimentations sur données artificielles et réelles, dont les résultats seront comparés aux méthodes du chapitre précédent.

\section{Modèle}
\label{chap2Model}

Dans cette section, nous formalisons le modèle de bandits avec profils d'action. Nous nous focalisons sur le cas où une seule action est sélectionnée à chaque pas de temps ($k=1$), que nous étendrons en fin de chapitre  au cas de la sélection multiple, adapté à notre besoin pour la collecte d'informations provenant de plusieurs sources simultanément. Considérons qu'à chaque utilisateur $i$ du réseau social, on associe un vecteur $\mu_{i} \in \mathbb{R}^{d}$. Ce vecteur, que l'on appellera profil, modélise certains attributs d'un utilisateur et peut contenir toute sorte de valeurs, selon le modèle et les informations disponibles sur chaque compte. Nous verrons dans la partie expérimentation un exemple de profil pouvant être utilisé. Afin de tirer profit de ces informations contextuelles, il est nécessaire de définir le modèle de récompense associé. Dans la section suivante, nous présentons un modèle de bandit contextuel linéaire existant, dont nous nous démarquerons ensuite pour notre problème de collecte basée sur des profils constants, mais cachés.

\subsection{Bandits avec profils d'actions connus}

Le cadre du bandit contextuel linéaire, que nous avons présenté dans la partie \ref{SectionBanditContextuel}, définit la récompense de chaque action comme une relation linéaire entre les profils et un vecteur de paramètre. Ce vecteur, inconnu à l'origine, doit être appris au fur et à mesure des itérations du processus. Concrètement, à chaque itération  $t\in \left\{1,..,T\right\}$, l'agent dispose d'un ensemble $\left\{ \mu_{1},..,\mu_{K} \right\} \subset \mathbb{R}^{d} $ de $K$ vecteurs. Il choisit ensuite une action $i_{t} \in \left\{1,...,K \right\}$ et reçoit la récompense associée $r_{i_{t},t} \in \left[ 0..1 \right]$. L'hypothèse de linéarité se traduit par l'existence d'un vecteur $\beta \in \mathbb{R}^{d}$ tel que $\forall t \in \left\{1,..,T\right\}, \forall i \in \left\{1,...,K \right\}: r_{i,t}=\mu_{i}^{\top}\beta+\eta_{i,t}$, où comme dans \cite{OFUL}, $\eta_{i,t}$ est un bruit $R$-sous-gaussien de moyenne nulle, c'est-à-dire $\forall \lambda \in \mathbb{R}: \mathbb{E}[e^{\lambda\eta_{i,t}}|\mathcal{H}_{t-1}] \leq e^{\lambda^{2}R^{2}/2}$ avec $\mathcal{H}_{t-1}=\left\{(i_{s},x_{i_{s},s},r_{i_{s},s})\right\}_{s=1..t-1}$ l'historique des choix passés. Soulignons qu'un bruit gaussien $\mathcal{N}(0,\sigma^{2})$ est sous-gaussien de constante $\sigma$ et qu'une variable uniforme dans l'intervalle $[-a,a]$ est aussi sous-gaussienne de constante $a$ (voir \cite{subGaussian} pour plus de résultats sur les variables aléatoires sous-gaussiennes).

Comme nous l'avons précédemment expliqué, la performance d'un algorithme de bandit peut se mesurer via la notion de pseudo-regret, dont nous rappelons les définitions ci-dessous.

\begin{defi}
Le pseudo-regret instantané d'un algorithme de bandit contextuel à un instant $t$, noté $reg_{t}$ est défini par:
\begin{equation}
reg_{t}=\mu_{i^{\star}}^{\top}\beta-\mu_{i_{t}}^{\top}\beta
\end{equation}
Avec  $\mu_{i^{\star}}=\argmax\limits_{\mu_{i},i=1..K}\mu_{i}^{\top}\beta$ représentant le profil de l'action optimale $i^{\star}$.
\end{defi}

Le but d'un algorithme de bandit est de minimiser le pseudo-regret cumulé au cours du temps, défini par  $\hat{R}_{T}=\sum\limits_{t=1}^{\top}reg_{t}$, avec une certaine probabilité, pour une séquence d'actions choisie $\left\{i_{1},...,i_{T}\right\}$.

Le problème formulé ci-dessus est bien connu dans la littérature et peut être directement résolu via des algorithmes existants tels que \texttt{OFUL}  \cite{OFUL} ou \texttt{LinUCB} \cite{banditCtxtLinUCB}, dont le regret cumulé peut être majoré. Le principe de ces algorithmes est d'utiliser un estimateur du paramètre de régression et de maintenir un ellipsoïde de confiance associé. Ce dernier permet d'obtenir un intervalle de confiance pour chaque récompense et offre ainsi la possibilité de définir des stratégies de type optimiste. De plus, il est important de noter que les profils et le paramètre de régression sont constants. Ce problème est donc aussi un problème de bandit stationnaire. En effet en notant $\mu_{i}^{\top}\beta=m_{i}$ on peut se ramener au modèle de la partie précédente. Cependant, la présence d'une structure sous-jacente entre les différentes actions, lorsqu'elle est exploitée, permet de définir des stratégies beaucoup plus performantes.

Afin d'illustrer cela, prenons par exemple l'algorithme \texttt{OFUL} qui à chaque instant $t$ et pour chaque action $i$ calcule le score $s_{i,t}=\mu_{i}^{\top}\hat{\beta}_{t-1}+\alpha_{t-1}||\mu_{i}||_{V_{t-1}^{-1}}$ où $\alpha_{t-1}$, $\hat{\beta}_{t-1}$ et $V_{t-1}$ sont des paramètres communs à tous les bras, définis dans la section \ref{sectionOFUL}. L'algorithme sélectionne ensuite l'action maximisant ce score. En se plaçant dans un cadre simple où $K=4$ et $d=2$, il est possible de représenter les scores à un instant $t$ (en fixant donc $\alpha_{t-1}$, $\hat{\beta}_{t-1}$ et $V_{t-1}$) dans un espace continu à deux dimensions, comme l'illustre la figure \ref{fig:scoreOFUL2D}. Dans cette figure, les zones de couleur verte correspondent à des valeurs de scores élevées tandis que les zones de couleur rouge correspondent à des scores faibles. Ainsi, les variations de couleurs reflètent la structure sous-jacente induite par le modèle. Dans ce cas précis,  l'algorithme \texttt{OFUL} sélectionnerait l'action 1, son profil étant situé dans une zone de l'espace plus prometteuse. En revanche, l'action 3 semble se situer dans une zone beaucoup moins avantageuse. Le fait d'utiliser un paramètre commun à toutes les actions permet de se projeter dans un espace où les récompenses sont structurées entre elles. De cette façon, les éléments appris sur une action nous renseignent également sur toutes les autres. L'apprentissage n'est donc plus individuel comme c'était le cas dans le chapitre précédent, mais mutualisé, ce qui permet d'explorer l'espace beaucoup plus efficacement. Imaginons qu'un grand nombre d'actions se situe dans la zone rouge de la figure. Dans ce cas, un algorithme de bandit classique devrait considérer ces dernières plusieurs fois avant de se rendre compte de leur mauvaise qualité. En revanche, un algorithme de bandit contextuel serait en mesure d'écarter ces actions beaucoup plus rapidement en les considérant comme appartenant à une zone de l'espace peu prometteuse.

        \begin{figureth}
			\includegraphics[width=0.9\linewidth]{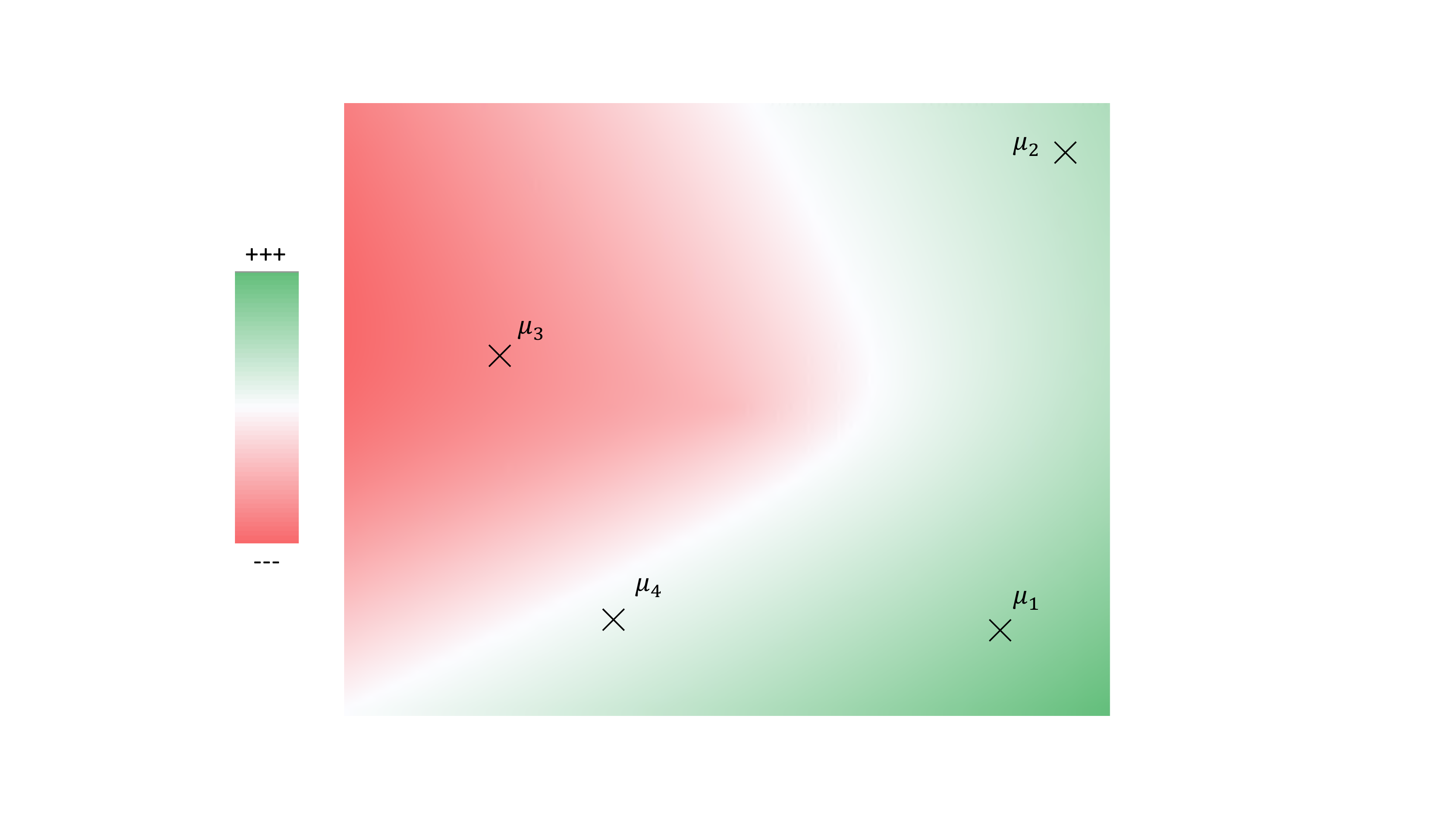}
        \caption[Score \texttt{OFUL} en deux dimensions]{Illustration  des scores de l'algorithme \texttt{OFUL} à un instant donné dans un espace à deux dimensions.}
			\label{fig:scoreOFUL2D}
		\end{figureth}

Dans la section suivante, nous introduisons un nouveau problème de bandit, basé sur la même hypothèse de linéarité, mais avec des contraintes sur l'observation des profils.

\subsection{Bandits avec profils d'actions inconnus}

Nous proposons un nouveau cadre, dans lequel l'ensemble des vecteurs de profils  $\left\{ \mu_{1},..,\mu_{K} \right\}$ n'est pas observé directement. A la place, à chaque itération $t$, l'agent dispose d'un sous-ensemble d'actions notées $\mathcal{O}_{t}$ tel que pour toutes actions $i \in \mathcal{O}_{t}$, un échantillon $x_{i,t}$ d'une variable aléatoire centrée sur $\mu_{i}$ est révélé. En considérant les mêmes hypothèses que précédemment, le problème peut être réécrit de la façon suivante:

\begin{align}
\label{chap2RewritePb}
\forall s\leq t :r_{i,s} &=\mu_{i}^{\top}\beta+\eta_{i,s} \nonumber \\
	   					 &=\hat{x}_{i,t}^{\top}\beta+(\mu_{i}-		\hat{x}_{i,t})^{\top}\beta+\eta_{i,s}  \nonumber\\
	   					 &=\hat{x}_{i,t}^{\top}\beta+\epsilon_{i,t}^{\top}\beta+\eta_{i,s}
\end{align}

 Où $\epsilon_{i,t}=\mu_{i}-\hat{x}_{i,t}$, $\hat{x}_{i,t}=\dfrac{1}{n_{i,t}}\sum\limits_{s \in{T_{i,t}^{obs}} } x_{i,s}$, avec $T_{i,t}^{obs}=\left\{ s\leq t, i\in \mathcal{O}_{s}\right\}$ et $n_{i,t}=|T_{i,t}^{obs}|$. Concrètement, $n_{i,t}$ correspond au nombre de fois où un échantillon associé à l'action $i$ a été délivré jusqu'au temps $t$ et  $\hat{x}_{i,t}$ correspond à la moyenne empirique des échantillons observés pour $i$ au temps $t$. Ainsi ce problème diverge des instances existantes du bandit contextuel traditionnel par les deux aspects principaux suivants:

\begin{enumerate}
\item Les profils ne sont pas visibles directement, on ne dispose que d'échantillons centrés sur ces derniers;
\item A chaque itération, on ne dispose pas d'échantillon pour tous les bras, seulement pour un sous-ensemble.
\end{enumerate}

Contrairement au bandit contextuel traditionnel de la section précédente, ici l'incertitude est double. Elle provient d'une part de l'estimateur du paramètre de régression $\beta$, mais aussi des échantillons observés. Ainsi un algorithme devra à la fois estimer $\beta$, mais aussi les vecteurs $\left\{ \mu_{1},..,\mu_{K} \right\}$. Pour ces derniers nous utiliserons la moyenne empirique comme estimateur, comme le suggère l'équation \ref{chap2RewritePb}.  Il est important de noter que dans l'écriture précédente, où l'on écrit la récompense à un instant $s\leq t$, le bruit dépend bien de $s$ et la moyenne empirique de $t$. D'après la loi des grands nombres, plus le nombre d'observations augmente, plus la moyenne empirique se rapproche de la vraie valeur du profil. Nous laissons les détails mathématiques nécessaires à la dérivation de notre algorithme à la section suivante. Nous illustrons la différence par rapport au cas précédent dans la figure  \ref{fig:scoreOFUL2DBis}\footnote{Nous insistons sur le fait qu'il s'agit uniquement d'une illustration permettant d'appréhender la différence avec le cas traditionnel sans incertitude sur les profils.}, qui représente par des cercles l'incertitude liée à l'approximation des profils par les moyennes empiriques des échantillons associés. Contrairement à la figure \ref{fig:scoreOFUL2D}, où les profils sont connus, on ne dispose ici que d'intervalles de confiance de différentes tailles, centrés sur les moyennes empiriques (représentées par des croix bleues verticales), dans lequel on sait que le profil se situe, avec une certaine probabilité. L'action la plus prometteuse est toujours l'action 1, dont le véritable profil (représenté par une croix noire) se situe dans la zone la plus verte de l'espace. Or, ce profil est inconnu, autrement dit, le véritable positionnement de cette croix n'est pas disponible. Une solution naïve consisterait alors à utiliser directement la moyenne empirique de chaque action pour déterminer son score. D'après la figure, ceci conduirait à sélectionner l'action 2, dont la moyenne empirique (croix bleue) se situe dans la zone la plus verte, ce qui constitue un choix sous-optimal par rapport à un algorithme qui connaîtrait les profils. Ceci est dû au fait que la moyenne empirique peut être plus ou moins éloignée du véritable profil. Nous proposons de prendre en compte cette incertitude supplémentaire dans le choix de l'action à sélectionner. Ainsi, en utilisant le principe d'optimisme devant l'incertain, un algorithme prenant en compte l'incertitude sur les profils sélectionnerait l'action 1, dont le cercle représentant l'incertitude du profil touche la zone la plus verte de l'espace.


        \begin{figureth}
			\includegraphics[width=0.9\linewidth]{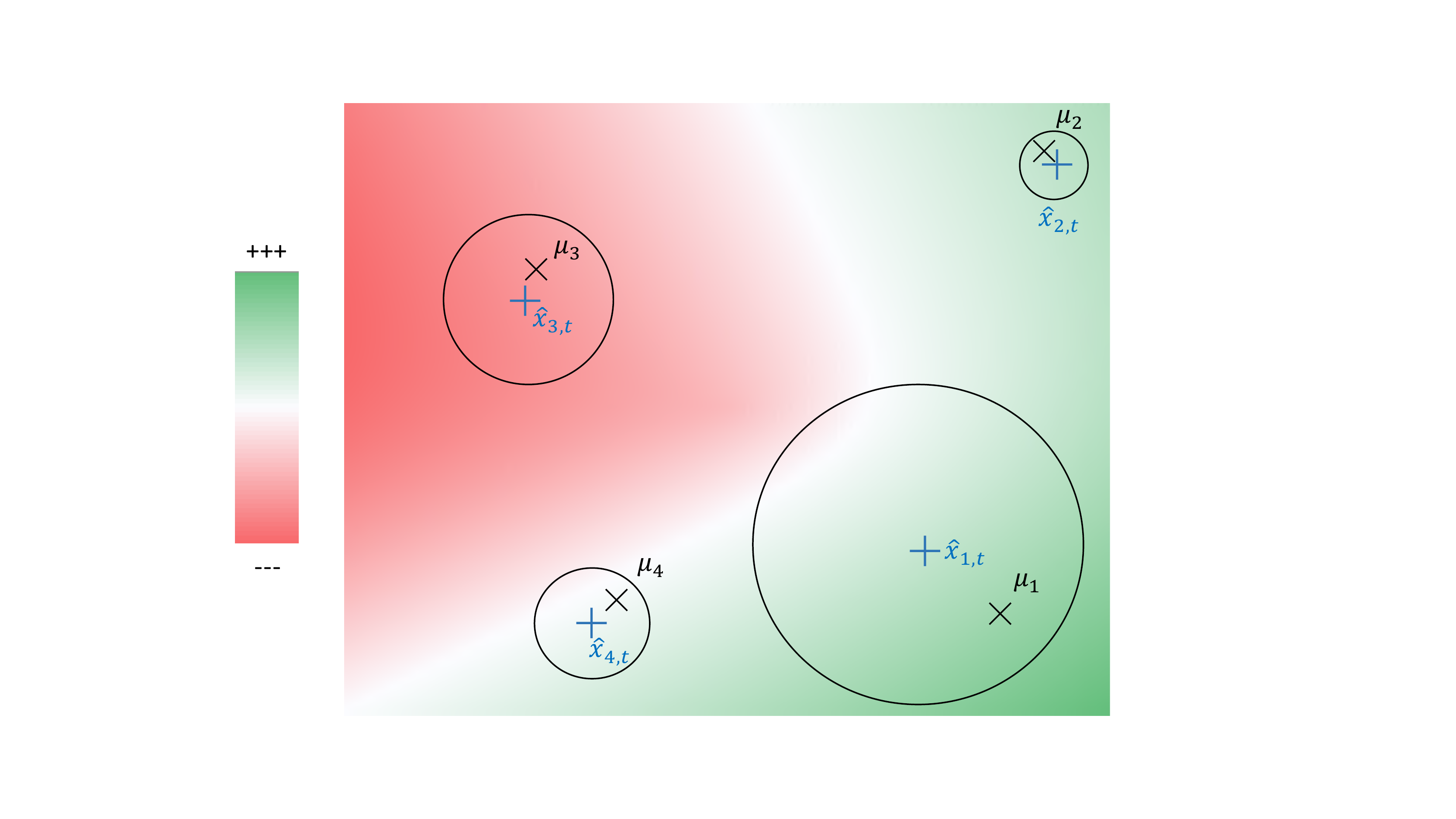}
        \caption[Score \texttt{OFUL} en deux dimensions avec incertitude sur les profils]{Illustration de l'incertitude supplémentaire apportée par la non-connaissance des profils.}
			\label{fig:scoreOFUL2DBis}
	\end{figureth}

	Dans la suite, nous proposons de définir un algorithme adapté à ce cas étape par étape, dans un contexte de processus de révélation des profils générique, que l'on déclinera pour les trois cas suivants: 
 \begin{itemize}
\item \textbf{Cas 1}: A chaque pas de temps $t$, toutes les actions livrent un échantillon c.-à-d.: $\forall t$, $\mathcal{O}_{t}=\left\{1,...,K \right\}$;
\item \textbf{Cas 2}: A chaque pas de temps $t$, chaque action possède une probabilité $p$ de livrer un échantillon. Dans ce cas, le contenu, mais aussi la taille de $\mathcal{O}_{t}$ varie au cours du temps;
\item \textbf{Cas 3}: A chaque pas de temps $t$, seule l'action sélectionnée à l'itération précédente livre un échantillon c.-à-d.: $\forall t$, $\mathcal{O}_{t}=i_{t-1}$.
\end{itemize}

Le premier cas correspond au cas le plus simple, où la seule différence avec un problème de bandit contextuel traditionnel provient du fait que les observations de contexte sont bruitées. Le second cas introduit une difficulté supplémentaire au sens où, à chaque instant, toutes les actions n'ont pas le même nombre d'observations, entraînant ainsi une différence d'incertitude entre ces dernières. Finalement, le dernier point, à notre sens le plus utile, s'intéresse au cas où un échantillon est observé uniquement pour l'action sélectionnée au temps précédent. Il apparaît que ce cas se rapproche le plus des conditions classiques du bandit stochastique, pour lequel aucune information n'est disponible à chaque instant, si ce n'est pour l'action qui a été sélectionnée. Dans cette configuration, l'utilisation de l'incertitude sur les profils dans la stratégie de sélection est indispensable pour obtenir une estimation relativement optimiste des utilités espérées pour les différentes actions, et ainsi être à même de garantir une convergence de l'algorithme vers les meilleures actions. 
Nous proposons dans la section suivante d'élaborer une politique permettant de prendre en compte l'incertitude sur les profils dans sa stratégie de sélection, afin de reconsidérer les actions dont les profils sont les moins bien connus.

\section{Algorithme}

Dans cette partie, nous dérivons une suite de propositions qui nous permettra de définir une politique de sélection adaptée à notre problème. Il s'agit dans un premier temps de définir un estimateur du paramètre de régression ainsi qu'un intervalle de confiance associé. Pour ce faire, nous utiliserons des résultats provenant de la théorie des processus auto normalisés étudiée dans \cite{PenaLaiShao2009}. Il s'agit ensuite de définir des estimateurs des profils d'actions, avec leurs propres intervalles de confiance. L'utilisation de ces derniers nous permettra de définir une borne supérieure de l'intervalle de confiance associé à chaque récompense et par conséquent de définir une politique optimiste.

\textbf{Notations:}  Etant donnée une matrice définie positive $A\in \mathbb{R}^{d\times d}$ on note dans la suite $\lambda_{min}(A)$ sa plus petite valeur propre.

\subsection{Régression et intervalle de confiance}

\begin{prop}
\label{SubGaussianProp}
Supposons que pour tout $i$, à chaque temps $t$, les échantillons $x_{i,t} \in \mathbb{R}^{d}$ sont iid d'une loi de moyenne $\mu_{i}\in \mathbb{R}^{d}$ , et qu'il existe un réel $L>0$ tel que $||x_{i,t}||\leq L$ et un réel $S>0$ tel que  $||\beta||\leq S$. Alors pour tout $i$ et tout $s\leq t$, la variable aléatoire $\epsilon_{i,t}^{\top}\beta+\eta_{i,s}$ est conditionnellement sous-gaussienne de constante 
$R_{i,t}=\sqrt{R^{2}+\dfrac{L^{2}S^{2}}{n_{i,t}}}$.
\end{prop}

\begin{preuve}
Disponible en annexe \ref{PreuveSubGauss}.
\end{preuve}

Cette proposition nous permettra dans la suite de construire un estimateur du paramètre $\beta$ et d'un intervalle de confiance associé. On se place à un instant $t$, après avoir observer les échantillons de profil $x_{i,t}$  des actions $i$ appartenant à $\mathcal{O}_{t}$, mais avant d'effectuer la sélection de l'action $i_{t}$. On dispose donc de l'ensemble d'observations suivantes: $\left\{r_{i_{s},s},\hat{x}_{i_{s},t}\right\}_{s=1..t-1}$.

Les notations vectorielles et matricielles suivantes sont utilisées:

\begin{itemize}
\item $\eta^{'}_{t-1}=(\eta_{i_s,s}+\epsilon_{i_{s},t}^{\top}\beta)_{s=1..t-1}^{\top}$ le vecteur des bruits de taille $t-1$. Par convention on prend $\eta^{'}_{0}=0$.
\item $X_{t-1}=(\hat{x}_{i_{s},t}^{\top})_{s=1..t-1}$ la matrice de taille $(t-1)\times d$ des moyennes empiriques des actions choisies, où la $s^{ieme}$ ligne correspond à la moyenne empirique au temps $t$ de l'action choisie au temps $s$. Il est important de noter que, contrairement au cas du bandit linéaire classique dans lequel à chaque instant on ajoute une ligne à cette matrice sans modifier les autres, ici dès qu'un bras révèle un échantillon de profil, sa moyenne empirique est modifiée et par conséquent toutes les lignes de $X_{t}$ associées le sont aussi. Nous verrons par la suite qu'il est possible de faire ces mises à jour de façon efficace. Par convention on prend $X_{0}=0_{d}$.
\item $Y_{t-1}=(r_{i_{s},s})_{s=1..t-1}^{\top}$ le vecteur des récompenses de taille $t-1$. Par convention on prend $Y_{0}=0$.
\item $A_{t-1}=diag(1/R_{i_{s},t})_{s=1..t-1}$ la matrice diagonale de taille $(t-1)\times (t-1)$ dont le $s^{ieme}$ élément diagonal vaut $1/R_{i_{s},t}$. Notons que, pour un bras spécifique, la valeur de ce coefficient augmente à mesure que son nombre d'observations augmente. Par convention on prend $A_{0}=1$.
\end{itemize}

Avec les notations précédentes, le problème de régression au temps $t$ peut s'écrire sous forme matricielle de la façon suivante:

\begin{equation}
\label{probReg}
Y_{t-1}=X_{t-1}\beta+\eta^{'}_{t-1}
\end{equation}

\begin{prop}
On note $\hat{\beta}_{t-1}$ l'estimateur des moindres carrés  $l^{2}$-régularisé de $\beta$ associé au problème défini dans l'équation \ref{probReg}, où chaque exemple est pondéré par le coefficient associé $1/R_{i_{s},t}$. On a:

\begin{align}
\label{regressor}
\hat{\beta}_{t-1} &= \arg\min_{\beta} \sum\limits_{s=1}^{t-1} \dfrac{1}{R_{i_{s},t}}(\beta^{\top}\hat{x}_{i_{s},t}- r_{i_{s},s})^{2} + \lambda ||\beta||^{2}\\
&=(X_{t-1}^{\top}A_{t-1}X_{t-1}+\lambda I)^{-1}X_{t-1}^{\top}A_{t-1}Y_{t-1}
\end{align}

Où $\lambda>0$ est la constante de régularisation.
\end{prop}

\begin{preuve}
En écrivant le problème de minimisation sous la forme matricielle suivante:

$\hat{\beta}_{t-1} = \argmin\limits_{\beta}(Y_{t-1}-X_{t-1}\beta)^{\top}A_{t-1}(Y_{t-1}-X_{t-1}\beta)+\lambda \beta^{\top}\beta$ et en annulant le gradient, on obtient directement le résultat souhaité.
\end{preuve}

Cet estimateur utilise les moyennes empiriques comme exemples d'apprentissages, cependant afin de prendre en compte l'incertitude associée à cette approximation, nous pondérons chaque exemple par un coefficient traduisant la confiance que l'on a en ce dernier. Ce coefficient, par définition, tendra vers une constante à mesure que le nombre d'observations augmente. Il nous permet également, conformément au théorème suivant, de définir un intervalle de confiance sur l'estimateur proposé.

\begin{theo}
\label{ConfInt}
Soit $V_{t-1}=\lambda I+ X_{t-1}^{\top}A_{t-1}X_{t-1}=\lambda I +\sum\limits_{s=1}^{t-1}\dfrac{\hat{x}_{i_{s},t}\hat{x}_{i_{s},t}^{\top}}{R_{i_{s},t}}$, alors avec les mêmes hypothèses que dans la proposition \ref{SubGaussianProp}, pour tout $0<\delta<1$, avec une probabilité au moins égale à $1-\delta$, pour tout $t \geq 0$:
\begin{equation}
\label{confIntTheta}
||\hat{\beta}_{t-1}-\beta||_{V_{t-1}} \leq   \sqrt{2\log\left(\dfrac{det(V_{t-1})^{1/2}det(\lambda I)^{-1/2}}{\delta}\right)} +\sqrt{\lambda}S=\alpha_{t-1}
\end{equation}
\end{theo}

\begin{preuve}
Disponible en annexe \ref{PreuveConfInt}.
\end{preuve}

Cet intervalle de confiance ressemble à celui utilisé dans l'algorithme \texttt{OFUL} \cite{OFUL}, cependant une différence notable vient de la définition de la matrice $V_{t}$ qui prend ici en compte une pondération. Un point  important du résultat du théorème précédent concerne le fait que l'inégalité est vraie uniformément pour tout $t$ avec une probabilité commune. Nous utiliserons ce résultat lorsque nous bornerons le pseudo-regret cumulé de notre algorithme. 

Le théorème suivant établit un intervalle pour les estimateurs des profils d'action.

\begin{theo}
\label{ConfIntProfils}
Pour tout $i$ et tout $t>0$ avec une probabilité au moins égale à $1-\delta/t^{2}$, on a:
\begin{equation}
\label{EqConfIntProfils}
||\hat{x}_{i,t}-\mu_{i}|| \leq Ld\sqrt{\dfrac{2}{n_{i,t}}\log\left( \dfrac{2dt^{2}}{\delta}\right)}  = \rho_{i,t,\delta} 
\end{equation}
\end{theo}

\begin{preuve}
Cette inégalité provient de l'application de l'inégalité de Hoeffding à chaque dimension séparément. Disponible en annexe \ref{PreuveConfIntProfils}.
\end{preuve}

Contrairement à l'intervalle de confiance associé au paramètre de régression, celui sur les profils n'est pas vrai de façon uniforme. En effet, pour le paramètre $\beta$, on dispose d'un intervalle de confiance pour $\hat{\beta}_{t}$ valable avec une certaine probabilité pour tous les temps $t$ simultanément. En revanche, pour les paramètres $\mu_{i}$ on dispose d'un intervalle de confiance pour $\hat{x}_{i,t}$ valable pour chaque temps séparément. Pour obtenir une probabilité uniforme (c.-à-d. tous les temps $t$ simultanément), nous utilisons le principe de la borne uniforme. Pour un $i$ donné, on a:

$\mathbb{P}(\forall t, ||\hat{x}_{i,t}-\mu_{i}||\leq  \rho_{i,t,\delta} )=1-\mathbb{P}(\exists t, ||\hat{x}_{i,t}-\mu_{i}|| \geq  \rho_{i,t,\delta} ) \geq 1-\sum\limits_{t}\mathbb{P}(||\hat{x}_{i,t}-\mu_{i}|| \geq  \rho_{i,t,\delta} ) \geq 1-\sum\limits_{t}\delta/t^{2}$. 

Ceci nous permet de justifier  la forme en $\delta/t^{2}$ permettant, une fois sommée sur tous les pas de temps, d'obtenir une probabilité uniforme. En effet, on a $\sum\limits_{t=2}^{\infty}\delta/t^{2} = \delta(\pi^{2}/6-1)\leq \delta$.

\begin{note}[Précision de l'inégalité de Hoeffding en plusieurs dimensions]

La figure \ref{fig:illustHoeffding} illustre l'effet provoqué par notre application de l'inégalité de Hoeffding à chaque dimension de façon individuelle. Le cercle représente l'hypothèse de travail de base, à savoir $||x_{i,t}||\leq L$. Cependant, en appliquant l'inégalité de Hoeffding à chaque dimension, il s'avère que nous majorons en quelque sorte cette hypothèse. Ainsi le carré illustre la nouvelle hypothèse, moins restrictive que la précédente. En effet, la zone à l'intérieur du carré et à l'extérieur du cercle représente une zone dans laquelle par hypothèse, on ne peut pas se trouver. Par conséquent, en pratique, on pourra prendre $\rho_{i,t,\delta} = \min\left( Ld\sqrt{\dfrac{2}{n_{i,t}}\log\left( \dfrac{2dt^{2}}{\delta}\right)},2L\right)$.
\end{note}

        \begin{figureth}
			\includegraphics[width=0.8\linewidth]{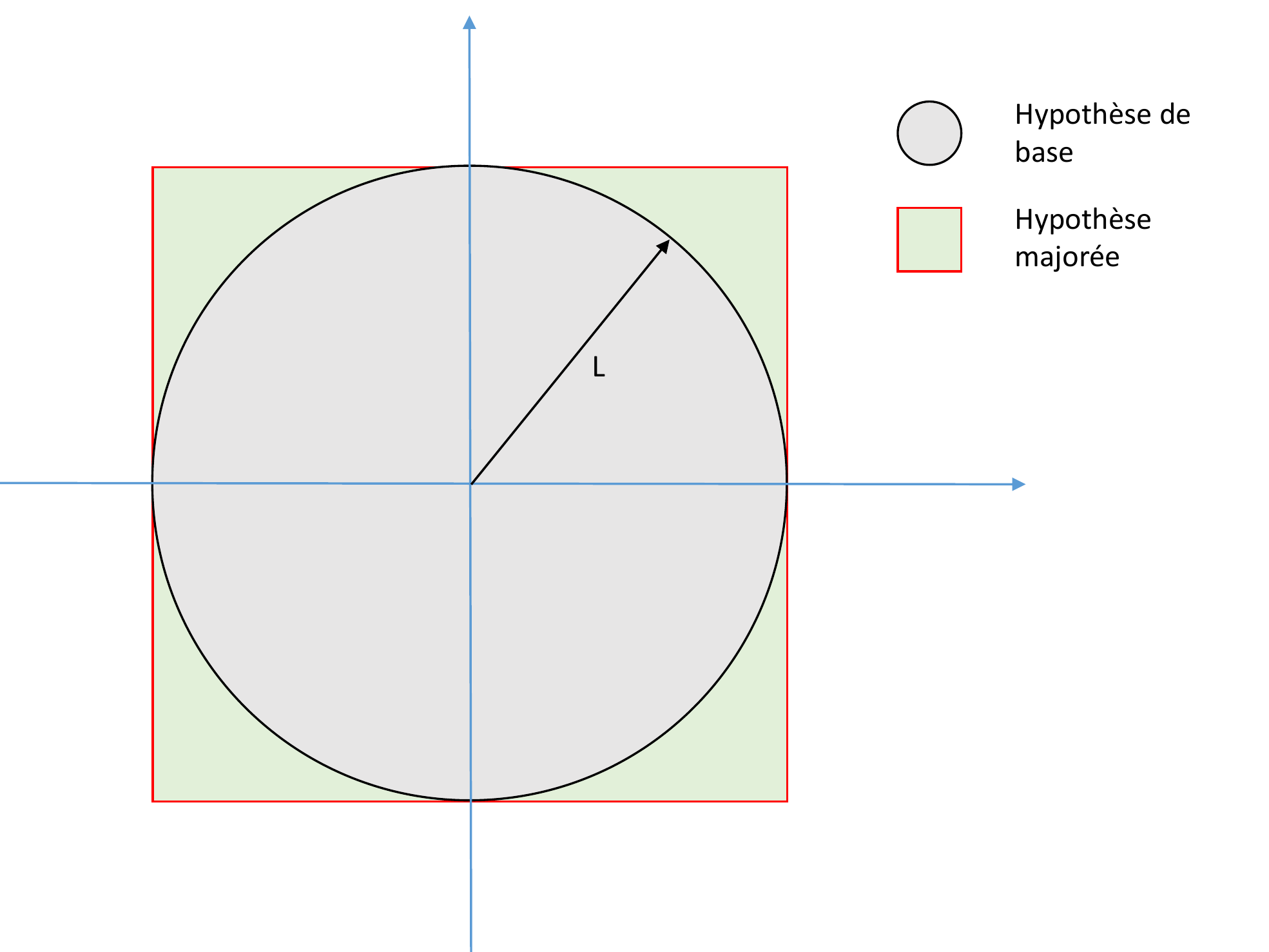}
        \caption[Illustration inégalité Hoeffding en plusieurs dimensions]{Illustration de l'application de l'inégalité de Hoeffding sur chaque dimension dans le cas $d=2$.}
			\label{fig:illustHoeffding}
	\end{figureth}

Nous sommes maintenant prêts à proposer un algorithme adapté à notre problème, ce qui est l'objet de la section suivante.

	\subsection{Présentation de l'algorithme}
 Le principe derrière les algorithmes de type UCB  est de maintenir un intervalle de confiance sur la récompense associée à chaque action et d'en sélectionner une de façon optimiste à chaque instant, c'est-à-dire celle avec la borne supérieure de l'intervalle de confiance de son utilité la plus élevée. D'une façon générale, plus serrés seront les intervalles de confiance que nous sommes capables de construire, meilleure sera la borne sur le regret. Dans le cas du bandit linéaire classique, cela revient directement à construire un bon estimateur du paramètre de régression $\beta$, comme nous l'avons fait dans la section précédente. Cependant dans notre cas, une source additionnelle d'incertitude est présente à cause de la variabilité des échantillons observés. Ainsi, à chaque instant $t$ il est nécessaire de définir une nouvelle borne supérieure de l'intervalle de confiance pour chaque récompense. Dans cette optique, nous proposons la politique  \texttt{SampLinUCB}, que nous détaillons dans l'algorithme \ref{algoSampLinUCB}. Il procède de la façon suivante:

 \begin{enumerate}
 \item Les paramètres communs $V$ et $b$ sont dans un premier temps initialisés (lignes 1 et 2).
 \item On initialise chaque action en la sélectionnant un fois: une récompense et un échantillon de contexte sont observés, ce qui permet d'initialiser les paramètres propres à chaque action (lignes 3 à 9). Notons que si cette initialisation est indispensable pour le cas 3, dans le cas 2 (et \textit{a fortiori} dans le cas 1) il n'est pas nécessaire de sélectionner une fois chaque action. En revanche, dans le cas 2, une action dont un échantillon de profil n'a jamais été observé ne peut pourra pas être sélecionnée.
 \item Ensuite, à chaque itération $t$, pour chaque  action $i$ appartenant à $\mathcal{O}_{t}$ (selon les cas 1, 2 ou 3 définis précédemment), on observe l'échantillon $x_{i,t}$  associé et on met à jour le nombre d'observation $n_{i}$, la moyenne empirique $\hat{x}_{i}$ ainsi que le paramètre d'incertitude $R_{i}$ (ligne 15). Les paramètres commun $V$ et $b$ sont également mis à jour en fonction de ces nouvelles valeurs en retranchant les anciennes (ligne 13) et en ajoutant les nouvelles (ligne 16). Notons que cette mise à jour a une emprunte mémoire faible.
 \item L'algorithme calcule ensuite le score $s_{i,t}$ (ligne 20) de chaque action en fonction de ses paramètres (voir l'équation \ref{scoreSampLinUCB} et la description des éléments du score ci-après), puis sélectionne l'action ayant le score le plus élevé (ligne 22), reçoit la récompense associée (ligne 23) et met à jour ses paramètres (lignes 24 à 26).
 \end{enumerate}

  \begin{algorithm}[h!]
 $V =\lambda I_{d\times d}$ (matrice identité de taille $d$)\; 
 $b = 0_{d}$ (vecteur nul de taille $d$)\;
 \For{$t=1..K$} {
 	Sélectionner $i_{t}=t$ et recevoir $r_{i_{t},t}$\;
    	$N_{i_{t}}=1$; $S_{i_{t}}=r_{i_{t},t}$\;
        Observer un échantillon de contexte $x_{i,t}$\;
        $n_{i_{t}}=1$; $\hat{x}_{i_{t}}=x_{i,t}$; $R_{i_{t}}=\sqrt{1+\dfrac{L^{2}S^{2}}{n_{i_{t}}}}$\;
    	$V=V+N_{i}\dfrac{\hat{x}_{i}\hat{x}_{i}^{\top}}{R_{i}}$; 
        $b= b+S_{i}\dfrac{\hat{x}_{i}}{R_{i}}$\;
 }
\For{$t=K+1..T$} {

	Recevoir $\mathcal{O}_{t}$\;
     \label{getnewObservationsSamplin}
	\For{$i  \in \mathcal{O}_{t}$} {
            $V=V-N_{i}\dfrac{\hat{x}_{i}\hat{x}_{i}^{\top}}{R_{i}}$; 
            $b = b-S_{i}\dfrac{\hat{x}_{i}}{R_{i}}$\;
        Observer $x_{i,t}$\;
        $n_{i}=n_{i}+1$; 
        $\hat{x}_{i}=\dfrac{(n_{i}-1)\hat{x}_{i}+x_{i,t}}{n_{i}}$; 
        $R_{i}=\sqrt{1+\dfrac{L^{2}S^{2}}{n_{i}}}$\; 
		$V=V+N_{i}\dfrac{\hat{x}_{i}\hat{x}_{i}^{\top}}{R_{i}}$; 
        $b= b+S_{i}\dfrac{\hat{x}_{i}}{R_{i}}$\;
       
	}
    	$\hat{\beta} = V^{-1}b$\;\label{updateThetaSamplin}
    \For{$i  \in \mathcal{K}$} {
		Calculer $s_{i,t}$ avec la formule \ref{scoreSampLinUCB} \;
	}
	Sélectionner $i_{t} = \argmax\limits_{i \in \mathcal{K}} \quad s_{i,t}$ \;
	Recevoir $r_{i_{t},t}$\;
    $N_{i_{t}}=N_{i_{t}}+1$\;
    $S_{i_{t}}=S_{i_{t}}+r_{i_{t},t}$\;
	$V= V+\dfrac{\hat{x}_{i_{t}}\hat{x}_{i_{t}}^{\top}}{R_{i_{t}}}$; 
    $b = b+r_{i_{t},t}\dfrac{\hat{x}_{i_{t}}}{R_{i_{t}}}$\;

	}
\caption{\texttt{SampLinUCB}}
\label{algoSampLinUCB}
\end{algorithm}

 Le score de sélection $s_{i,t}$ pour chaque action $i$ et à chaque instant $t$ est le suivant:
 
\begin{align}
\label{scoreSampLinUCB}
s_{i,t} =(\hat{x}_{i,t}+\overline{\epsilon}_{i,t})^{\top}\hat{\beta}_{t-1} + \alpha_{t-1} ||\hat{x}_{i,t}+\tilde{\epsilon}_{i,t}||_{V_{t-1}^{-1}}
\end{align}

avec:
\begin{itemize}

\item $\overline{\epsilon}_{i,t} =\dfrac{\rho_{i,t,\delta} \hat{\beta}_{t-1}}{||\hat{\beta}_{t-1}||}$;

\item $\tilde{\epsilon}_{i,t} = \dfrac{\rho_{i,t,\delta} \hat{x}_{i,t}}{\sqrt{\lambda}||\hat{x}_{i,t}||_{V_{t-1}^{-1}}}$;

\end{itemize}


Les vecteurs $\overline{\epsilon}_{i,t}$ et $\tilde{\epsilon}_{i,t}$ ont pour rôle de gérer l'incertitude sur les profils estimés. Intuitivement, ils autorisent l'algorithme à sélectionner des actions dont, soit le profil est dans une zone intéressante, soit l'incertitude sur le profil est grande, le but étant d'éliminer les actions étant les moins bonnes avec une forte probabilité (celles dont l'incertitude sur le profil ne permet pas d'atteindre une zone potentiellement intéressante, relativement à l'estimation courante du paramètre $\beta$). Nous proposons ci-dessous d'illustrer les différents paramètres de l'algorithme pour mieux en comprendre le fonctionnement.

\textbf{Illustration du fonctionnement de l'algorithme:} La figure \ref{fig:illustSamplLinUCB} représente une illustration des différentes grandeurs intervenant dans l'algorithme \texttt{SampLinUCB} pour une action donnée dans un espace à deux dimensions. Pour ne pas surcharger la figure, nous avons enlevé les indices de temps $t$ et d'action $i$. On représente les différents vecteurs par des flèches, ainsi la rouge représente $\hat{x}$, la  violette $\hat{x}+\tilde{\epsilon}$ et la verte $\hat{x}+\overline{\epsilon}$. Conformément à la définition de $\tilde{\epsilon}$,  $\hat{x}$ et  $\hat{x}+\tilde{\epsilon}$ sont colinéaires. En revanche, ce n'est pas le cas de $\hat{x}$ et  $\hat{x}+\overline{\epsilon}$, car $\overline{\epsilon}$ (qui correspond au vecteur joignant le bout de la flèche rouge à celui de la flèche verte) est colinéaire à $\hat{\beta}$. La projection de $\hat{x}+\overline{\epsilon}$ sur $\hat{\beta}$ (correspondant à $(\hat{x}+\overline{\epsilon})^{\top}\hat{\beta}$) a  une valeur plus élevée que la  projection de $\hat{x}$ sur $\hat{\beta}$ (correspondant à $\hat{x}^{\top}\hat{\beta}$). Par ailleurs, la valeur de $||\hat{x}+\tilde{\epsilon}||$ est toujours plus élevée que la valeur de $||\hat{x}||$. 
Ces deux phénomènes sont directement dus au fait que, d'une part, $\overline{\epsilon}$ est colinéaire à $\hat{\beta}$ et de même sens, et d'autre part, $\tilde{\epsilon}$  est colinéaire à $\hat{x}$ et de même sens. Ainsi, les vecteurs $\overline{\epsilon}$ et $\tilde{\epsilon}$ prenant en compte les incertitudes sur les profils ont toujours une contribution positive par rapport à un score ne les prenant pas en compte, ce qui traduit le fait que l'algorithme est optimiste sur le profils.

        \begin{figureth}
			\includegraphics[width=0.6\linewidth]{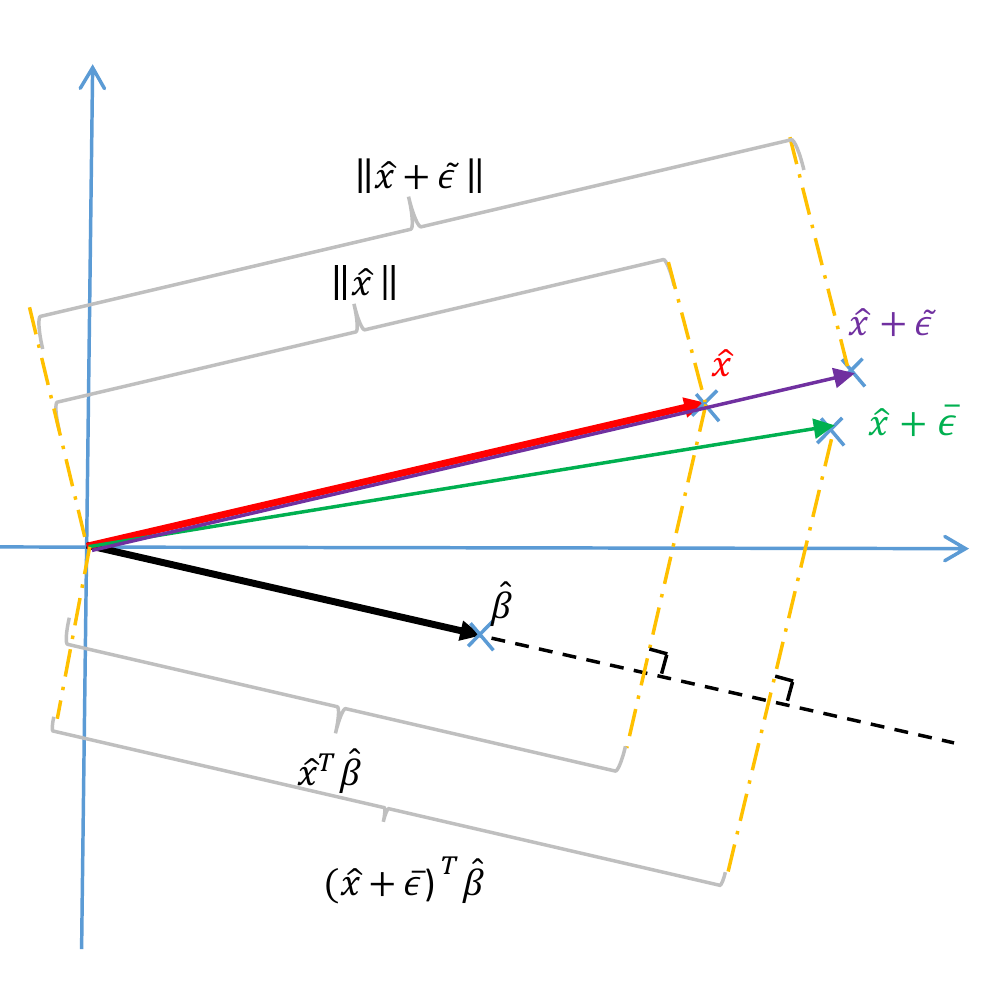}
        \caption[Illustration \texttt{SampLinUCB}]{Illustration du fonctionnement de l'algorithme \texttt{SampLinUCB} dans un espace à deux dimensions.}
			\label{fig:illustSamplLinUCB}
	\end{figureth}

    \begin{prop}
   Le score $s_{i,t}$ de l'équation \ref{scoreSampLinUCB} peut se réécrire de la façon suivante:
       \begin{align}
\label{scoreSampLinUCBBis}
s_{i,t} =\hat{x}_{i,t}^{\top}\hat{\beta}_{t-1} + \alpha_{t-1} ||\hat{x}_{i,t}||_{V_{t-1}^{-1}}+\rho_{i,t,\delta}\left(||\hat{\beta}_{t-1}||+\dfrac{\alpha_{t-1}}{\sqrt{\lambda}}\right)
		\end{align}
    \end{prop}

    \begin{preuve}
    \begin{align*}
s_{i,t} & =(\hat{x}_{i,t}+\overline{\epsilon}_{i,t})^{\top}\hat{\beta}_{t-1} + \alpha_{t-1} ||\hat{x}_{i,t}+\tilde{\epsilon}_{i,t}||_{V_{t-1}^{-1}}\\
		&= \hat{x}_{i,t}^{\top}\hat{\beta}_{t-1} + \dfrac{\rho_{i,t,\delta} \hat{\beta}_{t-1}^{\top}}{||\hat{\beta}_{t-1}||}\hat{\beta}_{t-1} + \alpha_{t-1}||\hat{x}_{i,t}+\dfrac{\rho_{i,t,\delta} \hat{x}_{i,t}}{\sqrt{\lambda}||\hat{x}_{i,t}||_{V_{t-1}^{-1}}}||_{V_{t-1}^{-1}} \\
        &=\hat{x}_{i,t}^{\top}\hat{\beta}_{t-1} +\rho_{i,t,\delta} ||\hat{\beta}_{t-1}|| +\alpha_{t-1}\left(1+\dfrac{\rho_{i,t,\delta}}{\sqrt{\lambda}||\hat{x}_{i,t}||_{V_{t-1}^{-1}}}\right) ||\hat{x}_{i,t}||_{V_{t-1}^{-1}} \\
        &=\hat{x}_{i,t}^{\top}\hat{\beta}_{t-1} + \alpha_{t-1} ||\hat{x}_{i,t}||_{V_{t-1}^{-1}}+\rho_{i,t,\delta}\left(||\hat{\beta}_{t-1}||+\dfrac{\alpha_{t-1}}{\sqrt{\lambda}}\right)
	\end{align*}
	\end{preuve}

Cette nouvelle forme du score associé à chaque action nous permet d'avoir un œil différent sur le comportement de l'algorithme à chaque instant. La première partie du score $\hat{x}_{i,t}^{\top}\hat{\beta}_{t-1} + \alpha_{t-1} ||\hat{x}_{i,t}||_{V_{t-1}^{-1}}$ ressemble fortement au score de l'algorithme \texttt{OFUL} classique ne prenant pas en compte d'incertitude sur les vecteurs de contexte. D'autre part, le second terme $\rho_{i,t,\delta}\left(||\hat{\beta}_{t-1}||+\dfrac{\alpha_{t-1}}{\sqrt{\lambda}}\right)$ est directement proportionnel au coefficient $\rho_{i,t,\delta}$ de chaque action, puisque les autres termes sont communs à tous les bras. On voit bien apparaître une "prime" donnée aux bras ayant été les moins observés. Notons que si l'on considère le cas 1, c'est à dire celui où on tous les bras délivrent un échantillon à chaque itération, alors le second terme est le même pour tous et peut donc être retiré du score de sélection. Ainsi dans ce cas, l'algorithme proposé revient à un algorithme \texttt{OFUL} qui utiliserait des moyennes empiriques en guise de vecteur de contexte, outre le mode de mise à jour des paramètres de régression, qui contiennent l'incertitude sur les  profils dans notre cas. Cependant, cette reformulation du score de sélection devient particulièrement intéressante lorsque l'on se penche sur le cas 3, c'est-à-dire celui où seule l'action choisie au temps précédent délivre un échantillon. Dans cette situation, la prime donnée aux bras peu observés traduit directement le fait que l'algorithme cherche à la fois à diminuer l'incertitude sur les paramètres de régression, et celle sur les profils des différentes actions. Sans ce terme supplémentaire, l'algorithme pourrait aisément se bloquer sur des actions sous-optimales (voir l'exemple proposé précédemment). Pour s'en convaincre, nous proposons d'analyser un scénario simple ci-dessous.
 
 \textbf{Exemple pour le cas 3: }
Considérons un scénario à deux actions $A$ et $B$ avec $\mu_{A}^{\top}\beta>\mu_{B}^{\top}\beta$, un espace des profils de dimension 2, des contextes dont toutes les composantes sont bornées (entre 0 et 1) et des récompenses bornées (entre 0 et 1). Le processus est initialisé de la façon suivante: à l'instant $t=1$, l'action $A$ est sélectionnée et la récompense $r_{A}$ est reçue (les indices de temps ont été enlevés pour alléger les notations). A l'instant $t=2$, conformément au cas étudié, on commence par observé un échantillon de contexte $x_{A}$ pour la dernière action sélectionnée puis l'action $B$ est sélectionnée et la récompense $r_{B}$ est reçue. La stratégie de sélection débute à l'instant $t=3$, pour lequel, on commence par observé un échantillon de contexte $x_{B}$ pour la dernière action sélectionnée. Considérons de plus que $x_{A}=(0,0)$ et $x_{B}=(1,1)$ (ce qui est une possibilité en raison du caractère aléatoire des échantillons). Si l'on considère un algorithme naïf n'utilisant que les moyennes empiriques (et pas les $\epsilon$), alors, on a pour chaque action un score de la forme $s_{i}=\hat{x}_{i}^{\top}\hat{\beta}+\alpha||\hat{x}_{i}||_{V^{-1}}$. 
On a donc $s_{A}=0$  car $\hat{x}_{A}=(0,0)$ d'après l'échantillon observé. On peut par ailleurs montrer que toutes les composante de $\hat{\beta}$ sont positives dans ce cas, ce qui entraîne $s_{B}>0$ car $\hat{x}_{B}=(1,1)$ d'après l'échantillon observé. Ainsi, à l'instant 3, c'est l'action $B$ qui serait sélectionnée. En répétant ce processus aux itérations suivantes, il apparaît que l'algorithme reserait bloquée sur l'action $B$, qui d'après l'hypothèse de départ est sous-optimale. Nous venons donc de trouver un cas dans lequel une application naïve de l'algorithme \texttt{OFUL} conduirait à un regret linéaire en raison d'une "mauvaise" initialisation due à l'aspect aléatoire des échantillons de profils observés.

Dans la section suivante, nous dérivons une borne du regret générique pour notre algorithme, puis, pour chacun des trois cas, nous proposons une borne spécifique. Nous finissons par étendre l'algorithme - ainsi que sa borne - au cas de la sélection multiple.

	\subsection{Regret}
   	\label{chap1RegSection}
    
    Le théorème ci-dessous établit une borne supérieure du regret cumulé de l'algorithme proposé dans la section précédente. La borne concerne le cas générique, où aucune hypothèse spécifique n'est faite sur le processus générant $\mathcal{O}_{t}$ à chaque itération $t$.
    
\begin{theo}[Borne générique]
\label{RegretSamplLinUCBGeneric}
En choisissant $\lambda \geq \max(1,2L^{2})$, avec une probabilité au moins égale à $1-3\delta$, le pseudo-regret cumulé de l'algorithme \texttt{SampLinUCB} est majoré par:

\begin{multline}
\label{cumulRegSamplLin}
  \hat{R}_{T} \leq C + 4Ld\left(\sqrt{\dfrac{d}{\lambda}\log\left(\dfrac{1+TL^{2}/\lambda}{\delta}\right)}+2S\right)\sqrt{\log\left( \dfrac{2dT^{2}}{\delta}\right)} \sum\limits_{t=1}^{T} \dfrac{2}{\sqrt{n_{i_{t},t}}} \\
  +  2\left(\sqrt{d\log\left(\dfrac{1+TL^{2}/\lambda}{\delta}\right)} + \sqrt{\lambda}S\right) \\
\times	 \sqrt{Td\left(\sqrt{R^{2}+L^{2}S^{2}} \log\left( 1 + \dfrac{TL^{2}}{\lambda d} \right)  + \dfrac{4L^{2}d^{2}}{\lambda}\log\left( \dfrac{2dT}{\delta}\right) \sum\limits_{t=1}^{T}  \dfrac{1}{n_{i_{t},t}}  \right)}
\end{multline}

\end{theo}

\begin{preuve}
La preuve complète de ce théorème est disponible en annexe \ref{preuveRegSampLinUCBGeneric}.
\end{preuve}

L'étude des facteurs dominants dans la borne précédente, associée aux trois cas d'études, permet d'établir le théorème suivant, où l'on a enlevé les dépendances en $L$, $\lambda$, $R$ et $S$ volontairement.

\begin{theo}[Borne pour les trois cas d'études]
\label{RegretSamplLinUCBDiffCases}
Pour les trois cas d'études proposés, la borne supérieure du pseudo-regret cumulé prend la forme suivante:
\begin{itemize}
\item Pour le cas 1, avec une probabilité au moins égale $1-3\delta$: 
\begin{equation}
\label{regCas1}
\hat{R}_{T} =\mathcal{O}\left(  d \sqrt{dT\log\left(\dfrac{T}{\delta}\right) \log\left(\dfrac{dT^{2}}{\delta}\right)} \right)
 \end{equation}
 
\item Pour le cas 2, avec une probabilité au moins égale $(1-3\delta)(1-\delta)$, et pour $T \geq 2\log(1/\delta)/p^{2}$:
\begin{equation}
\label{regCas2}
\hat{R}_{T} =\mathcal{O}\left(  d \sqrt{\dfrac{dT}{p}\log\left(\dfrac{T}{\delta}\right) \log\left(d\dfrac{T^{2}}{\delta}\right)} \right)
  \end{equation}
  Où $p$ est la probabilité pour chaque bras de délivrer un échantillon de profil à chaque itération.

\item Pour le cas 3, avec une probabilité au moins égale $1-3\delta$:
\begin{equation}
\label{regCas3}
\hat{R}_{T} =\mathcal{O}\left(  d \sqrt{dTK\log\left(\dfrac{T}{\delta}\right) \log\left(d\dfrac{T^{2}}{\delta}\right)} \right)
  \end{equation}

\end{itemize}

On rappelle que le symbole $\mathcal{O}$ traduit la relation "dominé par", autrement dit $f=\mathcal{O}(g)$ signifie qu'il existe une constante strictement positive $C$ telle qu' asymptotiquement on a: $|f|\leq C|g|$. 

\end{theo}

\begin{preuve}
Les trois preuves sont disponibles en annexe \ref{preuveRegSampLinUCBCas1}, \ref{preuveRegSampLinUCBCas2} et \ref{preuveRegSampLinUCBCas3} respectivement pour les cas 1, 2 et 3.
\end{preuve}

On remarque que dans les trois cas, la borne est bien sous-linéaire. On propose maintenant de faire une analyse qualitative des différents termes intervenant dans ces trois bornes.

\begin{itemize}
\item \textbf{Cas 1}:
Le facteur dominant de la borne du regret possède une forte dépendance en $d$, le nombre de dimensions de l'espace des profils. Une partie faible, d'un facteur $\sqrt{d}$, provient de l'incertitude liée à l'estimateur du paramètre de régression $\beta$. La plus grande part de la dépendance en $d$ vient cependant de l'incertitude liée à l'estimation des profils. En effet, cette dernière entraîne la présence d'un terme en $d\sqrt{\log(d)}$. Ceci est également valide pour les cas 2 et 3, qui possèdent des dépendances supplémentaires étudiées ci-après.
\item \textbf{Cas 2}:
La borne décrite dans l'équation \ref{regCas2} possède une dépendance en $p$. Plus la probabilité d'observer des échantillons est élevée, plus l'incertitude sera faible et plus l'algorithme sera performant. De plus, on note que cette borne n'est valide que pour un horizon $T$ supérieur à une valeur inversement proportionnelle à $p^{2}$. Intuitivement, ceci est dû au fait que plus $p$ est petit, plus un nombre élevé d'itérations est nécessaire pour s'assurer qu'un minimum d'observations a été effectué.
\item \textbf{Cas 3}:
La borne décrite dans l'équation \ref{regCas3} possède une dépendance en $\sqrt{K}$, le nombre total d'actions disponibles. Plus le nombre de bras augmente, plus la borne sera large. Ce facteur provient du fait que dans ce scénario, seule l'action sélectionnée à une itération révèle un échantillon de profil à l'étape suivante.
\end{itemize}

\begin{note}
Ces trois bornes ne sont pas optimales, au sens où elles sont plus larges que la borne inférieure du théorème \ref{infRegStoMP} valide pour tous les algorithmes de bandit stationnaire. En effet, il est important de rappeler que notre problème peut aussi être vu comme un problème stationnaire en remarquant que dans le modèle on a $\mu_{i}^{\top}\beta$ constant pour chaque action. En revanche, dans certains cas, en particulier lorsque le nombre d'actions devient grand, il peut être utile d'utiliser les informations de profils pour mieux explorer l'espace des récompenses. En effet, là où les algorithmes de bandit stationnaire classiques devront reconsidérer plusieurs fois les actions de mauvaise qualité, nous pensons que notre méthode peut tirer parti de la mutualisation effectuée dans l'apprentissage. Ceci fait l'objet d'expérimentations dans la section \ref{XPSampLinUCB}.
\end{note}


Dans la section suivante, nous proposons d'étendre le travail précédent au cas qui nous intéresse le plus, à savoir celui où plusieurs actions ($k>1$) sont sélectionnées simultanément à chaque tour.

\section{Extension au cas de la sélection multiple}

Cette courte section nous permet de faire le lien entre l'étude proposée ci-dessus et notre tâche de collecte d'information dans laquelle on a $k>1$. L'algorithme \texttt{SampLinUCB} (voir algorithme  \ref{algoSampLinUCB}) peut s'étendre aisément à ce scénario en spécifiant qu'à chaque instant $t$, plutôt que de sélectionner l'action $i$ ayant le meilleur score $s_{i,t}$ (voir équation \ref{scoreSampLinUCB}), la version étendue sélectionne les $k$ meilleures actions ordonnées selon leur score $s_{i,t}$. On rappelle que l'ensemble des actions choisies à est notée $\mathcal{K}_{t}$ au temps $t$. 

\begin{defi}
Le pseudo-regret instantané d'un algorithme de bandit contextuel  avec $k$ sélections à un instant $t$, noté $reg_{t}$; est défini par:
\begin{equation}
reg_{t}=\sum\limits_{i\in\mathcal{K}^{\star}}\mu_{i}^{\top}\beta-\sum\limits_{i\in\mathcal{K}_{t}}\mu_{i}^{\top}\beta
\end{equation}
Avec  $\mathcal{K}^{\star}$ l'ensemble des $k$ actions optimales, c.-à-d. ayant les plus grandes valeurs $\mu_{i}^{\top}\beta$.
\end{defi}

\begin{defi}
Le pseudo-regret associé, noté $\hat{R}_{T}$ est défini par:
\begin{equation}
\hat{R}_{T}=\sum\limits_{t=1}^{\top}reg_{t}
\end{equation}
\end{defi}

\begin{theo}[Borne générique avec sélection multiple]
\label{RegretSamplLinUCBGenericMult}
En choisissant $\lambda \geq \max(1,2L^{2})$, avec une probabilité au moins égale à $1-3\delta$, le pseudo-regret cumulé de l'algorithme \texttt{SampLinUCB} avec sélections multiples est majoré par:

\begin{multline}
\label{cumulRegSamplLinMult}
  \hat{R}_{T} \leq C + 4Ld\left(\sqrt{\dfrac{d}{\lambda}\log\left(\dfrac{1+TkL^{2}/\lambda}{\delta}\right)}+2S\right)\sqrt{\log\left( \dfrac{2dT^{2}}{\delta}\right)} \sum\limits_{t=1}^{T} \sum\limits_{i\in \mathcal{K}_{t}}\dfrac{2}{n_{i,t}} \\
  +  2\left(\sqrt{d\log\left(\dfrac{1+TkL^{2}/\lambda}{\delta}\right)} + \sqrt{\lambda}S\right) \\
\times	 \sqrt{Td\left(\sqrt{R^{2}+L^{2}S^{2}} \log\left( 1 + \dfrac{TkL^{2}}{\lambda d} \right)  + \dfrac{4L^{2}d^{2}}{\lambda}\log\left( \dfrac{2dT}{\delta}\right) \sum\limits_{t=1}^{T} \sum\limits_{i \in \mathcal{K}_{t}}\dfrac{1}{\sqrt{n_{i,t}}}  \right)}
\end{multline}

\end{theo}

\begin{preuve}
La preuve complète de ce théorème, dont les arguments sont très proches de la preuve pour le cas de la sélection unique, est disponible en annexe \ref{preuveRegSampLinUCBCasMP}.
\end{preuve}

Des bornes équivalentes pour les trois  cas peuvent être démontrées, à partir de la formulation ci-dessous et de la preuve du théorème \ref{RegretSamplLinUCBDiffCases}. 

Dans la section suivante, nous proposons d'expérimenter notre algorithme d'une part sur des données simulées, mais aussi sur des données réelles de réseau social, afin d'évaluer les performances empiriques de l'algorithme \texttt{SampLinUCB}. Pour le scénario de collecte de données, nous présenterons le modèle de contexte utilisé.

\section{Expérimentations}
\label{XPSampLinUCB}
Cette partie est divisée en deux sections. D'une part les expérimentations sur des données simulées, et d'autre part sur des données réelles.

	\subsection{Données artificielles}

    	\subsubsection{Protocole}
        
         \textbf{Génération des données:} Afin d'évaluer les performances de notre algorithme, nous l'expérimentation dans un premier temps dans un contexte de sélection simple ($k=1$) sur des données simulées. Pour cela, on fixe un horizon de $T=1000$ pas de temps, un ensemble de $K=50$ actions et $d=5$ dimensions. On tire ensuite un vecteur de régression $\beta$ aléatoirement dans $\left[-S/\sqrt{d}..S/\sqrt{d}\right]^{d}$, de façon à respecter la condition $||\beta||\leq S =1$. Pour chaque bras $i$ on tire un vecteur aléatoire $\mu_{i}$ de façon uniforme dans $\left[-L/\sqrt{d}..L/\sqrt{d}\right]^{d}$ avec $L=1$, de façon à être certain que $||\mu_{i}|| \leq L$. Ensuite, pour chaque itération $t \in \left\{1,...,T \right\}$ du processus, on procède de la façon suivante pour simuler les données:
    
    \begin{enumerate}
    	\item Pour chaque $i \in \left\{1,...,K \right\}$, on tire un échantillon $x_{i,t}$ suivant une loi normale centrée sur $\mu_{i}$, et de matrice de covariance $\sigma^{2}I$. Afin d'analyser l'influence du bruit sur les performances de \texttt{SampLinUCB}, nous testons différentes valeurs pour $\sigma \in \left\{0.1,0.5,1.0\right\}$.  De plus, pour respecter la condition $||x_{i,t}||\leq L=1$, tout en conservant le fait que les échantillons sont centrés sur $\mu_{i}$, la distribution normale est tronquée de façon symétrique autour de $\mu_{i}$. Ce procédé est illustré dans la figure \ref{fig:truncGauss} lorsque $d=1$, où les zones hachurées correspondent aux valeurs qui sont exclues. On représente à gauche un cas où $\mu_{i}>0$ et à droite un cas où $\mu_{i}<0$;
        \item Pour chaque $i \in \left\{1,...,K \right\}$, on tire une récompense $r_{i,t}$ selon une loi normale de moyenne $\mu_{i}^{\top}\beta$ et de variance $R^{2}=0.1$;
    \end{enumerate}

                	\begin{figureth}
		\begin{subfigureth}{0.45\textwidth}
			\includegraphics[width=\linewidth]{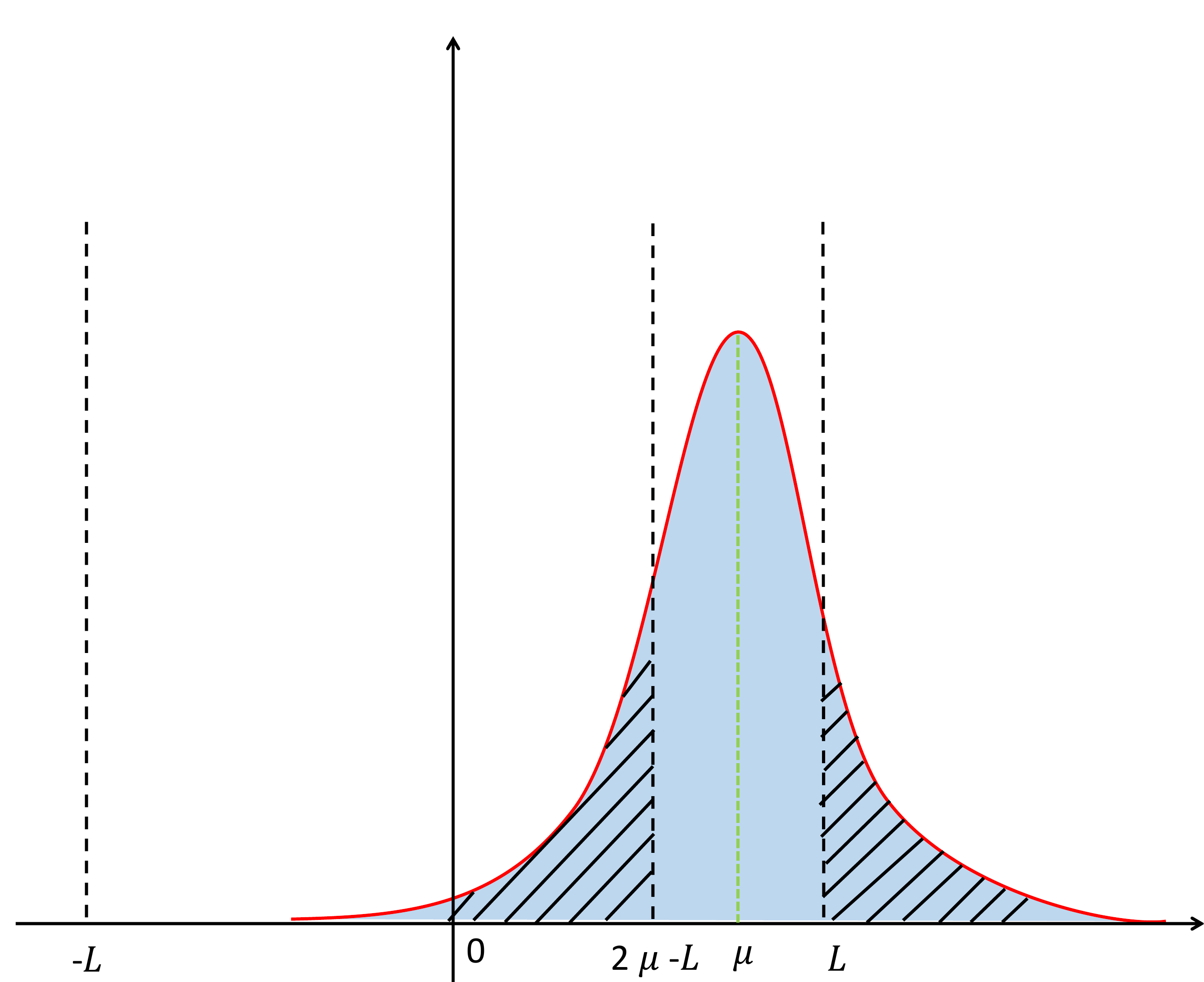}
			\caption{Cas où $\mu>0$}
		\end{subfigureth}
		\begin{subfigureth}{0.45\textwidth}
			\includegraphics[width=\linewidth]{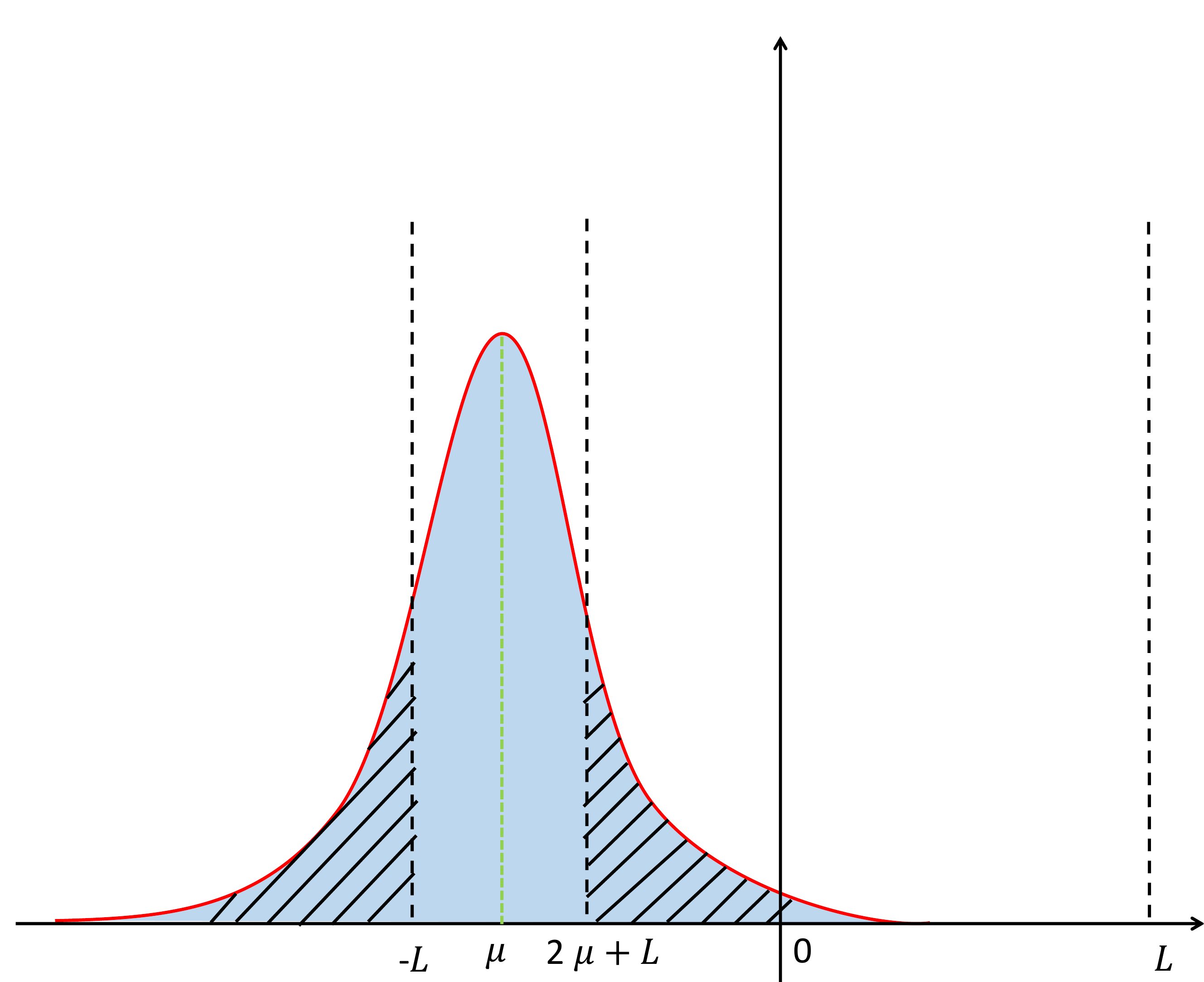}
			\caption{Cas où $\mu<0$}
		\end{subfigureth}
		\caption[Génération des échantillons de profils]{Processus de troncage des lois gaussiennes pour la génération des échantillons de profils bornés.}	
			\label{fig:truncGauss}
			\end{figureth}

    Nous avons généré 1000 jeux de données de cette façon et les résultats donnés plus loin correspondent à des moyennes.

	\textbf{Politiques testées:} Dans ce contexte de bandit stationnaire avec profil bruité, nous proposons de comparer \texttt{SampLinUCB} à \texttt{UCB}, \texttt{UCBV} et \texttt{MOSS}. 
    Soulignons que ces trois politiques n'utilisent aucune information extérieure autre que les récompenses observées. Par conséquent le seul bruit sur les données pour ces algorithmes provient de la loi normale de variance $R^{2}$. En revanche, une source de bruit supplémentaire pour notre algorithme \texttt{SampLinUCB} provient de la génération des échantillons de profils. Tout l'enjeu est de savoir si, bien qu'exploitée selon des échantillons bruités, la structure entre les actions peut être utilisée de façon à améliorer les performances des algorithmes de bandits stationnaires  classiques. Nous implémentons les trois scénarios présentés, en utilisant différentes valeurs de $p\in \left\{0, 0.05,0.5, 1 \right\}$. A noter que pour $p=0$ les actions sélectionnées au temps $t$ délivrent un échantillon de profil, ce qui correspond au cas 3 présenté précédemment.

    	\subsubsection{Résultats}

Les figures \ref{fig:regretVsTimeSimSampLinUCB}(a), \ref{fig:regretVsTimeSimSampLinUCB}(b) et \ref{fig:regretVsTimeSimSampLinUCB}(c) représentent l'évolution du regret des différentes politiques testées au cours du temps, pour des valeurs de $\sigma$  (bruit sur les profils) valant respectivement $\sigma=1.0$, $\sigma=0.5$ et $\sigma=0.1$. Notons que dans les trois représentations, les courbes de \texttt{UCB}, \texttt{UCBV} et \texttt{MOSS} sont identiques étant donné que leurs résultats ne sont pas influencés par le bruit sur les profils. On remarque premièrement que les politiques \texttt{UCB} et \texttt{UCBV} n'offrent pas de bons résultats. Il apparaît que ces deux politiques ont tendance à sur-explorer, du moins pendant la fenêtre de temps de 1000 itérations de l'expérimentation. De plus, dans ce cas précis, les moyennes des actions sont relativement bien réparties dans l'intervalle $\left[-1..1 \right]$ du fait du processus de simulation aléatoire des espérances, et ont toutes la même variance ($R^{2}=0.1$) . Ainsi, l'algorithme \texttt{UCBV} ne parvient pas à utiliser la variance pour identifier les meilleures actions de façon efficace. En revanche, les performances de la politique \texttt{MOSS} semblent bien plus élevées. Nous expliquons cela par le fait qu'il explore l'espace des récompenses de façon plus restreinte que ses deux concurrents, ce qui, dans ce cas précis, lui permet de s'orienter vers des bonnes actions plus rapidement. Cependant, \texttt{MOSS} nécessite de connaître l'horizon $T$, qui intervient directement dans la stratégie de sélection, e qui est relativement limitant en pratique dans le cadre d'utilisation en contexte réel. Par ailleurs, il est difficile d'extrapoler son comportement au-delà de l'horizon de l'expérience.

Penchons-nous maintenant sur les résultats offerts par l'algorithme \texttt{SampLinUCB} dans les divers scénarios. Afin d'isoler les différents effets, nous proposons d'étudier les deux aspects suivants:

\begin{itemize}
\item Influence de la probabilité d'observation des contextes (régi par $p$): La première chose que l'on remarque est que pour toutes les valeurs de bruit sur les profils, \texttt{SampLinUCB} offre de meilleures performances que ses compétiteurs, et ce pour toutes les valeurs de $p$, ce qui confirme la pertinence de notre approche. De plus, le cas où $p=1$, c'est-à-dire où tous les bras délivrent un échantillon de profil à chaque pas de temps, est le plus performant de tous, ce qui paraît cohérent puisque l'algorithme dispose de plus d'information que lorsque $p$ est plus faible. Vient ensuite le cas où $p=0.5$ qui, comme on aurait pu le penser intuitivement, est plus performant que lorsque $p=0.05$ et $p=0$. Ces deux derniers cas, dans lesquels le nombre d'échantillons de profil est beaucoup plus faible (en moyenne 2.5 lorsque $p=0.05$ et exactement 1 lorsque $p=0$) sont les plus intéressants. Nous remarquons que les performances de \texttt{SampLinUCB} sont plus élevées (sur l'horizon étudié) lorsque $p=0$ que lorsque $p=0.05$, ce qui s'explique par le fait que quand $p=0$, l'algorithme est actif non seulement dans l'exploration des récompenses, mais aussi dans l'exploration des profils.  Ceci lui permet donc de mieux s'orienter vers les actions optimales, même si la quantité d'information disponible (en terme de nombre d'échantillons) est moins importante. 

\item Influence de bruit sur les profils (régi par $\sigma$): D'une façon générale, le fait d'augmenter le bruit sur les échantillons de profils visibles par \texttt{SampLinUCB} a pour effet de dégrader ses performances. Ceci semble logique, puisque la convergence des moyenes empirique des différents échantillons (vers les véritables profils) est d'autant plus lente que le bruit est élevé. Lorsque $\sigma=0.1$, c'est à dire lorsque le bruit est le plus faible, les performances de \texttt{SampLinUCB} sont quasiment identiques pour toutes les valeurs de $p$, et toujours très largement supérieures à celles de \texttt{UCB}, \texttt{UCBV} et \texttt{MOSS}. Dans ce cas, les valeurs des échantillons sont très proches des vrais profils, et l'algorithme trouve très rapidement de bons estimateurs. Losque $\sigma$ augmente, les écarts de performances sont plus visibles et on voit très nettement que la probabilité $p$ impacte fortement les résultats. Lorsque $\sigma=0.5$, \texttt{SampLinUCB} est toujours bien meilleur que les autres, en revanche, lorsque $\sigma=1.0$, il apparaît que ses performances ne sont que  légèrement supérieures à celles de la politique \texttt{MOSS} pour $p=0.05$ et $p=0$. Nous expliquons cela par le fait que l'incertitude élevée due au bruit rend l'exploitation des échantillons de profils plus difficiles.
\end{itemize}

                	\begin{figureth}
               \begin{subfigureth}{0.5\textwidth}
			\includegraphics[width=\linewidth]{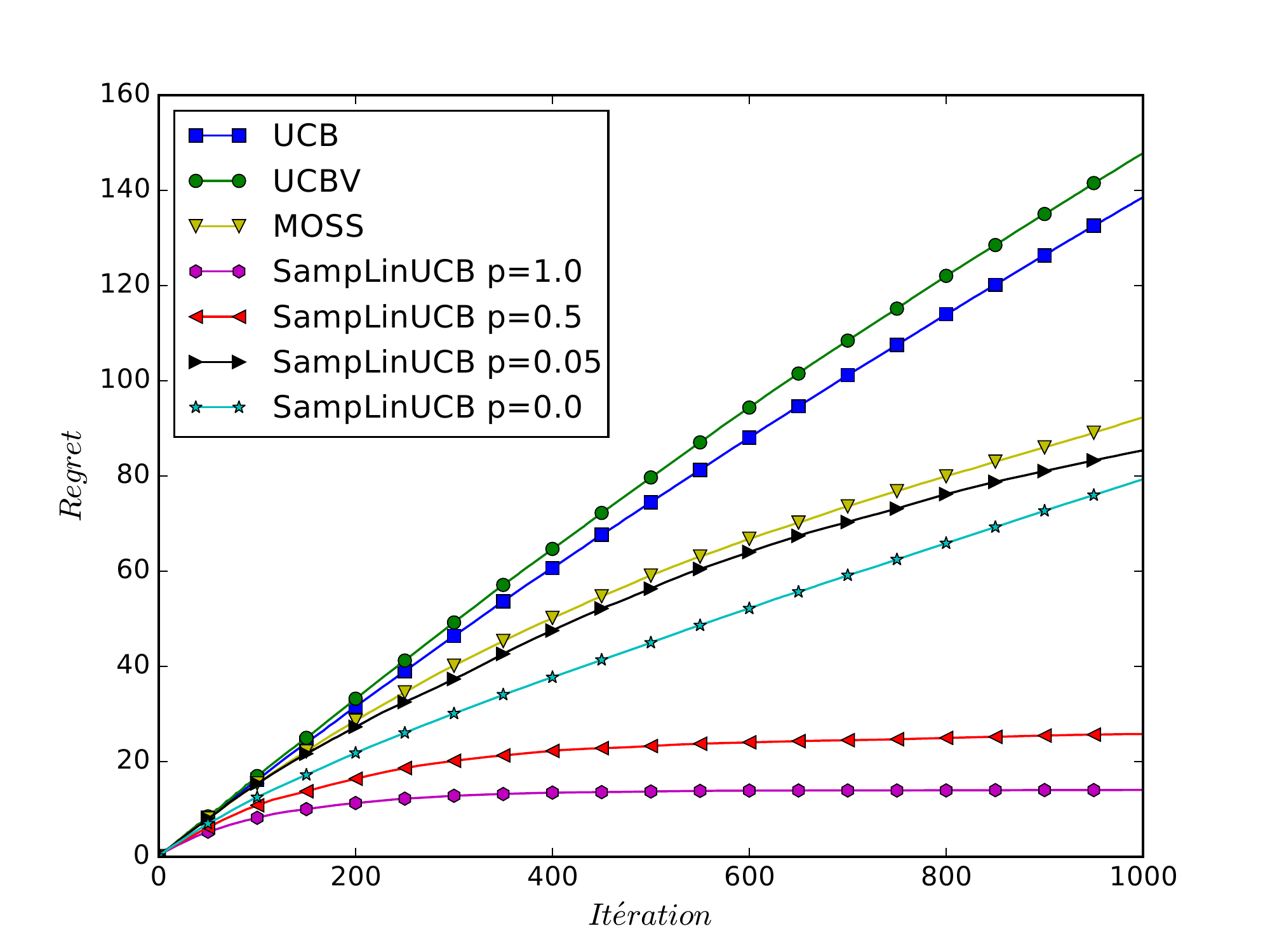}
			\caption{$\sigma=1.0$}
		\end{subfigureth}
		\begin{subfigureth}{0.5\textwidth}
			\includegraphics[width=\linewidth]{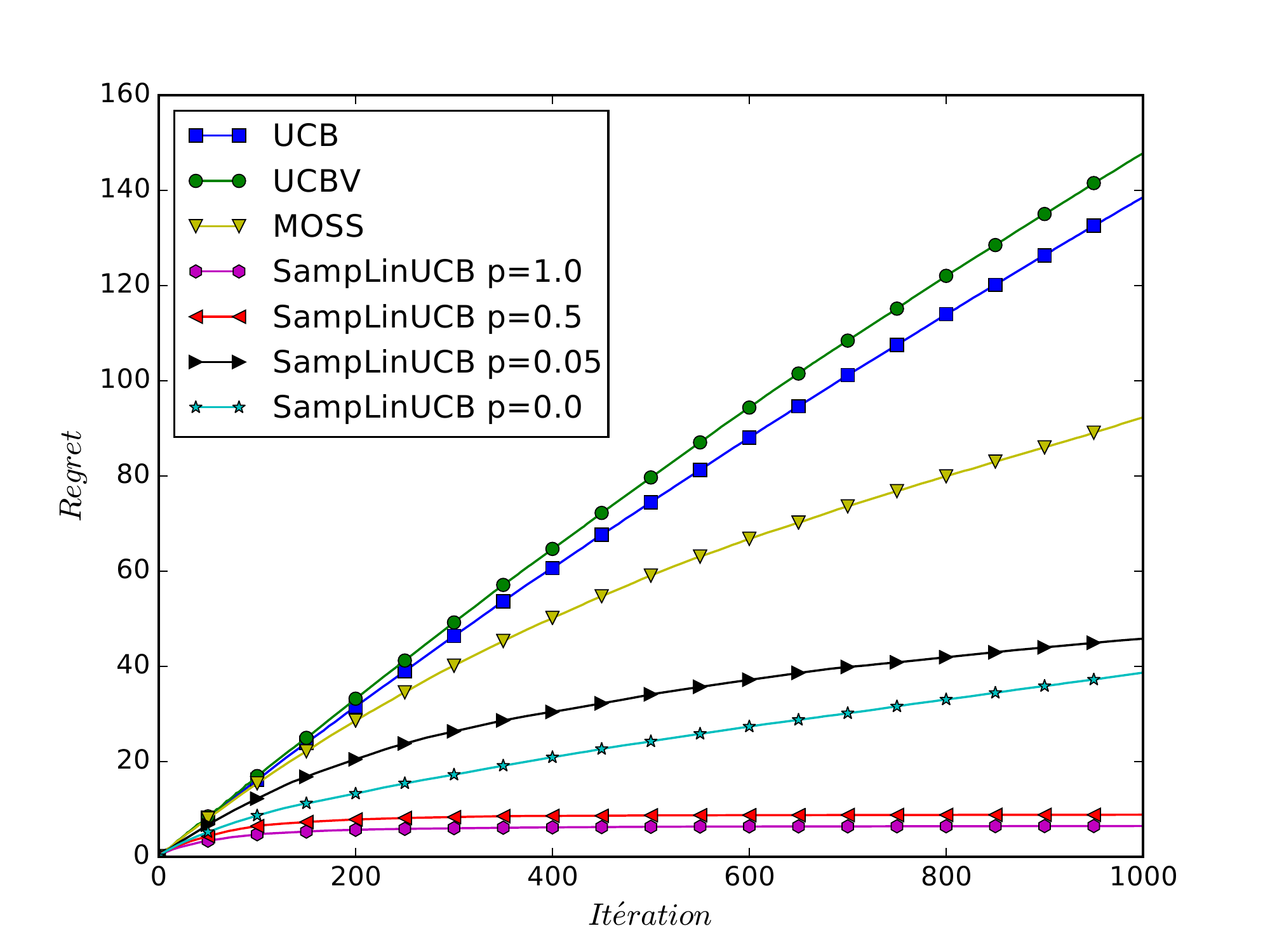}
			\caption{$\sigma=0.5$}
		\end{subfigureth}
        		\begin{subfigureth}{0.5\textwidth}
			\includegraphics[width=\linewidth]{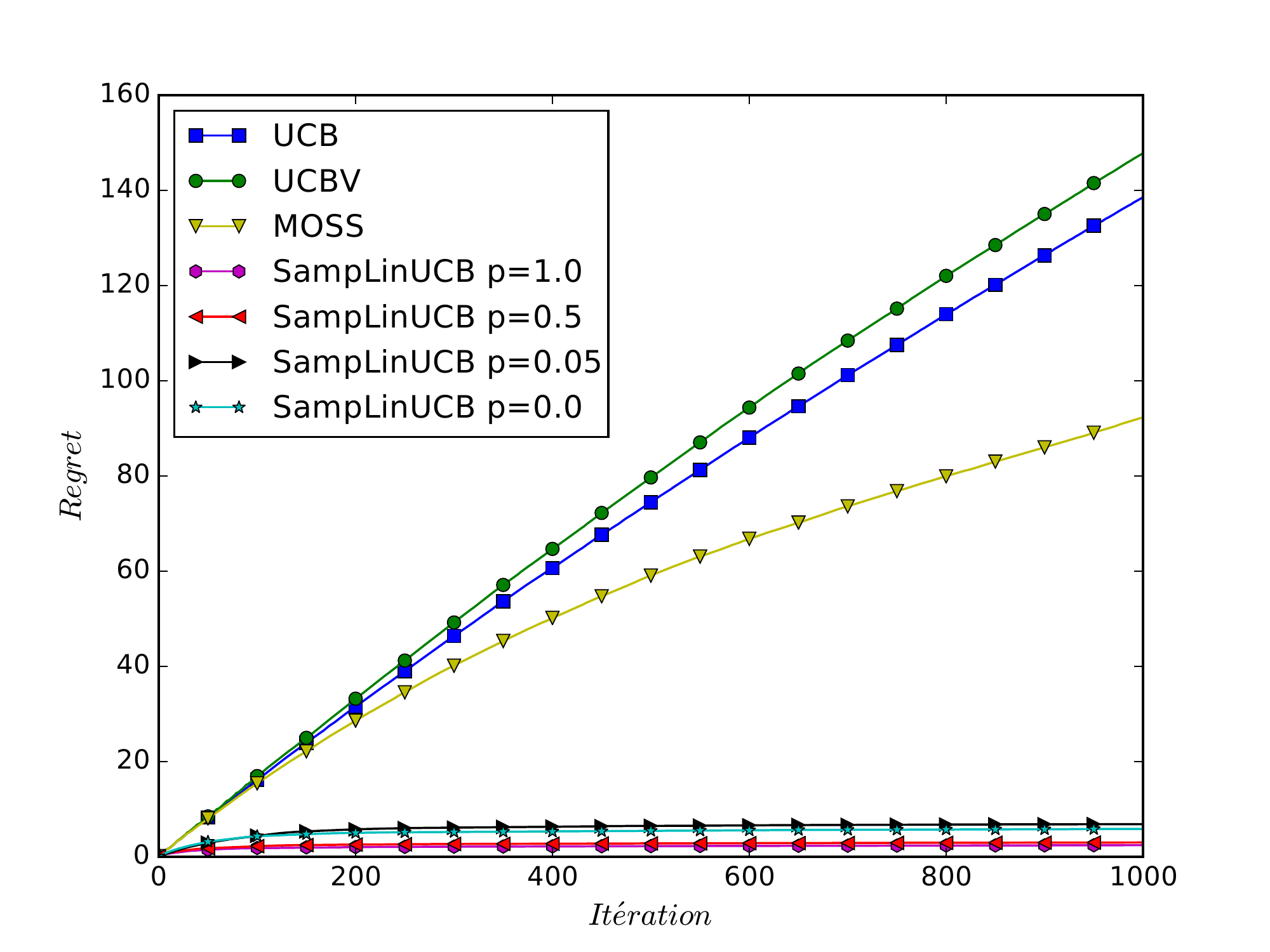}
			\caption{$\sigma=0.1$}
		\end{subfigureth}
		\caption[Regret simulation \texttt{SampLinUCB}]{Regret cumulé en fonction du temps pour l'expérience sur données artificielles et pour différents niveaux de bruit sur les échantillons de profils. Dans les trois représentations, les courbes de \texttt{UCB}, \texttt{UCBV} et \texttt{MOSS} sont identiques étant donné que leurs résultats ne sont pas influencés par le bruit sur les profils.}
			\label{fig:regretVsTimeSimSampLinUCB}
			\end{figureth}

En conclusion de cette analyse, il semble que l'algorithme \texttt{SampLinUCB}, bien que jouant avec une double incertitude - récompenses et profils - soit en mesure d'exploiter la structure sous-jacente de l'espace pour maximiser les récompenses récoltées au cours du temps.

	\subsection{Données réelles}
    
		\subsubsection{Modèle de profils}
        \label{modelProfilStat}

        Nous supposons que chaque utilisateur $i$ du réseau social est associé à un vecteur inconnu $\mu_{i}$ correspondant à son profil. Nous faisons l'hypothèse que ce dernier est une représentation de la moyenne des messages que chaque utilisateur a tendance à poster (ici, on entend poster au sens large, c'est à dire en incluant les interactions avec d'autres utilisateurs, à savoir les réponses et les reprises de messages). Pour notre application, le cas le plus réaliste est le cas 3, pour lequel on observe un échantillon de contexte au temps $t$ uniquement pour les utilisateurs écoutés au pas de temps $t-1$ c'est-à-dire $\mathcal{O}_{t}=\mathcal{K}_{t-1}$ . Par conséquent, si un compte $i$ appartient à $\mathcal{O}_{t}$, une façon de modéliser un échantillon de son profil $x_{i,t}$ est d'utiliser les messages $\omega_{i,t-1}$ récoltés au temps $t-1$. Afin d'avoir une représentation vectorielle des messages et nous permettre d'appliquer la méthode proposée dans ce chapitre, nous proposons d'utiliser la transformation LDA présentée au chapitre \ref{FormalisationTache} (section \ref{repMessage}), qui consiste à transformer chaque message $\omega_{i,t}$ en un vecteur $x_ {i,t}$ de l'espace des concepts. Nous rappelons que nous avons entraîné ce modèle avec $d=30 $. Pour résumer, en notant $F$ la fonction qui à un message associe sa représentation dans l'espace des concepts, l'échantillon de profil de l'utilisateur $i$ à l'instant $t$ est $x_{i,t}=F(\omega_{i,t-1})$.

        \subsubsection{Protocole}
        
        Nous utilisons le même protocole que celui présenté dans la partie \ref{protocole} du chapitre précédent, à savoir le modèle \textit{Topic+Influence} avec les trois thématiques \textit{Politique}, \textit{Religion} et \textit{Science}, et ce pour les trois bases de données collectées. On rappelle que l'on fixe $k$, le nombre d'utilisateurs écoutés à chaque période à 100. Nous nous comparons aux algorithmes \texttt{CUCBV}, \texttt{CUCB}, \texttt{UCB}-$\delta$ et \texttt{MOSS}. Finalement, on s'intéresse aux trois scénarios en simulant un processus délivrant des échantillons de profils de façon aléatoire pour différents $p\in \left\{0, 0.01,0.05, 0.1, 0.5,1 \right\}$. Comme dans les expérimentations précédentes, pour $p=0$, on ajoute les individus écoutés au temps $t$ à l'ensemble $\mathcal{O}_{t+1}$,  ce qui correspond au cas 3 présenté précédemment.

    	\subsubsection{Résultats}

    Les figures \ref{fig:XPStatCtxtUSRT}, \ref{fig:XPStatCtxtOGRT} et \ref{fig:XPStatCtxtBrexitRT} contiennent l'évolution du gain cumulé en fonction du temps respectivement pour les bases \textit{USElections}, \textit{OlympicGames} et \textit{Brexit} avec le modèle de récompense \textit{Topic+Influence} pour les trois thématiques. Afin de ne pas surcharger les graphiques, nous n'avons pas représenté les résultats pour l'algorithme \texttt{SampLinUCB} dans les cas $p=0.01,0.05,0.5$. Une première observation importante est que dans pratiquement tous les cas, notre algorithme \texttt{SampLinUCB} est plus performant que tous les autres, y compris \texttt{CUCBV}, qui était le plus performant d'une façon générale dans les expérimentations du chapitre précédent. Ceci permet d'affirmer que l'utilisation des profils sur les membres du réseau pour notre tâche de collecte d'information ciblée est pertinente. En effet, cette approche prenant en compte  un modèle des messages postés par chaque utilisateur permet de mieux s'orienter vers les "bons" profils. D'une façon générale, comme dans les expériences sur données simulées, les performances de notre approche augmentent avec $p$, pour atteindre un maximum lorsque $p=1$. Le point le plus important réside dans le fait que pour $p=0$, qui correspond au cas 3 décrit en début de chapitre, notre algorithme parvient à être plus performant que \texttt{CUCBV}. Ceci est très intéressant puisque, comme nous l'avons déjà mentionné, il s'agit du cas le plus réaliste, où l’on n’utilise rien d'autre que les informations données par les comptes écoutés.

        	\begin{figureth}
		\begin{subfigureth}{0.5\textwidth}
			\includegraphics[width=\linewidth]{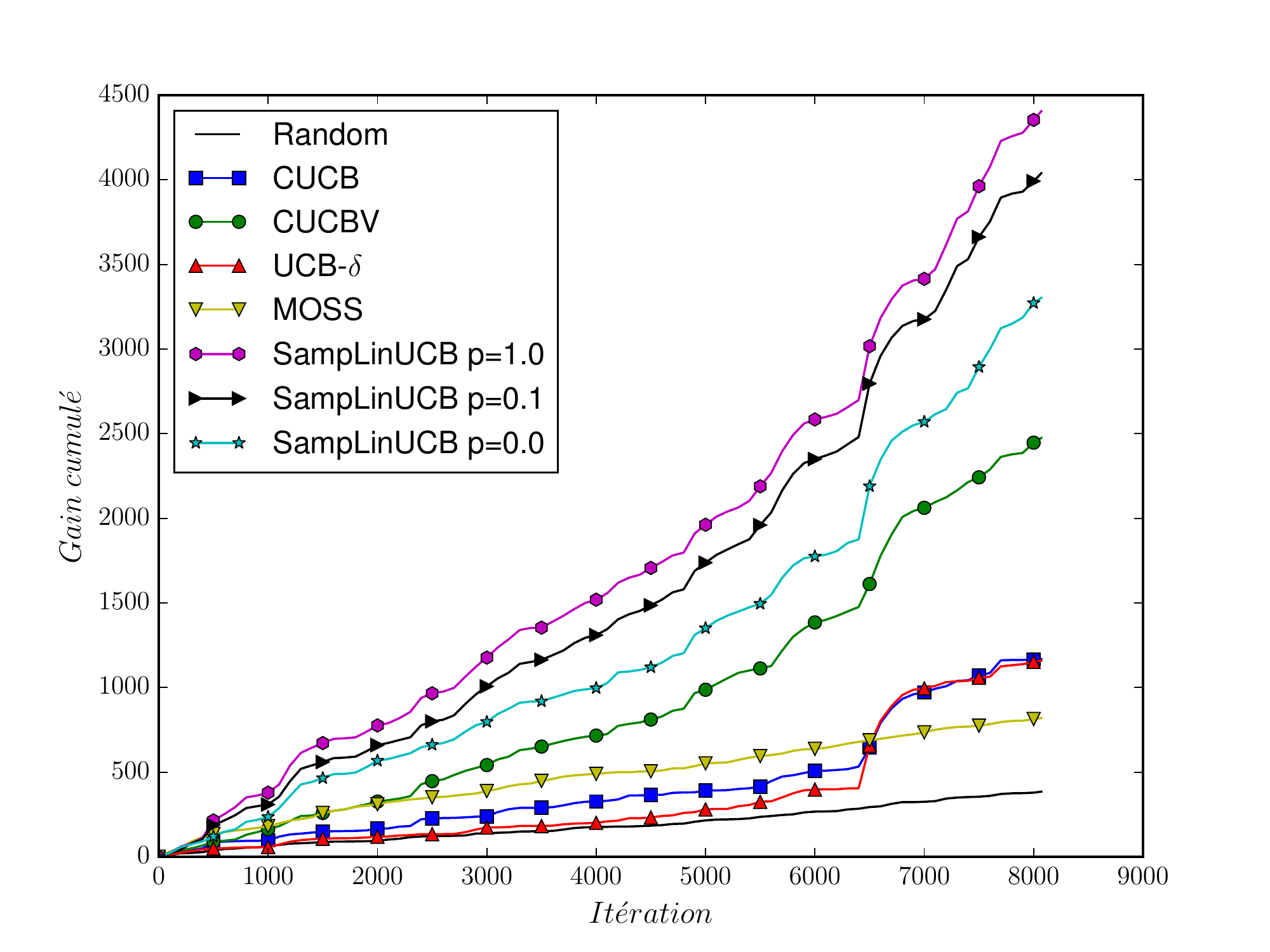}
			\caption{Politique}
		\end{subfigureth}
		\begin{subfigureth}{0.5\textwidth}
			\includegraphics[width=\linewidth]{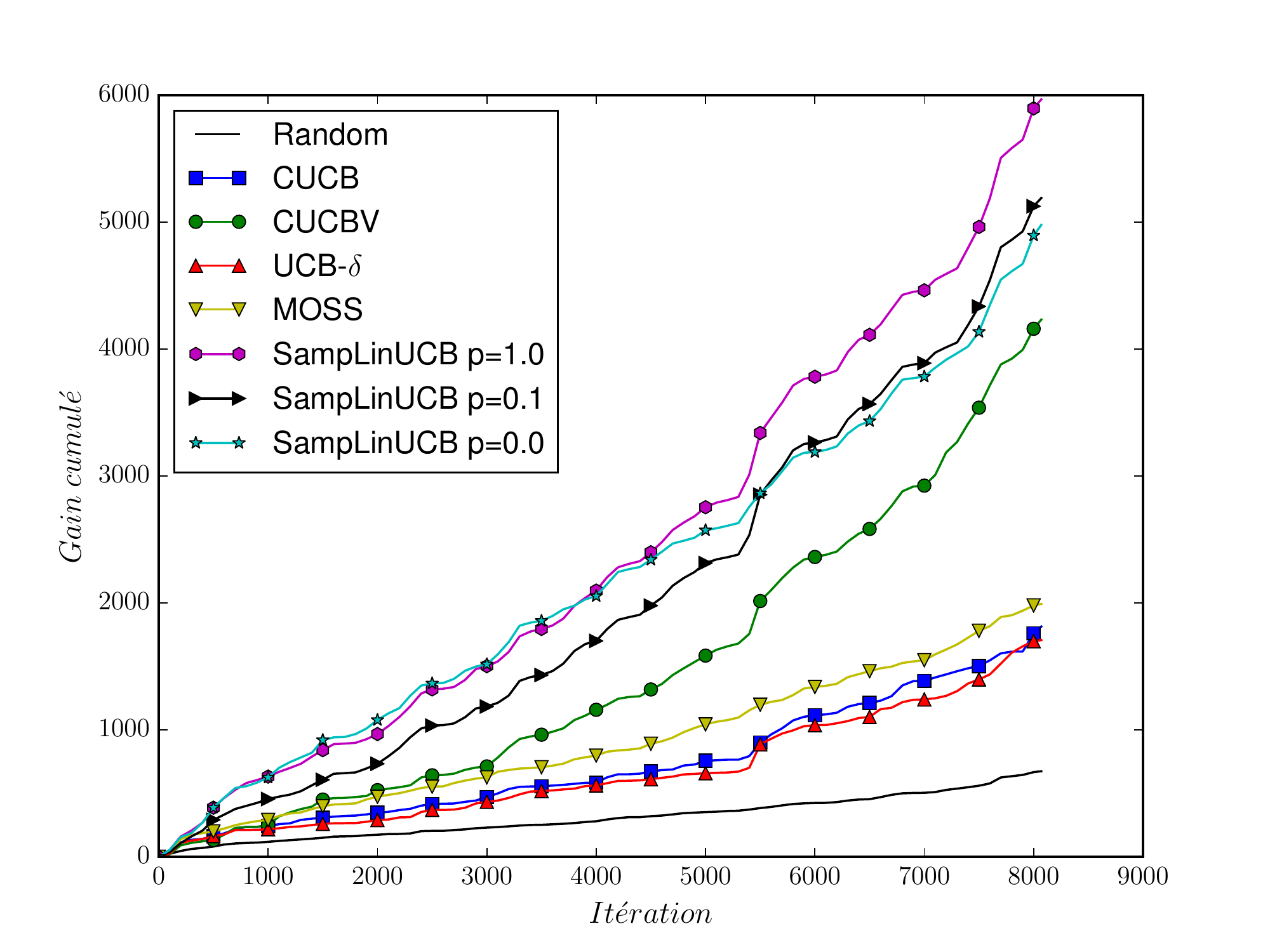}
			\caption{Religion}
		\end{subfigureth}
        \begin{subfigureth}{0.5\textwidth}
			\includegraphics[width=\linewidth]{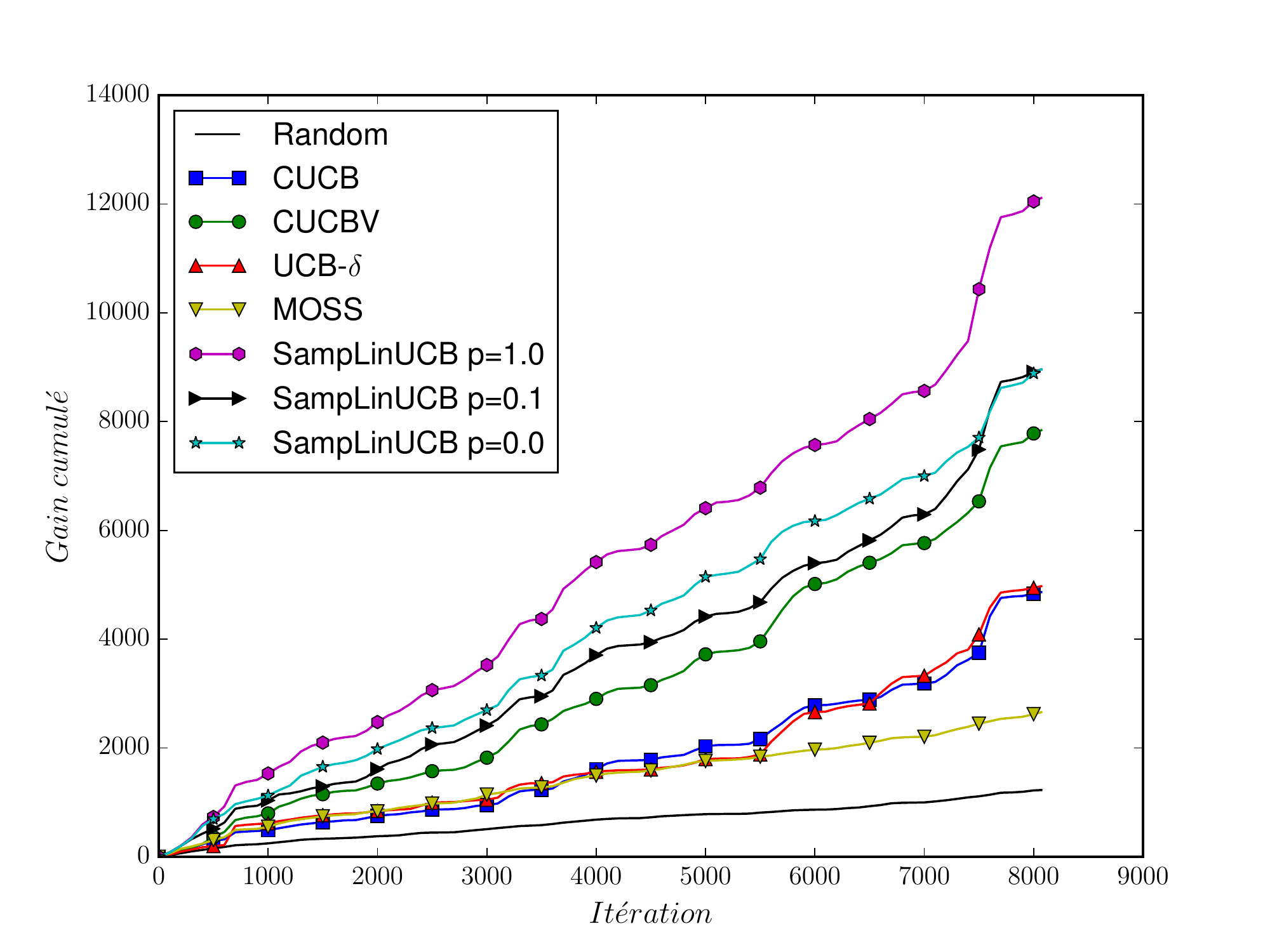}
			\caption{Science}
		\end{subfigureth}
		\caption[Gain vs temps base \textit{USElections} modèle statique.]{Evolution de la récompense cumulée en fonction du temps sur la base \textit{USElections} pour différentes politiques et le modèle \textit{Topic+Influence} avec différentes thématiques.}	
			\label{fig:XPStatCtxtUSRT}
	\end{figureth}

        	\begin{figureth}
		\begin{subfigureth}{0.5\textwidth}
			\includegraphics[width=\linewidth]{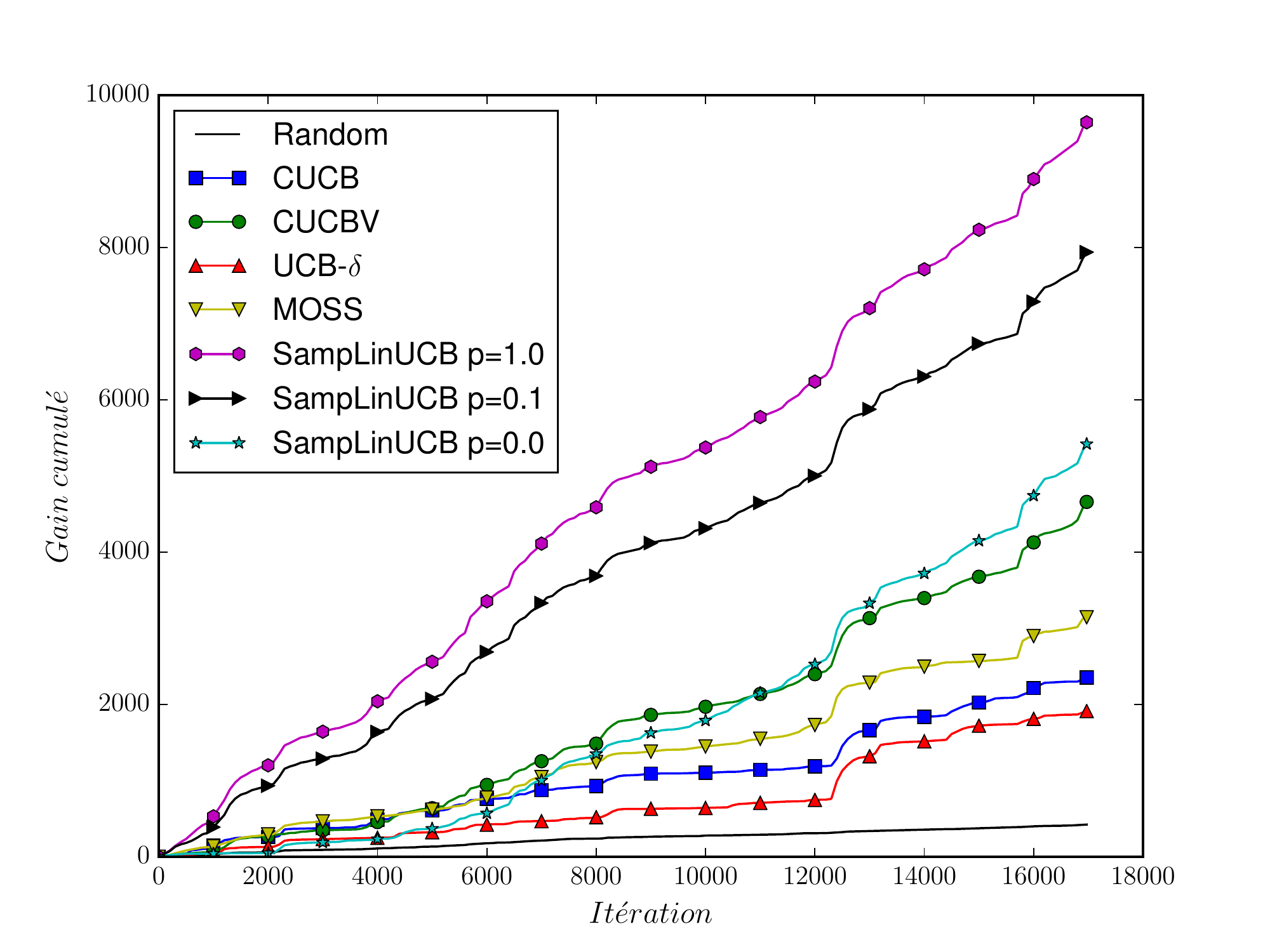}
			\caption{Politique}
		\end{subfigureth}
		\begin{subfigureth}{0.5\textwidth}
			\includegraphics[width=\linewidth]{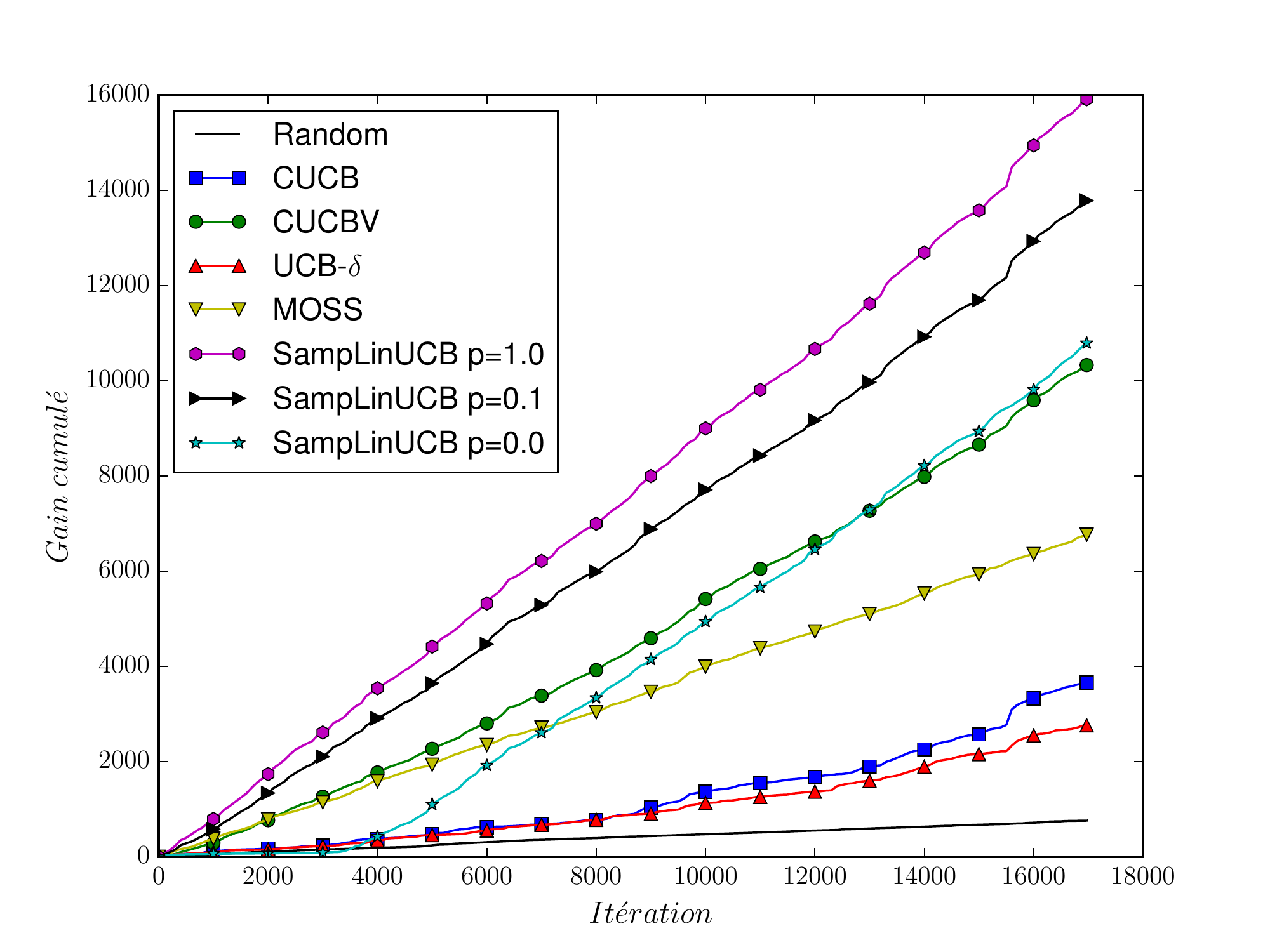}
			\caption{Religion}
		\end{subfigureth}
        \begin{subfigureth}{0.5\textwidth}
			\includegraphics[width=\linewidth]{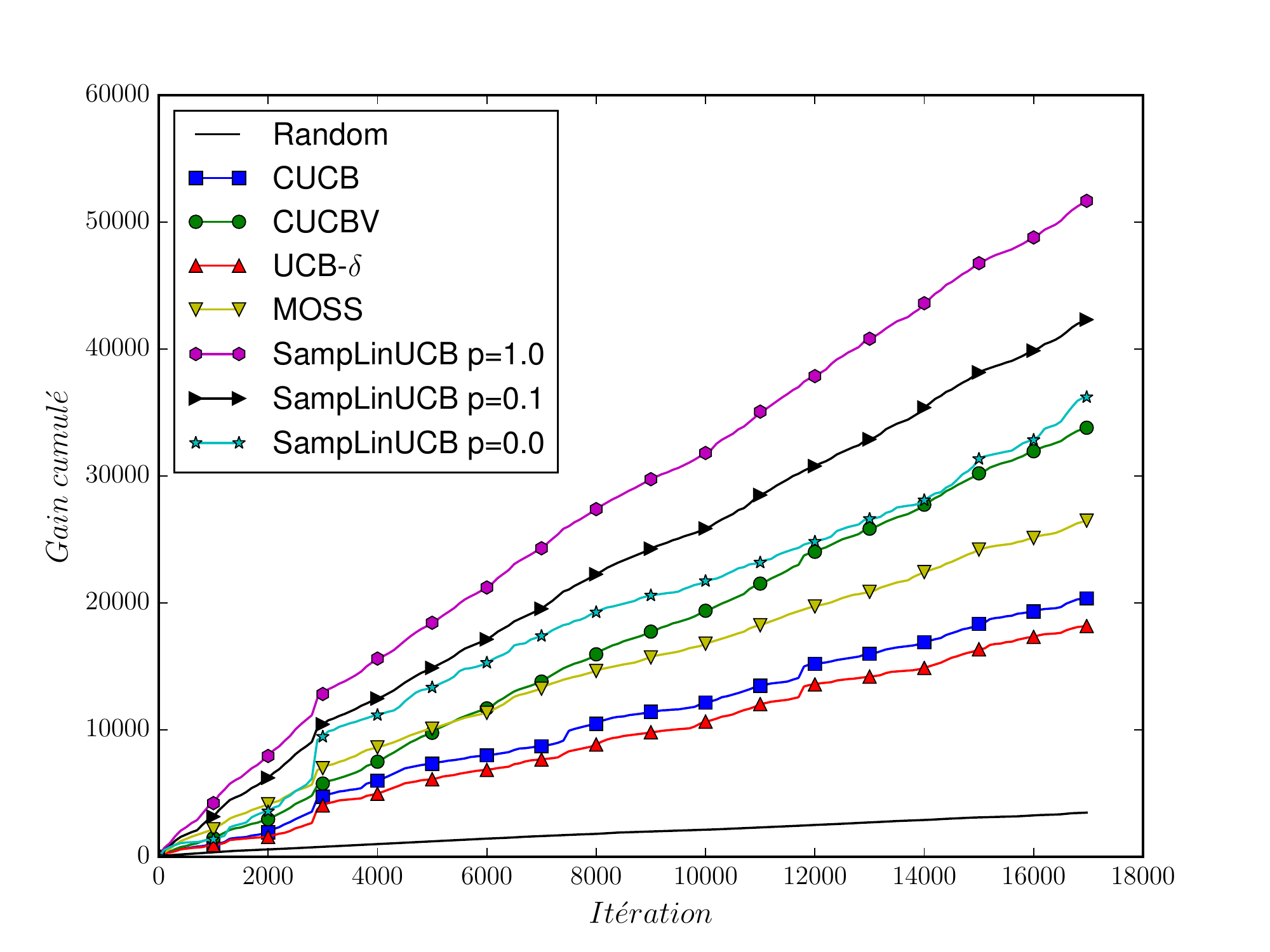}
			\caption{Science}
		\end{subfigureth}
		\caption[Gain vs temps base \textit{OlympicGames} modèle statique.]{Evolution de la récompense cumulée en fonction du temps sur la base \textit{OlympicGames} pour différentes politiques et le modèle \textit{Topic+Influence} avec différentes thématiques.}	
			\label{fig:XPStatCtxtOGRT}
	\end{figureth}

            	\begin{figureth}
		\begin{subfigureth}{0.5\textwidth}
			\includegraphics[width=\linewidth]{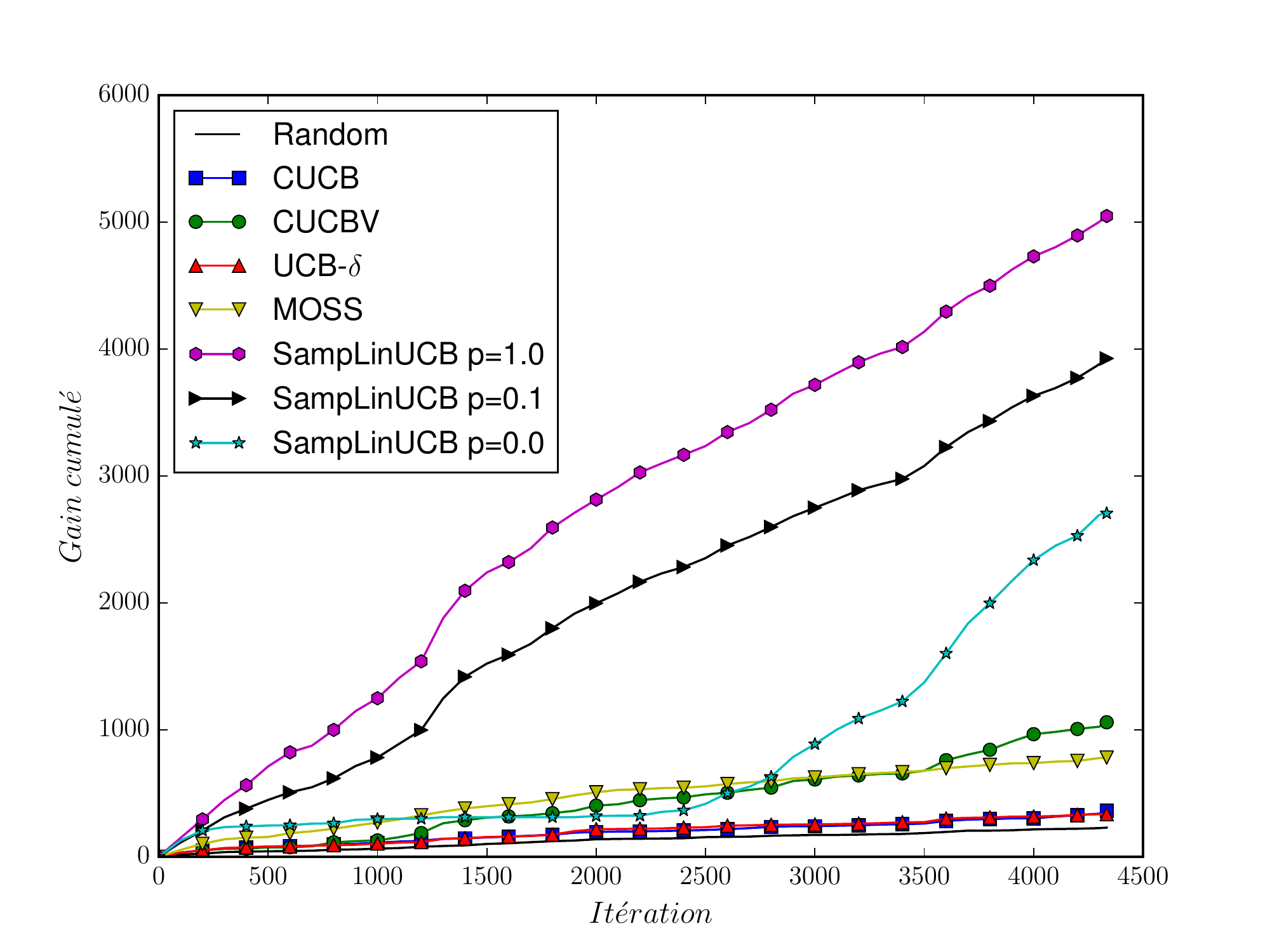}
			\caption{Politique}
		\end{subfigureth}
		\begin{subfigureth}{0.5\textwidth}
			\includegraphics[width=\linewidth]{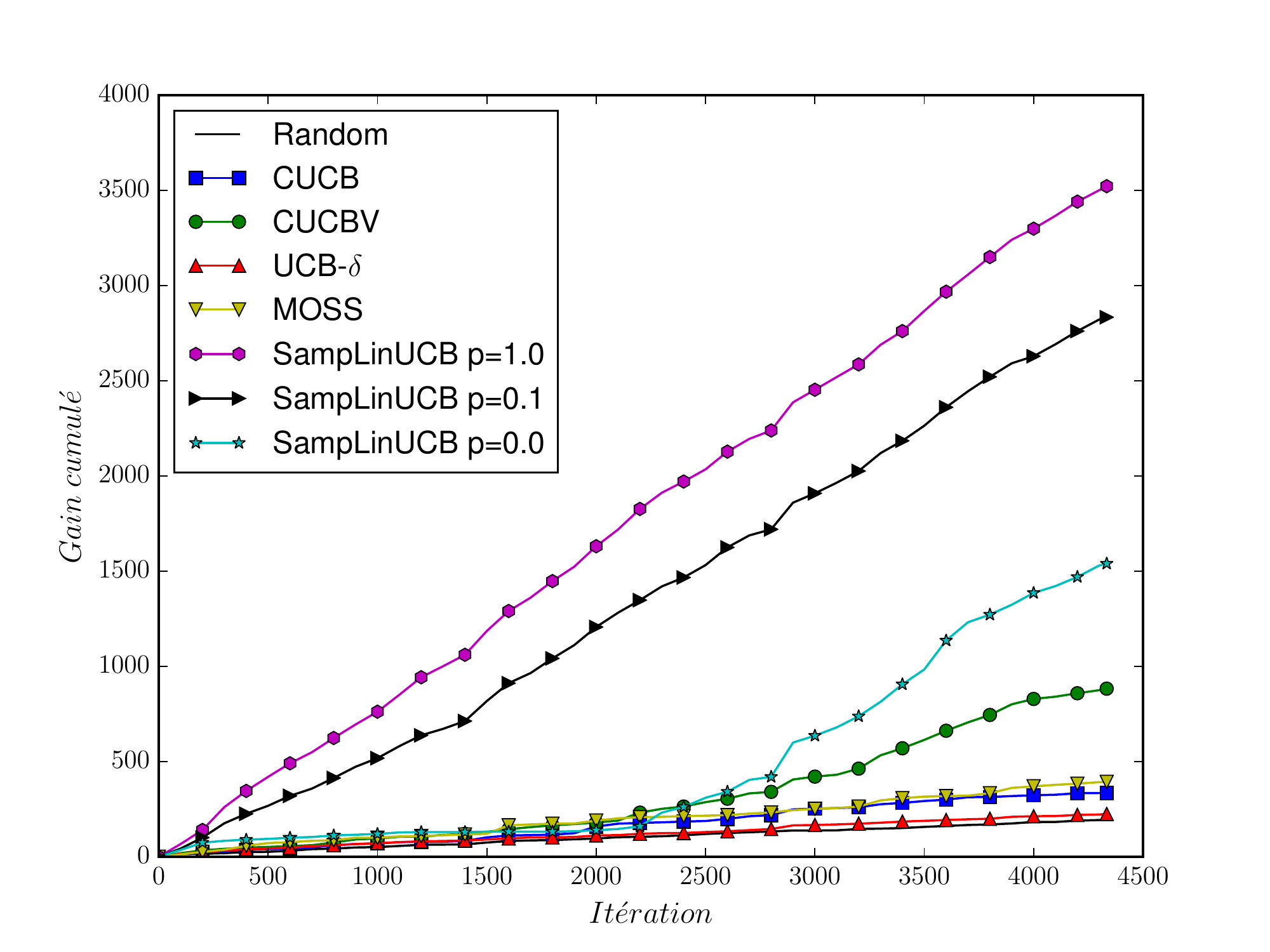}
			\caption{Religion}
		\end{subfigureth}
        \begin{subfigureth}{0.5\textwidth}
			\includegraphics[width=\linewidth]{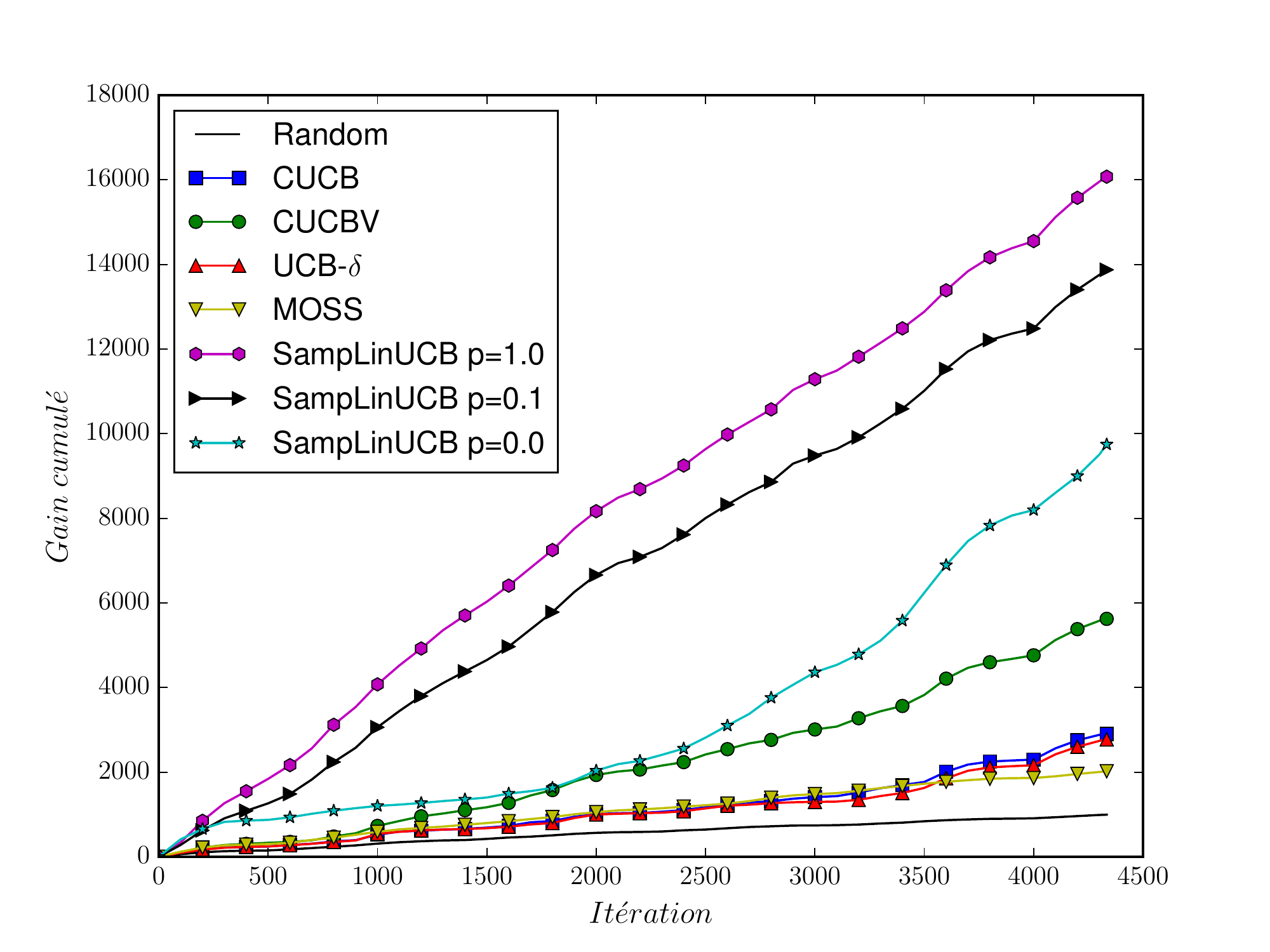}
			\caption{Science}
		\end{subfigureth}
		\caption[Gain vs temps base \textit{Brexit} modèle statique.]{Evolution de la récompense cumulée en fonction du temps sur la base \textit{Brexit} pour différentes politiques et le modèle \textit{Topic+Influence} avec différentes thématiques.}	
			\label{fig:XPStatCtxtBrexitRT}
	\end{figureth}

  On s'intéresse maintenant de plus près aux résultats propres à l'algorithme \texttt{SampLinUCB}. Les figures \ref{fig:XPStatCtxtUSRTFinal}, \ref{fig:XPStatCtxtOGRTFinal} et \ref{fig:XPStatCtxtBrexitRTFinal} représentent la récompense finale pour l'algorithme \texttt{SampLinUCB} sur les jeux de données \textit{USElections}, \textit{OlympicGames} et \textit{Brexit} avec le modèle de récompense \textit{Topic+Influence} pour les trois thématiques et pour différentes valeurs de $p$. Pour plus de clarté, les valeurs ont été normalisées par la plus grande, c'est-à-dire le gain final avec $p=1$. L'intuition générale se confirme bien puisque dans tous les cas, on observe que les performances se dégradent à mesure que $p$ diminue lorsque $p>0$. On remarque finalement que lorsque $p=0$, c'est-à-dire quand $\mathcal{O}_{t}=\mathcal{K}_{t-1}$  l'algorithme arrive la plupart du temps à battre le cas où $p=0.01$ et parfois le cas où $p=0.05$. Cela souligne la capacité de l'algorithme à être actif dans l'apprentissage des profils dans le cas 3: avec moins d'échantillons observés (100 sur 5000, ce qui correspondrait à $p=0.02$), il parvient à capturer plus de contenus utiles au cours du temps.

 	\begin{figureth}
		\begin{subfigureth}{0.3\textwidth}
			\includegraphics[width=\linewidth]{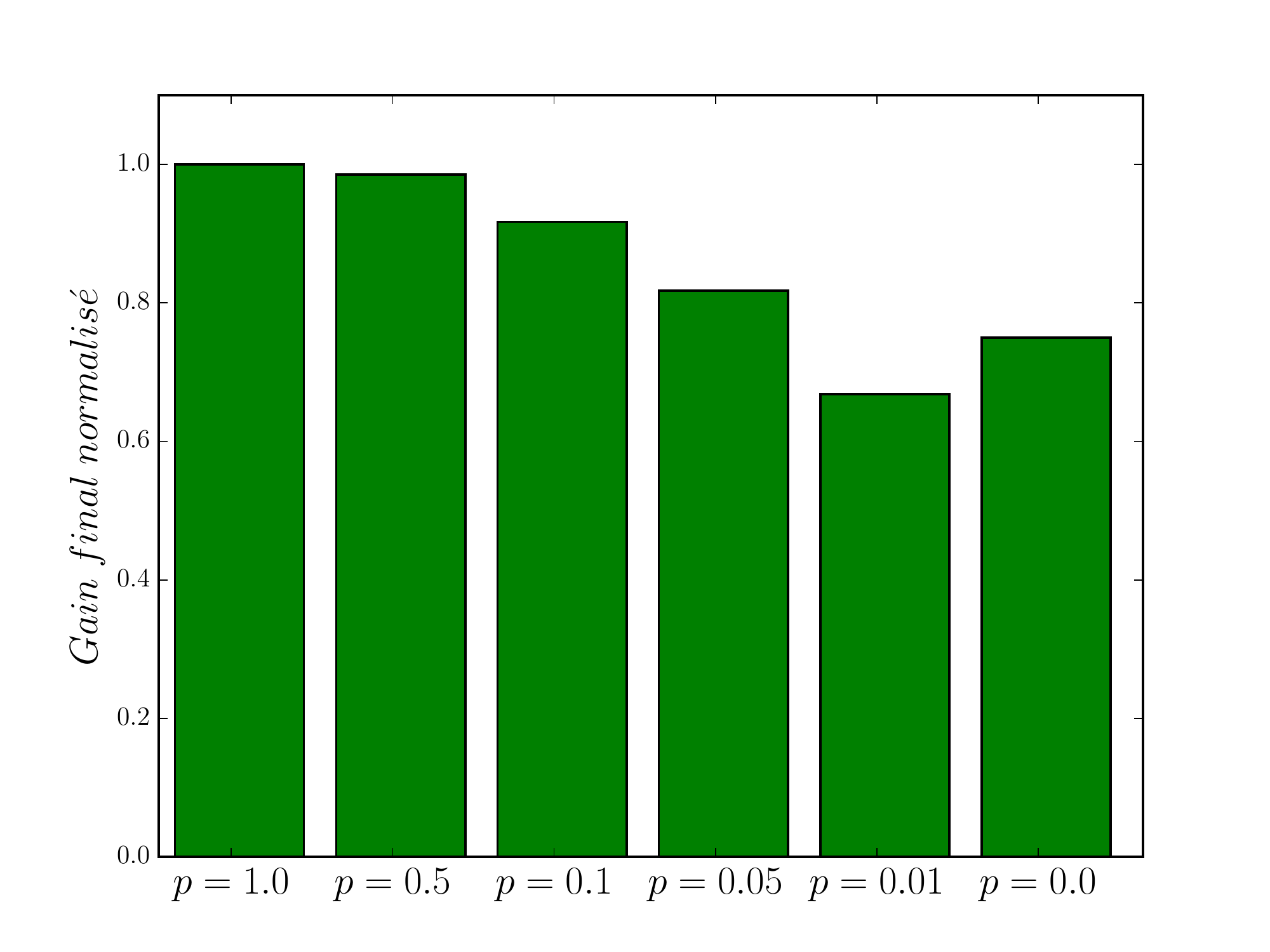}
			\caption{Politique}
		\end{subfigureth}
		\begin{subfigureth}{0.3\textwidth}
			\includegraphics[width=\linewidth]{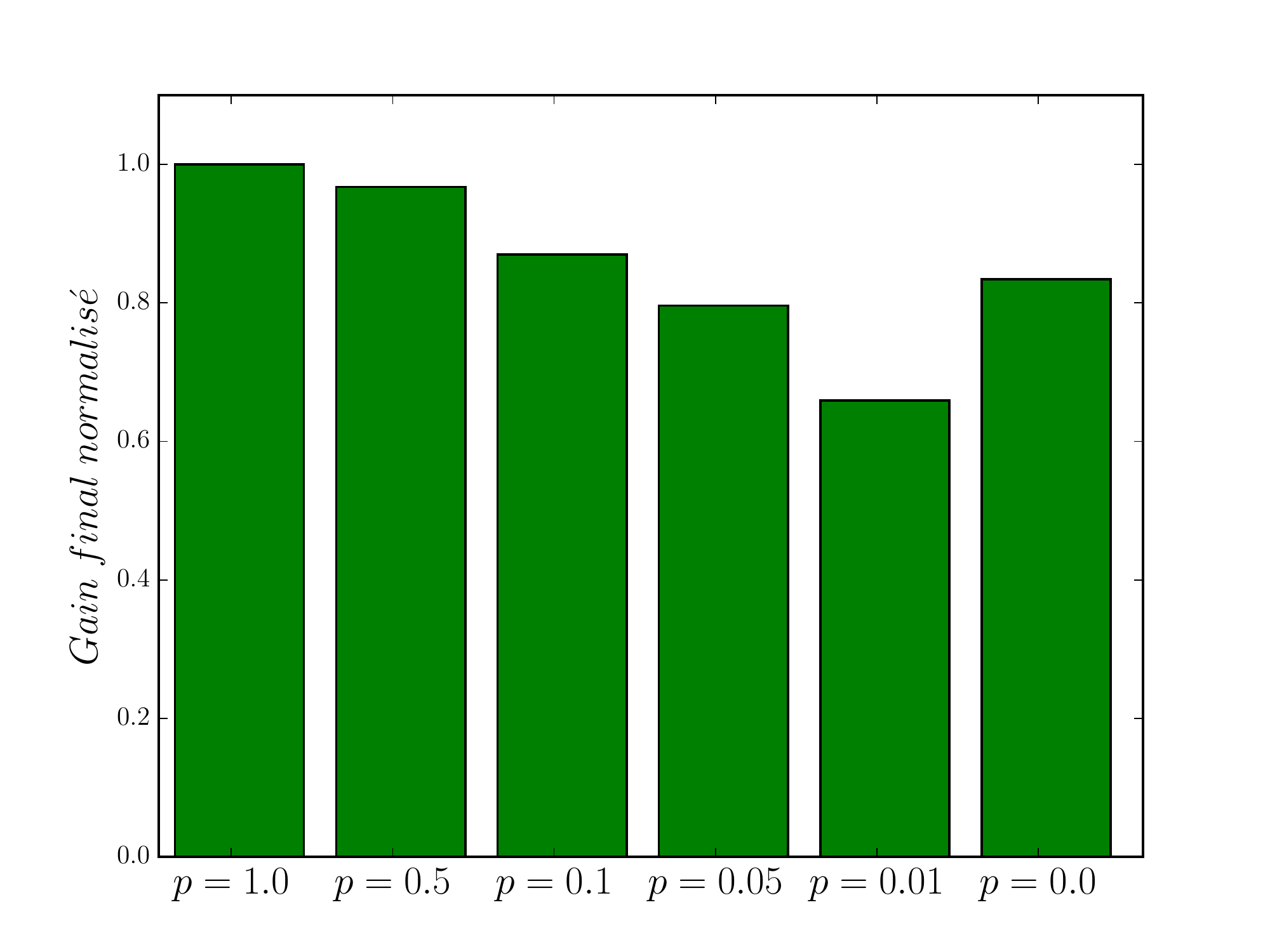}
			\caption{Religion}
		\end{subfigureth}
        \begin{subfigureth}{0.3\textwidth}
			\includegraphics[width=\linewidth]{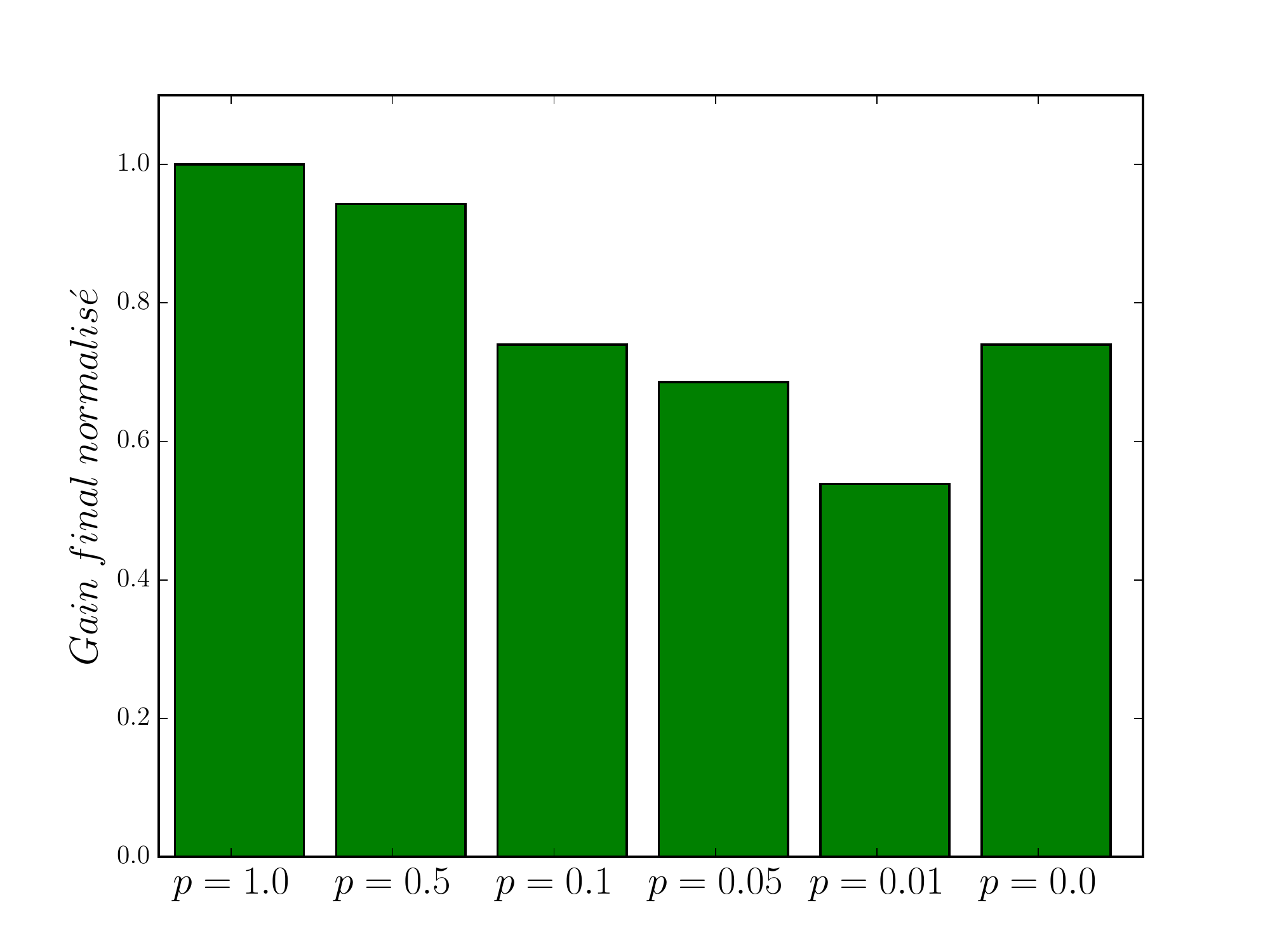}
			\caption{Science}
		\end{subfigureth}
		\caption[Gain final base \textit{USElections} algorithme \texttt{SampLinUCB}.]{Gain final pour l'algorithme \texttt{SampLinUCB} sur le jeu de données \textit{USElections} pour différentes valeurs de $p$.}	
			\label{fig:XPStatCtxtUSRTFinal}
	\end{figureth}

     \begin{figureth}
		\begin{subfigureth}{0.3\textwidth}
			\includegraphics[width=\linewidth]{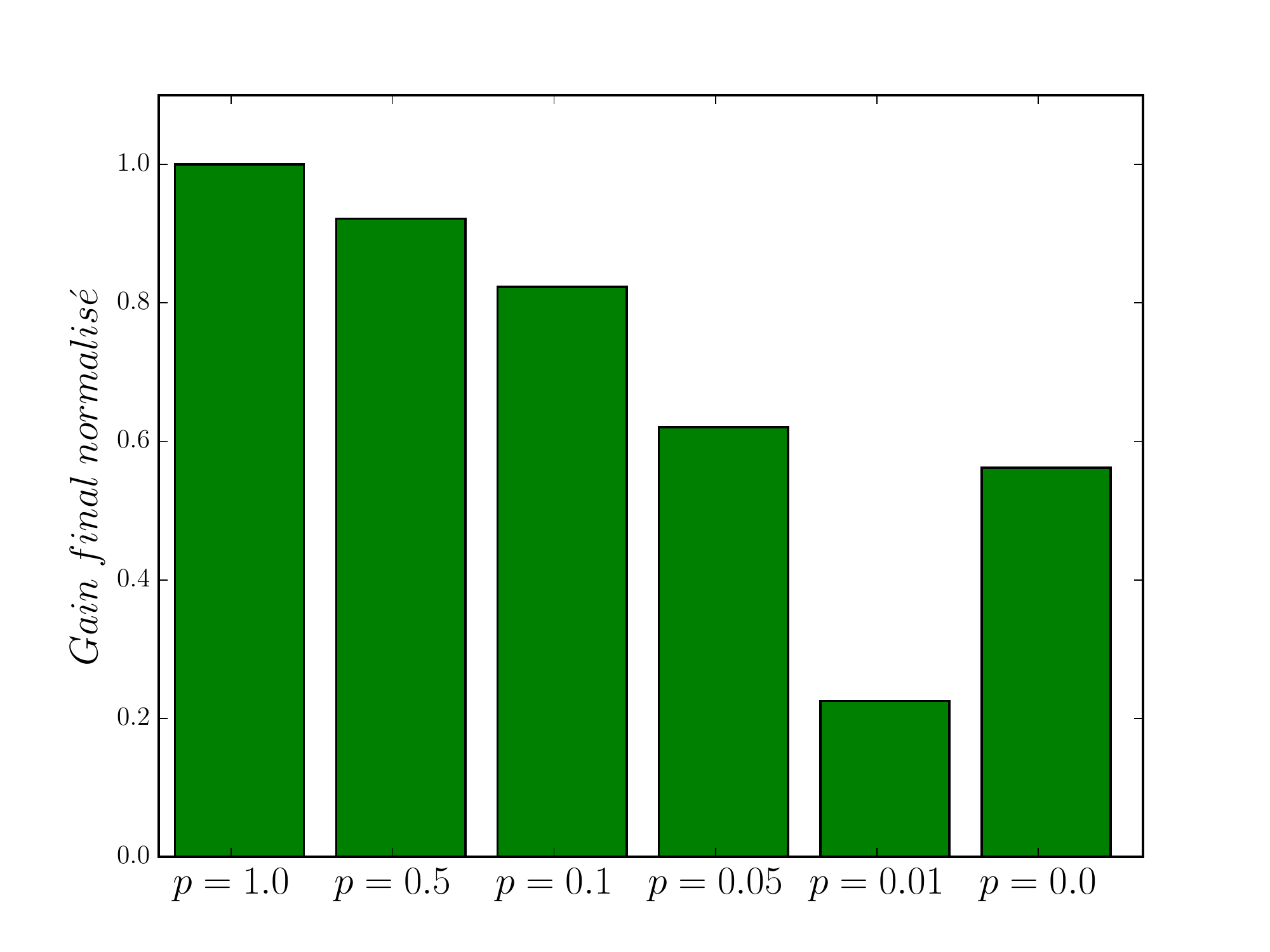}
			\caption{Politique}
		\end{subfigureth}
		\begin{subfigureth}{0.3\textwidth}
			\includegraphics[width=\linewidth]{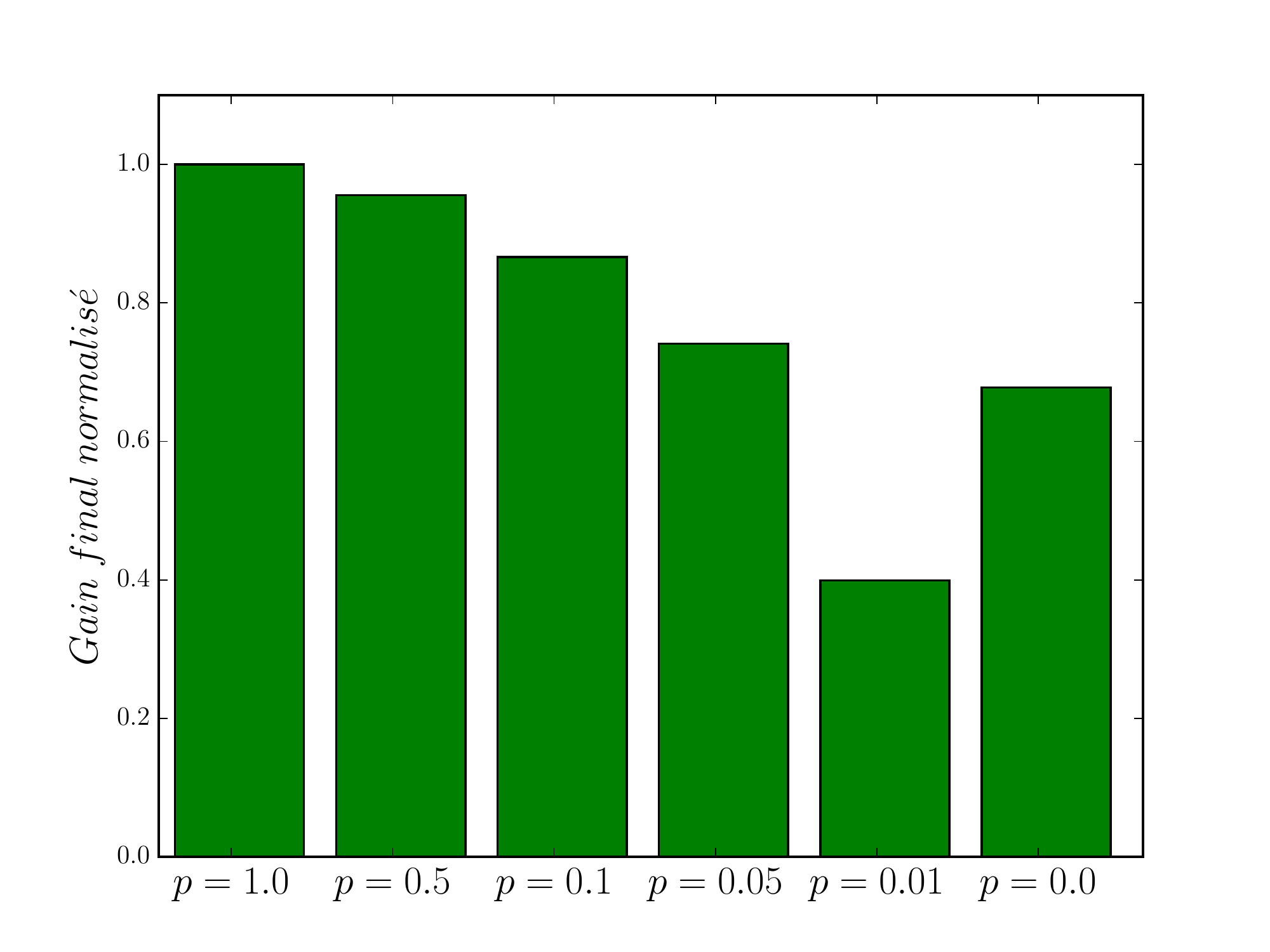}
			\caption{Religion}
		\end{subfigureth}
        \begin{subfigureth}{0.3\textwidth}
			\includegraphics[width=\linewidth]{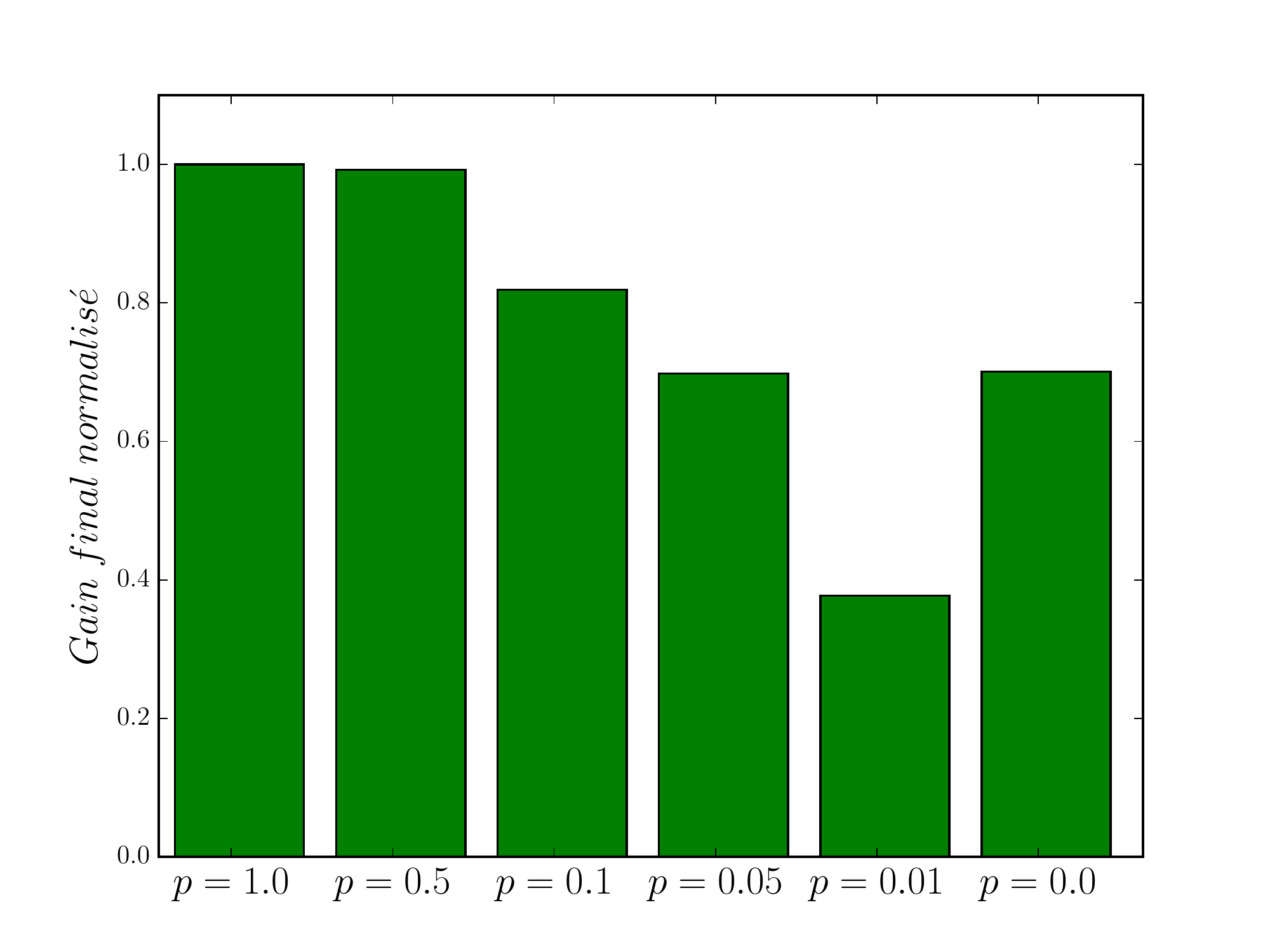}
			\caption{Science}
		\end{subfigureth}
		\caption[Gain final base \textit{OlympicGames} algorithme \texttt{SampLinUCB}.]{Gain final pour l'algorithme \texttt{SampLinUCB} sur le jeu de données \textit{OlympicGames} pour différentes valeurs de $p$.}	
			\label{fig:XPStatCtxtOGRTFinal}
	\end{figureth}

       \begin{figureth}
		\begin{subfigureth}{0.3\textwidth}
			\includegraphics[width=\linewidth]{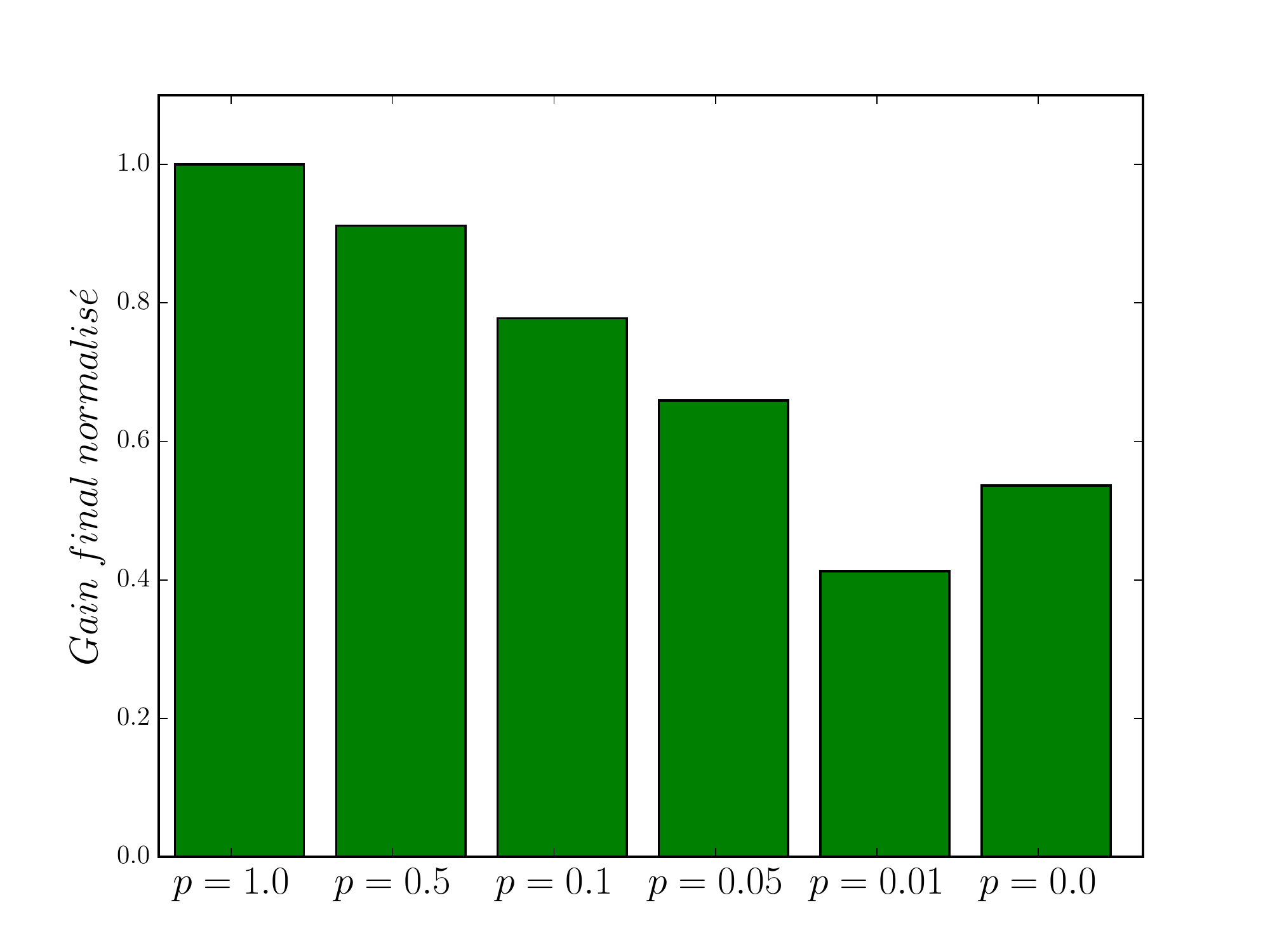}
			\caption{Politique}
		\end{subfigureth}
		\begin{subfigureth}{0.3\textwidth}
			\includegraphics[width=\linewidth]{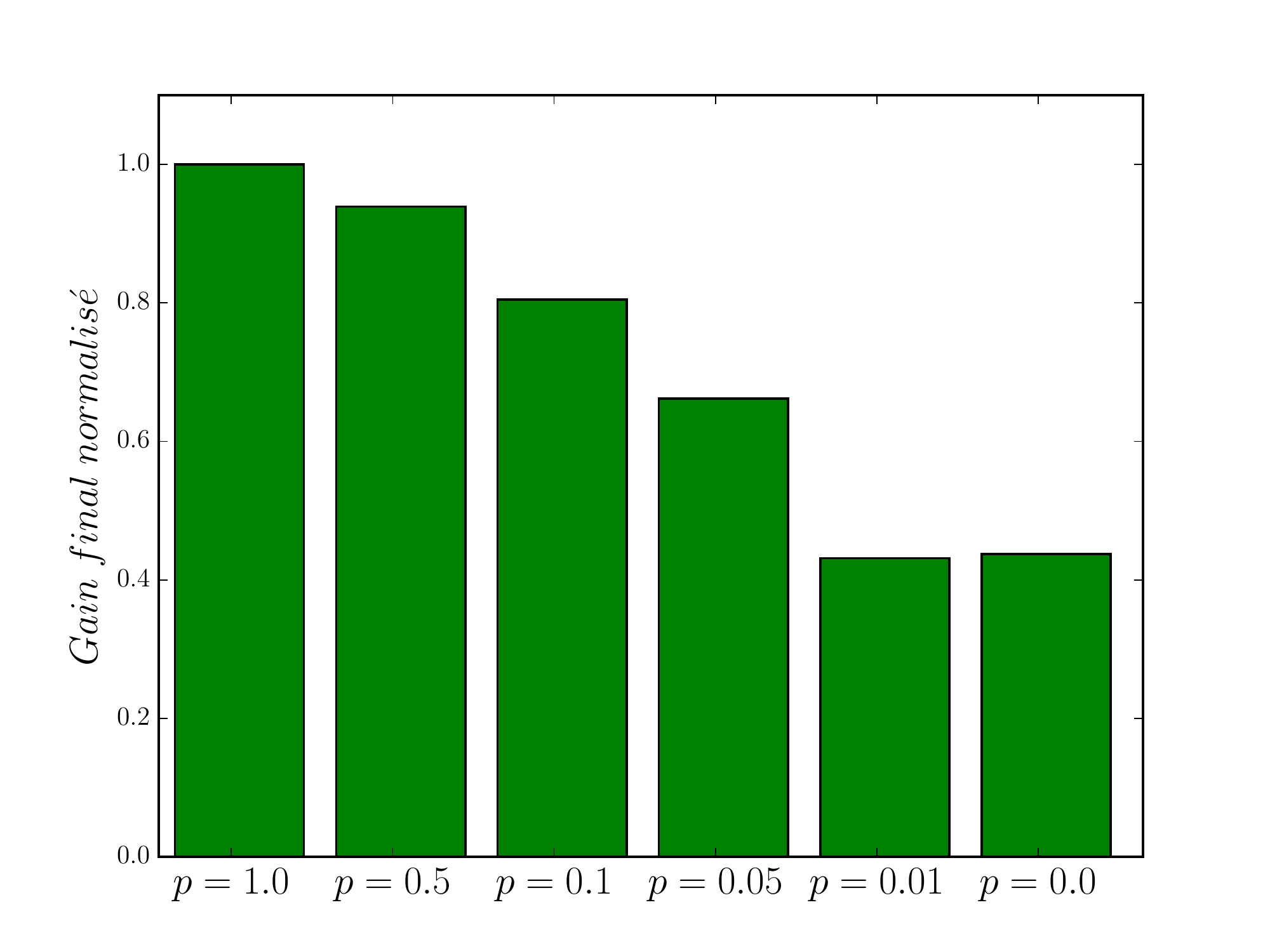}
			\caption{Religion}
		\end{subfigureth}
        \begin{subfigureth}{0.3\textwidth}
			\includegraphics[width=\linewidth]{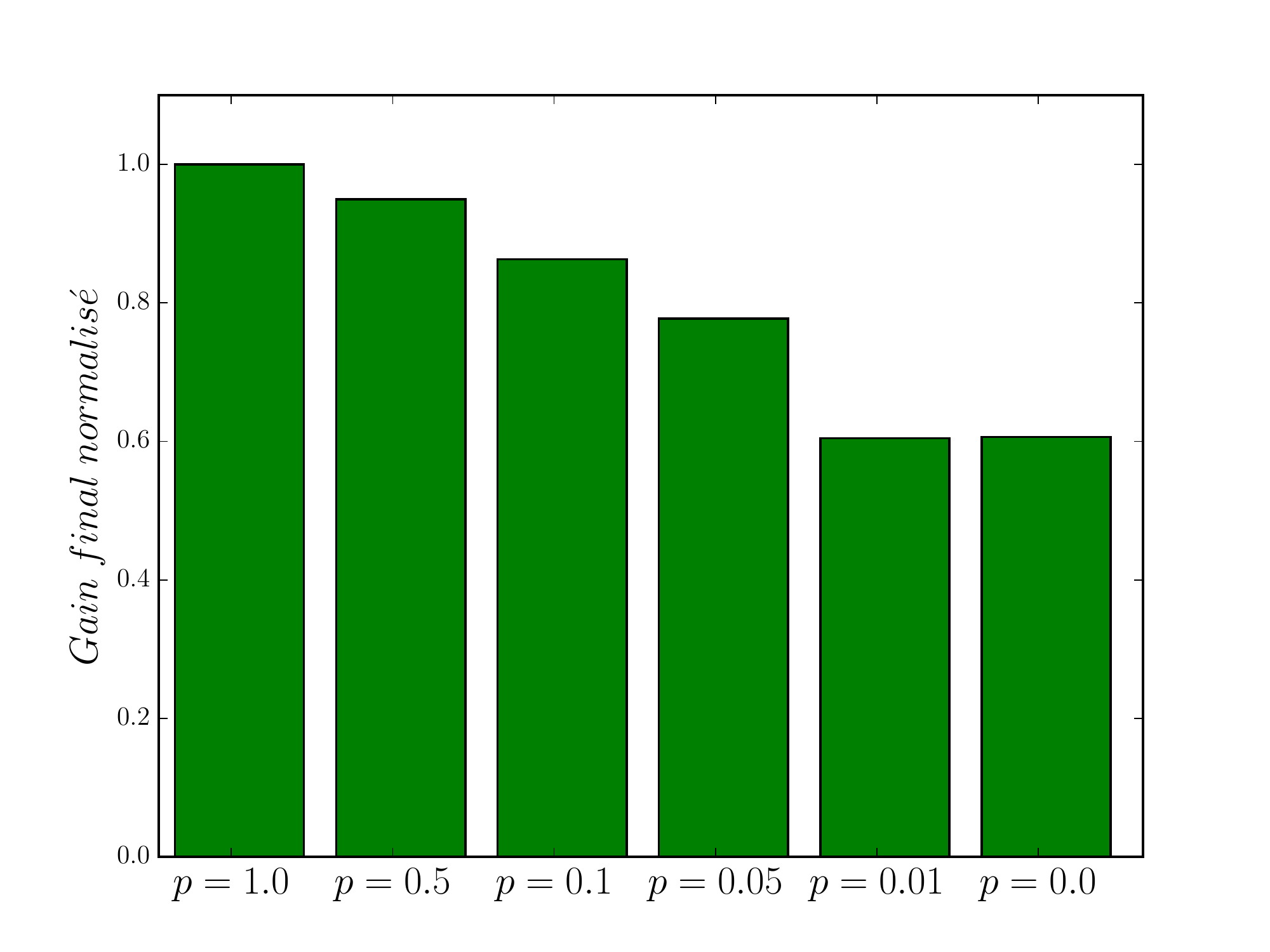}
			\caption{Science}
		\end{subfigureth}
		\caption[Gain final base \textit{Brexit} algorithme \texttt{SampLinUCB}.]{Gain final pour l'algorithme \texttt{SampLinUCB} sur le jeu de données \textit{Brexit} pour différentes valeurs de $p$.}	
			\label{fig:XPStatCtxtBrexitRTFinal}
	\end{figureth}

\section{Conclusion}

Dans ce chapitre, nous avons formalisé un nouveau problème de bandit dans lequel chaque action est associée à un profil inconnu \textit{a priori}, dans le but de mieux modéliser les utilisateurs d'un réseau social et d'améliorer les performances de notre processus de collecte d'information. Contrairement aux problèmes existants, dans notre cas l'agent décisionnel peut seulement observer des versions bruitées des profils, et ce, uniquement pour un nombre limité de comptes à chaque instant. Cette contrainte modélise le fait que dans un réseau social, seule une partie limitée de l'activité est observable via les API mises à notre disposition. Une fois le problème formalisé, nous avons dérivé un nouvel algorithme de type optimiste capable de tirer profit du modèle linéaire liant les vrais profils - non observables - aux récompenses des utilisateurs. D'une façon générale, cet algorithme maintient un intervalle de confiance à la fois sur l'estimateur du paramètre de régression, mais aussi sur les estimateurs des profils dans le but de construire un intervalle de confiance pour la récompense à proprement parler. Nous avons prouvé l'existence d'une borne supérieure sous-linéaire du regret associé pour trois scénarios distincts du processus déterminant le sous-ensemble de comptes délivrant des échantillons de profils à chaque itération, en particulier pour le cas le plus réaliste ou les seuls échantillons observables au temps $t$ sont ceux des comptes sélectionnés au temps $t-1$. Finalement, des expérimentations à la fois sur données simulées et réelles nous ont permis de confirmer la pertinence du modèle à la fois dans un cadre générique et pour notre tâche de collecte d'information sur un média social. En perspective, il aurait été intéressant d'étudier des cas où les échantillons de contextes suivent des lois moins régulières que celles utilisées dans les expérimentaions artificielles (par exemple des lois de type mélange), afin d'évaluer la robustesse de \texttt{SampLinUCB}. 

	\chapter{Modèle contextuel}
  \minitoc
  \newpage

\label{ModeleContextuelVariable}

La modélisation d'un profil pour chaque utilisateur, associée à une hypothèse de linéarité, nous a permis d'améliorer considérablement les performances de notre processus de collecte dans le chapitre précédent. Grâce à la présence d'un paramètre de régression commun à tous les utilisateurs, nous avons montré qu'il est possible de mutualiser l'apprentissage, permettant ainsi une exploration plus intelligente de l'espace des utilisateurs. L'idée était d'utiliser le "message moyen" - à une transformation LDA près - de chaque utilisateur pour exploiter des corrélations entre vocabulaire employé et récompenses espérées. Cependant, l'hypothèse de stationnarité sous-jacente peut être discutée, car il est possible que certains utilisateurs soient actifs sur plusieurs sujets complètement différents selon par exemple le moment de la journée. En supposant que les messages postés par un utilisateur pendant une période donnée puisse permettre de prévoir son utilité à venir, nous proposons dans ce chapitre un nouveau modèle pour tenir compte de ces variations et mieux exploiter les flux de données sous les contraintes associées à notre tâche.


\section{Modèle}

Dans cette section, nous formalisons le modèle mathématique du bandit contextuel ainsi que les hypothèses correspondantes.  On suppose qu'il existe un vecteur inconnu permettant d'estimer l'espérance de la récompense que l'on peut observer pour chaque action en fonction de son vecteur de contexte. On se place ici dans le cas où l'ensemble $\mathcal{K}$ des actions est fini et de taille $K$. Ainsi, à chaque pas de temps $t$, chaque action $i$ révèle un vecteur de contexte noté $x_{i,t} \in \mathbb{R}^{d}$ et l'agent sélectionne un sous-ensemble $\mathcal{K}_{t}$ de $k<K$ actions pour lesquelles il reçoit les récompenses associées, son but étant toujours de maximiser la somme des récompenses cumulées au cours du temps. L'hypothèse permettant de lier contexte et récompense que nous considérons suppose l'existence d'un vecteur de poids  $\beta \in \mathbb{R}^{d}$ inconnu tel que l'espérance de la récompense associée à $i$ selon le contexte $x_{i,t}$ est égale au produit scalaire des deux vecteurs à un instant $t$. Formellement, cela se traduit par:

 \begin{equation}
 \exists \beta \in \mathbb{R}^{d} \text{ tel que } \forall t \in \left\{1,..,T\right\}, \forall i \in \left\{1,..,K\right\}: \mathbb{E}[r_{i,t}|x_{i,t}]=x_{i,t}^{\top}\beta
 \end{equation}

Dans notre cas, le contexte $x_{i,t}$ de chaque utilisateur $i$ à chaque instant $t$ correspond au contenu publié dans la fenêtre de temps $t-1$. Contrairement au chapitre précédent, on utilise directement l'échantillon révélé par un utilisateur et non pas à une version bruitée d'un profil constant et inconnu. Dans la suite, nous supposons que $\beta$ est borné par une constante $S \in \mathbb{R}^{+*}$, c.-à-d. $||\beta|| \leq S $.

L'utilisation d'un vecteur de paramètres commun permet une exploration facilitée, comme précédemment, grâce à la généralisation des corrélations observées pour des utilisateurs à l'ensemble des utilisateurs. Comme nous l'avons déjà vu, dans notre cas, une difficulté majeure provient du fait que tous les contextes ne sont pas observables, contrairement aux problèmes de bandits contextuels traditionnels. En effet les contraintes des différentes API ne nous autorisent pas à voir un contexte pour chaque action et on ne peut donc pas utiliser des algorithmes de bandits contextuels existants tels que \texttt{LinUCB} \cite{banditCtxtLinUCB} ou encore \texttt{OFUL} \cite{OFUL} par exemple. À l'instar du chapitre précédent, plusieurs cas sont envisageables concernant le processus choisissant les actions pour lesquelles un contexte est observable. On note par la suite $\mathcal{O}_{t} \subset \mathcal{K}$ l'ensemble des utilisateurs pour lesquels un vecteur de contexte est disponible à l'itération $t$. Ainsi, le cas où $\mathcal{O}_{t}=\mathcal{K}$ correspond au bandit contextuel traditionnel.

Dans la section suivante, nous étudions dans un premier temps le cas le plus simple où tous les contextes sont observables. Ceci nous permettra de mettre en place les bases nécessaires à l'étude du cas plus complexe où une partie des contextes n'est pas accessible.

\section{Algorithmes}

Nous nous concentrons ici sur les algorithmes de types optimistes. Rappelons que le principe général, derrière toutes les politiques de ce type, consiste à construire  un intervalle de confiance sur les récompenses de chaque action. Pour ce faire, une des possibilités est de maintenir à jour des distributions \textit{a posteriori} sur les différents paramètres du modèle. Certaines hypothèses doivent être faites à la fois sur la vraisemblance des observations, mais aussi sur les distributions \textit{a priori} des paramètres. En particulier, l'utilisation des \textit{priors conjugués} permet d'obtenir des solutions exactes. Dans notre étude, nous utiliserons des distributions gaussiennes. Notons cependant que la dérivation d'un intervalle de confiance peut se faire à l'aide d'autres approches, comme ce fut le cas dans le chapitre précédent.

\subsection{Intervalles de confiance lorsque les contextes sont visibles}
\label{subseqcontextvixibles}

Construisons dans un premier temps un estimateur des paramètres dans le cas où tous les contextes seraient observables, avec les hypothèses suivantes :

\begin{itemize}
\item \textbf{Vraisemblance :} Les récompenses sont indépendamment  et identiquement distribuées selon les contextes observés: $r_{i,t} \sim \mathcal{N}(x_{i,t}^{\top}\beta,\sigma_{i}^2)$, avec $\sigma_{i}^2$ correspondant à la variance de la récompense $r_{i,t}$; 
\item \textbf{Prior :} Les paramètres inconnus sont normalement distribués: $\beta \sim \mathcal{N}(m_{0},v_{0}^{2} I)$, avec $m_{0}$ un vecteur de taille $d$, $I$ la matrice identité de taille $d$, et $v_{0}$ un paramètre permettant de contrôler la variance. 
\end{itemize}

\begin{note}
L'hypothèse de vraisemblance gaussienne est plus restrictive que celle du chapitre \ref{ModeleContextuelFixe} dans lequel un bruit sous-gaussien était suffisant. L'intérêt d'utiliser des gaussiennes est de conserver des formes analytiques sur les distributions des paramètres du problème.
\end{note}

Le but est de construire la distribution \textit{a posteriori} du paramètre $\beta$ à un instant $t$ connaissant les récompenses collectées, et les contextes associés, jusqu'à l'itération $t-1$. Pour alléger les notations, nous fixons $m_{0}=0$  (vecteur nul de taille $d$), $v_{0}=1$ et $\sigma_{i}=1$ pour tout  $i$. Cependant, tous les résultats peuvent être étendus à des cas plus complexes. Dans la pratique, lorsque nous n'avons pas d'information particulière sur les bras, toutes la valeurs des $\sigma_{i}$ sont prises égales à la même valeur.

Nous introduisons les notations matricielles condensées suivantes pour la suite de cette section :

\begin{itemize}
\item $\mathcal{T}_{i,t-1}$ l'ensemble des itérations où $i$ a été choisi durant les $t-1$ premières étapes: $\mathcal{T}_{i,t-1}=\left\{s \leq t-1 , i \in \mathcal{K}_{s}\right\}$. On a $|\mathcal{T}_{i,t-1}|=N_{i,t-1}$; 
\item $c_{i,t-1}$ le vecteur de récompenses obtenues par le bras $i$ avant l'itération $t$: $c_{i,t-1}=(r_{i,s})_{s\in \mathcal{T}_{i,t-1}}$ ;
\item $D_{i,t-1}$ la matrice de taille $N_{i,t-1}\times d$, composée des contextes observés  pour le bras $i$, aux itérations antérieures à $t$ où il a été choisi: $D_{i,t-1}=(x_{i,s}^{\top})_{s\in \mathcal{T}_{i,t-1}}$.
\end{itemize}

Les deux propositions suivantes expriment les distributions \textit{a posteriori} des paramètres du modèle à un instant $t$, avant que la sélection des actions n'ait été effectuée, c'est-à-dire en utilisant l'historique des choix effectués jusqu'au temps $t-1$.

\begin{prop}
\label{propDistBetaChap3}
La distribution \textit{a posteriori} du paramètre $\beta$ à l'étape $t$, lorsque tous les contextes sont visibles, respecte:

\begin{equation}
\label{eqBetaChap3}
\beta \sim \mathcal{N}\left(\hat{\beta}_{t-1},V_{t-1}^{-1}\right)
\end{equation}

Où: 

\begin{align}
\hat{\beta}_{t-1}= V_{t-1}^{-1}b_{t-1}& &b_{t-1}= \sum\limits_{i=1}^{K}D_{i,t-1}^{\top}c_{i,t-1}& & V_{t-1}=I+\sum\limits_{i=1}^{K}D_{i,t-1}^{\top}D_{i,t-1}
\end{align}

\end{prop}

\begin{preuve}
Il s'agit d'un résultat classique de regression que nous avons démontré dans la section \ref{TSLinBanditEtatArt} de l'état de l'art.
\end{preuve}

Tous ces paramètres peuvent être mis à jour de façon peu coûteuse à mesure que de nouveaux exemples d'apprentissage arrivent (la complexité de la mise à jour des paramètres est constante sur l'ensemble du processus). 

\begin{prop}
 La valeur espérée $\mathbb{E}[r_{i,t}|x_{i,t}]$, de la récompense de l’utilisateur $i$ à l’itération $t$ sachant son contexte $x_{i,t}$, suit la distribution:
\begin{equation}
\label{distAllSeen}
\mathbb{E}[r_{i,t}|x_{i,t}] \sim 
\mathcal{N}\left(x_{i,t}^{\top}\hat{\beta}_{t-1}, x_{i,t}^{\top}V_{t-1}^{-1}x_{i,t}\right)
\end{equation}
\end{prop}

\begin{preuve}
D'après le modèle, on sait que $\mathbb{E}[r_{i,t}|x_{i,t}]=x_{i,t}^{\top}\beta$. Or $\beta$ est une variable aléatoire gaussienne de moyenne $\hat{\beta}_{t-1}$ et matrice de covariance $V_{t-1}^{-1}$ d'après la proposition précédente, d'où le résultat annoncé.

\end{preuve}

\begin{theo}
\label{theoUCBICWSM}
Pour tout $0<\delta <1$ et $x_{i,t} \in \mathbb{R}^{d}$, en notant $\alpha = \Phi^{-1}(1-\delta/2)$ \footnote{$\Phi^{-1}$ est la fonction de répartition inverse d'une loi normale.}, pour chaque action $i$, après $t$ itérations:
\begin{equation}
P\left( | \mathbb{E}[r_{i,t}|x_{i,t}] -  x_{i,t}^{\top}\hat{\beta}_{t-1}|  \leq   \alpha \sqrt{x_{i,t}^{\top}V_{t-1}^{-1}x_{i,t}} \right)  \geq 1-\delta 
\label{probObs}
\end{equation}

\end{theo}

\begin{preuve}
En utilisant l'équation \ref{distAllSeen}, la variable aléatoire $\dfrac{\mathbb{E}[r_{i,t}|x_{i,t}] -  x_{i,t}^{\top}\hat{\beta}_{t-1}}{\sqrt{x_{i,t}^{\top}V_{t-1}^{-1}x_{i,t}}}$ suit une loi normale de moyenne 0 et de variance 1, d'où le résultat proposé.
\end{preuve}

Cette formule peut directement être utilisée pour définir une borne supérieure de l'intervalle de confiance (\textit{UCB - Upper Confidence Bound}) de la récompense espérée pour chaque utilisateur dont le contexte a été observé à l’itération $t$:

\begin{equation}
s_{i,t} = x_{i,t}^{\top}\hat{\beta}_{t-1} + \alpha\sqrt{x_{i,t}^{\top}V_{t-1}^{-1}x_{i,t}}
\label{obsScoreChap3}
\end{equation}
 
Ainsi les approches de type UCB étant des approches optimistes, l'idée est alors de sélectionner à chaque instant $t$ les $k$ utilisateurs ayant les $k$ bornes $s_{i,t}$ les plus élevées. 
Le score précédent pourrait directement être utilisé pour notre tâche de collecte d'information si les contextes de tous les utilisateurs étaient disponibles, ce qui n'est pas le cas, hormis lorsque $\mathcal{O}_{t}=\mathcal{K}$. Dans la section qui suit, nous étudions le cas des contextes manquants, qui nécessite un traitement particulier.

\subsection{Prise en compte des utilisateurs dont le contexte n'est pas visible} 

Dans le cas qui nous intéresse, c'est-à-dire lorsque tous les contextes ne peuvent pas être observés, l'application directe du modèle précédent pose deux difficultés majeures:

\begin{enumerate}
\item Les scores $s_{i,t}$, permettant de sélectionner les utilisateurs que l'on souhaite écouter, ne peuvent pas être calculés directement selon l'équation \ref{obsScoreChap3} lorsque les contextes $x_{i,t}$ ne sont pas observés (c.-à-d. pour les utilisateurs qui ne sont pas dans $\mathcal{O}_{t}$). Une approche naïve pourrait consister à éliminer les utilisateurs sans contexte observé des choix possibles. Néanmoins, cela peut conduire à écarter de nombreux comptes potentiellement intéressants, et selon le processus d'alimentation des contexte considéré, il est alors possible de s'enfermer dans des stratégies de sélection peu efficaces (par exemple, si $\mathcal{O}_{t}=i_{t-1}$, aucun changement d'utilisateur écouté ne peut être effectué). Il est donc primordial de définir une méthode permettant de donner un score aux utilisateurs dont le contexte est inconnu.
\item  Pour effectuer l'apprentissage des paramètres du modèle de la section précédente, nous devons disposer à un instant $t$ d'un ensemble de couple $(x_{i,s},r_{i,s})_{i\in \mathcal{K}_{s},s=1..t-1}$. Les paramètres du problèmes ne peuvent alors être mis à jour que pour les utilisateurs appartenant à $\mathcal{K}_{t} \cap \mathcal{O}_{t}$. Or rien ne garantit des intersections non vides entre ces deux ensembles. Afin d'assurer un apprentissage efficace, il est nécessaire de définir une méthode permettant d'apprendre à chaque itération, pour toutes les récompenses récoltées, même si aucun contexte n'y est associé.
\end{enumerate}

Pour pallier ces deux problèmes, l'idée générale que nous proposons est d'effectuer des hypothèses probabilistes sur les distributions des contextes.  Dans la section suivante, nous définissons le nouveau modèle probabiliste adopté. Nous verrons que l'apprentissage de ce dernier, contrairement au cas précédent ne peut pas être fait de façon analytique. Pour cette raison, nous utiliserons une approche variationnelle \cite{Bishopvar}, permettant d'approximer la véritable distribution des paramètres. Finalement, nous proposerons une borne supérieure de l'intervalle de confiance de la récompense espérée de chaque utilisateur, qui nous permettra de définir un algorithme de type UCB, spécifiquement adapté à notre cas d'étude.

\subsubsection{Modèle probabiliste et apprentissage}

Considérons le contexte $x_{i,t}$ d'un utilisateur $i$ au temps $t$ comme une variable aléatoire. Lorsque ce contexte n'est pas observé, nous supposons qu'il suit une certaine distribution. Il s'agit d'un modèle gaussien défini pour chaque utilisateur, dans lequel les contextes sont normalement distribués (vraisemblance), selon une moyenne et une variance qui leur est propre. 
 Considérons les hypothèses suivantes:
 \begin{itemize}
 \item \textbf{Vraisemblance :} Pour chaque utilisateur $i$, les contextes sont supposés provenir d'une loi gaussienne multivariée: $x_{i,t} \sim \mathcal{N}(\mu_{i},\tau_{i}^{-1}I)$, de moyenne $\mu_{i}$ et de précision $\tau_{i}^{-1}I$. Nous utilisons ici la précision, qui sera plus simple à traiter d'un point de vue calculatoire;
  \item \textbf{Priors :} Pour chaque utilisateur $i$, on suppose que l'espérance de ses contextes est également une variable aléatoire gaussienne de moyenne nulle et de précision $\tau_{i}^{-1}I$: $\mu_{i}\sim \mathcal{N}(0,\tau_{i}^{-1}I)$. La précision $\tau_{i}$ est supposée suivre une loi Gamma de paramètres $a_{0}$ et $b_{0}$: $\tau_{i}\sim Gamma(a_{0},b_{0})$.
 \end{itemize}
 
 La figure \ref{fig:genProcessPlate} représente le modèle probabiliste considéré sous forme graphique. Les carrés à bords droits représentent des valeurs fixées, les cercles représentent des variables aléatoires et les carrés à bords ronds signifient la répétition des variables aléatoires qu'ils contiennent. On rappelle que nous avons fixé $\sigma_{i}=1$ pour tout $i$, $v_{0}=1$ et $m_{0}=0$ (vecteur nul de taille $d$).

\begin{figureth}
      \includegraphics[width=0.65\linewidth]{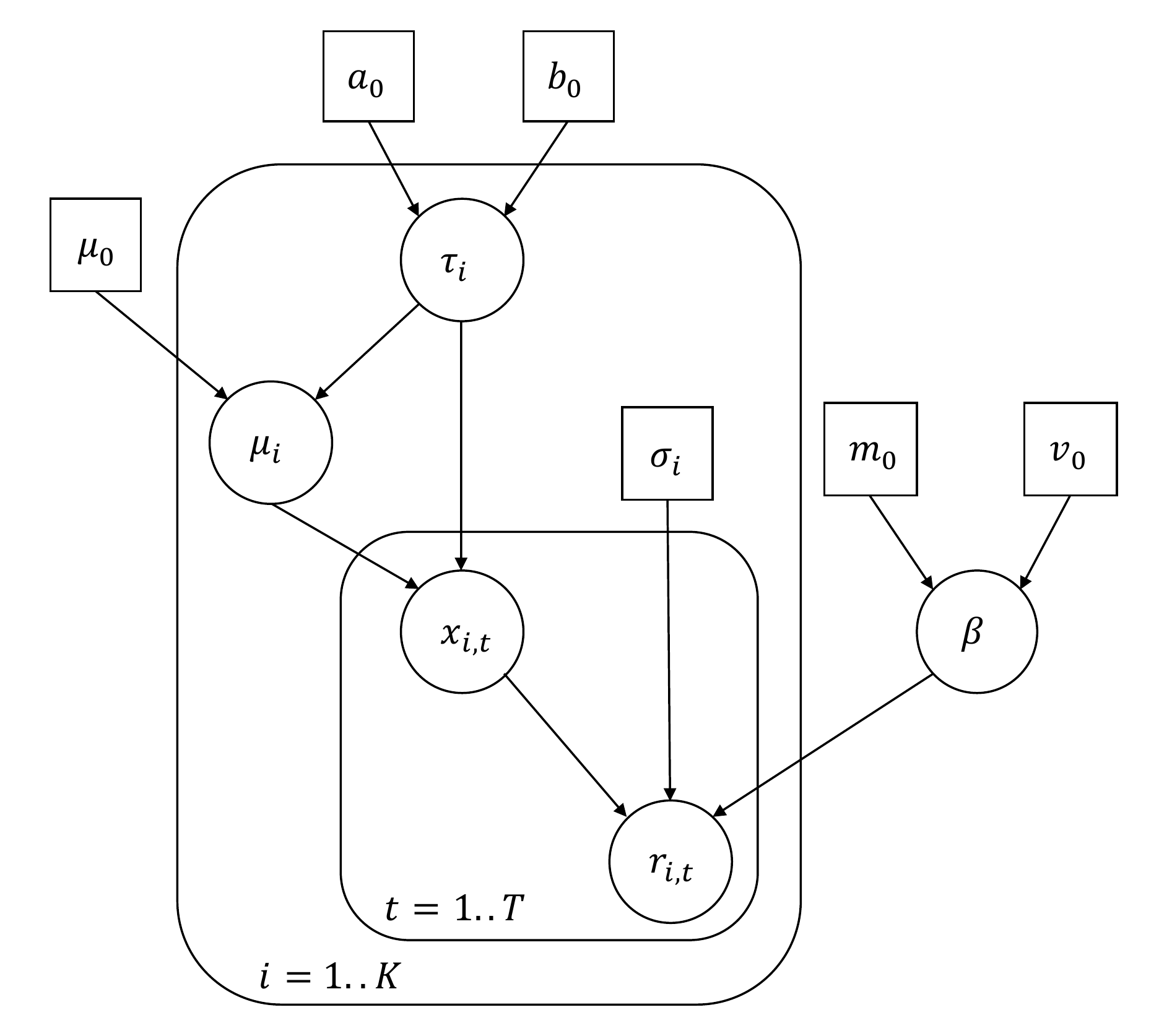}
      \caption[Modèle contextuel génératif]{Représentation graphique du modèle contextuel génératif.}   
      \label{fig:genProcessPlate}
\end{figureth}

\begin{note}
Nous avons choisi d'utiliser des matrices de précision sphériques (c.-à-d. diagonales avec des composantes identiques), cependant, d'autres possibilités sont envisageables. Nous aurions également pu considéré des matrices diagonales, avec des valeurs de variances différentes pour chaque composante des contextes (et nécessitant une distribution Gamma a priori sur chaque composante), ou encore des matrices non diagonales (et nécessitant une distribution normal-Wishart a priori sur la matrice de précision des contextes), mais au prix d'un coût calculatoire plus important.
\end{note}

Nous disposons désormais d'un modèle décrivant la dynamique des récompenses, mais aussi des contextes au cours du temps. Rappelons que dans la section précédente, seuls $\beta$ et les $r_{i,t}$ étaient considérés comme des variables aléatoires. Dans le cas présent, tous les $x_{i,t}$, $\mu_{i}$ et  $\tau_{i}$ sont également des variables aléatoires. Si certaines sont directement observées, d'autres ne sont pas visibles et, comme précédemment, notre but est de dériver leurs distributions à un instant $t$, en fonction des observations faites jusqu'au temps $t-1$. Les observations sont constituées des récompenses et contextes, à savoir $(r_{i,s})_{i\in \mathcal{K}_{s},s=1..t-1}$ et $(x_{i,s})_{i\in \mathcal{O}_{s},s=1..t-1}$. D'autre part, les variables cachées dont la distribution doit être calculée sont $\beta$,  $(\mu_{i})_{i=1..K}$, $(\tau_{i})_{i=1..K}$ et les contextes non observés ayant générés des récompenses $(x_{i,s})_{i\in \mathcal{K}_{s}\cap\bar{\mathcal{O}_{s}},s=1..t-1}$. Notons que les récompenses non observées liées à des contextes observés ne sont pas prises en considérations ici car elles n'influencent pas les variables qui nous intéressent. Formellement, il s'agit de calculer la distribution conditionnelle suivante:

\begin{align}
\label{distAtrouver}
p\left(\beta, (\mu_{i})_{i=1..K}, (\tau_{i})_{i=1..K}, (x_{i,s})_{i\in \mathcal{K}_{s}\cap\bar{\mathcal{O}_{s}},s=1..t-1} | (r_{i,s})_{i\in \mathcal{K}_{s},s=1..t-1},(x_{i,s})_{i\in \mathcal{O}_{s},s=1..t-1} \right)
\end{align}

Le calcul de cette distribution en appliquant directement le théorème de Bayes posent de nombreuses difficultés calculatoires, et il est malheuresement impossible d'en obtenir une forme analytique. Pour pallier ce problème, nous proposons d'adopter une démarche variationnelle, qui nous permettra de dériver les distributions souhaitées. Nous présentons ci-dessous le principe  général de cette approche, que nous appliquerons ensuite à notre cas d'étude.

\paragraph{Principe de l'approche variationnelle: }
\label{principeVariationalMethode}
Rappelons dans un premier temps le principe de cette approche. Il s'agit d'un ensemble de techniques utilisant le calcul variationnel, permettant d'approximer des distributions \textit{a posteriori} dans le cas où ces dernières n'ont pas de forme analytique. Il s'agit d'une alternative aux méthodes dites MCMC (Monte Carlo Markov Chain) dont le Gibbs sampling fait partie. En particulier, alors que les techniques de Monte Carlo fournissent une approximation numérique de la véritable distribution \textit{a posteriori} (via un ensemble d'échantillons), la méthode bayésienne variationnelle fournit une solution analytique localement optimale à une approximation de la véritable distribution \textit{a posteriori}.

L'idée de l'approche variationnelle est d'approximer la véritable distribution \textit{a posteriori} par une distribution d'une forme plus simple. On considère un ensemble de variables observées noté $X$, un ensemble de variables non observées noté $Z$. La distribution \textit{a posteriori} de $Z$ sachant $X$ , noté $p(Z|X)$ est approximée par une distribution dite variationnelle notée $q(Z)$. Le but est de trouver une distribution $q$ la plus proche possible de $p$ en minimisant une distance entre $p$ et $q$. On choisit comme distance la divergence de Kullback–Leibler de $p$ par rapport à $q$, notée $KL(p,q)$, dont on rappelle la définition ci-dessous:
 
 \begin{equation}
 \label{eqDivKLVar}
KL(q,p)=\int q(Z) \log\left(\dfrac{q(Z)}{p(Z|X)}\right) \, \mathrm{d}Z
\end{equation}

On note que $q$ et $p$ sont inversées par rapport à ce à quoi l’on pourrait s'attendre. Cette utilisation de la divergence de Kullback–Leibler est conceptuellement similaire à ce qui est fait dans l'algorithme EM (Expectation-Maximisation). Comme le précisent les auteurs de \cite{varMessPassing}, le fait de minimiser $KL(q,p)$ (divergence dite "exclusive") peut mener à une distribution $q$ ignorant certains modes de $p$. En revanche, minimiser $KL(p,q)$ (divergence dite "inclusive") peut mener à une distribution $q$ donnant du poids à des régions où la distribution $p$ est absente. A titre informatif, l'algorithme EP (Expectation-Propagation) \cite{Minka2001}, qui constitue une autre méthode d'inférence, est basé sur la minimisation de la divergence "inclusive".

Après quelques calculs, notamment détaillés dans \cite{varMessPassing}, le terme de l'équation \ref{eqDivKLVar} peut se réécrire sous la forme qui suit:

 \begin{equation}
\log p(X) = KL(q,p) + F(q)
\end{equation}

où $F(q)$ est l'énergie libre et vaut $F(q)=\int q(Z) \log\left(\dfrac{p(X,Z)}{q(Z)}\right) \, \mathrm{d}Z$.

Etant donnée que la log-vraisemblance $\log p(X)$ est fixe par rapport à $q$, maximiser le terme $F(q)$ revient à minimiser la divergence $KL(q,p)$. Ainsi, si l'on choisit bien $q$, on obtient non seulement une forme analytique d'une approximation de la véritable distribution \textit{a posteriori} $p(Z|X)$, mais aussi une borne inférieure de la log-vraisemblance des données $F(q)$.

En pratique, il est d'usage d'utiliser une hypothèse supplémentaire sur la forme de la distribution $q$. Cette hypothèse est appelée \textit{mean field assumption} et consiste à factoriser la distribution $q$ selon une partition des variables $Z$. Si l'on considère une partition $Z_{1},...,Z_{M}$ de l'ensemble des variables $Z$, alors l'hypothèse se traduit par:

\begin{equation}
q(Z)=\prod\limits_{i=1}^{M}q_{i}(Z_{i})
\end{equation}

On peut démontrer à l'aide du calcul variationnel que la distribution optimale pour chacune des composantes $q_{j}$, notée  $q_{j}^{\star}$ satisfait:

\begin{equation}
 q_{j}^{\star} (Z_{j})= \dfrac{e^{\mathbb{E}_{i\neq j}[\log p(Z,X)]}}{ 
 \int  e^{\mathbb{E}_{i\neq j}[\log p(Z,X)]} \, \mathrm{d}Z_{j}}
\end{equation}

où  $\mathbb{E}_{i\neq j}[\log p(Z,X)]$ est l'espérance du logarithme de la probabilité jointe des données et des variables latentes, prise selon toutes les variables sauf $j$. En pratique, il est plus commode de travailler avec les logarithmes, c'est-à-dire:

\begin{equation}
\label{varOptSol}
\log q_{j}^{\star} (Z_{j})= \mathbb{E}_{i\neq j}[\log p(Z,X)] +constante
\end{equation}

La constante dans l'expression ci-dessus correspond au terme de normalisation de la distribution, qui ne dépend pas de la valeur de $Z_j$. Elle peut être réinjectée \textit{a posteriori} lorsqu’une forme (gaussienne ou autre) est reconnue dans le numérateur. Cette formulation crée des dépendances circulaires entre les paramètres des distributions des variables présentes dans les différentes partitions. Cela suggère naturellement d'utiliser  un algorithme itératif, dans lequel les espérances (et éventuellement les moments d'ordre supérieurs) des variables latentes sont initialisées d'une certaine manière, puis les paramètres de chaque partition sont calculés tour à tour en utilisant les distributions courantes des autres partitions, jusqu'à convergence de l'algorithme.

\paragraph{Application à notre cas: }

Nous introduisons les nouvelles  notations ensemblistes suivantes, qui serviront pas la suite pour dériver les paramètres des différentes distributions:

\begin{itemize}
\item Pour tout $i$, l'ensemble des itérations jusqu'au temps $t-1$, telles que $i$ a été sélectionné en ayant eu un contexte observé  est noté $\mathcal{A}_{i,t-1}=\left\{s \text{ tel que } 1\leq s\leq t-1 \text{ et } i\in \mathcal{K}_{s}\cap\mathcal{O}_{s} \right\}$;
\item Pour tout $i$, l'ensemble des itérations jusqu'au temps $t-1$, telles que $i$ a été sélectionné sans avoir eu un contexte observé est noté $\mathcal{B}_{i,t-1}=\left\{s \text{ tel que } 1\leq s\leq t-1 \text{ et } i\in \mathcal{K}_{s}\cap\bar{\mathcal{O}_{s}} \right\}$;
\item Pour tout $i$, l'ensemble des itérations  jusqu'au temps $t-1$,  telles que le contexte de $i$ a été observé est noté $\mathcal{C}_{i,t-1}=\left\{s \text{ tel que } 1\leq s\leq t-1 \text{ et } i\in \mathcal{O}_{s} \right\}$;
\end{itemize}

En utilisant ces notations, la distribution \textit{a posteriori} des différents paramètres, décrite dans l'équation \ref{distAtrouver}, peut se réécrire:

\begin{align}
p\left(\beta, (\mu_{i})_{i=1..K}, (\tau_{i})_{i=1..K}, (x_{i,s})_{i=1..K,s\in \mathcal{B}_{i,t-1}} | (r_{i,s})_{i=1..K, s\in \mathcal{A}_{i,t-1}\cup\mathcal{B}_{i,t-1}},(x_{i,s})_{i=1..K,s \in \mathcal{C}_{i,t-1}}\right)
\end{align}

\textbf{Factorisation: } Afin de se placer dans un cadre favorable à l'inférence variationnelle, on suppose la factorisation suivante à un instant $t$:

\begin{equation}
q\left(\beta, (\mu_{i})_{i=1..K}, (\tau_{i})_{i=1..K}, (x_{i,s})_{i=1..K,s\in \mathcal{B}_{i,t-1}}\right)=q_{\beta}(\beta)\prod\limits_{i=1}^{K}\left(q_{\mu_{i}}(\mu_{i})q_{\tau_{i}}(\tau_{i})\prod\limits_{s\in \mathcal{B}_{i,t-1}}q_{x_{i,s}}(x_{i,s})\right)
\end{equation}

Grâce à cette hypothèse et en utilisant la méthode variationnelle présentée plus haut, nous pouvons désormais calculer les distributions de chacun des paramètres. Les propositions \ref{distribVarBeta}, \ref{distribVarXit}, \ref{distribVarMui} et \ref{distribVarTaui} établissent ces distributions respectivement pour les variables $\beta$, $x_{i,s}$, $\mu_{i}$ et $\tau_{i}$. Les preuves sont disponibles en annexes \ref{PreuveVarContextualBeta}, \ref{PreuveVarContextualXit}, \ref{PreuveVarContextualMui} et \ref{PreuveVarContextualTaui}.

\begin{prop}
\label{distribVarBeta}
La distribution variationnelle de $\beta$ après $t-1$ itérations est une gaussienne de moyenne $\hat{\beta}_{t-1}$ et de matrice de covariance $V_{t-1}^{-1}$:

\begin{equation}
\beta \sim \mathcal{N}(\hat{\beta}_{t-1}, V_{t-1}^{-1})
\end{equation}

avec

$V_{t-1}=I+\sum\limits_{i=1}^{K} \left[ \sum\limits_{s\in \mathcal{A}_{i,t-1}}x_{i,s}x_{i,s}^{\top} +\sum\limits_{s\in \mathcal{B}_{i,t-1}} \mathbb{E}[x_{i,s}x_{i,s}^{\top}] \right]$ 

$\hat{\beta}_{t-1}=V_{t-1}^{-1}\left(\sum\limits_{i=1}^{K}\left[ \sum\limits_{s\in \mathcal{A}_{i,t-1}}x_{i,s}r_{i,s} +\sum\limits_{s\in \mathcal{B}_{i,t-1}} \mathbb{E}[x_{i,s}]r_{i,s} \right]\right)$
\end{prop}

On remarque que l'espérance sur les $x_{i,s}$ n'est calculée que lorsque l'action a été sélectionnée mais que son contexte \textbf{n'a pas} été observé. Par ailleurs, le calcul de $\mathbb{E}[x_{i,s}]$ et $\mathbb{E}[x_{i,s}x_{i,s}^{\top}]$ est effectué en utilisant le fait que $\mathbb{E}[xx^{\top}]=Var(x)+\mathbb{E}[x]\mathbb{E}[x]^{\top}$ et la proposition \ref{distribVarXit}.

\begin{prop}
\label{distribVarXit}
Pour tout $i$ et pour tout $s\in \mathcal{B}_{i,t-1}$, la distribution variationnelle de $x_{i,s}$ est une gaussienne de moyenne $\hat{x}_{i,s}$ et de matrice de covariance $W_{i,s}^{-1}$:

\begin{equation}
x_{i,s}\sim \mathcal{N}(\hat{x}_{i,s}, W_{i,s}^{-1})
\end{equation}

avec

$W_{i,s}= \mathbb{E}[\beta\beta^{\top}]+\mathbb{E}[\tau_{i}]I$

$\hat{x}_{i,s}=W_{i,s}^{-1}(\mathbb{E}[\beta]r_{i,s}+\mathbb{E}[\tau_{i}]\mathbb{E}[\mu_{i}])$
\end{prop}

Les expressions de $\mathbb{E}[\mu_{i}]$ et $\mathbb{E}[\tau_{i}]$ sont calculées selon les propositions \ref{distribVarMui} et \ref{distribVarTaui}. De plus, le calcul de $\mathbb{E}[\beta\beta^{\top}]$ s'effectue en utilisant la proposition \ref{distribVarBeta}.

\begin{prop}
\label{distribVarMui}
Pour tout $i$, la distribution variationnelle de $\mu_{i}$ après $t-1$ itérations est une gaussienne de moyenne $\hat{\mu}_{i,t-1}$ et de matrice de covariance $\Sigma_{i,t-1}^{-1}$:

\begin{equation}
\mu_{i}\sim \mathcal{N}(\hat{\mu}_{i,t-1}, \Sigma_{i,t-1}^{-1})
\end{equation}

avec

$\Sigma_{i,t-1}=  (1+n_{i,t-1})E[\tau_{i}]I$ 

$\hat{\mu}_{i,t-1}=\dfrac{\sum\limits_{s\in \mathcal{C}_{i,t-1}}x_{i,s} +\sum\limits_{s\in \mathcal{B}_{i,t-1}} \mathbb{E}[x_{i,s}]}{1+n_{i,t-1}}$
\end{prop}

et $n_{i,t-1}=|\mathcal{B}_{i,t-1}|+|\mathcal{C}_{i,t-1}|$.

\begin{prop}
\label{distribVarTaui}
Pour tout $i$, la distribution variationnelle de $\tau_{i}$ après $t-1$ itérations suit une loi gamma de paramètres $a_{i,t-1}$ et $b_{i,t-1}$: 

\begin{equation}
\tau_{i}\sim Gamma(a_{i,t-1},b_{i,t-1})
\end{equation}

avec

$a_{i,t-1}= a_{0}+\dfrac{d(1+n_{i,t-1})}{2}$ 

$b_{i,t-1}=b_{0}+\dfrac{1}{2}  \left[ (1+n_{i,t-1})\mathbb{E}[\mu_{i}^{\top}\mu_{i}]-2\mathbb{E}[\mu_{i}]^{\top}\left( \sum\limits_{s\in \mathcal{C}_{i,t-1}}x_{i,s} +\sum\limits_{s\in \mathcal{B}_{i,t-1}}\mathbb{E}[x_{i,s}] \right) \right.$

$\qquad \qquad \qquad ~~~~~~~~~ \left.    + \sum\limits_{s\in \mathcal{C}_{i,t-1}}x_{i,s}^{\top}x_{i,s} + \sum\limits_{s\in \mathcal{B}_{i,t-1}}\mathbb{E}[x_{i,s}^{\top}x_{i,s}]      \right] $

\end{prop}

On a $\mathbb{E}[\tau_{i}]=a_{i,t-1}/b_{i,t-1}$, ce qui permet de calculer les paramètres dans les propositions \ref{distribVarXit} et \ref{distribVarMui}. De plus, le calcul de $\mathbb{E}[x_{i,s}^{\top}x_{i,s}] $ et $\mathbb{E}[\mu_{i}^{\top}\mu_{i}]$  s'effectue en remarquant que $\mathbb{E}[x^{\top}x]=Trace(Var(x))+\mathbb{E}[x]^{\top}\mathbb{E}[x]$ et les propositions \ref{distribVarXit} et \ref{distribVarMui}.

\vspace{0.5cm}

Les quatre propositions précédentes fournissent les distributions des différentes variables du problème. Il apparaît très clairement que les paramètres de chacune d'elles sont liés à tous les autres, d'où la nécessité d'une procédure itérative, dont le but est de converger vers une solution stable. Nous proposons d'expliciter cette procédure dans l'algorithme \ref{varPassLinear}. Cet algorithme prend en entrée un entier $nbIt$  qui correspond au nombre d'itérations à effectuer. Plus il est élevé, plus la solution trouvée sera précise. Dans la pratique, il est également possible de surveiller la convergence des différents paramètres et d'arrêter la procédure lorsque les variations de ces derniers ne dépassent pas un certain seuil d'une itération à la suivante.

\begin{algorithm}[!ht]
  \KwIn{$nbIt$ (nombre d'itérations)}
    Initialiser les paramètres de façon aléatoire\;
  \For{$It  = 1..nbIt$} {
      Calculer $\hat{\beta}_{t-1}$ et $V_{t-1}$  selon la proposition \ref{distribVarBeta}\;
    \For{$i =1.. K$} {
     Calculer $\hat{\mu}_{i,t-1}$ et $\Sigma_{i,t-1}$ selon la proposition \ref{distribVarMui} \;
         Calculer $a_{i,t-1}$ et $b_{i,t-1}$ selon la proposition \ref{distribVarTaui} \;
         \For{$s  \in \mathcal{B}_{i,t-1}$} {
          Calculer  $\hat{x}_{i,s}$ et $W_{i,s}$ selon la proposition \ref{distribVarXit} \;
         }
    }
  }
\caption{\texttt{Processus itératif variationnel pour le modèle contextuel}}
\label{varPassLinear}
\end{algorithm}

\textbf{Note sur la complexité :} 
La complexité de l'algorithme d'inférence variationnelle augmente avec le nombre d'itérations $t$. En effet, pour chaque $i$, à un instant $t$, on voit clairement qu'il est nécessaire de traiter tous les contextes cachés $x_{i,s}$ pour tous les temps passés $s\in \mathcal{B}_{i,t-1}$. Cette croissance, potentiellement très rapide (selon processus délivrant les contextes), peut  mener à des problèmes de mémoire et n'est pas compatible avec la nature intrinsèquement "en ligne" des problèmes de bandits. Afin de s'affranchir de cet inconvénient, nous proposons d'utiliser une approximation des algorithmes se restreignant à une fenêtre de temps limitée dans le passé pour le calcul des distributions des $x_{i,s}$ lorsque $s\in \mathcal{B}_{i,t-1}$. Soit $S$ la taille de cette fenêtre, alors à un instant $t$, nous considérons l'algorithme précédent avec $\mathcal{B}_{i,t-1}\backslash \mathcal{B}_{i,t-1-S}$ au lieu de $\mathcal{B}_{i,t-1}$, rendant ainsi la complexité constante avec le temps. Le fait de considérer une fenêtre de temps influe sur les distributions de tous les paramètres, car chacun d'eux fait intervenir $\mathcal{B}_{i,t-1}$. Il est toutefois important de noter que cela n'a pas pour effet d'effacer entièrement l'influence des événements passés avant l'instant $t-1-S$. En effet pour l'apprentissage de $\beta$, les sommes $\sum\limits_{s\in \mathcal{A}_{i,t-1}}x_{i,s}x_{i,s}^{\top}$ et $\sum\limits_{s\in \mathcal{A}_{i,t-1}}x_{i,s}r_{i,s}$ peuvent être mise à jour sans avoir besoin de conserver tous les éléments en mémoire, comme cela est fait dans \texttt{LinUCB}. Une remarque similaire est possible pour l'apprentissage des paramètres $\mu_{i}$ et $\tau_{i}$ qui fait intervenir l'ensemble $\mathcal{C}_{i,t-1}$.

\subsubsection{Algorithme de bandit}

Nous disposons désormais d'une méthode permettant, à chaque instant $t$, de calculer la distribution des paramètres du problème en fonction des choix effectués jusqu'au temps $t-1$. Nous nous intéressons désormais à la politique de sélection que l'on peut y associer. Rappelons qu'à l'instant $t$, le but est de sélectionner $k$  actions (ensemble $\mathcal{K}_{t}$) parmi l'ensemble des actions disponibles (ensemble $\mathcal{K}$). Afin d'effectuer cette sélection, nous utilisons une approche de type UCB, qui choisit à chaque instant les actions dont la borne supérieure de l'intervalle de confiance est la plus élevée. La première étape consiste donc à déterminer une borne pour chaque action (comme dans la section \ref{subseqcontextvixibles}). Dans le cas présent, à un instant $t$ pour une action $i$, il existe deux possibilités:

\begin{itemize}
\item Le contexte de $i$ \textbf{est disponible} (c.-à-d. $i\in \mathcal{O}_{t}$): dans ce cas, l'utilisation de l'hypothèse de linéarité ($\mathbb{E}[r_{i,t}|x_{i,t}]=x_{i,t}^{\top}\beta$) et de la distribution variationnelle gaussienne de $\beta$ définie en proposition \ref{distribVarBeta} permet de considérer $\mathbb{E}[r_{i,t}|x_{i,t}]$ comme une variable aléatoire gaussienne de moyenne $x_{i,t}^{\top}\hat{\beta}_{t-1}$ et de variance $x_{i,t}^{\top}V_{t-1}^{-1}x_{i,t}$. Grâce à cela, il est directement possible de déterminer un intervalle de confiance comme nous l'avons précédemment effectué dans le théorème \ref{theoUCBICWSM}. Finalement, pour chaque action dont le contexte est visible, le score UCB associé est donc défini par la même formule que dans l'équation \ref{obsScoreChap3}. Il est important de noter cependant que les valeurs des paramètres $\hat{\beta}_{t-1}$ et $V_{t-1}^{-1}$ ne sont pas les mêmes et doivent être appris de façon itérative via l'approche variationnelle proposée plus haut.

\item Le contexte de $i$ \textbf{n'est pas disponible} (c.-à-d. $i\notin \mathcal{O}_{t}$): dans ce cas, il n'est pas possible d'appliquer la méthode précédente, car le vecteur $x_{i,t}$ n'est pas observable. Dans la suite, nous proposons de dériver un intervalle de confiance de l’espérance des récompenses lorsque le contexte associé n'est pas observé, ce qui nous permettra finalement d'aboutir à un algorithme de type UCB mettant en jeu des intervalles de confiance de même niveau que ceux considérés pour les contextes observés.
\end{itemize}

Rappelons que notre modèle considère que pour chaque utilisateur $i$, il existe une distribution sous-jacente responsable des contextes observés. Ainsi, lorsque le contexte n'est pas observé à un instant $t$ pour une action $i$, nous proposons de prendre en considération l'espérance de la récompense de l'action selon la distribution de contexte associé.

\begin{prop}
A un instant $t$, la récompense moyenne de l'action $i$ prise selon la distribution de ses contextes  peut s'écrire de la façon suivante:
\begin{align*}
\mathbb{E}[r_{i,t}]&=\int_{x_{i,t}} \mathbb{E}[r_{i,t}|x_{i,t}]p(x_{i,t}) \, \mathrm{d}x_{i,t}\\
  &=\int_{x_{i,t}} x_{i,t}^{\top}\beta p(x_{i,t}) \, \mathrm{d}x_{i,t}\\
    &=\beta^{\top} \int_{x_{i,t}} x_{i,t} p(x_{i,t}) \, \mathrm{d}x_{i,t}\\
    &=\beta^{\top} \mathbb{E}[x_{i,t}] \\
    &=\beta^{\top} \mu_{i}\\
    &= \beta^{\top}\hat{\mu}_{i,t-1} +\beta^{\top}(\mu_{i}-\hat{\mu}_{i,t-1})
\end{align*}

où $\hat{\mu}_{i,t-1}$ correspond à l'espérance de $\mu_{i}$ (voir proposition \ref{distribVarMui})
\end{prop}

Cette écriture fait apparaître les termes $\beta^{\top}(\mu_{i}-\hat{\mu}_{i,t-1})$ et $\beta^{\top}\hat{\mu}_{i,t-1} $, dont nous allons dériver un intervalle de confiance dans les deux propositions qui suivent.

\begin{prop}
\label{ICterm1}
Pour tout $0<\delta_{1}<1$, à un instant $t$ et pour tout $i$, on a:
\begin{align}
\mathbb{P}\left(|\beta^{\top}\hat{\mu}_{i,t-1} -\hat{\beta}_{t-1}^{\top}\hat{\mu}_{i,t-1}| \leq  \alpha_{1} \sqrt{\hat{\mu}_{i,t-1}^{\top}V_{t-1}^{-1}\hat{\mu}_{i,t-1}}\right)= 1-\delta_{1}
\end{align}

où $\alpha_{1}=\Phi^{-1}(1-\delta_{1}/2)$.
\end{prop}

\begin{preuve}
D'après la proposition \ref{distribVarBeta}, à un instant $t$, $\beta$ suit une loi gaussienne de moyenne $\hat{\beta}_{t-1}$ et de variance $V_{t-1}^{-1}$. Par conséquent $\beta^{\top}\hat{\mu}_{i,t-1}$ suit une loi gaussienne de moyenne $\hat{\beta}_{t-1}^{\top}\hat{\mu}_{i,t-1}$ et de variance $\hat{\mu}_{i,t-1}^{\top}V_{t-1}^{-1}\hat{\mu}_{i,t-1}$, donc~ $\mathbb{P}\left(|\beta^{\top}\hat{\mu}_{i,t-1}-\hat{\beta}_{t-1}^{\top}\hat{\mu}_{i,t-1} | \leq  \alpha_{1} \sqrt{\hat{\mu}_{i,t-1}^{\top}V_{t-1}^{-1}\hat{\mu}_{i,t-1}}\right)= 1-\delta_{1}$. Ce qui prouve le résultat annoncé.
\end{preuve}

\begin{prop}
\label{ICterm2}
Pour tout $0<\delta_{2}<1$, à un instant $t$ et pour tout $i$, on a:
\begin{align}
\mathbb{P}\left(|\beta^{\top}(\mu_{i}-\hat{\mu}_{i,t-1}) | \leq \alpha_{2}\sqrt{\dfrac{b_{i,t-1}}{a_{i,t-1}(1+n_{i,t-1})}} \right)\geq 1-\delta_{2}
\end{align}
avec $\alpha_{2}=S\sqrt{\Psi^{-1}(1-\delta_{2})}$, $\Psi^{-1}$ la fonction de répartition inverse de la loi $\chi^{2}$ à $d$ degrés de liberté, et où $n_{i,t-1}$, $a_{i,t-1}$ et $b_{i,t-1}$ sont définis dans les propositions \ref{distribVarMui} et \ref{distribVarTaui}.
\end{prop}

\begin{preuve}
On utilise dans un premier temps le fait que $|\beta^{\top}(\mu_{i}-\hat{\mu}_{i,t-1}) | \leq S||\mu_{i}-\hat{\mu}_{i,t-1}||$, où l'on rappelle que $S$ est une borne supérieure de $\beta$. 
D'après la proposition \ref{distribVarMui}, à un instant $t$, $\mu_{i}$ suit une loi gaussienne de moyenne $\hat{\mu}_{i,t-1}$ et de variance $\Sigma_{i,t-1}^{-1}$. Donc la varible aléatoire $\mu_{i}-\hat{\mu}_{i,t-1}$ suit une loi gaussienne de moyenne nulle et de variance $\Sigma_{i,t-1}^{-1}$. Toujours d'après la proposition \ref{distribVarMui}, la matrice $\Sigma_{i,t-1}^{-1}$ est diagonale et chaque composante de la diagonale sont égales. En effet on a $\Sigma_{i,t-1}^{-1}=((1+n_{i,t-1})E[\tau_{i}])^{-1}I$ que nous écrivons temporairement $\sigma_{i,t}^{2} I$. Par conséquent, toutes les  composantes du vecteur $\dfrac{\mu_{i}-\hat{\mu}_{i,t-1}}{\sigma_{i,t}}$ suivent une loi normale $\mathcal{N}(0,1)$. Par définition de la loi $\chi^{2}$, $\dfrac{||\mu_{i}-\hat{\mu}_{i,t-1}||^{2}}{\sigma_{i,t}^{2}}$ suit donc une loi du $\chi^{2}$ à $d$ degrés de liberté. Notons $\Psi$ la fonction de répartition de cette loi. On a donc 
$\mathbb{P}\left(\dfrac{||\mu_{i}-\hat{\mu}_{i,t-1}||^{2}}{\sigma_{i,t}^{2}} \leq \eta \right)= \Psi(\eta)$ où, de façon équivalente $\mathbb{P}\left(||\mu_{i}-\hat{\mu}_{i,t-1}|| \leq \sqrt{\eta}\sigma_{i,t} \right)= \Psi(\eta)$. Finalement, comme $|\beta^{\top}(\mu_{i}-\hat{\mu}_{i,t-1}) | \leq S||\mu_{i}-\hat{\mu}_{i,t-1}||$, 
on a $\mathbb{P}\left(|\beta^{\top}(\mu_{i}-\hat{\mu}_{i,t-1}) | \leq S\sqrt{\eta}\sigma_{i,t} \right) \geq \Psi(\eta)$. On pose maintenant $\Psi(\eta)=1-\delta_{2}$, et on note $\Psi^{-1}$ la fonction de répartition inverse de la loi $\chi^{2}$. On a donc $\mathbb{P}\left(|\beta^{\top}(\mu_{i}-\hat{\mu}_{i,t-1}) | \leq S\sigma_{i,t}\sqrt{\Psi^{-1}(1-\delta_{2})} \right) \geq 1-\delta_{2}$. Par ailleurs, d'après la proposition \ref{distribVarTaui}, on a $\sigma_{i,t}=\sqrt{\dfrac{1}{(1+n_{i,t-1})E[\tau_{i}]}}=\sqrt{\dfrac{b_{i,t-1}}{a_{i,t-1}(1+n_{i,t-1})}}$, d'où le résultat annoncé.

\end{preuve}

\begin{theo}
\label{theoUCBICWSMBis}
Soit $0<\delta_1<1$ et $0<\delta_2<1$. Alors à un instant $t$ et pour tout $i$, on a: 
\begin{multline}
P\left(|\mathbb{E}[r_{i,t}]-\hat{\beta}_{t-1}^{\top}\hat{\mu}_{i,t-1}| \leq   \alpha_{1} \sqrt{\hat{\mu}_{i,t-1}^{\top}V_{t-1}^{-1}\hat{\mu}_{i,t-1}} +   \alpha_{2}\sqrt{\dfrac{b_{i,t-1}}{a_{i,t-1}(1+n_{i,t-1})}}\right) \geq  
1-\delta_{1}-\delta_{2} \\
\end{multline}
 où $\alpha_{1}=\Phi^{-1}(1-\delta_{1}/2)$ et $\alpha_{2}=S\sqrt{\Psi^{-1}(1-\delta_{2})}$
\end{theo}

\begin{preuve}
La preuve de ce résultat vient directement des résultats des propositions \ref{ICterm1} et\ref{ICterm2}. On a:

$P\left(|\mathbb{E}[r_{i,t}]-\hat{\beta}_{t-1}^{\top}\hat{\mu}_{i,t-1}| >  \alpha_{1} \sqrt{\hat{\mu}_{i,t-1}^{\top}V_{t-1}^{-1}\hat{\mu}_{i,t-1}} +   \alpha_{2}\sqrt{\dfrac{b_{i,t-1}}{a_{i,t-1}(1+n_{i,t-1})}}  \right)$

$=P\left(|\beta^{\top}\hat{\mu}_{i,t-1} +\beta^{\top}(\mu_{i}-\hat{\mu}_{i,t-1})-\hat{\beta}_{t-1}^{\top}\hat{\mu}_{i,t-1}| >  \alpha_{1} \sqrt{\hat{\mu}_{i,t-1}^{\top}V_{t-1}^{-1}\hat{\mu}_{i,t-1}} +   \alpha_{2}\sqrt{\dfrac{b_{i,t-1}}{a_{i,t-1}(1+n_{i,t-1})}}  \right)$

$\leq \mathbb{P}\left(|\beta^{\top}\hat{\mu}_{i,t-1} - \hat{\beta}_{t-1}^{\top}\hat{\mu}_{i,t-1} |+|\beta^{\top}(\mu_{i}-\hat{\mu}_{i,t-1}) | > \alpha_{1} \sqrt{\hat{\mu}_{i,t-1}^{\top}V_{t-1}^{-1}\hat{\mu}_{i,t-1}}+\alpha_{2}\sqrt{\dfrac{b_{i,t-1}}{a_{i,t-1}(1+n_{i,t-1})}}\right)$

$\leq \mathbb{P}\left(\left\{|\beta^{\top}\hat{\mu}_{i,t-1} - \hat{\beta}_{t-1}^{\top}\hat{\mu}_{i,t-1} | > \alpha_{1} \sqrt{\hat{\mu}_{i,t-1}^{\top}V_{t-1}^{-1}\hat{\mu}_{i,t-1}} \right\}  \text{ } \cup \text{ } \left\{|\beta^{\top}(\mu_{i}-\hat{\mu}_{i,t-1}) | >\alpha_{2}\sqrt{\dfrac{b_{i,t-1}}{a_{i,t-1}(1+n_{i,t-1})}}\right\}\right)$

$\leq\delta_{1}+\delta_{2}$

d'après les résultats des propositions \ref{ICterm1} et\ref{ICterm2} et l'inégalité de Boole. En prenant la contraposée, on obtient le résultat annoncé.

\end{preuve}

Ainsi, l'inégalité ci-dessus nous donne une borne relativement serrée de l'intervalle de confiance associé à la récompense espérée de l’utilisateur $i$, d'où nous pouvons dériver un nouveau score de sélection. À chaque itération $t$, si le contexte de $i$ n'est pas observé, son score noté $s_{i,t}$ est tel que:

\begin{equation}
s_{i,t}=\hat{\mu}_{i,t-1}^{\top}\hat{\beta}_{t-1}+\alpha_{1} \sqrt{\hat{\mu}_{i,t-1}^{\top}V_{t-1}^{-1}\hat{\mu}_{i,t-1}} +   \alpha_{2}\sqrt{\dfrac{b_{i,t-1}}{a_{i,t-1}(1+n_{i,t-1})}}
\label{scoreNotObsMeanChap3}
\end{equation}

La première partie du score ressemble fortement à un score \texttt{LinUCB} \cite{banditCtxtLinUCB} dans lequel on utilise $\hat{\mu}_{i,t-1}$ au lieu de $x_{i,t}$. De façon intuitive, il s'agit d'approximer le contexte $x_{i,t}$ par un terme proche de sa moyenne empirique lorsque celui-ci n'est pas visible. Cependant, cette approximation a un prix, et a pour effet d'ajouter un terme supplémentaire dans l'exploration. Ce terme correspond à l'incertitude sur les contextes. Nous proposons d'effectuer une analyse de ce terme d'exploration afin de comprendre son comportement général. Pour cela, on étudie $\dfrac{b_{i,t-1}}{a_{i,t-1}(1+n_{i,t-1})}$, vu comme une fonction de $n_{i,t-1}$. Le terme $n_{i,t-1}$ étant au moins égal au nombre de fois où le contexte de l'utilisateur $i$ a été observé jusqu'au temps $t$, nous pouvons facilement l'interpréter. En observant la formule pour $a_{i,t-1}$ dans la proposition \ref{distribVarTaui} on peut considérer que le terme dominant $a_{i,t-1}(1+n_{i,t-1})$ évolue en $n_{i,t-1}^{2}$. D'autre part, on remarque  qu'en considérant toutes les espérances comme des valeurs finies que $b_{i,t-1}$ évolue en $n_{i,t-1}$. Ainsi, d'une façon générale, le terme d'exploration supplémentaire évolue en $1/\sqrt{n_{i,t-1}}$, qui diminue à mesure que le nombre d'observations du contexte de l'utilisateur augmente. Intuitivement, cela traduit la diminution de l'incertitude entre la moyenne empirique et l'espérance. Finalement, le score des utilisateurs dont le contexte n'est pas observé tend vers un score de type \texttt{LinUCB} \cite{banditCtxtLinUCB} dans lequel on utilise $\hat{\mu}_{i,t-1}$ au lieu de $x_{i,t}$ quand son contexte a été observé un nombre suffisant de fois.

\textbf{Bilan :} Nous avons dérivé un intervalle de confiance de la récompense espérée de chaque utilisateur $i$ à un instant $t$, que son contexte soit observé ou non. Si l'on souhaite avoir une probabilité égale pour les deux cas (contexte observé et non observé), il suffira de choisir les paramètres $\delta$, $\delta_{1}$ et $\delta_{2}$ tels que $\delta=\delta_{1}+\delta_{2}$ dans les équations \ref{obsScoreChap3} et \ref{scoreNotObsMeanChap3}. Les différentes combinaisons de $\delta_{1}$ et $\delta_{2}$ respectant cette condition permettrons de jouer sur l'exploration. Nous sommes donc en mesure d'élaborer une approche de type UCB, que nous désignons par \texttt{HiddenLinUCB} et que nous décrivons dans l'algorithme \ref{HiddenLinUCB}.

\textbf{Regret de l'algorithme \texttt{HiddenLinUCB} :} Notons qu'une borne supérieure du pseudo-regret ne peut être prouvée théoriquement en raison du fait que le pseudo-regret d'une stratégie de bandit contextuel considère les véritables valeurs de contextes. Or l'utilisation d'une valeur moyenne dans l'algorithme \texttt{HiddenLinUCB} pour les actions dont le contexte n'est pas observé, entraîne une erreur qui ne peut être réduite au fil des itérations. En effet,  $\hat{\mu}_{i,t-1}$ n'est pas un estimateur asymptotique de $x_{i,t}$  (au sens où $\lim_{t\to\infty}||\hat{\mu}_{i,t-1}-x_{i,t}||= 0$ avec une forte probabilité). En revanche, comme nous allons le voir dans les expérimentations, la modélisation des contextes cachés effectuée par l'algorithme \texttt{HiddenLinUCB} permet en pratique d'obtenir de meilleurs performances que des algorithmes ne les utilisant pas.

\begin{algorithm}[p]
 \KwIn{$k$, $T$, $\alpha$,$\alpha_1$,$\alpha_2$}
 $\mathcal{K} = \emptyset$\; 
  $V_{0} = I_{d\times d}$ (matrice identité de taille $d$)\; 
  $\hat{\beta}_{0} = 0_{d}$ (vecteur nul de taille $d$)\;
  \For{$t=1..T$} {
  Recevoir $\mathcal{O}_{t}$\;  \label{getnewObservationsBis}
  $\mathcal{K} = \mathcal{O}_{t} \cup \mathcal{K}$\; \label{feedArmSetBis}
    \For{$i  \in \mathcal{K}$} {
    \eIf {$i  \in \mathcal{O}_{t}$}{
          \If {$i$ est nouveau}{
              $\mathcal{A}_{i}= \emptyset$,  
              $\mathcal{B}_{i}= \emptyset$, 
              $\mathcal{C}_{i}= \emptyset$\; 
                Initialiser $\hat{\mu}_{i}$, $\Sigma_{i}$, $a_{i}$ et $b_{i}$ aléatoirement\;
            }
        Observer $x_{i,t}$\; \label{beginObserveAndUpdateBis}
        $\mathcal{C}_{i}=\mathcal{C}_{i} \cup \left\{t\right\} $ \;
        $s_{i,t} = x_{i,t}^{\top}\hat{\beta}_{t-1} + \alpha\sqrt{x_{i,t}^{\top}V_{t-1}^{-1}x_{i,t}}$ \;
        } 
        {
        $s_{i,t}=\hat{\mu}_{i,t-1}^{\top}\hat{\beta}_{t-1}+\alpha_{1} \sqrt{\hat{\mu}_{i,t-1}^{\top}V_{t-1}^{-1}\hat{\mu}_{i,t-1}} +   \alpha_{2}\sqrt{\dfrac{b_{i,t-1}}{a_{i,t-1}(1+n_{i,t-1})}}$\;
        }
    }
    Sélectionner les $k$ actions ayant les scores les plus élevés:\\ 
    $\qquad\mathcal{K}_{t} \leftarrow \argmax\limits_{\hat{\mathcal{K}} \subseteq \mathcal{K}, |\hat{\mathcal{K}}|=k} \quad \sum\limits_{i \in \hat{\mathcal{K}}} s_{i,t}$ \; \label{userSelectionBis}
  \For{$i \in \mathcal{K}_{t}$} {
    \eIf {$i  \in \mathcal{O}_{t}$}{
          $\mathcal{A}_{i}=\mathcal{A}_{i} \cup \left\{t\right\}$
        }
        {
        $\mathcal{B}_{i}=\mathcal{B}_{i} \cup \left\{t\right\}$\;
        Initialiser $\hat{x}_{i,t}$ et $W_{i,t}$ aléatoirement\;
        }
  }
    Exécuter le processus d'inférence variationnelle décrit dans l'algorithme \ref{varPassLinear} pour mettre à jour les distributions des variables aléatoires\;

    }
    
\caption{\texttt{HiddenLinUCB}}
\label{HiddenLinUCB}
\end{algorithm}

\section{Expérimentations}

\label{XPChap3}

  \subsection{Données artificielles}
    
    \subsubsection{Protocole}
        
         \textbf{Génération des données:} Afin d'évaluer les performances de notre algorithme nous l'expérimentation dans un premier temps sur des données simulées. Pour cela, on fixe un horizon de $T=5000$ pas de temps, un ensemble de $K=50$ actions et $d=10$ dimensions. On tire ensuite un vecteur de régression $\beta$ aléatoirement dans $\left[-S/\sqrt{d}..S/\sqrt{d}\right]^{d}$, de façon à respecter la condition $||\beta||\leq S =1$. Pour chaque bras $i$ on tire un réel $\tau_{i}$ selon une loi gamma de paramètres $a_{0}=2$ et $b_{0}=1$, et un vecteur aléatoire $\mu_{i}$ de dimension $d$, depuis une loi normale $\mathcal{N}(0,\tau_{i}^{-1}I)$. Ensuite, pour chaque itération $t \in \left\{1,...,T \right\}$  on procède de la façon suivante pour simuler les données:
    
    \begin{enumerate}
      \item Pour chaque $i \in \left\{1,...,K \right\}$ on tire un échantillon $x_{i,t}$ suivant une loi normale $\mathcal{N}(\mu_{i},\tau_{i}^{-1}I)$;
        \item Pour chaque $i \in \left\{1,...,K \right\}$, on tire une récompense $r_{i,t}$ selon une normale $\mathcal{N}(x_{i,t}^{\top}\beta,1)$;
    \end{enumerate}

1000 jeux de données ont été générés de cette manière, les résultats présentés correspondent à une moyenne sur ces différents corpus.
   
  \textbf{Politiques testées:} Les algorithmes de bandits contextuels existants ne sont pas bien adaptés pour traiter notre problème de bandit avec contextes cachés. En effet, ces derniers ne peuvent pas prendre les actions dont le contexte n'est pas disponible à un instant donné dans le processus de sélection. Contrairement au cas du chapitre précédent, ici, notre problème ne peut pas se voir directement comme un problème de bandit stationnaire. Cependant, les algorithmes de bandits stationnaires peuvent être en mesure de détecter certaines tendances. En revanche, ils ne peuvent pas exploiter les contextes observés. 
  Nous choisissons de nous comparer aux algorithmes suivants : \texttt{CUCB}, \texttt{CUCBV}, \texttt{MOSS} et un algorithme \texttt{Thompson Sampling} avec récompense gaussienne décrite dans l'état de l'art, et initialement proposé par \cite{banditSimpleThompsonAnalysis2}. Soulignons que ces trois politiques n'utilisent aucune information extérieure autre que les récompenses observées. 
  Nous implémentons également la politique \texttt{HiddenLinUCB} proposée dans ce chapitre et décrite dans l'algorithme \ref{HiddenLinUCB}. On fixe $\delta=0.05$, $\delta{1}=\delta{2}=0.025$ et $S$, la taille de la fenêtre de temps à 100. Par ailleurs, nous implémentons les trois scénarios présentés au chapitre précédent pour le processus générant l'ensemble $\mathcal{O}_{t}$ à chaque itération. On rappelle que dans le cas 1 on a $\mathcal{O}_{t}=\mathcal{K}$, dans le cas 2, chaque action a une probabilité $p$ d'être dans $\mathcal{O}_{t}$ à chaque pas de temps et dans le cas 3, $\mathcal{O}_{t}=\mathcal{K}_{t-1}$. 
  On compare différentes valeurs de $p\in \left\{0,0.1,0.2, 0.5,0.7, 1 \right\}$. On rappelle que lorsque $p=0$, les actions sélectionnées au temps $t$ sont ajoutées à $\mathcal{O}_{t+1}$.  Lorsque $p=1$, le processus d'inférence variationnel n'est plus nécessaire car tout les contextes sont visibles, ce qui correspond à l'algorithme \texttt{LinUCB}. Pour l'ensemble des politiques, on considère que l'on sélectionne $k=5$ actions à chaque itération.

  \subsubsection{Résultats}
    
    Nous proposons de mesurer les performances de nos algorithmes en termes de regret cumulé dont nous rappelons la définition:
    \begin{equation*}
\sum\limits_{t=1}^{T} \left(\sum\limits_{i\in \mathcal{K}_{t}^{\star}} r_{i,t} - \sum\limits_{i\in \mathcal{K}_{t}} r_{i,t}\right)
    \end{equation*}  
    où $\mathcal{K}_{t}^{\star}$ est  l'ensemble des $k$ actions ayant la plus grande valeur de $x_{i,t}^{\top}\beta$. La figure \ref{fig:regretVsTimeSimHiddenLinUCB} représente l'évolution du regret cumulé au cours du temps pour les différents algorithmes testés, où nous avons retiré l'algorithme \texttt{Random} en raison de ses trop mauvais résultats. 
    On remarque que les deux politiques \texttt{CUCB} et \texttt{MOSS} offrent des performances pratiquements égales, tandis que \texttt{CUCBV} et \texttt{Thompson Sampling} sont moins performants. Il semble que dans ce cas, l'utilisation de la variance empirique par \texttt{CUCBV} dans l'exploration ne soit pas efficace. Plus intéressant est le fait que dans toutes les configurations (c.-à-d. pour n'importe quelle valeur de $p$) l'algorithme \texttt{HiddenLinUCB} offre de meilleurs résultats que les politiques de bandit stationnaire, toutes confondues. Il est donc possible de tirer partie du modèle contextuel dans le processus de sélection, même lorsque tous les contextes ne sont pas visibles. On note d'une façon générale que les performances se dégradent à mesure que $p$ diminue. Ceci semble cohérent, étant donné que plus $p$ est grand, plus le nombre  de contextes observés est élevé. De plus, quand $p$ augmente, la qualité des estimateurs augmente aussi. En revanche lorsque $p$ est petit, on dispose de moins d'exemples pour effectuer l'apprentissage des différents paramètres. En outre, les performances de l'algorithme \texttt{HiddenLinUCB} s'avèrent meilleures pour le cas $p=0$ que pour le cas $p=0.1$. En ne faisant pas appel à un quelconque processus aléatoire externe délivrant les contextes, l'algorithme \texttt{HiddenLinUCB p=0} définit lui même ceux qui sont observés, puisque les contextes obtenus correspondent alors aux actions sélectionnées précédemment. Cette exploration active sur les contextes permet une identification plus efficace des bras les plus intéressants, pour un ratio de contextes observés identique (1/10).

               \begin{figureth}
      \includegraphics[width=0.9\linewidth]{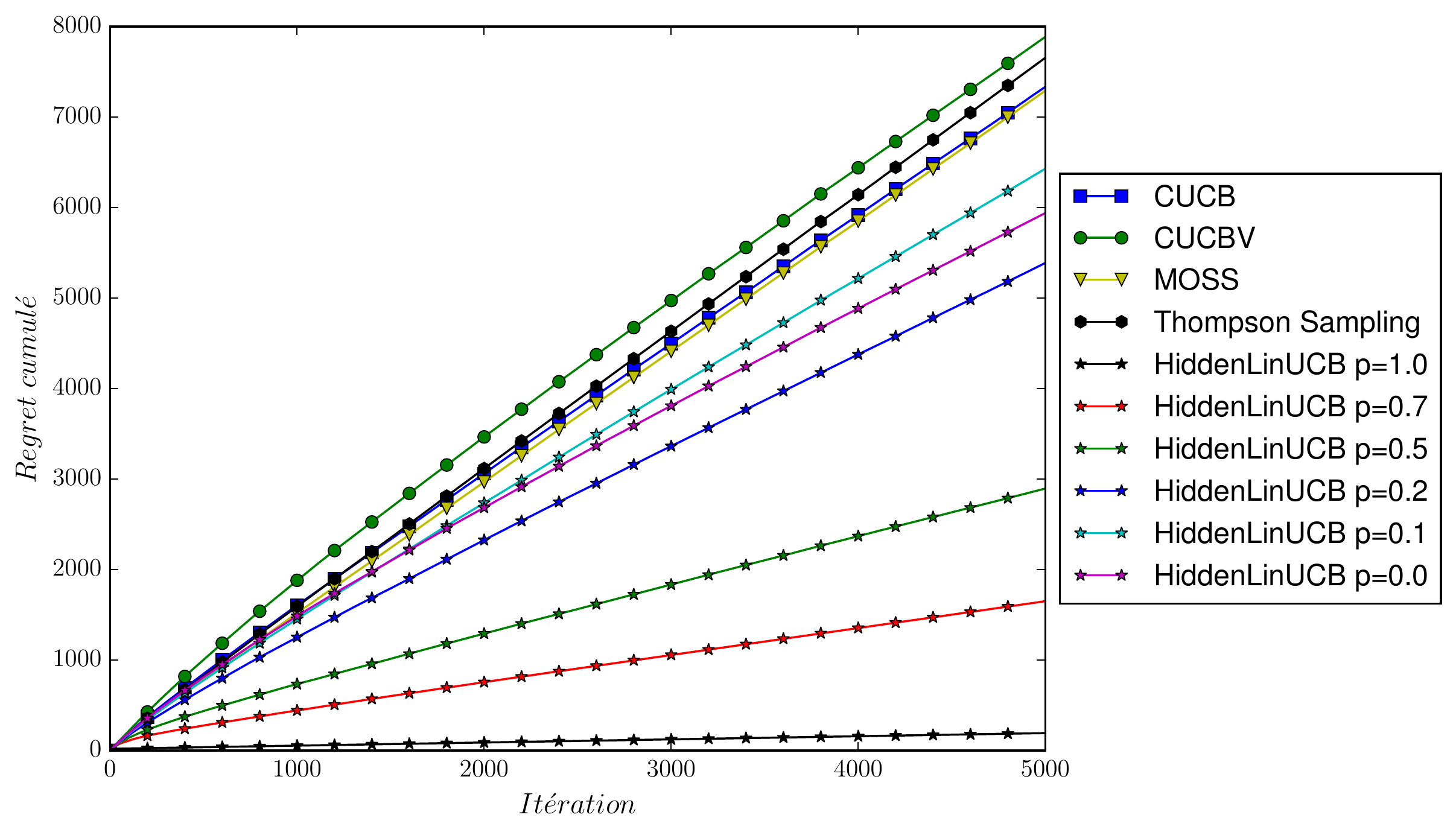}
        \caption[Regret simulation \texttt{HiddenLinUCB}]{Regret cumulé en fonction du temps pour l'expérience sur données artificielles.}
      \label{fig:regretVsTimeSimHiddenLinUCB}
      \end{figureth}

L'algorithme \texttt{HiddenLinUCB} obtient de très bons résultats quand les données utilisées respectent bien les hypothèses du modèle. Dans la section qui suit, nous expérimentons notre approche sur des données réelles, afin d'en mesurer l'efficacité lorsque les données ne suivent plus exactement les hypothèses ayant servies à dériver l'algorithme.

  \subsection{Données réelles}

Nous proposons d'expérimenter l'algorithme proposé dans ce chapitre dans le cadre de notre tâche de collecte d'information sur Twitter. Nous nous évaluerons à la fois sur des jeux de données hors-ligne, mais aussi dans une expérimentation en conditions réelles.

    \subsubsection{Hors ligne}

    \paragraph{Modèle de contexte}

On se place dans le même cadre que dans le chapitre précédent (c.f. \ref{modelProfilStat}) dans lequel, à un instant $t$, nous utilisions les messages postés au temps $t-1$ comme échantillons de profils à l'instant $t$ (pour les utilisateurs appartenant à $\mathcal{O}_{t}$). 
Formellement nous avions $x_{i,t}=F(\omega_{i,t-1})$ où $\omega_{i,t-1}$ représente les messages du compte $i$ à l'instant $t-1$ et $F$ est une transformation nous permettant de réduire la dimension de l'espace des profils. Nous proposons d'utiliser cette même représentation pour modéliser le contexte d'un utilisateur à un instant donné.

    \paragraph{Protocole}
        
        Nous utilisons le même protocole que celui présenté dans la partie \ref{protocole} du chapitre précédent, à savoir le modèle \textit{Topic+Influence} avec les trois thématiques \textit{Politique}, \textit{Religion} et \textit{Science}, et ce pour les trois bases de données collectées (toujours avec $k=100$). Nous comparons la politique \texttt{HiddenLinUCB} à tous les algorithmes testés jusqu'ici, à savoir \texttt{CUCB}, \texttt{CUCBV}, \texttt{UCB}-$\delta$, \texttt{MOSS} et \texttt{SampLinUCB}. On fixe $\delta=0.05$, $\delta_{1}=\delta_{2}=0.025$ et une fenêtre de temps de $S=10$ itérations, permettant de réduire la complexité du processus variationnel (voir remarque précédente). Pour les valeurs de $a_{0}$ et $b_{0}$, nous avons utilisé le jeu de données \textit{USElections} et la thématique politique comme ensemble de validation, ce qui a conduit à prendre $a_{0}=10$ et $b_{0}=1$\footnote{L'utilisation d'une valeur de $a_{0}$ plus élevée que $b_{0}$ a pour effet de réduire l'exploration sur les contextes.}. Finalement, pour les algorithmes nécessitant un processus de génération de l'ensemble $\mathcal{O}_{t}$, c'est-à-dire \texttt{SampLinUCB} et \texttt{HiddenLinUCB} on simule le processus pour différents $p\in \left\{0, 0.01,0.05, 0.1, 0.5,1 \right\}$. Comme précédemment, lorsque $p=0$, les individus écouté au temps $t$ sont ajoutés à l'ensemble $\mathcal{O}_{t+1}$ (ce qui correspond au cas 3)  \footnote{Cet ajout aurait également pu être effectué pour $p> 0$, ce qui sera fait dans l'expérimentation en ligne qui suit, mais on souhaite ici isoler les comportements pour une analyse plus fine des différents cas.}. Notons que lorsqu’ un utilisateur $i$ est dans $\mathcal{O}_{t}$, la même valeur de $x_{i,t}$ est délivrée aux deux algorithmes (\texttt{SampLinUCB} et \texttt{HiddenLinUCB}), la différence se situe dans la manière dont ils s'en servent dans leur politique de sélection.

      \paragraph{Résultats}

Le nombre d'algorithmes et de configurations étant élevé, la représentation graphique de l'évolution de la récompense cumulée en fonction du temps sur une même courbe n'est pas envisageable. Nous proposons plutôt une représentation condensée dans laquelle, pour chaque algorithme on illustre la valeur finale de la récompense cumulée. Ainsi les figures \ref{fig:finalRewardDBUS}, \ref{fig:finalRewardDBOG} et \ref{fig:finalRewardDBBrexit} représentent cette grandeur respectivement pour les bases \textit{USElections}, \textit{OlympicGames} et \textit{Brexit} selon les trois thématiques. Pour plus de lisibilité, nous utilisons une même couleur pour représenter les algorithmes dans lesquels la probabilité $p$ est la même. Premièrement, lorsque chaque contexte est observable ($p=1$), notre approche contextuelle  correspond à un algorithme  \texttt{LinUCB} classique et fonctionne mieux que les approches \texttt{CUCB}, \texttt{CUCBV}, \texttt{MOSS} et \texttt{UCB}-$\delta$. Ce résultat montre que nous sommes en mesure de mieux anticiper les utilisateurs qui vont être les plus pertinents à l'étape suivante, compte tenu de leur activité courante. Cela confirme aussi le comportement non stationnaire des utilisateurs. Considérer les contextes permet également de converger plus rapidement vers les utilisateurs intéressants puisque tous les comptes partagent le même paramètre $\beta$. En outre, les résultats montrent que, même pour de faibles probabilités d'observation des contextes $p$, notre politique contextuelle se comporte beaucoup mieux que les approches non contextuelles, ce qui valide empiriquement notre approche : il est possible de tirer parti de l'information contextuelle, même si une grande partie de cette information est cachée. En donnant la possibilité de sélectionner des utilisateurs même si leur contexte est inconnu, nous permettons à l'algorithme de compléter sa sélection en choisissant les utilisateurs dont le contexte moyen correspond à un profil appris. Si aucun utilisateur dans $\mathcal{O}_{t} $ ne semble actuellement pertinent, l'algorithme peut alors compter sur les différents estimateurs associés aux utilisateurs pour effectuer sa sélection. A titre d'exemple, pour la base \textit{USElections} et la thématique science, le nombre moyen d'utilisateurs sélectionnés pour lesquels le contexte a été observé à chaque pas de temps est de 43 pour $p=0.1$ et 58 pour $p=0.5$, ce qui confirme que l'utilisation de contextes intéressants lorsque ceux-ci sont disponibles, tout en conservant une probabilité de sélection non négligeable pour les utilisateurs de qualité dont le contexte n'est pas connu. Par ailleurs, nous remarquons que les résultats obtenus avec $p=0$ (correspondant a un ratio de contexte observés de 1/50) sont meilleurs que ceux obtenus avec $p=0.01$ (ratio de 1/100), et dans certains cas, que ceux obtenus avec  $p=0.05$ (ratio de 1/20), ce qui confirme que  le terme d'exploration utilisé, favorisant une exploration active des différents contextes, permet de s'orienter vers des utilisateurs prometteurs. 
Continuons cette analyse par une comparaison avec la méthode \texttt{SampLinUCB} du chapitre précédent, basée sur l'estimation des profils des utilisateurs. Il apparaît que pour une même valeur de $p$, \texttt{HiddenLinUCB} permet d'obtenir de meilleurs résultats que \texttt{SampLinUCB}. Ce résultat est plus prononcé sur les bases \textit{USElections} et \textit{Brexit} que sur la base \textit{OlympicGames}. Il semble que l'approche non stationnaire soit particulièrement pertinente dans un réseau social puisque les utilisateurs y ont une attitude très variable de façon générale, pouvant être mieux captée par le modèle contextuel utilisé. 


             \begin{figureth}
    \begin{subfigureth}{0.5\textwidth}
      \includegraphics[width=\linewidth]{DBUS_PoliticFinalRwdNew.pdf}
      \caption{Politique}
    \end{subfigureth}
    \begin{subfigureth}{0.5\textwidth}
      \includegraphics[width=\linewidth]{DBUS_ReligionFinalRwdNew.pdf}
      \caption{Religion}
    \end{subfigureth}
        \begin{subfigureth}{0.5\textwidth}
      \includegraphics[width=\linewidth]{DBUS_ScienceFinalRwdNew.pdf}
      \caption{Science}
    \end{subfigureth}
    \caption[Récompense finale pour tous les algorithmes testés sur la base \textit{USElections}.]{Récompense finale pour tous les algorithmes testés sur la base \textit{USElections} et le modèle \textit{Topic+Influence} avec différentes thématiques.} 
      \label{fig:finalRewardDBUS}
        \end{figureth}
                
                           \begin{figureth}
    \begin{subfigureth}{0.5\textwidth}
      \includegraphics[width=\linewidth]{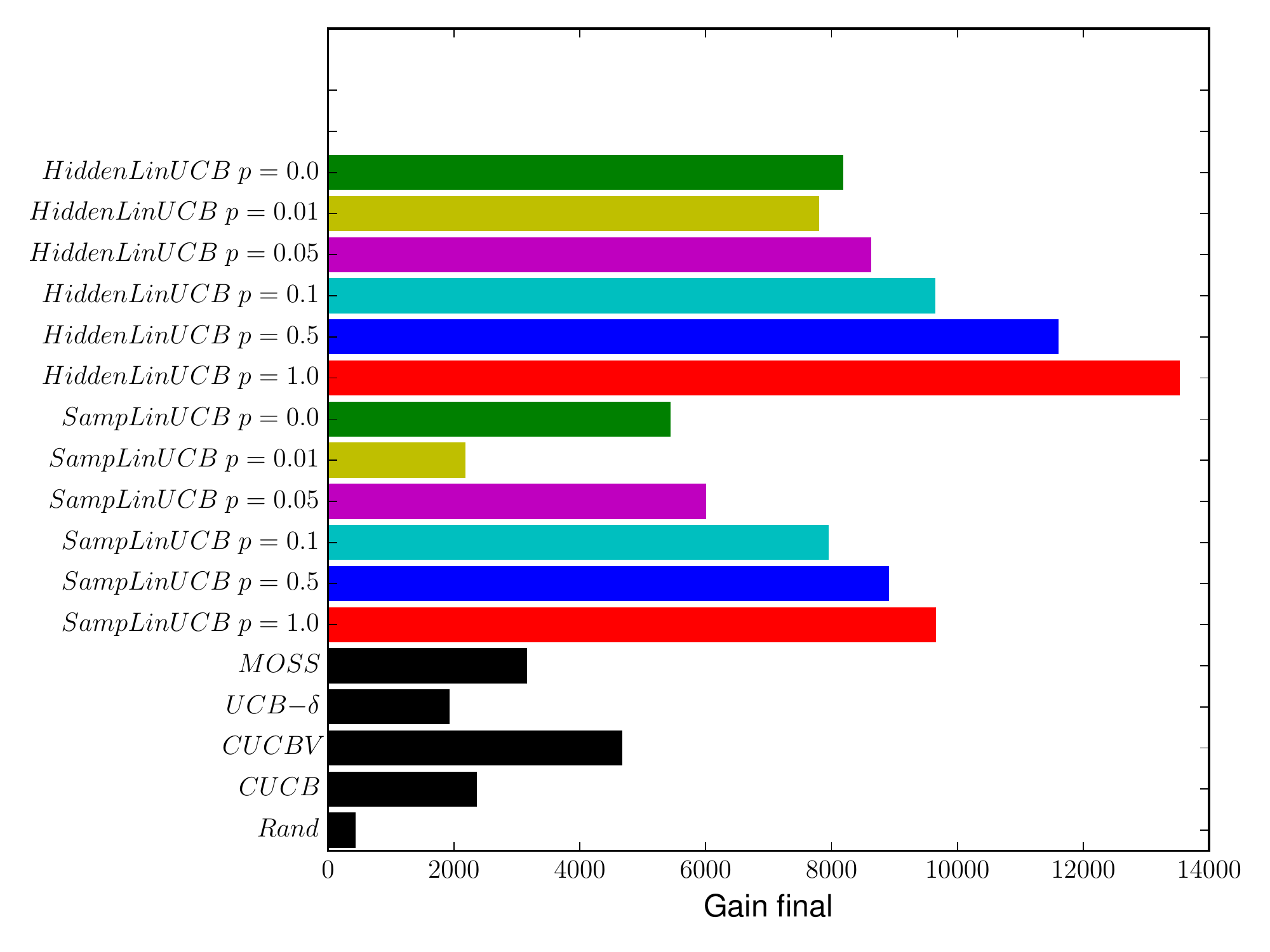}
      \caption{Politique}
    \end{subfigureth}
    \begin{subfigureth}{0.5\textwidth}
      \includegraphics[width=\linewidth]{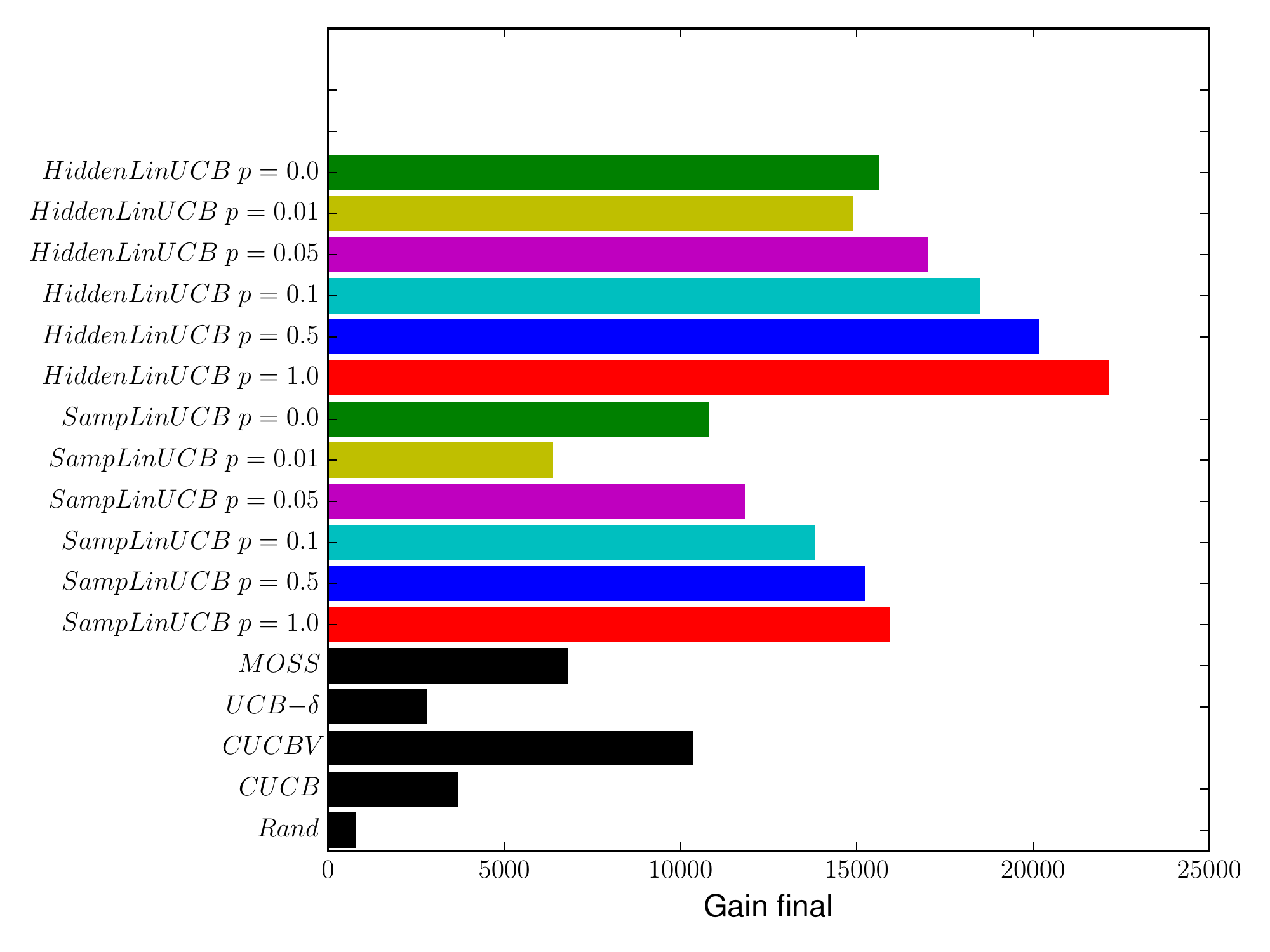}
      \caption{Religion}
    \end{subfigureth}
        \begin{subfigureth}{0.5\textwidth}
      \includegraphics[width=\linewidth]{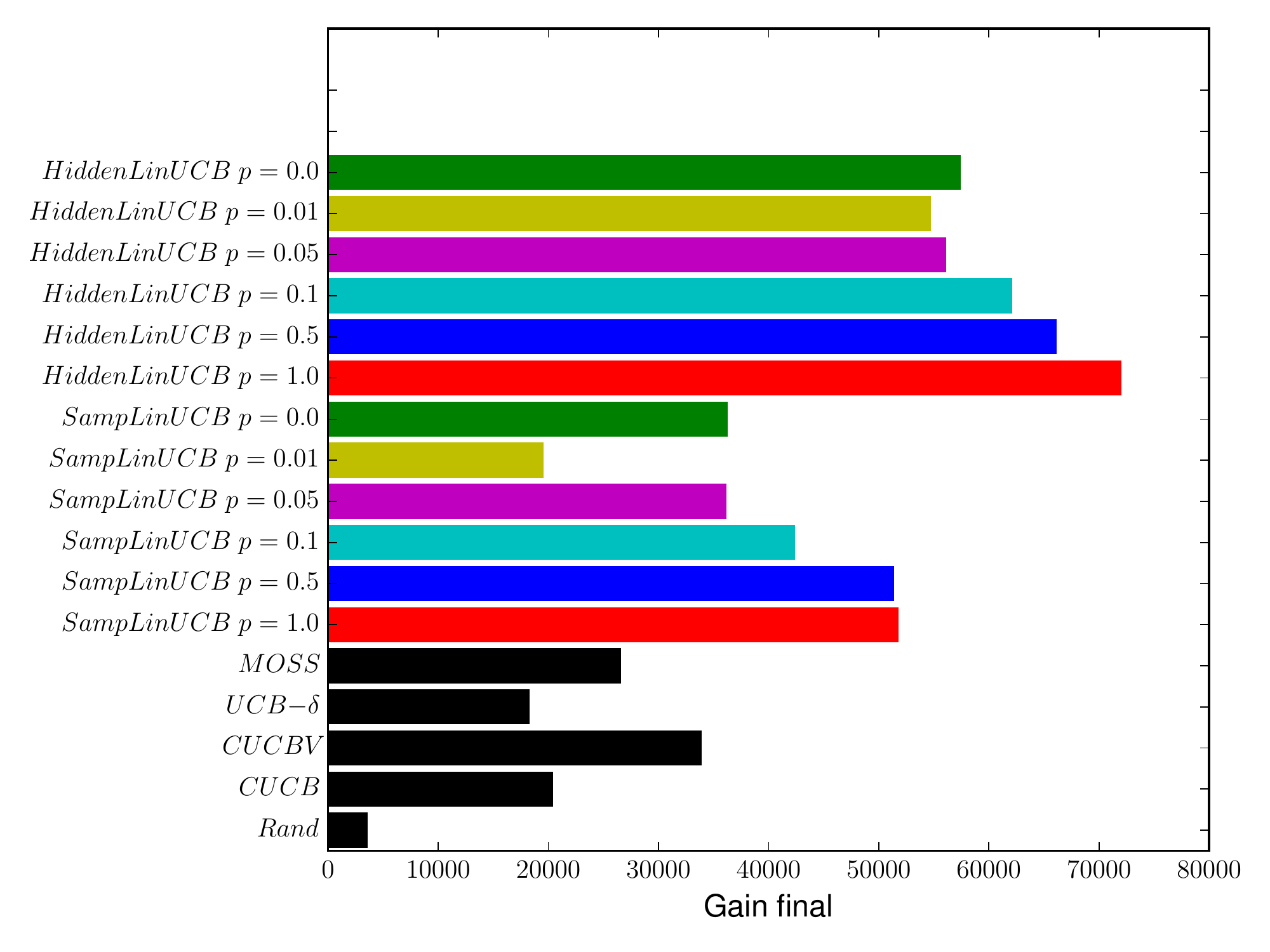}
      \caption{Science}
    \end{subfigureth}
    \caption[Récompense finale pour tous les algorithmes testés sur la base \textit{OlympicGames}.]{Récompense finale pour tous les algorithmes testés sur la base \textit{OlympicGames} et le modèle \textit{Topic+Influence} avec différentes thématiques.} 
      \label{fig:finalRewardDBOG}
        \end{figureth}

           \begin{figureth}
    \begin{subfigureth}{0.5\textwidth}
      \includegraphics[width=\linewidth]{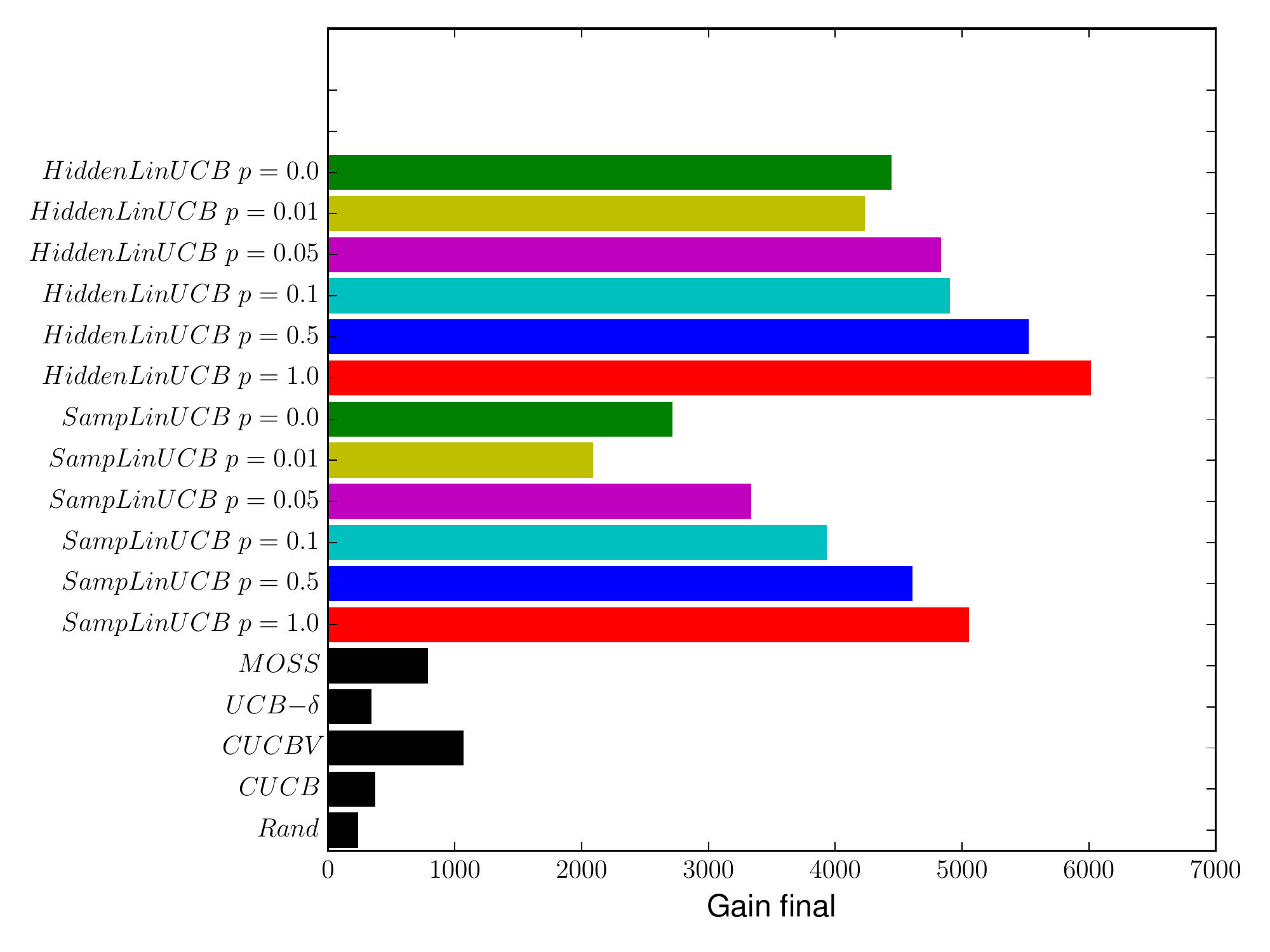}
      \caption{Politique}
    \end{subfigureth}
    \begin{subfigureth}{0.5\textwidth}
      \includegraphics[width=\linewidth]{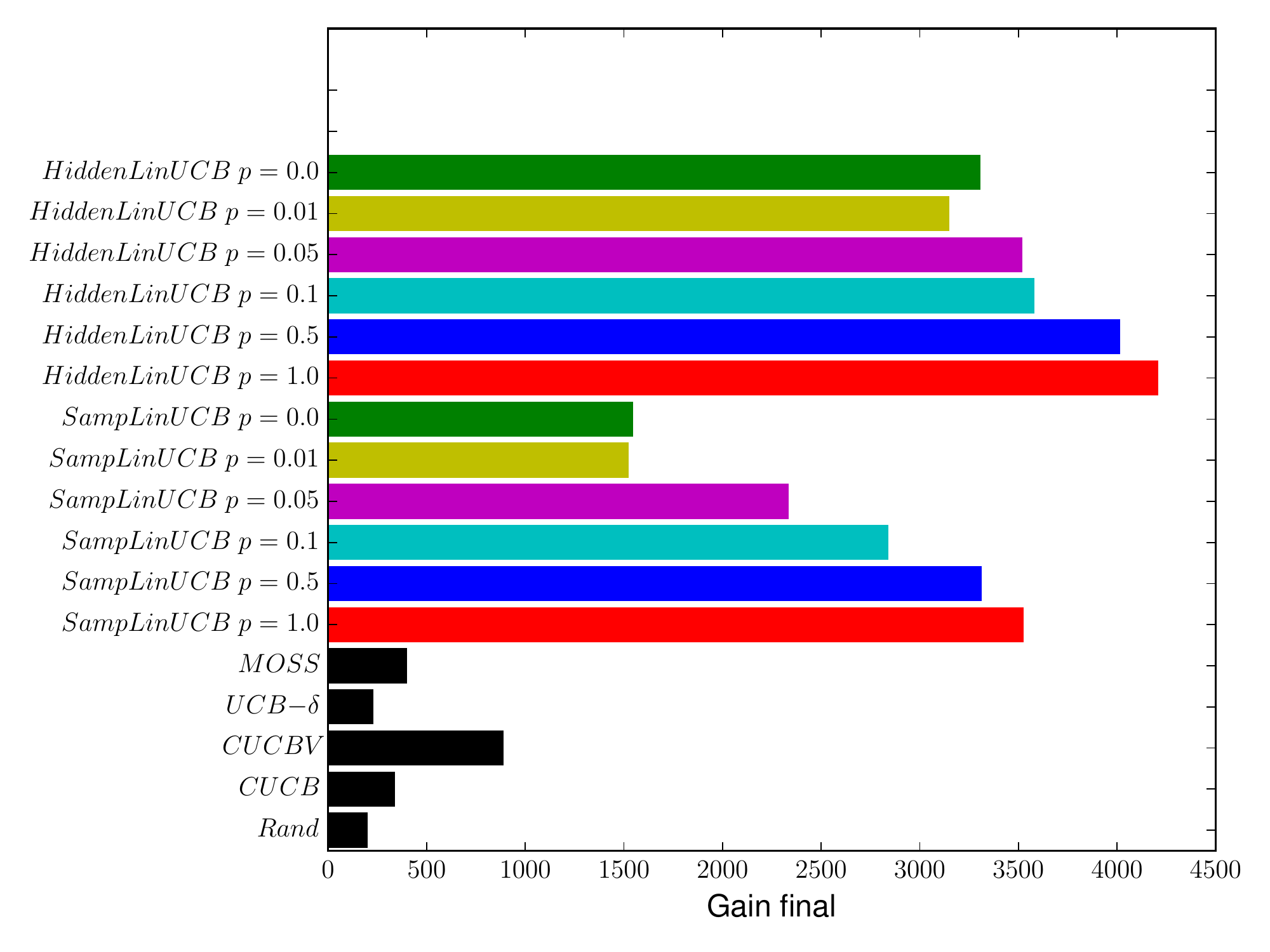}
      \caption{Religion}
    \end{subfigureth}
        \begin{subfigureth}{0.5\textwidth}
      \includegraphics[width=\linewidth]{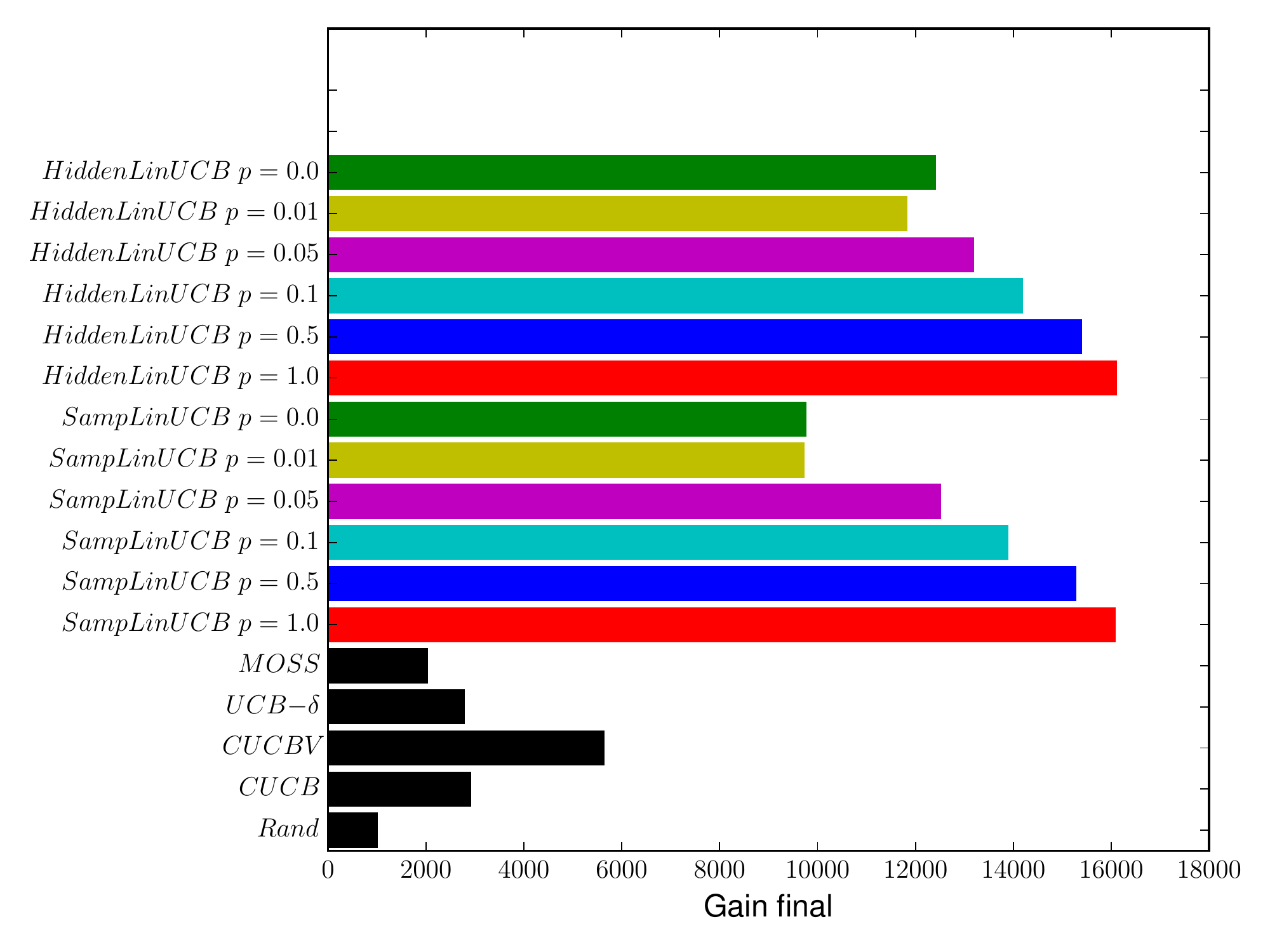}
      \caption{Science}
    \end{subfigureth}
    \caption[Récompense finale pour tous les algorithmes testés sur la base \textit{Brexit}.]{Récompense finale pour tous les algorithmes testés sur la base \textit{Brexit} et le modèle \textit{Topic+Influence} avec différentes thématiques.} 
      \label{fig:finalRewardDBBrexit}
        \end{figureth}

\subsubsection{En ligne}

      \paragraph{Protocole}

Nous considérons notre tâche de collecte dans une expérimentation en ligne sur Twitter prenant en compte l'ensemble du réseau social ainsi que les contraintes des différentes API. Comme nous l'avons déjà présenté, l'API \textit{Follow} \textit{Streaming} de Twitter nous offre la possibilité d'écouter un total de 5000 utilisateurs du réseau social simultanément. À l'instar de l'expérimentation en ligne du chapitre \ref{FormalisationTache}, nous utiliserons cette dernière pour récupérer les contenus produits par les comptes dans $\mathcal{K}_{t}$. Dans le cas où l'on ne dispose d'aucune autre source d'information, cette API sert également à récupérer les contextes de certains utilisateurs, ce qui correspond au scénario où $\mathcal{O}_{t}=\mathcal{K}_{t-1}$. Comme nous l'avons vu dans les expérimentations hors-ligne, de façon générale, l'augmentation du nombre de contextes visibles à chaque itération permet d'améliorer les résultats. Dans cette optique, rendre disponible plus de contextes semble être pertinent. En plus des 5000 comptes que l'on peut écouter en même temps, Twitter propose également une API \textit{Sample} \textit{streaming}, qui renvoie en temps réel 1\% de tous les \textit{tweets} publics. Les éléments renvoyés par cette API sont les mêmes, peu importe la connexion utilisée. Ainsi, multiplier les comptes pour récupérer plus de données via cette API est inutile. Nous utilisons cette dernière pour découvrir de nouveaux utilisateurs, mais aussi pour récolter les contextes (activités) d'un grand nombre d'utilisateurs actifs à un moment donné. Concrètement, l'utilisation de cette seconde source d'information nous permet d'avoir plus d'utilisateurs dans $\mathcal{O}_{t}$ à chaque itération. Afin d'expliciter le fonctionnement du processus de collecte en temps réel dans ce cas, nous proposons une illustration dans la figure \ref{fig:processLiveCtxt}, où trois itérations sont représentées. Au début de chaque étape, la politique de sélection des sources choisit un ensemble d’utilisateurs à suivre parmi tous les utilisateurs connus du système, selon des observations et des connaissances fournies par un module d'apprentissage correspondant à la politique en question. Ensuite, les messages postés par les $k=5000$ profils sélectionnés sont collectés via l'API \textit{Follow streaming}. Comme nous pouvons le voir dans la partie centrale du schéma, après avoir suivi les utilisateurs de $\mathcal{K}_t$ pendant l'itération courante $t$, les messages collectés sont analysés et le résultat - traduisant la pertinence - est renvoyé au module d'apprentissage. En parallèle, à chaque itération, l'activité courante de certains utilisateurs est capturée par l'API \textit{Sample streaming}. Cela nous offre la possibilité d'enrichir la base des utilisateurs potentiels ${\cal K}$, mais aussi de construire l'ensemble  des utilisateurs dont on connaît le contexte. Chaque utilisateur ayant publié au moins un message parmi ceux qui ont été collectés avec l'API \textit{Sample streaming} à l'étape $t$ sont inclus dans $\mathcal{O}_{t+1}$. À cet ensemble sont ajoutés les membres de $\mathcal{K}_{t}$, puisque leur activité a été suivie pendant l'étape $t$.

\begin{figureth}
      \includegraphics[width=0.8\linewidth]{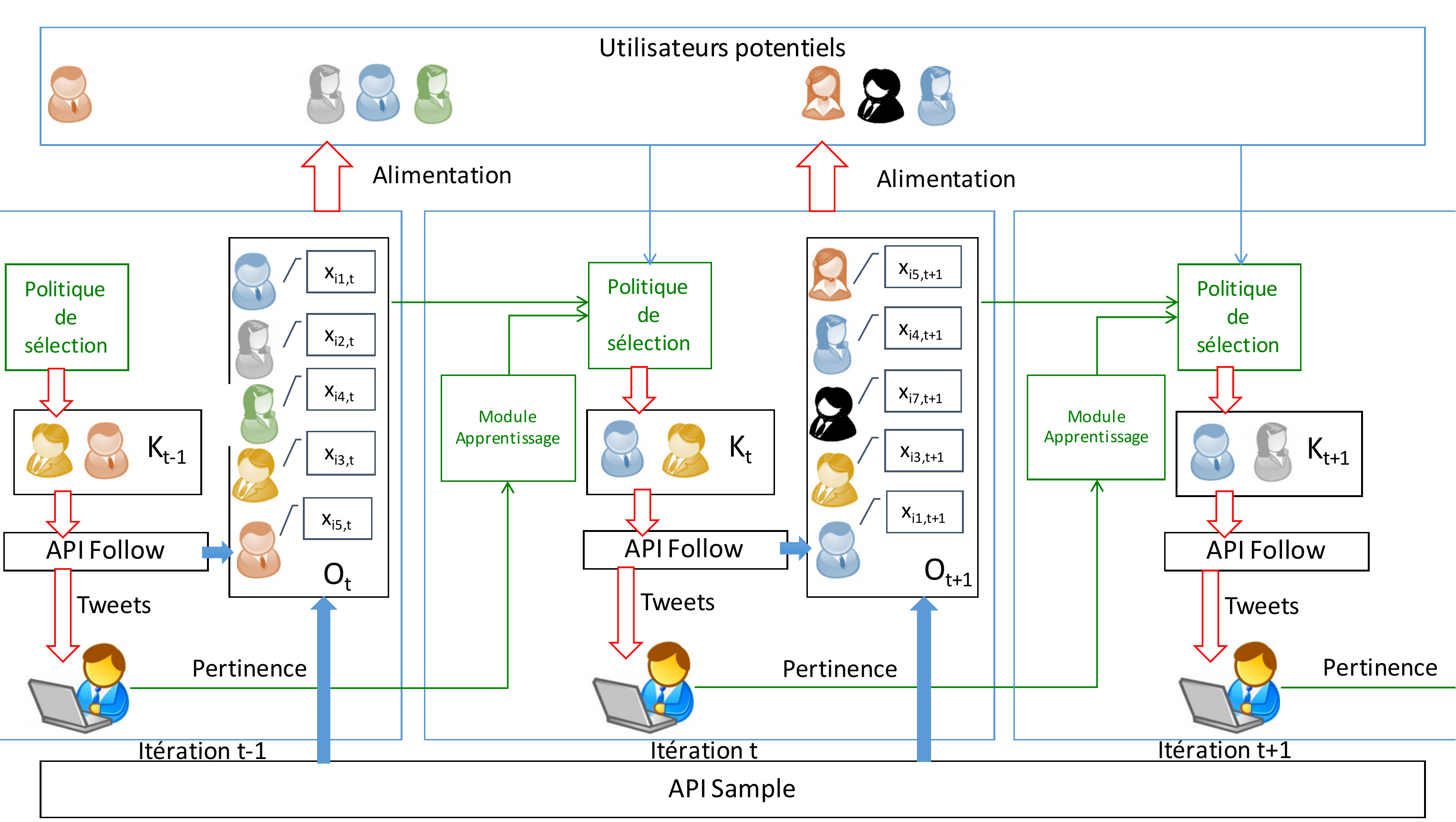}
      \caption[Illustration du système]{Illustration du système.}   
      \label{fig:processLiveCtxt}
\end{figureth}

Dans cette expérience, on fixe la durée d'une période d'écoute à $\mathcal{L}=15$ minutes et l'horizon à deux semaines, ce qui correspond à $T=1344$. Nous choisissons la fonction de récompense \textit{Topic+Influence} avec la thématique politique. Comme précédemment, compte tenu des restrictions de Twitter, chaque expérience nécessite un compte développeur, ce qui limite le nombre de politiques que nous pouvons tester en parallèle. Nous avons choisi de tester les quatre stratégies suivantes : notre approche contextuelle \texttt{HiddenLinUCB}, l'algorithme \texttt{CUCBV}, une politique \texttt{Random} et une autre appelée \texttt{Static}, qui suit les mêmes 5000 comptes, sélectionnés \textit{a priori}, à chaque étape du processus. Les 5000 comptes suivis par l'approche \texttt{Static} ont été choisis en sélectionnant les 5000 utilisateurs ayant cumulé le plus fort volume de récompenses selon les messages collectés par l'API \textit{Sample} sur une période de 24 heures. Pour information, certains comptes célèbres tels que \textit{@dailytelegraph}, \textit{@Independent} ou \textit{@CNBC} en faisaient partie. Remarquons enfin que $p$, la probabilité pour chaque utilisateur de révéler son activité n'est pas un paramètre, mais est contraint par la tâche.

      \paragraph{Résultats}

La partie gauche de la figure \ref{fig:liveContextPolitic} représente l'évolution du gain cumulé en fonction du temps (représenté en termes de nombre d'itérations). On remarque le très bon comportement de notre algorithme dans un scénario réel, car la somme des récompenses qu'il accumule croît beaucoup plus rapidement que celle des autres politiques, surtout après les 500 premières itérations. Après ces premières itérations, notre algorithme semble avoir acquis une bonne connaissance sur les distributions de récompenses en fonction des contextes observés sur les différents utilisateurs du réseau. Afin d'analyser le comportement des politiques expérimentées pendant les premières itérations de la capture, nous représentons aussi à droite de la figure \ref{fig:liveContextPolitic} un zoom des mêmes courbes sur les 150 premiers pas de temps.
Au début, la politique \texttt{Static} est plus performante que toutes les autres, ce qui peut s'expliquer par deux raisons: premièrement les algorithmes de bandits utilisés ont besoin de sélectionner tous les utilisateurs au moins une fois pour initialiser les scores et deuxièmement les utilisateurs faisant partie de l'ensemble \texttt{Static} sont supposés être des sources relativement pertinentes étant donné la façon dont ils ont été choisis. Vers l'itération 80, aussi bien l'algorithme \texttt{CUCBV} que l'algorithme contextuel deviennent meilleurs que \texttt{Static}, ce qui correspond au moment où ils  commencent à pouvoir exploiter de ``bons'' comptes identifiés. Ensuite, une période d'environ 60 itérations est observée (approximativement entre les itérations 80 et 140), au cours de laquelle l'algorithme \texttt{CUCBV} et notre approche contextuelle ont des performances comparables. Cette période s'explique par le fait que notre approche nécessite un certain nombre d'itérations pour apprendre des corrélations efficaces entre contextes et récompenses afin d'en tirer avantage. Enfin, après cette période d'initialisation, on peut remarquer un changement significatif de pente pour la courbe correspondant à notre algorithme, ce qui souligne sa faculté à être bien plus réactif dans un environnement dynamique. De plus, le volume de récompenses collectées avec notre approche sur la période d'expérimentation représente plus du double de ce qui a été collecté avec l'approche état de l'art \texttt{CUCBV}. Enfin, il est à noter que le nombre de fois que chaque utilisateur a été sélectionné par notre algorithme est mieux réparti qu'avec un algorithme stationnaire comme \texttt{CUCBV}, ce qui confirme la pertinence d'adopter une approche dynamique. En conclusion, que ce soit en ligne ou hors ligne, notre approche se révèle très efficace pour la sélection dynamique d'utilisateurs pour la collecte de données ciblée.

             \begin{figureth}
    \begin{subfigureth}{0.45\textwidth}
      \includegraphics[width=\linewidth]{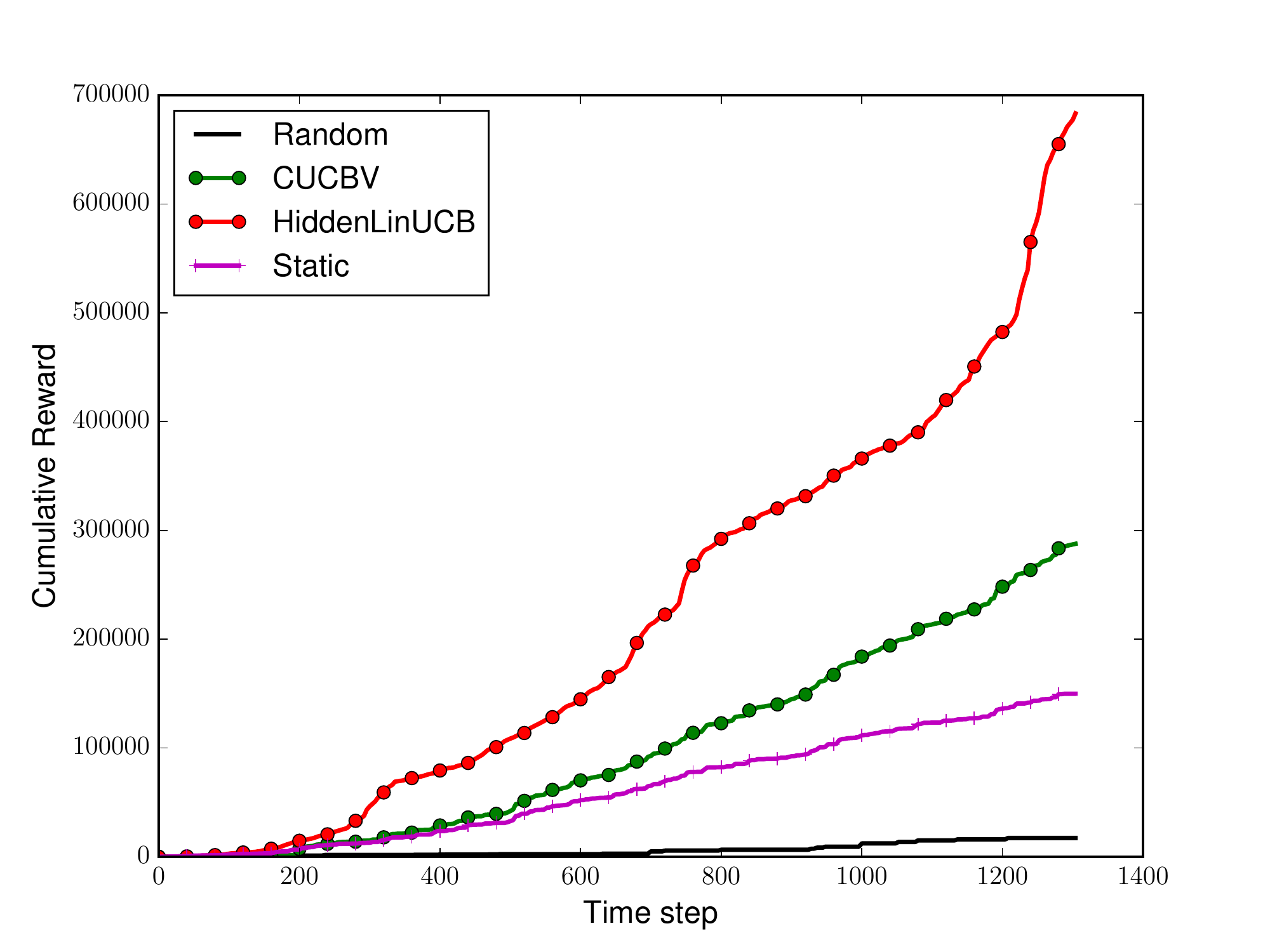}
      \caption{Récompense cumulée sur toute la durée de la collecte.}
    \end{subfigureth}
    \begin{subfigureth}{0.45\textwidth}
      \includegraphics[width=\linewidth]{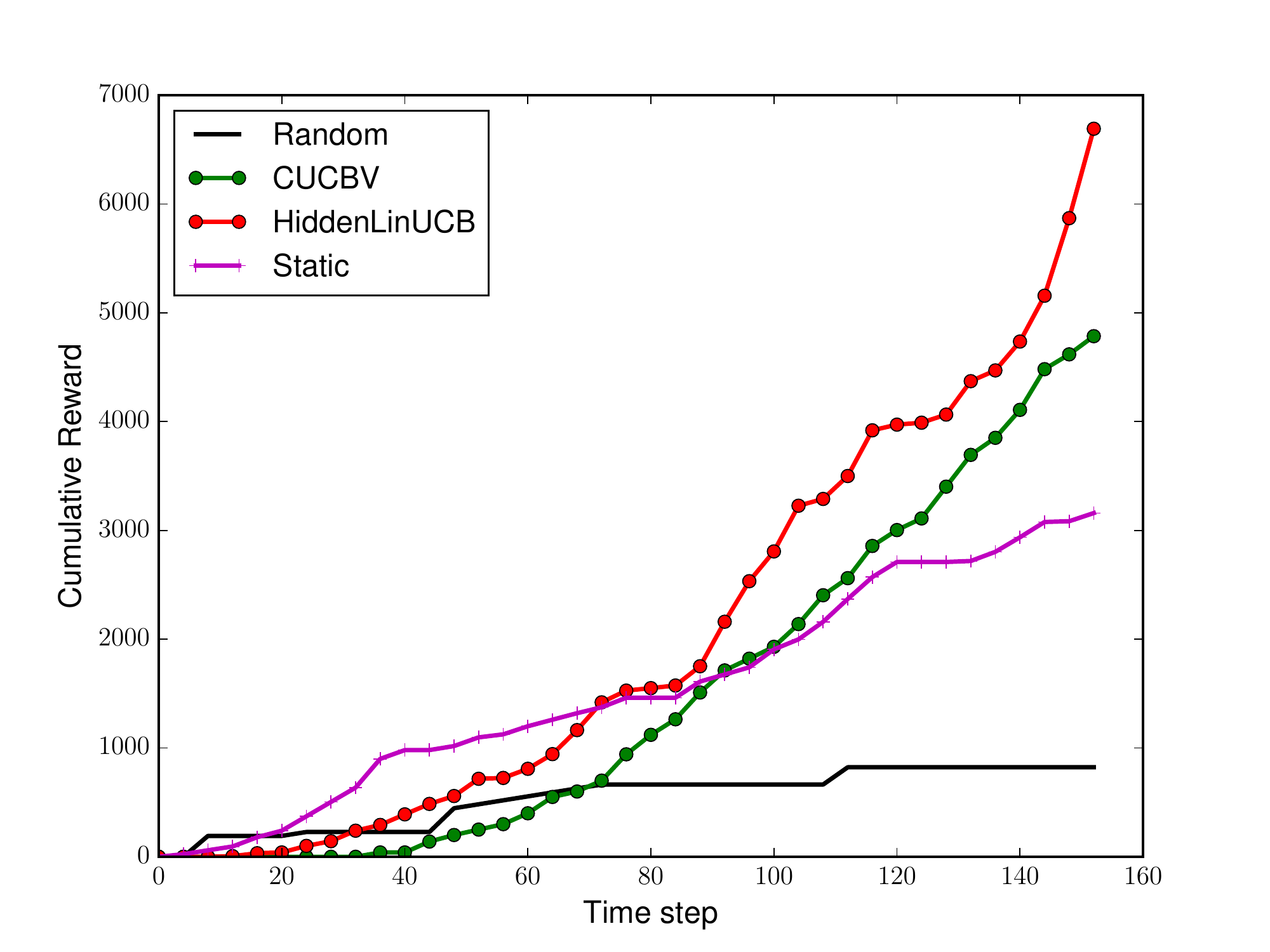}
      \caption{Agrandissement sur les 150 premiers pas de temps.}
    \end{subfigureth}
    \caption[Récompense cumulée en fonction du temps dans l'expérience en ligne]{Récompense cumulée en fonction du temps dans l'expérience en ligne pour différentes politiques. A gauche se trouve une version agrandie des 150 premiers pas de temps.}  
      \label{fig:liveContextPolitic}
        \end{figureth}

\section{Conclusion}

Dans ce chapitre, nous avons formalisé la tâche de collecte de données sur les réseaux sociaux comme une instance spécifique du problème de bandit contextuel, dans laquelle, à cause des restrictions imposées par les médias sociaux, seule une partie des contextes est observable à chaque itération. Pour résoudre cette tâche, nous avons proposé un nouvel algorithme de bandit faisant intervenir un processus d'inférence variationnelle et permettant de prendre en compte les utilisateurs dont le contexte n'est pas disponible. L'approche proposée permet de définir un intervalle de confiance pour les récompenses de chaque utilisateurs et donc de définir un stratégie de type UCB. Les résultats expérimentaux ont montré la validité de cette méthode dans un cadre réel. Il aurait également été intéressant d'utiliser des modèles de contextes non gaussiens afin d'obtenir une meilleure modélisation des messages. Cependant, des approximations supplémentaires et des calculs plus lourds auraient été nécessaires pour dériver les distributions des différents paramètres. 
Dans le chapitre qui suit, nous introduisons un modèle dans lequel nous modélisons des relations temporelles entre les différents utilisateurs.	
    \chapter{Modèles récurrents}
  \minitoc
  \newpage
    
    \label{ModeleRelationnel}

L'approche du chapitre précédent a permis d'améliorer les performances de notre système de collecte en temps réel en utilisant des contextes décisionnels associés à chaque utilisateur. Cependant, le fait qu'une majorité de ces contextes ne soit pas accessible en raison des restrictions imposées pas les API de Twitter, constitue une contrainte forte. Nous proposons dans ce chapitre une approche ne nécessitant pas d'information extérieure, mais permettant de modéliser des relations entre utilisateurs, afin de mieux capter les variations d'utilité des récompenses associées, malgré les données manquantes. Le principe de ce modèle est de prendre en compte des dépendances temporelles afin de les exploiter dans la stratégie de sélection des différentes actions. Ceci nous amène à considérer un nouveau modèle de bandit de type récurrent dans lequel on introduit des transitions d'une itération à la suivante. Dans cette optique, deux types d'approches sont alors proposés: un premier modèle considérant des dépendances linéaires entre les récompenses de périodes d'écoute successives, et un second, faisant intervenir des hypothèses de transitions entre des états latents du système afin de capturer les variations d'utilité des différents comptes. Nous verrons en particulier que cette seconde approche possède à la fois l'avantage de pouvoir modéliser des dépendances à plus long terme, mais aussi de détecter des schémas de dépendance temporelle plus complexes.

\section{Modèle relationnel récurrent}

\label{modRecurrent}


\subsection{Hypothèses et notations}

On rappelle que le problème du bandit avec sélection multiple procède de la façon suivante: à chaque itération $t \in \left\{1,...,T\right\}$, un sous-ensemble $\mathcal{K}_{t} \subset \mathcal{K}$ d'actions de taille $k$ est sélectionné. Pour chaque action $i \in \mathcal{K}_{t}$, l'agent décisionnel reçoit une récompense $r_{i,t} \in \mathbb{R}$. Le choix de $\mathcal{K}_{t}$  à chaque instant est effectué grâce à une politique de sélection se basant sur l'historique des choix et des récompenses passés. Pour dériver une politique efficace, c'est à dire permettant de récolter un maximum de récompenses au cours du temps, certaines hypothèses sont nécessaires. Par exemple, dans le  chapitre \ref{ModeleStationnaire} nous avons fait une hypothèse de stationnarité, tandis que dans le chapitre \ref{ModeleContextuelVariable} nous avons supposé qu'il existait une certaine structure dans l'espace des récompenses, avec notamment l'utilisation d'un contexte décisionnel. De nouvelles hypothèses sont émises dans ce chapitre afin de considérer des dépendances temporelles entre les utilités observées de chaque utilisateur.

Dans un premier temps, nous proposons d'étudier une relation de récurrence directe entre les récompenses. Concrètement, l'espérance de la récompense de l'action $i$ au temps $t$ est supposée être une fonction de la somme pondérée des récompenses au temps $t-1$ plus un terme de biais visant à modéliser une qualité intrinsèque, indépendamment des relations. Nous formalisons le problème de la façon suivante:

\begin{align}
\label{modelVect}
  \forall i \in \left\{1,...,K\right\}, \exists \theta_{i} \in \mathbb{R}^{K+1}  ~\text{tel que}~:  
  \forall t \in \left\{2,...,T\right\}: \mathbb{E}[r_{i,t}|R_{t-1}]=\theta_{i}^{\top}R_{t-1}
\end{align}

Avec $R_{t}=(r_{1,t},...,r_{K,t},1)^{\top} \in \mathbb{R}^{K+1}$ le vecteur des récompenses au temps $t$ concaténé avec une valeur constante égale à 1 qui nous permet de modéliser un biais. Les vecteurs $\theta_{i}$ sont propres à chaque action et déterminent le poids des unes sur les autres entre deux itérations consécutives. Par ailleurs $R_{1}$ est un vecteur aléatoire que nous spécifierons par la suite. 
Dans le cadre de notre tâche de collecte d'information, ce genre de modélisation revient à considérer que l'information pertinente se propage sur le réseau selon des relations d'influence et/ou de communication entre utilisateurs.

\begin{note}
Nous insistons sur la différence avec le type de dépendances généralement étudiées (type graphe), que l'on pourrait qualifier de structurelles, en opposition aux dépendances temporelles que nous souhaitons prendre en compte. Plutôt que de on modéliser des corrélations ou des similarités entre les récompenses des différentes actions à un instant donné, nous cherchons ici à modéliser des phénomènes de propagation d'utilité d'un pas de temps à un autre. En outre, à la différence de la plupart des méthodes existantes, nous nous plaçons dans un cadre où le graphe de dépendances est inconnu a priori.
\end{note}

En vue de dériver un algorithme de bandit, nous spécifions, comme au chapitre précédent, des distributions pour les récompenses respectant l'équation \ref{modelVect}, mais aussi pour les paramètres du problème, c'est-à-dire les différents vecteurs $\theta_{i}$. Dans cette optique, nous proposons le modèle probabiliste gaussien suivant\footnote{Il aurait également été possible de définir une distribution \textit{a priori} sur les paramètres  variance $\sigma^{2}$ et $\alpha^{2}$. Cependant, nous verrons que la complexité due au fait que de nombreuses composantes du vecteur de récompense sont manquantes rend le problème suffisamment complexe à résoudre. Il aurait également été possible de considérer une variance $\sigma_{i}$ propre à chaque utilisateur $i$, ce qui permettrait de prendre en compte des variabilités différentes d'un utilisateur à l'autre. En l'absence d'information, on choisit la même valeur $\sigma_{i}=\sigma$ pour chacun. }:

\begin{itemize}
\item \textbf{Vraisemblance : } $\forall i \in \left\{1,...,K\right\}, \exists \theta_{i} \in \mathbb{R}^{K+1} ~\text{tel que}~ \forall t \in \left\{2,...,T\right\}$:   $r_{i,t}=\theta_{i}^{\top}R_{t-1} + \epsilon_{i,t}$, où $\epsilon_{i,t} \sim \mathcal{N}(0,\sigma^{2})$ (bruit gaussien de moyenne $0$ et de variance $\sigma^{2}$). 
\item \textbf{Prior sur les paramètres: } $\forall i \in \left\{1,...,K\right\}$:  $\theta_{i} \sim \mathcal{N}(0,\alpha^{2}I)$ (bruit gaussien de dimension $K+1$, de moyenne  $0$ et de matrice de covariance  $\alpha^{2}I$, avec $I$ la matrice identité de taille $K+1$).
\item \textbf{Prior au temps 1: } $\forall i \in \left\{1,...,K\right\}$: $r_{i,1} \sim \mathcal{N}(\mu_{i},\sigma^{2})$ avec $\mu_{i}\in \mathbb{R}$.
\end{itemize}

Notons que l'on a défini une distribution \textit{a priori} sur le vecteur des récompenses initiales $R_{1}$, ce dernier ne pouvant pas être modélisé par une somme pondérée des récompenses précédentes.

\begin{note}[Lien avec les processus autorégressifs]
D'une façon générale, un processus autorégressif vectoriel ("\textit{Vector autoregression}" ou \textit{VAR}) d'ordre $p \in \mathbb{N}^{\star}$ et de dimension $n$ suppose que les valeurs de $n$ séries différentes au temps $t$ sont des combinaisons linéaires des valeurs des séries aux pas de temps $t-1, t-2,...t-p$, plus un bruit. Le modèle présenté ici s'apparente donc à un processus autorégressif vectoriel de dimension $K$ et  d'ordre 1. Lorsque les différents poids du modèle sont considérés comme des variables aléatoires, on se situe dans le cadre du "\textit{Bayesian Vector autoregression}" \cite{bayesianVectorAutoregression}. Une difficulté supplémentaire dans notre cas est qu'un certain nombre de valeurs au temps $t-1$ sont manquantes, en raison du processus décisionnel de bandit associé ne permettant d'observer que $k$ récompenses à chaque itération.
\end{note}


\textbf{Note sur  de la cohérence du modèle:} 
En notant $A$ la matrice de taille $(K+1) \times (K+1)$ dont la ligne $i$ vaut $\theta_{i}$ pour $1\leq i\leq K$ et la ligne $K+1$ vaut $(0,..,0,1)$ on a: $\forall t \in \left\{1,...,T\right\}: \mathbb{E}[R_{t}]=A^{t}\mathbb{E}[R_{0}]$. Ainsi, selon les valeurs des $\theta_{i}$, le problème peut diverger, ce que nous souhaitons éviter. On a:    $||\mathbb{E}[R_{t}]|| \leq ||A^{t}|| ||\mathbb{E}[R_{0}]|| \leq ||A||^{t} ||\mathbb{E}[R_{0}]||$ où   $||A||$ est la norme spectrale, c'est-à-dire la racine carrée de la plus grande valeur propre de la matrice semi-définie positive $A^{\top}A$: $||A||=\sqrt{\lambda_{max}(A^{\top}A)}$. Donc on a: 
$||\mathbb{E}[R_{t}]|| \leq  (\sqrt{\lambda_{max}(A^{\top}A)})^{t}||\mathbb{E}[R_{0}]||$. Pour s'assurer d'un modèle non divergent, on doit alors avoir $\lambda_{max}(A^{\top}A) \leq 1$, ce qui garantit $\lim\limits_{t \rightarrow +\infty} ||\mathbb{E}[R_{t}]|| < +\infty$.


\textbf{Illustration d'un scénario simple:} 
La figure \ref{fig:RecurrentModelK3T100} représente une simulation du modèle sur un cas simple où l'a choisi $K=3$, $T=100$, $\mu_{i}=0$ et $\sigma=0.1$. Cette illustration a pour but de montrer le comportement apparemment chaotique d'un tel modèle. En effet bien que nous ayons incorporé un biais de valeur différente pour chaque action, il apparaît que les influences d'un pas de temps à l'autre ont plus d'importance, si bien qu'aucune des trois actions n'apparaît meilleure que les autres sur l'ensemble des pas de temps.

                \begin{figureth}
      \includegraphics[width=0.6\linewidth]{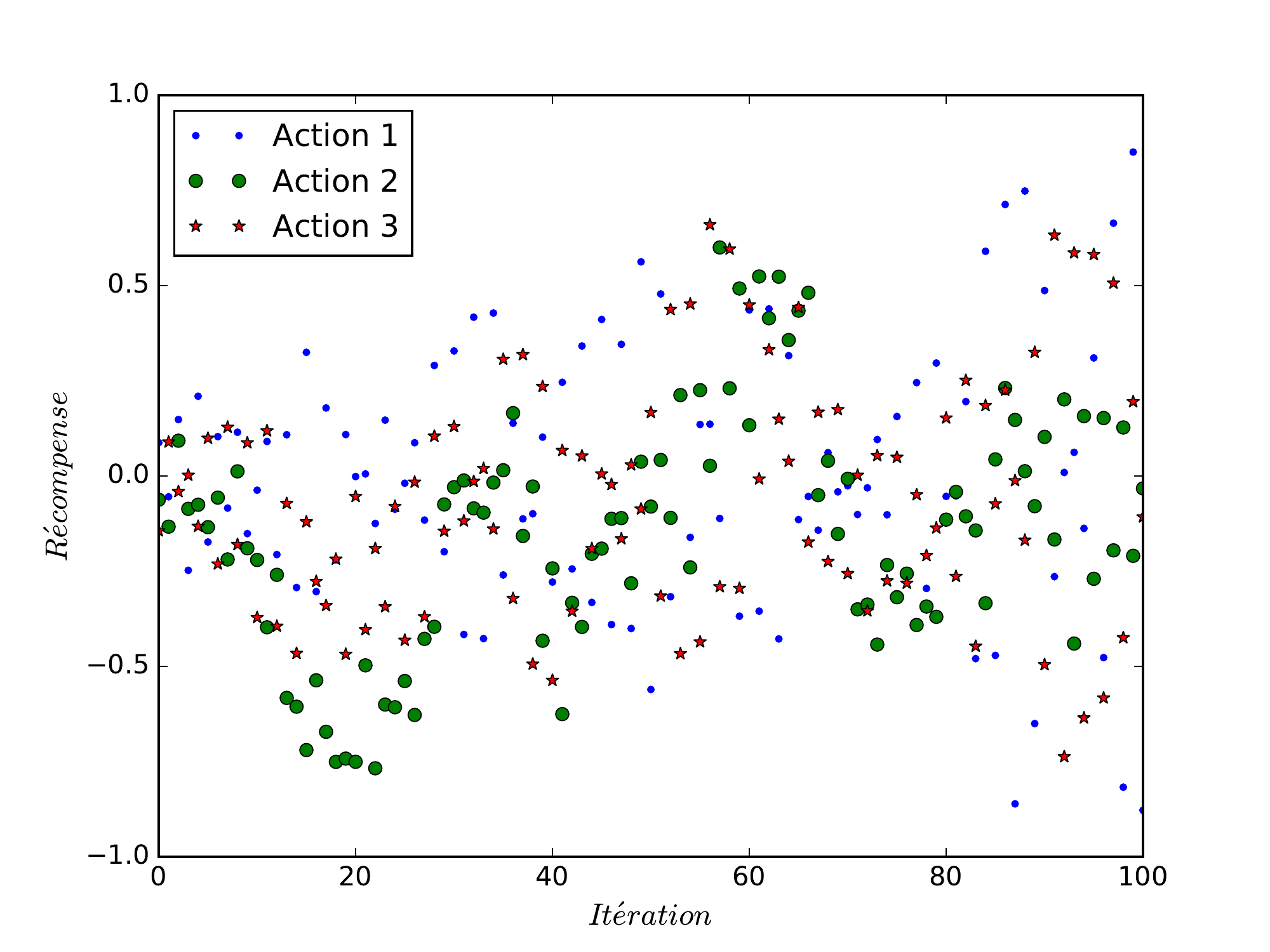}
      \caption[Simulation modèle récurrent $K=3$ et $T=100$ avec récompenses réelles.]{Simulation du modèle récurrent $K=3$ et $T=100$.}
      \label{fig:RecurrentModelK3T100}
      \end{figureth}

\subsection{Algorithme}
\label{sectionAlgorModelRec}

Nous proposons dans ce chapitre d'utiliser une approche de type Thompson sampling. Rappelons que le principe de celle-ci est de produire un échantillon de la valeur espérée de chaque récompense à partir de distributions \textit{a posteriori}, apprises au fur et à mesure, puis de sélectionner l'action ayant la plus forte valeur échantillonnée. Dans notre cas, à chaque itération $t\geq 2$, on doit donc effectuer un échantillonnage de la variable aléatoire $\theta_{i}^{\top}R_{t-1}$, que l'on notera $\tilde{r}_{i,t}$, pour chaque action $i$\footnote{Au premier pas de temps, lorsque $t=1$ on effectue un échantillonnage selon la distribution \textit{a priori}.}. Cet échantillonnage doit être effectué en utilisant l'historique des choix $\mathcal{H}_{t-1}$ et les $k$ actions ayant les plus fortes valeurs de $\tilde{r}_{i,t}$ sont sélectionnées. Si les récompenses des actions non sélectionnées étaient révélées à chaque itération, c'est-à-dire si l'on avait accès à toutes les composantes de $R_{t-1}$, nous serions dans un problème de bandit contextuel classique. Cependant au temps $t$, pour tout $s \in \left\{1..t-1\right\}$ et tout $i \notin \mathcal{K}_{s}$, la composante $r_{i,s}$ est manquante, et doit donc être traitée comme une variable aléatoire. 
D'un point de vue formel, l'échantillonnage de chaque $\tilde{r}_{i,t}$ doit être effectué à partir de la distribution \textit{a posteriori} suivante:

\begin{equation}
p((r_{i,s})_{s=1..t-1,i\notin \mathcal{K}_{s}},(\theta_{i})_{i=1..K}| (r_{i,s})_{s=1..t-1,i\in \mathcal{K}_{s}})
\end{equation}

Cependant, cette distribution est très complexe et ne peut être échantillonnée directement en raison de la complexité provenant de l'aspect récurrent du problème. Pour surmonter cette difficulté, nous proposons d'approximer cette distribution à l'aide d'une approche variationnelle. Cette méthode, que nous avons utilisée au chapitre \ref{ModeleContextuelVariable} pour modéliser les contextes cachés, nous permettra d'obtenir des distributions analytiques sur les différents paramètres. 

\begin{note}
Il aurait également été possible d'utiliser un algorithme MCMC type Gibbs sampling, permettant d'échantillonner des variables aléatoires multivariées à partir de distributions jointes complexes. Cet algorithme procède de façon itérative en échantillonnant successivement chaque variable selon sa distribution conditionnelle par rapport à toutes les autres. Si l'on répète ce processus suffisamment de fois, alors il est possible de montrer que l'échantillon final provient de la distribution jointe originale. Chacune des deux méthodes d'inférence (Gibbs sampling / Variationnelle) possède des avantages et des inconvénients qui selon le cas d'usage orienteront le choix de l'utilisateur. Nous proposons un comparatif de certaines propriétés dans le tableau 
\ref{tab:compGibbsVSVar}\footnote{Données extraites du site \texttt{http://infernet.azurewebsites.net}}. Dans le cas présent, nous avons favorisé l'inférence variationnelle en raison du fait qu'il est plus commode de suivre sa convergence, mais aussi pour sa rapidité de convergence.
\end{note}

              \begin{tableth}
        \begin{tabular}{|c|c|c|}
                  \hline
                    \textbf{Propriété} & \textbf{Gibbs sampling} & \textbf{Inférence variationnelle}\\
                    \hline
                    Déterministe & Non & Oui \\
                        &   & \\
                    \hline
                    Résultat exact & Oui, mais au bout  & Non\\
                          & d'un nombre infini d'itérations & \\
                     \hline
                    Garantie de convergence & Oui, mais très souvent  & Oui\\
                      & difficile à suivre  & \\
                     \hline
                    Efficacité & Peu efficace  & Souvent le plus\\
                    &   &efficace \\
                     \hline
         \end{tabular}
      \caption[Avantages et inconvénients Gibbs sampling vs variationnelle.]{Propriétés des méthodes d'inférences: Gibbs sampling vs variationnelle.}
    \label{tab:compGibbsVSVar}
      \end{tableth}


La méthode d'inférence variationnelle nécessite de supposer une factorisation des différentes variables aléatoires cachées dont nous souhaitons obtenir la distribution. Dans cette optique, la factorisation suivante est supposée:

\begin{equation}
q\left((\theta_{i})_{i=1..K}, (r_{i,s})_{s=1..t-1,i\notin \mathcal{K}_{s}}\right)=\prod_{i=1}^{K}q_{\theta_{i}}(\theta_{i})\prod_{s=1}^{t-1}\prod_{i\notin \mathcal{K}_{s}} q_{r_{i,s}}(r_{i,s})
\end{equation}

Grâce à cette hypothèse, nous pouvons désormais calculer les distributions de chacun des paramètres. Les propositions \ref{propThetaReal}  et  \ref{propRwdReal} établissent ces distributions respectivement pour les variables  $(\theta_{i})_{i=1..K}$ et  $(r_{i,s})_{s=1..t-1,\notin \mathcal{K}_{s}}$. Les preuves sont disponibles en annexes \ref{proofLin1Params} et \ref{proofLin1Rwds}.

\begin{prop}
\label{propThetaReal}

Pour $t\geq 3$ , on note $D_{t-1}=((R_{s},1)^{\top})_{s=1..t-1}$ la matrice de taille $(t-1)\times (K+1)$ où la ligne $s$ correspond au vecteur de récompense au temps $s$ avec 1 à la fin, $D_{1..t-2}$ la matrice de taille $(t-2)\times K$ composée des $t-2$ premières lignes de $D_{t-1}$ et $D_{i:2..t-1}$ le vecteur de taille $t-2$ correspondant à la colonne $i$ et les $t-2$ dernières lignes de $D_{t-1}$ (c.-à-d. le vecteur de récompenses du bras $i$ du temps $1$ au temps $t-2$). Alors, pour tout $t \geq 2  $, et pour tout $i$, la distribution conditionnelle de $\theta_{i}$ est donnée par:
\begin{equation}
\theta_{i}  \sim \mathcal{N}(A_{i,t-1}^{-1}b_{i,t-1}, A_{i,t-1}^{-1})
\label{dist2}
\end{equation}
Avec: 
\begin{itemize}
\item $A_{i,t-1}=\dfrac{\mathbb{E}[D_{1..t-2}^{\top}D_{1..t-2}]}{\sigma^{2}}+\dfrac{I}{\alpha^{2}}$
\item $b_{i,t-1}=\dfrac{\mathbb{E}[D_{1..t-2}^{\top}]}{\sigma^{2}}\mathbb{E}[D_{i:2..t-1}]$ 
\end{itemize}

Et par convention $b_{i,1}=0$ et $A_{i,1}= \dfrac{I}{\alpha^{2}}$.

De plus, 
$\mathbb{E}[D_{1..t-2}^{\top}D_{1..t-2}]=\sum\limits_{s=1}^{t-2} \mathbb{E}[(R_{s},1)]\mathbb{E}[(R_{s},1)]^{\top}+Var((R_{s},1))$, les valeurs de $\mathbb{E}[(R_{s},1)]$ et $Var((R_{s},1))$ étant déterminées grâce à la proposition \ref{propRwdReal}.
\end{prop}

\begin{prop}
\label{propRwdReal}
On note $\Theta$ la matrice de taille $K \times (K+1)$ dont la ligne $i$ vaut $\theta_{i}$ et $\beta_{j}$ la  $j^{eme}$ colonne de $\Theta$. 
Pour $t\geq 2$ et $1 \leq s \leq t-1$, la distribution variationnelle de  $r_{i,s}$ est donnée par:

\begin{equation}
 r_{i,s} \sim \mathcal{N}(\mu_{i,s}, \sigma_{i,s}^{2})
\label{dist1}
\end{equation}
Avec: 
\begin{itemize}
\item si $s=1$: $\mu_{i,s}=\dfrac{\mathbb{E}[\beta_{i}]^{\top}\mathbb{E}[R_{s+1}] + \mu_{i} - \sum\limits_{j=1,j \neq i}^{K+1} \mathbb{E}[\beta_{i}^{\top}\beta_{j}]\mathbb{E}[r_{j,s}]  }{1+\mathbb{E}[\beta_{i}^{\top}\beta_{i}]}$ 
, 
$\sigma_{i,s}^{2}=\dfrac{\sigma^{2}}{1+\mathbb{E}[\beta_{i}^{\top}\beta_{i}]}$ 

\item si $s=t-1$: $\mu_{i,s}=\mathbb{E}[\theta_{i}]^{\top}\mathbb{E}[R_{s-1}] $ ,$\sigma_{i,s}^{2}=\sigma^{2}$

\item si $1\leq s<t-1$: 
$\mu_{i,s}=\dfrac{\mathbb{E}[\beta_{i}]^{\top}\mathbb{E}[R_{s+1}] + \mathbb{E}[\theta_{i}]^{\top}\mathbb{E}[R_{s-1}] - \sum\limits_{j=1,j \neq i}^{K+1} \mathbb{E}[\beta_{i}^{\top}\beta_{j}]\mathbb{E}[r_{j,s}]  }{1+\mathbb{E}[\beta_{i}^{\top}\beta_{i}]}$ 
, 
$\sigma_{i,s}^{2}=\dfrac{\sigma^{2}}{1+\mathbb{E}[\beta_{i}^{\top}\beta_{i}]}$
\end{itemize}

où par convention on a pris $r_{j,K+1}=1$ pour tout $j$.  

De plus, $\mathbb{E}[\beta_{i}^{\top}\beta_{j}]$ peut être calculé de la façon suivante: 
$\mathbb{E}[\beta_{i}^{\top}\beta_{j}]=\sum\limits_{l=1}^{K} Var(\theta_{l})_{i,j}+\mathbb{E}[\theta_{l}]_{i}\mathbb{E}[\theta_{l}]_{j}$, où $Var(\theta_{l})_{i,j}$ désigne l'élément $(i,j)$ de la matrice de covariance de $\theta_{l}$ et $\mathbb{E}[\theta_{l}]_{i}$ désigne l'élement $i$ de la moyenne de $\theta_{l}$, tous deux définis dans la proposition \ref{propThetaReal}.

\end{prop}

Dans les propositions précédentes, lorsque une récompense a été observée, on utilise la valeur associée au lieu de l'espérance. Concrètement, la composante $i$ de $\mathbb{E}[R_{s}]$ est égale à $r_{i,s}$  si $i \in \mathcal{K}_{s}$  et à $\mathbb{E}[r_{i,s}]$ sinon. 
De plus, $Var(R_{s})$ correspond à une matrice diagonale de taille $K$ dont l'élément $(i,i)$ vaut 0 si $i \in \mathcal{K}_{s}$ et $\sigma_{i,s}^{2}$ dans le cas opposé. Finalement, $Var((R_{s},1))$ correspond à la matrice $Var(R_{s})$ à laquelle on a ajouté une colonne et une ligne de 0.

Les deux propositions précédentes fournissent les distributions des différentes variables du problème. Il apparaît très clairement que les paramètres de chacune d'elles sont liés à tous les autres, d'où la nécessité d'une procédure itérative, dont le but est de converger vers une solution stable. Nous proposons d'expliciter cette procédure dans l'algorithme \ref{varPassRecLinear}. Cet algorithme prend en entrée un entier $nbIt$  qui correspond au nombre d'itérations à effectuer.

\begin{algorithm}[!ht]
  \KwIn{$nbIt$ (nombre d'itérations)}
  \For{$It  =1..nbIt$} {  
    \For{$i  =1..K$} {
        Calculer $A_{i,t-1}$ et $b_{i,t-1}$  selon la proposition \ref{propThetaReal}\;
         \For{$s  =1..t-1$} {
         \If{$i\notin \mathcal{K}_{s}$}{
         Calculer $\mu_{i,s}$ et $\sigma_{i,s}^{2}$ selon la proposition  \ref{propRwdReal} \;
         }          
         }
    }
  }
\caption{\texttt{Processus itératif variationnel pour le modèle relationnel récurrent}}
\label{varPassRecLinear}
\end{algorithm}

Nous disposons désormais de toutes les distributions variationnelles dont nous avons besoin pour dériver un algorithme de Thompson sampling adapté à notre problème de bandit relationnel récurrent. Cet algorithme, que nous désignons par \texttt{Recurrent Relational Thompson Sampling}, est  décrit dans  l'algorithme  \ref{recurrentTS}. A chaque itération $t$ (sauf pour $t=1$ où l'on utilise uniquement les distributions \textit{a priori} du modèle), l'algorithme fait appel au processus itératif (algorithme \ref{varPassRecLinear}) permettant d'obtenir des distributions approchées pour les variables $\theta_{i}$ et les récompenses non observées à l'itération précédente. Une fois ces distributions obtenues, l'algorithme effectue un échantillonnage des variables en vue d'obtenir un échantillon $\tilde{r}_{i,t}$ de $\theta_{i}^{\top}R_{t-1}$ pour toutes les actions $i$. Il sélectionne ensuite les $k$ actions ayant les plus fortes valeurs de $\tilde{r}_{i,t}$  et récolte les récompenses associées.

\begin{algorithm}[!ht]
\For{$t  =1..T$} {
  \eIf{$t=1$}{
    Pour tout $i$, échantillonner $\tilde{r}_{i,1}$ selon les distributions \textit{a priori} du modèle de récompense\;
  }
  {
    Exécuter l'algorithme variationnel \ref{varPassRecLinear} pour obtenir les distributions de chaque $\theta_{i}$ et des $r_{i,s}$ manquants\;
    \For{$i  =1..K$} {
      Echantillonner $\theta_{i}$\;
         \If{$i\notin \mathcal{K}_{t-1}$}{
            Echantillonner $\tilde{r}_{i,t-1}$\;
         }          
         Calculer $\tilde{r}_{i,t}=\theta_{i}^{\top}R_{t-1}$\;
    }

  }
   $\qquad\mathcal{K}_{t} \leftarrow \argmax\limits_{\hat{\mathcal{K}} \subseteq \mathcal{K}, |\hat{\mathcal{K}}|=k} \quad \sum\limits_{i \in \hat{\mathcal{K}}} \tilde{r}_{i,t}$ \;
  \For{$i  \in \mathcal{K}_{t}$} {
    Recevoir $r_{i,t}$ \;
    }
}
\caption{\texttt{Recurrent Relational Thompson Sampling}}
\label{recurrentTS}
\end{algorithm}

\textbf{Complexité :}
La complexité de l'algorithme proposé, que l'on notera de façon condensée \texttt{Recurrent TS}, augmente de façon linéaire avec le nombre d'itérations $t$. Ceci est dû à la nécessité de calculer les distributions de chaque récompense manquante depuis le début. Plus précisément, on a $K(K+1)+(t-1)(K-k)$ variables aléatoires dont on doit trouver les paramètres au temps $t$. Cela prend en compte les $K-k$ récompenses manquantes du temps $1$ au temps $t-1$ et les $K$ vecteurs de $K+1$ variables. Ce taux de croissance très rapide, pouvant mener à des problèmes de mémoire, n'est pas compatible avec la nature intrinsèquement "en ligne" des problèmes de bandits. Afin de nous affranchir de cet inconvénient, nous proposons d'utiliser une approximation se restreignant à une fenêtre de temps limité du passé. Soit $S$ la taille de cette fenêtre, alors au lieu de considérer toutes les récompenses manquantes du temps $1$ au temps $t-1$, l'algorithme se limite aux itérations moins anciennes que $S$ pas de temps (de $t-S-1$ à $t-1$), rendant ainsi la complexité constante avec le temps. En revanche, les facteurs $K(K+1)$ et $(K-k)$ ne peuvent pas être réduits simplement. Cela provient du fait que les algorithmes essayent d'apprendre tous les paramètres du modèle. Lorsque le nombre d'actions $K$ augmente, il apparaît donc que la méthode proposée ne peut plus être envisageable. Dans la suite, nous proposons un   modèle de récompenses faisant intervenir des transitions entre couches cachées et permettant de réduire considérablement le nombre de paramètres à apprendre.

\section{Modèle récurrent à états cachés}

\subsection{Hypothèses et notations}

Dans le modèle présenté ici, on suppose à chaque itération l'existence d'un état caché du système $h_{t}\in \mathbb{R}^{d}$, qui sous-tend la génération des récompenses à chaque itération $t$. Par ailleurs, nous supposons  qu'il existe une transformation linéaire permettant de passer de l'état $h_{t}$ à l'état $h_{t+1}$. Formellement, on a: 

\begin{equation}
\label{hiddenRel}
\exists \Theta \in \mathbb{R}^{d\times d}, \forall t \in \left\{2,...,T\right\}: \mathbb{E}[h_{t}|h_{t-1}]=\Theta h_{t-1}
\end{equation}

\begin{equation}
\label{genRel}
\forall i \in \left\{1,...,K\right\}, \exists W_{i} \in \mathbb{R}^{d}, \exists b_{i} \in \mathbb{R} ~\text{tels que}~: 
\forall t \in \left\{1,...,T\right\}: \mathbb{E}[r_{i,t}|h_{t}]=W_{i}^{\top}h_{t}+b_{i}
\end{equation}

Avec $h_{t}\in \mathbb{R}^{d}$ l'état caché associé à l'itération $t$, $\Theta$ la matrice carré de dimension $d$ constituée des poids du modèle récurrent (permettant de passer d'un état au suivant), $W_{i}\in \mathbb{R}^{d}$ le vecteur de poids liant l'état caché à la récompense de l'action $i$ à l'instant $t$, et $b_{i}$ un terme de biais spécifique à chaque action $i$.

L'équation \ref{hiddenRel} décrit l'évolution du système entre deux itérations successives.  Chaque composante du vecteur d'état à un instant donné est le résultat d'une combinaison linéaire des composantes du vecteur d'état au temps précédent. Le modèle de récompense est quant à lui décrit par l'équation \ref{genRel}. Il fait intervenir la couche cachée du temps courant pondérée par un vecteur $W_{i}$, plus un biais $b_{i}$ pour chaque action $i$.  Le modèle est illustré sur trois itérations dans la figure \ref{fig:genmod}, où l'on distingue les dépendances temporelles entre états. Bien qu'à un instant $t$, l'état courant ne dépend que de l'état précédent, ce modèle est capable de capter des dépendances à plus long terme en raison de la relation de récurrence entre états cachés.

              \begin{figureth}    
            \includegraphics[width=0.5\linewidth]{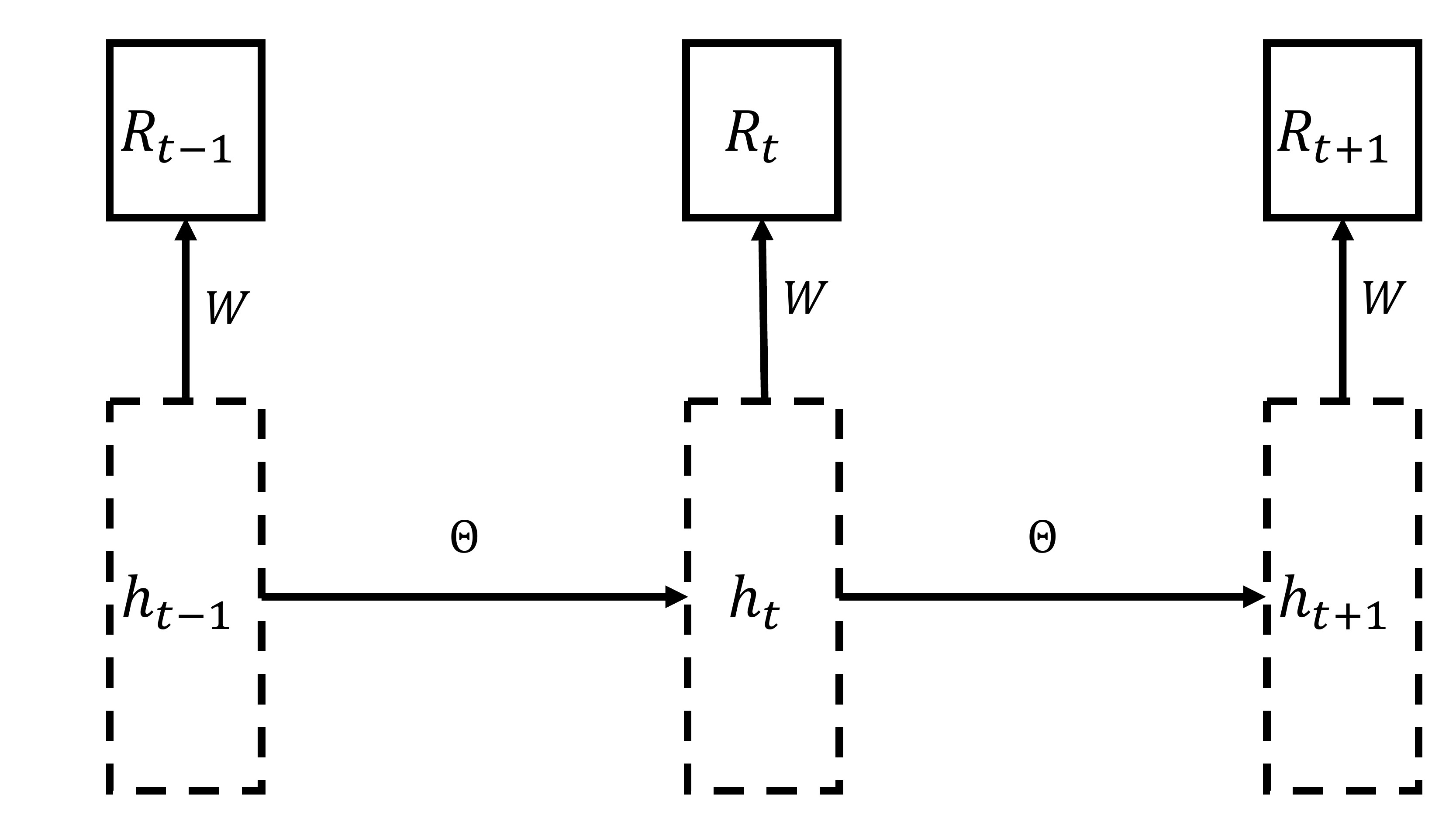}
      \caption[Modèle récurrent à états cachés.]{Modèle récurrent à états cachés.}   
      \label{fig:genmod}
  \end{figureth}

 Comme précédemment, nous considérons un modèle génératif basé sur des ditributions gaussiennes:

\begin{itemize}
\item \textbf{Vraisemblance 1: } $\exists \Theta \in \mathbb{R}^{d\times d}, \forall t \in \left\{2,...,T\right\}: h_{t}=\Theta h_{t-1} +\epsilon_{t}$, où $\epsilon_{t} \sim \mathcal{N}(0,\delta^{2}I)$.
\item \textbf{Vraisemblance 2: } $\forall i \in \left\{1,...,K\right\}, \exists W_{i} \in \mathbb{R}^{d}, \exists b_{i} \in \mathbb{R}$ \text{ tels que }: $\forall t \in \left\{1,...,T\right\}$ : $r_{i,t}=W_{i}^{\top}h_{t}+b_{i}$ $\epsilon_{i,t} \sim \mathcal{N}(0,\sigma^{2})$.
\item \textbf{Prior au temps 1: } $h_{1} \sim \mathcal{N}(0,\delta^{2}I)$, où $I$ désigne la matrice identité de taille $d$.
\item \textbf{Prior sur les paramètres 1: } $\forall i \in \left\{1,...,d\right\}$:  $\theta_{i} \sim \mathcal{N}(0,\alpha^{2}I)$, où $\theta_{i}$ désigne la ligne $i$ de $\Theta$.
\item \textbf{Prior sur les paramètres 2: } $\forall i \in \left\{1,...,K\right\}$:  $(W_{i}, b_{i})  \sim \mathcal{N}(0,\gamma^{2}I)$, avec $(W_{i}, b_{i})$ le vecteur aléatoire de taille $d+1$ issu de la concaténation de $W_{i}$ et $b_{i}$, et $I$ la matrice identité de taille $d+1$. 
\end{itemize}

Pour ne pas allourdir les écritures, on  utilise la même notation pour désigner les matrices identité de taille $d$ et $d+1$.

Le modèle probabiliste utilisé, avec la présence de variables observables, générées par un état latent non observable au sein duquel des transitions temporelles s'effectuent, nous rappelle le formalisme des modèles de Markov cachés ("Hidden Markov Model" - HMM) \cite{HiddenMarkovModel}. Dans les HMM classiques, les états possibles du système sont discrets et en nombre fini, ce qui n'est pas notre cas, puisque nous utilisons des états continus (les composantes de $h_{t}$ suivent des lois gaussiennes). Lorsque les états sont continus, que les distributions sont gaussiennes, et que les transitions entre états sont linéaires, ce type de modèle est appelé système dynamique linéaire (\textit{"Linear Dynamical System"}). De plus, lorsque les matrices de transition (notée $\Theta$) et d'émission (notée $(W_{i})_{i=1..K}$) sont connues, ce modèle correspond au Filtre de Kalman \cite{KalmanFilter}. Les filtres de Kalman ont généralement pour but de retrouver l'état d'un système physique (position, vitesse, etc.) en fonction de mesures bruitées, lorsqu’ un modèle de la dynamique de ce système est disponible. Le modèle dynamique, que décrit la matrice $\Theta$, provient généralement  d'équations physiques, tandis que le modèle d'observation décrit par $(W_{i})_{i=1..K}$, dépend des capteurs disponibles. Dans le cas qui nous intéresse ici, ces paramètres ne sont pas connus et sont considérés comme des variables aléatoires. Ce problème entre dans le cadre générique des \textit{"Bayesian Linear Dynamical Systems"}, pour lequel une approche variationnelle permettant d'approximer les distributions des différents paramètres a été étudiée dans \cite{Variationalmethods}. L'application de la méthode très générique proposée par l'auteur à notre tâche n'étant pas directe, nous dériverons l'ensemble des distributions nécessaires à notre cas spécifique.

\begin{note}[Lien avec un modèle de bandit existant]
Cette remarque a pour but de clarifier la différence entre notre modèle et un modèle pouvant paraître proche, dans lequel un processus de Markov (MDP) régit l'évolution des différents états des actions \cite{restlessBandit2}. Dans cette approche, les auteurs supposent que chaque action $i$  possède un ensemble fini de $k_{i}$ états, dont les transitions sont gouvernées par une matrice   $P_{i}$ de taille $k_{i} \times k_{i}$. De plus, à chaque état $j\in \left\{1,...,k_{i}\right\}$ correspond une espérance (inconnue) $\mu_{i,j}$. La première différence réside dans le fait que, dans leur approche, les auteurs utilisent des états (et des espérances de récompenses) discrets, contrairement à la représentation continue proposée ici.  De plus, dans leur cas, chaque action possède un ensemble d'états qui lui est propre, et l'évolution d'un état à l'autre se fait de façon indépendante entre les actions, ce qui constitue une seconde différence avec notre modèle. En effet, nous utilisons un état "partagé" par toutes les actions. Enfin, le modèle proposé dans \cite{restlessBandit2} ne permet pas d'encoder des dépendances temporelles entre états, car les probabilités de transition sont constantes.
\end{note}

\subsection{Algorithme}
\label{sectionalgoHiddenState}

Notre but est de dériver un algorithme de Thompson sampling, ce qui se traduit par la nécessité d'échantillonner la variable aléatoire  $W_{i}^{\top}\Theta h_{t-1}+b_{i}$ pour chaque action $i$ à l'instant $t$ en utilisant l'historique des choix effectués. Formellement, cet échantillonnage doit être effectué selon la distribution suivante:

\begin{equation}
p(h_{t-1},\Theta,W,b| (r_{i,s})_{s=1..t-1,i\in \mathcal{K}_{s}})
\end{equation}

Comme précédemment, cette distribution n'est pas calculable directement et nous proposons d'adopter une méthode variationnelle afin d'approximer les distributions des différents paramètres.


Dans notre cas, l'ensemble des variables observées est celui des récompenses des $k$ actions sélectionnées entre le temps $1$ et le temps $t-1$. L'ensemble des variables latentes est constitué des couches cachées entre le temps $1$ et le temps $t-1$ et des paramètres, à savoir $(h_{1},...,h_{t-1},\Theta,W,b)$. On rappelle que notre but est de trouver une approximation de la distribution des paramètres sachant les données observées. Commençons par factoriser l'ensemble des variables latentes de la façon suivante:

\begin{equation}
q(h_{1},...,h_{t-1},\Theta,W,b)=\prod\limits_{s=1}^{t-1}q_{h_{s}}(h_{s})\prod\limits_{i=1}^{K}q_{W_{i},b_{i}}(W_{i},b_{i})\prod\limits_{j=1}^{d}q_{\theta_{j}}(\theta_{j})
\end{equation}

 où pour tout $j \in \left\{1,...,d \right\}$, $\theta_{j}$ représente la ligne $j$ de la matrice $\Theta$. Cette factorisation séparant les couches cachées d'un pas de temps à l'autre, mais aussi les paramètres propres à chaque action nous semblent être la plus naturelle étant donné le modèle probabiliste décrit plus haut.

 Les propositions \ref{genModHiddenVar}, \ref{genModThetaVar} et \ref{genModWVar} qui suivent, dont nous fournissons les preuves en annexes \ref{proofLin1HiddenLayerVar}, \ref{proofLin1HiddenParamsThetaVar} et \ref{proofLin1HiddenParamsVar}, établissent les distributions variationnelles des différentes variables en se basant sur la factorisation précédente. Il est important de noter que chaque distribution fait appel aux paramètres des autres, il s'agit donc d'un modèle itératif.

 \begin{prop}
\label{genModHiddenVar}
On note  $W$ la matrice de taille  $K \times d$ dont la ligne $i$ vaut  $W_{i}$ et $b$ le vecteur des biais de taille $K$. De plus au temps $s$, la sous-matrice (resp. le sous-vecteur) composée des lignes d'indexe $i\in \mathcal{K}_{s}$ de $W$ (resp. de $b$) est notée $W_{s}$ (resp. $b_{s}$). Finalement le vecteur de taille $k$ des récompenses observées au temps $s$ est noté $R_{s}$. 
A un instant $t\geq 2$, pour tout $s$ tel que $s\leq t-1$, la distribution variationnelle de $h_{s}$ est donnée par: 

\begin{equation}
 h_{s}\sim \mathcal{N}(F_{s}^{-1}g_{s}, F_{s}^{-1})
\label{distHs}
\end{equation}

 Avec:

\begin{itemize}
\item si $s=1$: 
$F_{s}=\dfrac{I}{\delta^{2}} +  \dfrac{\mathbb{E}[\Theta ^{\top}\Theta]}{\delta^{2}} + \dfrac{\mathbb{E}[W_{s}^{\top}W_{s}]}{\sigma^{2}}$ et 
$g_{s}= \dfrac{\mathbb{E}[\Theta]^{\top} \mathbb{E}[h_{s+1}]}{\delta^{2}}  + \dfrac{\mathbb{E}[W_{s}]^{\top}\mathbb{E}[R_{s}]-\mathbb{E}[W_{s}^{\top}b_{s}]}{\sigma^{2}}$

\item si $s=t-1$: 
$F_{s}=\dfrac{I}{\delta^{2}}  + \dfrac{\mathbb{E}[W_{s}^{\top}W_{s}]}{\sigma^{2}}$ et 
$g_{s}= \dfrac{\mathbb{E}[\Theta] \mathbb{E}[h_{s-1}]}{\delta^{2}}  + \dfrac{\mathbb{E}[W_{s}]^{\top}\mathbb{E}[R_{s}]-\mathbb{E}[W_{s}^{\top}b_{s}]}{\sigma^{2}}$

\item si $1<s<t-1$: 
$F_{s}=\dfrac{I}{\delta^{2}} +  \dfrac{\mathbb{E}[\Theta ^{\top}\Theta]}{\delta^{2}} + \dfrac{\mathbb{E}[W_{s}^{\top}W_{s}]}{\sigma^{2}}$ et 
$g_{s}= \dfrac{\mathbb{E}[\Theta] \mathbb{E}[h_{s-1}]}{\delta^{2}}  + \dfrac{\mathbb{E}[W_{s}]^{\top}\mathbb{E}[R_{s}]-\mathbb{E}[W_{s}^{\top}b_{s}]}{\sigma^{2}} +\dfrac{\mathbb{E}[\Theta]^{\top} \mathbb{E}[h_{s+1}]}{\delta^{2}}$

Où:

$\qquad \mathbb{E}[\Theta ^{\top}\Theta]=\sum\limits_{i=1}^{d}\mathbb{E}[\theta_{i}\theta_{i}^{\top}]=\sum\limits_{i=1}^{d}(\mathbb{E}[\theta_{i}]\mathbb{E}[\theta_{i}]^{\top}+Var(\theta_{i}))$;

$\qquad \mathbb{E}[W_{s}^{\top}W_{s}]=\sum\limits_{i\in\mathcal{K}_{s}}\mathbb{E}[W_{i}W_{i}^{\top}]=\sum\limits_{i\in\mathcal{K}_{s}}(\mathbb{E}[W_{i}]\mathbb{E}[W_{i}]^{\top}+Var(W_{i}))$;

$\qquad \mathbb{E}[W_{s}^{\top}b_{s}]=\sum\limits_{i\in\mathcal{K}_{s}} (\mathbb{E}[W_{i}]\mathbb{E}[b_{i}] + Cov(W_{i},b_{i}))$

 \end{itemize}

\end{prop}

\begin{prop}
\label{genModThetaVar}
Pour $t\geq 3$, on note $D_{t-1}=(h_{s}^{\top})_{s=1..t-1}$ la matrice de taille $(t-1)\times d$ des  états cachés jusqu'au temps $t-1$. 
Alors, pour tout $t\geq 2$ la distribution variationnelle de $\theta_{i}$ suit une loi gaussienne telle que :

\begin{equation}
 \theta_{i}\sim \mathcal{N}(A_{i,t-1}^{-1}b_{i,t-1}, A_{i,t-1}^{-1})
\label{distTheta}
\end{equation}

Avec: 

\begin{itemize}
\item $A_{i,t-1}=\dfrac{I}{\alpha^{2}}+\dfrac{\mathbb{E}[D_{1..t-2}^{\top}D_{1..t-2}]}{\sigma^{2}}$
\item $b_{i,t-1}=\dfrac{\mathbb{E}[D_{1..t-2}]^{\top}}{\sigma^{2}}\mathbb{E}[D_{i:2..t-1} ]$ 
\end{itemize}

Où par convention $b_{i,1}=0$ et $A_{i,1}= \dfrac{I}{\alpha^{2}}$. Lorsque $t\geq 3$ on utilisera:

$\qquad \mathbb{E}[D_{1..t-2}^{\top}D_{1..t-2}]=\sum\limits_{s=1}^{t-2} (\mathbb{E}[h_{s}]\mathbb{E}[h_{s}]^{\top}+Var(h_{s}))$

\end{prop}

\begin{prop}
\label{genModWVar}
Pour chaque action $i$  on note $\mathcal{T}_{i,t-1}$ l'ensemble des itérations où elle a été choisie jusqu'au temps $t-1$, c'est-à-dire $\mathcal{T}_{i,t-1}=\left\{s \text{ tel que } i\in \mathcal{K}_{s}  \text{ pour } 1\leq s\leq t-1\right\}$. On note également $M_{i,t-1}=((h_{s},1)^{\top})_{s\in \mathcal{T}_{i,t-1}}$  et $c_{i,t-1}=(r_{i,s})_{s\in \mathcal{T}_{i,t-1}}$. 
Pour tout $i$ et tout  $t\geq 1$,  la distribution variationnelle de $(W_{i},b_{i})$ suit une loi gaussienne telle que:

\begin{equation}
(W_{i},b_{i})\sim \mathcal{N}(V_{i,t-1}^{-1}v_{i,t-1}, V_{i,t-1}^{-1})
\label{distWb}
\end{equation}

\begin{itemize}
\item $V_{i,t-1}=\dfrac{I}{\gamma^{2}}+\dfrac{\mathbb{E}[M_{i,t-1}^{\top}M_{i,t-1}]}{\sigma^{2}}$
\item $v_{i,t-1}=\dfrac{\mathbb{E}[M_{i,t-1}]^{\top}c_{i,t-1}}{\sigma^{2}}$ 
\end{itemize}

Où par convention $v_{i,1}=0$ et $V_{i,1}= \dfrac{I}{\gamma^{2}}$ et lorsque $t\geq 2$ on utilisera:

$\qquad \mathbb{E}[M_{i,t-1}^{\top}M_{i,t-1}]=\sum\limits_{s \in\mathcal{T}_{i,t-1}}(\mathbb{E}[(h_{s},1)]\mathbb{E}[(h_{s},1)]^{\top}+Var((h_{s},1)))$
\end{prop}

 On notera que la matrice de covariance $Var((h_{s},1))$ correspond à la matrice $Var(h_{s})$ à laquelle on a ajouté une colonne et une ligne finale de 0. De plus, le terme $Cov(W_{i},b_{i})$ de la proposition \ref{genModHiddenVar} s'extrait simplement de la matrice $Var((W_{i},b_{i}))$. 
 
  Nous disposons maintenant de toutes les informations nécessaires pour calculer les paramètres des distributions des variables qui nous intéressent de façon itérative. Le processus itératif variationnel associé est décrit dans l'algorithme \ref{varPass}.

 \begin{algorithm}[!ht]
  \KwIn{$nbIt$ (nombre d'itérations)}
  \For{$It =1..nbIt$} {
    \For{$s =1..t-1$} {
          Calculer $g_{s}$ et $F_{s}$ selon la proposition \ref{genModHiddenVar} \;
  }
  \For{$i=1..d$} {
        Calculer $b_{i,t-1}$ et $A_{i,t-1}$ selon la proposition \ref{genModThetaVar} \;
    }
    \For{$i =1..K$} {
      Calculer $v_{i,t-1}$ et $V_{i,t-1}$ selon la proposition \ref{genModWVar} \;
    }
  }
\caption{\texttt{Processus itératif variationnel pour le modèle récurrent à états cachés}}
\label{varPass}
\end{algorithm}

L'algorithme de Thompson sampling \ref{varRecurrentTS} utilise cette procédure variationnelle afin d'être à même d'échantillonner une variable $W_{i}^{\top}\Theta h_{t-1}+b_{i}$ pour chaque action $i$ à l'instant $t$ et d'effectuer la sélection des meilleures actions. Il procède de façon similaire au précédent modèle, en faisant appel à la procedure variationnelle à chaque itération pour recalculer les paramètres des distributions en fonctions des nouvelles observations de récompenses. 
Un des atouts de cette méthode par rapport à celle du modèle précédent est que sa complexité est beaucoup moins élevée et croît moins vite avec le nombre d'actions $K$. En effet, ici, à chaque instant $t$, on doit considérer $d^{2}+d(t-1)+K(d+1)$ paramètres. Au lieu de grandir quadratiquement avec le nombre d'action, la complexité augmente avec le carré de l'espace latent, d'où l'intérêt de prendre $d<K$. 
La méthode utilisée précédemment, qui consiste à se restreindre à une fenêtre temporelle de taille $S$ est également utilisable pour réduire la complexité en temps. Cependant, elle peut être à la source d'une perte d'information. En effet, en ne regardant que $S$ pas de temps en arrière, il est possible que l'algorithme oublie les informations acquises avant. Dans cette optique, nous proposons d'introduire un concept de mémoire. Cette méthode simple consiste à utiliser les valeurs des paramètres calculés au temps $t-1-S$ comme \textit{prior}. Par exemple pour $\theta_{i}$, au lieu de prendre $A_{i,t-1}=\dfrac{I}{\alpha^{2}}+\dfrac{\mathbb{E}[D_{t-2-S..t-2}^{\top}D_{t-2-S..t-2}]}{\sigma^{2}}$, il suffit de prendre $A_{i,t-1}=A_{i,t-1-S}+\dfrac{\mathbb{E}[D_{t-2-S..t-2}^{\top}D_{t-2-S..t-2}]}{\sigma^{2}}$. On peut procéder ainsi également pour les paramètres $(W_{i},b_{i})$. En revanche, les couche cachées en dehors de la fenêtres de temps ne sont pas reconsidérées.

 \begin{algorithm}[!ht]
\For{$t  =1..T$} {
  \eIf{$t=1$}{
    Pour tout $i$, échantillonner $\tilde{r}_{i,1}$ selon les distributions \textit{a priori}  du modèle de récompense\;
  }
  {
    Exécuter le processus d'inférence variationnelle selon l'algorithme \ref{varPass} afin de calculer les paramètres de chaque distribution \;
    Echantillonner $h_{t-1}$ et $\Theta$ \;
    \For{$i =1..K$} {
      Echantillonner $(W_{i},b_{i})$ \;
        $\tilde{r}_{i,t}=W_{i}^{\top}\Theta h_{t-1}+b_{i}$\;
    }
  }
     $\qquad\mathcal{K}_{t} \leftarrow \argmax\limits_{\hat{\mathcal{K}} \subseteq \mathcal{K}, |\hat{\mathcal{K}}|=k} \quad \sum\limits_{i \in \hat{\mathcal{K}}} \tilde{r}_{i,t}$ \;
  \For{$i  \in \mathcal{K}_{t}$} {
    Recevoir $r_{i,t}$ \;
    }
}
\caption{\texttt{Recurrent State Thompson Sampling }}
\label{varRecurrentTS}
\end{algorithm}


\section{Expérimentations}

Dans cette section nous proposons dans un premier temps des expérimentations sur données artificielles, générées selon le modèle relationnel défini en début de chapitre. Nous verrons que l'algorithme de Thompsons Sampling proposé offre des performances supérieures aux algorithmes de l'état de l'art. Dans le cas où le nombre d'actions devient trop grand, nous étudierons l'algorithme utilisant une couche cachée et verrons que ce dernier permet d'obtenir de bons résultats. Dans un second temps, nous étudierons les performances de notre algorithme dans un scénario simulé où les données seront générées d'une façon différente, en faisant intervenir des périodicités. Nous verrons à nouveau que l'algorithme utilisant une couche cachée est en mesure de retrouver la structuration temporelle des récompenses. Finalement, nous étudierons les approches proposées dans ce chapitre sur des données réelles, où nous aborderons en particulier les difficultés rencontrées pour apprendre notre modèle sur ces données.

    \subsection{Données artificielles relationnelles}
    
      \subsubsection{Protocole}

        \textbf{Génération des données:} Afin d'évaluer les performances des méthodes décrites précédemment, nous les expérimentons sur des données simulées selon le processus relationnel de la section \ref{modRecurrent}. Nous fixons $\sigma=\alpha=\gamma=\delta=1$ pour les valeurs des différentes variances, et $\mu_{i}=0$  pour la moyenne \textit{a priori} de chaque action $i$. Nous générons ensuite les paramètres $\Theta$ en prenant soin de vérifier que la condition sur les valeurs propres de la matrice des paramètres $\Theta$ est bien respectée pour éviter que le problème diverge (voir remarque en début de chapitre). Les récompenses sont générées selon le modèle probabiliste décrit dans la section \ref{modRecurrent} selon les deux scénarios suivant:
        
        \begin{itemize}
      \item  \textbf{XP1: } on considère un nombre d'actions relativement faible de $K=30$ et un horizon temporel de $T=1000$ itérations. Nous testons différentes valeurs de $k \in \left\{ 1, 2, ..., 29 \right\}$. Cette expérimentation a pour but d'évaluer les performances de nos approches récurrentes lorsque le nombre d'actions est raisonnable.
\item \textbf{XP2: }  on se place dans un cas où $K=200$ et  $T=10000$ avec des récompenses réelles. Nous testons également plusieurs valeurs de $k\in \left\{ 10, 20, ..., 190 \right\}$. La version originale sans couche cachée de l'algorithme n'est plus envisageable ici en raison du trop grand nombre de paramètres. Le but est donc de tester la version utilisant une couche générative cachée en présence de multiples bras. Les données utilisées ici sont générées selon un modèle linéaire classique comme dans le premier scénario.
      \end{itemize}

Pour chacune de ces deux séries d'expérimentations, 100 jeux de données sont générés. Les résultats sont présentés correspondent à des moyennes sur ces différents corpus.

\textbf{Politiques testées:} 
A notre connaissance , il n'existe pas d'algorithme adapté à notre problème de bandit récurrent. Cependant, afin d'évaluer les performances de nos politiques, on se compare à divers algorithmes de l'état de l'art (en plus d'une politique \texttt{Random}):
\begin{itemize}
\item Algorithmes de bandit stochastique: nous proposons de se comparer aux deux politiques optimistes \texttt{UCB} et \texttt{UCBV} et à un algorithme de Thompson sampling pour des récompenses suivant des gaussiennes, que nous appelons \texttt{TS classique}. Ce dernier est décrit dans l'état de l'art dans l'algorithme \ref{TSGauss}. Toutes ces politiques, bien que n'étant pas conçues pour les récompenses non stationnaires comme celles étudiées ici, peuvent néanmoins être en mesure de capter des tendances stationnaires pour les différentes actions.

\item  Algorithmes de bandit non stationnaire: nous nous comparons aux algorithmes \texttt{Discount UCB} et \texttt{Sliding Window UCB}, spécifiquement conçus pour des problèmes de bandit non stationnaire  (présenté dans l'état de l'art en section \ref{nonstatBanditEtatArt}). Nous adaptons ces algorithmes au cas de la sélection multiple en sélectionnant les $k$ actions avec le meilleur score à chaque itération. On rappelle que l'algorithme  \texttt{Discount UCB} utilise un facteur de \textit{discount} pour prendre en compte les variations dans le temps tandis que l'algorithme \texttt{Sliding Window UCB} utilise une fenêtre glissante. Dans les remarques 3 et 9 de \cite{nonStationaryBanditAbrupt}, les auteurs proposent des valeurs optimisées pour ces deux paramètres, 
à savoir pour l'algorithme \texttt{Discount UCB}, un facteur de \textit{discount} égal à $1-\sqrt{\gamma_{T}/T}/4$, et pour l'algorithme \texttt{Sliding Window UCB}, une fenêtre de temps égale à la partie entière de $2\sqrt{T\log(T)/\gamma_{T}}$, avec $\gamma_{T}$ le nombre de changements de moyenne des actions. Dans le scénario que nous étudions ici, sans hypothèses supplémentaires on a $\gamma_{T}=T$.
\end{itemize}

Nous implémentons également deux autres politiques non réalistes, au sens où elles ne sont pas envisageable dans un cas réel: 
\begin{itemize}
\item  Algorithme de bandit contextuel: à titre informatif, nous implémentons un algorithme de Thompson sampling pour bandit contextuel traditionnel (voir algorithme \ref{TSContextLin}), que nous notons \texttt{TS contextuel}, dans lequel \textbf{aucune} composante du vecteur de récompense ne serait manquante. Pour cela, à la fin de chaque itération, on met à disposition de l'algorithme le vecteur des récompenses complet. Notons que cette politique n'est pas réaliste dans la configuration que nous avons choisie puisque l'on n'a normalement pas accès au contexte complet. De plus, cette politique n'est pas testée pour $K=200$ (\textbf{XP2}) en raison de la complexité d'inverser 200 matrices de dimension 200 à chaque itération. 

\item Algorithme optimal linéaire: finalement, nous nous comparons à un oracle qui connaîtrait la véritable valeur des paramètres  ayant permis de générer les données, et dans lequel aucune composante du vecteur de récompense ne serait manquante. Cet algorithme, noté \texttt{Optimal contextuel}, sélectionne à chaque itération les $k$ actions ayant la plus forte valeur de $\theta_{i}^{\top}R_{t-1}$.
\end{itemize}

Pour nos algorithmes, nous choisissons une fenêtre de temps de taille $S=200$ itérations et un total $nbIt=10$ itérations du processus variationnel à chaque instant. Dans la suite, \texttt{Recurrent TS} désigne notre modèle récurrent à relations directes, et \texttt{Recurrent TS d=valeur} désigne notre modèle à couche cachée, où $d$ désigne la dimension de l'espace latent. Pour \textbf{XP1} nous testons $d\in \left\{2, 4,8 \right\}$, et pour \textbf{XP2} nous testons $d\in \left\{ 5, 10, 20,30 \right\}$.

        \subsubsection{Résultats}
        
        Nous présentons les résultats en termes de gain cumulé final (c'est-à-dire au temps $t=T$) pour chaque politique en fonction du nombre d'actions sélectionnées $k$. Les figures \ref{fig:RecurrentTSK30} et \ref{fig:RecurrentTSK200} illustrent ces valeurs respectivement pour les scénarios \textbf{XP1}  et \textbf{XP2}. Tout d'abord, on note que l'aspect en forme de cloche des courbes est dû au fait que les récompenses peuvent être négatives. Par conséquent, la récompense cumulative finale peut diminuer même si le nombre d'actions sélectionnées augmente. Tel qu'attendu, la politique \texttt{Optimal contextuel} obtient les meilleurs résultats, suivie par \texttt{TS contextuel} (présent dans \textbf{XP1}), qui, malgré ses observations de toutes les récompenses à chaque itération doit estimer les paramètres du modèle au fur et à mesure. \texttt{TS contextuel} constitue en quelques sorte une borne supérieure de ce que l'on peut espérer obtenir avec \texttt{Recurrent TS}. A noter que plus le nombre $k$ d'actions sélectionnées augmente, plus \texttt{Recurrent TS} observe des composantes de contexte, ce qui explique le rapprochement des ses résultats de ceux de cette borne supérieure. Nous remarquons en outre dans la figure \ref{fig:RecurrentTSK30}, que quel que soit ce nombre $k$, \texttt{Recurrent TS} obtient de meilleurs résultats que tous les autres algorithmes. Cela signifie que ce dernier est capable de reconstruire correctement les poids du modèle. L'utilisation de couches cachées permet également d'obtenir de bonnes performances, qui sont toutefois moins élevées que la version n'en utilisant pas. Ceci est cohérent étant donné le processus de génération des données qui correspond exactement au modèle recherché par \texttt{Recurrent TS}. Notons que nous n'avons pas représenté la courbe de \texttt{Recurrent TS d=8}, celle-ci n'étant que très légèrement au dessus  la courbe de \texttt{Recurrent TS d=4}. De plus, il n'y a aucune amélioration avec les politiques \texttt{Discount UCB} et \texttt{Sliding Window UCB} par rapport aux politiques traditionnelles telles que \texttt{UCB}. Dans \textbf{XP1}, \texttt{Sliding Window UCB} est même moins performant qu'une stratégie aléatoire. Ceci s'explique par le fait que ces deux méthodes sont conçues pour le cas où les récompenses changent brusquement, comme c'est le cas ici, mais seulement un nombre restreint de fois pendant la durée de l'expérience, ce qui n'est pas vérifié ici puisque des changements se produisent à chaque itération. 
        
        Dans la seconde expérience (\textbf{XP2}), l'algorithme \texttt{Recurrent TS} avec couche cachée obtient également de meilleures performances que n'importe quelle autre politique, ce qui nous permet de confirmer que notre algorithme est capable d'extraire des relations entre les actions depuis un espace de dimension réduite. Enfin, ses performances augmentent avec le nombre de dimensions de l'espace caché. Cependant, même si le taux d'amélioration entre $d=5$ et $d=10$ (ou entre $d=10$ et $d=20$) est élevé, nous remarquons qu'il diminue ensuite et atteint une limite lorsque $d=30$.

        \begin{figureth}    
        \includegraphics[width=0.8\linewidth]{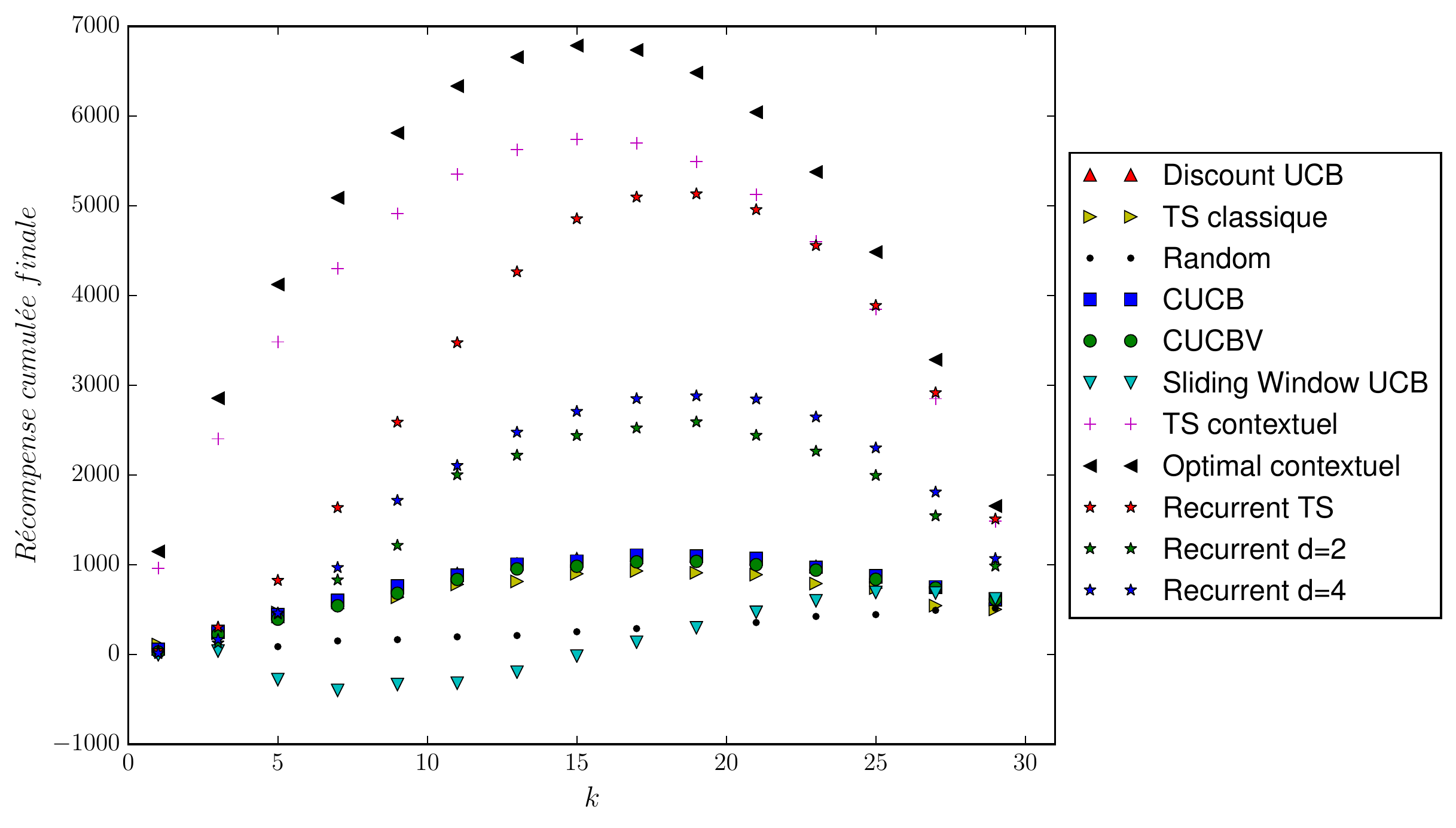}
      \caption[Gain final simulation \texttt{Recurrent TS} $K=30$ données artificielles relationnelles.]{Gain cumulé final  en fonction du nombre de bras sélectionnés à chaque itération sur données artificielles avec $K=30$.}   
      \label{fig:RecurrentTSK30}
  \end{figureth}

          \begin{figureth}    
            \includegraphics[width=0.8\linewidth]{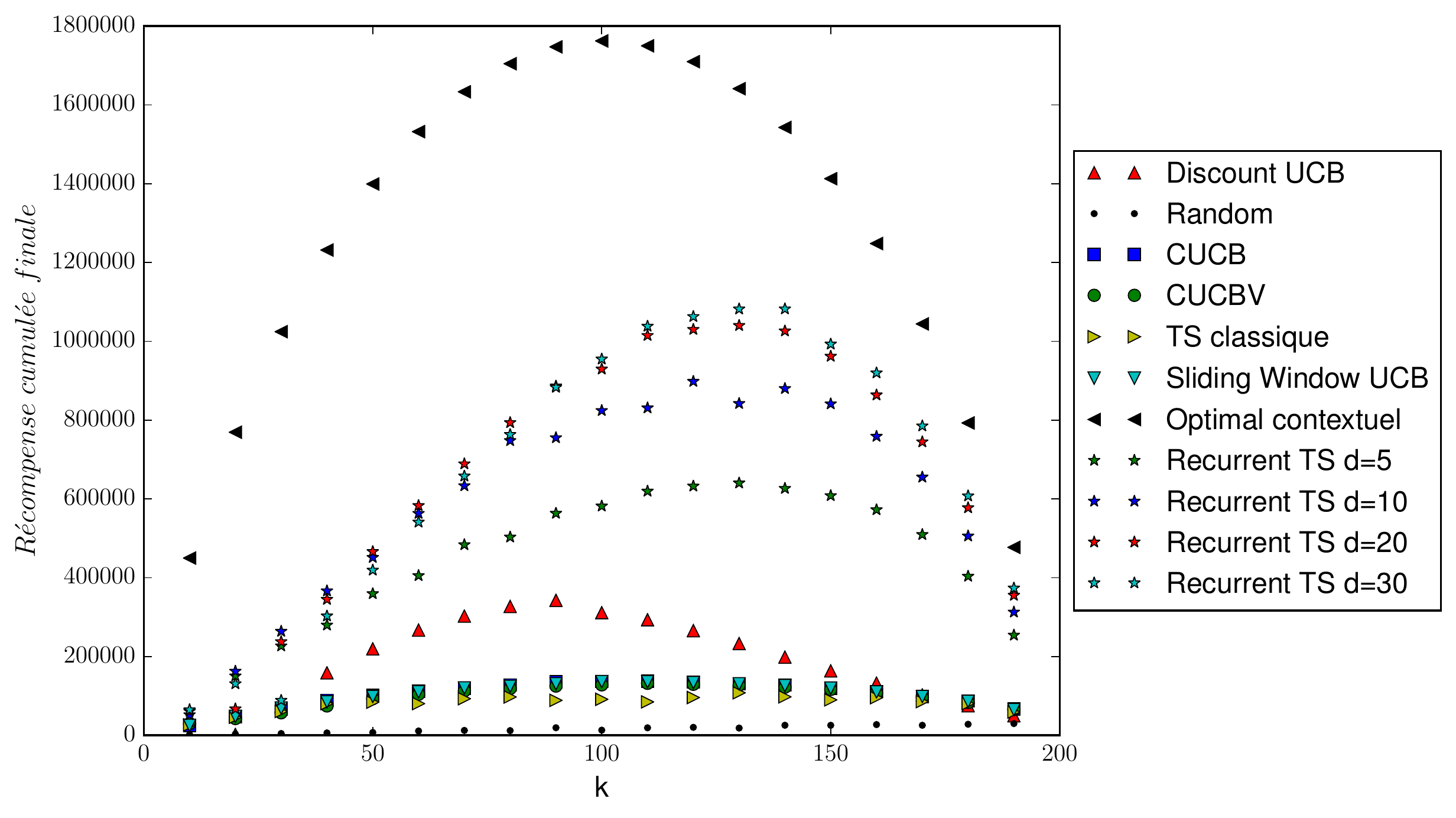}
      \caption[Gain final simulation \texttt{Recurrent TS} $K=200$ données artificielles relationnelles.]{Gain cumulé final  en fonction du nombre de bras sélectionnés à chaque itération sur données artificielles avec $K=200$.}   
      \label{fig:RecurrentTSK200}
      \end{figureth}

    \subsection{Données artificielles périodiques}
        
        \subsubsection{Protocole} 
        
        Nous venons de voir que les algorithmes proposés dans ce chapitre sont en mesure de retrouver les relations existantes entre les différentes actions lorsque les données ont été générées selon un modèle relationnel récurrent. Nous proposons dans cette section d'étudier le comportement de notre algorithme dans un scénario ou les données ne sont pas directement générées par le modèle de la section \ref{modRecurrent}. 
        
\textbf{Génération des données:} Le modèle étudié considère une périodicité dans le comportement des différentes actions. On se place sur un horizon de $T=10000$ itérations et on prend un ensemble de $K=200$ actions. On découpe l'horizon en 100 cycles de 100 itérations chacun. On découpe chaque cycle en 4 périodes de 25 itérations. Pour chaque action $i$, on tire aléatoirement une moyenne $\mu_{i}$ entre 0 et 1, et un entier $j_{i}\in \left\{1,2,3,4\right\}$. Les récompenses sont ensuite générées de la façon suivante: à chaque temps $t\in \left\{1,...,T\right\}$, on calcule $j$, la période dans laquelle se trouve $t$. Pour chaque action $i$, si $j_{i}=j$ on génère une récompense $r_{i,t}$ selon une loi gaussienne de moyenne $\mu_{i}$ et de variance 1. Si $j_{i}\neq j$ on fixe une récompense nulle, c'est-à-dire $r_{i,t}=0$. Afin d'illustrer ce processus, nous représentons les récompenses générées pour un sous-ensemble de 7 actions sur 500 itérations dans la figure \ref{fig:SimProcessK200}, où l'on voit bien apparaître les différentes fenêtres pour chaque action. 
Nous souhaitons étudier dans quelle mesure le modèle avec couche cachée est capable de retrouver une structure de ce type dans les données sans avoir besoin de modéliser explicitement les différentes périodes. 

        \begin{figureth}    
        \includegraphics[width=0.6\linewidth]{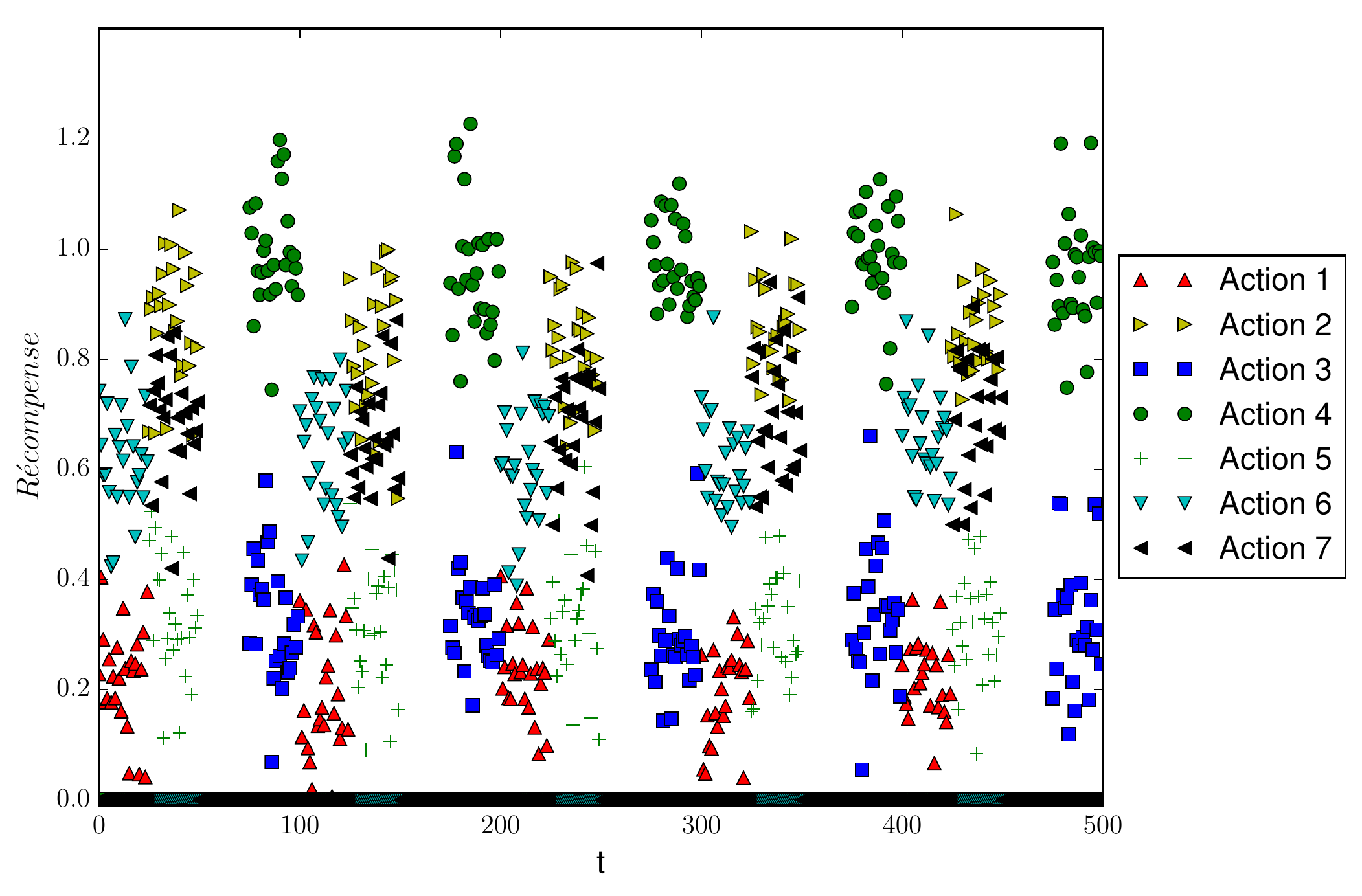}
      \caption[Echantillon de données artificielles périodiques.]{Illustration des récompenses produites au cours du temps pour 7 actions (parmi 200), sur 500 itérations.}   
      \label{fig:SimProcessK200}
  \end{figureth}

\textbf{Politiques testées:} 
Nous proposons d'étudier les performances de notre algorithme par rapport aux politiques décrites plus haut, à savoir: \texttt{Random}, \texttt{UCB}, \texttt{UCBV}, \texttt{TS classique}, \texttt{Discount UCB} et \texttt{Sliding Window UCB}. Pour les algorithmes \texttt{Discount UCB} et \texttt{Sliding Window UCB}, étant donné que le nombre de changements d'espérance pour chaque action sur l'horizon de l'expérience vaut  $\gamma_{T}=200$, on fixe la valeur du facteur de \textit{discount} à 0.96 et celle de la fenêtre glissante à 42, afin de respecter les recommandations de la section précédente. Pour l'algorithme avec couche cachée, nous choisissons une fenêtre de temps de taille $S=100$ itérations et un total $nbIt=10$. Nous testons différentes valeurs de $d\in \left\{ 1, 2 ,4 \right\}$. De plus, afin de valider l'utilité d'utiliser une mémoire (voir note sur la complexité de l'algorithme dans la section \ref{sectionalgoHiddenState}) pour l'apprentissage des distributions  nous testons différents cas, à savoir: \\
- avec mémoire sur les $(W_{i},b_{i})$ et avec mémoire sur les $\theta_{j}$;\\
- avec mémoire sur les $(W_{i},b_{i})$ et sans mémoire sur les $\theta_{j}$;\\
- sans mémoire sur les $(W_{i},b_{i})$ et sans mémoire sur les $\theta_{j}$;\\
- sans mémoire sur les $(W_{i},b_{i})$ et avec mémoire sur les $\theta_{j}$.

                \subsubsection{Résultats}

         \textbf{Résultats préliminaires:} 
        Avant d'expérimenter les différentes politiques, nous souhaitons étudier le comportement du modèle en mesurant sa capacité à retrouver les différents cycles présents dans les données. La figure \ref{fig:hiddelLayersEvolution} représente l'évolution de la valeur moyenne des couches cachées sur 500 itérations  et pour différents nombres de passes de l'algorithme itératif variationnel, en considérant l'ensemble des données. On observe une évolution qui tend à se stabiliser au bout de la dixième passe. Il apparaît clairement qu'une certaine structure est découverte par le modèle, des motifs périodiques se détachant nettement sur les différentes dimensions. Nous souhaitons maintenant vérifier que dans le scénario du bandit qui nous intéresse, notre approche est en mesure de récolter de bonnes récompenses.

            \begin{figureth}
    \begin{subfigureth}{0.33\textwidth}
      \includegraphics[width=\linewidth]{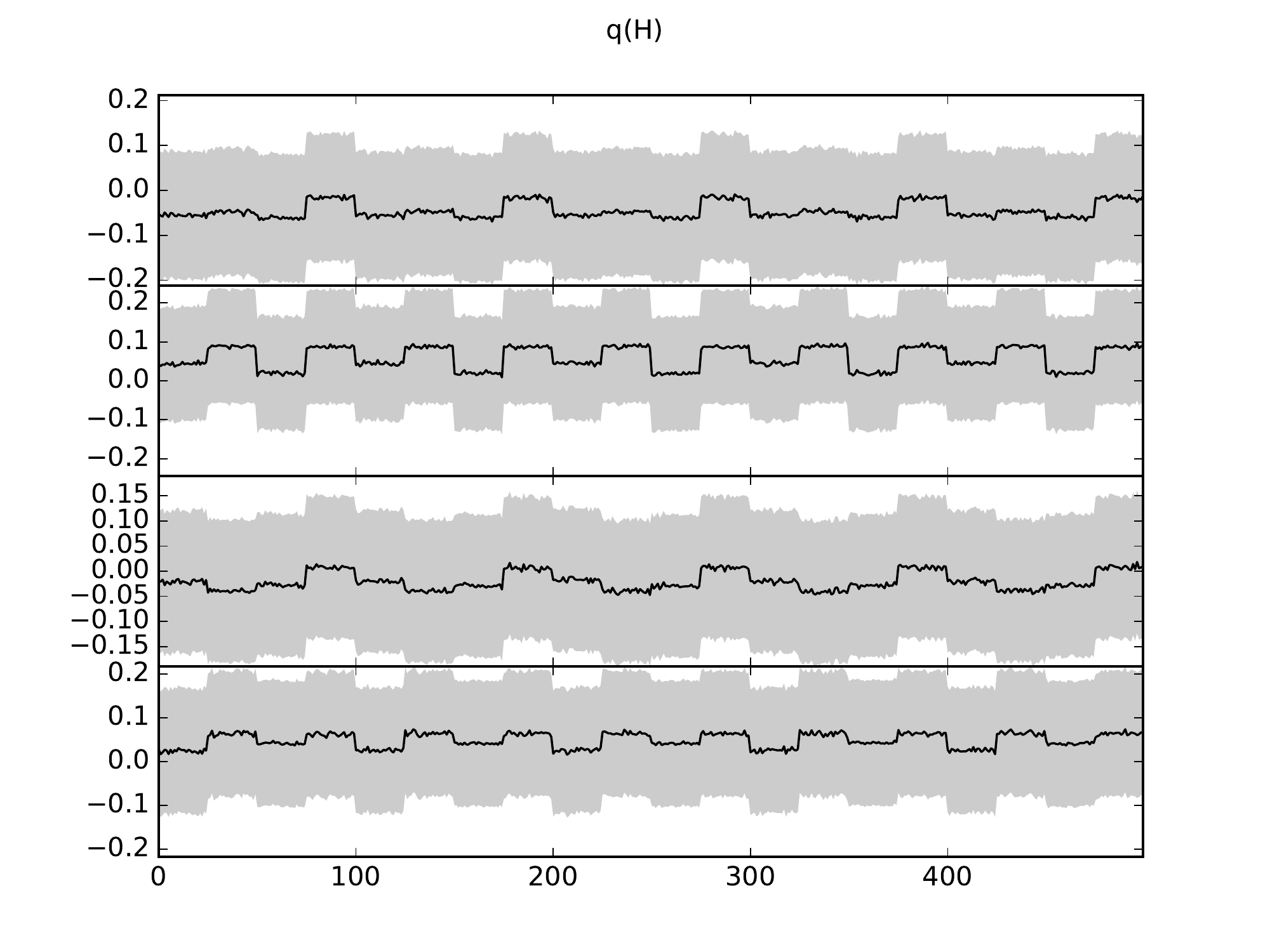}
      \caption{1 passe}
    \end{subfigureth}
    \begin{subfigureth}{0.33\textwidth}
      \includegraphics[width=\linewidth]{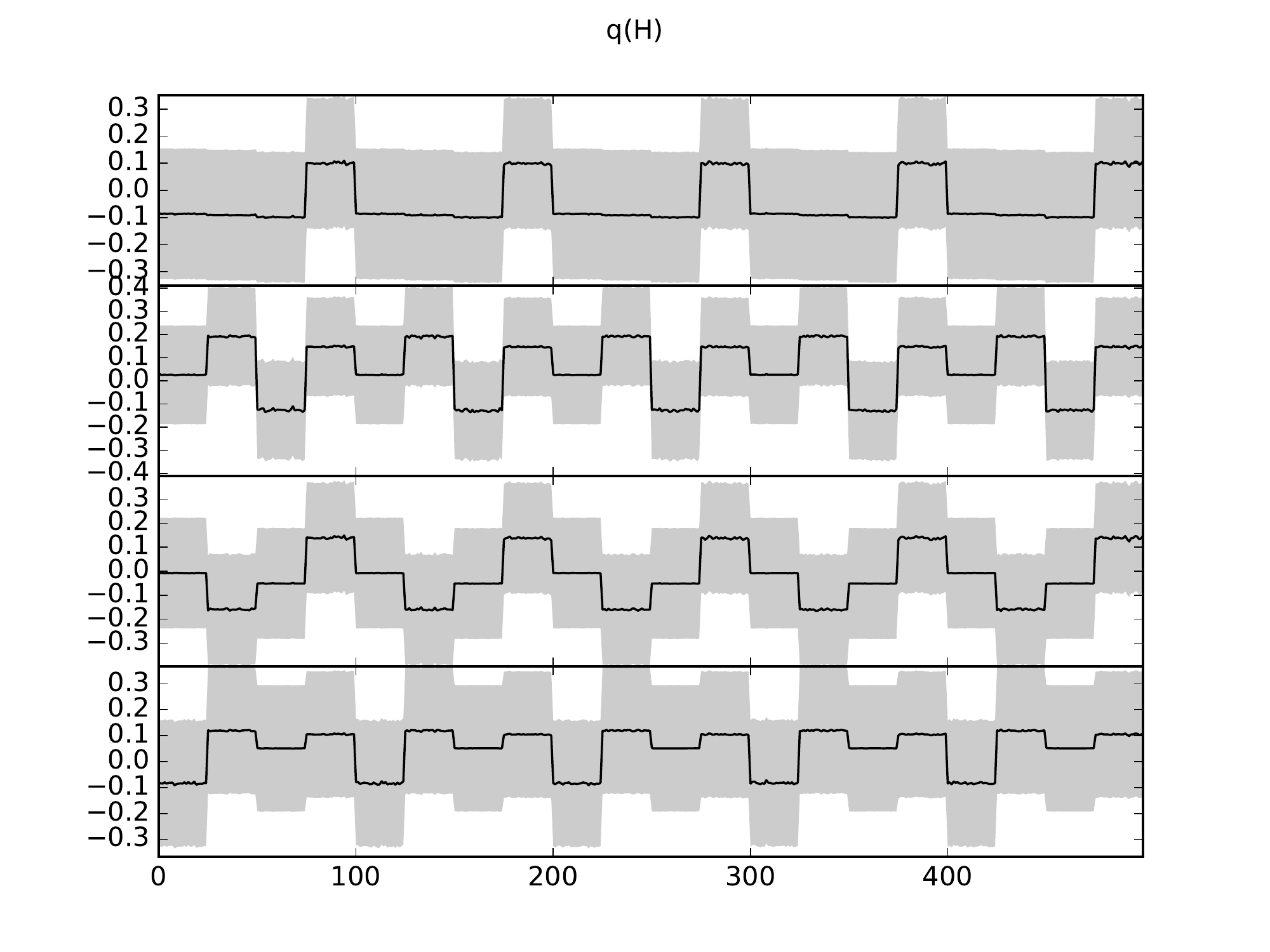}
      \caption{2 passes}
    \end{subfigureth}
        \begin{subfigureth}{0.33\textwidth}
      \includegraphics[width=\linewidth]{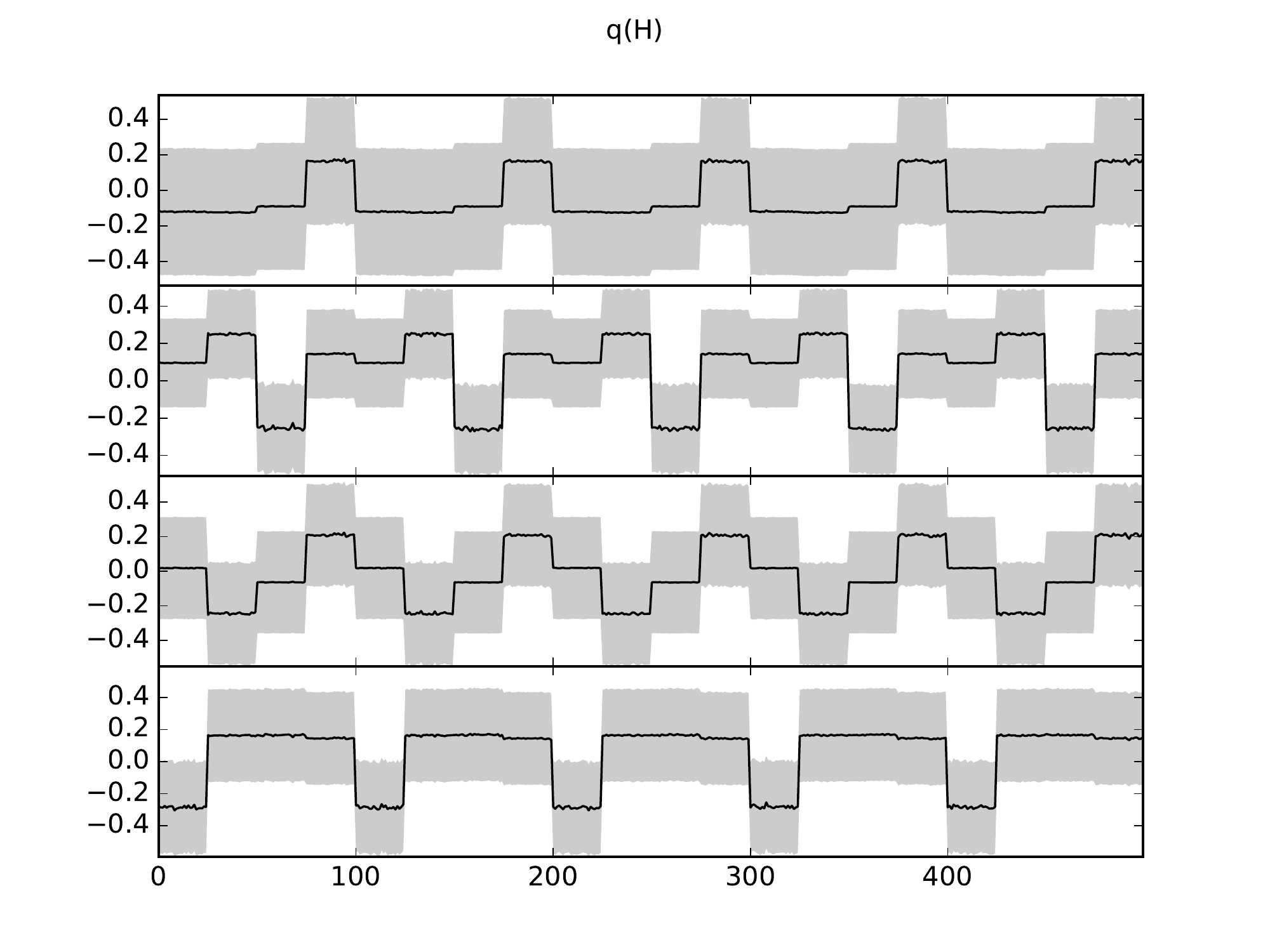}
      \caption{3 passes}
    \end{subfigureth}
           \begin{subfigureth}{0.33\textwidth}
      \includegraphics[width=\linewidth]{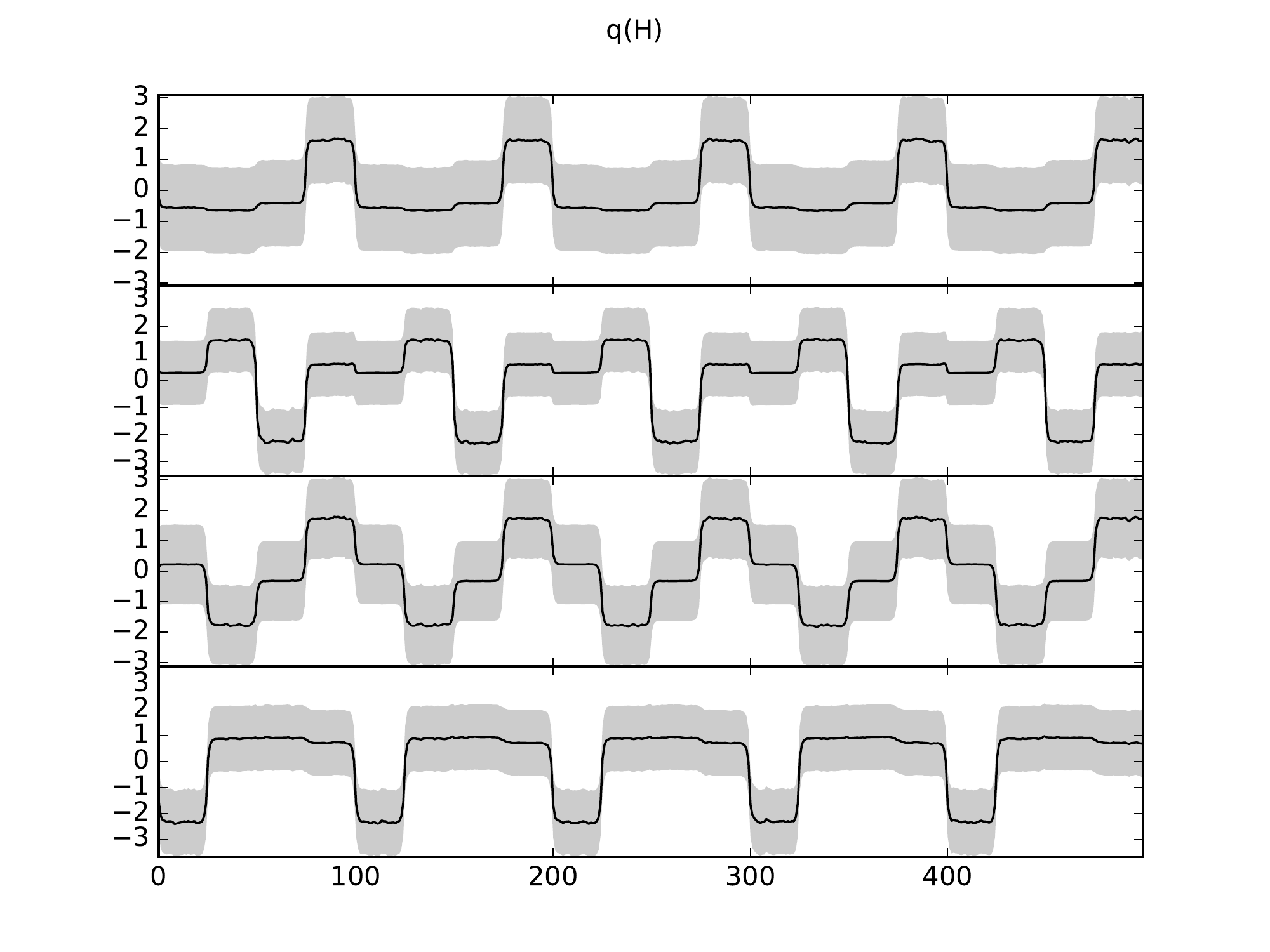}
      \caption{10 passes}
    \end{subfigureth}
    \caption[Couches cachées avec $K=200$ et données artificielles périodiques: première expérimentation.]{Evolution des valeurs des différentes couches cachées au cours du temps pour $d=4$ sur un horizon de 500 itérations.}
          \label{fig:hiddelLayersEvolution}
    \end{figureth}

        \textbf{Performances des différentes politiques:}  Premièrement, il est clairement apparu que l'utilisation d'une mémoire est importante dans ce scénario. En effet, nous avons observé que la version de l'algorithme \texttt{Recurrent TS} offrant les meilleurs résultats est celle qui utilise à la fois la mémoire sur les $(W_{i},b_{i})$ et les $\theta_{j}$. Ceci est illustré dans la figure \ref{fig:effectMemory}, où l'on représente le gain final obtenu par la politique \texttt{Recurrent TS d=4} lorsque $k=50$, dans les quatre configurations proposées plus haut ("true" signifie que la mémoire a été activée et "false" signifie que non). Il est également apparu  que la mémoire sur les $(W_{i},b_{i})$ a plus d'impact que celle sur les $\theta_{j}$. Ceci est dû au fait que l'apprentissage des paramètres $\theta_{j}$ est mutualisé entre tous les bras, tandis que $(W_{i},b_{i})$ est propre à chaque action $i$. Ainsi, si une action n'est pas sélectionnée pendant $S$ itérations et que la mémoire sur les $(W_{i},b_{i})$ n'est pas active, toute l'information la concernant est perdue.
        
                \begin{figureth}    
        \includegraphics[width=0.6\linewidth]{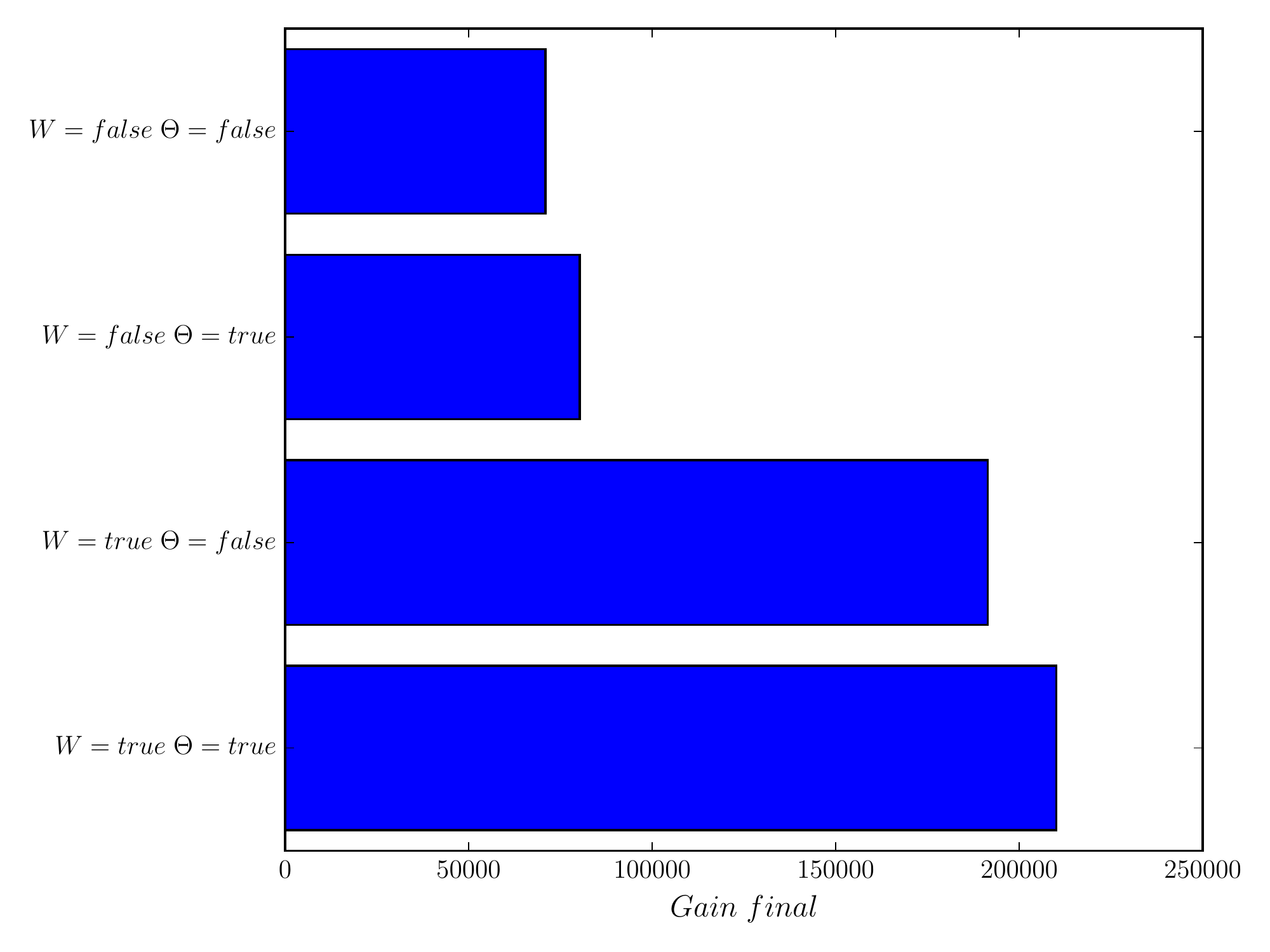}
      \caption[Effet de la mémoire sur les performances de \texttt{Recurrent TS}.]{Effet de la mémoire sur les performances de \textit{Recurrent TS}: représentation du gain final obtenu pour les quatre configurations proposées avec $k=50$.}   
      \label{fig:effectMemory}
  			\end{figureth}

        Nous présentons maintenant les résultats en terme de gain cumulé final (c'est-à-dire au temps $t=T$) pour chaque politique dans la figure \ref{fig:RecurrentTSK200Bis} en fonction du nombre d'actions sélectionnées $k$  (pour l'algorithme avec couches cachées, on illustre la version avec mémoire). La première chose que l'on remarque est la présence d'une sorte de plateau à $k=50$. Ceci est dû au fait que l'on a 200 actions au total et 4 périodes. Comme chaque action n'a une récompense non nulle que pendant une des quatre périodes, à chaque itération on a en moyenne 150 actions qui ont une récompense nulle. 
        On remarque que les politiques de bandit non stationnaire \texttt{Discount UCB} et \texttt{Sliding Window UCB} ne se démarquent des politiques de bandit stationnaire \texttt{UCB}, \texttt{UCBV} et \texttt{TS classique} que lorsque $k$ est suffisamment grand (au-delà de 100). Toutes ces politiques apparaissent moins performantes que la politique \texttt{Recurrent TS}, du moins lorsque $d=2$ et $d=4$. Lorsque $d=1$ les résultats sont plus mitigés, mais nous arrivons tout de même à être performants sur une plage de $k$ entre 50 et 100. De plus, les performances de \texttt{Recurrent TS} sont plus élevées avec $d=4$ qu'avec $d=2$. Nous avons cependant observé qu'augmenter le nombre de dimensions au-delà de 4 ne permet pas d'améliorer les performances de l'algorithme. Ceci s'explique par le fait que 4 variables sont suffisantes pour reconstruire la structure périodique des récompenses (comme le suggère la figure \ref{fig:hiddelLayersEvolution}). 

        \begin{figureth}    
        \includegraphics[width=0.8\linewidth]{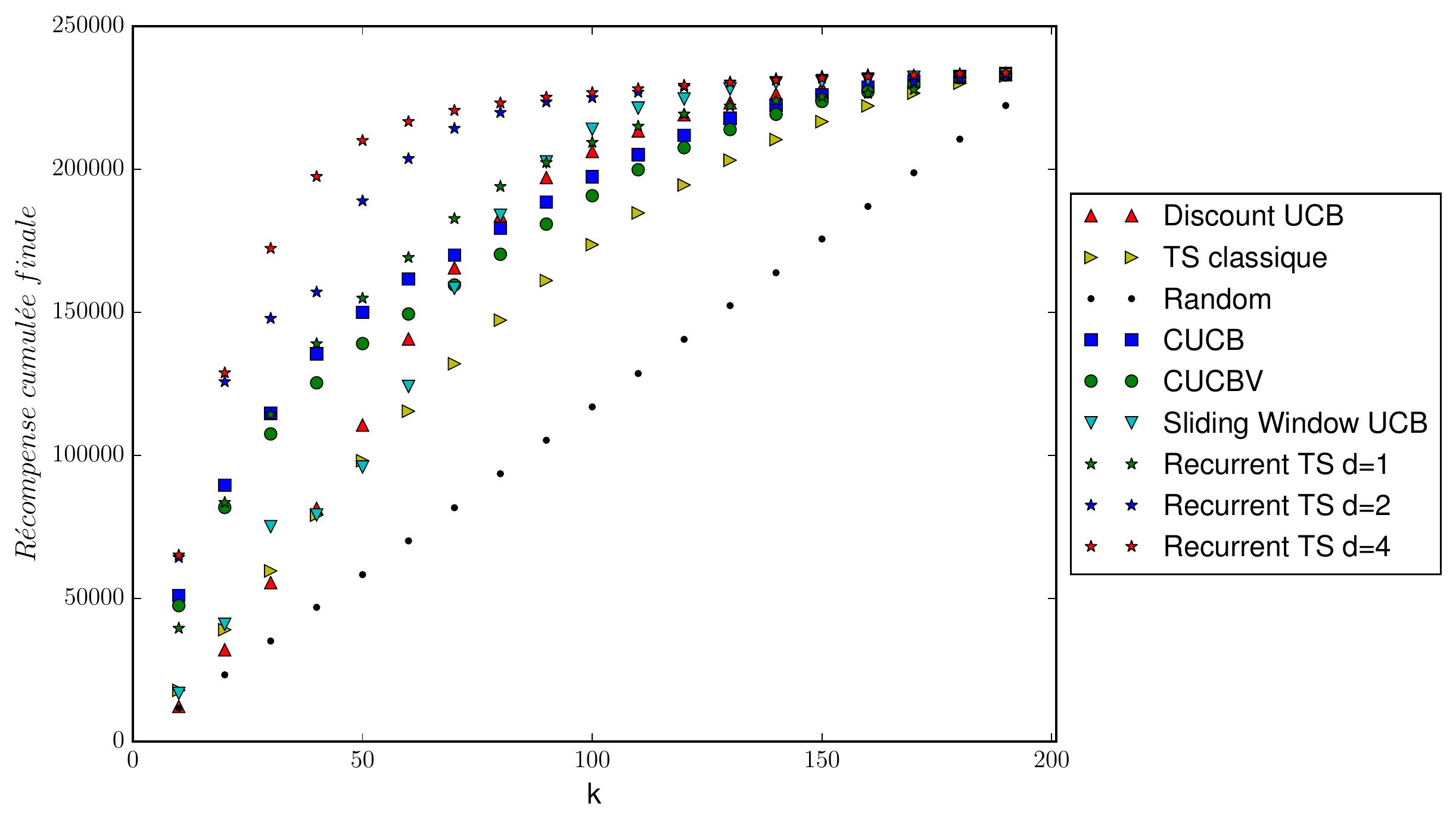}
      \caption[Gain final simulation \texttt{Recurrent TS} $K=200$ données artificielles périodiques: première expérimentation.]{Gain cumulé final  en fonction du nombre de bras sélectionnés sur données artificielles avec $K=200$ et des récompenses  périodiques.}  
      \label{fig:RecurrentTSK200Bis}
  \end{figureth}

\begin{note}[Expérimentation avec des cycles plus complexes]
Afin de nous assurer de la capacité du modèle à détecter des variations de récompenses périodiques, nous avons également expérimenté celui-ci sur des données où chacune des 200 actions possède une dynamique plus singulière. Plutôt que d'avoir un socle commun de 4 périodes de 25 itérations, nous autorisons chaque action $i$ à avoir son propre cycle de $per_{i}$ périodes de $it_{i}$ itérations. Ces valeurs sont tirées aléatoirement pour chaque action, respectivement dans les ensembles $\left\{2,3,4,5\right\}$ et $\left\{10,20,30,40\right\}$. On représente l'évolution des valeurs des couches cachées pour $d=10$ dans la figure \ref{fig:XPSimPeriodRandomHiddenState}. On voit très nettement apparaître des formes cycliques de différentes tailles, ce qui nous pousse à penser que le modèle est en mesure de retrouver une certaine structure dans les données. La dynamique apparaît cependant plus complexe à détecter, car beaucoup plus de paramètres sont en jeu. Finalement, la figure \ref{fig:RecurrentTSK200BisBis} représente les récompenses finales obtenue par différentes politiques en fonction de $k$. A nouveau, notre algorithme permet d'obtenir de meilleurs résultats que ses compétiteurs lorsque $d$ est suffisamment élevé (à partire de $d=4$).
\end{note}

      \begin{figureth}
			\includegraphics[width=0.7\linewidth]{HiddenAfter21StepsBis.pdf}
		\caption[Couches cachées avec $K=200$ et données artificielles périodiques: seconde expérimentation.]{Evolution des valeurs des différentes couches cachées au cours du temps pour $d=10$ sur un horizon de 1000 itérations lorsque les périodes des différentes actions peuvent être différentes et se chevaucher.}
        	\label{fig:XPSimPeriodRandomHiddenState}
		\end{figureth}

      	\begin{figureth}		
        \includegraphics[width=0.8\linewidth]{RecurrentTSSimulationK200T10000BisBis.pdf}
			\caption[Gain final simulation \texttt{Recurrent TS} $K=200$ données artificielles  périodiques: seconde expérimentation]{Gain cumulé final  en fonction du nombre de bras sélectionnés sur données artificielles avec $K=200$ et des récompenses périodiques (avec chevauchement des périodes).}	
			\label{fig:RecurrentTSK200BisBis}
	\end{figureth}

\begin{note}[Modèle relationnel sur données cycliques]
Nous terminons ce jeu d'expérimentations par un test de notre approche relationnelle (c.à.d. sans couche cachée), uniquement utilisable lorsque le nombre d'actions est faible, sur des données cycliques. Pour cela, nous générons des données périodiques de la même façon que précédemment (sans chevauchement, avec 4 périodes de 25 itérations), avec $K=30$ et $T=1000$. Le gain final cumulé en fonction du nombre de bras sélectionnés est présenté dans la figure \ref{fig:RecurrentTSK30Period4} pour différentes politiques. Si comme précédemment, notre approche utilisant des états cachés est la plus performante, il apparaît que l'approche sans couche cachée ne parvient pas à surpasser les performances des algorithmes de bandit stationnaire tel que \texttt{CUCB}. Cela s'explique par le fait que le modèle que l'approche relationnelle cherche à reconstruire considère chaque récompense comme un somme pondérée de toutes les précédentes, ce qui n'est pas le cas ici étant donné le processus de génération des données utilisé. 
\end{note}

       	\begin{figureth}		
        \includegraphics[width=0.8\linewidth]{RecurrentTSSimulationK30T1000Period4.pdf}
			\caption[Gain final simulation \texttt{Recurrent TS} données artificielles  périodiques.]{Gain cumulé final  en fonction du nombre de bras sélectionnés sur les données artificielles avec $K=30$ et des récompenses périodiques (sans chevauchement des périodes).}	
			\label{fig:RecurrentTSK30Period4}
	\end{figureth}

\subsection{Données réelles}

\subsubsection{Etude préliminaire}

Face au nombre élevé d'utilisateurs ($K=5000$) dans les différents jeux de données utilisés précédemment et au temps d'exécution des différentes expérimentations, nous avons fait le choix de n'utiliser que 500 utilisateurs sur les 5000 présents au total dans chaque base. Ces 500 comptes sont sélectionnés en amont de façon aléatoire et uniforme. De plus, au vu des résultats obtenus sur données artificielles, qui montrent que d'une façon générale les performances se détériorent à mesure que le $k$ diminue, nous choisissons de fixer le nombre d'utilisateurs pouvant être sélectionnés simultanément à 50. Nous sommes donc dans une situation où $K=500$ et $k=50$, c'est-à-dire où un dixième des actions peuvent être sélectionnées à chaque itération. De nombreux essais ont été effectués en utilisant un grand ensemble de jeux de paramètres, mais ne nous ont pas permis d'obtenir de meilleures performances que des algorithmes de bandit stationnaire (comme \texttt{UCB}, \texttt{UCBV} ou \texttt{TS classique}) pour les récompenses utilisées dans les chapitres précédents. En vue d'obtenir de meilleurs résultats, nous avons décidé d'augmenter la taille de la fenêtre d'écoute, précédemment d'une valeur de 100 secondes. En effet, cette valeur nous est apparue relativement faible dans un contexte ou l'on souhaite capter un certain nombre de régularités grâce au modèle récurrent. Dans cette optique, nous avons  augmenté la période d'écoute pour la fixer à 30 minutes en vue de lisser les récompenses produites par les différents comptes. Grâce à cette première étape, les cycles potentiels dans les comportements des utilisateurs auront plus de chance d'être détectés par le modèle. Après de nombreux essais, nous ne sommes toujours pas parvenus à atteindre des performances satisfaisantes. Afin de mieux comprendre ces résultats, nous proposons d'étudier un cas, à savoir le jeu de données \textit{OlympicGames} et la thématique \textit{Science}\footnote{Les résultats présentés ici sont également valables pour les autres bases de données et les autres thématiques.}. 

Afin de mesurer la capacité maximale du modèle récurrent avec couches cachées à récolter des récompenses élevées au cours du temps, nous considérons une politique optimale, sélectionnant à chaque itération $t$ les 50 utilisateurs ayant les plus fortes valeurs $W_{i}^{\top}\Theta h_{t-1} + b_{i}$, où les valeurs des différents paramètres sont apprises en amont sur le jeu de données, selon différents pourcentages de données disponibles. Nous testons différentes valeurs de $d \in \left\{ 5,10,20\right\}$ et nous comparons à un modèle optimal n'utilisant que des biais (ce qui correspond à la politique optimale stationnaire). Nous fixons les écart-types des vraisemblances du modèle à $\sigma=\delta=0.1$ et les écart-types des distributions \textit{a priori} à $\alpha=\gamma=1.0$, ces valeurs offrant les meilleures résultats en moyenne. Il est ainsi possible de mesurer dans quelle mesure l'algorithme \texttt{Recurrent TS} peut offrir de meilleures performances que des algorithmes de bandit stationnaire selon les conditions expérimentales. Intuitivement, la part de données disponibles aura une influence directe sur la qualité des paramètres appris en amont. Le tableau \ref{tab:rwdOptScienceOGK500T1800} reporte les valeurs de récompenses finales obtenues par ces différentes politiques, où $d=0$ désigne la politique n'utilisant que des biais. Premièrement, nous remarquons que lorsque toutes les données sont disponibles (ligne 100\%), les performances augmentent avec le nombre de dimensions $d$. Par ailleurs, toujours dans cette configuration, le surplus de récompense obtenu avec $d=20$  par rapport à la politique optimale stationnaire ($d=0$) est d'environ 9\%, ce qui peut paraître assez faible. Lorsque le pourcentage de données disponibles diminue, nous observons que les performances des politiques utilisant des couches cachées se dégradent considérablement. Ce phénomène est d'autant plus marqué que le nombre de dimensions est grand. En revanche, les performances de la politique stationnaire sont beaucoup plus stables. Ceci semble cohérent puisque cette dernière a beaucoup moins de paramètres à apprendre. Finalement, lorsque seulement 10\% des données sont utilisées, ce qui correspond en moyenne à 50 (500$\times$0.1) valeurs par itération, l'utilisation d'un modèle avec couches cachées ne semble plus pertinente puisque l'on obtient des résultats pratiquement identiques à ceux d'un modèle stationnaire plus simple. Or, ce cas correspond à notre scénario, dans lequel $k=50$. Ainsi, les meilleurs résultats pouvant être obtenus avec notre algorithme ne sont pas supérieurs aux meilleurs résultats pouvant être obtenus avec un algorithme de bandit traditionnel. Cette analyse nous permet d'apporter une première explication au fait que nous ne sommes pas parvenus à obtenir des performances satisfaisantes précédemment. 

                 \begin{tableth}
        \begin{tabular}{|c|c|c|c|c|}
                  \hline
                    \textbf{\% données utilisées} & \textbf{Opt. d=0 (uniq. biais)} & \textbf{Opt. d=5}& \textbf{Opt. d=10}& \textbf{Opt. d=20}\\
                    \hline
                    100\% & 50378 & 51056 & 53735 & 55220 \\
                    \hline
                    80\% & 50304 & 51051  & 53264 & 54365 \\
                     \hline
                    50\% & 50201 & 50754 & 52007 & 52376 \\
                     \hline
                    30\% & 50123 & 50341 & 50609 & 512016 \\
                     \hline
                    20\% & 50048 & 49642 & 50019 & 50213 \\
                     \hline
                    10\% & 49923 & 49629 & 48976 & 49327\\
                    \hline
         \end{tabular}
      \caption[Récompense finale politiques optimales avec différentes valeurs de $d$, $k=50$ avec \textit{Science}+\textit{OlympicGames}.]{Récompense finale obtenue avec les politiques optimales pour différentes valeurs de $d$ et $k=50$ selon le pourcentage de données accessibles lors de l'apprentissage pour la thématique \textit{Science} et le jeu de données \textit{OlympicGames}.}
    \label{tab:rwdOptScienceOGK500T1800}
      \end{tableth}

Pour mieux comprendre les raisons de ce phénomène, nous représentons dans la figure \ref{fig:HiddenScienceK500T1800} les valeurs des différentes couches  cachées (après convergence) sur 500 itérations lorsque $d=5$ et que toutes les données sont disponibles. Bien que certains pics soient visibles par endroits, les valeurs prises par $h_{t}$ sont nulles la plupart du temps. Ainsi nous avons $W_{i}^{\top}\Theta h_{t-1} + b_{i} \approx b_{i}$ et la majorité des éléments appris par le modèle se situe dans les termes de biais. Cette remarque nous conforte dans le fait que l'utilisation du modèle proposé dans ce chapitre sur ces données n'est pas pertinente. 
        
Au vu de ces résultats, nous proposons dans la section suivante un autre modèle de récompense pour évaluer le modèle sur des données réelles possédant une structure plus explicite.

              \begin{figureth}    
        \includegraphics[width=0.6\linewidth]{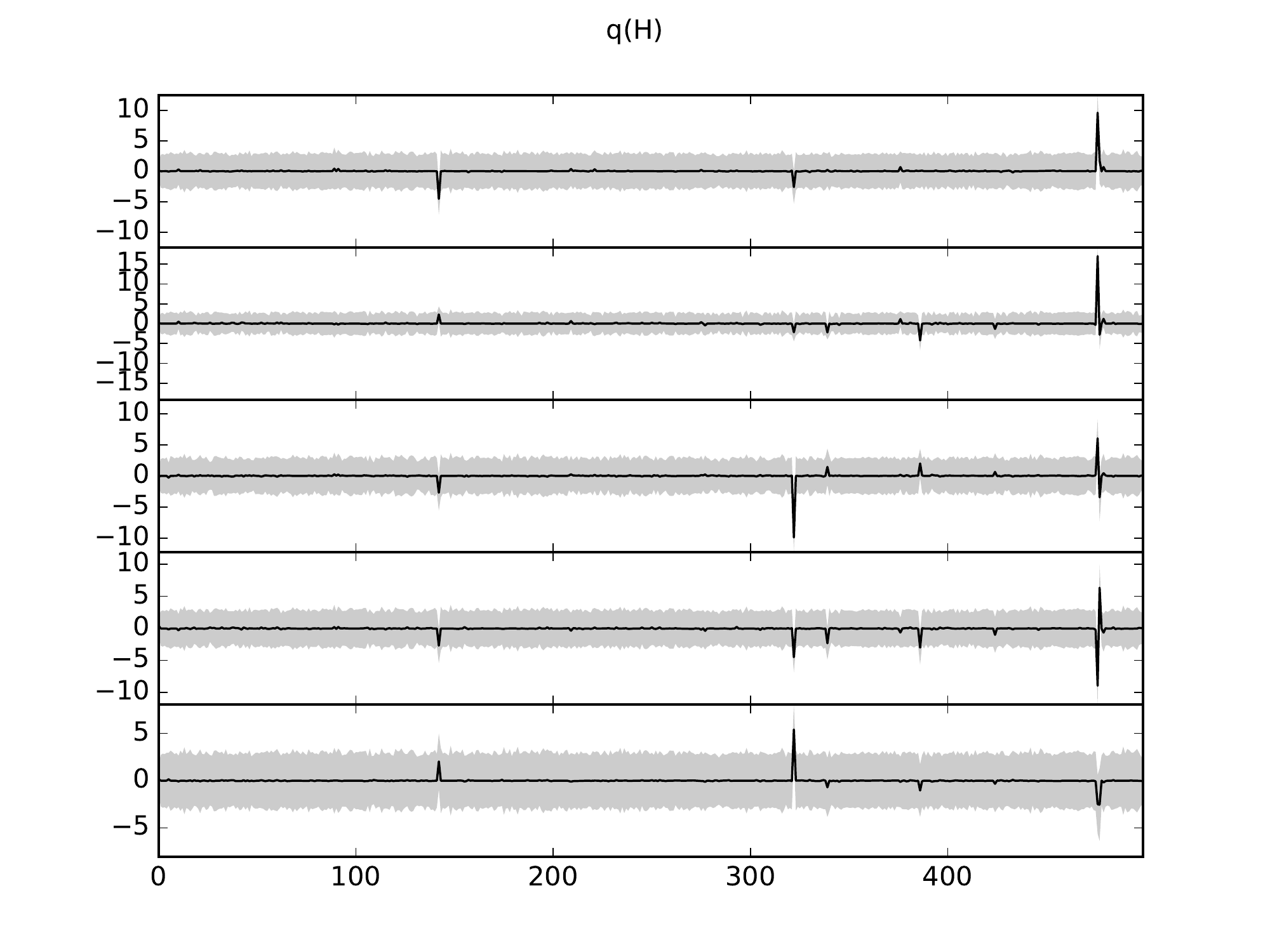}
      \caption[Evolution couches cachées \textit{Science} + \textit{OlympicGames} ]{Evolution des valeurs des différentes couches cachées au cours du temps pour $d=5$ sur un horizon de 500 itérations, la thématique \textit{Science} et le jeu de données \textit{OlympicGames}.}  
      \label{fig:HiddenScienceK500T1800}
      \end{figureth}

\subsubsection{Seconde expérimentation}

Nous utilisons le jeu de données \textit{OlympicGames}, et comme précédemment un sous-ensemble de $K=500$ utilisateurs et des fenêtres d'écoute de 30 minutes.  Nous choisissons également de fixer le nombre d'utilisateurs pouvant être sélectionnés simultanément à $k=50$.

\textbf{Modèle de récompense: }Nous proposons un modèle de récompense plus simple dans lequel un utilisateur produit une récompense égale à 1 s'il publie au moins un contenu durant la période d'écoute considérée, indépendamment du contenu. Lorsqu'un utilisateur ne publie pas, la récompense associée vaut 0. L'objectif est alors de retrouver des cycles d'activité, afin d'identifier à chaque itération les utilisateurs les plus susceptibles des produire du contenu. Ce type de collecte permet l'observation de plus de récompenses non nulles, ce qui permet d'espérer un meilleur apprentissage des dynamiques en jeu.

\textbf{Résultats préliminaires: }Le tableau \ref{tab:rwdOptisTalkOG} présente les résultats d'une étude similaire à celle menée dans la section précédente sur les politiques optimales. Nous remarquons également que les performances des politiques optimales sont d'autant plus élevées que le nombre de couches cachées est grand lorsque toutes les données sont disponibles pour l'apprentissage des paramètres. Dans ce cas, on atteint une amélioration d'environ 30\% en utilisant notre modèle par rapport à une politique optimale stationnaire. Bien que ces résultats se détériorent lorsque le pourcentage de données visibles baisse (sauf pour $d=0$ qui est toujours stable), nous remarquons que, contrairement à l'étude précédente, même lorsque l'on n'a que 10\% de données disponibles, l'utilité d'utiliser le modèle récurrent est notable. Nous remarquons aussi que dans ce cas, l'utilisation d'un nombre $d$ élevé n'est pas la meilleure solution. En effet, il apparaît que les modèles avec $d=5$ et $=10$ sont plus stables que le modèle avec $d=20$ quand le nombre de données utilisées diminue. Nous expliquons cela par le fait que ce dernier possède trop de paramètres à apprendre. Les résultats obtenus favorisent l'idée que l'algorithme \texttt{Recurrent TS} pourrait fonctionner plus efficacement que ses compétiteurs stationnaires.

                 \begin{tableth}
        \begin{tabular}{|c|c|c|c|c|}
                  \hline
                    \textbf{\% données utilisées} & \textbf{Opt. d=0 (uniq. biais)} & \textbf{Opt. d=5}& \textbf{Opt. d=10}& \textbf{Opt. d=20}\\
                    \hline
                    100\% & 30039 & 41866 & 42999 & 43795  \\
                    \hline
                    80\% & 30017 & 41813  & 42846 &  43560 \\
                     \hline
                    50\% & 29900 & 41569 & 42448 & 42746  \\
                     \hline
                    30\% & 29723 & 40748 & 41429 &  41357 \\
                     \hline
                    20\% & 29688 & 39968 & 40085 &  39588 \\
                     \hline
                    10\% & 29271  & 38015 & 37459 &  36024 \\
                    \hline
         \end{tabular}
         \caption[Récompense finale politiques optimales avec différentes valeurs de $d$, $k=50$ avec le nouveau modèle de récompense.]{Récompense finale obtenue avec les politiques optimales pour différentes valeurs de $d$ et $k=50$ selon le pourcentage de données accessibles lors de l'apprentissage pour le nouveau modèle de récompense et le jeu de données \textit{OlympicGames}.}
    \label{tab:rwdOptisTalkOG}
      \end{tableth}

        La figure \ref{fig:HiddenisTalkT1800US} représente l'évolution des couches cachées pour $d=5$ sur 500 itérations et après convergence du modèle. Non seulement les valeurs ne sont pas nulles, mais on voit clairement apparaître des cycles de périodicité différentes. Ce dernier élément nous conforte dans l'idée que les données possèdent une certaine structure, dont l'algorithme \texttt{Recurrent TS} pourrait tirer parti. Nous proposons donc de tester cette approche dans un scénario de bandit.
        
                \begin{figureth}    
        \includegraphics[width=0.6\linewidth]{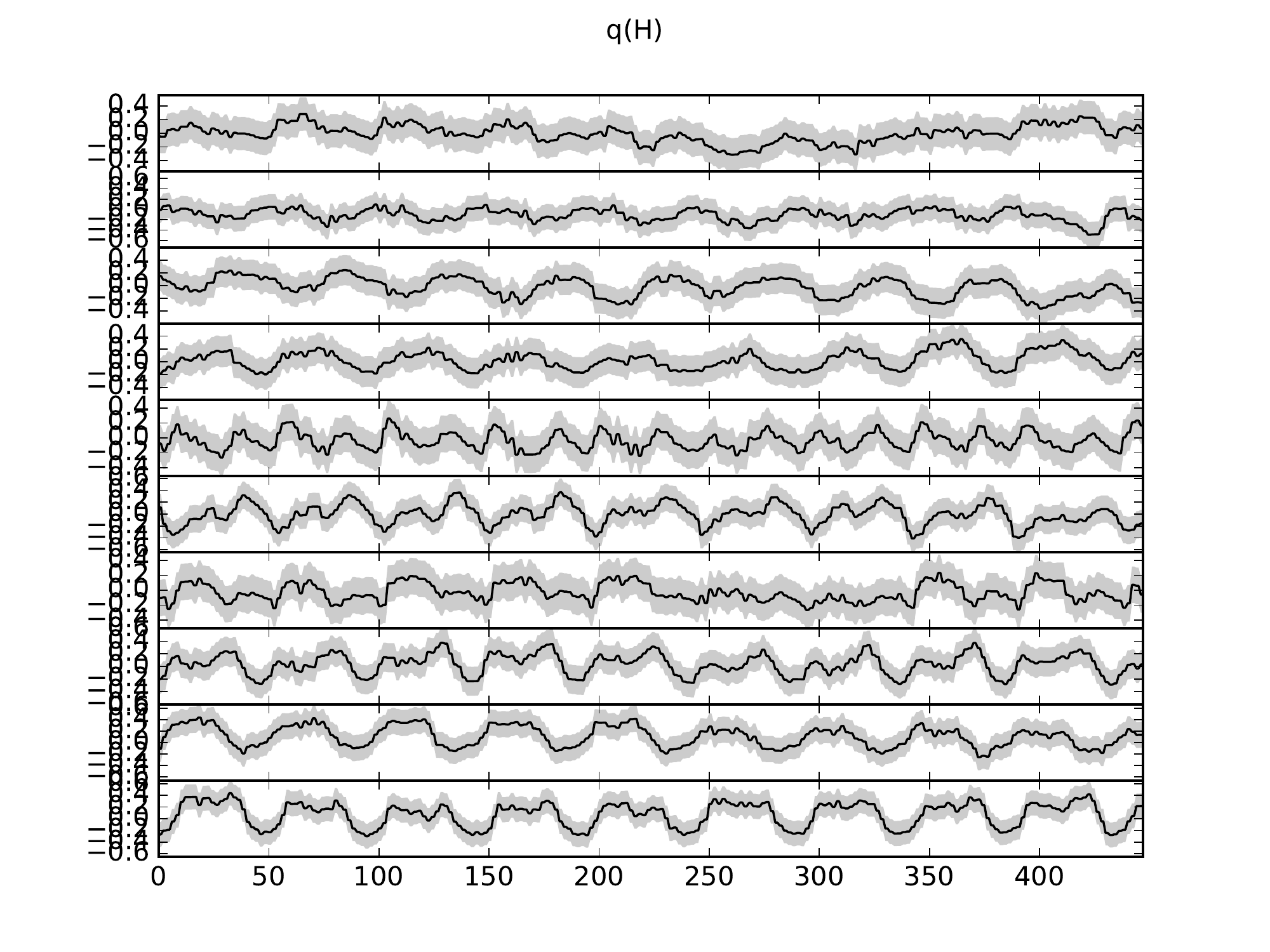}
      \caption[Evolution couches cachées pour le nouveau modèle de récompense et le jeu de données \textit{OlympicGames}]{Evolution des valeurs des différentes couches cachées au cours du temps pour $d=5$ sur un horizon de 500 itérations pour le nouveau modèle de récompense et le jeu de données \textit{OlympicGames}.}   
      \label{fig:HiddenisTalkT1800US}
      \end{figureth}

\textbf{Politiques testées: }En vue de mesurer les performances de l'approche proposée dans ce chapitre, nous nous comparons à divers algorithmes de bandits stationnaires et non stationnaires, à savoir: \texttt{Random}, \texttt{CUCB}, \texttt{CUCBV}, \texttt{TS classique}, \texttt{Discount UCB} et \texttt{Sliding Window UCB}. 
 Pour chacun de ces algorithmes, nous avons fixer les différents paramètres - lorsqu'ils en ont - à leurs valeurs optimales, c'est-à-dire celle offrant la meilleure récompense cumulée \textit{a posteriori}. En particulier pour \texttt{TS classique}, des valeurs de variance de la vraisemblance relativement faible de l'ordre de 0.01, favorisant l'exploitation, semblent mieux adaptées. Pour \texttt{Sliding Window UCB}, la taille de la fenêtre glissante optimale est de 200 itérations, et pour \texttt{Discount UCB}, la valeur optimale se situe proche de 1, correspondant à un \texttt{UCB} classique. Pour l'approche \texttt{Recurrent TS}, nous fixons les écarts-types des vraisemblances du modèle à $\sigma=\delta=0.1$ et les écarts-types des distributions \textit{a priori} à $\alpha=\gamma=1.0$ (pour les mêmes raisons que dans la section précédente). Nous utilisons par ailleurs la fonction de mémoire, conformément aux résultats observés sur données artificielles, avec une fenêtre glissante de 100 itérations.

      \textbf{Résultats: }La figure \ref{fig:resisTalkOGT1800K500} représente l'évolution du gain cumulé en fonction du temps pour les différentes politiques testées. Nous observons clairement que la politique \texttt{Recurrent TS} offre de meilleurs résultats que toutes ses compétitrices pour $d=5$ et $d=10$. En revanche, lorsque $d=20$, les performances se dégradent considérablement, pour atteindre le niveau des approches stationnaires, qui obtiennent toutes des résultats comparables (avec un léger avantage pour l'algorithme \texttt{CUCB}). L'algorithme  de bandit non stationnaire \texttt{Sliding Window UCB} est quant à lui le plus efficace\footnote{Nous ne représentons pas explicitement \texttt{Discount UCB} puisqu'il est équivalent à \texttt{UCB} lorsque le facteur de \textit{discount} vaut 1.}.  
      Finalement, les résultats obtenus dans un scénario de bandit confirme ceux de l'étude préliminaire puisque notre modèle est en mesure de surpasser toutes les autres approches sur ce type de problème lorsque l'on considère un nombre de dimensions limité. Plus d'itérations seraient sans doute nécessaires à l'apprentissage  des paramètres lorsque $d>10$.
      
              \begin{figureth}    
        \includegraphics[width=0.8\linewidth]{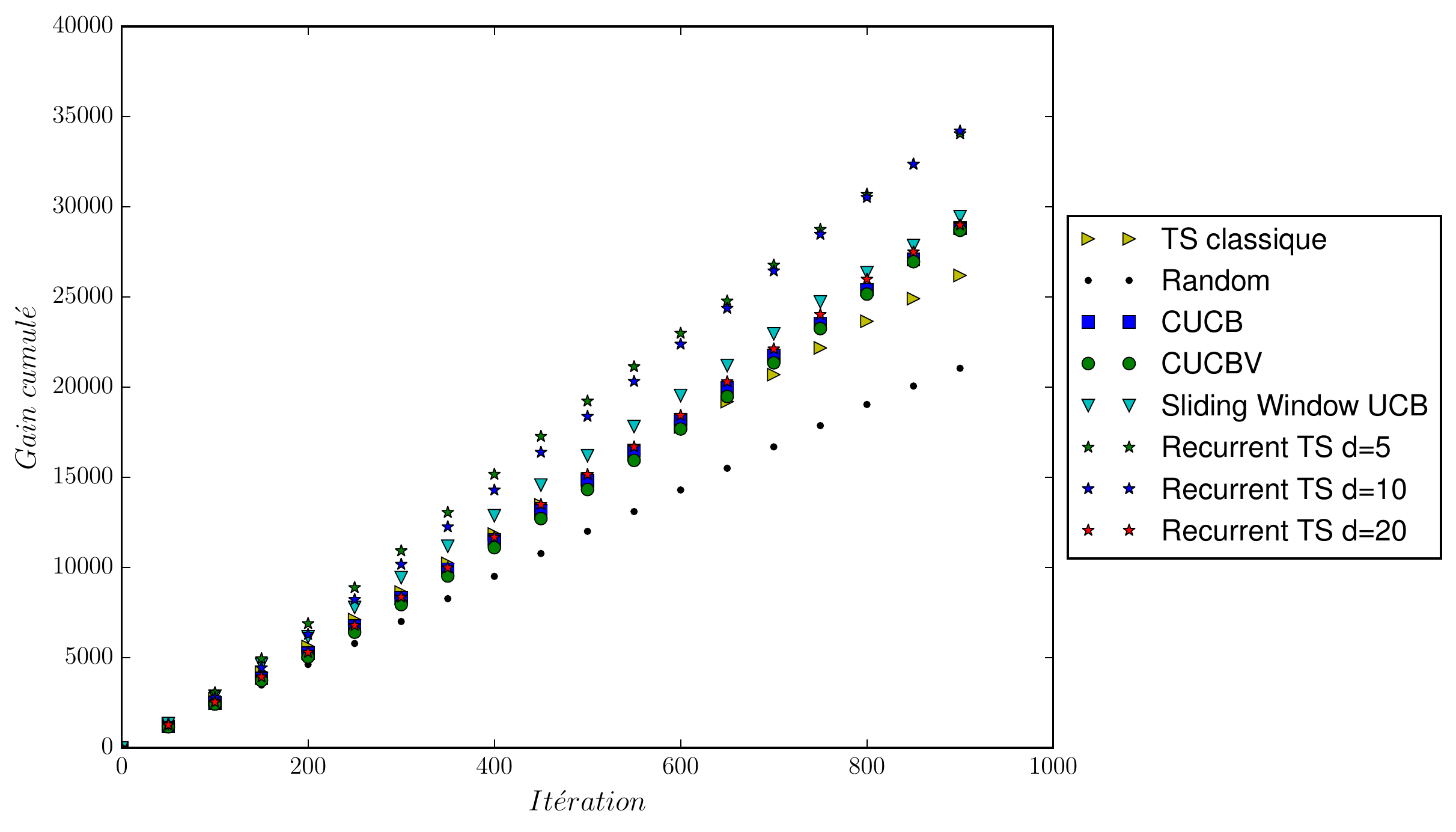}
      \caption[Gain cumulé  en fonction du temps pour le scénario de collecte sur le jeu de données \textit{OlympicGames} avec le nouveau modèle de récompense.]{Gain cumulé en fonction du temps pour le scénario de collecte sur le jeu de données \textit{OlympicGames} avec le nouveau modèle de récompense. }  
      \label{fig:resisTalkOGT1800K500}
      \end{figureth}

\section{Conclusion}

Dans ce chapitre, nous avons proposé un premier modèle de bandit récurrent dans lequel chaque action a la possibilité d'agir sur les autres d'une itération à l'autre, en vue de modéliser des phénomènes d'influence entre utilisateurs. Nous avons proposé un algorithme de Thompson sampling  pour résoudre cette tâche, basé sur une approximation variationnelle des distributions des paramètres du modèle. Nous avons ensuite étudié un modèle utilisant une représentation intermédiaire cachée, au sein de laquelle on modélise des transitions entre états successifs, en vue de capter des dépendances temporelles plus complexes. Nous avons montré que ce dernier offre en outre l'avantage de pouvoir être utilisé lorsque le nombre d'actions est élevé. Les diverses expérimentations effectuées sur données artificielles ont révélé l'efficacité des diverses méthodes. En particulier, lorsque le nombre d'actions devient grand, l'utilisation du modèle comportant des représentations cachées est en mesure de capter divers types de structures au sein des récompenses. Ce modèle semble en outre bien adapté à la détection de structures périodiques. Nous avons ensuite expérimenté notre méthode sur des données réelles de réseaux sociaux. Les mauvaises performances de cette approche en considérant les récompenses utilisées dans les chapitres précédent nous ont menés à conduire une étude plus précise afin d'apporter des éléments explicatifs. Finalement, une modification des récompenses considérées a permis de montrer la validité de notre approche dans un contexte de collecte d'information en temps réel. Pour améliorer les performances dans un cadre plus large, il serait envisageable d'inclure dans le modèle une prise en compte de contextes observés (tels que dans le chapitre précédent), afin de disposer de plus d'information pour estimer les données manquantes du problème de collecte.

	\chapter{Conclusions et perspectives}

\section{Conclusions}
 
Au cours de ce travail de thèse, nous avons étudié le problème de la collecte de données en temps réel dans les médias sociaux. En raison des différentes limitations imposées par ces médias, mais aussi de la quantité très importante de données, nous sommes partis du principe que la collecte de la totalité des contenus produits  n'est pas envisageable. Par conséquent, pour être en mesure de récolter des informations pertinentes, relativement à un besoin prédéfini, il est nécessaire de se focaliser sur un sous-ensemble des données existantes. En considérant chaque utilisateur comme une source de données pouvant être écoutée à chaque itération d'un processus de capture, nous avons modélisé la tâche de collecte comme un problème de bandit, dans lequel la qualité de l'information produite par un utilisateur correspond à une récompense. La notion de qualité peut être définie de diverses façons, selon le type d'information devant être récolté, et dépend de chaque application. Afin de cadrer l'étude, nous avons proposé divers modèles de récompense prenant en compte à la fois la thématique, mais aussi les réactions suscitées (\textit{Retweet} par exemple) par un contenu pour en mesurer sa qualité.

Nous avons proposé plusieurs modèles de bandit visant à identifier en temps réel les utilisateurs les plus pertinents. Dans une première contribution (chapitre \ref{ModeleStationnaire}), le cas du bandit dit stochastique, dans lequel chaque utilisateur est associé à une distribution de probabilité stationnaire, a été étudié. Ce travail a permis de montrer la pertinence d'utiliser le formalisme du bandit pour notre tâche de collecte en temps réel. De plus, nous avons proposé un nouvel algorithme ainsi qu'une analyse théorique du regret associé. Par la suite, nous avons étudié deux modèles de bandit contextuel, l'un stationnaire (chapitre \ref{ModeleContextuelFixe}) et l'autre non stationnaire (chapitre \ref{ModeleContextuelVariable}), dans lesquels l'utilité de chaque utilisateur peut être estimée de façon plus efficace en supposant une certaine structure, permettant ainsi de mutualiser l'apprentissage. En particulier, la première approche introduit la notion de profil, qui correspond au contenu moyen produit par chaque compte. Ces derniers n'étant pas accessibles, à cause des contraintes imposées par les médias sociaux, nous avons défini un nouveau problème de bandit dans lequel l'agent décisionnel doit effectuer une double exploration, mêlant incertitude sur les profils et incertitude sur les paramètres de régression considérés. Une analyse théorique de l'algorithme associé nous a permis de qualifier une borne supérieure de son regret. La seconde approche prend en compte l'activité d'un utilisateur à un instant donné pour prédire son comportement futur. Dans cette approche, une méthode d'inférence variationnelle a été utilisée afin de prendre en compte les limitations des APIs, et permettant une estimation des contextes manquants. Pour finir, nous nous sommes intéressés à la modélisation de dépendances temporelles entres  utilités des différents utilisateurs. Deux types d'approches ont été proposées: un premier modèle considérant des dépendances linéaires entre récompenses de périodes d'écoute successives, et un second, faisant intervenir des hypothèses de transitions entre des états latents du système, afin de capturer les variations d'utilité des différents comptes. Pour chacune des approches étudiées, nous avons mené des expérimentations sur des données artificielles, mais aussi sur divers jeux de données réelles provenant de Twitter. Les modèles proposés aux chapitres \ref{ModeleStationnaire}, \ref{ModeleContextuelFixe} et \ref{ModeleContextuelVariable}, en modélisant les dynamiques en jeu de plus en plus finement, ont permis d'offrir des performances croissantes sur les jeux de données et modèles de récompenses proposés. Pour le modèle récurrent du chapitre \ref{ModeleRelationnel}, nous avons défini un modèle de récompense alternatif, répondant mieux aux hypothèses du modèle, afin de valider cette approche sur données réelles. 

\section{Perspectives}

Les travaux effectués dans cette thèse ouvrent la voie à différentes perspectives, dont nous en énumérons certaines ci-dessous:	

\begin{itemize}

\item \textbf{Modèle hybride: } Il aurait été intéressant de combiner l'ensemble des modèles proposés dans un unique modèle, prenant en compte à la fois des profils moyens, des contextes instantanés et des dépendances temporelles. Concrètement, l'utilité espérée d'un utilisateur à un instant donné pourrait se modéliser par une fonction des différentes composantes de chaque modèle. Dans ces conditions, l'élaboration d'un modèle probabiliste gaussien (du même type que ceux proposés aux chapitres \ref{ModeleContextuelVariable} et \ref{ModeleRelationnel}) permettrait de définir un algorithme de Thompson sampling utilisant des approximations variationnelles des véritables distributions \textit{a posteriori} des paramètres. Ce type de modèle, bien que nécessitant probablement des approximations supplémentaires pour l'apprentissage des paramètres, permettrait de modéliser d'une façon plus réaliste le comportement d'un utilisateur sur un média social, et donc d'améliorer les performances du processus de collecte.

\item \textbf{Modèle récurrent: }Le modèle récurrent linéaire proposé dans le dernier chapitre pourrait être étendu en utilisant des relations temporelles plus complexes d'un pas de temps à l'autre, à l'instar des nombreux travaux existants sur les réseaux de neurones récurrents. Ceux-ci permettraient en outre de capter des relations plus abstraites, qu'un modèle linéaire peut difficilement modéliser. La difficulté majeure d'appliquer ces approches à des problèmes de bandit est double. D'une part, une grande partie de l'information permettant d'effectuer un apprentissage efficace est manquante, de par la nature du problème qui ne permet d'observer qu'un sous-ensemble restreint de la donnée. D'autre part, pour répondre au besoin d'exploration du processus de décision, il est nécessaire de maintenir soit des intervalles de confiance, soit des distributions sur les différents paramètres, afin de dériver les politiques respectivement optimistes et de Thompson sampling associées. Ceci peut s'avérer complexe lorsque des réseaux de neurones sont impliqués en raison des formes prises par les différentes fonctions d'activation. Un certain nombre de travaux, utilisant entre autres des méthodes variationnelles, existent à ce sujet et permettent d'approximer les distributions des paramètres (voir \cite{varAutoEnco} par exemple). Nous pensons que l'étude de ces méthodes pour des problèmes de bandit avec récompenses non stationnaires possédant des corrélations temporelles constituerait un travail intéressant. 

\item \textbf{Modèles de récompenses alternatifs: }Finalement, les récompenses utilisées tout au long de ce manuscrit sont relativement simples et pourraient également être étendues en vue de récolter des informations de nature plus complexe. Dans cette optique, il aurait été intéressant d'utiliser des fonctions de récompenses prenant en compte la diversité des contenus produits par les utilisateurs écoutés (voir \cite{banditCtxtCombLinUCB} pour des exemples de fonctions). La principale difficulté d'utiliser ce type de fonction vient de la combinatoire complexe qu'elles engendrent. En effet, la récompense d'un ensemble d'actions n'est alors plus égale à la somme des récompenses des actions individuelles, ce qui ne permet plus d'utiliser les algorithmes proposés dans ce manuscrit directement.


\end{itemize}

	\begin{appendix}
		\pagenumbering{Roman}
			\chapter*{Annexes}

\addcontentsline{toc}{chapter}{Annexes}
\renewcommand{\thesection}{\Alph{section}}

\section{Liste des publications}
\label{listPublis}

\begin{itemize}
\item \bibentry{ecml16}

\item \bibentry{icwsm16}

\item \bibentry{coria16}

\item \bibentry{dn16}

\item \bibentry{ictai15}

\item \bibentry{icwsm15}

\item \bibentry{coria15}

\item \bibentry{dn15}
\end{itemize}

\section{Preuve borne supérieure du regret pour l’algorithme \texttt{CUCBV}}
\label{PreuveRegCUCBV}

Nous rappelons les notations suivantes:

\begin{itemize}
\item $\mathcal{K}$ l'ensemble complet des $K$ actions possibles;
\item $\mu_{i}$ l'espérance de la récompense associée à l'action $i$;
\item $\sigma_{i}$ l'écart type de la récompense associée à l'action $i$;
\item On ordonne les actions de la façon suivante: $\forall i, \mu_{i} > \mu_{i+1}$;
\item $\mathcal{K}^{\star}$ est l'ensemble des $k$ actions ayant la plus forte espérance: $\mathcal{K}^{\star}=\left\{1,..,k\right\}$; 
\item $\overline{\mu}^{\star}$ est la moyenne des espérances des actions dans $\mathcal{K}^{\star}$, $\overline{\mu}^{\star}=\dfrac{1}{k}\sum\limits_{i=1}^{k}\mu_{i}$
\item $\Delta_i$ est défini comme la différence entre $\overline{\mu}^{\star}$, la moyenne des espérances des actions dans $\mathcal{K}^{\star}$, et $\mu_{i}$: $\Delta_i = \overline{\mu}^{\star}-\mu_{i}$; 
\item $\underline{\mu}^{\star}$ est la plus faible espérance dans $\mathcal{K}^{\star}$: $\underline{\mu}^{\star}=\min\limits_{i\in\mathcal{K}^{\star} }\mu_{i}=\mu_{k}$ 
\item $\delta_{i}$ correspond à la différence entre $\underline{\mu}^{\star}$, la plus faible espérance dans $\mathcal{K}^{\star}$, et $\mu_i$: $\delta_{i}=\underline{\mu}^{\star} - \mu_i$;
\item $N_{i,t}$ est le nombre de fois qu'une action $i$ a été choisie dans les $t$ premiers pas de temps.
\item $\hat{\mu}_{i,t}$ est la moyenne empirique associée à l'action $i$ après dans les $t$ premiers pas de temps;
\item $V_{i,t}$ est la variance empirique associée à l'action $i$ après dans les $t$ premiers pas de temps;
\end{itemize}

\subsection{Expression du regret}
\label{PreuveRegCUCBVFirstPart}

 En utilisant l'ordonnancement des actions selon leur espérance on a:

\begin{equation*}
\hat{R}_{T}  = \mathbb{E}\left[\sum\limits_{t=1}^{T} \left(\sum\limits_{j=1}^k \mu_{j}-\sum\limits_{i \in \mathcal{K}_t} \mu_i \right)\right]
=\mathbb{E}\left[T\sum\limits_{j=1}^k\mu_{j} -\sum\limits_{i=1}^{K} N_{i,T}  \mu_{i}\right]
\end{equation*}


De plus, indépendamment de l'action choisie:
 \begin{equation*}
 \sum_{i=1}^{K} N_{i,T} =kT
 \end{equation*}
 
 Car on sélectionne exactement $k$ actions distinctes à chaque itération. Donc:
 
  \begin{align*}
 \hat{R}_{T}&=\mathbb{E}\left[T\sum\limits_{j=1}^k\mu_{j} -\sum\limits_{i=1}^{K} N_{i,T}  \mu_{i}\right]\\
 			&=\mathbb{E}\left[\sum_{i=1}^{K} \dfrac{N_{i,T}}{k}\sum\limits_{j=1}^k\mu_{j} -\sum\limits_{i=1}^{K} N_{i,T}  \mu_{i}\right]\\
 		&=\mathbb{E}\left[\sum_{i=1}^{K}  N_{i,T}\left( \sum_{j=1}^k \frac{\mu_{j}}{k} -\mu_{i}\right)\right]\\
 			&=\sum\limits_{i=1}^{K} \mathbb{E}[N_{i,T}] \Delta_i
  \end{align*}

\subsection{Borne du regret}
\label{PreuveRegCUCBVSecondPart}

L'étude de la borne supérieure du regret cumulé moyen commence par la séparation des contributions des actions optimales et des actions non optimales: 

  \begin{align*}
  \hat{R}_{T}=\sum\limits_{i=1}^{K} \mathbb{E}[N_{i,T}] \Delta_i=\sum\limits_{i=1}^{k} \mathbb{E}[N_{i,T}] \Delta_i + \sum\limits_{i=k+1}^{K} \mathbb{E}[N_{i,T}] \Delta_i
    \end{align*}

Etant donné que l'on ne peut pas choisir une action plusieurs fois dans un même pas de temps, nous savons que $\mathbb{E}[N_{i,T}]\leq T$ d'où:
 
 \begin{align*}
\hat{R}_{T} \leq T \sum\limits_{i=1}^{k} \Delta_i + \sum\limits_{i=k+1}^{K} \mathbb{E}[N_{i,T}] \Delta_i
\end{align*}

En remarquant aussi que $\sum_{i=1}^{k}\Delta_i = 0$, nous obtenons une première borne supérieure du regret ne dépendant que des actions sous-optimales:

 \begin{align*}
 \hat{R}_{T} \leq  \sum\limits_{i=k+1}^{K} \mathbb{E}[N_{i,t}] \Delta_i
 \end{align*}

Considérant l'algorithme \texttt{CUCBV}, avec l'hypothèse que toutes les actions sont connues on sait qu'après $t_{0}=\lceil K/k \rceil$, chaque action est choisie au moins une fois (ceci est dû  à cause de l'initialisation des scores à $+ \infty$).

Dans la suite, on note le score du bras $i$ à l'intant $t$ de la façon suivante:

 \begin{align*}
 G_{i,N_{i,t-1},t}=\widehat{\mu}_{i,t-1}+\sqrt{\frac{2a\log(t) V_{i,t-1}}{N_{i,t-1}}}+3bc\frac{\log(t)}{N_{i,t-1}}
 \end{align*}
 
Afin de limiter les difficultés de notations, on se limite au cas où $a=c=1$. En pratique, on pourrait effectuer la démonstration dans un cadre plus générique avec d'autres constantes (c.f \cite{banditUCBV}).

Donc à chaque pas de temps $t\geq t_{0} $, pour les actions sélectionnées on obtient:

\begin{align*}
\forall i \in \mathcal{K}_{t}:N_{i,t-1}>0, \exists j \in \mathcal{K}^{\star} ~\text{tel que}~ N_{j,t-1}>0 ~\text{et}~  G_{i,N_{i,t-1},t} \geq G_{j,N_{j,t-1},t}
\end{align*}

Etudions maintenant le terme $N_{i,T}$. Soit $u>1$:

\begin{align*}
N_{i,T}&=\sum\limits_{t=1}^{T} \mathbf{1}_{\left\{i \in \mathcal{K}_t\right\}} \\
	&\leq 1 + \sum\limits_{t=1+t_{0}}^{T} \mathbf{1}_{\left\{i \in \mathcal{K}_t\right\}} \\
	&\leq 1+u+\sum\limits_{t=1+u+t_{0}}^{T} \mathbf{1}_{\left\{N_{i,t-1}\geq u ; \ i \in \mathcal{K}_t\right\}}\\
    &\leq 1+u+\sum\limits_{t=1+u+t_{0}}^{T} \mathbf{1}_{\left\{
  N_{i,t-1}\geq u ; \exists j \in \mathcal{K}^{\star} ~\text{tel que}~ N_{j,t-1}>0 ~\text{et}~
   G_{i,N_{i,t-1},t} \geq G_{j,N_{j,t-1},t}
\right\}}
\end{align*}

Par ailleurs, pour tout $\gamma \in \mathbb{R}$, on a:

\begin{align*}
&\mathbf{1}_{\left\{
  N_{i,t-1}\geq u ; \exists j \in \mathcal{K}^{\star} ~\text{tel que}~ N_{j,t-1}>0 ~\text{et}~
   G_{i,N_{i,t-1},t} \geq G_{j,N_{j,t-1},t}
\right\}}\\
&\leq \mathbf{1}_{\left\{ \exists s: u\leq s\leq t-1 ~\text{tel que}~  G_{i,s,t} > \gamma \right\}} + \mathbf{1}_{\left\{ \exists j \in \mathcal{K}^{\star}, \exists s_{j}: 1\leq s_{j}\leq t-1 ~\text{tel que}~ G_{j,s_{j},t} \leq \gamma \right\}}
\end{align*}

En choisissant $\gamma=\underline{\mu}^{\star}$ on obtient:

\begin{align*}
\mathbb{E}[N_{i,T}] &\leq 1+u+\sum\limits_{t=u+1+t_{0}}^{T}  \mathbb{P}(\exists s: u\leq s\leq t-1 ~\text{tel que}~  G_{i,s,t} > \underline{\mu}^{\star}) \\ &\qquad \qquad\qquad+\sum\limits_{t=u+1+t_{0}}^{T}\mathbb{P}(\exists j \in \mathcal{K}^{\star}, \exists s_{j}: 1\leq s_{j}\leq t-1 ~\text{tel que}~ G_{j,s_{j},t} \leq \underline{\mu}^{\star})\\
& \leq 1+u+\sum\limits_{t=u+1+t_{0}}^{T}  \sum\limits_{s=u}^{t-1} \mathbb{P}(G_{i,s,t} > \underline{\mu}^{\star}) + \sum\limits_{t=u+1+t_{0}}^{T}\sum\limits_{j=1}^{k} \mathbb{P}(\exists s_{j}: 1\leq s_{j}\leq t-1 ~\text{tel que}~ G_{j,s_{j},t} \leq \underline{\mu}^{\star})
\end{align*}

Dans la suite, nous allons majorer chacun de ces deux termes. Nous utiliserons les notations suivantes:

\begin{itemize}
\item $\mathbb{P}_{j,t}^{1} = \mathbb{P}(\exists s_{j}: 1\leq s_{j}\leq t-1 ~\text{tel que}~ G_{j,s_{j},t} \leq \underline{\mu}^{\star})$ avec $j \in \mathcal{K}^{\star}$
\item $\mathbb{P}_{i,s,t}^{2}=\mathbb{P}(G_{i,s,t} > \underline{\mu}^{\star})$
\end{itemize}

\subsubsection{Contribution du premier terme}

Etant donné que $\forall j \in \mathcal{K}^{\star}$ on a $\underline{\mu}^{\star}=\mu_{k} \leq \mu_{j}$, on obtient directement:

\begin{align*}
\mathbb{P}_{j,t}^{1}
\leq \mathbb{P}(\exists s_{j}: 1\leq s_{j}\leq t-1 ~\text{tel que}~ G_{j,s_{j},t} \leq \mu_{j})
\end{align*}

En utilisant les résultats du \textbf{Théorème 1} dans \cite{banditUCBV}, on obtient: 

\begin{align*}
\mathbb{P}(G_{j,s_{j},t} \leq \mu_{j}) \leq \beta(\log(t),t)
\end{align*}

Simultanément $\forall s_{j} \in \left\{1,2, ...,t\right\}$, où $\beta(\log(t),t)$ est de l'ordre de $e^{-\log(t)}=1/t$. 

Donc on peut majorer $\mathbb{P}_{j,t}^{1}$ par:

\begin{align*}
\mathbb{P}_{j,t}^{1} \leq  \leq \beta(\log(t),t)
\end{align*}

Finalement:
 
\begin{align*}
\sum\limits_{t=u+1+t_{0}}^{T}\sum\limits_{j=1}^{k} \mathbb{P}(\exists s_{j}: 1\leq s_{j}\leq t-1 ~st~ G_{j,s_{j},t} \leq \underline{\mu}^{\star})\leq k \sum\limits_{t=u}^{\top} \beta(\log(t),t)
\end{align*}

\subsubsection{Contribution du second terme}

D'après la preuve du \textbf{Théorème 3} dans \cite{banditUCBV} on peut conclure qu'en prenant $u$ le plus petit entier inférieur à $\lceil 8\left(\dfrac{\sigma_{i}^{2}}{\delta_{i}^{2}} + \dfrac{2}{\delta_{i}}\right)\log(T) \rceil$, pour tout $s\leq u \leq t-1$ et $t \geq 2$:

\begin{align}
\sqrt{\frac{2\ln(t) (\sigma_{i}^2+\delta_{i}/2)}{s}} +3\frac{\log(t)}{s} \leq  \dfrac{\delta_{i}}{2}
\end{align}

Par ailleurs, pour tout $s\geq u$ et $t\geq 2,$ en utilisant l'équation précédente:

\begin{align*}
\mathbb{P}(G_{i,s,t} > \underline{\mu}^{\star})
&=\mathbb{P}\left( \widehat{\mu}_{i,s}+\sqrt{\frac{2\ln(t) \widehat{\sigma}_{i,s}^{2}}{s}}+3\frac{\ln(t)}{s} > \underline{\mu}^{\star}\right)\\
&=\mathbb{P}\left( \widehat{\mu}_{i,s}+\sqrt{\frac{2\ln(t) \widehat{\sigma}_{i,s}^{2}}{s}}+3\frac{\ln(t)}{s} > \delta_{i}+\mu_{i}\right)\\
&\leq \mathbb{P}\left( \widehat{\mu}_{i,s}+\sqrt{\frac{2\ln(t) (\sigma_{i}^2+\delta_{i}/2)}{s}}+3\frac{\ln(t)}{s} > \delta_{i}+\mu_{i}\right)
+ \mathbb{P}\left(\widehat{\sigma}_{i,s}^{2} \geq \sigma_{i}^2+\delta_{i}/2\right)\\
&\leq \mathbb{P}\left( \widehat{\mu}_{i,s}-\mu_{i} >\delta_{i}/{2}\right)
+ \mathbb{P}\left(\widehat{\sigma}_{i,s}^{2} - \sigma_{i}^2 \geq \delta_{i}/2\right) \leq 2e^{-s\delta_{i}^{2}/(8\sigma_{i}^{2}+4\delta_{i}/3)} \\
\end{align*}

Comme précisé dans \cite{banditUCBV}, à la dernière étape, l'inégalité de Bernstein est appliquée deux fois.
En sommant ces probabilités, on arrive à:

\begin{align*}
\sum\limits_{s=u}^{t-1} \mathbb{P}(G_{i,s,t} > \underline{\mu}^{\star})
\leq 2 \sum\limits_{s=u}^{t-1} e^{-s\delta_{i}^{2}/(8\sigma_{i}^{2}+4\delta_{i}/3)}
\leq \left(\dfrac{24\sigma_{i}^{2}}{\delta_{i}^{2}} + \dfrac{4}{\delta_{i}}\right)\dfrac{1}{T}
\end{align*}

Donc: 
\begin{align*}
\sum\limits_{t=u+1+t_{0}}^{T}\sum\limits_{s=u}^{t-1} \mathbb{P}(G_{i,s,t} > \underline{\mu}^{\star})
\leq \sum\limits_{t=1}^{T} \left(\dfrac{24\sigma_{i}^{2}}{\delta_{i}^{2}} + \dfrac{4}{\delta_{i}}\right)\dfrac{1}{T}
\leq \left(\dfrac{24\sigma_{i}^{2}}{\delta_{i}^{2}} + \dfrac{4}{\delta_{i}}\right)
\end{align*}

\subsubsection{Conclusion}

\begin{align*}
\mathbb{E}[N_{i,T}] \leq
  1 +  \left(1+8\left(\dfrac{\sigma_{i}^{2}}{\delta_{i}^{2}} + \dfrac{2}{\delta_{i}}\right)\right)\log(T) 
+\left(\dfrac{24\sigma_{i}^{2}}{\delta_{i}^{2}} + \dfrac{4}{\delta_{i}}\right)
+k\sum\limits_{t=u}^{T} \beta(\log(t),t)
\end{align*}

Et étant donné que $\dfrac{\sigma_{i}^{2}}{\delta_{i}^{2}} + \dfrac{2}{\delta_{i}}\geq 2 $:

\begin{align*}
\mathbb{E}[N_{i,T}] \leq 
  1 +  \left(1+8\left(\dfrac{\sigma_{i}^{2}}{\delta_{i}^{2}} + \dfrac{2}{\delta_{i}}\right)\right)\log(T) 
+\left(\dfrac{24\sigma_{i}^{2}}{\delta_{i}^{2}} + \dfrac{4}{\delta_{i}}\right)
+k\sum\limits_{t=16\log(T)}^{T} \beta(\log(t),t)
\end{align*}

On arrive à:

$\hat{R}_{T} \leq
 \sum\limits_{i \notin \mathcal{K}^{\star}} \Big(   1 +  \left(1+8\left(\dfrac{\sigma_{i}^{2}}{\delta_{i}^{2}} + \dfrac{2}{\delta_{i}}\right)\right)\log(T)
				+\left(\dfrac{24\sigma_{i}^{2}}{\delta_{i}^{2}} + \dfrac{4}{\delta_{i}}\right) +k\sum\limits_{t=16\log(T)}^{T} \beta(\log(t),t) \Big)\Delta_i$

En utilisant le fait que $\beta(\log(t),t)=\alpha \frac{1}{t}$ et $\sum\limits_{t=1}^{n}\frac{1}{T}\leq \log(T)+\gamma+\frac{1}{2T} \leq \log(T)+\gamma+\frac{1}{2}$, où $\alpha$ est un réel, et $\gamma$ est la constante d'Euler, on obtient finalement:




\begin{align*}
\hat{R}_{T} \leq \log(T)\sum\limits_{i \notin \mathcal{K}^{\star}} \left(C+8\left(\dfrac{\sigma_{i}^{2}}{\delta_{i}^{2}} + \dfrac{2}{\delta_{i}}\right)\right)\Delta_i + D
\end{align*}

Avec $C=1+k\alpha$ et $D=\sum\limits_{i \notin \mathcal{K}^{\star}}\left(1+\dfrac{24\sigma_{i}^{2}}{\delta_{i}^{2}} + \dfrac{4}{\delta_{i}}+k\alpha\left(\gamma+\frac{1}{2}\right)\right)\Delta_i$

\section{Borne supérieure du regret pour l’algorithme \texttt{SampLinUCB}}

\subsection{Preuve de la proposition \ref{SubGaussianProp}}
\label{PreuveSubGauss}
Les deux lemmes suivants viennent de la définition des variables aléatoires sous-gaussiennes.
\begin{lem}
\label{lem1}
Soit $X$ une variable aléatoire sous-gaussienne centrée,  alors les points suivants sont équivalents:
\begin{itemize}
\item Condition de Laplace  : $\exists R>0, \forall \lambda \in \mathbb{R}, \mathbb{E}[e^{\lambda X}]\leq e^{R^{2}\lambda^{2}/2}$
\item Queue sous-gaussienne: $\exists R>0, \forall \gamma>0, P(|X|\geq \gamma)\leq 2e^{-\gamma^{2}/(2R^{2})} $
\end{itemize}
\end{lem}

\begin{lem}
\label{lem2}
Soient $X_{1}$ et $X_{2}$ deux variables aléatoires sous-gaussiennes de constante respective $R_{1}$ et $R_{2}$. Soient $\alpha_{1}$ et $\alpha_{2}$ deux réels. Alors la VA $\alpha_{1}X_{1}+\alpha_{2}X_{2}$ est aussi sous-gaussienne de constante $\sqrt{\alpha_{1}^{2}R_{1}^{2}+\alpha_{2}^{2}R_{2}^{2}}$. 

\end{lem}

\begin{lem}
\label{lem3}
Supposons que pour tout $i$, et à chaque temps $t$, les échantillons $x_{i,t} \in \mathbb{R}^{d}$ sont iid de moyenne $\mu_{i} \in \mathbb{R}^{d}$, et que $||x_{i,t}||\leq L$ et $||\beta||\leq S$ . Alors pour tout $i$, et à chaque temps $t$, $\epsilon_{i,t}^{\top}\beta$ est sous-gaussien, de constante $\dfrac{LS}{\sqrt{n_{i,t}}}$. On rappelle que $\epsilon_{i,t}=\mu_{i}-\hat{x}_{i,t}$.
\end{lem}

\begin{preuve}

En utilisant l'inégalité de Cauchy-Schwarz, pour tout $i$, et à chaque temps $t$ on a: $|x_{i,t}^{\top}\beta|\leq ||\beta||||\hat{x}_{i,t}||\leq LS$. Donc, étant donné que pour tout  $i$, les  $x_{i,t}$ sont  iid et  $\mathbb{E}[x_{i,t}]=\mu_{i}$, on peut appliquer l'inégalité de Hoeffding à la variable aléatoire $x_{i,t}^{\top}\beta$ de moyenne $\mu_{i}^{\top}\beta$.

\begin{align*}
\forall \gamma>0,\mathbb{P}\left( |\beta^{\top}\hat{x}_{i,t}-\beta^{\top}\mu_{i}| > \gamma \right) = \mathbb{P}\left( |\beta^{\top}\epsilon_{i,t}| > \gamma \right) \leq 2e^{-\frac {n_{i,t}\gamma^{2}}{2S^{2}L^{2}}}
\end{align*}
En appliquant le lemme  \ref{lem1} on obtient le résultat souhaité avec $T_{i,t}\gamma^{2}/(2S^{2}L^{2}) = 1 /(2R^{2})$
\end{preuve}

On utilise finalement le lemme  \ref{lem2} avec la somme des $\beta^{\top}\epsilon_{i,t}$ et $\eta_{i,s}$ pour prouver la proposition \ref{SubGaussianProp}, qui établit que la variable aléatoire  $\epsilon_{i,t}^{\top}\beta+\eta_{i,s}$ est conditionnellement sous-gaussien de constante 
$R_{i,t}=\sqrt{R^{2}+\dfrac{L^{2}S^{2}}{n_{i,t}}}$.

\subsection{Preuve du théorème \ref{ConfInt}}
\label{PreuveConfInt}

Pour alléger les notations, on enlève la dépendance en $t$ dans $A$ et $X$. On a:
\begin{align*}
\hat{\beta}_{t-1} &= \arg\min_{\beta} \sum\limits_{s=1}^{t-1} \dfrac{1}{R_{i_{s},t}}(\beta^{\top}\hat{x}_{i_{s},t}- r_{i_{s},s})^{2} + \lambda ||\beta||^{2} \\
				 &= (X^{\top}AX+\lambda I)^{-1}X^{\top}AY \\
				 &= (X^{\top}AX+\lambda I)^{-1}X^{\top}A(X\beta+\eta^{'}) \\
				 &= (X^{\top}AX+\lambda I)^{-1}X^{\top}A\eta^{'} + (X^{\top}AX+\lambda I)^{-1} (X^{\top}AX+\lambda I)\beta \\ & \ \ \ - (X^{\top}AX+\lambda I)^{-1}\lambda I \beta\\
				 &= (X^{\top}AX+\lambda I)^{-1}X^{\top}A\eta^{'} + \beta - \lambda (X^{\top}AX+\lambda I)^{-1} \beta
\end{align*}

En utilisant une méthode identique à celle de \cite{OFUL} on arrive a:
\begin{align*}
||\hat{\beta}_{t-1}-\beta||_{V_{t-1}} &\leq ||X^{\top}A\eta^{'}||_{V_{t-1}^{-1}}+\lambda||\beta||_{V_{t-1}^{-1}}
\end{align*}

Avec $V_{t-1}=\lambda I+ X^{\top}AX$, qui est bien définie positive puisque $\lambda >0$. Etant donné que  $||\beta||\leq S$  et  $||\beta||_{V_{t-1}^{-1}}^{2}\leq ||\beta||^{2}/\lambda_{min}(V_{t-1}) \leq ||\beta||^{2}/\lambda$ on arrive a:

\begin{align*}
||\hat{\beta}_{t-1}-\beta||_{V_{t-1}} &\leq ||X^{\top}A\eta^{'}||_{V_{t-1}^{-1}}+ \sqrt{\lambda}S
\end{align*}

En utilisant le théorème 1 de \cite{OFUL} et le fait que  $\dfrac{\eta^{'}_{s}}{R_{i_{s},t}}$ est sous-gaussien de constante 1 (grâce à la proposition  \ref{SubGaussianProp}), pour tout $\delta>0$, avec une probabilité au moins égale à $1-\delta$, pour tout $t \geq 0$:

\begin{align*}
||X^{\top}A\eta^{'}||_{V_{t-1}^{-1}} &= ||\sum\limits_{s=1}^{t-1} \dfrac{\eta^{'}_{s}}{R_{i_{s},t}}\hat{x}_{i_{s},t}||_{V_{t-1}^{-1}} \\
								&\leq   \sqrt{2\log\left(\dfrac{det(V_{t-1})^{1/2}det(\lambda I)^{-1/2}}{\delta}\right)}\\
								&\leq  \sqrt{d\log\left(\dfrac{1+tL^{2}/\lambda}{\delta}\right)}
\end{align*}

Les principaux arguments de cette preuve viennent de la théorie des processus auto normalisés décrite dans \cite{PenaLaiShao2009}.

\subsection{Preuve du théorème \ref{ConfIntProfils}}
\label{PreuveConfIntProfils}
Etant donné l'hypothèse selon laquelle $||x_{i,t}||\leq L$, on peut appliquer Hoeffding à chaque dimension $j\in \left[1..d\right]$, de telle sorte que $|x_{i,t}^{j}|\leq L$:

\begin{equation*}
\forall \gamma>0: ~\mathbb{P}\left( |\hat{x}_{i,t}^{j}-\mu_{i}^{j}|  > \gamma/d \right) \leq 2e^{-\frac {n_{i,t}\gamma^{2}}{2L^{2}d^{2}}}
\end{equation*}

En utilisant le fait que  $||\hat{x}_{i,t}-\mu_{i}|| \leq \sum\limits_{i=1}^{d}|\hat{x}_{i,t}^{i}-\mu_{i}^{i}|$ la propriété de la borne uniforme:

\begin{equation*}
\mathbb{P}\left( ||\hat{x}_{i,t}-\mu_{i}||  \leq \gamma \right)\geq 1-2de^{-\frac {n_{i,t}\gamma^{2}}{2L^{2}d^{2}}}
\end{equation*}

Finalement en choisissant un $\gamma$ approprié, pour tout $i$ et tout temps $t>0$ avec une probabilité au moins égale à $1-\delta/t^{2}$:

\begin{equation*}
||\hat{x}_{i,t}-\mu_{i}|| \leq Ld\sqrt{\dfrac{2}{n_{i,t}}\log\left( \dfrac{2dt^{2}}{\delta}\right)}  = \rho_{i,t,\delta} 
\end{equation*}

\subsection{Preuve du théorème \ref{RegretSamplLinUCBGeneric}}
\label{preuveRegSampLinUCBGeneric}




\begin{lem}
\label{lem}
Pour tout $i$ et $t>0$ avec une probabilité au moins égale $1-\delta/t^{2}-\delta$:

\begin{align*}
0\leq \hat{\beta}_{t-1}^{\top}(\hat{x}_{i,t}+\overline{\epsilon}_{i,t}) + \alpha_{t-1} ||\hat{x}_{i,t}+\tilde{\epsilon}_{i,t}||_{V_{t-1}^{-1}}-\beta^{\top}\mu_{i} 
							& \leq 2\alpha_{t}||\hat{x}_{i,t}||_{V_{t-1}^{-1}} + 4\sqrt{d}(\alpha_{t-1}/\sqrt{\lambda}+S)\rho_{i,t,\delta} 
\end{align*}

\end{lem}

\begin{preuve}

Supposons que l'inégalité du théorème    \ref{ConfIntProfils}  est valide, alors:

\begin{itemize}
\item $||\hat{x}_{i,t}-\mu_{i}||_{V_{t-1}^{-1}} \leq ||\hat{x}_{i,t}-\mu_{i}||/\sqrt{\lambda} \leq \rho_{i,t,\delta}/\sqrt{\lambda}$ donc $||\mu_{i}||_{V_{t-1}^{-1}} \leq || \hat{x}_{i,t}||_{V_{t-1}^{-1}} + \rho_{i,t,\delta}/\sqrt{\lambda} = || \hat{x}_{i,t} + \tilde{\epsilon}_{i,t} ||_{V_{t-1}^{-1}}$, avec $\tilde{\epsilon}_{i,t}=  \rho_{i,t,\delta} \hat{x}_{i,t}/(\sqrt{\lambda}|| \hat{x}_{i,t}||_{V_{t-1}^{-1}})$.
\item $|\hat{\beta}_{t-1}^{\top}(\hat{x}_{i,t}-\mu_{i})|\leq \hat{\beta}_{t-1}^{\top}\overline{\epsilon}_{i,t}$, avec $\overline{\epsilon}_{i,t}=\rho_{i,t,\delta}\hat{\beta}_{t-1}/||\hat{\beta}_{t-1}||$.
\end{itemize}

En utilisant ces deux résultats et la propriété de la borne uniforme,  on peut montrer le lemme annoncé:

Première partie:
\begin{align*}
 & \hat{\beta}_{t-1}^{\top}(\hat{x}_{i,t}+\overline{\epsilon}_{i,t}) + \alpha_{t-1} ||\hat{x}_{i,t}+\tilde{\epsilon}_{i,t}||_{V_{t-1}^{-1}}-\beta^{\top}\mu_{i} \\
=& (\hat{\beta}_{t-1}-\beta)^{\top}\mu_{i} + \alpha_{t-1} ||\hat{x}_{i,t}+\tilde{\epsilon}_{i,t}||_{V_{t-1}^{-1}} - \hat{\beta}_{t-1}^{\top}(\mu_{i}-\hat{x}_{i,t})+\hat{\beta}_{t-1}^{\top}\overline{\epsilon}_{i,t}\\
\geq &-||\hat{\beta}_{t-1}-\beta||_{V_{t-1}}||\mu_{i}||_{V_{t-1}^{-1}}+ \alpha_{t-1} ||\hat{x}_{i,t}+\tilde{\epsilon}_{i,t}||_{V_{t-1}^{-1}} + \hat{\beta}_{t-1}^{\top}(\hat{x}_{i,t}-\mu_{i}+\overline{\epsilon}_{i,t})\\
\geq &-\alpha_{t-1}||\mu_{i}||_{V_{t-1}^{-1}}+ \alpha_{t-1} ||\hat{x}_{i,t}+\tilde{\epsilon}_{i,t}||_{V_{t-1}^{-1}} + \hat{\beta}_{t-1}^{\top}(\hat{x}_{i,t}-\mu_{i}+\overline{\epsilon}_{i,t})\\
\geq &-\alpha_{t-1} ||\hat{x}_{i,t}+\tilde{\epsilon}_{i,t}||_{V_{t-1}^{-1}}+ \alpha_{t-1} ||\hat{x}_{i,t}+\tilde{\epsilon}_{i,t}||_{V_{t-1}^{-1}} + \hat{\beta}_{t-1}^{\top}(\hat{x}_{i,t}-\mu_{i}+\overline{\epsilon}_{i,t})\\	
\geq & 0
\end{align*}

Seconde partie: en remarquant que $||\overline{\epsilon}_{i,t}||_{V_{t-1}^{-1}} \leq ||\overline{\epsilon}_{i,t}||/\sqrt{\lambda} = \rho_{i,t,\delta}/\sqrt{\lambda} $ et $||\tilde{\epsilon}_{i,t}||_{V_{t-1}^{-1}} = \rho_{i,t,\delta}/\sqrt{\lambda}$:

\begin{align*}
& \hat{\beta}_{t-1}^{\top}(\hat{x}_{i,t}+\overline{\epsilon}_{i,t}) + \alpha_{t-1} ||\hat{x}_{i,t}+\tilde{\epsilon}_{i,t}||_{V_{t-1}^{-1}}-\beta^{\top}\mu_{i} \\
=& (\hat{\beta}_{t-1}-\beta)^{\top}\mu_{i} + \alpha_{t-1} ||\hat{x}_{i,t}+\tilde{\epsilon}_{i,t}||_{V_{t-1}^{-1}} - \hat{\beta}_{t-1}^{\top}(\mu_{i}-\hat{x}_{i,t})+\hat{\beta}_{t-1}^{\top}\overline{\epsilon}_{i,t}\\
\leq & ||\hat{\beta}_{t-1}-\beta||_{V_{t-1}}||\mu_{i}||_{V_{t-1}^{-1}}+ \alpha_{t-1} ||\hat{x}_{i,t}+\tilde{\epsilon}_{i,t}||_{V_{t-1}^{-1}} + \hat{\beta}_{t-1}^{\top}(\hat{x}_{i,t}-\mu_{i}+\overline{\epsilon}_{i,t})\\
\leq  & 2\alpha_{t-1}||\hat{x}_{i,t}+\tilde{\epsilon}_{i,t}||_{V_{t-1}^{-1}} + 2 ||\hat{\beta}_{t-1}||_{V_{t-1}}||\overline{\epsilon}_{i,t}||_{V_{t-1}^{-1}}\\
\leq & 2\alpha_{t-1}||\hat{x}_{i,t}||_{V_{t-1}^{-1}} +  2\alpha_{t-1}||\tilde{\epsilon}_{i,t}||_{V_{t-1}^{-1}} +  2(\alpha_{t-1}+S\sqrt{\lambda}) ||\overline{\epsilon}_{i,t}||_{V_{t-1}^{-1}}\\
\leq &2\alpha_{t-1}||\hat{x}_{i,t}||_{V_{t-1}^{-1}} + 4(\alpha_{t-1}/\sqrt{\lambda}+S)\rho_{i,t,\delta}
\end{align*}

\end{preuve}

\begin{prop}
\label{lemApp}
Pour tout $t$, avec une probabilité au moins égale $1-\delta/t^{2}-\delta$, le regret instantané de l'algorithme \texttt{SampLinUCB}, noté $reg_{t}=\beta^{\top}\mu_{i^{\star}}-\beta^{\top}\mu_{i_{t}}$ est majoré par:

\begin{align*}
reg_{t} \leq  2\alpha_{t-1}||\hat{x}_{i_{t},t}||_{V_{t-1}^{-1}} + 4(\alpha_{t-1}/\sqrt{\lambda}+S)\rho_{i_{t},t,\delta} = reg_{t}^{(1)}+reg_{t}^{(2)}
\end{align*}
\end{prop}

\begin{preuve}
Etant donné la politique de sélection de \texttt{SampLinUCB} et la première inégalité de lemme précédent, on a pour tout $t$:

 $s_{i_{t},t}\geq s_{i^{\star},t} \geq \beta^{\top}\mu_{i^{\star}}$.
 
 D'autre part, la seconde inégalité du lemme précédent prouve que pour tout $t$

$s_{i_{t},t} \leq \beta^{\top}\mu_{i_{t}} + 2\alpha_{t-1}||\hat{x}_{i,t}||_{V_{t-1}^{-1}} + 4(\alpha_{t-1}/\sqrt{\lambda}+S)\rho_{i,t,\delta}$

Ce qui conclut la preuve.

\end{preuve}

\textbf{Preuve du théorème}

On utilise d'une part le fait que  $\sum\limits_{t=2}^{\infty}\delta_{t} = \delta(\pi^{2}/6-1)\leq \delta$ et la propriété de la borne uniforme, et d'autres parts le fait que dans  le théorème \ref{ConfInt} la borne est uniforme. Ainsi, avec une probabilité au moins égale à $1-2\delta$:

\begin{align*}
\sum\limits_{t=1}^{T} reg_{t}^{(2)} &\leq C +\sum\limits_{t=2}^{T}  4(\alpha_{t-1}/\sqrt{\lambda}+S)\rho_{i_{t},t,\delta}\\
&\leq C + \sum\limits_{t=2}^{T} 4(\alpha_{t-1}/\sqrt{\lambda}+S)Ld\sqrt{\dfrac{2}{n_{i,t}}\log\left( \dfrac{2dt^{2}}{\delta}\right)} \\							
& \leq C + 4Ld(\alpha_{T}/\sqrt{\lambda}+S)\sqrt{\log\left( \dfrac{2dT^{2}}{\delta}\right)} \sum\limits_{t=2}^{T} \dfrac{2}{\sqrt{n_{i_{t},t}}}
\end{align*}

D'autre part:

\begin{align*}
\sum\limits_{t=1}^{T} reg_{t}^{(1)} &\leq \sum\limits_{t=1}^{T} 2\alpha_{t-1}||\hat{x}_{i_{t},t}||_{V_{t-1}^{-1}}^{2} \\
									&\leq \sqrt{T\sum\limits_{t=1}^{T}4\alpha_{t-1}^{2}||\hat{x}_{i_{t},t}||_{V_{t-1}^{-1}}^{2}}\\
								    &\leq 2\alpha_{T} \sqrt{T\sum\limits_{t=1}^{T}||\hat{x}_{i_{t},t}||_{V_{t-1}^{-1}}^{2}}
\end{align*}

Il faut maintenant majorer le terme $\sum\limits_{t=1}^{T}||\hat{x}_{i_{t},t}||_{V_{t-1}^{-1}}^{2}$. 

Pour cela, on introduit la grandeur suivante:

\begin{equation*}
\nu_{i,t,\delta}=Ld\sqrt{2/(n_{i,t})\log\left( 2dT/\delta\right)}
\end{equation*}

En utilisant à nouveau Hoeffding, avec une probabilité au moins égale à $1-\delta/T$ on a pour $s \leq t-1$:

\begin{equation*}
||\hat{x}_{i_{s},t}|| \leq ||\mu_{i_{s}}||+\nu_{i_{s},t,\delta}
\end{equation*}

En définissant:

\begin{equation*}
\check{\epsilon}_{i,t}=\nu_{i,t,\delta} \mu_{i}/||\mu_{i}||
\end{equation*}

on a, pour $s\leq t-1$: 

\begin{equation*}
1/\sqrt{R_{i_{s},s}}||\mu_{i_{s}}-\check{\epsilon}_{i_{s},s}|| \leq 1/\sqrt{R_{i_{s},t}}||\hat{x}_{i_{s},t}||
\end{equation*}

On arrive ainsi à:

\begin{equation*}
V_{t-1}=\lambda I+\sum\limits_{s=1}^{t-1} \dfrac{1}{R_{i_{s},t}} \hat{x}_{i_{s},t}\hat{x}_{i_{s},t}^{\top}  \geq \lambda I + \sum\limits_{s=1}^{t-1}\dfrac{1}{R_{i_{s},s}} (\mu_{i_{s}}-\check{\epsilon}_{i_{s},s})(\mu_{i_{s}}-\check{\epsilon}_{i_{s},s})^{\top} = W_{t-1}
\end{equation*}

Ce qui signifie que pour tout vecteur $x$: $||x||_{V_{t-1}^{-1}} \leq ||x||_{W_{t-1}^{-1}}$.

On définit maintenant:

\begin{equation*}
\hat{\epsilon}_{i,t}=\nu_{i,t,\delta} \mu_{i}/(\sqrt{\lambda}||\mu_{i}||_{W_{t-1}^{-1}})
\end{equation*}

de telle sorte que pour $s\leq t-1$:

\begin{equation*}
||\hat{x}_{i_{s},t}||_{W_{t-1}^{-1}} \leq ||\mu_{i_{s}}+\hat{\epsilon}_{i_{s},s}||_{W_{t-1}^{-1}}
\end{equation*}
et
\begin{equation*}
||\hat{\epsilon}_{i_{s},s}||_{W_{t-1}^{-1}}=\nu_{i_{s},s,\delta}/\sqrt{\lambda}
\end{equation*}


Finalement, en utilisant la propriété de la borne uniforme et le fait que  $\sum\limits_{t=1}^{T} \dfrac{\delta}{T}=\delta$, avec une probabilité au moins égale à $1-\delta$:

\begin{align*}
\sum\limits_{t=1}^{T} ||\hat{x}_{i_{t},t}||_{V_{t-1}^{-1}}^{2}  &\leq  \sum\limits_{t=1}^{T} ||\hat{x}_{i_{t},t}||_{W_{t-1}^{-1}}^{2}\\ &\leq\sum\limits_{t=1}^{T} ||  \mu_{i_{t}} +\hat{\epsilon}_{i_{t},t}||_{W_{t-1}^{-1}}^{2} \leq \sum\limits_{t=1}^{T} ||  \mu_{i_{t}} +\hat{\epsilon}_{i_{t},t} -\check{\epsilon}_{i_{t},t}   +\check{\epsilon}_{i_{t},t} ||_{W_{t-1}^{-1}}^{2}\\ 
&\leq \sum\limits_{t=1}^{T} ||\mu_{i_{t}}-\check{\epsilon}_{i_{t},t}||_{W_{t-1}^{-1}}^{2} +\sum\limits_{t=1}^{T} ||\hat{\epsilon}_{i_{t},t}||_{W_{t-1}^{-1}}^{2}+\sum\limits_{t=1}^{T} ||\check{\epsilon}_{i_{t},t}||_{W_{t-1}^{-1}}^{2}\\ 
&\leq \sum\limits_{t=1}^{T} ||\mu_{i_{t}}-\check{\epsilon}_{i_{t},t}||_{W_{t-1}^{-1}}^{2} + \dfrac{2}{\lambda}\sum\limits_{t=1}^{T} \nu_{i_{t},t,\delta}^{2} \\ 
&\leq \sum\limits_{t=1}^{T} ||\mu_{i_{t}}-\check{\epsilon}_{i_{t},t}||_{W_{t-1}^{-1}}^{2} +\dfrac{4L^{2}d^{2}}{\lambda}\log\left( \dfrac{2dT}{\delta}\right) \sum\limits_{t=1}^{T}  \dfrac{1}{n_{i_{t},t}}
\end{align*}

D'autre part:

\begin{align*}
det(W_{T}) &=det(W_{T-1}+\dfrac{1}{R_{i_{T},T}} (\mu_{i_{T}}-\check{\epsilon}_{i_{T},T})(\mu_{a_{T}}-\check{\epsilon}_{i_{T},T})^{\top})\\
&=det(W_{T-1})det(I+\dfrac{1}{R_{i_{T},T}}W_{T-1}^{-1/2}(\mu_{i_{T}}-\check{\epsilon}_{i_{T},T})(W_{T-1}^{-1/2}(\mu_{i_{T}}-\check{\epsilon}_{i_{T},T}))^{\top})\\
&=det(W_{T-1})(1+\dfrac{1}{R_{i_{T},T}}||\mu_{i_{T}}-\check{\epsilon}_{i_{T},T}||_{W_{T-1}^{-1}}^{2}) \\
&=det(\lambda I) \prod\limits_{t=1}^{T} (1+\dfrac{1}{R_{i_{t},t}}||\mu_{i_{t}}-\check{\epsilon}_{i_{t},t}||_{W_{t-1}^{-1}}^{2})
\end{align*}

Où l'on a utilisé le fait que les valeurs propres de $I + xx^{\top}$  valent 1 sauf une qui vaut $1+||x||^{2}$ et correspond au vecteur propre $x$.

Etant donné que par hypothèse $\lambda > \max(1,2L^{2})$, on a:

\begin{equation*}
||\mu_{i_{t}}-\check{\epsilon}_{i_{t},t}||_{W_{t}^{-1}}^{2} \leq ||x||^{2}/\lambda \leq 2L^{2}/ \lambda \leq 1
\end{equation*}

Donc en utilisant le fait que  $x\leq 2\log(1+x)$ lorsque $1 \leq x \leq 1$, on obtient:

\begin{align*}
2\log\left(\dfrac{det(W_{T})}{det(\lambda I)}\right) &\geq \sum\limits_{t=1}^{T} \dfrac{1}{R_{i_{t},t}}||\mu_{i_{t}}-\check{\epsilon}_{i_{t},t}||_{W_{t-1}^{-1}}^{2} \\
&\geq  \min\limits_{t=1..T}\left(\dfrac{1}{R_{i_{t},t}}\right) \sum\limits_{t=1}^{T} ||\mu_{i_{t}}-\check{\epsilon}_{i_{t},t}||_{W_{t-1}^{-1}}^{2}\\
&\geq  1/\sqrt{R^{2}+L^{2}S^{2}} \sum\limits_{t=1}^{T} ||\mu_{i_{t}}-\check{\epsilon}_{i_{t},t}||_{W_{t-1}^{-1}}^{2}
\end{align*}

Comme dans le lemme 11 de \cite{OFUL} on a également:
\begin{align*}
\log\left(\dfrac{det(W_{T})}{det(\lambda I)}\right) \leq d \log\left( 1 + \dfrac{TL^{2}}{\lambda d} \right)
\end{align*}

Ce qui nous donne:

\begin{align*}
\sum\limits_{t=1}^{T} ||\mu_{i_{t}}-\check{\epsilon}_{i_{t},t}||_{W_{t-1}^{-1}}^{2} 
\leq
\sqrt{R^{2}+L^{2}S^{2}} d \log\left( 1 + \dfrac{TL^{2}}{\lambda d} \right)
\end{align*}

Finalement comme dans le lemme 10 de \cite{OFUL}, l'inégalité trace-téterminant donne: 

\begin{align*}
\alpha_{T}\leq \sqrt{d\log\left(\dfrac{1+TL^{2}/\lambda}{\delta}\right)} + \sqrt{\lambda}S
\end{align*}

En rassemblant ces résultats, on arrive à la borne générique proposée dans le théorème.

\subsection{Preuve du théorème \ref{RegretSamplLinUCBDiffCases}}

\subsubsection{Cas 1:}
\label{preuveRegSampLinUCBCas1}
D'une part, on a:

\begin{align*}
\sum\limits_{t=1}^{T}  \dfrac{1}{n_{i_{t},t}} =  \sum\limits_{t=1}^{T}  \dfrac{1}{t} \leq 1+ \log(T)
\end{align*}

Et d'autre part:

\begin{align*}
\sum\limits_{t=1}^{T}  \dfrac{1}{\sqrt{n_{i_{t},t}}} = \sum\limits_{t=1}^{T}  \dfrac{1}{\sqrt{t}}
 													 \leq \int_0^T \dfrac{1}{\sqrt{t}} \, \mathrm dt
 													 \leq 2\sqrt{T}
                                                     \end{align*}
                                                  
En introduisant ces deux résultats dans la borne du théorème  \ref{RegretSamplLinUCBDiffCases} et en remarquant que le terme dominant vient de $\sqrt{T}$, on arrive au résultat annoncé.

\subsubsection{Cas 2:}
\label{preuveRegSampLinUCBCas2}

\begin{lem}
\label{lem6}
 $ \forall i, \forall t \geq \ceil{2\log(1/\delta)/p^{2}}$,avec une probabilité au moins égale à $1-\delta$: 

\begin{equation*}
n_{i,t} \geq \dfrac{tp}{2}
\end{equation*}

\end{lem}

\begin{preuve}
En Hoeffding, pour tout $\epsilon>0$:

\begin{equation*}
\mathbb{P}(n_{i,t} \geq tp -\epsilon) \geq 1- e^{-2\epsilon^{2}/t}
\end{equation*}

En prenant $\epsilon = tp/2$ on arrive à 
\begin{equation*}
\mathbb{P}(n_{i,t} \geq tp/2) \geq 1- e^{-tp^{2}/2} 
\end{equation*}

Si $t \geq 2\log(1/\delta)/p^{2}$, alors $1- e^{-tp^{2}/2} \geq 1-\delta$, ce qui prouve le lemme.

Dans la suite on note $u=ceil(2\log(1/\delta)/p^{2})$.

\end{preuve}

\textbf{Preuve principale:} 
D'une part grâce au lemme \ref{lem6}, avec une probabilité au moins égale à $1-\delta$:
\begin{align*}
 \sum\limits_{t=1}^{T}  \dfrac{1}{n_{i_{t},t}}& = \sum\limits_{t=1}^{u}\dfrac{1}{n_{i_{t},t}} +\sum\limits_{t=u+1}^{T}\dfrac{1}{n_{i_{t},t}}  \\
 												   &\leq u + \dfrac{2}{p}\sum\limits_{t=u+1}^{T} \dfrac{1}{t}\\
 												   &\leq u + \dfrac{2}{p} \int_u^T \dfrac{1}{t} \, \mathrm dt\\
 												   &\leq  u +  \dfrac{2\log(T)}{p}
\end{align*}

D'autre part, toujours grâce au lemme \ref{lem6}, avec une probabilité au moins égale à $1-\delta$:
 
 \begin{align*}
 \sum\limits_{t=1}^{T}  \dfrac{1}{\sqrt{n_{i_{t},t}}} &=\sum\limits_{t=1}^{u}\dfrac{1}{\sqrt{n_{i_{t},t}}} +\sum\limits_{t=u+1}^{T}\dfrac{1}{\sqrt{n_{i_{t},t}}}\\
& \leq u + \sqrt{\dfrac{2}{p}}\sum\limits_{t=u+1}^{T} \dfrac{1}{\sqrt{t}}\\
&\leq u + \sqrt{\dfrac{2}{p}} \int_u^T \dfrac{1}{\sqrt{t}} \, \mathrm dt \\
&\leq   u +  2\sqrt{\dfrac{2T}{p}}
 \end{align*}

En utilisant la borne générique du théorème \ref{RegretSamplLinUCBDiffCases} on arrive au résultat annoncé.

\subsubsection{Cas 3:}
\label{preuveRegSampLinUCBCas3}

On décompose $T=\floor{n/K}K+r$ avec $r<K$.

La somme $\sum\limits_{t=1}^{T} \dfrac{1}{n_{i_{t},t}}$ est maximale quand chaque bras est observé exactement $\floor{T/K}$ fois pendant les  $\floor{T/K}K$ premières itérations. Donc:

\begin{align*}
\sum\limits_{t=1}^{T} \dfrac{1}{n_{i_{t},t}} &\leq \sum\limits_{i=1}^{K}\sum\limits_{t=1}^{\floor{T/K}+1} \dfrac{1}{t}\\
& \leq K\sum\limits_{t=1}^{\ceil{T/K}}\dfrac{1}{t}\\
&\leq  1+ \log(\ceil{T/K})
\end{align*}

Avec le même argument on arrive à:
 
 \begin{align*}
 \sum\limits_{t=1}^{T} \dfrac{1}{\sqrt{n_{i_{t},t}}} &\leq K\sum\limits_{t=1}^{\ceil{T/K}}\dfrac{1}{\sqrt{t}}\\ 
 &\leq 2K\sqrt{\ceil{T/K}}
 \end{align*}

Finalement, en remarquant que  $K\log(\ceil{T/K}) \sim K\log(n/K)$ et $K\sqrt{\ceil{T/K}} \sim \sqrt{Kn}$ et en utilisant la borne générique du théorème \ref{RegretSamplLinUCBDiffCases} on prouve le résultat annoncé.

\subsection{Preuve du théorème \ref{RegretSamplLinUCBGenericMult}}
\label{preuveRegSampLinUCBCasMP}

On suit une démarche similaire à celle présentée dans  \cite{banditCtxtCombLinUCB} pour le cas particulier de la somme: étant donné que l'on considère que la récompense d'un ensemble de bras est égale à la somme des récompenses individuelles qui le compose, on a: 

\begin{equation*}
reg_{t}=\sum\limits_{i\in\mathcal{K}^{\star}}\mu_{i}^{\top}\beta-\sum\limits_{i\in\mathcal{K}_{t}}\mu_{i}^{\top}\beta
\end{equation*}
Avec  $\mathcal{K}^{\star}$ l'ensemble des $k$ actions optimales, c.-à-d. ayant les plus grandes valeurs $\mu_{i}^{\top}\beta$.

On utilise ensuite le fait que pour tout $t$:
\begin{equation*}
\sum\limits_{i \in \mathcal{K}_{t}} s_{i,t} \geq \sum\limits_{i \in \mathcal{K}_{\star}} s_{i^{\star},t}
\end{equation*}

Ce qui conduit à:

\begin{equation*}
reg_{t} \leq \sum\limits_{i \in \mathcal{K}_{t}} 2\alpha_{t-1}||\hat{x}_{i,t}||_{V_{t-1}^{-1}} + 4\sqrt{d}(\alpha_{t-1}/\sqrt{\lambda}+S)\rho_{i,t,\delta}
\end{equation*}

Où la matrice $V_{t-1}$ est définie en considérant les $k$ exemples d'apprentissage disponibles à chaque itération: $V_{t-1}=\lambda I+ \sum\limits_{s=1}^{t-1}\sum\limits_{i \in \mathcal{K}_{s}} \dfrac{1}{R_{i,t}}\hat{x}_{i,t}\hat{x}_{i,t}^{\top}$. Le fait que l'on ajoute $k$ terme à la matrice $V$ à chaque itération, se traduit deux façon distinctes:
\begin{itemize}
\item D'une part sur l'ellipsoide de confiance associée au paramètre $\beta$. Dans ce cas, on a:

$\alpha_{T}\leq \sqrt{d\log\left(\dfrac{1+kTL^{2}/\lambda}{\delta}\right)} + \sqrt{\lambda}S$;

\item D'autre part dans la majoration de $\sum\limits_{t=1}^{T}\sum\limits_{i_{t} \in \mathcal{K}_{t} } ||\mu_{i_{t}}-\check{\epsilon}_{i_{t},t}||_{W_{t-1}^{-1}}^{2}$. On a:

$\sum\limits_{t=1}^{T} \sum\limits_{i_{t} \in \mathcal{K}_{t} } ||\mu_{i_{t}}-\check{\epsilon}_{i_{t},t}||_{W_{t-1}^{-1}}^{2} \leq  \sqrt{R^{2}+L^{2}S^{2}} d \log\left( 1 + \dfrac{TkL^{2}}{\lambda d}\right)$
\end{itemize}

En utilisant ensuite les mêmes méthodes que précédemment, le regret cumulé prend une forme similaire, oùles $k$ actions sélectionnées à chaque itération apparaissent de façon explicite dans les deux termes: $\sum\limits_{t=1}^{T} \sum\limits_{i \in \mathcal{K}_{t}}\dfrac{1}{n_{i,t}}$ et $\sum\limits_{t=1}^{T} \sum\limits_{i \in \mathcal{K}_{t}}\dfrac{1}{\sqrt{n_{i,t}}}$.

\section{Distributions variationnelles pour le modèle contextuel et l'algorithme \texttt{HiddenLinUCB}}
\label{PreuveVarContextual}

On rappelle dans un premier temps les notations utilisées:

\begin{itemize}
\item Pour tout $i$, l'ensemble des itérations jusqu'au temps $t-1$ telles que $i$ a été sélectionné en ayant eu un contexte observé  est noté $\mathcal{A}_{i,t-1}=\left\{s \text{ tel que } 1\leq s\leq t-1 \text{ et } i\in \mathcal{K}_{s}\cap\mathcal{O}_{s} \right\}$;
\item Pour tout $i$, l'ensemble des itérations jusqu'au temps $t-1$ telles que $i$ a été sélectionné sans avoir eu un contexte observé est noté $\mathcal{B}_{i,t-1}=\left\{s \text{ tel que } 1\leq s\leq t-1 \text{ et } i\in \mathcal{K}_{s}\cap\bar{\mathcal{O}_{s}} \right\}$;
\item Pour tout $i$, l'ensemble des itérations jusqu'au temps $t-1$ telles que le contexte de $i$ a été observé est noté $\mathcal{C}_{i,t-1}=\left\{s \text{ tel que } 1\leq s\leq t-1 \text{ et } i\in \mathcal{O}_{s} \right\}$;
\end{itemize}

Dans cette section, nous fournissons les preuves des ditributions variationnelles des différentes varibales aléatoires du modèle. Commençons par écrire la probabilité jointe du modèle complet, qui nous sevira ensuite dans chacun des cas. Celle ci s'écrit de la façon suivante: 

\begin{align*}
p\left(\underbrace{\beta, (\mu_{i})_{i=1..K}, (\tau_{i})_{i=1..K}, (x_{i,s})_{i=1..K,s\in \mathcal{B}_{i,t-1}}}_{Z} \underbrace{(r_{i,s})_{i=1..K, s\in \mathcal{A}_{i,t-1}\cup\mathcal{B}_{i,t-1}},(x_{i,s})_{i=1..K,s \in \mathcal{C}_{i,t-1}}}_{X}\right)
\end{align*}

$Z$ représente les variables cachées tandis que $X$ représente les variables observées. On rappelle de plus que l'on utilise la factorisation suivante pour les distributions des variables cachées:

\begin{align*}
q(\beta, (\mu_{i})_{i=1..K}, (\tau_{i})_{i=1..K}, (x_{i,s})_{i=1..K,s\in \mathcal{B}_{i,t-1}}=q_{\beta}(\beta)\prod\limits_{i=1}^{K}\left(q_{\mu_{i}}(\mu_{i})q_{\tau_{i}}(\tau_{i})\prod\limits_{s\in \mathcal{B}_{i,t-1}}q_{x_{i,s}}(x_{i,s})\right)
\end{align*}

\subsection{Preuve de la proposition \ref{distribVarBeta}}
\label{PreuveVarContextualBeta}

Dans cette preuve, on désigne par $C$ les termes qui ne dépendent pas de $\beta$. La distribution variationnelle $q_{\beta}^{\star}(\beta)$ de $\beta$ est déterminée en utilisant:

\begin{align*}
\log q_{\beta}^{\star}(\beta)=\mathbb{E}_{\beta^{\backslash}}[\log p(Z,X)]
\end{align*}

où $\mathbb{E}_{\beta^{\backslash}}$ désigne l'espérance prise selon toutes les variables sauf $\beta$. 

\begin{align*}
\log q_{\beta}^{\star}(\beta)&=\mathbb{E}_{\beta^{\backslash}}[\log p((r_{i,s})_{i=1..K, s\in \mathcal{A}_{i,t-1}\cup\mathcal{B}_{i,t-1}} | \beta,  (x_{i,s})_{i=1..K,s\in \mathcal{A}_{i,t-1}},  (x_{i,s})_{i=1..K,s\in \mathcal{B}_{i,t-1}} )+\log p(\beta)]+C\\
&=-\dfrac{1}{2}\mathbb{E}_{\beta^{\backslash}}\left[\sum\limits_{i=1}^{K} \left[\sum\limits_{s\in \mathcal{A}_{i,t-1}} (r_{i,s}-x_{i,s}^{\top}\beta)^{2} + \sum\limits_{s\in \mathcal{B}_{i,t-1}} (r_{i,s}-x_{i,s}^{\top}\beta)^{2}\right] +\beta^{\top}\beta\right] +C\\
&=-\dfrac{1}{2} \mathbb{E}_{\beta^{\backslash}}\left[\beta^{\top}\left(I+\sum\limits_{i=1}^{K} \left[ \sum\limits_{s\in \mathcal{A}_{i,t-1}}x_{i,s}x_{i,s}^{\top} +\sum\limits_{s\in \mathcal{B}_{i,t-1}} x_{i,s}x_{i,s}^{\top} \right]\right)\beta -2\beta^{\top}\left(\sum\limits_{i=1}^{K}\left[ \sum\limits_{s\in \mathcal{A}_{i,t-1}}x_{i,s}r_{i,s} +\sum\limits_{s\in \mathcal{B}_{i,t-1}}x_{i,s}r_{i,s} \right]\right) \right] +C
\end{align*}

En prenant l'espérance selon les $x_{i,s}$ lorsque $s\in \mathcal{B}_{i,t-1}$ (qui sont les seules VA autres que $\beta$ présentes dans cette expression), et en posant:

$V_{t-1}=I+\sum\limits_{i=1}^{K} \left[ \sum\limits_{s\in \mathcal{A}_{i,t-1}}x_{i,s}x_{i,s}^{\top} +\sum\limits_{s\in \mathcal{B}_{i,t-1}} \mathbb{E}[x_{i,s}x_{i,s}^{\top}] \right]$ 

$\hat{\beta}_{t-1}=V_{t-1}^{-1}\left(\sum\limits_{i=1}^{K}\left[ \sum\limits_{s\in \mathcal{A}_{i,t-1}}x_{i,s}r_{i,s} +\sum\limits_{s\in \mathcal{B}_{i,t-1}} \mathbb{E}[x_{i,s}]r_{i,s} \right]\right)$

On obtient: 
\begin{align*}
\log q_{\beta}^{\star}(\beta)&= -\dfrac{1}{2}(\beta-\hat{\beta}_{t-1})^{\top}V_{t-1}(\beta-\hat{\beta}_{t-1}) +C
\end{align*}

Ceci correspond à une distribution gaussienne multidimensionnelle de moyenne $\hat{\beta}_{t-1}$ et de matrice de covariance $V_{t-1}^{-1}$.

\subsection{Preuve de la proposition \ref{distribVarXit}}
\label{PreuveVarContextualXit}

Soit une action $i$ et $s\in \mathcal{B}_{i,t-1}$. 
Dans cette preuve, on désigne par $C$ les termes qui ne dépendent pas de $x_{i,s}$. La distribution variationnelle $q_{x_{i,s}}^{\star}(x_{i,s})$ de $x_{i,s}$ est déterminée en utilisant:

\begin{align*}
\log q_{x_{i,s}}^{\star}=\mathbb{E}_{x_{i,s}^{\backslash}}[\log p(Z,X)]
\end{align*}

où $\mathbb{E}_{x_{i,s}^{\backslash}}$ désigne l'espérance prise selon toutes les variables sauf $x_{i,s}$.

\begin{align*}
\log q_{x_{i,s}}^{\star}(x_{i,s})&=\mathbb{E}_{x_{i,s}^{\backslash}}[\log p(r_{i,s} | \beta,x_{i,s})+\log p(x_{i,s}|\mu_{i},\tau_{i})]+C\\
&=-\dfrac{1}{2}\mathbb{E}_{x_{i,s}^{\backslash}}\left[(r_{i,s}-x_{i,s}^{\top}\beta)^{2} +\tau_{i}(x_{i,s}-\mu_{i})^{\top}(x_{i,s}-\mu_{i})\right] +C\\
&=-\dfrac{1}{2}\mathbb{E}_{x_{i,s}^{\backslash}}\left[x_{i,s}^{\top}(\beta\beta^{\top}+\tau_{i}I)x_{i,s}-2x_{i,s}^{\top}(\beta r_{i,s}+\mu_{i}\tau_{i})\right] +C\\
\end{align*}

En prenant l'espérance selon les $\beta$, $\mu_{i}$ et $\tau_{i}$ lorsque $s\in \mathcal{B}_{i,t-1}$ (qui sont les seules VA autres que $x_{i,s}$ présentes dans cette expression), et en posant:

$W_{i,s}= \mathbb{E}[\beta\beta^{\top}]+\mathbb{E}[\tau_{i}]I$

$\hat{x}_{i,s}=W_{i,s}^{-1}(\mathbb{E}[\beta]r_{i,s}+\mathbb{E}[\tau_{i}]\mathbb{E}[\mu_{i}])$

On obtient:

\begin{align*}
\log q_{x_{i,s}}^{\star}(x_{i,s})&=-\dfrac{1}{2}(x_{i,s}-\hat{x}_{i,s})^{\top}W_{i,s}(x_{i,s}-\hat{x}_{i,s}) +C
\end{align*}

Ceci correspond à une distribution gaussienne multidimensionnelle de moyenne $\hat{x}_{i,s}$ et de matrice de covariance $W_{i,s}^{-1}$.

\subsection{Preuve de la proposition \ref{distribVarMui}}
\label{PreuveVarContextualMui}

Soit $i$ une action. Dans cette preuve, on désigne par $C$ les termes qui ne dépendent pas de $\mu_{i}$. La distribution variationnelle $q_{\mu_{i}}^{\star}(\mu_{i})$ de $\mu_{i}$ est déterminée en utilisant:

\begin{align*}
\log q_{\mu_{i}}^{\star}(\mu_{i})=\mathbb{E}_{\mu_{i}^{\backslash}}[\log p(Z,X)]
\end{align*}

où $\mathbb{E}_{\mu_{i}^{\backslash}}$ désigne l'espérance prise selon toutes les variables sauf $\mu_{i}$. 

\begin{align*}
\log q_{\mu_{i}}^{\star}(\mu_{i})&=\mathbb{E}_{\mu_{i}^{\backslash}}[\log p((x_{i,s})_{s\in \mathcal{C}_{i,t-1}},(x_{i,s})_{s\in \mathcal{B}_{i,t-1}}|\mu_{i},\tau_{i})+\log p(\mu_{i}|\tau_{i})]+C\\
&=-\dfrac{1}{2}\mathbb{E}_{\mu_{i}^{\backslash}}\left [\sum\limits_{s\in \mathcal{C}_{i,t-1}}\tau_{i}(x_{i,s}-\mu_{i})^{\top}(x_{i,s}-\mu_{i})+\sum\limits_{s\in \mathcal{B}_{i,t-1}}\tau_{i}(x_{i,s}-\mu_{i})^{\top}(x_{i,s}-\mu_{i}) +\tau_{i}\mu_{i}^{\top}\mu_{i}  \right]+C\\
&=-\dfrac{1}{2}\mathbb{E}_{\mu_{i}^{\backslash}}\left [\tau_{i}\left(\mu_{i}^{\top}\mu_{i} +\sum\limits_{s\in \mathcal{C}_{i,t-1}} \mu_{i}^{\top}\mu_{i}  +\sum\limits_{s\in \mathcal{B}_{i,t-1}} \mu_{i}^{\top}\mu_{i} -2\mu_{i}^{\top}\left(\sum\limits_{s\in \mathcal{C}_{i,t-1}}x_{i,s}+\sum\limits_{s\in \mathcal{B}_{i,t-1}}x_{i,s}\right) \right)  \right]+C\\
&=-\dfrac{1}{2}\mathbb{E}_{\mu_{i}^{\backslash}}\left [\tau_{i}\left(\mu_{i}^{\top}\mu_{i}(1+n_{i,t-1})-2\mu_{i}^{\top}\left(\sum\limits_{s\in \mathcal{C}_{i,t-1}}x_{i,s}+\sum\limits_{s\in \mathcal{B}_{i,t-1}}x_{i,s}\right) \right)  \right]+C\\
\end{align*}
Avec $n_{i,t-1}=|\mathcal{B}_{i,t-1}|+|\mathcal{C}_{i,t-1}|$. On a donc:

\begin{align*}
\log q_{\mu_{i}}^{\star}(\mu_{i})&=-\dfrac{1}{2}\mathbb{E}_{\mu_{i}^{\backslash}}\left [\tau_{i}(1+n_{i,t-1})\left(\mu_{i}^{\top}\mu_{i}-2\mu_{i}^{\top}\left(\dfrac{\sum\limits_{s\in \mathcal{C}_{i,t-1}}x_{i,s}+\sum\limits_{s\in \mathcal{B}_{i,t-1}}x_{i,s}}{1+n_{i,t-1}}\right) \right)  \right]+C\\
\end{align*}

En prenant l'espérance selon les $x_{i,s}$ lorsque $s\in \mathcal{B}_{i,t-1}$ et $\tau_{i}$ (qui sont les seules VA autres que $\mu_{i}$ présentes dans cette expression), et en posant:

$\Sigma_{i,t-1}=  (1+n_{i,t-1})E[\tau_{i}]I$ 

$\hat{\mu}_{i,t-1}=\dfrac{\sum\limits_{s\in \mathcal{C}_{i,t-1}}x_{i,s} +\sum\limits_{s\in \mathcal{B}_{i,t-1}} \mathbb{E}[x_{i,s}]}{1+n_{i,t-1}}$

On obtient: 
\begin{align*}
\log  q_{\mu_{i}}^{\star}(\mu_{i})&= -\dfrac{1}{2}(\mu_{i}-\hat{\mu}_{i,t-1})^{\top}\Sigma_{i,t-1}(\mu_{i}-\hat{\mu}_{i,t-1}) +C
\end{align*}

Ceci correspond à une distribution gaussienne multidimensionnelle de moyenne $\hat{\mu}_{i,t-1}$ et de matrice de covariance $\Sigma_{i,t-1}^{-1}$.

\subsection{Preuve de la proposition \ref{distribVarTaui}}
\label{PreuveVarContextualTaui}

Soit $i$ une action. Dans cette preuve, on désigne par $C$ les termes qui ne dépendent pas de $\tau_{i}$. La distribution variationnelle $q_{\tau_{i}}^{\star}(\tau_{i})$ de $\tau_{i}$ est déterminée en utilisant:

\begin{align*}
\log q_{\tau_{i}}^{\star}(\tau_{i})=\mathbb{E}_{\tau_{i}^{\backslash}}[\log p(Z,X)]
\end{align*}

où $\mathbb{E}_{\tau_{i}^{\backslash}}$ désigne l'espérance prise selon toutes les variables sauf $\tau_{i}$.

\begin{align*}
\log q_{\tau_{i}}^{\star}(\tau_{i})&=\mathbb{E}_{\tau_{i}^{\backslash}}[\log p((x_{i,s})_{s\in \mathcal{C}_{i,t-1}},(x_{i,s})_{s\in \mathcal{B}_{i,t-1}}|\mu_{i},\tau_{i})+\log p(\mu_{i}|\tau_{i})+\log p(\tau_{i})]+C\\
\end{align*}

On rappelle que $p(\tau_{i})$ correspond à la loi gamma de paramètres $b_{0}$ et $b_{0}$, qui a pour densité: $p(\tau_{i})=\dfrac{b_{0}^{b_{0}}\tau_{i}^{a_{0}-1}e^{-\tau_{i}b_{0}}}{\Gamma(a_{0})}$, où $\Gamma(a_{0})$ est la constante de normalisation. Par ailleurs, $\tau_{i}$ apparaît dans la densité des loi gaussiennes en dehors de l'exponentielle. En effet pour une loi gaussienne de dimension $d$, de moyenne $m$ et de matrice de covariance $\tau_{i} I$, la densité en $x$ vaut: $f(x;m,\alpha)=\dfrac{\tau_{i}^{d/2}}{(2\pi)^{d/2}}e^{-\tau_{i}/2(x-m)^{\top}(x-m)}$. On a donc:

\begin{multline*}
\log q_{\tau_{i}}^{\star}(\tau_{i}) = \mathbb{E}_{\tau_{i}^{\backslash}}\left [-\dfrac{1}{2}\left(\sum\limits_{s\in \mathcal{C}_{i,t-1}}\tau_{i}(x_{i,s}-\mu_{i})^{\top}(x_{i,s}-\mu_{i})+\sum\limits_{s\in \mathcal{B}_{i,t-1}}\tau_{i}(x_{i,s}-\mu_{i})^{\top}(x_{i,s}-\mu_{i}) +\tau_{i}\mu_{i}^{\top}\mu_{i}  \right)   \right.\\
\left. +\sum\limits_{s\in \mathcal{C}_{i,t-1}} \dfrac{d}{2}\log \tau_{i} +\sum\limits_{s\in \mathcal{B}_{i,t-1}} \dfrac{d}{2}\log \tau_{i} +  \dfrac{d}{2}\log \tau_{i} +(a_{0}-1)\log \tau_{i} -b_{0}\tau_{i} \right] +C
\end{multline*}

\begin{multline*}
\log q_{\tau_{i}}^{\star}(\tau_{i}) = \mathbb{E}_{\tau_{i}^{\backslash}}\left [ -\dfrac{\tau_{i}}{2}\left((n_{i,t-1}+1) \mu_{i}^{\top}\mu_{i}-2\mu_{i}^{\top}(\sum\limits_{s\in \mathcal{C}_{i,t-1}}x_{i,s}+\sum\limits_{s\in \mathcal{B}_{i,t-1}}x_{i,s})+\sum\limits_{s\in \mathcal{C}_{i,t-1}}x_{i,s}^{\top}x_{i,s}+\sum\limits_{s\in \mathcal{B}_{i,t-1}}x_{i,s}^{\top}x_{i,s}\right)  \right. \\
\left. +(a_{0}-1)\log \tau_{i} -b_{0}\tau_{i}+\dfrac{d(n_{i,t-1}+1)}{2} \log \tau_{i}   \right] +C
\end{multline*}

En prenant l'espérance selon les $x_{i,s}$ lorsque $s\in \mathcal{B}_{i,t-1}$ et $\mu_{i}$ (qui sont les seules VA autres que $\tau_{i}$ présentes dans cette expression), et en posant:

$a_{i,t-1}= a_{0}+\dfrac{d(1+n_{i,t-1})}{2}$ 

$b_{i,t-1}=b_{0}+\dfrac{1}{2}  \left[ (1+n_{i,t-1})\mathbb{E}[\mu_{i}^{\top}\mu_{i}]-2\mathbb{E}[\mu_{i}]^{\top}\left( \sum\limits_{s\in \mathcal{C}_{i,t-1}}x_{i,s}   + \sum\limits_{s\in \mathcal{B}_{i,t-1}}\mathbb{E}[x_{i,s}] \right)   + \sum\limits_{s\in \mathcal{C}_{i,t-1}}x_{i,s}^{\top}x_{i,s} + \sum\limits_{s\in \mathcal{B}_{i,t-1}}\mathbb{E}[x_{i,s}^{\top}x_{i,s}]      \right] $

\begin{align*}
\log  q_{\tau_{i}}^{\star}(\tau_{i})&= (a_{i,t-1}-1 )\log \tau_{i}-b_{i,t-1}\tau_{i}+C
\end{align*}

Ceci correspond à une distribution gamma de paramètres $a_{i,t-1}$ et $b_{i,t-1}$.

\section{Distributions variationnelles pour les modèles récurrents et l'algorithme \texttt{Recurrent TS}}
\label{PreuveRelational}

\subsection{Modèle relationnel}
\label{relationnalLinearProof}

\textbf{Note importante} pour simplifier les notations on n'incorpore pas le terme de biais dans cette démonstration. Cependant les résultats sont directement transposables en ajoutant un terme constant égal à 1 à la fin de chaque vecteur de récompense.

Etant donné un ensemble de $K$ bras, on rappelle les hypothèses:

\begin{itemize}
\item \textbf{Vraisemblance : } $\forall i \in \left\{1,...,K\right\}, \exists \theta_{i} \in \mathbb{R}^{K} ~\text{tel que}~ \forall t \in \left\{2,...,T\right\}$:   $r_{i,t}=\theta_{i}^{\top}R_{t-1} + \epsilon_{i,t}$, où $\epsilon_{i,t} \sim \mathcal{N}(0,\sigma^{2})$ (bruit gaussien de moyenne $0$ et de variance $\sigma^{2}$). 
\item \textbf{Prior sur les paramètres: } $\forall i \in \left\{1,...,K\right\}$:  $\theta_{i} \sim \mathcal{N}(0,\alpha^{2}I)$ (bruit gaussien de dimension $K$, de moyenne  $0$ et de matrice de covariance  $\alpha^{2}I$).
\item \textbf{Prior sur $R_{1}$ : } $R_{1} \sim \mathcal{N}(\mu,\sigma^{2}I)$ où $\mu \in \mathbb{R}^{K}$ ($\forall i \in \left\{1,...,K\right\}$: $r_{i,1} \sim \mathcal{N}(\mu_{i},\sigma^{2})$).
\end{itemize}

La distribution jointe des différentes variables du modèle est la suivante:

\begin{align*}
p\left(\underbrace{(\theta_{i})_{i=1..K}, (r_{i,s})_{i=1..K,i\notin \mathcal{K}_{s},s=1..t-1}}_{Z}, \underbrace{(r_{i,s})_{i=1..K,i\in \mathcal{K}_{s},s=1..t-1}}_{X}\right)
\end{align*}

où $Z$ représente les variables cachées et $X$ les variables observées. Notre but est de déterminer les distributions variationnelles des différentes variables cachées selon la factorisation suivante:

\begin{align*}
q\left((\theta_{i})_{i=1..K}, (r_{i,s})_{i=1..K,i\notin \mathcal{K}_{s},s=1..t-1}\right)=\prod_{i=1}^{K}q(\theta_{i})\prod_{s=1}^{t-1}\prod_{i=1, i\notin \mathcal{K}_{s}}^{K} q(r_{i,s})
\end{align*}

\subsubsection{Paramètres}
\label{proofLin1Params}

Pour $t\geq 3$ , on note $D_{t-1}=(R_{s}^{\top})_{s=1..t-1}$ la matrice de taille $(t-1)\times (K+1)$ où la ligne $s$ correspond au vecteur de récompense au temps $s$, $D_{1..t-2}$ la matrice de taille $(t-2)\times K$ composée des $t-2$ premières lignes de $D_{t-1}$ et $D_{i:2..t-1}$ le vecteur de taille $t-2$ correspondant à la colonne $i$ et les $t-2$ dernières lignes de $D_{t-1}$ (c.-à-d. le vecteur de récompenses du bras $i$ du temps $1$ au temps $t-2$). 

La distribution variationnelle $q_{\theta_{i}}^{\star}(\theta_{i})$ de $\theta_{i}$ est déterminée de la façon suivante:

\begin{align*}
\log q_{\theta_{i}}^{\star}(\theta_{i})&=\mathbb{E}_{\theta_{i}^{\backslash}}[\log p(Z,X)]\\
&=\mathbb{E}_{\theta_{i}^{\backslash}}\left[-\dfrac{1}{2\sigma^{2}}\sum\limits_{v=1}^{t-2}    ( r_{i,v+1} - \theta_{i}^{\top}R_{v}  )^{2} -\dfrac{1}{2\alpha^{2}}\theta_{i}^{\top} \theta_{i}\right] +C\\
&=\mathbb{E}_{\theta_{i}^{\backslash}}\left[-\dfrac{1}{2\sigma^{2}}  (D_{i:2..t-1} - D_{1..t-2} \theta_{i} )^{\top} (D_{i:2..t-1} - D_{1..t-2} \theta_{i}) -\dfrac{1}{2\alpha^{2}} \theta_{i}^{\top} \theta_{i} \right] +C\\
&=\mathbb{E}_{\theta_{i}^{\backslash}}\left[ -\dfrac{1}{2}\left( \theta_{i}^{\top}(  \dfrac{D_{1..t-2}^{\top}D_{1..t-2}}{\sigma^{2}}  +  \dfrac{I}{\alpha^{2}})\theta_{i}  -2 \theta_{i}^{\top}  \dfrac{D_{1..t-2}^{\top} D_{i:2..t-1} }{\sigma^{2}}      \right)\right] +C\\
\end{align*}

En prenant l'espérance selon les paramètres les récompenses n'ayant pas été observées, et en posant: 

$A_{i,t-1}=\dfrac{\mathbb{E}[D_{1..t-2}^{\top}D_{1..t-2}]}{\sigma^{2}}+\dfrac{I}{\alpha^{2}}$

$b_{i,t-1}=\dfrac{\mathbb{E}[D_{1..t-2}^{\top}]}{\sigma^{2}}\mathbb{E}[D_{i:2..t-1}]$ 

On obtient: $\log q_{\theta_{i}}^{\star}(\theta_{i})=-\dfrac{1}{2}( \theta_{i} -  A_{i,t-1}^{-1}b_{i,t-1})^{\top}A_{i,t-1}(\theta_{i} -  A_{i,t-1}^{-1}b_{i,t-1})+C$

Ce qui correspond à une distribution gaussienne de moyenne $A_{i,t-1}^{-1}b_{i,t-1}$ et de matrice de covariance  $A_{i,t-1}^{-1}$.

\subsubsection{Récompenses}
\label{proofLin1Rwds}

On note $\Theta=(\theta_{i}^{\top})_{i=1..K}$ la matrice de taille $K \times K$  dont la ligne $i$ vaut $\theta_{i}$ et on note $\beta_{i}$ la colonne $i$ de $\Theta$. 
On fixe  $i \in \left\{1,...,K\right\}$ et  $s \in \left\{1,...,t-1\right\}$ pour $t\geq 2$. Le cas $t=1$ ne nécessite pas de traitement particulier puisque il s'agit de la distribution du prior. On étudie les trois distributions suivante pour $r_{i,s}$ en fonction cas suivants selon la valeur de $s$.

\begin{itemize}
\item  \textbf{Cas 1: $1<s<t-1$}

 La distribution variationnelle $q_{r_{i,s}}^{\star}(r_{i,s})$ de $r_{i,s}$ est déterminée de la façon suivante:

\begin{align*}
\log q_{r_{i,s}}^{\star}(r_{i,s})&=\mathbb{E}_{r_{i,s}^{\backslash}}[\log p(Z,X)]\\
&=\mathbb{E}_{r_{i,s}^{\backslash}}\left[\log p(  (r_{j,s+1})_{j=1..K}  | (r_{j,s})_{j=1..K}  , (\theta_{j})_{j=1..K})+ \log      p(r_{i,s} |  (r_{j,s-1})_{j=1..K}, (\theta_{j})_{j=1..K} )\right] +C\\
&=\mathbb{E}_{r_{i,s}^{\backslash}}\left[\sum\limits_{j=1}^{K} \log p( r_{j,s+1} |  (r_{l,s})_{l=1..K} , \theta_{j})    +\log  p(r_{i,s} |  (r_{l,s-1})_{l=1..K}, \theta_{i})\right] +C\\
&=-\dfrac{1}{2\sigma^{2}}\mathbb{E}_{r_{i,s}^{\backslash}}\left[\sum\limits_{j=1}^{K}   \left( r_{j,s+1} - \sum\limits_{l=1}^{K}\theta_{j,l}r_{l,s} \right)^{2} + \left( r_{i,s} - \sum\limits_{l=1}^{K}\theta_{i,l}r_{l,s-1} \right)^{2} \right] +C\\
\end{align*}
où $C$ désigne tout terme qui ne dépend pas de $r_{i,s}$.

On note $A=\sum \limits_{j=1}^{K}    \left( r_{j,s+1} - \sum\limits_{l=1}^{K}\theta_{j,l}r_{l,s} \right)^{2}+ \left( r_{i,s} - \sum\limits_{l=1}^{K}\theta_{i,l}r_{l,s-1} \right)^{2}$ 

En développant le carré, on obtient:

\begin{align*}
 A& =\sum \limits_{j=1}^{K}  \theta_{j,i}^{2}r_{i,s}^{2} - 2r_{i,s}\left(\sum \limits_{j=1}^{K}\theta_{j,i} r_{j,s+1}  - \sum \limits_{j=1}^{K}\theta_{j,i} \sum\limits_{l=1,l\neq i}^{K}  \theta_{j,l}r_{l,s}   \right)   + r_{i,s}^{2} - 2r_{i,s}\left(  \sum\limits_{l=1}^{K} \theta_{i,l}r_{l,s-1} \right) + C\\
 &=r_{i,s}^{2}\left(1+ \sum \limits_{j=1}^{K} \theta_{j,i}^{2} \right ) - 2r_{i,s}\left(  \sum \limits_{j=1}^{K} \theta_{j,i} r_{j,s+1}  -  \sum \limits_{j=1}^{K} \theta_{j,i} \sum\limits_{l=1,l\neq i}^{K} \theta_{j,l}r_{l,s} +\sum\limits_{l=1}^{K} \theta_{i,l}r_{l,s-1}  \right) + C\\
 &=r_{i,s}^{2}\left(1+ \sum \limits_{j=1}^{K} \theta_{j,i}^{2} \right ) - 2r_{i,s}\left(  \sum \limits_{j=1}^{K} \theta_{j,i} r_{j,s+1}  -  \sum\limits_{l=1,l\neq i}^{K} r_{l,s}\sum \limits_{j=1}^{K} \theta_{j,i}\theta_{j,l}  +\sum\limits_{l=1}^{K} \theta_{i,l}r_{l,s-1}  \right) + C
\end{align*}

On note $R_{s}=(r_{1,s},...,r_{K,s})^{\top}$, ce qui nous donne:

\begin{align*}
A&=r_{i,s}^{2}(1+\beta_{i}^{\top}\beta_{i})-2r_{i,s} (\beta_{i}^{\top}R_{s+1} -  \sum\limits_{l=1,l\neq i}^{K}r_{l,s} \beta_{i}^{\top}\beta_{l} + \theta_{i}^{\top}R_{s-1})+C\\
& =r_{i,s}^{2}(1+\beta_{i}^{\top}\beta_{i})-2r_{i,s} (\beta_{i}^{\top}R_{s+1} - \beta_{i}^{\top} \sum\limits_{l=1,l\neq i}^{K} \beta_{l}r_{l,s} + \theta_{i}^{\top}R_{s-1})+C\\
\end{align*}


En prenant l'espérance selon les paramètres $\theta_{i}$ et les récompenses n'ayant pas été observées sauf $r_{i,s}$, et en posant: 

$\mu_{i,s}=\dfrac{\mathbb{E}[\beta_{i}]^{\top}\mathbb{E}[R_{s+1}] + \mathbb{E}[\theta_{i}]^{\top}\mathbb{E}[R_{s-1}] - \sum\limits_{l=1,l \neq i}^{K} \mathbb{E}[\beta_{i}^{\top}\beta_{l}]\mathbb{E}[r_{l,s}]  }{1+\mathbb{E}[\beta_{i}^{\top}\beta_{i}]}$

$\sigma_{i,s}^{2}=\dfrac{\sigma^{2}}{1+\mathbb{E}[\beta_{i}^{\top}\beta_{i}]}$
 
On obtient: 
\begin{align*}
\log q_{r_{i,s}}^{\star}(r_{i,s})&=-\dfrac{1}{2\sigma_{i,s}^{2}} \left( r_{i,s} -  \mu_{i,s} \right)^{2}+C\\
\end{align*}

Ceci correspond à une distribution gaussienne de moyenne $\mu_{i,s}$ et de variance  $\sigma_{i,s}^{2}$.

\item \textbf{Cas 2: $s=1$}

\begin{align*}
\log q_{r_{i,s}}^{\star}(r_{i,s})&=\mathbb{E}_{r_{i,s}^{\backslash}}[\log p(Z,X)]\\
&=\mathbb{E}_{r_{i,s}^{\backslash}}\left[\log p((r_{j,s+1})_{j=1..K}  | (r_{j,s})_{j=1..K},(\theta_{j})_{j=1..K} )  + \log p(r_{i,s}  |  \mu_{i}) \right] +C\\
&=-\dfrac{1}{2\sigma^{2}}\mathbb{E}_{r_{i,s}^{\backslash}}\left[\sum\limits_{j=1}^{K}   \left( r_{j,s+1} - \sum\limits_{l=1}^{K}\theta_{j,l}r_{l,s} \right)^{2} + \left( r_{i,s} - \mu_{i} \right)^{2} \right] +C\\
\end{align*}

On retrouve une forme similaire au cas 1 ci-dessus.

En prenant l'espérance selon les paramètres $\theta_{i}$ et les récompenses n'ayant pas été observées sauf $r_{i,s}$, et en posant:

$\mu_{i,s}=\dfrac{\mathbb{E}[\beta_{i}]^{\top}\mathbb{E}[R_{s+1}] + \mu_{i} - \sum\limits_{l=1,l \neq i}^{K} \mathbb{E}[\beta_{i}^{\top}\beta_{l}]\mathbb{E}[r_{l,s}]  }{1+\mathbb{E}[\beta_{i}^{\top}\beta_{i}]}$

$\sigma_{i,s}^{2}=\dfrac{\sigma^{2}}{1+\mathbb{E}[\beta_{i}^{\top}\beta_{i}]}$

On obtient: 
\begin{align*}
\log q_{r_{i,s}}^{\star}(r_{i,s})&=-\dfrac{1}{2\sigma_{i,s}^{2}} \left( r_{i,s} -  \mu_{i,s} \right)^{2}+C\\
\end{align*}

Ce qui correspond à une distribution gaussienne de moyenne $\mu_{i,s}$ et de variance  $\sigma_{i,s}^{2}$.

\item \textbf{Cas 3: $s=t-1$}

\begin{align*}
\log q_{r_{i,s}}^{\star}(r_{i,s})=\mathbb{E}_{r_{i,s}^{\backslash}}[\log p(Z,X)]
=\mathbb{E}_{r_{i,s}^{\backslash}}\left[\log p(r_{i,t}  | (r_{j,s-1})_{j=1..K} ,\theta_{i} )\right] +C=-\dfrac{1}{2\sigma^{2}}\mathbb{E}_{r_{i,s}^{\backslash}}\left[ \left( r_{i,s}-\theta_{i}^{\top}R_{s-1}\right)^{2}\right] +C
\end{align*}

En prenant l'espérance selon les paramètres $\theta_{i}$ et les récompenses n'ayant pas été observées sauf $r_{i,s}$, et en posant: 

$\mu_{i,s}=\mathbb{E}[\theta_{i}]^{\top}\mathbb{E}[R_{s-1}] $ 

$\sigma_{i,s}^{2}=\sigma^{2}$

On obtient: $\log q_{r_{i,s}}^{\star}(r_{i,s})=-\dfrac{1}{2\sigma_{i,s}^{2}} \left( r_{i,s} -  \mu_{i,s} \right)^{2}+C$

Ce qui correspond à une distribution gaussienne de moyenne $\mu_{i,s}$ et de variance  $\sigma_{i,s}^{2}$.

\end{itemize}

\subsection{Modèle à états cachés}
\label{relationnalLinearHiddenProofVar}

\begin{itemize}
\item \textbf{Vraisemblance 1: } $\exists \Theta \in \mathbb{R}^{d\times d}, \forall t \in \left\{2,...,T\right\}: h_{t}=\Theta h_{t-1} +\epsilon_{t}$, où $\epsilon_{t} \sim \mathcal{N}(0,\delta^{2}I)$.
\item \textbf{Vraisemblance 2: } $\forall i \in \left\{1,...,K\right\}, \exists W_{i} \in \mathbb{R}^{d}, \exists b_{i} \in \mathbb{R}$ \text{ tels que }: $\forall t \in \left\{1,...,T\right\}$ : $r_{i,t}=W_{i}^{\top}h_{t}+b_{i}$ $\epsilon_{i,t} \sim \mathcal{N}(0,\sigma^{2})$.
\item \textbf{Prior au temps 1: } $h_{1} \sim \mathcal{N}(0,\delta^{2}I)$, où $I$ désigne la matrice identité de taille $d$.
\item \textbf{Prior sur les paramètres 1: } $\forall i \in \left\{1,...,d\right\}$:  $\theta_{i} \sim \mathcal{N}(0,\alpha^{2}I)$, où $\theta_{i}$ désigne la ligne $i$ de $\Theta$.
\item \textbf{Prior sur les paramètres 2: } $\forall i \in \left\{1,...,K\right\}$:  $(W_{i}, b_{i})  \sim \mathcal{N}(0,\gamma^{2}I)$, avec $(W_{i}, b_{i})$ le vecteur aléatoire de taille $d+1$ issu de la concaténation de $W_{i}$ et $b_{i}$, et $I$ la matrice identité de taille $d+1$. 
\end{itemize}

Notre but est de déterminer les distributions variationnelles des différentes variables cachées selon la factorisation suivante:

\begin{align*}
q(h_{1},...,h_{t-1},\Theta,W,b)=\prod\limits_{s=1}^{t-1}q_{h_{s}}(h_{s})\prod\limits_{i=1}^{K}q_{W_{i},b_{i}}(W_{i},b_{i})\prod\limits_{j=1}^{d}q_{\theta_{j}}(\theta_{j})
\end{align*}

\subsubsection{Couches cachées}
\label{proofLin1HiddenLayerVar}

On note  $W$ la matrice de taille  $K \times d$ dont la ligne $i$ vaut  $W_{i}$ et $b$ le vecteur des biais de taille $K$. De plus au temps $s$, la sous-matrice (resp. le sous-vecteur) composée des lignes d'index $i\in \mathcal{K}_{s}$ de $W$ (resp. de $b$) est notée $W_{s}$ (resp. $b_{s}$). Finalement le vecteur de taille $k$ des récompenses observées au temps $s$ est noté $R_{s}$. Soit $1 \leq s \leq t-1$. On s'intéresse au trois cas suivant:

\begin{itemize}
\item \textbf{Cas 1: $1 < s< t-1$}

\begin{align*}
\log q_{h_{s}}^{\star} (h_{s})&= \mathbb{E}_{h_{s}^{\backslash}}[ \log p(h_{s+1}|h_{s},\Theta) +\log p(h_{s}|h_{s-1},\Theta)+ \log p(R_{s} |h_{s},W_{s},b_{s})] + C \\
&= \mathbb{E}_{h_{s}^{\backslash}}\left[-\dfrac{1}{2}  \left( \dfrac{\left( h_{s+1} - \Theta h_{s} \right)^{\top}\left( h_{s+1} - \Theta h_{s} \right)}{\delta^{2}}   +  \dfrac{\left( h_{s} - \Theta h_{s-1} \right)^{\top}\left( h_{s} - \Theta h_{s-1} \right)}{\delta^{2}} \right. \right.\\
&\left. \left. \qquad+\dfrac{ \left( R_{s} - (W_{s}h_{s}+b_{s}) \right)^{\top}\left( R_{s} - (W_{s}h_{s}+b_{s} )\right)}{\sigma^{2}} \right) \right] + C\\
&=\mathbb{E}_{h_{s}^{\backslash}}\left[-\dfrac{1}{2}  \left(     h_{s}^{\top}\left(  \dfrac{I}{\delta^{2}} +  \dfrac{\Theta ^{\top}\Theta}{\delta^{2}} + \dfrac{W_{s}^{\top}W_{s}}{\sigma^{2}}   \right) h_{s}     -2h_{s}^{\top}\left(  \dfrac{\Theta h_{s-1}}{\delta^{2}}  + \dfrac{\Theta^{\top} h_{s+1}}{\delta^{2}}  + \dfrac{W_{s}^{\top}(R_{s}-b_{s})}{\sigma^{2}}   \right) \right) \right] + C\\
\end{align*}

En utilisant l'indépendance des variables aléatoire selon la partition choisie, on peut exprimer l'expression ci-dessus sous la forme:

\begin{align*}
\log  q_{h_{s}}^{\star} (h_{s})&= -\dfrac{1}{2} \left( h_{s}^{\top}F_{s} h_{s} - 2h_{s}^{\top}g_{s}\right) +C\\
&= -\dfrac{1}{2}  (h_{s}-F_{s}^{-1}g_{s})^{\top}F_{s}(h_{s}-F_{s}^{-1}g_{s})+C
\end{align*}

Avec: 

$F_{s}=\dfrac{I}{\delta^{2}} +  \dfrac{\mathbb{E}[\Theta ^{\top}\Theta]}{\delta^{2}} + \dfrac{\mathbb{E}[W_{s}^{\top}W_{s}]}{\sigma^{2}}$

$g_{s}= \dfrac{\mathbb{E}[\Theta] \mathbb{E}[h_{s-1}]}{\delta^{2}}  + \dfrac{\mathbb{E}[W_{s}]^{\top}\mathbb{E}[R_{s}]-\mathbb{E}[W_{s}^{\top}b_{s}]}{\sigma^{2}} +\dfrac{\mathbb{E}[\Theta]^{\top} \mathbb{E}[h_{s+1}]}{\delta^{2}}$

On reconnaît  bien une loi gaussienne de moyenne $F_{s}^{-1}g_{s}$ et de matrice de covariance $F_{s}^{-1}$.  
On remarque de plus que $\Theta ^{\top}\Theta=\sum\limits_{i=1}^{d}\theta_{i} \theta_{i}^{\top}$ d'où $\mathbb{E}[\Theta^{\top}\Theta]=\sum\limits_{i=1}^{d}\mathbb{E}[\theta_{i}\theta_{i}^{\top}]=\sum\limits_{i=1}^{d}\mathbb{E}[\theta_{i}]\mathbb{E}[\theta_{i}]^{\top}+Var(\theta_{i})$. Le même raisonnement est appliqué pour $\mathbb{E}[W_{s}^{\top}W_{s}]$.

\item \textbf{Cas 2: $s=1$}

On effectue un calcul similaire excepté qu'on ne considère pas d'état caché au temps $s-1$. On trouve ainsi: 

\begin{align*}
\log  q_{h_{s}}^{\star} (h_{s})&= -\dfrac{1}{2} \left( h_{s}^{\top}F_{s} h_{s} - 2h_{s}^{\top}g_{s}\right) +C\\
&= -\dfrac{1}{2}  (h_{s}-F_{s}^{-1}g_{s})^{\top}F_{s}(h_{s}-F_{s}^{-1}g_{s})+C
\end{align*}

Avec: 

$F_{s}=\dfrac{I}{\delta^{2}} +  \dfrac{\mathbb{E}[\Theta ^{\top}\Theta]}{\delta^{2}} + \dfrac{\mathbb{E}[W_{s}^{\top}W_{s}]}{\sigma^{2}}$ 

$g_{s}= \dfrac{\mathbb{E}[\Theta]^{\top} \mathbb{E}[h_{s+1}]}{\delta^{2}}  + \dfrac{\mathbb{E}[W_{s}]^{\top}\mathbb{E}[R_{s}]-\mathbb{E}[W_{s}^{\top}b_{s}]}{\sigma^{2}}$

\item \textbf{Cas 3: $s=t-1$}

On effectue un calcul simialaire excepté qu'on ne considère pas d'état caché au temps $s+1$.

On trouve ainsi: 

\begin{align*}
\log  q_{h_{s}}^{\star} (h_{s})&= -\dfrac{1}{2} \left( h_{s}^{\top}F_{s} h_{s} - 2h_{s}^{\top}g_{s}\right) +C\\
&= -\dfrac{1}{2}  (h_{s}-F_{s}^{-1}g_{s})^{\top}F_{s}(h_{s}-F_{s}^{-1}g_{s})+C
\end{align*}

Avec:

$F_{s}=\dfrac{I}{\delta^{2}}  + \dfrac{\mathbb{E}[W_{s}^{\top}W_{s}]}{\sigma^{2}}$ 

$g_{s}= \dfrac{\mathbb{E}[\Theta] \mathbb{E}[h_{s-1}]}{\delta^{2}}  + \dfrac{\mathbb{E}[W_{s}]^{\top}\mathbb{E}[R_{s}]-\mathbb{E}[W_{s}^{\top}b_{s}]}{\sigma^{2}}$

\end{itemize}

\subsubsection{Paramètres $\theta_{i}$}
\label{proofLin1HiddenParamsThetaVar}

Pour $t\geq 3$, on note $D_{t-1}=(h_{s}^{\top})_{s=1..t-1}$ la matrice de taille $(t-1)\times d$ des  états cachés jusqu'au temps $t-1$. 
On a:
\begin{align*}
\log  q_{\theta_{i}}^{\star} (\theta_{i})&= -\dfrac{1}{2}   \left(\theta_{i}-A_{i,t-1}^{-1}b_{i,t-1}\right)^{\top}A_{i,t-1}\left(\theta_{i}-A_{i,t-1}^{-1}b_{i,t-1}\right)  +cte\\
\end{align*}

Avec:

$A_{i,t-1}=\dfrac{\mathbb{E}[D_{1..t-2}^{\top}D_{1..t-2}]}{\sigma^{2}}+\frac{I}{\alpha^{2}}$

$b_{i,t-1}=\dfrac{\mathbb{E}[D_{1..t-2}]^{\top}}{\sigma^{2}}\mathbb{E}[D_{i:2..t-1} ]$

On reconnaît  bien une loi gaussienne de moyenne $A_{i,t-1}^{-1}b_{i,t-1}$ et de matrice de covariance $A_{i,t-1}^{-1}$. 
On remarque de plus que $D_{0..t-2}^{\top}D_{0..t-2}=\sum\limits_{s=0}^{t-2} h_{s}h_{s}^{\top}$ 
et donc $\mathbb{E}[D_{0..t-2}^{\top}D_{0..t-2}]=\sum\limits_{s=0}^{t-2} \mathbb{E}[h_{s}]\mathbb{E}[h_{s}]^{\top}+Var(h_{s})$.

\subsubsection{Paramètres $W$ et $b$}
\label{proofLin1HiddenParamsVar}

Pour chaque action $i$  on note $\mathcal{T}_{i,t-1}$ l'ensemble des itérations où elle a été choisie jusqu'au temps $t-1$, c'est-à-dire $\mathcal{T}_{i,t-1}=\left\{s \text{ tel que } i\in \mathcal{K}_{s}  \text{ pour } 1\leq s\leq t-1\right\}$. On note également $M_{i,t-1}=((h_{s},1)^{\top})_{s\in \mathcal{T}_{i,t-1}}$  et $c_{i,t-1}=(r_{i,s})_{s\in \mathcal{T}_{i,t-1}}$. 

On a :
\begin{align*}
\log  q_{(W_{i},b_{i})}^{\star} ((W_{i},b_{i}))&= -\dfrac{1}{2}   \left((W_{i},b_{i})-V_{i,t-1}^{-1}v_{i,t-1}\right)^{\top}V_{i,t-1}\left((W_{i},b_{i})-V_{i,t-1}^{-1}v_{i,t-1}\right)  +cte\\
\end{align*}

Avec:

$V_{i,t-1}=\dfrac{I}{\gamma^{2}}+\dfrac{\mathbb{E}[M_{i,t-1}^{\top}M_{i,t-1}]}{\sigma^{2}}$ 

 $v_{i,t-1}=\dfrac{\mathbb{E}[M_{i,t-1}]^{\top}c_{i,t-1}}{\sigma^{2}}$ 

On reconnaît  bien une loi gaussienne de moyenne $V_{i,t-1}^{-1}v_{i,t-1}$ et de matrice de covariance $V_{i,t-1}^{-1}$. 
On remarque de plus que $M_{i,t-1}^{\top}M_{i,t-1}=\sum\limits_{s \in\mathcal{T}_{i,t-1}}(h_{s},1)(h_{s},1)^{\top}$ donc  $ \mathbb{E}[M_{i,t-1}^{\top}M_{i,t-1}]=\sum\limits_{s \in\mathcal{T}_{i,t-1}}\mathbb{E}[(h_{s},1)]\mathbb{E}[(h_{s},1)]^{\top}+Var((h_{s},1))$.

	\end{appendix}

\bibliographystyle{apalike}
\bibliography{Biblio.bib}

\end{document}